\documentclass[aps,prx,10pt,onecolumn,amsmath,amssymb,nofootinbib,superscriptaddress,longbibliography]{revtex4-2}
\usepackage[T1]{fontenc}
\usepackage[utf8]{inputenc}
\usepackage{lmodern}
\usepackage{bm,graphicx,booktabs}
\usepackage{microtype}
\usepackage{xcolor}
\usepackage{ragged2e}
\usepackage{hyperref}
\hypersetup{colorlinks=true,linkcolor=blue,citecolor=blue,urlcolor=blue,
 pdftitle={Basis-set error bounds for first-quantised quantum simulation of chemistry},
 pdfauthor={Dominic W. Berry}}
\newtheorem{theorem}{Theorem}[section]
\newtheorem{lemma}[theorem]{Lemma}
\newtheorem{proposition}[theorem]{Proposition}
\newtheorem{corollary}[theorem]{Corollary}
\newtheorem{remark}[theorem]{Remark}
\newenvironment{proofx}[1][Proof]{\par\medskip\noindent\textit{#1.}\ }{\hfill$\square$\par\medskip}
\newcommand{\T}{\mathbb{T}}
\newcommand{\R}{\mathbb{R}}
\newcommand{\Z}{\mathbb{Z}}
\newcommand{\ind}[1]{\mathbf{1}\!\left[#1\right]}
\newcommand{\norm}[1]{\left\lVert #1 \right\rVert}

\newcommand{\HBO}{H_{\mathrm{per}}}
\newcommand{\Hnbo}{H_{\mathrm{nBO}}}
\newcommand{\Ttot}{T_{\mathrm{tot}}}
\newcommand{\Pp}{P'}
\newcommand{\Hgrid}{H_{\mathrm{grid}}}
\newcommand{\E}{\mathcal E}

\newcommand{\ket}[1]{|#1\rangle}
\newcommand{\ran}{\operatorname{ran}}

\makeatletter
\let\l@subsection\@gobbletwo
\let\l@subsubsection\@gobbletwo
\makeatother
\begin{document}
\title{Basis-set error bounds for first-quantised quantum simulation of chemistry}
\author{Dominic W. Berry}\email{dominic.berry@mq.edu.au}
\affiliation{School of Mathematical and Physical Sciences, Macquarie University, NSW 2109, Australia}
\affiliation{School of Electrical Engineering and Computer Science, University of Queensland, QLD 4072, Australia}

\author{Arkin Tikku}
\affiliation{School of Chemistry, University of Sydney, NSW 2006, Australia}

\author{Sophia Simon}
\affiliation{School of Electrical Engineering and Computer Science, University of Queensland, QLD 4072, Australia}

\author{Ryan Babbush}
\affiliation{Google Quantum AI, Los Angeles, CA, USA}

\begin{abstract}
In quantum simulation of electronic structure it is often expected that the estimate of the ground energy converges to the continuum basis limit at a rate of $\mathcal O(N^{-1})$ for a basis of size $N$. 
This has been found in analytical and numerical studies, but without a rigorous treatment of the error in the quantities considered in quantum algorithms.
Here we show the $\mathcal O(N^{-1})$ scaling of the error for a wide range of cases needed for quantum simulation of chemistry.
We show the error in the ground energy scales as $\mathcal O(N^{-1})$ for plane-wave as well as spatial grid bases.
For the spatial grid we consider both a periodic system and a non-periodic system with open boundaries.
For time evolution, our bound on the error in the state itself is weaker than $\mathcal O(N^{-1})$, but we show that the error in observables scales as $\widetilde{\mathcal O}(N^{-1})$ in important cases.
In particular, we may consider a smoothed observable that describes the position of particles.
We show that this error scales as $\mathcal O(N^{-5/3}+tN^{-1})$ for time $t$, again both for plane-wave and spatial grid bases.
For the non-periodic grid case we use a condition that the true evolution would not result in significant leakage past the boundaries, which introduces basis error unrelated to the resolution of the discretisation.
We also show the error in the kinetic energy of a particle scales as $\mathcal O((t+\log N)/N)$ for the plane-wave and grid bases.
Together, these results show that $\mathcal O(N^{-1})$ scaling of the error holds for many of the relevant cases for quantum algorithms (with a trivial log factor for kinetic energy).
\end{abstract}
\maketitle
\begingroup
\makeatletter
\renewcommand*\l@section[2]{\@dottedtocline{1}{0pt}{1.8em}{#1}{#2}}
\makeatother
\tableofcontents
\endgroup

\section{Introduction}
\label{sec:intro}

Quantum simulation offers a way to calculate molecular
energies~\cite{AspuruGuzik2005} and chemical
dynamics~\cite{Kassal2008} without storing the many-particle
wavefunction classically.
There are two main approaches to representing the system.
First, one may use a compact representation of the system with a
small number of carefully selected orbitals, in which case the
parameters in the Hamiltonian are typically precomputed classically
and input to the block encoding, using methods such as double
factorisation~\cite{vonBurg2021}, tensor
hypercontraction~\cite{Lee2021THC}, or a combination of the
two~\cite{Low2025}.
Second, one may use a much larger basis such as plane waves, or
discretise space on a grid, where the parameters of the Hamiltonian
may be calculated on the fly within the block
encoding~\cite{Kivlichan2017RealSpace,Babbush2018Materials,Babbush2019,SuBerryWiebeRubinBabbush2021,Childs2022RealSpace}.
In that case, it is convenient to use first quantisation, which
means that each particle requires a register that scales only
logarithmically with the number $N$ of basis functions per particle.
Nevertheless, the complexity still depends polynomially on $N$,
even if that dependence may be made
sublinear~\cite{Babbush2019,SuBerryWiebeRubinBabbush2021}.

Determining a minimum sufficient basis size is therefore essential
for obtaining accurate resource estimates.
To choose a sufficient basis size, we need to bound the error introduced by restricting the continuous problem to a finite basis. A bound on the basis-set error in terms of $N$ therefore allows us to choose $N$ such that the basis-set error stays within its allocated share of the total error tolerance.
For the electronic structure problem, a basis-set error of $\mathcal O(N^{-1})$ is commonly assumed and expected to hold for most reasonable choices of basis function such as plane waves or nuclei-centred Gaussians~\cite{Gruneis2013,Hattig2012,HarlKresse2008,Shepherd2012,Helgaker1997,Klopper1995,Halkier1998}.

The supporting literature contains both analytic and numerical
results. Classical partial-wave analyses relate slow convergence
to the electron--electron cusp. Schwartz~\cite{Schwartz1962} and
Kutzelnigg and Morgan~\cite{KutzelniggMorgan1992} derived asymptotic
contributions to atomic correlation energies in perturbation theory,
while Hill~\cite{Hill1985} obtained an inverse-cubic truncation error
for variational helium calculations. These results concern an
angular-momentum cutoff and do not directly give bounds on the
number of plane waves. For a plane-wave basis, Shepherd
\emph{et al.}~\cite{Shepherd2012} derived an $\mathcal O(N^{-1})$
correlation-energy error within second-order M{\o}ller--Plesset
(MP2) theory for the uniform electron gas, and presented
supporting numerical results from coupled-cluster and full
configuration interaction quantum Monte Carlo calculations.
Masios \emph{et al.}~\cite{Masios2024} subsequently analysed this
scaling in coupled-cluster doubles theory using a diagrammatic
decomposition.

For quantum algorithms targeting the ground energy, the required
basis-set bound concerns the difference between the exact
ground-state eigenvalues of the finite-basis and continuum
Hamiltonians. The MP2 and coupled-cluster analyses describe
convergence to the complete-basis limits of approximate methods,
which does not establish the same convergence rate for the exact
eigenvalue. The Monte Carlo calculations provide numerical evidence
for that rate, but not a general bound. Dupuy~\cite{Dupuy2019}
proved a leading plane-wave eigenvalue error of order $N^{-1}$
for a periodic one-electron Coulomb Hamiltonian at fixed cell size;
this result does not include electron--electron interactions.
A principal aim of the present work is to establish the
$\mathcal O(N^{-1})$ ground-state eigenvalue bound for interacting
many-particle Coulomb Hamiltonians in a fixed periodic cell,
thereby providing a rigorous basis for its use in
quantum-algorithm complexity estimates.

For a fixed periodic cell, we prove an $\mathcal O(N^{-1})$ ground-energy error for Galerkin discretisation, in which the Hamiltonian is projected onto the plane-wave basis. The result includes electronic spin and applies with either fixed or quantum nuclei.
A similar result was also recently given by Fang and Zhang~\cite{FangZhang2026}, which was developed independently and concurrently.
That result was for the Born-Oppenheimer approximation, where all nuclei are in fixed positions.
They used the result to give simulation algorithm complexities explicitly in terms of $\epsilon$ without $N$.
Another important case is the error in the wavefunction under time evolution.
It might be hoped that the $L^2$ error also scales as $\mathcal O(N^{-1})$, but this has not been established.

Instead, we find the bound on the wavefunction error $\mathcal O(N^{-5/6}+t/N)$ under time evolution, which is slightly weaker than the desired $\mathcal O(N^{-1})$ scaling.
Fang and Zhang gave a different result for the wavefunction error, a scaling as $\mathcal O((1+t)N^{-2/3}+t^{2/3}N^{-4/9})$ (see Theorem 11 of that work).
This difference is in part due to a different regularity assumption.
Fang and Zhang consider a condition of ordinary $H^2$ Sobolev regularity, which means that the wavefunction and its first and second derivatives are square-integrable.
In contrast, we assume the stronger condition $\ket{\psi_0}\in D(H^2)$, which means that both $H\ket{\psi_0}$ and $H^2\ket{\psi_0}$ must be well-defined and have finite norm.
This is a reasonable condition for the type of initial states to be prepared.
(It will hold for a finite superposition of eigenstates.)

However, in practice, what is important is the result of observables measured on the time-evolved state.
For example, in Ref.~\cite{Eklund2026Dynamics} one would aim to measure positions of particles to determine the progress of chemical reactions.
Another application is measurement of the kinetic energy of a particle to estimate stopping power \cite{Rubin2024Stopping}.
In these cases we are able to show that the error scales close to $\mathcal O(N^{-1})$ with the same $\ket{\psi_0}\in D(H^2)$ condition.

More specifically, we consider an observable to determine if a particle is in a region.
Sharp boundaries on the region tend to induce larger errors, so we smooth the boundaries, which is a reasonable choice for the intended application.
With this condition, the discretisation error in the expectation value is $\mathcal O(N^{-5/3}+tN^{-1})$ for all $t\ge0$.
For the case of kinetic energy, we simply consider error in the expectation value, and show that it scales as $\mathcal O((t+\log N)/N)$.
While this is not strictly $\mathcal O(N^{-1})$, it can be given as $\widetilde{\mathcal O}(N^{-1})$.

As well as considering the plane-wave basis for these cases (ground-energy and observable error), we consider a spatial grid with a suitably truncated interaction kernel, and consider both the periodic case and the non-periodic case for isolated systems.
The non-periodic case is important for cases like in Ref.~\cite{Eklund2026Dynamics}, where there is a single reacting system and we should not have periodic images.
For the periodic grid case we obtain similar results as for plane waves, both for ground-energy and observable errors.
For the non-periodic case we show that, provided the initial state and exact solution are appropriately localised in the finite region, the scaling of the error is again similar to the plane-wave case (though with an extra term for leakage).

Fang and Zhang~\cite{FangZhang2026} also consider isolated systems, but need to increase the size of a periodic region to control the error from periodic images.
As a result, they give $N=\mathcal O(\epsilon^{-2})$ as a sufficient basis size for isolated-system ground-energy estimation (see Eq.~(13) of that work).
In contrast, for our approach, for isolated-system ground energies with clamped nuclei in the
specified scalar exchange sector, our range-truncated interaction
gives a sufficient basis size
$N=\mathcal O(\epsilon^{-1}\log^3(1/\epsilon))$
under the free-space and finite-volume localisation assumptions
of Proposition~\ref{prop:free}.
For dynamical observables at fixed time, a sufficient basis size is
$\widetilde{\mathcal O}(\epsilon^{-1})$ if the wavefunction and
shell-gradient leakage decay exponentially with cell size
throughout the evolution, with uniform cutoff-domain bounds.

Section~\ref{sec:Main_results} fixes conventions and summarises the bounds. Sections~\ref{sec:energy-overview}--\ref{sec:kinetic-overview} develop the energy, smooth-observable and kinetic-energy arguments. Section~\ref{sec:free-overview} treats isolated systems, and Sec.~\ref{sec:conclusion} gives the conclusions. Detailed statements and proofs appear in Appendices~\ref{app:bo}--\ref{app:source}.

\section{Main results}
\label{sec:Main_results}

\subsection{Setup}
\paragraph{Hamiltonian} 
Let $\eta$ be the number of electrons and $\eta_{\rm tot}$ the total number of quantum particles.
We use bold symbols for three-dimensional spatial vectors and
an arrow over a bold symbol for an ordered collection of such
vectors. Thus $\bm{x}_j$ is the position of particle $j$, while
$\vec{\bm{x}}=(\bm{x}_1,\ldots,\bm{x}_{\eta_{\rm tot}})$
is the full spatial configuration. We use the same convention
for momenta and Fourier indices.

Particle $j$ has position $\bm{x}_j$, mass $m_j$ and charge $q_j$; any clamped nucleus $\ell$ has position $\bm{R}_\ell$ and charge $\zeta_\ell$. We consider the following Hamiltonian in the continuum, which allows for both quantum and fixed classical nuclei.
\begin{equation}
 H=T+V_{\mathrm{tot}},\qquad
 T=\sum_{j=1}^{\eta_{\rm tot}}-\frac{\nabla^2_j}{2m_j},\qquad
 V_{\mathrm{tot}}=\sum_{i<j}q_iq_j w(\bm{x}_i-\bm{x}_j)
                  +\sum_{j,\ell}q_j\zeta_\ell w(\bm{x}_j-\bm{R}_\ell),
 \label{eq:overview-hamiltonian}
\end{equation}
where $T$ is the total kinetic energy of the quantum particles and
$V_{\mathrm{tot}}$ is their Coulomb potential energy, including
interactions with any clamped nuclei.
The function $w(\bm{x})$ is the zero-mean periodic Coulomb kernel (see  Eq.~\ref{eq:w} for a definition).
For the isolated systems of Section~\ref{sec:free-overview},
we replace $w(\bm{x})$ by $1/\|\bm{x}\|$ and denote the free-space
Hamiltonian by $H_\infty$.

We use $U$ for electron--nucleus attraction, $V$ for electron--electron repulsion and $V_{\mathrm{nuc}}$ for nucleus--nucleus repulsion; $V_{\mathrm{tot}}$ denotes the sum of the potential terms included in $H$. In the Born--Oppenheimer (BO) case, only the electrons are quantum particles, so $V_{\mathrm{tot}}=U+V$ and the nuclear repulsion is an additive energy constant. In the fully quantum, non-Born--Oppenheimer (non-BO) case, the nuclei are included among the $\eta_{\rm tot}$ particles, $V_{\mathrm{tot}}=U+V+V_{\mathrm{nuc}}$, and the clamped-nucleus sum is absent.
We also allow hybrid models in which some nuclei are held fixed
and others are treated quantum mechanically. In this case,
$\eta_{\rm tot}$ counts the electrons and quantum nuclei, while
the clamped-nucleus sum includes only the fixed nuclei.
Interactions involving a quantum nucleus are included in
$V_{\mathrm{tot}}$; repulsion between two fixed nuclei contributes
only an additive energy constant.

\paragraph{Plane-wave basis and spatial resolution}
Our bounds compare the continuum Coulomb Hamiltonian with finite Galerkin plane-wave and spatial-grid representations. We describe their spatial resolution using both a momentum cutoff $K$ and a one-particle basis size $N$. Throughout the periodic analysis we use atomic units and a cubic computational cell $\T=\R^3/L\Z^3$ of volume $\Omega=L^3$.
For simplicity we take $N$ to be a cube of an odd number.

For each particle, we define the plane-wave basis 
\begin{align}
\mathbb{K}=\left\{\frac{2\pi \bm{p}}{\Omega^{1/3}}: \bm{p}\in G\right\}\quad\text{with}\quad G = \left[ -\frac{N^{1/3} - 1}{2}, \, \frac{N^{1/3} - 1}{2} \right]^3 \cap \mathbb{Z}^3 \, 
,
\end{align}
with momentum cutoff $K=\max_{\bm{k}\in \mathbb{K}} \norm{\bm{k}}_{\infty}=\Theta( (N/\Omega)^{1/3}) $. Here, $\|\bm{k}\|_\infty=\max_{\alpha=1,2,3}|k_\alpha|$ is the largest magnitude of any component $k_\alpha$ of the momentum vector $\bm{k}\in \mathbb{K}$. Thus the truncated basis is a cube in momentum space, and the momentum cutoff $K$ bounds each component rather than the Euclidean magnitude $\|\bm{k}\|$. The appendices specify whether a momentum tail is radial or cubical. Since $\{\|\bm{k}\|_\infty>\lambda\}\subseteq\{\|\bm{k}\|>\lambda\}$, a Euclidean tail bound also controls the corresponding cubical tail.

A discrete Fourier transform of a finite plane-wave basis gives
a spatial-grid basis, also called a plane-wave dual basis.
We use a potential diagonal in this grid basis, with entries
obtained by sampling a finite Fourier sum, as in the
quantum-simulation construction of
Babbush \emph{et al.}~\cite{Babbush2018Materials}.
In the momentum representation, transfers can wrap around the
register, with indices reduced modulo the number of grid points
in each direction. The resulting Hamiltonian therefore differs
from Galerkin projection.

For periodic systems, we truncate the Fourier series of $w$.
For isolated systems, we first truncate the spatial range of
$1/\|\bm{x}\|$ and periodise the resulting kernel, choosing
the truncation so that it agrees with the free-space interaction
throughout the retained region~\cite{FustiMolnarPulay2002}.
We then truncate its Fourier series.
In both cases, the finite Fourier sum remains finite when two
particles occupy the same grid point. This includes
electron--nucleus coincidences and coincidences of opposite-spin
electrons, which are allowed by fermionic antisymmetry.
Our bounds include the errors from Fourier truncation and
momentum wrapping; see Appendix~\ref{app:grid} for the grid
construction and Section~\ref{sec:free-overview} for the
additional finite-cell comparison needed for isolated systems.

\paragraph{Spin}
We include both position and spin when imposing the usual fermionic
exchange symmetry. The electronic wavefunction depends on each
electron's position and one of two spin values, and changes sign
when the positions and spins of any two electrons are exchanged
together. Exchanging positions alone need not change its sign.
The Hamiltonian is independent of spin.
The two spin states give $2N$ one-particle spin orbitals, without changing any basis-set exponent. For other species we likewise allow a finite internal-spin space and impose the appropriate exchange statistics. The Hamiltonian $H$ and spatial projector act as the identity on spin and preserve the resulting exchange-symmetry sector. The observables also act as the identity on spin, although a labelled one-particle observable need not preserve the exchange sector. We therefore allow its auxiliary test states to lie in the full particle space when estimating an expectation.

\subsection{Exact and discretised quantum dynamics}
\label{sec:exactdisc}

Let $P$ denote the joint projection of all particle registers onto the plane-wave basis, such that $H_N=PHP$ is the discrete Hamiltonian represented using $N$ spatial basis functions per particle. We then compare the exact evolution with the evolution of its projection:
\begin{equation}
 \ket{\psi(t)}=e^{-iHt}\ket{\psi_0},\qquad
 \ket{\psi_N(t)}=e^{-iH_Nt}P\ket{\psi_0}.
 \label{eq:overview-evolution}
\end{equation}
We do not renormalise after projection. Even when $\|\ket{\psi_0}\|=1$, the Galerkin state has norm $\|\ket{\psi_N(t)}\|=\|P\ket{\psi_0}\|\le1$. Our observable estimates therefore compare the quadratic forms $\langle\psi(t)|O|\psi(t)\rangle$ and $\langle\psi_N(t)|O|\psi_N(t)\rangle$, without dividing by the state norms. Corollary~\ref{cor:norm} gives the correction for a normalised Galerkin state and a bounded observable.

This convention separates two effects that will recur in the proofs. At a given time, $\ket{\psi(t)}-P\ket{\psi(t)}$ is the part of the exact state that the basis cannot represent. The difference $P\ket{\psi(t)}-\ket{\psi_N(t)}$ is instead an evolution error within the basis. These two retained states agree initially, but the projected exact state continues to receive amplitude from discarded modes through the potential. Its evolution equation therefore contains the additional term $PV_{\mathrm{tot}}(1-P)\ket{\psi(t)}$, which is absent from Galerkin evolution. We derive its contribution to the error in Eq.~\eqref{eq:overview-duhamel}. We will estimate the two contributions separately and then combine them in the measured expectation.

\begin{table}[!tb]
\caption{Principal discretisation-error bounds.
For periodic systems, the cell size is fixed.
For non-periodic systems, $L$ is the computational cell side length
and $N=\Theta((LK)^3)$, where $K$ is the physical momentum cutoff.
Dynamical bounds require the stated squared-Hamiltonian-domain
regularity; the periodic smooth-observable bounds also require
the stated derivative bounds and rapid decay of the observable's
Fourier tails. All entries are asymptotic upper bounds for
sufficiently large cutoffs.}
\label{tab:rates}
\begin{ruledtabular}
\begin{tabular}{
    p{0.26\textwidth}
    p{0.37\textwidth}
    p{0.27\textwidth}
}
Quantity and discretisation
& Error bound
& Scope and reference
\\
\hline
\multicolumn{3}{l}{\emph{Periodic systems: fixed cell}}
\\[3pt]

Ground-state energy, Galerkin
& $\mathcal O(N^{-1})$
& BO and non-BO;
  Theorems~\ref{thm:main} and~\ref{thm:nbo}
\\[5pt]

Ground-state energy, truncated-kernel grid
& $\mathcal O(N^{-1})$
& Theorem~\ref{thm:grid-energy}
\\[5pt]

Smooth observable, Galerkin
& $\mathcal O(N^{-5/3}+tN^{-1})$
& $t\ge0$;
  Theorem~\ref{thm:dyn}
\\[5pt]

Smooth observable, truncated-kernel grid
& $\mathcal O(N^{-5/3}+tN^{-1})$
& $t\ge0$;
  Theorem~\ref{thm:grid-dyn-sharp}
\\[5pt]

Particle kinetic energy, Galerkin and truncated-kernel grid
& $\mathcal O\bigl((t+\log N)/N\bigr)$
& $t\ge0$;
  Theorem~\ref{thm:kinetic-log}
\\[7pt]

\hline
\multicolumn{3}{l}{\emph{Non-periodic systems: finite-cell grid approximation}}
\\[3pt]

Ground-state energy
& $\mathcal O(L^3N^{-1}+e^{-\gamma L})$
& Clamped nuclei, scalar exchange sector;
  free-space and finite-cell localisation assumptions;
  Proposition~\ref{prop:free}
\\[5pt]

Smooth, compactly supported position observable
& $\begin{gathered}
     \mathcal O\bigl(L^5N^{-5/3}+tL^3N^{-1}
     {}+(1+t)\epsilon_{\rm leak}\bigr)
   \end{gathered}$
& $t\ge0$;
  leakage and cutoff regularity;
  Proposition~\ref{prop:free-dyn}
\\[5pt]

Particle kinetic energy
& $\begin{gathered}
     \mathcal O\bigl(L^3N^{-1}[t
     {}+\log(N^{1/3}/L)]
     {}+(1+t)\epsilon_{\rm leak}\bigr)
   \end{gathered}$
& $t\ge0$;
  leakage and cutoff regularity;
  Proposition~\ref{prop:free-kinetic-log}
\\
\end{tabular}
\end{ruledtabular}

\smallskip
\parbox{\textwidth}{\justifying\footnotesize
For non-periodic-system energies, $\gamma>0$ is determined by the
assumed localisation rates. Choosing the cell size and momentum
cutoff gives the sufficient basis size
$N=\mathcal O(\epsilon^{-1}\log^3(1/\epsilon))$
for energy error at most $\epsilon$.
For non-periodic-system dynamics, $\epsilon_{\rm leak}$ bounds the
wavefunction outside the retained interior and its gradient on
the cutoff transition shell throughout the evolution interval. The stated regularity bounds are held fixed;
the non-periodic-system kinetic bound displays its cell-size dependence explicitly.
\par}
\end{table}

Table~\ref{tab:rates} collects the main results.
For the periodic ground-energy bounds, the proofs derive the
needed eigenfunction regularity from the Hamiltonian.
For the dynamical bounds, we use $D(H)$ to denote the domain of $H$:
the states on which $H$ is defined and produces a vector of finite
norm. We assume $\ket{\psi_0}\in D(H^2)$, meaning that both
$\ket{\psi_0}$ and $H\ket{\psi_0}$ belong to $D(H)$.
Thus $H\ket{\psi_0}$ and $H^2\ket{\psi_0}$ are well-defined and
have finite norm. We require
$\|\ket{\psi_0}\|+\|H\ket{\psi_0}\|+\|H^2\ket{\psi_0}\|$
to remain bounded as the basis cutoff increases.
We refer to this sum as the Hamiltonian graph-norm budget.
For the Coulomb Hamiltonians considered here, membership in $D(H)$
also ensures square-integrable first and second spatial derivatives.
Our assumption therefore gives this regularity for both the initial
wavefunction and the wavefunction obtained by applying $H$ to it.
It does not require four square-integrable spatial derivatives:
the singular potential makes repeated application of $H$ different
from repeated application of the Laplacian.
The constants may depend on the fixed cell, particle numbers,
charges, masses and the stated norm bounds, but not on the basis
cutoff. The detailed statements specify the dependence on the
observable and any cutoff threshold.

We separate these spatial errors from the errors incurred when implementing a quantum algorithm. The dynamical bounds assume exact evolution under the specified finite Hamiltonian. Any errors introduced by a product formula, block encoding, qubitisation procedure or arithmetic oracle must be budgeted separately. Similarly, an energy bound controls the lowest eigenvalue of the finite Hamiltonian, not the probability of obtaining it by phase estimation. That probability also depends on the prepared state's overlap with the desired eigenspace.

The accompanying Lean development~\cite{BerryPlaneWave2026,deMoura2021,Mathlib2020}
checks the periodic ground-energy and observable error bounds,
including the logarithmic kinetic-energy and stopping-power estimates,
together with the non-periodic-system comparisons stated here. It constructs
the periodic Coulomb Hamiltonians and their propagators and verifies
the error estimates for those operators. The proofs use only the
standard logical axioms recorded by the release's dependency checks.
The repository provides the source code and build instructions;
\texttt{docs/FORMALISATION.md} gives the correspondence between
the manuscript results and the Lean declarations.

For non-periodic systems, the formal statements take the specified
free-space states or solutions and their localisation or leakage
bounds as inputs. They cover families of interaction profiles with
uniformly bounded total variation and include the physical zero modes
in the cell and grid Hamiltonians. The cutoff-domain hypothesis is
retained for a general spatial cutoff and proved for the stated
compatible constructions. Formal verification establishes the error
bounds under these hypotheses; establishing the required localisation
or leakage for a particular physical system remains a separate task.

\section{Ground-state energy convergence}
\label{sec:energy-overview}

We obtain the $\mathcal O(N^{-1})$ Galerkin ground-energy bound by controlling the part of an exact ground state omitted by the basis. The key is that the variational principle makes the energy error quadratic in this omitted part, provided we measure it with the appropriate kinetic-energy weight. Throughout this argument, the cell and system parameters are fixed.

Let $\ket{\psi}$ be a normalised ground state in the chosen exchange sector, with $H\ket{\psi}=E\ket{\psi}$, and let $E_N$ be the lowest eigenvalue of $H_N$ in the same sector. Write $\ket{\delta}:=(1-P)\ket{\psi}$ for the discarded component. Restricting the variational space cannot lower the ground energy, so $E_N\ge E$. For an upper bound, we use $P\ket{\psi}/\|P\ket{\psi}\|$ as a trial state whenever $P\ket{\psi}\ne0$. To see why its energy error is quadratic in $\ket{\delta}$, expand $P\ket{\psi}=\ket{\psi}-\ket{\delta}$:
\begin{equation}
 \langle\psi|P(H-E)P|\psi\rangle
 =(\langle\psi|-\langle\delta|)(H-E)(\ket{\psi}-\ket{\delta})
 =\langle\delta|(H-E)|\delta\rangle.
\end{equation}
All terms containing $\ket{\psi}$ vanish by the eigenvalue equation and self-adjointness. Since $\|P\ket{\psi}\|^2=1-\|\ket{\delta}\|^2$, the Rayleigh quotient therefore gives
\begin{equation}
 0\le E_N-E\le
 \frac{\langle\delta|(H-E)|\delta\rangle}{1-\|\ket{\delta}\|^2}.
 \label{eq:overview-variational}
\end{equation}
The Coulomb potential satisfies a quadratic-form bound: its expectation on a state vector is bounded in magnitude by a constant times that vector's kinetic energy plus its squared norm. In particular,
\begin{equation}
 |\langle\delta|V_{\mathrm{tot}}|\delta\rangle|
 \le C\bigl(\langle\delta|T|\delta\rangle+\|\ket{\delta}\|^2\bigr).
\end{equation} It therefore controls the numerator by a constant times $\|(1+T)^{1/2}\ket{\delta}\|^2$, as in Proposition~\ref{prop:sandwich}. This weighted norm includes the kinetic energy of the discarded component: a small wavefunction norm alone would not control its energy at high momentum. We therefore need momentum-tail estimates with and without kinetic weighting.

Ordinary second-order Sobolev regularity would not yet give the desired rate. It yields $\|\ket{\delta}\|=\mathcal O(K^{-2})$ and a bounded $\|(1+T)\ket{\delta}\|$, so their product would control the energy error only by $\mathcal O(K^{-2})$. We need an additional power of $K^{-1}$ in that product. The Coulomb interaction and the eigenvalue equation supply it; we do not impose an extra derivative assumption on the eigenfunction.

We first estimate the high-momentum content of $V_{\mathrm{tot}}\ket{\psi}$, rather than that of $\ket{\psi}$ directly. For each particle register, Lemma~\ref{lem:tail} bounds its norm above momentum $K$ by $\mathcal O(K^{-1/2})$. The proof separates small and large Fourier transfers, where a transfer is the change in a particle's momentum caused by a potential term. A transfer of magnitude at most $K/2$ can produce momentum above $K$ only from an input already above $K/2$; the second-order Sobolev regularity obtained from the Hamiltonian controls this contribution. For larger transfers, the decay of the Coulomb Fourier coefficients is the decisive input. At fixed cell size their magnitude is proportional to $\|\bm{k}\|^{-2}$, so the squared $L^2$ norm of the part above $K/2$ has the scaling
\begin{equation}
 \sum_{\|\bm{k}\|>K/2}\|\bm{k}\|^{-4}
 =\mathcal O(K^{-1}).
\end{equation}
A three-dimensional shell has a number of modes proportional to its radius squared; this is why summing the fourth inverse power leaves one inverse power of $K$. Taking the square root gives the potential-tail factor $K^{-1/2}$.

We must still control multiplication of this kernel by the state. For a pair interaction we use the three-dimensional relative coordinate in which the kernel is singular and hold the other coordinates fixed. Second-order Sobolev regularity gives the needed supremum bound on that slice, which we combine with the kernel's $L^2$ bound before integrating over the spectator coordinates. Equivalently, the Fourier argument fixes the spectator momenta and the pair's total momentum. The point of the slice construction is that its dimension stays three as particles are added. We never ask for a supremum estimate in the full many-particle configuration space.

We then return to the eigenvalue equation. Since the Fourier projector commutes with $T$, its discarded component satisfies $(T-E)\ket{\delta}=-(1-P)V_{\mathrm{tot}}\ket{\psi}$. On the discarded modes, the kinetic energy is at least a fixed mass-dependent multiple of $K^2$. Once $K$ is sufficiently large relative to $|E|$ and the masses, we can divide by $T-E$ and gain two powers of $K^{-1}$. Combining this with the potential-tail estimate gives
\begin{equation}
 \|\ket{\delta}\|=\mathcal{O}(K^{-5/2}),\qquad
 \|(1+T)\ket{\delta}\|=\mathcal{O}(K^{-1/2}),
 \label{eq:overview-tails}
\end{equation}
with the mass factors tracked in the detailed proofs. The first bound controls the discarded amplitude; the second controls the same component with a full kinetic-energy weight. The half-power weight required in Eq.~\eqref{eq:overview-variational} follows by Cauchy--Schwarz:
\begin{equation}
 \|(1+T)^{1/2}\ket{\delta}\|^2
 \le\|\ket{\delta}\|\,\|(1+T)\ket{\delta}\|
 =\mathcal{O}(K^{-3}),
\end{equation}
while $\|\ket{\delta}\|^2=\mathcal O(K^{-5})$ ensures that the denominator in Eq.~\eqref{eq:overview-variational} is bounded away from zero for sufficiently large $K$. We thus obtain $E_N-E=\mathcal O(K^{-3})=\mathcal O(N^{-1})$ at fixed cell size.

The argument does not require a non-degenerate ground state or a spectral-gap estimate: we compare energies through a trial state, rather than identifying a particular numerical eigenvector. When the nuclei are quantum particles, the kinetic energy includes their masses, and we apply the tail restrictions and Hardy estimates register by register. Here a Hardy estimate bounds the $L^2$ norm after multiplication by a Coulomb singularity using the first spatial derivatives and norm of the state, with the other particle coordinates held fixed before integration. The cutoff exponent is unchanged; the explicit mass factors and ground-energy dependence are retained in Theorems~\ref{thm:main} and~\ref{thm:nbo}. The constants are not optimised.

Two scope distinctions remain important. The projection estimate in Eq.~\eqref{eq:overview-tails} measures how well the retained basis represents the exact eigenfunction; it does not compare that eigenfunction with a selected Galerkin eigenvector. Such a comparison requires an additional eigenspace argument and, in general, spectral separation. Nor does the energy upper bound establish sharpness. The proposed matching one-electron lower bound in Appendix~\ref{app:bo} still requires the additional local-expansion and eigenvalue-equivalence ingredients stated there.

\section{Smooth observables and spatial grids}
\label{sec:dynamics-overview}

\subsection{Momentum tails and smooth-observable error}

We next consider the error in a time-dependent expectation value, rather than in a ground energy.
Let $O=\chi(\bm{x}_j)$ be a smooth one-particle position observable,
where $0\le\chi\le1$. We assume that its Fourier coefficients
are absolutely summable, with tails decaying faster than any
inverse power of momentum, and that its gradient and Laplacian
are bounded. These conditions are stated quantitatively in
Appendix~\ref{app:dynamics}.
We assume $\ket{\psi_0}\in D(H^2)$, with the state and observable data fixed as the cutoff increases. We distinguish the effect of discarding the high-momentum part of the exact state from the error accumulated by evolving within the retained basis. The norms $\|\ket{\psi(t)}\|$, $\|H\ket{\psi(t)}\|$ and $\|H^2\ket{\psi(t)}\|$ are conserved. We can therefore apply the dynamical tail estimates of Appendix~\ref{app:dynamics} uniformly in time, including $\|(1-P)\ket{\psi(t)}\|=\mathcal O(K^{-5/2})$.

The role of $D(H^2)$ can be seen by comparing this argument with the eigenstate calculation. We now write $T\ket{\psi}=H\ket{\psi}-V_{\mathrm{tot}}\ket{\psi}$ in place of $T\ket{\psi}=E\ket{\psi}-V_{\mathrm{tot}}\ket{\psi}$. Since $H\ket{\psi}\in D(H)$ has second-order Sobolev regularity, its high-momentum tail is $\mathcal O(K^{-2})$. The potential term has the larger tail $\mathcal O(K^{-1/2})$ established above. Thus the high-momentum part of $T\ket{\psi}$ is $\mathcal O(K^{-1/2})$, and division by the kinetic weight $T\ge cK^2$ on discarded modes again gives $\mathcal O(K^{-5/2})$ for the state. Applying the Hamiltonian regularity estimate to both $\ket{\psi}$ and $H\ket{\psi}$ makes these constants depend only on the conserved graph-norm budget. This is the reason for the squared-Hamiltonian domain assumption.

To identify the accumulated error, compare the projected exact state $P\ket{\psi(t)}$ with the Galerkin state $\ket{\psi_N(t)}$. They start from the same state $P\ket{\psi_0}$, but their equations differ by the potential's action on the discarded modes. Since the kinetic operator $T$ and the Fourier projector $P$ commute, Duhamel's formula gives
\begin{equation}
 P\ket{\psi(t)}-\ket{\psi_N(t)}
 =-i\int_0^t e^{-iH_N(t-s)}PV_{\mathrm{tot}}(1-P)\ket{\psi(s)}\,ds.
 \label{eq:overview-duhamel}
\end{equation}
The factor $PV_{\mathrm{tot}}(1-P)\ket{\psi(s)}$ is the additional term in the projected Schr\"odinger equation: the potential transfers amplitude from discarded modes back into the retained basis. It is often called an inhomogeneous or source term, because it drives the difference between the projected exact state and the Galerkin state. To estimate its norm, put $\ket{\delta(s)}=(1-P)\ket{\psi(s)}$. Hardy's inequality controls multiplication by the Coulomb potential with a half-power kinetic weight. Interpolating between the two tail norms gives
\begin{equation}
 \|PV_{\mathrm{tot}}\ket{\delta(s)}\|
 \le C\|(1+T)^{1/2}\ket{\delta(s)}\|
 \le C\bigl(\|\ket{\delta(s)}\|\,\|(1+T)\ket{\delta(s)}\|\bigr)^{1/2}
 =\mathcal O(K^{-3/2}).
\end{equation}
Taking norms directly in Eq.~\eqref{eq:overview-duhamel} would therefore give only an $\mathcal O(tK^{-3/2})$ bound on the difference of these states. To obtain a better observable bound, we keep this omitted-mode contribution inside the inner products that define the expectation value.

The expectation difference has exactly the pairings we need. For two states $\ket{u},\ket{v}$ and self-adjoint $O$,
\begin{equation}
 \langle u|O|u\rangle-\langle v|O|v\rangle
 =(\langle u|-\langle v|)O\ket{u}+\langle v|O(\ket{u}-\ket{v}).
\end{equation}
We apply this identity with $\ket{u}=P\ket{\psi(t)}$ and $\ket{v}=\ket{\psi_N(t)}$. Thus it suffices to pair the retained-state error with the observable applied to either endpoint; we do not need to control every possible direction of that error. For either final-time test state $\ket{v}=P\ket{\psi(t)}$ or $\ket{v}=\ket{\psi_N(t)}$, we apply the observable, project, and evolve backwards under the Galerkin Hamiltonian:
\begin{equation}
 \ket{\varphi_v(s)}:=e^{iH_N(t-s)}P(O\ket{v}),\qquad 0\le s\le t.
\end{equation}
Both choices of $\ket{v}$ have a uniformly bounded kinetic graph norm $\|(1+T)\ket{v}\|$. Here we use the cutoff-independent comparison of the kinetic and discrete Hamiltonian graph norms, together with conservation of $\|H_N\ket{\psi_N(t)}\|$. To see why applying the observable preserves the required bound, commute it through the kinetic operator:
\begin{equation}
 [T,O]\ket{v}=-\frac{1}{m_j}\nabla\chi(\bm{x}_j)\cdot\nabla_j \ket{v}
         -\frac{1}{2m_j}(\Delta\chi)(\bm{x}_j)\ket{v}.
\end{equation}
Only the bounded first and second derivatives of $\chi$ occur. The resulting unnormalised test state $P(O\ket{v})$ therefore has a bounded discrete Hamiltonian graph norm. Backward evolution conserves that norm, and the cutoff-independent estimate
\begin{equation}
 \|(1+T)\ket{z}\|
 \le C\bigl(\|H_N\ket{z}\|+\|\ket{z}\|\bigr),
 \qquad P\ket{z}=\ket{z}.
\end{equation}
returns the desired kinetic graph bound for $\ket{\varphi_v(s)}$ (Lemma~\ref{lem:smear}). This backward evolution of the test state acts on the full retained particle space: applying $\chi(\bm{x}_j)$ need not preserve an exchange-symmetry sector. The potential-tail estimate applied to this backward state therefore gives $\|(1-P)V_{\mathrm{tot}}\ket{\varphi_v(s)}\|=\mathcal O(K^{-1/2})$.

We now move the propagator in Eq.~\eqref{eq:overview-duhamel} to the test side of the inner product and use symmetry of $V_{\mathrm{tot}}$. This pairs the discarded state with $V_{\mathrm{tot}}\ket{\varphi_v(s)}$. Only the high-momentum part of the latter can contribute, because $(1-P)\ket{\psi(s)}$ is orthogonal to the retained basis. Consequently,
\begin{equation}
 \begin{aligned}
 \big|(\langle\psi(t)|P-\langle\psi_N(t)|)O\ket{v}\big|
 &\le\int_0^t\|(1-P)\ket{\psi(s)}\|\,
                 \|(1-P)V_{\mathrm{tot}}\ket{\varphi_v(s)}\|\,ds\\
 &=\mathcal O\!\left(tK^{-5/2}K^{-1/2}\right)
 =\mathcal O(tK^{-3}).
 \end{aligned}
\end{equation}
This is the pairing estimate of Proposition~\ref{prop:duality}. The improvement comes from testing the error against the observable, not from a stronger bound on the full wavefunction error.

We must also account for replacing $\ket{\psi(t)}$ by $P\ket{\psi(t)}$ in the expectation value. With all states in the following identity evaluated at time $t$, we have
\begin{equation}
 \begin{aligned}
 \langle\psi|O|\psi\rangle-\langle\psi|POP|\psi\rangle
 &=2\operatorname{Re}\langle\psi|(1-P)O|\psi\rangle
 -\langle\psi|(1-P)O(1-P)|\psi\rangle.
 \end{aligned}
\end{equation}
Smoothness makes the first term quadratic in small tails. A Fourier mode of $\chi$ with transfer at most $K/2$ can produce output beyond the cutoff only from input with momentum above $K/2$ in the observed register; restrictions on the other registers commute with $O$. 
The assumed decay of the Fourier tails of $\chi$ makes the
contribution from transfers larger than $K/2$ decay faster than
any inverse power of $K$. Consequently,
$\|(1-P)O\ket{\psi(t)}\|=\mathcal O(K^{-5/2})$.
Pairing this tail with $(1-P)\ket{\psi(t)}$ gives $\mathcal O(K^{-5})$, while the last term is at most $\|O\|\,\|(1-P)\ket{\psi(t)}\|^2=\mathcal O(K^{-5})$. Combining the projection and evolution contributions proves the $\mathcal O(K^{-5}+tK^{-3})$ estimate of Theorem~\ref{thm:dyn} for all $t\ge0$.

\subsection{Spatial-grid discretisation and Fourier aliasing}

We then ask whether a position-grid implementation has the same accuracy. In collocation, the potential is diagonal with entries obtained by sampling its values on the grid. A discrete Fourier transform changes the basis, but does not make this operator equal to the Galerkin compression. Fourier transfers that differ by a grid period are indistinguishable on the sample points and are folded into the same retained register. This is aliasing. In particular, a sampled pair interaction conserves total momentum only modulo the register size, rather than exactly.

To control this difference, we use a register with momentum cutoff $2K$ and truncate the interaction kernel to transfers with componentwise cutoff $2K$. Let $P'$ project onto this larger register. Its full difference set extends to $4K$, so truncating the kernel removes some transfers that would be present in $P'HP'$. Lemma~\ref{lem:grid-structure} separates those dropped transfers from the aliases:
\begin{equation}
 H_{\rm grid}=P'HP'-D+R.
\end{equation}
Here $D$ contains the Galerkin transfers between $2K$ and $4K$ that we omitted, and $R$ contains the additional aliased transfers caused by sampling. For an aliased transfer, the output momentum index would lie outside the retained register, so the grid operation reduces it modulo the number of grid points in that coordinate and places it back inside the register. This is the operation called wrapping in Appendix~\ref{app:grid}. The bandwidth choice forces at least one endpoint of such a transfer to have momentum above the inner cutoff $K$ in a register where this modular reduction occurs. It does not eliminate coupling between the inner and outer regions. Rather, it allows us to bound the aliasing terms using the momentum tails of the states on which they act.

A one-coordinate example makes the distinction between omitted and aliased transfers explicit. Take an inner register with indices $-2,\ldots,2$ and the larger grid register $-4,\ldots,4$, which has period nine. With the convention that a transfer $\nu$ sends input $q$ to $q-\nu$, the transfer $\nu=1$ sends $q=-4$ to $-5$, which is reduced modulo nine to $+4$. Even a small transfer can therefore be aliased, but both endpoints in this example are outside the inner register. The allowed transfer $\nu=4$ instead sends $q=-1$ to $-5$, again reduced modulo nine to $+4$: aliasing can couple the inner and outer regions. A Galerkin transfer from $+4$ to $-4$ uses $\nu=8$ and is dropped by the kernel cutoff $|\nu|\le4$. The three-dimensional construction follows the same modular rule in each coordinate.

The energy comparison must account for both $D$ and $R$. Unlike Galerkin restriction, the grid modification does not impose a variational ordering relative to the continuum Hamiltonian. We instead use a two-sided comparison: a ground state of each finite Hamiltonian is a trial state for the other. Appendix~\ref{app:grid} first establishes cutoff-independent energy and graph-norm bounds for these states, then derives their momentum tails and bounds the expectations of the omitted-transfer operator $D$ and the aliasing operator $R$. This gives the absolute estimate $|E_{\rm grid}-E|=\mathcal O(K^{-3})$ in Theorem~\ref{thm:grid-energy}. Establishing the energy budgets before using the tail estimates prevents their constants from depending circularly on an uncontrolled grid eigenvalue.

We evolve the projected initial state as $\ket{\psi_g(t)}:=e^{-iH_{\rm grid}t}P'\ket{\psi_0}$, again without renormalising, and implement the observable by the grid-diagonal operator $O_g$,
whose entries are $\chi$ evaluated at the grid points of register $j$. Thus the grid estimate compares $\langle\psi(t)|O|\psi(t)\rangle$ with $\langle\psi_g(t)|O_g|\psi_g(t)\rangle$. Sampling the observable introduces its own aliases: $O_g$ need not equal $P'OP'$, even when $\chi$ has only low Fourier modes. Lemma~\ref{lem:grid-obs} bounds the resulting quadratic-form error on the projected exact state by $\mathcal O(K^{-5})$ using its momentum tails.

We can now identify the three contributions to the dynamical grid error. First replace the exact state by $P'\ket{\psi(t)}$ in the continuum observable; this is the projection contribution. Next replace $P'OP'$ by $O_g$ on that projected state; this is the sampling contribution. Finally compare $P'\ket{\psi(t)}$ with $\ket{\psi_g(t)}$ under the same observable $O_g$. The first two contributions are $\mathcal O(K^{-5})$. For the third, the Hamiltonian decomposition gives the following Duhamel formula for the retained-state error
\begin{equation}
 \begin{aligned}
 P'\ket{\psi(t)}-\ket{\psi_g(t)}
 =-i\int_0^t e^{-iH_{\rm grid}(t-s)}
 \bigl[&P'V_{\mathrm{tot}}(1-P')\ket{\psi(s)}
       +(D-R)P'\ket{\psi(s)}\bigr]ds.
 \end{aligned}
\end{equation}
The first term in square brackets is the familiar return from discarded continuum modes. The second is the change in the retained dynamics caused by omitting transfers and introducing aliases. We must therefore bound the inner product of each of these three contributions---the discarded-mode term, the $D$ term and the $R$ term---with the backward-evolved grid test state.

\subsection{Regularity of the backward-evolved grid states}

The grid argument also needs tails for its backward test states,
\begin{equation}
 \ket{\varphi_{g,v}(s)}:=e^{iH_{\rm grid}(t-s)}O_g\ket{v},
 \qquad \ket{v}\in\{P'\ket{\psi(t)},\ket{\psi_g(t)}\},\quad 0\le s\le t.
\end{equation}
A bounded first kinetic graph norm would give only a $K^{-2}$ tail for these vectors. That was enough for the Galerkin argument, where we took the high-momentum tail of the potential applied to the test state. The inner products involving $D$ and $R$ in the finite register also require a tail of the test state itself. A $K^{-2}$ tail would leave $K^{-1/2}K^{-2}=K^{-5/2}$ in a dropped-transfer pairing, short of the desired $K^{-3}$. We therefore need a stronger finite-grid estimate.

Kernel truncation supplies
\begin{equation}
 \|T^2\ket{\varphi_{g,v}(s)}\|=\mathcal O(K^{3/2}),
\end{equation}
uniformly in $s$ and $t$ (Lemmas~\ref{lem:grid-data2} and~\ref{lem:grid-obs2}). The origin of the $K^{3/2}$ factor is useful to keep in view. Moving a kinetic weight through a Fourier transfer with indices reduced modulo the register size introduces a quadratic momentum factor. This cancels the inverse-square decay of the band-limited Coulomb coefficient, leaving bounded coefficients on $\mathcal O(K^3)$ transfers. Their squared sum is $\mathcal O(K^3)$, and the associated slice estimate therefore carries a factor $K^{3/2}$. The weighted discrete estimates convert this count into a comparison with the squared grid Hamiltonian, whose norm on the evolving state is conserved. The projected exact states have the same $K^{3/2}$ budget by summing their weighted momentum tails over shells. Multiplication by the sampled smooth observable preserves it with a constant involving the fourth absolute Fourier moment of $\chi$, namely $\sum_{\bm{\nu}\in\mathbb Z^3}(1+\norm{\bm{k}_{\bm{\nu}}}^4)|\widehat\chi(\bm{\nu})|$, where $\bm{k}_{\bm{\nu}}=2\pi\bm{\nu}/L$ and $\widehat\chi(\bm{\nu})$ is the corresponding Fourier coefficient. The fourth-power weight accounts for the momentum factors introduced when the kinetic operator acts twice.

This is a bound on the norm after applying the squared kinetic operator in a finite register. It is allowed to grow with $K$ and does not assert ordinary fourth-order regularity of the continuum state. On a shell where one register momentum is of order $K$, the squared kinetic weight is of order $K^4$.
For any fixed particle register $j$, let
$\mathbf1_{\text{shell}}$ project onto the retained plane-wave
configurations satisfying $K<\|\bm{k}_j\|_\infty\le2K$,
where $\bm{k}_j$ is the momentum in that register.
Then
\begin{equation}
 \|\mathbf1_{\text{shell}}\ket{\varphi_{g,v}(s)}\|
 \le CK^{-4}\|T^2\ket{\varphi_{g,v}(s)}\|
 =\mathcal O(K^{-5/2}),
\end{equation}
which is the stronger backward tail needed for the grid comparison.
The same $\mathcal O(K^{-5/2})$ scaling holds for the retained
modes with $\|\bm{k}_j\|_\infty>K/2$.

We now apply the omitted-transfer and aliasing estimates to each error term paired with the backward-evolved test state. Dropped transfers have the Coulomb coefficient factor $K^{-1/2}$ and a state-tail factor $K^{-5/2}$, giving $K^{-3}$. For aliased transfers, we split according to the Fourier transfer entering the kernel coefficient, which need not equal the difference between the represented input and output momenta after modular reduction. When this Fourier transfer exceeds the componentwise cutoff $K$, the smaller Coulomb coefficients give the same $K^{-3}$ estimate with one tail. Within that cutoff, the coefficients need not be small, but modular reduction forces both coupled momenta above $K$ in any register where it occurs. These pairings contain two tails and are bounded by $K(K^{-5/2})^2=K^{-4}$. The coefficient bound and the momentum geometry therefore supply complementary ways of making the aliases small.

After pairing with the backward-evolved test state, each term in the Duhamel integrand is now at most $\mathcal O(K^{-3})$. Integrating over time and restoring the two $\mathcal O(K^{-5})$ projection and sampling contributions gives the grid bound $\mathcal O(K^{-5}+tK^{-3})$ of Theorem~\ref{thm:grid-dyn-sharp}. The constants also depend on a weighted sum of the magnitudes
of the observable's Fourier coefficients, with fourth-power
momentum weights. This sum is finite because the assumed Fourier
tails decay faster than any inverse power of momentum.

The use of $2K$ for the register cutoff is a convention for stating the effective resolution, not a projection or filter applied to the state during evolution. Starting with a fixed register, we may instead take $K$ to be half its cutoff. The one-particle mode count changes by only a fixed factor, so at fixed cell size the common observable rate is $\mathcal O(N^{-5/3}+tN^{-1})$. The essential design choice is the interaction bandwidth relative to the register size.

For comparison, Proposition~\ref{prop:fullkernel} keeps the maximal band-limited kernel, with transfers up to $4K$ on the same register. Its absolute ground-energy error is bounded by $\mathcal O(K^{-3/2})$, improving to $\mathcal O(K^{-2})$ under the cyclic correlation summability condition in Eq.~\eqref{eq:fullkernel-correlation}. This condition bounds the sum of the absolute overlaps between a ground state and all its nonzero cyclic momentum shifts, using opposite shifts on the two particles of a quantum pair and single-register shifts for clamped-nucleus interactions. Here cyclic shifts reduce momentum indices modulo the register size, as in the aliasing example above. That bound must hold uniformly in $K$ for both selected ground states, of $P'HP'$ and of the full-kernel grid Hamiltonian, including the single-particle correlations for any clamped-nucleus interactions. These are upper bounds, not lower bounds showing that this method must perform worse. Nor is that maximal band-limited kernel the same as direct sampling of the singular, untruncated Coulomb function. The latter requires a prescription at singular grid points and a separate error analysis.

\section{Kinetic energy and stopping-power estimates}
\label{sec:kinetic-overview}

We now estimate a particle's kinetic energy, including the projectile
energy used to determine stopping power. 
The continuum kinetic-energy operator for particle $j$ is
$T_j=-\nabla_j^2/(2m_j)$. In the momentum representation,
it acts by multiplication by $\|\bm{k}_j\|^2/(2m_j)$.
Unlike the smooth position
observables considered above, this operator is unbounded: an
error carried by a small amplitude at high momentum can still change
its expectation appreciably. Here $j$ labels the particle whose kinetic energy is measured; for stopping power, that particle is the projectile. 
We nevertheless obtain, for both Galerkin
and truncated-kernel grid evolution,
\begin{equation}
 \big|\langle\psi(t)|T_j|\psi(t)\rangle
       -\langle\psi_{\rm d}(t)|T_j|\psi_{\rm d}(t)\rangle\big|
 \le C\mathcal D^2 K^{-3}[1+t+\log(2+K)],\qquad t\ge0,
 \label{eq:overview-kinetic}
\end{equation}
where $\mathcal D=\|\ket{\psi_0}\|+\|H\ket{\psi_0}\|+\|H^2\ket{\psi_0}\|$ and
$\ket{\psi_0}\in D(H^2)$. 
Here $\ket{\psi_{\rm d}(t)}$ denotes either the Galerkin state
$\ket{\psi_N(t)}$, evolved from $P\ket{\psi_0}$, or the grid state
$\ket{\psi_g(t)}$, evolved from $P'\ket{\psi_0}$, with the
corresponding definition of $K$.
Neither initial projection is
renormalised. At fixed cell size, $N=\Theta(K^3)$ for both methods, so
the bound is $\mathcal O((t+\log N)/N)$.

The proof separates two ways of restricting a state. The Fourier
projector restricts individual particle momenta and defines the
discretisation. A second restriction, used only in the proof, selects
low eigenvalues of the full Hamiltonian. These Hamiltonian spectral
projectors commute with their respective evolutions, whereas a momentum
projector need not. We compare the two dynamics on the selected
Hamiltonian-energy spaces, use their momentum tails to make the
consistency error---the difference between applying the finite Hamiltonian
and projecting the exact Hamiltonian image---small, and finally bound the kinetic contribution of
the components omitted by this additional restriction. The simulation
itself uses the original finite Hamiltonian and projected initial state.

\subsection{Projection and the kinetic-energy pairing}

We first isolate the error caused by the finite momentum register.
For Galerkin evolution let $\ket{\delta}=(1-P)\ket{\psi(t)}$. Since $P$ and $T_j$
are diagonal in the same Fourier basis, the cross terms vanish:
\begin{equation}
 \langle\psi|T_j|\psi\rangle-\langle\psi|PT_jP|\psi\rangle
 =\langle\delta|T_j|\delta\rangle
 \le\|\ket{\delta}\|\,\|(1+T)\ket{\delta}\|
 \le C\mathcal D^2K^{-3}.
 \label{eq:overview-kinetic-projection}
\end{equation}
The last inequality combines the ordinary $K^{-5/2}$ state tail
with its $K^{-1/2}$ kinetic-weighted tail. The same argument applies at
the grid register cutoff $2K$. Thus the projection contribution already
has the desired inverse-basis-size rate.

To control the error between retained states, we use the exact identity
\begin{equation}
 \langle u|T_j|u\rangle-\langle v|T_j|v\rangle
 =(\langle u|-\langle v|)T_j\ket{u}+\langle v|T_j(\ket{u}-\ket{v}).
 \label{eq:overview-kinetic-pairing}
\end{equation}
The existing graph estimates bound the norms of $T_j\ket{u}$ and $T_j\ket{v}$ for
the states we compare. Consequently
\begin{equation}
 |\langle u|T_j|u\rangle-\langle v|T_j|v\rangle|
 \le (\|T_j\ket{u}\|+\|T_j\ket{v}\|)\|\ket{u}-\ket{v}\|
 \le C\mathcal D\|\ket{u}-\ket{v}\|.
\end{equation}
Thus a retained-state error of order
$\mathcal D K^{-3}[1+t+\log(2+K)]$ gives the desired kinetic error,
even though $T_j$ has no cutoff-independent operator norm. The new spectral comparison
provides that bound on the low-energy components. On the grid the
remaining high-energy components require an additional weighted
pairing, which we explain after establishing the low-energy estimate.

\subsection{Dynamics restricted to low Hamiltonian energies}

We explain the grid comparison first, since it includes the
truncated-transfer and aliasing errors as well as the contribution from discarded modes. Write $G=\Hgrid$ for the grid Hamiltonian acting on the
finite-dimensional space retained by $P'$, and define
\begin{equation}
 \mathcal C_g=GP'-P'H
 =-D_{\mathrm{drop}}P'+R_{\mathrm{wrap}}P'-P'V_{\mathrm{tot}}(1-P').
\end{equation}
Here $D_{\mathrm{drop}}$ contains the potential transfers removed by truncating the
kernel, and $R_{\mathrm{wrap}}$ contains the aliased transfers introduced by sampling.
These are the operators $D$ and $R$ of Eq.~\eqref{eq:grid-structure};
the subscripts identify their omitted-transfer and aliasing roles. The operator $\mathcal C_g$ measures the difference between applying the grid Hamiltonian after projection and projecting the result of the continuum Hamiltonian.
To estimate the effect of this residual on an observable, we must
control both the state to which it is applied and the state against
which its output is paired. We first select low Hamiltonian energies
on each side and then estimate the residual between those spaces.

Choose a fixed positive energy shift $c_0$, large enough that $\mathsf H=H+c_0\ge1$ and $\mathsf H_g=G+c_0\ge1$ for every admissible cutoff. The continuum
and grid graph estimates compare their norms with that of
$\mathcal E=1+T$, uniformly in the cutoff. Choose a sufficiently small
fixed $a>0$ and put
\begin{equation}
 \Lambda=aK^2,\qquad
 S=\ind{\mathsf H\le\Lambda},\qquad
 S_g=\ind{\mathsf H_g\le2\Lambda}.
\end{equation}
Here $\mathbf1[A\le b]$ denotes the spectral projector onto
energies of the self-adjoint operator $A$ at most $b$.
Thus $S$ and $S_g$ select low energies of the shifted
continuum and grid Hamiltonians, respectively.
The two ceilings differ by a factor of two so that a component below
the exact ceiling and one above the grid ceiling have an energy
separation of at least $\Lambda$. We have chosen this separation
ourselves; it imposes no condition on physical eigenvalue gaps.
The graph and tail estimates give the following norm bound on the Hamiltonian difference restricted to low continuum energies
\begin{equation}
 \|\mathcal C_g\mathsf H^{-1}S\|\le CK^{-1/2}.
\end{equation}
This bound still allows the output to occupy arbitrary grid energies.
The sharper comparison below restricts the output as well and gains
the additional decay needed for the observable estimate.

The first step is to show that the inverse-weighted states
$\ket{u_{\rm inv}}=\mathsf H^{-1}S\ket{f}$ and
$\ket{v_{\rm inv}}=\mathsf H_g^{-1}S_g\ket{h}$ have unusually
small momentum tails. Here $\ket{f}$ and $\ket{h}$ are arbitrary unit vectors
in the continuum and grid spaces. The comparison between the kinetic and Hamiltonian graph norms gives
$\|\mathcal E \ket{u_{\rm inv}}\|+\|\mathcal E \ket{v_{\rm inv}}\|\le C$: applying an inverse
Hamiltonian supplies the first graph norm needed for the
potential-tail estimate. If $Q_{j,\theta K}$ selects momenta
$\|\bm{k}_j\|_\infty>\theta K$ for a fixed $\theta>0$, then
\begin{equation}
 \|Q_{j,\theta K}\mathsf H^{-1}S\|
 +\|Q_{j,\theta K}\mathsf H_g^{-1}S_g\|
 \le CK^{-5/2}.
 \label{eq:overview-spectral-tail}
\end{equation}
To see the origin of this power, the kinetic energy on that momentum
region is at least $b_{\rm kin}K^2$, for a fixed $b_{\rm kin}>0$ determined by the masses and $\theta$, whereas the selected Hamiltonian energies
are at most $2aK^2$. The Schr\"odinger equation and the Coulomb
potential-tail estimate therefore give an inequality of the form
\begin{equation}
 b_{\rm kin}K^2 \mathcal X\le(2aK^2+c_0)\mathcal X+CK^{-1/2},
\end{equation}
where $\mathcal X$ is either operator norm in
Eq.~\eqref{eq:overview-spectral-tail}. Choosing $a$ sufficiently small
lets us absorb the first term on the right. Dividing the remaining
$K^{-1/2}$ potential tail by $K^2$ gives $K^{-5/2}$. For the Hamiltonian term in this estimate, write $Q=Q_{j,\theta K}$ and $\mathcal B=\mathsf H^{-1}S$. Since $\mathcal B$ is a spectral function of $H$ with input restricted by $S$,
\begin{align}
 QH\mathcal B&=Q\mathcal BHS,\\
 \|QH\mathcal B\|&\le(\Lambda+c_0)\|Q\mathcal B\|.
\end{align}
The grid argument has the same form with $2\Lambda+c_0$. Thus the factor multiplying the tail norm comes from the restricted input Hamiltonian; we never commute the momentum projector $Q$ through $H$.

We now test $\mathcal C_g$ between these inverse-weighted states.
For the exterior term, the ordinary state tail contributes
$K^{-5/2}$ and the potential applied to the test state contributes
$K^{-1/2}$. The omitted-transfer and aliasing terms obey a similar estimate: a nonzero large transfer forces at least one of its input and output momenta to have magnitude proportional to $K$. The
$K^{-1/2}$ coefficient-tail bound and the $K^{-5/2}$ state tail at
that endpoint therefore give the same product. Aliasing terms with both endpoint momenta high
have the smaller bound $K(K^{-5/2})^2$. Together these estimates give
\begin{equation}
 \|S_g\mathsf H_g^{-1}\mathcal C_g\mathsf H^{-1}S\|
 \le CK^{-3}.
 \label{eq:overview-low-consistency}
\end{equation}
The estimate is uniform over the kernel class in Eq.~\eqref{eq:kernelclass}, including its smooth momentum-transfer truncations, and extends to the range-truncated kernel families used in Appendix~\ref{app:free}: their nonzero Fourier coefficients are bounded by a fixed multiple of the Coulomb coefficients, and only a uniformly bounded number of transfer indices can reduce to the same retained index. These two properties preserve the potential-tail and aliasing estimates. The latter bound concerns the number of aliased transfers, not the number of particles or basis states.

For $\ket{\psi_{\rm low}(t)}=e^{-iHt}S\ket{\psi_0}$, compare the full retained vectors
$\ket{x(t)}=P'\ket{\psi_{\rm low}(t)}$ and $\ket{y(t)}=e^{-iGt}P'S\ket{\psi_0}$.
They agree at time zero. Their low grid-energy difference
$S_g(\ket{x(t)}-\ket{y(t)})$ is a Duhamel integral driven by the Hamiltonian difference $S_g\mathcal C_g\ket{\psi_{\rm low}(t)}$. 
To use Eq.~\eqref{eq:overview-low-consistency}, we divide the
input and output shifted Hamiltonian energies into bands
$[2^j,2^{j+1})$, indexed by integers $j$. Each successive band
doubles both endpoints of the preceding band. This division
into scales related by powers of two is called a dyadic
decomposition.
For comparable energies, the
inverse weights have a bounded ratio and direct integration gives a
term proportional to $t$. At each fixed band offset the blocks have
orthogonal inputs and outputs, so this term is not multiplied by the
number of bands. For separated energies, we solve an operator
Sylvester equation $B_1X-XB_2=\mathcal L$, where $B_1,B_2$ are the two
Hamiltonians restricted to the chosen bands and $\mathcal L$ is the corresponding block of the Hamiltonian-difference operator in the Duhamel integral. This identity expresses the oscillatory time integral
as a difference of endpoint terms. Its inverse
energy separation replaces the time factor, and summing the remaining
bands costs $\log(2+K)$. Lemma~\ref{lem:kinetic-dyadic} gives the
complete operator argument. Since
$S\ket{\psi_0}=\mathsf H^{-2}S\mathsf H^2\ket{\psi_0}$, the two Hamiltonian
images of the initial state supply precisely the inverse input weights
needed there. We obtain
\begin{equation}
 \|S_g(\ket{x(t)}-\ket{y(t)})\|
 \le C\mathcal D K^{-3}[t+\log(2+K)].
 \label{eq:overview-low-flow}
\end{equation}
The first graph norms of $\ket{x}$ and $\ket{y}$ are bounded by $C\mathcal D$.
We next combine this low-energy difference bound with control of its
high-energy part, and then use the kinetic pairing in
Eq.~\eqref{eq:overview-kinetic-pairing}.

\subsection{High-energy contributions to the kinetic-energy error}

We now compare the full retained states, rather than only their
low grid-energy components. Write
\begin{equation}
 \ket{x(t)}=P'e^{-iHt}S\ket{\psi_0},\qquad
 \ket{y(t)}=e^{-iGt}P'S\ket{\psi_0},
\end{equation}
so that Eq.~\eqref{eq:overview-low-flow} bounds
$\|S_g(\ket{x(t)}-\ket{y(t)})\|$. We will estimate the high grid-energy part
of this same difference inside the kinetic pairing. This lets us use
momentum-tail bounds for $\ket{x}$ and $\ket{y}$ themselves, without needing a
momentum-tail bound after applying the Hamiltonian projector $S_g$.

The grid projection of a low exact-energy state can have grid energies
above $2\Lambda$. To bound this contribution, put
\begin{equation}
 X_{\perp}=(1-S_g)P'S\mathsf H^{-1}S,\qquad
 F=(1-S_g)\mathcal C_g\mathsf H^{-1}S.
\end{equation}
The identity $\mathsf H_gP'-P'\mathsf H=\mathcal C_g$ gives
$\mathsf H_gX_{\perp}-X_{\perp}\mathsf H=F$. The output energy exceeds $2\Lambda$
and the input energy is at most $\Lambda$, so the decaying integral
\begin{equation}
 X_{\perp}=\int_0^\infty e^{-s\mathsf H_g}F e^{s\mathsf H}S\,ds
\end{equation}
has norm at most $\|F\|/\Lambda$. The norm bound on $\mathcal C_g\mathsf H^{-1}S$ established above gives $\|F\|\le CK^{-1/2}$, and hence
$\|X_{\perp}\|\le CK^{-5/2}$. Applying this bound to
$\mathsf H e^{-iHt}S\ket{\psi_0}$ controls $(1-S_g)\ket{x(t)}$. For $\ket{y(t)}$,
we apply it at time zero and use that $S_g$ commutes with the grid
evolution. Both contributions are at most $C\mathcal D K^{-5/2}$,
so
\begin{equation}
 \|(1-S_g)(\ket{x(t)}-\ket{y(t)})\|\le C\mathcal D K^{-5/2}.
\end{equation}

The additional power needed for kinetic energy comes from the vector
against which this difference is paired. For either $\ket{w}=\ket{x(t)}$ or
$\ket{w}=\ket{y(t)}$, the state-tail estimates give
$\|Q_{j,\lambda}\mathcal E \ket{w}\|\le
C\mathcal D\lambda^{-1/2}$ for $1\le\lambda\le K$.
On a shell where the largest register momentum is of order
$\lambda$, applying $T_j$ contributes
at most $C\lambda^2$. Thus the squared contribution to
$\|\mathcal E T_j\ket{w}\|^2$ is at most
$C\mathcal D^2\lambda^3$. Summing over disjoint shells whose
radii double is a geometric sum, bounded by $C\mathcal D^2K^3$.
The finitely supported grid state has no momenta above $2K$; its
last shell is controlled by the tail estimate at $K$.
The same graph-norm comparison therefore gives
\begin{equation}
 \|\mathsf H_gT_j\ket{w}\|\le C\|\mathcal E T_j\ket{w}\|
 \le C\mathcal D K^{3/2}.
\end{equation}
On the grid-energy space above $2\Lambda$, an inverse Hamiltonian
has norm at most $(2\Lambda)^{-1}$. Consequently
\begin{equation}
 \|(1-S_g)T_j\ket{w}\|\le(2\Lambda)^{-1}\|\mathsf H_gT_j\ket{w}\|
 \le C\mathcal D K^{-1/2}.
\end{equation}
This step uses the spectral support of the test vector after projection;
it does not assume that $T_j$ commutes with $S_g$.

We can now split each pairing in
Eq.~\eqref{eq:overview-kinetic-pairing} into the two orthogonal
grid-energy spaces. For $\ket{w}=\ket{x(t)}$ or $\ket{w}=\ket{y(t)}$,
\begin{align}
 |(\langle x|-\langle y|)T_j\ket{w}|
 &\le \|S_g(\ket{x}-\ket{y})\|\,\|T_j\ket{w}\|
    +\|(1-S_g)(\ket{x}-\ket{y})\|\,\|(1-S_g)T_j\ket{w}\|\\
 &\le C\mathcal D^2K^{-3}[1+t+\log(2+K)].
\end{align}
The high-energy term is $O(\mathcal D^2K^{-3})$; the logarithm
comes entirely from the low-energy dynamical comparison.

Finally, the exact initial component removed by $S$ has norm at most
$\Lambda^{-2}\|\mathsf H^2\ket{\psi_0}\|\le C\mathcal D K^{-4}$.
Unitarity and the first graph bounds convert its restoration into a
kinetic-energy error of order $\mathcal D^2K^{-4}$ in both
state evolutions. Adding the Fourier projection error from
Eq.~\eqref{eq:overview-kinetic-projection} proves the grid bound.

For Galerkin evolution the consistency residual is
$H_NP-PH=-PV_{\mathrm{tot}}(1-P)$. Its inverse-weighted norm on
the selected exact-energy space is $O(K^{-1})$. Dividing by the
same artificial energy separation therefore gives an omitted
retained-state contribution of order $\mathcal D K^{-3}$.
Combining this with the low-energy comparison yields
\begin{equation}
 \|P\ket{\psi(t)}-\ket{\psi_N(t)}\|
 \le C\mathcal D K^{-3}[1+t+\log(2+K)].
\end{equation}
The ordinary kinetic pairing now suffices for the Galerkin kinetic
bound. This is an error between retained states: the full continuum
state still has its separate $K^{-5/2}$ Fourier tail.
Appendix~\ref{app:source} proves both kinetic bounds in
Theorem~\ref{thm:kinetic-log}.

\subsection{Complementary direct and capped bounds}

We give two complementary bounds: a direct Duhamel estimate
and an estimate obtained by temporarily capping the kinetic-energy
observable. Here capping means replacing its momentum-space value
$\|\bm{k}_j\|^2/(2m_j)$ by
$\min\{\|\bm{k}_j\|^2,\kappa^2\}/(2m_j)$, so that it cannot exceed
$\kappa^2/(2m_j)$. This modifies the observable used in the proof,
not the basis or the Hamiltonian evolution. We separately bound
the error from this replacement to recover a bound for the original
kinetic energy. These arguments isolate simpler mechanisms and,
for the capped estimate, give sublinear time growth.
We state them for the Galerkin state.

In these estimates we retain the notation $\Xi$ of
Proposition~\ref{prop:dyn-reg} for the propagated regularity budget.
For the fixed system, $\Xi=C_0m_{\max}\mathcal D$, so it
differs from the initial graph-norm budget $\mathcal D$ above only
by a fixed factor. Here $m_{\max}=\max_j m_j$ is the largest moving-particle mass, and $C_0$ is the cutoff-independent constant that converts the Hamiltonian graph budget into a bound on $\|(1+T)\ket{\psi(t)}\|+\|(1+T)H\ket{\psi(t)}\|$. The expectations below are consequently
quadratic in the same initial data budget.

The accumulated error requires more care because applying $T_j$ also increases the high-momentum weight of the backward test states. The direct state-norm bound from Eq.~\eqref{eq:overview-duhamel} gives $\mathcal O(\Xi^2(K^{-3}+tK^{-3/2}))$. To improve it, we keep the contribution $PV_{\mathrm{tot}}(1-P)\ket{\psi(s)}$ inside the expectation pairing, as for smooth observables. For either endpoint $\ket{v}=P\ket{\psi(t)}$ or $\ket{v}=\ket{\psi_N(t)}$, define
\begin{equation}
 \ket{\varphi_v(s)}:=e^{iH_N(t-s)}T_j\ket{v},\qquad 0\le s\le t.
\end{equation}
We need only its half-power kinetic norm: Lemma~\ref{lem:phi} gives $\|(1+T)^{1/2}\ket{\varphi_v(s)}\|\le C\Xi K^{1/2}$. This follows by summing the momentum-shell tails of $T_j\ket{v}$ and propagating the corresponding fractional Hamiltonian norm. Hardy's inequality then bounds the omitted-mode contribution paired with the backward-evolved test state by
\begin{equation}
 \begin{aligned}
 |\langle\delta(s)|V_{\mathrm{tot}}P|\varphi_v(s)\rangle|
 &=|\langle\delta(s)|V_{\mathrm{tot}}|\varphi_v(s)\rangle|\\
 &\le C\|\ket{\delta(s)}\|\,\|(1+T)^{1/2}\ket{\varphi_v(s)}\|\\
 &\le C\Xi^2K^{-5/2}K^{1/2}=C\Xi^2K^{-2}.
 \end{aligned}
\end{equation}
Time integration therefore gives $\mathcal O(\Xi^2(K^{-3}+tK^{-2}))$ for all $t\ge0$ (Theorem~\ref{thm:src1}). This estimate uses every momentum transfer in $V_{\mathrm{tot}}$ and retains the exponent $K^{-2}$ without an arbitrarily small power loss.

For long times we can reduce the growth of the bound by temporarily capping the observable. We use $\mu_\kappa(\bm{k}_j)=\min\{\|\bm{k}_j\|^2,\kappa^2\}/(2m_j)$, with $\kappa$ between a fixed lower threshold and $K/2$. The cap changes each kinetic expectation by at most $C\Xi^2\kappa^{-3}$. To estimate its evolution error, we now use the capped backward state $\ket{\varphi_{\kappa,v}(s)}=e^{iH_N(t-s)}\mu_\kappa \ket{v}$. Lemma~\ref{lem:capsmear} gives the endpoint bound $\|(1+T)\mu_\kappa \ket{v}\|\le C\Xi\kappa^{3/2}$. The discrete graph-norm comparison and Hamiltonian conservation used in Section~\ref{sec:dynamics-overview} propagate it to $\|(1+T)\ket{\varphi_{\kappa,v}(s)}\|\le C\Xi\kappa^{3/2}$, uniformly in $s$ and $t$. This permits the potential-tail estimate used in Section~\ref{sec:dynamics-overview}: $\|(1-P)V_{\mathrm{tot}}\ket{\varphi_{\kappa,v}(s)}\|\le C\Xi\kappa^{3/2}K^{-1/2}$. Pairing this tail with $\ket{\delta(s)}$ therefore gives
\begin{equation}
 |\langle\delta(s)|V_{\mathrm{tot}}P|\varphi_{\kappa,v}(s)\rangle|
 \le C\Xi^2K^{-5/2}\kappa^{3/2}K^{-1/2}
 =C\Xi^2\kappa^{3/2}K^{-3}.
\end{equation}
After time integration, the estimate for the original, uncapped observable is
\begin{equation}
 C\Xi^2\bigl(K^{-3}+\kappa^{-3}+t\kappa^{3/2}K^{-3}\bigr).
\end{equation}
The two $\kappa$-dependent terms have opposite preferences. Raising the cap makes the observable replacement more accurate, but makes the backward state harder to control. For $t>0$, their balance is
\begin{equation}
 \kappa^{-3}=t\kappa^{3/2}K^{-3}
 \quad\Longleftrightarrow\quad
 \kappa^{9/2}=K^3/t,
 \qquad \kappa=(K^3/t)^{2/9}.
\end{equation}
Substituting this value into either term gives $t^{2/3}K^{-2}$. The balanced value is usable only when it lies between the fixed lower threshold and $K/2$. Outside that interval we clamp the cap to the appropriate endpoint; for very long times the uniform bound on the kinetic expectations is already sufficient. Theorem~\ref{thm:mult-cap} carries out these endpoint cases, including $t=0$, without extending the balance formula beyond its range. Combining its capped estimate with the stronger direct Duhamel estimate of Theorem~\ref{thm:src1} gives
\begin{equation}
 \mathcal{O}\!\left(K^{-3}+\min\{tK^{-2},\,t^{2/3}K^{-2}\}\right),
 \qquad t\ge0,
 \label{eq:overview-capped}
\end{equation}
with the common factor $\Xi^2$ understood. The cap is only a proof device; the certified expectation still uses $T_j$.

\subsection{Projection error in the kinetic-energy rate}
\label{sec:projection-flux-overview}

We next estimate how discarding high-momentum components changes the expression for the rate of kinetic-energy transfer. This comparison uses the exact state and its momentum projection; it does not yet compare the exact and Galerkin evolutions as in Eq.~\eqref{eq:overview-kinetic}. For $\ket{u},\ket{v}\in D(T)$, define the sesquilinear form
\begin{equation}
 \mathfrak G(\ket{u},\ket{v}):=i\bigl(\langle V_{\mathrm{tot}}u\,|\,T_jv\rangle
                         -\langle T_ju\,|\,V_{\mathrm{tot}}v\rangle\bigr).
\end{equation}
Here $\langle Bu|Cv\rangle$ denotes the inner product of
$B\ket{u}$ and $C\ket{v}$.
The expression is conjugate-linear in its first state vector and linear in its second. Evaluated twice on the same evolving state, $\mathfrak G(\ket{\psi},\ket{\psi})$ gives the derivative of the kinetic expectation along exact evolution, and likewise along Galerkin evolution when evaluated on the Galerkin state. Writing it through inner products ensures that we use only the displayed square-integrable factors; we do not require an everywhere-defined force operator. We integrate the change in this expression caused by momentum projection:

\begin{equation}
 \mathcal F_K(t):=\int_0^t
  \bigl[\mathfrak G(\ket{\psi(s)},\ket{\psi(s)})
       -\mathfrak G(P\ket{\psi(s)},P\ket{\psi(s)})\bigr]ds.
\end{equation}
This quantity is called integrated projection flux in Appendix~\ref{app:flux}; here it measures a difference of kinetic-energy-rate expressions, rather than a probability current across a spatial boundary. Theorem~\ref{thm:flux3} proves $|\mathcal F_K(t)|\le C_\varepsilon\Xi^2(1+t)K^{-3+\varepsilon}$ for every fixed $\varepsilon>0$, all $t\ge0$ and sufficiently large admissible cutoffs. The evolution equation for the projected exact state contains the additional term $PV_{\mathrm{tot}}(1-P)\ket{\psi}$, so its kinetic derivative is not simply $\mathfrak G(P\ket{\psi},P\ket{\psi})$. This is why a bound on $\mathcal F_K(t)$ alone does not establish the full Galerkin kinetic-energy error.

Appendix~\ref{app:flux} proves this bound by finite-matrix identities in a larger momentum box. Let $M$ be an integer cutoff on each Cartesian momentum-index component in each particle register and let $B_M$ project onto the larger momentum box. Define
\begin{equation}
 \ket{u_M(s)}:=B_M\ket{\psi(s)},\qquad T_M:=B_MTB_M,\qquad
 V_M:=B_MV_{\mathrm{tot}}B_M.
\end{equation}
The box projection is a proof device, distinct from the simulation projector $P$. The state projected into this larger box satisfies an inhomogeneous Schr\"odinger equation,
\begin{equation}
 i\ket{\dot u_M}=(T_M+V_M)\ket{u_M}+\ket{f_M},\qquad
 \ket{f_M}:=B_MV_{\mathrm{tot}}(1-B_M)\ket{\psi}.
\end{equation}
Here $\ket{f_M}$ is the amplitude returned by the potential from modes outside the larger box. For a self-adjoint matrix $J$ on the box, differentiating $\langle u_M|J|u_M\rangle$ and integrating gives
\begin{equation}
 \begin{aligned}
 \int_0^t\langle u_M|i[T_M,J]|u_M\rangle\,ds
 ={}&[\langle u_M|J|u_M\rangle]_0^t
   -\int_0^t\langle u_M|i[V_M,J]|u_M\rangle\,ds\\
 &-i\int_0^t\bigl(\langle f_M|J|u_M\rangle
                    -\langle u_M|J|f_M\rangle\bigr)\,ds.
 \end{aligned}
\end{equation}
We choose $J$ so that the kinetic commutator reproduces the force contribution to be eliminated. This identity then replaces its time integral by endpoint terms, a potential commutator and an explicitly retained forcing error. Working with finite matrices avoids assuming that a singular continuum force or the auxiliary matrix $J$ preserves an unbounded-operator domain. The auxiliary box is chosen as a function of the basis cutoff, and the exterior tails control its forcing and boundary errors.

The final terms describing repeated transfers through the same Coulomb interaction require a cancellation before we take absolute values.
An interaction channel is one fixed particle-pair Coulomb term, or the
sum of the clamped-nucleus potentials acting on one particle register.
All fixed centres acting on that register are grouped together.
Each channel transfers a three-dimensional
relative momentum while preserving the spectator momenta and, for a
pair term, the pair's total momentum. We call a sequence of transfers
that leaves the low-momentum region and returns to it an excursion.
For the return term below, $\vec{\bm{b}},\vec{\bm{c}}$ are the low-momentum endpoints and
$\vec{\bm{d}},\vec{\bm{f}}$ are the two intermediate channel momenta. We pair this path
with its reflection
\begin{equation}
 (\vec{\bm{d}},\vec{\bm{f}})\longmapsto(\vec{\bm{b}}+\vec{\bm{c}}-\vec{\bm{d}},\,\vec{\bm{b}}+\vec{\bm{c}}-\vec{\bm{f}}).
\end{equation}
Each interaction channel has Fourier matrix elements that
depend only on the transferred momentum. Together with
Hermitian reversal, this makes the products of three Coulomb
coefficients identical in the two oriented sums, including
the phases from off-grid nuclei.
Subtraction therefore acts on the inverse kinetic denominators. Their difference gains a factor bounded by $C(\|\vec{\bm{b}}\|+\|\vec{\bm{c}}\|)/\|\vec{\bm{f}}-\vec{\bm{b}}\|^3$. If $A(\vec{\bm{c}},\vec{\bm{b}})$ denotes the coefficient obtained by summing these ordered return paths from momentum $\vec{\bm{b}}$ to momentum $\vec{\bm{c}}$, the resulting estimate has the form
\begin{equation}
 |A(\vec{\bm{c}},\vec{\bm{b}})-\overline{A(\vec{\bm{b}},\vec{\bm{c}})}|
 \le CK^{-3}(1+\|\vec{\bm{b}}\|+\|\vec{\bm{c}}\|)\log^q(2+K),
\end{equation}
where $q$ is a fixed finite logarithmic power and the fixed lattice spacing is absorbed into the constant. 
The unmatched reflected paths are those for which only one
orientation satisfies the cutoff restrictions. They lie in
layers near the faces of the simulation or auxiliary momentum
box, of width $\mathcal O(1+\|\vec{\bm{b}}+\vec{\bm{c}}\|)$. The slab estimates count
momentum-transfer lattice points in these layers: one Cartesian
component is restricted to a short interval near a box face,
while the other components range over the momentum shell being
summed. Combining these counts with the decay of the summand
gives the weighted lattice bounds in
Eq.~\eqref{eq:flux-reflection-slabs}. Their contribution is
included in the displayed bound on
$|A(\vec{\bm{c}},\vec{\bm{b}})-\overline{A(\vec{\bm{b}},\vec{\bm{c}})}|$.
Thus it is the anti-Hermitian combination, rather than either oriented kernel separately, that gains the extra inverse power of the cutoff. The weighted estimates for each interaction channel control the factor $1+\|\vec{\bm{b}}\|+\|\vec{\bm{c}}\|$ when this coefficient is paired with the state. No reality assumption on the Coulomb matrix is used.

Appendix~\ref{app:flux} applies the finite-matrix identity above through three successive stages, replacing the remaining time-integrated force contributions by endpoint and commutator terms. These stages bound both the endpoint terms and the remaining integrands by $K^{-3}$ times fixed logarithmic powers and the squared data budget. Time integration supplies the factor $1+t$, and absorbing the logarithms into $K^\varepsilon$ gives the stated flux rate. This argument is independent of the Hamiltonian spectral comparison used above for the full kinetic-energy error.

\subsection{Discretisation error in stopping-power estimates}

For a fixed positive reference speed $v_j$ and a window duration
$\tau>0$, define the signed secant estimator on the exact
state evolution by
\begin{equation}
 \widehat S_\tau=
 \frac{\langle T_j(\tau)\rangle-\langle T_j(0)\rangle}{v_j\tau}.
\end{equation}
The conventional positive stopping power for energy loss has the
opposite sign, with the same absolute discretisation error. For
either Galerkin or truncated-kernel grid evolution, we add the errors
at the two endpoints and divide by $v_j\tau$. Applying
Eq.~\eqref{eq:overview-kinetic} at both $t=0$ and $t=\tau$
gives the claimed dependence on the window duration after absorbing
a factor of two into the constant. Corollary~\ref{cor:stopping-log}
therefore gives
\begin{equation}
 |\Delta\widehat S_\tau|
 \le\frac{C\mathcal D^2}{v_j}K^{-3}
       \left[1+\frac{1+\log(2+K)}{\tau}\right],
 \qquad \tau>0.
 \label{eq:overview-stopping-log}
\end{equation}
The window here starts at the common initial time zero. A shifted
window on the same state evolutions must retain the time dependence of
both endpoint errors. This estimate controls the discretisation error
of the stated finite-window functional; the bias of a secant or fitted
slope as an estimate of instantaneous stopping power is separate.

\section{Non-periodic systems and finite-volume error}
\label{sec:free-overview}

For a non-periodic system we must control two approximations: replacing free space by a finite cell, and representing the resulting cell Hamiltonian in a finite basis. Increasing the momentum cutoff controls only the second error. We therefore first choose a finite-volume interaction that agrees exactly with the free-space Coulomb interaction on a retained region, and then use localisation to compare states supported predominantly in that region. Throughout this comparison, $L$ denotes the cell side length and $K$ the physical momentum cutoff; both may vary. We use the free-space setup of Appendix~\ref{app:free}, with fixed moving-particle masses $m_j\ge1$ in atomic units. This lower mass bound is used below when placing the concrete mass-weighted cutoff inside the retained region.

\subsection{Range-truncated Coulomb interaction in a finite cell}

We obtain exact agreement by truncating the \emph{range} of the Coulomb interaction before periodising it. Fix a transition width $\rho>0$ and a smooth radial profile $0\le\theta_{L,\rho}\le1$ that equals one for $r\le L/2-\rho$ and zero for $r\ge L/2$, with uniformly bounded total variation, meaning that $\int_0^{L/2}|\theta_{L,\rho}'(r)|\,dr$ has a bound independent of $L$. This controls the total change of the radial profile and hence its nonzero Fourier coefficients. We periodise $\theta_{L,\rho}(r)/r$ to define the cell kernel $g_L$. If the clamped nuclei lie within radius $R_0$, the retained product region is
\begin{equation}
 \mathcal B_L:=\{\vec{\bm{x}}:\ \|\bm{x}_j\|<R_L\text{ for every moving particle }j\},
 \qquad R_L:=L/4-R_0-\rho.
\end{equation}
The radius is chosen to keep every interacting separation inside the unmodified part of the kernel. Indeed,
\begin{equation}
 \|\bm{x}_i-\bm{x}_j\|\le2R_L<L/2-\rho,\qquad
 \|\bm{x}_j-\bm{R}_\ell\|\le R_L+R_0<L/2-\rho.
\end{equation}
Thus $g_L$ gives exactly the same pair and nuclear interactions as $1/r$ throughout this product region. Restricting each particle coordinate is what controls all the pair separations simultaneously. This spatial range truncation is separate from the momentum-transfer truncation used to implement the grid Hamiltonian. The latter retains transfers up to $2K$ on a register of cutoff $2K$, as in Section~\ref{sec:dynamics-overview}.

The physical cell Hamiltonian $H_L$ retains the zero Fourier coefficient of $g_L$, which is the constant part of the interaction. In each pair term this coefficient contributes a configuration-independent scalar multiplied by the two charges. Omitting it gives the operator $H_L^{(0)}$ used in intermediate
formal proofs and changes the Hamiltonian by the known scalar in
Eq.~\eqref{eq:free-zero-mode}:
\begin{equation}
 H_L=H_L^{(0)}+C_{\rm zm}(L)I.
\end{equation}
We restore this scalar in both the continuum cell energy and its finite-basis approximation. Otherwise the omitted shift could leave an $\mathcal O(L^{-1})$ contribution to the energy error. For expectation values under time evolution, the shift produces only a common phase and cancels.

Restoring the scalar is therefore essential for the energy comparison below: an uncorrected algebraic shift could dominate the exponentially small localisation error. For dynamics, the same restoration changes neither the measured expectation nor its discretisation error.

\subsection{Ground-state localisation and finite-cell energy error}

We first consider ground energies with clamped nuclei, in a specified scalar exchange sector. Here the wavefunctions are scalar-valued functions of the spatial coordinates, with the prescribed exchange sign for permutations within each species. There is no spin coordinate in this sector. Its ground energy is therefore the minimum in that stated spatial sector; we do not identify it with the full spin-inclusive ground energy. Let $E_\infty$ be the minimum free-space energy in that sector, including both kinetic and Coulomb potential energy, attained by a normalised ground state with the exponential localisation of Eq.~\eqref{eq:free-agmon-data}. Its rate $\alpha>0$ means that the probability mass (the squared wavefunction norm) outside a product region of radius $r$ is $\mathcal O(e^{-2\alpha r})$.

We first approximate the supplied ground state in the energy-form norm by smooth states in the same sector. We take the form norm after adding a fixed positive energy shift that makes the Hamiltonian strictly positive. For the Coulomb Hamiltonian, convergence in this norm is equivalent to convergence of both the state and its first spatial derivatives in $L^2$. This controls both norm and energy, so the smooth near-minimisers are derived from the ground state rather than assumed as extra data. We then choose complementary cutoffs $X,Y$ with $X^2+Y^2=1$, both invariant under the particle permutations defining the sector. Multiplication by either cutoff therefore preserves the exchange signs, which is necessary when we use the same-sector energy lower bound on the complementary piece. The construction in Eq.~\eqref{eq:free-IMS-pair} supplies this invariance together with uniform derivative bounds. The retained cutoff $X$ is supported in $\mathcal B_L$ and equals one a fixed distance inside its boundary.

The IMS localisation identity in Appendix~\ref{app:free} is the tool that controls the energy cost of this cut. For a state vector with position wavefunction $v(\vec{\bm{x}})$, the free-space energy quadratic form is
\begin{equation}
 \mathfrak q(\ket{v}) :=
 \sum_j\frac{1}{2m_j}\int \|\nabla_j v(\vec{\bm{x}})\|^2\,d\vec{\bm{x}}
 +\int V_\infty(\vec{\bm{x}})|v(\vec{\bm{x}})|^2\,d\vec{\bm{x}},
\end{equation}
where $V_\infty$ is the total free-space Coulomb potential and $\vec{\bm{x}}$ denotes the full spatial configuration. This expression equals $\langle v|H_\infty|v\rangle$ on the operator domain but also defines the energy for vectors in the larger form domain, where the displayed integrals are finite. The localisation identity states
\begin{equation}
 \begin{aligned}
 \mathfrak q(\ket{v})&=\mathfrak q(X\ket{v})+\mathfrak q(Y\ket{v})-\mathcal E_{\rm loc}(\ket{v}),\\
 \mathcal E_{\rm loc}(\ket{v})&=
 \sum_j\frac1{2m_j}\int
   \bigl(\|\nabla_jX\|^2+\|\nabla_jY\|^2\bigr)|v(\vec{\bm{x}})|^2\,d\vec{\bm{x}}.
 \end{aligned}
\end{equation}
Here $v(\vec{\bm{x}}):=\langle \vec{\bm{x}}|v\rangle$ is the position amplitude, with
$\vec{\bm{x}}$ denoting the full spatial configuration.
The potential splits exactly because the squared cutoffs sum to one. The extra term comes from differentiating them in the kinetic energy. Their gradients are uniformly bounded and supported in the outer transition layer. If a normalised trial state has mass at most $\tau<1$ in that outer region, then $\mathcal E_{\rm loc}(\ket{v})\le C\tau$ and $\|X\ket{v}\|^2\ge1-\tau$.

We can now see why the cut state remains a good trial state. If $\mathfrak q(\ket{v})\le E_\infty+\delta_{\rm trial}$, the lower bound $\mathfrak q(Y\ket{v})\ge E_\infty\|Y\ket{v}\|^2$ gives
\begin{equation}
 \mathfrak q(X\ket{v})
 \le E_\infty(1-\|Y\ket{v}\|^2)+\delta_{\rm trial}+C\tau
 =E_\infty\|X\ket{v}\|^2+\delta_{\rm trial}+C\tau.
\end{equation}
Dividing by the retained norm shows that the normalised retained piece has energy at most
\begin{equation}
 E_\infty+\frac{\delta_{\rm trial}+C\tau}{1-\tau}.
\end{equation}
The denominator accounts for its retained norm. We take $\delta_{\rm trial}\downarrow0$ and use the free-state localisation. If $w_0$ is the fixed width of the cutoff layer, the relevant tail begins at $R_L-w_0$. Since $R_L=L/4-R_0-\rho$, its mass satisfies
\begin{equation}
 \tau=\mathcal O(e^{-2\alpha(R_L-w_0)})
      =\mathcal O(e^{-\alpha L/2}),
\end{equation}
with the fixed offsets absorbed into the constant. The retained state can be placed in the cell without changing its norm or Coulomb interaction. The variational principle then gives an upper comparison for the cell ground energy $E_L$,
\begin{equation}
 E_L\le E_\infty+C_+e^{-\alpha L/2}.
\end{equation}
Thus localisation of the free-space ground state supplies a good cell trial state; it does not by itself exclude a lower-energy cell state elsewhere.

To obtain the reverse comparison, Proposition~\ref{prop:free} assumes a normalised ground eigenvector of the range-truncated cell Hamiltonian, with ordinary second-order Sobolev regularity, for each sufficiently large $L$, in the same scalar sector.
The finite-cell localisation assumption in
Proposition~\ref{prop:free} controls its localisation
relative to the smaller product region $\mathcal B_{L/2}$,
uniformly in $L$.
The distance from that reference region to the outer cutoff layer is $R_L-w_0-R_{L/2}=L/8-w_0$. The localisation rate $\beta>0$ in this assumption therefore gives an outer mass $\mathcal O(e^{-2\beta(L/8-w_0)})=\mathcal O(e^{-\beta L/4})$. Applying the same localisation argument to this cell state, and placing the retained piece back in free space, gives $E_\infty\le E_L+C_-e^{-\beta L/4}$. The two directions use different supplied localisation data.

We have now compared the continuum energies in both directions. The slower of the two exponential rates controls their absolute difference. Neither direction uses a claim that free-space localisation automatically forces localisation of a cell ground state; the latter is the separate content of the finite-cell localisation assumption.

We then add the grid error $|E_{N,L}-E_L|\le CK^{-3}$. Here $E_{N,L}$ is the lowest eigenvalue of the physical grid Hamiltonian in the same scalar exchange sector, at cell size $L$ and one-particle basis size $N$. The physical-unit estimates in Appendix~\ref{app:free} make its constant and momentum threshold independent of $L$ for this family. A fixed-cell estimate alone would not suffice when the cell grows. This uniformity requires more than rescaling the Coulomb coefficients: the Sobolev slice bounds, relative operator bounds, grid regularity and observable moments must all remain controlled. For example, the physical Sobolev sum obeys
\begin{equation}
 L^{-3}\sum_{\bm{\nu}\in\mathbb Z^3}(1+\|2\pi\bm{\nu}/L\|^2)^{-2}=\mathcal O(1).
\end{equation}
The increasing number of modes in a fixed physical momentum region is cancelled by the volume normalisation $L^{-3}$. This is one of the estimates that allows the fixed-cell argument to be used uniformly along the growing-cell family. Combining the two comparisons gives the bound of Proposition~\ref{prop:free},
\begin{equation}
 |E_{N,L}-E_\infty|\le CK^{-3}+C'e^{-\gamma L},\qquad
 \gamma=\tfrac12\min\{\alpha,\beta/2\}.
\end{equation}
For the grid, $K=2\pi m/L$ is the effective physical cutoff and the actual register extends to $2K$. Its spatial mode count per particle is $N=(4m+1)^3$. Thus
\begin{equation}
 m=\frac{LK}{2\pi},\qquad N=\Theta((LK)^3)
\end{equation}
in the admissible large-cutoff regime. Increasing the cell while keeping $N$ fixed would reduce the physical momentum resolution. To control both errors, we instead choose $L=\mathcal O(\log(1/\epsilon))$ and a sufficient $K=\mathcal O(\epsilon^{-1/3})$, above the fixed thresholds, so each contribution is at most a chosen fraction of $\epsilon$. Substituting these choices into $(LK)^3$, rounding $m$ upwards and enforcing its admissible minimum gives
\begin{equation}
 N=\mathcal O\!\left(\epsilon^{-1}\log^3(1/\epsilon)\right).
\end{equation}
This is a sufficient basis size under the stated hypotheses.
The formal proof constructs the smooth approximation and comparisons;
it does not establish free-space localisation, the finite-cell
localisation assumption, or the supplied range-truncated cell
ground eigenvectors.

\subsection{Spatial localisation and grid approximation of free-space dynamics}

For time evolution we instead compare a supplied strong free-space solution $\ket{\psi_\infty(t)}$ with a cell state and then a grid state. Here ``strong'' means that the time-dependent state has an $L^2$ time derivative and obeys the Schr\"odinger equation $i\partial_t\ket{\psi_\infty}=H_\infty\ket{\psi_\infty}$ as an $L^2$ equality. In the weak formulation of Proposition~\ref{prop:free-dyn}, the time-dependent state, its square-integrable Hamiltonian image and its weak spatial gradient are supplied data. The weak gradient is defined by integration by parts against smooth compactly supported test functions, so its use does not require a pointwise differentiable wavefunction at particle collisions. This gives the equation that we differentiate after cutting off the state; we are not constructing a free-space propagator at this step. We assume $\ket{\psi_0}\in D(H_\infty^2)$ and measure a fixed smooth observable $f(\bm{x}_j)$, where $f\in C_c^\infty(\mathbb R^3)$ and $0\le f\le1$. Thus $f$ is infinitely differentiable and vanishes outside a fixed bounded region of the observed particle's position space. To place the state in the cell, choose a real $C^2$ cutoff $0\le X_L\le1$ compactly supported inside the retained region and preserving any imposed exchange sector. It equals one on an inner region, and its first and second derivatives have uniform bounds and vanish outside a transition shell. Let $\mathcal T_L$ denote the isometric placement of a supported function in the centred cell, and set
\begin{equation}
 \ket{\varphi(t)}:=\mathcal T_L(X_L\ket{\psi_\infty(t)}),\qquad
 \ket{\varphi_{0,L}}:=\mathcal T_L(X_L\ket{\psi_0}).
\end{equation}
The spatially truncated state is not an exact solution for $H_L$, because the kinetic operator differentiates $X_L$. Since the potentials agree on its support, the additional term in its Schr\"odinger equation comes entirely from differentiating the cutoff:
\begin{equation}
 \begin{aligned}
 i\partial_t\ket{\varphi(t)}
 &=H_L\ket{\varphi(t)}+\mathcal T_L([X_L,T]\ket{\psi_\infty(t)}),\\
 [X_L,T]\ket{\psi}
 &=\sum_j\frac1{m_j}\left(\nabla_jX_L\cdot\nabla_j\ket{\psi}
                         +\tfrac12\Delta_jX_L\,\ket{\psi}\right).
 \end{aligned}
\end{equation}
The two terms in the cutoff commutator explain the two kinds of leakage we must control. The term containing $\Delta_jX_L$ needs the wavefunction to be small where the cutoff changes; the term containing $\nabla_jX_L$ also needs its gradient to be small there.
A small wavefunction norm on the cutoff transition shell
controls the term containing $\Delta_jX_L$. The term containing
$\nabla_jX_L$ also requires a small gradient norm on that shell.
We must also control the spatial gradient of the exact state where the cutoff changes. The leakage hypothesis of Proposition~\ref{prop:free-dyn} bounds both the wavefunction norm outside the inner region and the gradient norm on the transition shell by $\epsilon_{\rm leak}\le1$, at every time in $[0,t]$. It therefore bounds the norm of this additional term by $C\epsilon_{\rm leak}$, and Duhamel's formula gives
\begin{equation}
 \|\ket{\varphi(t)}-e^{-iH_Lt}\ket{\varphi_{0,L}}\|
 \le Ct\epsilon_{\rm leak}.
\end{equation}
To pass from the state-norm difference to an expectation, we use
\begin{equation}
 |\langle u|O|u\rangle-\langle v|O|v\rangle|
 \le\|O\|(\|\ket{u}\|+\|\ket{v}\|)\|\ket{u}-\ket{v}\|.
\end{equation}
The norms of the cut and cell states are bounded by the initial data, so this gives an expectation error at most $Ct\epsilon_{\rm leak}$. Cutting the exact state introduces one more contribution. Since $f$ and $X_L$ are commuting multiplication operators,
\begin{equation}
 \begin{aligned}
 \langle\psi_\infty|f|\psi_\infty\rangle
 -\langle\psi_\infty|X_LfX_L|\psi_\infty\rangle
 &=\langle\psi_\infty|f(1-X_L^2)|\psi_\infty\rangle\\
 &\le\|\mathbf1_{\mathrm{inner}^c}\ket{\psi_\infty}\|^2
 \le\epsilon_{\rm leak}^2.
 \end{aligned}
\end{equation}
Here all states are evaluated at the same time. This estimate uses the full configuration-space leakage. The support of the one-particle observable restricts the observed particle, but places no restriction on the other particles; we cannot replace that leakage assumption by localisation of the measured coordinate alone.

To apply the grid estimate of Section~\ref{sec:dynamics-overview} to the cell state, we need the cut initial state to satisfy
\begin{equation}
 \ket{\varphi_{0,L}}\in D(H_L^2),\qquad
 \sup_{L\ge L_0}\sum_{j=0}^2\|H_L^j\ket{\varphi_{0,L}}\|<\infty,
\end{equation}
where $L_0>0$ is a fixed lower threshold for the cell side length, chosen sufficiently large that the spatial-cutoff construction and the geometric conditions of Proposition~\ref{prop:free-dyn} hold for every $L\ge L_0$.
This is the cutoff-domain hypothesis (H$_{\rm dom}$). It controls not only the cut state but also its first and second cell-Hamiltonian images, with one bound for the whole family of cells. Membership in $D(H_L^2)$ separately for each $L$ would not supply the uniform dynamical constant we need. The physical-unit operator estimates and Fourier moments of the fixed observable $f$ are uniform in $L$, so this graph bound makes the torus-to-grid error uniform in $L$ as well. We use the grid Hamiltonian constructed from $g_L$ and set $\ket{\psi_g(t)}:=e^{-iH_{\rm grid}t}P'\ket{\varphi_{0,L}}$, without renormalising either the cutoff or the projection. With $O_g$ the diagonal grid observable whose entries are $f$ at the physical grid points, combining the two comparisons gives
\begin{equation}
 \begin{split}
 &\big|\langle\psi_\infty(t)|f(\bm{x}_j)|\psi_\infty(t)\rangle
       -\langle\psi_g(t)|O_g|\psi_g(t)\rangle\big|
\le C(K^{-5}+tK^{-3})+C(1+t)\epsilon_{\rm leak},
 \qquad t\ge0.
 \end{split}
\end{equation}
Here the cell and momentum cutoffs exceed the fixed thresholds in Proposition~\ref{prop:free-dyn}, and $C$ depends on the stated system, observable, cutoff and graph-norm data, but not on $L$, $K$, $t$ or $\epsilon_{\rm leak}$.

\subsection{Kinetic energy and stopping power for a non-periodic system}
\label{sec:free-kinetic-overview}

The same spatial localisation also lets us compare a particle's kinetic
energy with its value on the non-periodic-system grid. The bounded-observable
inequality used above cannot be applied with a cutoff-independent norm
for $T_j$. We instead use the kinetic pairing from
Section~\ref{sec:kinetic-overview}:
\begin{equation}
 |\langle u|T_j|u\rangle-\langle v|T_j|v\rangle|
 \le\|\ket{u}-\ket{v}\|(\|T_j\ket{u}\|+\|T_j\ket{v}\|).
\end{equation}
The important distinction is that the two kinetic norms may remain
bounded even when no small kinetic norm has been proved for $\ket{u}-\ket{v}$.
This is precisely the information supplied by the global Hamiltonian
graph bounds and the existing norm-level leakage comparison.

To apply the pairing, let
$\mathcal D_\infty=\|\ket{\psi_0}\|+\|H_\infty\ket{\psi_0}\|
+\|H_\infty^2\ket{\psi_0}\|$, and let $D_*$ bound the three cell
graph norms in (H$_{\rm dom}$).
Hamiltonian graph conservation bounds
$\|T_j\ket{\psi_\infty(t)}\|$ by $C\mathcal D_\infty$.
The product rule
\begin{equation}
 T_j(X_L\ket{u})=X_LT_j\ket{u}-\frac1{m_j}\nabla_jX_L\cdot\nabla_j\ket{u}
                    -\frac1{2m_j}(\Delta_jX_L)\ket{u}
\end{equation}
gives the same bound for the cut state, because the cutoff's first
and second derivatives are uniformly bounded. Its norm difference
from the free state is at most $\epsilon_{\rm leak}$, so cutting
off the kinetic expectation costs at most
$C\mathcal D_\infty\epsilon_{\rm leak}$.
The Duhamel norm error between the spatially truncated state placed in the cell and the cell
state is $Ct\epsilon_{\rm leak}$, as above. Their kinetic norms
are bounded by $C\mathcal D_\infty$ and $CD_*$, respectively, so this
second comparison costs
$C(\mathcal D_\infty+D_*)t\epsilon_{\rm leak}$.
Thus the same wavefunction and shell-gradient leakage conditions
suffice; no additional assumption on derivatives in the whole exterior
region is needed.

For the cell-to-grid step, we apply the spectral argument of
Theorem~\ref{thm:kinetic-log} in physical units. Its constants remain
uniform in $L$ for the range-truncated reference interaction. In particular,
the high-transfer coefficient norm contributes
$L^{-3/2}K^{-1/2}$, while the three-dimensional slice estimate for
state coefficients contributes $L^{3/2}$; these volume factors cancel.
The number of transfer indices that alias to a given retained index, and the constants comparing the kinetic and Hamiltonian graph norms, are also bounded uniformly in $L$.
Finally, the number of dyadic energy bands is set by the physical
ceiling $O(K^2)$, rather than by the number of modes in the cell.
Proposition~\ref{prop:free-kinetic-log} verifies these points and gives
\begin{align}
 &\left|\langle\psi_\infty(t)|T_j|\psi_\infty(t)\rangle
       -\langle\psi_g(t)|T_j|\psi_g(t)\rangle\right|
\le CD_*^2K^{-3}[1+t+\log(2+K)]
       +C(\mathcal D_\infty+D_*)(1+t)\epsilon_{\rm leak}.
 \label{eq:overview-free-kinetic}
\end{align}
This holds for all $t\ge0$ for which leakage is controlled throughout
$[0,t]$, with the same unnormalised spatial cutoff and projected
initial state. 
Here the measured grid kinetic-energy operator is diagonal in
the retained momentum basis, with eigenvalue
$\|\bm{k}_j\|^2/(2m_j)$ on a basis state whose particle-$j$
momentum is $\bm{k}_j$. Thus we use the exact kinetic-energy
eigenvalue for each retained mode. There is no additional
observable-sampling error.

Since $N=\Theta((LK)^3)$, the basis term in
Eq.~\eqref{eq:overview-free-kinetic} is
\begin{equation}
 \mathcal O\!\left(\frac{L^3}{N}
       \left[1+t+\log\left(2+\frac{N^{1/3}}L\right)\right]\right).
\end{equation}
It is $\mathcal O(\log N/N)$ at fixed $L,t$, while leakage remains a
separate error. When choosing a growing cell, we must retain the
factor $L^3$ and supply a leakage estimate for that choice; initial
$D(H_\infty^2)$ regularity alone does not determine the required cell
size.

For a signed stopping-power secant over $[0,\tau]$, with $\tau>0$ and
fixed reference speed $v_j$, Corollary~\ref{cor:free-stopping-log}
divides the two endpoint errors by $v_j\tau$:
\begin{align}
 |\Delta\widehat S_\tau|
 &\le\frac{CD_*^2}{v_j}K^{-3}
           \left[1+\frac{1+\log(2+K)}{\tau}\right]
+\frac{C(\mathcal D_\infty+D_*)}{v_j}
                       (1+\tau^{-1})\epsilon_{\rm leak}.
 \label{eq:overview-free-stopping}
\end{align}
This is an error bound for the finite-window functional, separate
from its bias as a model of instantaneous stopping power.
Both comparisons retain the same uniform cutoff-domain condition
(H$_{\rm dom}$); the kinetic-energy estimate has introduced no
additional condition on the cutoff.

\subsection{Spatial cutoffs preserving the squared Hamiltonian domain}

We next explain how the specified cutoffs supply (H$_{\rm dom}$) from the initial assumption $\ket{\psi_0}\in D(H_\infty^2)$. This step is necessary because an arbitrary smooth multiplier need not preserve the squared domain of a Coulomb Hamiltonian. The first commutator differentiates the cutoff, as in the spatially truncated evolution equation above. Commuting a second time can also differentiate the Coulomb potential. To identify the problem, put $\ket{u}=\ket{\psi_0}$ and $B_L=[T,X_L]$, with the opposite sign to the evolution commutator $[X_L,T]$. The formal second-image identity to be justified is
\begin{equation}
 H_\infty^2(X_L\ket{u})
 =X_LH_\infty^2\ket{u}+2B_LH_\infty \ket{u}
       +[T,B_L]\ket{u}+[V_\infty,B_L]\ket{u},
\end{equation}
where $V_\infty$ is the total free-space potential. The differential operator $B_L$ is first order, and $[T,B_L]$ is at most second order. The potentially more singular term is $[V_\infty,B_L]$. For a quantum-particle pair it contains
\begin{equation}
 [V_{ij},B_L]
 =\left(\frac{\nabla_iX_L}{m_i}-\frac{\nabla_jX_L}{m_j}\right)
       \cdot\nabla_iV_{ij}.
\end{equation}
Here $V_{ij}=q_iq_j/\|\bm{x}_i-\bm{x}_j\|$ is the free-space potential for the indicated quantum-particle pair, acting by multiplication. The derivative of the Coulomb potential has inverse-square magnitude. We therefore design the mass-weighted gradient difference to vanish linearly when $\bm{x}_i=\bm{x}_j$. This recovers one power of the separation, leaving an inverse-distance singularity, of the same order as the Coulomb potential, that the available Hardy estimates can control.

For fully quantum electrons and nuclei, we choose
\begin{equation}
 Q_m(\vec{\bm{x}}):=\sum_j m_j\|\bm{x}_j\|^2,\qquad
 X_L(\vec{\bm{x}}):=F\!\left(\frac{Q_m(\vec{\bm{x}})}{(L/8)^2}\right),
\end{equation}
where $F$ is a fixed smooth function with values in $[0,1]$, equal to one for arguments at most one and zero for arguments at least two. On its support, $Q_m\le2(L/8)^2$. Since $m_j\ge1$, this implies
\begin{equation}
 \|\bm{x}_j\|\le\sqrt2\,L/8<R_L
\end{equation}
for every moving particle and all sufficiently large $L$. Thus this particular cutoff fits inside the retained product region. The mass weights are also chosen to match the inverse masses in the kinetic operator. Indeed,
\begin{equation}
 \frac{\nabla_iX_L}{m_i}-\frac{\nabla_jX_L}{m_j}
 =\frac{2}{(L/8)^2}F'\!\left(\frac{Q_m}{(L/8)^2}\right)(\bm{x}_i-\bm{x}_j).
\end{equation}
Substituting this gradient difference into the pair commutator gives the exact cancellation
\begin{equation}
 [V_{ij},B_L]
 =-\frac{2q_iq_jF'(Q_m/(L/8)^2)}{(L/8)^2\|\bm{x}_i-\bm{x}_j\|}.
\end{equation}
Only the original inverse-distance Coulomb singularity remains. The regularity bridge in Lemma~\ref{lem:free-regularity-bridge} supplies second-order Sobolev regularity for $\ket{u}$ and $H_\infty \ket{u}$, together with bounds on the inverse-distance weighted state norms that appear in Hardy's inequality, such as $\|\ket{u}/\|\bm{x}_i-\bm{x}_j\|\|$. The second-image expression uses at most two ordinary derivatives of $\ket{u}$ and one of $H_\infty \ket{u}$. Derivatives through order four fall on the smooth cutoff, not on the initial state. This explains concretely why the argument needs $D(H_\infty^2)$ but no ordinary $H^4$ assumption.

A square-integrable formal expression alone would not prove domain membership. Lemma~\ref{lem:free-nbo-domain} establishes the weak identities using regularised Coulomb kernels, proves that $B_L\ket{u}$ lies in the Hamiltonian domain, and transplants the cut state and its first Hamiltonian image successively to the cell. These steps justify the displayed second image and give the uniform squared-domain bound. Corollary~\ref{cor:free-dyn-nbo} consequently gives the dynamical estimate without an independent cutoff-domain assumption. The mass-weighted region restricts centre-of-mass position as well as internal separation, so the comparison keeps the full centre-of-mass motion; it does not require that motion to be removed.

With clamped nuclei, the cutoff must also avoid creating a singularity when a moving particle approaches a fixed centre. For electrons with a common mass $m_e$, we use
\begin{equation}
 X_L^{\rm BO}(\vec{\bm{x}}):=\prod_jq(16\bm{x}_j/L),
\end{equation}
where the map $q:\mathbb R^3\to[0,1]$ equals one on the unit ball and vanishes outside the ball of radius two. At $\bm{x}_i=\bm{x}_j$, the common bump and equal masses make the two mass-weighted gradients coincide. Their difference therefore vanishes linearly with the separation: the mean-value estimate in Eq.~\eqref{eq:free-bo-cancellation} gives
\begin{equation}
 \left\|\frac{\nabla_iX_L^{\rm BO}}{m_e}
       -\frac{\nabla_jX_L^{\rm BO}}{m_e}\right\|
 \le CL^{-2}\|\bm{x}_i-\bm{x}_j\|.
\end{equation}
This again reduces the pair commutator to an inverse-distance singularity. For sufficiently large $L$, the derivative in each electron coordinate is also zero near every nucleus, regardless of the positions of the other electrons. The nuclear term in that commutator is proportional to
\begin{equation}
 -\frac{q_j\zeta_\ell}{m_j}
 \frac{\nabla_jX_L\cdot(\bm{x}_j-\bm{R}_\ell)}{\|\bm{x}_j-\bm{R}_\ell\|^3}.
\end{equation}
Coordinate-wise flatness makes it zero near the singularity. Away from a fixed neighbourhood of the nucleus, the inverse-square factor is bounded, so the remaining term is harmless. The product cutoff thus has two separate jobs: its common bump and equal masses cancel the additional singularity at particle coincidences, while flatness in each coordinate removes the fixed-centre singularities. Corollary~\ref{cor:free-dyn-bo} therefore supplies the same uniform graph bound. These two constructions require no ordinary $H^4$ Sobolev regularity. For arbitrary admissible cutoffs, however, the general result retains (H$_{\rm dom}$); domain preservation is proved for the specified compatible constructions, not for every smooth cutoff.

\subsection{Extension to spin-dependent states}

Finally, Theorem~\ref{thm:free-dyn-spin} extends the smooth-observable comparison to finite internal spin. Proposition~\ref{prop:free-kinetic-log} gives the kinetic-energy comparison in the same Hilbert direct-sum norm. We use the Hilbert-space data norm $\|\ket{\Psi_0}\|+\|H_\infty\ket{\Psi_0}\|+\|H_\infty^2\ket{\Psi_0}\|$ and bound the sums of squared component leakage norms at each time. Writing $\sigma$ for a spin configuration and $\ket{\Psi_\sigma}$
for its spatial component, the direct-sum norm is $\|\ket{\Psi}\|^2=\sum_\sigma\|\ket{\Psi_\sigma}\|^2$. Since the spatial operators act identically on each component, the same operator estimates hold in this norm. The quadratic data budgets also combine without a component count: the triangle inequality in the space of spin-indexed norm lists gives
\begin{equation}
 \sum_\sigma\left(\sum_{j=0}^2\|H_\infty^j\ket{\Psi_{0,\sigma}}\|\right)^2
 \le\left(\sum_{j=0}^2\|H_\infty^j\ket{\Psi_0}\|\right)^2.
\end{equation}
We sum the squared leakage norms at each time before applying the time integral. Thus the assumption controls the whole spin state at that time; it need not bound a sum of separately chosen componentwise time maxima. These two Hilbert-space estimates yield the same dynamical constant without a factor depending on the number of spin configurations. No antisymmetry is required for this extension, so it also applies to states in the physical position--spin exchange sectors. This extension concerns dynamics; the ground-energy comparison above remains in its specified scalar exchange sector. The exact free-space solution and its leakage remain inputs, and the comparison does not construct the canonical free-space propagator. In particular, membership in $D(H_\infty^2)$ does not establish small leakage, and measured boundary occupancy of the simulated state does not certify the required leakage of the exact state.

\section{Conclusion}
\label{sec:conclusion}

In this work we have established rigorous basis-set error bounds
for first-quantised quantum simulations of many-particle Coulomb
systems. For a fixed periodic cell, the ground-state energy error
is $\mathcal O(N^{-1})$ for both plane-wave Galerkin discretisation
and spatial-grid discretisation with the stated interaction
truncation, where $N$ counts spatial basis functions per particle.
The results apply with either clamped or quantum nuclei and
justify the inverse-basis-size scaling commonly assumed in
quantum resource estimates.

We have also bounded errors in observables after time evolution,
which do not follow directly from energy convergence. For initial
states in $D(H^2)$, the error in a smooth one-particle position
observable is $\mathcal O(N^{-5/3}+t/N)$, while the error in an
individual particle's kinetic energy is
$\mathcal O((t+\log N)/N)$. Both rates hold for Galerkin and
truncated-kernel grid evolution for all $t\ge0$, with the stated
regularity data fixed. At fixed time, the kinetic-energy result
therefore gives inverse-basis-size convergence up to a logarithm.

For isolated systems, the range-truncated grid comparisons
separate basis discretisation from the error caused by restricting
the system to a finite cell. The ground-energy result gives a
sufficient basis size
$\mathcal O(\epsilon^{-1}\log^3(1/\epsilon))$
under the stated localisation assumptions, for clamped nuclei in
the specified scalar exchange sector. The dynamical results retain
the stated cutoff-domain condition and leakage assumptions on the
exact solution, with constants uniform in cell size when expressed
in terms of the physical momentum cutoff. The kinetic estimates
also control discretisation error in finite-window stopping-power
calculations.

These bounds provide the spatial-error estimates needed to connect
basis size with the accuracy of energy and dynamical calculations.
Further work could determine whether the logarithm in the kinetic
bound is necessary, sharpen the dependence on particle number,
charges and masses, and establish the localisation and leakage
estimates for specific physical systems. Combining these results
with bounds on state preparation, finite-Hamiltonian simulation,
arithmetic and measurement would give complete accuracy guarantees
for quantum chemistry algorithms.

\begin{acknowledgments}
DWB worked on this project under a sponsored research agreement with Google Quantum AI.
This project is supported by Australian Research Council Discovery Project DP260102543.
Derivations in this work were developed with Gemini 3, Fable 5, and GPT-6 Astra.
Lean code was developed with Opus 5.5.
\end{acknowledgments}

\bibliographystyle{apsrev4-2}

\appendix

\section{Periodic Born--Oppenheimer ground-state energy}
\label{app:bo}
\label{sec:setting}

We prove that restricting each electron to $N$ spatial plane waves changes
the periodic Born--Oppenheimer ground-state energy by at most
$\mathcal O(N^{-1})$. The nuclei remain fixed throughout this appendix.
Our starting point is the exact ground state: we project it into the
finite basis and use the projected state as a variational trial state.
The eigenvalue equation cancels the terms linear in the discarded
component, leaving an energy error controlled by its squared $H^1$ norm.
Here the $H^1$ norm measures both the wavefunction norm and its first
spatial derivatives, so it controls the kinetic energy of the discarded
component as well as its probability weight.

The main task is therefore to estimate the high-momentum part of the
ground state with this derivative weight. The standard $H^2$ regularity
of Coulomb eigenfunctions does not by itself give the desired rate.
We obtain the additional decay by first estimating the high-momentum
part of the potential applied to the state, and then using the eigenvalue
equation to convert that estimate into bounds on the state itself.
We keep the spin variables throughout, since imposing antisymmetry on
spatial coordinates alone would describe only the fully spin-polarised
problem.

\subsection{Hamiltonian and Fourier projection}

We work in atomic units on the periodic cell
$\T:=\R^3/L\Z^3$, of side length $L=\Omega^{1/3}$ and volume $\Omega$.
The positions of the $\eta$ electrons are $\bm{x}_j\in\T$, with
$j\in\{1,\ldots,\eta\}$. The $L_{\mathrm{nuc}}$ clamped nuclei have
positions $\bm{R}_\ell\in\T$ and positive charges $\zeta_\ell$, with
$\ell\in\{1,\ldots,L_{\mathrm{nuc}}\}$; we write
$\lambda_\zeta:=\sum_{\ell=1}^{L_{\mathrm{nuc}}}\zeta_\ell$ for their
total charge. Positivity lets us use this total when summing the
magnitudes of the nuclear contributions.

We first specify the Fourier convention, so that the finite matrix and
the momentum transfers used in the proof have consistent signs.
Following Ref.~\cite{SuBerryWiebeRubinBabbush2021}, Eqs.~(5)--(6), the
orthonormal one-electron spatial basis is indexed by $\bm{p}\in\Z^3$:
\begin{equation}
\phi_{\bm{p}}(\bm{x}):=\langle \bm{x}|\bm{p}\rangle=\frac{1}{\sqrt{\Omega}}e^{-i\bm{k}_{\bm{p}}\cdot \bm{x}},
\qquad \bm{k}_{\bm{p}}=\frac{2\pi \bm{p}}{\Omega^{1/3}}.
\label{eq:pw}
\end{equation}
For a multiplication operator we instead write its Fourier expansion as
$f(\bm{x})=\sum_{\bm{\nu}\in\Z^3}\hat f(\bm{\nu})e^{i\bm{k}_{\bm{\nu}}\cdot \bm{x}}$. Consequently,
\begin{equation}
 e^{i\bm{k}_{\bm{\nu}}\cdot \bm{x}}\phi_{\bm{q}}(\bm{x})=\phi_{\bm{q}-\bm{\nu}}(\bm{x}),\qquad
 \langle \bm{p}|f|\bm{q}\rangle=\hat f(\bm{q}-\bm{p}).
 \label{eq:fourier-sign}
\end{equation}
Thus an input index $\bm{q}$ and output index $\bm{p}=\bm{q}+\bm{\tau}$ differ by a signed
transfer $\bm{\tau}$, whose coefficient is $\hat f(-\bm{\tau})$. The Coulomb
kernel below is even, so its coefficients obey
$\hat w(-\bm{\tau})=\hat w(\bm{\tau})$ and we may reverse this transfer sign
without changing the coefficient. For general multipliers we retain the
sign in Eq.~\eqref{eq:fourier-sign}, including any complex phase.

We retain a cube of one-electron Fourier indices,
\begin{equation}
\bm{p} \in G := \left[-\frac{N^{1/3}-1}{2},\,\frac{N^{1/3}-1}{2}\right]^3
\cap\Z^3.
\label{eq:G}
\end{equation}
For this cube to contain exactly $N$ points, we take $N=n^3$ with
$n$ an odd positive integer. We call these sizes admissible and take
all limits in $N$ along this sequence. The sizes $n=2^{n_p}-1$ used
in Ref.~\cite{SuBerryWiebeRubinBabbush2021} satisfy this condition.
Here $N$ counts spatial plane waves for one electron, rather than
many-electron configurations or spin orbitals.

The periodic interaction is specified by removing the zero Fourier
mode from the Coulomb kernel:
\begin{equation}
w(\bm{x}):=\frac{4\pi}{\Omega}\sum_{\bm{\nu}\in\Z^3\setminus\{0\}}
\frac{e^{i\bm{k}_{\bm{\nu}}\cdot \bm{x}}}{\norm{\bm{k}_{\bm{\nu}}}^2}.
\label{eq:w}
\end{equation}
This convention fixes the mean of $w$ to zero. We interpret the series
in $L^2(\T)$, as justified in Proposition~\ref{prop:F1}; it retains
the usual $1/\|\bm{x}\|$ singularity near the origin. On
$L^2(\T^\eta;\mathcal V_\eta)$, where
$\mathcal V_\eta:=(\mathbb C^2)^{\otimes\eta}$ is the $2^\eta$-dimensional
electronic spin space, the Hamiltonian acts identically on each spin
component:
\begin{equation}
\HBO:=-\frac12\sum_{j=1}^{\eta}\Delta_j
-\sum_{j=1}^{\eta}\sum_{\ell=1}^{L_{\mathrm{nuc}}}
 \zeta_\ell w(\bm{x}_j-\bm{R}_\ell)
+\sum_{1\le i<j\le\eta}w(\bm{x}_i-\bm{x}_j).
\label{eq:Hper}
\end{equation}
The three terms are the electronic kinetic energy $T$, the
 electron--nucleus attraction $U$, and the electron--electron repulsion
$V$, respectively. We write $V_{\mathrm{tot}}:=U+V$. The nuclear
repulsion $C(\vec{\bm{R}})$ in Eq.~(8) of
Ref.~\cite{SuBerryWiebeRubinBabbush2021} is constant for fixed nuclear
positions and has been absorbed into the energy convention. Adding
it back would shift the continuum and finite-basis energies by the
same amount, leaving their difference unchanged.

To identify the finite matrix used in the error bound, let
$\vec{\bm{p}}=(\bm{p}_1,\ldots,\bm{p}_\eta)\in(\Z^3)^\eta$ and form the product basis
$\ket{\vec{\bm{p}}}:=\bigotimes_{j=1}^\eta\ket{\bm{p}_j}$. Its position amplitude is
$\Phi_{\vec{\bm{p}}}(\bm{x}_1,\ldots,\bm{x}_\eta):=\langle \bm{x}_1,\ldots,\bm{x}_\eta|\vec{\bm{p}}\rangle
=\prod_{j=1}^\eta\phi_{\bm{p}_j}(\bm{x}_j)$.
We project every spatial coordinate onto $G$ and leave spin unchanged:
\begin{equation}
P=P_N:=P_N^{\mathrm{sp}}\otimes I_{\mathcal V_\eta},\qquad
\mathrm{ran}\,P_N^{\mathrm{sp}}
=\mathrm{span}\{\ket{\vec{\bm{p}}}:\bm{p}_j\in G\text{ for every }j\}.
\label{eq:P}
\end{equation}
Here $\mathrm{ran}\,P$ denotes the image of the orthogonal projector,
namely the set of states retained in the finite basis. Before imposing
fermionic antisymmetry, its dimension is $N^\eta2^\eta$.
The restriction of $P\HBO P$ to this image is the Galerkin, or
Rayleigh--Ritz, matrix: its entries are the exact continuum matrix
elements between retained basis states.

We can see the agreement with Eqs.~(10), (11), and (13) of
Ref.~\cite{SuBerryWiebeRubinBabbush2021} directly from
Eq.~\eqref{eq:fourier-sign}. The one-electron kinetic entries are
$\norm{\bm{k}_{\bm{p}}}^2/2$. A nuclear term transfers one momentum index, while
a pair term transfers the two indices in opposite directions. The
zero transfer is absent because Eq.~\eqref{eq:w} omits $\bm{\nu}=0$.
Finally, both the input and output indices of a projected matrix
element must lie in $G$. The restrictions $\bm{p}+\bm{\nu}\in G$ and
$\bm{q}-\bm{\nu}\in G$ in the cited formulas are precisely these output
conditions, after the permitted relabelling of the even Coulomb
kernel's transfer index. Thus this is a projection of the continuum
Hamiltonian, with the Fourier phases fixed as above.

The physical ground energy is obtained after imposing electronic
antisymmetry. Since the same spatial projector is applied to each
electron, $P$ commutes with simultaneous permutations of positions
and spins, and preserves
\begin{equation}
 \mathcal H_-:=\Lambda^\eta\big(L^2(\T)\otimes\mathbb C^2\big).
 \label{eq:spin-sector}
\end{equation}
The exterior-power notation $\Lambda^\eta$ specifies states that
change sign when two complete one-electron variables are exchanged.
Writing $\boldsymbol\sigma=(\sigma_1,\ldots,\sigma_\eta)
\in\{\uparrow,\downarrow\}^\eta$, these variables are the pairs
$(\bm{x}_j,\sigma_j)$. A single spatial component with its spin labels
held fixed need not be antisymmetric under exchange of positions.
The smaller scalar space $\Lambda^\eta L^2(\T)$, with aligned spins
adjoined, describes the fully spin-polarised sector. We do not make
that restriction here. Because the Hamiltonian is independent of
spin, its matrix elements are the spatial matrix elements above
multiplied by a Kronecker delta in the spin indices, as in the
spectator-spin convention of
Ref.~\cite{SuBerryWiebeRubinBabbush2021}.

Let $E$ be the lowest eigenvalue of $\HBO$ on $\mathcal H_-$, and let
$\ket{\psi}$ be any corresponding normalised eigenfunction. We do not
assume that this eigenvalue is non-degenerate. Let $E_N$ be the
lowest eigenvalue of $P\HBO P$ on
$\mathrm{ran}\,P\cap\mathcal H_-$, whenever this finite sector is
nonzero; our sufficiently large cutoffs will ensure this. We measure
the spatial resolution by
\begin{equation}
K:=\frac{\pi(N^{1/3}-1)}{L}\;\propto\;N^{1/3}.
\label{eq:K}
\end{equation}
This cutoff bounds each Cartesian momentum component:
$\bm{p}\in G$ if and only if $\norm{\bm{k}_{\bm{p}}}_\infty\le K$.
The proportionality in Eq.~\eqref{eq:K} refers to asymptotic scaling
at fixed $L$. More explicitly, for admissible $N=n^3\ge27$, we have
$n-1\ge2n/3$ and hence
\begin{equation}
 K^{-3}\le\left(\frac{3L}{2\pi}\right)^3N^{-1}.
\end{equation}
This is the conversion from a momentum-cutoff bound to a basis-size
bound that we use below.

For the estimates, use the momentum and spin basis
$\ket{\vec{\bm{p}},\boldsymbol\sigma}:=\ket{\vec{\bm{p}}}\otimes\ket{\boldsymbol\sigma}$.
For any state $\ket{u}$, define its momentum amplitudes by
\begin{equation}
 u_{\vec{\bm{p}},\boldsymbol\sigma}:=\langle\vec{\bm{p}},\boldsymbol\sigma|u\rangle,
 \qquad
 \ket{u}=\sum_{\vec{\bm{p}},\boldsymbol\sigma}
 u_{\vec{\bm{p}},\boldsymbol\sigma}\ket{\vec{\bm{p}},\boldsymbol\sigma}.
\end{equation}
The amplitude of $B\ket{u}$ is
$\langle\vec{\bm{p}},\boldsymbol\sigma|B|u\rangle$. When spin labels are
suppressed, $u_{\vec{\bm{p}}}$ denotes the vector
$(u_{\vec{\bm{p}},\boldsymbol\sigma})_{\boldsymbol\sigma}$ of these amplitudes;
$\|u_{\vec{\bm{p}}}\|^2$ is its squared norm. In position space we write
$u(\vec{\bm{x}},\boldsymbol\sigma):=\langle \vec{\bm{x}},\boldsymbol\sigma|u\rangle$,
where $\vec{\bm{x}}=(\bm{x}_1,\ldots,\bm{x}_\eta)$, and retain function notation for
pointwise values and slices. With spin labels suppressed, $u(\vec{\bm{x}})$
is the corresponding spin vector. Differential operators acting on
a ket act on its position amplitude.
We use the total squared momentum
$\|\vec{\bm{k}}\|^2:=\sum_{j=1}^\eta\norm{\bm{k}_{\bm{p}_j}}^2$ and define the Sobolev
norms by
\begin{equation}
 \norm{\ket{u}}_{H^s}^2=\sum_{\vec{\bm{p}},\boldsymbol\sigma}
 (1+\|\vec{\bm{k}}\|^2)^s|u_{\vec{\bm{p}},\boldsymbol\sigma}|^2.
 \label{eq:spin-sobolev}
\end{equation}
For $s=1$ this includes all first derivatives, and for $s=2$ it
controls derivatives through second order. Unqualified $L^2$ norms
also include the spin sum. When spin labels are omitted, inner
products contract the spin components and squared norms sum them.
For example, let $\ket{u_\sigma}$ denote the spatial state for a
fixed spin configuration $\sigma$. An estimate
$\norm{L\ket{u_\sigma}}\le C\norm{B\ket{u_\sigma}}$ with the same constant for
each component gives
\begin{equation}
 \sum_\sigma\norm{L\ket{u_\sigma}}^2
 \le C^2\sum_\sigma\norm{B\ket{u_\sigma}}^2.
\end{equation}
Bilinear estimates extend by Cauchy--Schwarz in the spin index.
Consequently, the Hardy and Sobolev estimates used below have the
same constants for our finite spin space; no factor $2^\eta$ is
introduced. We understand all subsequent spatial function spaces
with this fixed spin factor.

The component omitted by the finite basis is
\begin{equation}
\ket{\delta}:=(1-P)\ket{\psi}.
\end{equation}
We will control the energy error by estimating this component, without
having to select or compare a particular eigenvector of the finite
matrix.

\subsection{Ground-energy theorem}

The following theorem gives the convergence rate at fixed cell size
and fixed nuclear configuration.
\begin{theorem}[Basis-set error of the BO ground-state energy]
\label{thm:main}
Fix $\eta$, $L$, $\{\bm{R}_\ell, \zeta_\ell\}$. There exist constants
$C$ and $N_0$, depending on these data and on $E$ --- polynomially on
$\eta$ and $\lambda_\zeta$ --- but not on $N$, such that for all
admissible $N \ge N_0$,
\begin{equation}
0 \;\le\; E_N - E \;\le\; C\, K^{-3} \;\le\; C'\, N^{-1}.
\label{eq:main}
\end{equation}
\end{theorem}

We prove the theorem in four steps. We first establish that the
Coulomb Hamiltonian has a ground state in $H^2$, together with a
bound on this norm. We then use $P\ket{\psi}$ as a trial state to reduce
the energy error to $\norm{\ket{\delta}}_{H^1}^2$, and reduce that norm to
one-electron momentum tails. The remaining step is to improve those
tails using the eigenvalue equation. The key input to this last step
is Lemma~\ref{lem:tail}, which bounds the high-momentum part of
$V_{\mathrm{tot}}\ket{\psi}$. In this way the stronger decay is a consequence
of the Hamiltonian, rather than an additional regularity assumption.

\subsection{Preliminaries: self-adjointness and a priori regularity}
\label{sec:prelim}

Before projecting an eigenfunction, we establish its existence and
the regularity needed for the Fourier estimates. The periodic
interaction has the same local singularity as the ordinary Coulomb
potential, plus a bounded remainder. This lets us use the
three-dimensional Hardy inequality to control each interaction by
spatial derivatives, and then apply the standard self-adjointness
results of Refs.~\cite{Kato1951,ReedSimonII}.

\begin{proposition}[Structure of the periodic kernel]
\label{prop:F1}
(i) $w \in L^2(\T)$. (ii) There exist $\chi \in C_c^\infty(\R^3)$ with
$\chi \equiv 1$ near $0$ and support in a ball contained in the open
fundamental domain, and $B \in L^\infty(\T)$, such that
$w(\bm{x}) = \chi(\bm{x})/\|\bm{x}\| + B(\bm{x})$ on the fundamental domain.
\end{proposition}

Here a fundamental domain is one representative cell for the periodic
coordinates. The notation $C_c^\infty$ specifies a smooth function
with compact support, so $\chi$ localises the singularity without
reaching the cell boundary. The condition $B\in L^\infty$ means that
the remainder is bounded apart from a possible set of measure zero.

\begin{proofx}
We first verify square integrability from the Fourier coefficients.
Parseval's identity with the normalisation in Eq.~\eqref{eq:w} gives
\begin{equation}
 \norm{w}_{L^2(\T)}^2
 =\Omega\sum_{\bm{\nu}\ne0}
 \left(\frac{4\pi}{\Omega\norm{\bm{k}_{\bm{\nu}}}^2}\right)^2
 =\frac{16\pi^2}{\Omega}\sum_{\bm{\nu}\ne0}\norm{\bm{k}_{\bm{\nu}}}^{-4}<\infty.
\end{equation}
The sum converges because in three dimensions the number of lattice
points in a shell grows quadratically in its radius, while the
summand decays as the fourth inverse power.

For the local decomposition, we first identify the equation solved by
$w$. Let $\delta_{\mathrm{per}}$ denote the unit point mass at the
origin of the torus. Applying $-\Delta$ to Eq.~\eqref{eq:w} multiplies
each nonzero Fourier coefficient by $\norm{\bm{k}_{\bm{\nu}}}^2$, and gives
\begin{equation}
 -\Delta w=4\pi\big(\delta_{\mathrm{per}}-\Omega^{-1}\big).
\end{equation}
This identity is understood in the distributional sense, that is,
after integration against a smooth periodic test function. The
constant term subtracts the zero Fourier mode of the point mass.
It is the uniform neutralising background associated with our
zero-mean convention for the potential.

We compare this equation with the local Coulomb identity
$-\Delta(1/\|\bm{x}\|)=4\pi\delta_0$ in three dimensions, where $\delta_0$
is the unit point mass at the origin in $\R^3$. Choose a smooth
cutoff $\chi$ equal to one near the origin and supported strictly
inside the fundamental cell. Multiplying $1/\|\bm{x}\|$ by this cutoff
makes it a periodic function without changing its singularity at
the origin. The product rule gives
\begin{equation}
 -\Delta\left(\frac{\chi}{\|\bm{x}\|}\right)
 =4\pi\delta_{\mathrm{per}}+g,\qquad
 g:=-2\nabla\chi\cdot\nabla\left(\frac1{\|\bm{x}\|}\right)
       -\frac{\Delta\chi}{\|\bm{x}\|}.
\end{equation}
The derivatives of $\chi$ are supported away from both the origin
and the cell boundary. Thus $g$ is a smooth periodic function:
the apparent inverse powers of $\|\bm{x}\|$ occur only where $\|\bm{x}\|$ is
bounded away from zero.

Now set $B:=w-\chi/\|\bm{x}\|$. Both terms are in $L^2(\T)$, since
$\|\bm{x}\|^{-2}$ is locally integrable in three dimensions. Their singular
point-mass terms cancel, leaving
\begin{equation}
 -\Delta B=-4\pi\Omega^{-1}-g.
\end{equation}
For each nonzero Fourier index this implies
$\hat B(\bm{\nu})=-\hat g(\bm{\nu})/\norm{\bm{k}_{\bm{\nu}}}^2$.
Smoothness of $g$ makes its Fourier coefficients decay faster than
any inverse power, as follows by repeated integration by parts.
The nonzero coefficients of $B$ therefore have an absolutely
convergent Fourier series. Its zero coefficient is finite because
$B\in L^2(\T)$. This shows that $B$ is bounded and proves the
required decomposition. In particular, we have obtained the unit
Coulomb singularity directly from the Fourier definition of $w$.
\end{proofx}

The bounded part of this decomposition is harmless for the domain of
the Hamiltonian. We now control the singular part in one three-dimensional
coordinate at a time; this avoids requiring a Hardy estimate in the
full $3\eta$-dimensional configuration space.

\begin{proposition}[Self-adjointness and $H^2$ regularity]
\label{prop:F2}
$\HBO$ is self-adjoint on $H^2(\T^{\eta})$, bounded below, with compact
resolvent; in particular the ground state on $\mathcal{H}_-$ exists.
Each multiplication operator $w(\bm{x}_i - \bm{x}_j)$ and $w(\bm{x}_j - \bm{R}_\ell)$ is
infinitesimally bounded, both in operator and in form sense, relative to
$-\Delta = -\sum_j \Delta_j$. Every normalised eigenfunction with
eigenvalue $E$ satisfies
\begin{equation}
\norm{\ket{\psi}}_{H^2} \le C_0 \big(1 + |E|\big),
\label{eq:H2bound}
\end{equation}
with $C_0$ polynomial in $\eta$ and $\lambda_\zeta$.
\end{proposition}

Compact resolvent means that, for a shift $z$ outside the
spectrum, the inverse $(\HBO-z)^{-1}$ maps every bounded sequence
of states to a sequence with a convergent subsequence. Its role
here is to ensure a discrete spectrum with finite-dimensional
eigenspaces. Infinitesimal operator boundedness allows an
arbitrarily small coefficient of $\norm{\Delta\ket{u}}$ when bounding
the norm of a potential-applied state, at the cost of a larger
coefficient of $\norm{\ket{u}}$. The form version instead bounds the
potential expectation by an arbitrarily small multiple of
$\norm{\nabla\ket{u}}^2$, plus a multiple of $\norm{\ket{u}}^2$.

\begin{proofx}
Let $\ket{u}\in H^2(\T^\eta)$. For an electron--electron term, hold all
coordinates except $\bm{x}_i$ fixed and set $\bm{x}=\bm{x}_i-\bm{x}_j$. Apply the
three-dimensional Hardy inequality
\begin{equation}
 \norm{v/\|\bm{x}\|}_{L^2(\R^3)}\le2\norm{\nabla v}_{L^2(\R^3)}
\end{equation}
to $v(\bm{x})=\chi(\bm{x})u(\bm{x}+\bm{x}_j,\vec{\bm{x}}')$, where $\vec{\bm{x}}'$ collects the fixed
coordinates. The cutoff gives a compactly supported function that
can be extended by zero outside the cell. Differentiating $\chi u$
produces a derivative of $u$ and a bounded-cutoff term. After
integrating over the other coordinates and using
Proposition~\ref{prop:F1} to include $B$, we obtain
\begin{equation}
 \norm{w(\bm{x}_i-\bm{x}_j)\ket{u}}
 \le C_L\big(\norm{\nabla_i\ket{u}}+\norm{\ket{u}}\big).
\end{equation}
The same argument applies to $w(\bm{x}_i-\bm{R}_\ell)$ with the nucleus held
fixed. All these steps hold componentwise in spin and then sum as in
Eq.~\eqref{eq:spin-sobolev}.

The Fourier interpolation inequality
$\norm{\nabla_i\ket{u}}\le\varepsilon'\norm{\Delta\ket{u}}
+C_{\varepsilon'}\norm{\ket{u}}$ now turns this first-derivative bound
into an arbitrarily small relative bound with respect to the total
Laplacian $\Delta:=\sum_j\Delta_j$. Thus, for every
$\varepsilon>0$,
\begin{equation}
\norm{w(\bm{x}_i-\bm{x}_j)\ket{u}}
\le\varepsilon\norm{\Delta\ket{u}}+C_\varepsilon\norm{\ket{u}},
\qquad \ket{u}\in H^2(\T^\eta).
\label{eq:KR}
\end{equation}
This is what infinitesimal operator boundedness means here: the
coefficient of $\norm{\Delta\ket{u}}$ can be made as small as desired,
at the cost of increasing the coefficient of $\norm{\ket{u}}$.
For the corresponding quadratic-form bound, Cauchy--Schwarz and the
first-derivative estimate give
\begin{equation}
 |\langle u|w(\bm{x}_i-\bm{x}_j)|u\rangle|
 \le C_L\norm{\ket{u}}\big(\norm{\nabla_i\ket{u}}+\norm{\ket{u}}\big)
 \le\varepsilon\norm{\nabla_i\ket{u}}^2+C_\varepsilon\norm{\ket{u}}^2.
\end{equation}
The same estimates hold for every nuclear term before multiplying by
its charge. Summing the pair terms and the charge-weighted nuclear
terms preserves the arbitrarily small relative bound, because their
total weight is bounded by
$\eta(\eta-1)/2+\eta\lambda_\zeta$. The interpolation and summation
constants can therefore be chosen polynomially in these parameters.

The Kato--Rellich theorem now gives self-adjointness of $T+V_{\mathrm{tot}}$
on the domain of $T=-\Delta/2$, namely $H^2(\T^\eta)$.
The form bound also gives a lower bound on the energy. Moreover,
Eq.~\eqref{eq:KR}, after summation and absorption of a relative
bound smaller than one, makes the Hamiltonian graph norm
$\norm{\ket{u}}+\norm{\HBO\ket{u}}$ equivalent to the $H^2$ norm.
The embedding of $H^2(\T^\eta)$ into $L^2(\T^\eta)$ is compact
on a fixed torus, so the resolvent is compact. Restricting to the
closed antisymmetric sector preserves these properties; the
spectrum there is discrete and bounded below, and its lowest
energy is an eigenvalue.

Finally, we extract the quantitative regularity estimate used in the
rest of the proof. Choose the relative bound in the summed potential
estimate so that
\begin{equation}
 \norm{V_{\mathrm{tot}}\ket{u}}\le\tfrac12\norm{T\ket{u}}+C\norm{\ket{u}}.
\end{equation}
For a normalised eigenfunction, $T\ket{\psi}=(E-V_{\mathrm{tot}})\ket{\psi}$,
and hence
\begin{equation}
 \norm{T\ket{\psi}}\le|E|+\norm{V_{\mathrm{tot}}\ket{\psi}}
 \le|E|+\tfrac12\norm{T\ket{\psi}}+C.
\end{equation}
Moving the half of $\norm{T\ket{\psi}}$ to the left gives
$\norm{T\ket{\psi}}\le2(|E|+C)$. Since
$\norm{\ket{\psi}}_{H^2}\lesssim\norm{\ket{\psi}}+\norm{\Delta\ket{\psi}}$ and
$\Delta\ket{\psi}=-2T\ket{\psi}$, this proves Eq.~\eqref{eq:H2bound} with the
stated polynomial dependence.
\end{proofx}

\subsection{Variational energy comparison}
\label{sec:sandwich}

We now connect Fourier approximation to the quantity of interest,
$E_N-E$. The finite-basis variational space is a subspace of the
continuum space, so its ground energy cannot be lower. For an upper
bound, we use the projected exact ground state as a trial state.
The important feature of this choice is that the eigenvalue equation
removes all terms linear in the discarded component $\ket{\delta}$.

\begin{proposition}
\label{prop:sandwich}
Suppose $\norm{P\ket{\psi}}^2 \ge \tfrac12$ (verified in
Sec.~\ref{sec:bootstrap} for $N \ge N_0$). Then
\begin{equation}
0 \;\le\; E_N - E \;\le\; 2\, \langle\delta|\HBO-E|\delta\rangle
\;\le\; 2 C_1 \norm{\ket{\delta}}_{H^1}^2 ,
\label{eq:sandwich}
\end{equation}
with $C_1$ depending polynomially on $\eta$, $\lambda_\zeta$, and
$|E|$.
\end{proposition}

\begin{proofx}
Every normalised state in $\mathrm{ran}\,P\cap\mathcal H_-$ is also
an admissible state for the continuum variational problem on
$\mathcal H_-$. Taking the infimum over the smaller set therefore
gives $E_N\ge E$.

For the opposite inequality, $P\ket{\psi}$ remains in the antisymmetric
sector because $P$ preserves that sector. The hypothesis ensures
that it is nonzero, so its normalisation is an admissible finite-basis
trial state and
\begin{equation}
 E_N-E\le
 \frac{\langle\psi|P(\HBO-E)P|\psi\rangle}{\norm{P\ket{\psi}}^2}.
\end{equation}
By Proposition~\ref{prop:F2}, $\ket{\psi}\in H^2$; the Fourier projector
preserves this space, so $\ket{\delta}=(1-P)\ket{\psi}\in H^2$ as well. We can
therefore use the eigenvalue equation and self-adjointness directly
in the following expansion:
\begin{align}
 \langle\psi|P(\HBO-E)P|\psi\rangle
 &=\big(\langle\psi|-\langle\delta|\big)(\HBO-E)\big(\ket{\psi}-\ket{\delta}\big)\\
 &=\langle\psi|\HBO-E|\psi\rangle
   -\langle\psi|\HBO-E|\delta\rangle
   -\langle\delta|\HBO-E|\psi\rangle
   +\langle\delta|\HBO-E|\delta\rangle.
\end{align}
The first and third terms vanish because $(\HBO-E)\ket{\psi}=0$.
The second vanishes after moving $\HBO-E$ onto $\ket{\psi}$ by
self-adjointness. Thus
\begin{equation}
 \langle\psi|P(\HBO-E)P|\psi\rangle
 =\langle\delta|\HBO-E|\delta\rangle.
\end{equation}
This is the quadratic cancellation: the trial-state energy error
depends only on a quadratic expression in the discarded component.
Dividing by $\norm{P\ket{\psi}}^2\ge1/2$ gives the first upper bound in
Eq.~\eqref{eq:sandwich}.

It remains to control this quadratic expression by the $H^1$ norm.
Separating its kinetic, potential, and energy-shift contributions,
we have
\begin{equation}
 \langle\delta|\HBO-E|\delta\rangle
 \le\tfrac12\norm{\nabla\ket{\delta}}^2
   +|\langle\delta|V_{\mathrm{tot}}|\delta\rangle|
   +|E|\norm{\ket{\delta}}^2,
 \qquad
 \norm{\nabla\ket{\delta}}^2:=\sum_j\norm{\nabla_j\ket{\delta}}^2.
\end{equation}
The Hardy estimate in the proof of Proposition~\ref{prop:F2} gives,
for each pair kernel,
\begin{equation}
 |\langle\delta|w(\bm{x}_i-\bm{x}_j)|\delta\rangle|
 \le C\big(\norm{\ket{\delta}}\norm{\nabla_i\ket{\delta}}
             +\norm{\ket{\delta}}^2\big)
 \le C\norm{\ket{\delta}}_{H^1}^2.
\end{equation}
The nuclear kernels satisfy the same estimate, with their charges
included when we sum. Combining the finitely many terms gives
$\langle\delta|\HBO-E|\delta\rangle\le C_1\norm{\ket{\delta}}_{H^1}^2$
with the stated polynomial dependence, proving the second upper
bound.
\end{proofx}

We have reduced the theorem to proving the tail estimate
\begin{equation}
\norm{\ket{\delta}}_{H^1}^2\le C_3\,\eta^2K^{-3}
\label{eq:target}
\end{equation}
and checking that the projection retains at least half the norm.
Both will follow from bounds that restrict only one electron's
momentum at a time.

\subsection{Reduction to one-particle momentum tails}
\label{sec:reduction}

A Fourier configuration is discarded by $P$ if at least one of its
$\eta$ spatial indices lies outside $G$. We can bound this event by
summing the single-electron events. Since leaving the cube
$\norm{\bm{k}_{\bm{p}_j}}_\infty\le K$ also implies leaving the Euclidean ball
$\norm{\bm{k}_{\bm{p}_j}}_2\le K$, the indicators satisfy
\begin{equation}
 \ind{\exists j:\bm{p}_j\notin G}
 \le\sum_{j=1}^\eta\ind{\norm{\bm{k}_{\bm{p}_j}}_\infty>K}
 \le\sum_{j=1}^\eta\ind{\norm{\bm{k}_{\bm{p}_j}}_2>K}.
\end{equation}
This replacement enlarges the discarded region and therefore gives
an upper bound. From now on, unqualified momentum norms in tail
restrictions are Euclidean norms.

For $j,m\in\{1,\ldots,\eta\}$, define
\begin{align}
 t_j(K)&:=\norm{\ind{\norm{\bm{k}_{\bm{p}_j}}>K}\ket{\psi}},\\
 s_{j,m}(K)&:=\norm{\ind{\norm{\bm{k}_{\bm{p}_j}}>K}\norm{\bm{k}_{\bm{p}_m}}\ket{\psi}}.
\end{align}
Here the indicators and momentum weights act diagonally in the
momentum basis, and the norms include all spin components.
The quantity $t_j$ measures the amplitude with electron $j$ above
the cutoff. The quantity $s_{j,m}$ measures the same amplitude with
a first-derivative weight in electron $m$; the electron with high
momentum and the differentiated electron need not be the same.
This distinction matters because the $H^1$ norm includes derivatives
in every electron coordinate.

Multiplying the indicator inequality by the appropriate squared
Fourier weights and summing gives
\begin{align}
 \norm{\ket{\delta}}^2
 &\le\sum_j t_j(K)^2\le\eta t(K)^2,\\
 \norm{\ket{\delta}}_{H^1}^2
 &\le\sum_j\left[t_j(K)^2+\sum_m s_{j,m}(K)^2\right]
 \le\eta\left[t(K)^2+\sum_m s_m(K)^2\right],
 \label{eq:red}
\end{align}
where $t(K):=\max_j t_j(K)$ and
$s_m(K):=\max_j s_{j,m}(K)$. We use maxima to obtain uniform bounds,
rather than assume permutation symmetry of individual spin
components. In the following proofs we may therefore choose any
restricted electron and call it particle $1$. The estimates are
uniform in that choice and in the derivative index $m$, so they
bound all the quantities in Eq.~\eqref{eq:red}.

\begin{remark}[Other reducing sectors]
The same variational comparison and tail reduction apply in any
nonzero closed sector preserved by the Hamiltonian and every $P_N$,
with the operators restricting to that sector. In that case $E$
denotes its ground energy. The arguments require that projection
keeps the trial state in the chosen sector; they do not require a
particular exchange symmetry. This is an extension of the analytic
argument, rather than an additional formal sector theorem.
\end{remark}

We can now identify precisely why the regularity established so far
is insufficient. On the region $\norm{\bm{k}_{\bm{p}_j}}>K$,
$1+\|\vec{\bm{k}}\|^2\ge K^2$, so
\begin{equation}
 1\le K^{-2}(1+\|\vec{\bm{k}}\|^2),\qquad
 \norm{\bm{k}_{\bm{p}_m}}\le(1+\|\vec{\bm{k}}\|^2)^{1/2}
                  \le K^{-1}(1+\|\vec{\bm{k}}\|^2).
\end{equation}
Multiplying these pointwise inequalities by $\|\psi_{\vec{\bm{p}}}\|$ and taking
the Fourier square-summation norm yields
\begin{equation}
 t(K)\le K^{-2}\norm{\ket{\psi}}_{H^2},\qquad
 s_m(K)\le K^{-1}\norm{\ket{\psi}}_{H^2}.
 \label{eq:apriori}
\end{equation}
For large $K$, substitution into Eq.~\eqref{eq:red} gives only
$\norm{\ket{\delta}}_{H^1}^2=\mathcal O(K^{-2})$, and hence an energy
bound of order $N^{-2/3}$. Our target in Eq.~\eqref{eq:target}
requires one further inverse power of $K$ in the squared norm.
We next estimate the potential-applied state, which will let us
obtain that improvement from the eigenvalue equation.
\subsection{Potential-tail estimate}
\label{sec:lemma}

We now estimate the part of $V_{\mathrm{tot}}\ket{\psi}$ with a large momentum
in at least one specified electron coordinate. Here
$V_{\mathrm{tot}}=U+V$ is the sum of the electron--nucleus and
electron--electron potentials in Eq.~\eqref{eq:Hper}. We need a bound
uniform in the choice of electron, so we define
\begin{equation}
\Lambda(K):=\max_{1\le j\le\eta}
\norm{\ind{\norm{\bm{k}_{\bm{p}_j}}>K}V_{\mathrm{tot}}\ket{\psi}}.
\end{equation}
The indicator acts diagonally in the momentum basis, and the norm
includes all momentum and spin components. Our aim is to show that this quantity decays as $K^{-1/2}$.
The eigenvalue equation will then supply the additional powers of
$K^{-1}$ needed to bound the discarded state and its derivatives.

\begin{lemma}[High-momentum content of $V_{\mathrm{tot}}\ket{\psi}$]
\label{lem:tail}
For $K \ge 1$,
\begin{equation}
\Lambda(K) \;\le\; C_{\mathrm{pot}}\, K^{-1/2},
\label{eq:lambda}
\end{equation}
with $C_{\mathrm{pot}}$ proportional to $\norm{\ket{\psi}}_{H^2}$ and polynomial in
$\eta$ and $\lambda_\zeta$ (with $\Omega$-dependent numerical factors).
\end{lemma}

\begin{proofx}
Fix the electron whose output momentum is restricted and call it
particle $1$. Let $Q_1(K)$ be the orthogonal Fourier projector onto
$\norm{\bm{k}_{\bm{p}_1}}>K$. Thus we must bound
$\norm{Q_1(K)V_{\mathrm{tot}}\ket{\psi}}$ independently of which electron
was designated as particle $1$. We separate interactions that leave
this electron's momentum unchanged from those that can change it.
For the latter, we split the kernel according to the size of the
momentum transfer. This separates two reasons why the output can have
large momentum: either the input already had large momentum, or the
interaction supplied a large transfer.

\emph{Interactions not involving particle $1$.}
Consider first $w(\bm{x}_i-\bm{x}_j)$ with $2\le i<j\le\eta$, or
$w(\bm{x}_j-\bm{R}_\ell)$ with $j\ge2$, leaving the nuclear charge outside the
estimate. Such a multiplier commutes with $Q_1(K)$ because it is
independent of $\bm{x}_1$. For the pair term write $w_{ij}:=w(\bm{x}_i-\bm{x}_j)$;
the same estimate below holds for $w(\bm{x}_j-\bm{R}_\ell)$. Write $\ket{\psi_{>K}}:=Q_1(K)\ket{\psi}$. The Hardy estimate
used in Proposition~\ref{prop:F2} bounds multiplication by each Coulomb
kernel from $H^1$ to $L^2$, and hence
\begin{equation}
 \norm{Q_1(K)(w_{ij}\ket{\psi})}
 =\norm{w_{ij}\ket{\psi_{>K}}}
 \le C\norm{\ket{\psi_{>K}}}_{H^1}
 \le CK^{-1}\norm{\ket{\psi}}_{H^2}.
\end{equation}
For the last step, every retained Fourier coefficient of $\ket{\psi_{>K}}$
has $\|\vec{\bm{k}}\|^2>K^2$, so
$(1+\|\vec{\bm{k}}\|^2)\le K^{-2}(1+\|\vec{\bm{k}}\|^2)^2$. This contribution already decays
faster than the required $K^{-1/2}$ for $K\ge1$. Summing these
interactions introduces at most a factor of order
$\eta^2+\eta\lambda_\zeta$, with the nuclear charges counted as
weights rather than as a number of terms.

\emph{Interactions involving particle $1$: small transfers.}
For each remaining interaction, write $w=w_<+w_>$, where $w_<$
contains the Fourier modes with $0<\norm{\bm{k}_{\bm{\nu}}}\le K/2$ and $w_>$
contains those with $\norm{\bm{k}_{\bm{\nu}}}>K/2$. Multiplication by
$w_<(\bm{x}_1-\bm{y})$ changes the first electron's momentum by at most $K/2$,
whether $\bm{y}=\bm{x}_j$ is another electron position or $\bm{y}=\bm{R}_\ell$ is fixed.
The triangle inequality therefore excludes an output above $K$ from
an input at or below $K/2$:
\begin{equation}
 Q_1(K)w_<(\bm{x}_1-\bm{y})\ket{\psi}
 =Q_1(K)w_<(\bm{x}_1-\bm{y})Q_1(K/2)\ket{\psi}.
\end{equation}
Since $Q_1(K)$ is a contraction in $L^2$, this identity gives
\begin{equation}
 \norm{Q_1(K)(w_<\ket{\psi})}
 \le\norm{w_<}_{L^\infty}\,t(K/2).
\end{equation}
We bound the kernel by the sum of the magnitudes of its Fourier
coefficients:
\begin{equation}
 \norm{w_<}_{L^\infty}
 \le\frac{4\pi}{\Omega}
 \sum_{0<\norm{\bm{k}_{\bm{\nu}}}\le K/2}\norm{\bm{k}_{\bm{\nu}}}^{-2}
 \le c_\Omega K.
\end{equation}
In three dimensions a shell of radius $r$ has order $r^2$ lattice
points, which cancels the $r^{-2}$ coefficient and leaves growth
linear in the radius. Applying Eq.~\eqref{eq:apriori} at $K/2$, we obtain
\begin{equation}
 \norm{Q_1(K)(w_<\ket{\psi})}
 \le c_\Omega K(K/2)^{-2}\norm{\ket{\psi}}_{H^2}
 =4c_\Omega K^{-1}\norm{\ket{\psi}}_{H^2}.
\end{equation}
Thus small transfers also give the stronger decay $K^{-1}$.

\emph{Interactions involving particle $1$: large transfers.}
A large transfer can create high output momentum even from a low-momentum
input. We control this contribution using the decay of the kernel's
Fourier coefficients, and may discard the output projector when taking
an upper bound. Parseval's identity gives
\begin{equation}
 \norm{w_>}_{L^2(\T)}^2
 =\frac{16\pi^2}{\Omega}
 \sum_{\norm{\bm{k}_{\bm{\nu}}}>K/2}\norm{\bm{k}_{\bm{\nu}}}^{-4}
 \le c_\Omega' K^{-1}.
\end{equation}
The exponent follows by comparing the three-dimensional lattice tail
with $\int_{K/2}^\infty r^2r^{-4}\,dr$, which is proportional to
$K^{-1}$. The constants here and in the preceding lattice sum may
depend on the fixed cell.

To turn this kernel bound into a bound on $w_>\ket{\psi}$, we need a supremum
estimate on the state only in the three coordinates of particle $1$.
We hold $\vec{\bm{x}}':=(\bm{x}_2,\ldots,\bm{x}_\eta)$ fixed and apply the embedding
$H^2(\T)\hookrightarrow L^\infty(\T)$ to the function
$\bm{x}_1\mapsto\psi(\bm{x}_1,\vec{\bm{x}}')$. In this slice $\bm{y}$ is fixed, even when
$\bm{y}=\bm{x}_j$, and translating the periodic kernel does not change its
$L^2$ norm. H\"older's inequality in $\bm{x}_1$, followed by integration in
$\vec{\bm{x}}'$, therefore yields
\begin{align}
\norm{w_>(\bm{x}_1-\bm{y})\ket{\psi}}^2
&=\int d\vec{\bm{x}}'\int d\bm{x}_1\,|w_>(\bm{x}_1-\bm{y})|^2\|\psi(\bm{x}_1,\vec{\bm{x}}')\|^2
\\
&\le\norm{w_>}_{L^2(\T)}^2
 \int d\vec{\bm{x}}'\,\norm{\psi(\cdot,\vec{\bm{x}}')}_{L^\infty_{\bm{x}_1}}^2
\\
&\le C_S^2\norm{w_>}_{L^2(\T)}^2
 \int d\vec{\bm{x}}'\,\norm{\psi(\cdot,\vec{\bm{x}}')}_{H^2_{\bm{x}_1}}^2
\\
&\le C_S^2\norm{w_>}_{L^2(\T)}^2\norm{\ket{\psi}}_{H^2}^2.
\label{eq:sliceSobolev}
\end{align}
Here $C_S$ is the three-dimensional Sobolev-embedding constant. The
last step follows directly from the Fourier weights:
\begin{equation}
 \int d\vec{\bm{x}}'\,\norm{\psi(\cdot,\vec{\bm{x}}')}_{H^2_{\bm{x}_1}}^2
 =\sum_{\vec{\bm{p}}}(1+\norm{\bm{k}_{\bm{p}_1}}^2)^2\|\psi_{\vec{\bm{p}}}\|^2
 \le\sum_{\vec{\bm{p}}}(1+\|\vec{\bm{k}}\|^2)^2\|\psi_{\vec{\bm{p}}}\|^2.
\end{equation}
These displayed slice calculations can first be read for one spin
component and then summed over spin; $\|\psi_{\vec{\bm{p}}}\|^2$ denotes the
squared norm of the spin vector when all components are included.
Thus we use a three-dimensional embedding regardless of the number
of electrons, and introduce no factor from the number of spin states.
Taking square roots in Eq.~\eqref{eq:sliceSobolev} gives
\begin{equation}
 \norm{w_>(\bm{x}_1-\bm{y})\ket{\psi}}
 \le C_S(c_\Omega')^{1/2}K^{-1/2}\norm{\ket{\psi}}_{H^2}.
\end{equation}
There are $\eta-1$ electron--electron terms involving particle $1$,
and its electron--nucleus terms have total charge weight
$\lambda_\zeta$. Combining their bounds with the two $K^{-1}$
contributions above proves Eq.~\eqref{eq:lambda}, uniformly in the
chosen electron. The slowest decay in this estimate comes from the
large-transfer part of the Coulomb kernel. This identifies the source
of the exponent in the upper bound; it does not establish a matching
lower bound.
\end{proofx}

\subsection{Eigenvalue bootstrap}
\label{sec:bootstrap}

We have now bounded the high-momentum part of $V_{\mathrm{tot}}\ket{\psi}$
by $K^{-1/2}$. To finish the theorem, we use the kinetic term in the
eigenvalue equation to convert this into faster decay of $\ket{\psi}$.
This is the bootstrap step: the $H^2$ estimate supplied the potential
tail bound, and the equation in turn improves the state tail beyond
what that initial regularity bound alone implied.

\begin{proofx}[Completion of the proof of Theorem~\ref{thm:main}]
For every momentum and spin configuration, the eigenvalue equation reads
\begin{equation}
 \tfrac12\|\vec{\bm{k}}\|^2\psi_{\vec{\bm{p}},\boldsymbol\sigma}
 =E\psi_{\vec{\bm{p}},\boldsymbol\sigma}
 -\langle\vec{\bm{p}},\boldsymbol\sigma|V_{\mathrm{tot}}|\psi\rangle.
 \label{eq:evF}
\end{equation}
We again choose an arbitrary restricted electron and call it
particle $1$. On $\norm{\bm{k}_{\bm{p}_1}}>K$, the total momentum obeys
$\|\vec{\bm{k}}\|>K$. Dividing by the kinetic multiplier therefore gains two
inverse powers of $K$ for the state itself. If we also insert the
first-derivative multiplier of any electron $m$, we retain one
inverse power, since $\norm{\bm{k}_{\bm{p}_m}}\le\|\vec{\bm{k}}\|$:
\begin{equation}
 \frac{2}{\|\vec{\bm{k}}\|^2}\le\frac{2}{K^2},\qquad
 \frac{2\norm{\bm{k}_{\bm{p}_m}}}{\|\vec{\bm{k}}\|^2}\le\frac{2}{K}.
 \label{eq:weights}
\end{equation}
These bounds hold for every $m$, including the restricted electron.
They control the total derivative weight without requiring the
other electrons' momenta to be small.

Divide Eq.~\eqref{eq:evF} by $\|\vec{\bm{k}}\|^2/2$, restrict to the chosen
high-momentum region, and take the Fourier square-summation norm.
The triangle inequality and Eq.~\eqref{eq:weights} give, uniformly
in the choice of restricted electron,
\begin{align}
 t(K)&\le\frac{2}{K^2}\big(|E|t(K)+\Lambda(K)\big),
 \label{eq:boot1}\\
 s_m(K)&\le\frac{2}{K}\big(|E|t(K)+\Lambda(K)\big).
 \label{eq:boot2}
\end{align}
The right side still contains the unknown tail $t(K)$, but its
coefficient is small at sufficiently large momentum. Specifically,
if $K^2\ge4|E|$, then $2|E|/K^2\le1/2$. Moving that term to the
left of Eq.~\eqref{eq:boot1} gives
\begin{equation}
 \left(1-\frac{2|E|}{K^2}\right)t(K)
 \le\frac{2\Lambda(K)}{K^2}.
\end{equation}
The parenthesis is at least $1/2$, and Lemma~\ref{lem:tail} therefore
implies, also taking $K\ge1$,
\begin{equation}
 t(K)\le\frac{4\Lambda(K)}{K^2}
 \le4C_{\mathrm{pot}}K^{-5/2}.
 \label{eq:tfinal}
\end{equation}
This is the improved amplitude tail. Substituting it and the same
potential-tail bound into Eq.~\eqref{eq:boot2} yields
\begin{equation}
 s_m(K)\le\frac{2}{K}
 \Big(4|E|C_{\mathrm{pot}}K^{-5/2}
          +C_{\mathrm{pot}}K^{-1/2}\Big)
 \le C_2K^{-3/2}
 \label{eq:sfinal}
\end{equation}
for $K\ge\max(1,2|E|^{1/2})$. For example, the same condition
$4|E|/K^2\le1$ lets us take $C_2=4C_{\mathrm{pot}}$ on this range.
The derivative tail has thus gained the half power of $K^{-1}$
that was missing from Eq.~\eqref{eq:apriori}.

We now assemble the bounds for all electrons. Equation~\eqref{eq:red}
gives
\begin{equation}
 \norm{\ket{\delta}}_{H^1}^2
 \le\eta\big[16C_{\mathrm{pot}}^2K^{-5}
                 +\eta C_2^2K^{-3}\big]
 \le C_3\eta^2K^{-3},
\end{equation}
where the last step uses $K\ge1$ and $\eta\ge1$. This proves
Eq.~\eqref{eq:target}. The norm estimate from the same reduction
also verifies that our trial state is usable:
\begin{equation}
 \norm{P\ket{\psi}}^2=1-\norm{\ket{\delta}}^2
 \ge1-\eta t(K)^2
 \ge1-16\eta C_{\mathrm{pot}}^2K^{-5}.
\end{equation}
In particular it is at least $1/2$ once
$K^5\ge32\eta C_{\mathrm{pot}}^2$. Choosing the cutoff above this
threshold and the earlier thresholds validates all hypotheses of
Proposition~\ref{prop:sandwich}.

Finally, applying that proposition and the explicit relation between
$K$ and admissible $N$ following Eq.~\eqref{eq:K}, we obtain
\begin{equation}
 0\le E_N-E\le2C_1C_3\eta^2K^{-3}\le C'N^{-1}
 \qquad(\text{admissible }N\ge N_0).
\end{equation}
The constants in the potential-tail estimate are proportional to
$\norm{\ket{\psi}}_{H^2}$, which Proposition~\ref{prop:F2} bounds in terms
of $|E|$ and polynomially in the particle and charge parameters.
All subsequent combinations preserve this dependence. Taking $N_0$
large enough for the stated cutoff conditions completes the proof.
\end{proofx}

\subsection{Sharpness and scope}
\label{sec:remarks}

The argument establishes an upper bound without assuming a spectral
gap or a more detailed local expansion of the eigenfunction. We
finish by separating this proved statement from two related questions:
whether its exponent is optimal, and what the same proof says about
approximating the state.

\begin{remark}[Expected sharpness, and what it requires]
\label{rem:sharp}
We expect the energy error to be $\Theta(1/N)$ already for one
electron and one nucleus in the periodic cell. To discuss positivity
and simplicity, fix one electron spin component. The two spin blocks
are identical, so this scalar restriction has the same ground energy
and Galerkin energy error as the spin-inclusive problem. A matching
lower bound would require two additional ingredients, which we
invoke here without a complete proof.

The first ingredient is a sufficiently precise local expansion at
the nucleus. Writing the nuclear charge as $Z$ and placing the
nucleus at the origin, the expected form is
\begin{equation}
 \psi(\bm{x})=\psi(0)(1-Z\|\bm{x}\|)+r(\bm{x}),\qquad\psi(0)\ne0.
\end{equation}
We would need enough control of the remainder $r$ to show that its
Fourier coefficients decay strictly faster than the contribution
from the singular $\|\bm{x}\|$ term, whose leading decay is
$\norm{\bm{k}}^{-4}$. A local expansion is used here with a smooth cutoff
around the nucleus, so that it can be separated from the regular
part elsewhere in the periodic cell. Kato's cusp condition
\cite{Kato1957} alone fixes a radial derivative; it does not by
itself give the remainder estimate required for this Fourier
asymptotic. The structural regularity theory of Ref.~\cite{FHOS},
specialised to one electron, supplies expansions of the relevant
type, but the Fourier conclusion needs its own justification.

In particular, we need a nonzero leading coefficient along enough
of the actual Fourier lattice modes. An informal asymptotic along
a set of spatial directions of full measure is not yet this lattice
statement. If the required lattice asymptotic is established, then
we can bound the squared $H^1$ tail from below using the modes with
$\norm{\bm{k}}>\sqrt{3}K$, all of which lie outside the retained cube.
Their contribution is bounded below by a positive constant times
$\sum_{\norm{\bm{k}}>\sqrt{3}K}\norm{\bm{k}}^{2}\norm{\bm{k}}^{-8}$, which scales as
$K^{-3}$ in three dimensions. Equivalently,
\begin{equation}
 \mathrm{dist}_{H^1}(\ket{\psi},\mathrm{ran}\,P)
 :=\inf_{\ket{v}\in\mathrm{ran}\,P}\norm{\ket{\psi}-\ket{v}}_{H^1}
 \ge c'K^{-3/2}.
\end{equation}
The cubical cutoff changes constants in this comparison, not the
power. Here the Fourier projection is also the best approximation
in the $H^1$ norm, since that norm is diagonal in the same basis.

The second ingredient relates this approximation lower bound to an
eigenvalue lower bound. For a simple isolated eigenvalue, the
Babu\v{s}ka--Osborn theory \cite{BabuskaOsborn} gives the two-sided
comparison
\begin{equation}
 E_N-E\asymp
 \mathrm{dist}_{H^1}(\ket{\psi},\mathrm{ran}\,P)^2.
\end{equation}
The notation $a\asymp b$ means that $c\,b\le a\le C\,b$
for some positive constants $c$ and $C$. Here the constants
may depend on the fixed system but are independent of the
basis cutoff, and the comparison holds for all sufficiently
large cutoffs.
The two-sided eigenvalue bound uses a shifted energy form:
adding a sufficiently large multiple of the squared norm makes
the form comparable to the squared $H^1$ norm.
It also uses simplicity and
spectral isolation of the eigenvalue. For the scalar periodic
hydrogen ground state, simplicity follows from positivity-improving
arguments and isolation from the compact resolvent.
The positivity property used here is that imaginary-time evolution
$e^{-t\HBO}$, for $t>0$, sends every nonzero nonnegative scalar
function to a function that is strictly positive almost everywhere.

Granting both ingredients gives $E_N-E\ge c''K^{-3}$, and hence a
matching $N^{-1}$ lower bound at fixed cell size. This explains the
expected sharpness, but we do not claim to have proved the required
Fourier-lattice asymptotic here. The upper bound of
Theorem~\ref{thm:main} is independent of both additional ingredients.
\end{remark}

\begin{remark}[Byproduct: gap-free Fourier truncation error]
\label{rem:trunc}
The improved amplitude tail also gives a state-approximation bound.
Combining Eqs.~\eqref{eq:red} and~\eqref{eq:tfinal}, we obtain
\begin{equation}
 \norm{(1-P)\ket{\psi}}
 \le\sqrt{\eta}\,t(K)
 \le4\sqrt{\eta}\,C_{\mathrm{pot}}K^{-5/2}
 =\mathcal O(N^{-5/6}).
\end{equation}
No spectral gap enters this estimate. Its meaning is the error made
by Fourier truncation of the exact eigenfunction:
\begin{equation}
 \norm{(1-P)\ket{\psi}}
 =\inf_{\ket{v}\in\mathrm{ran}\,P}\norm{\ket{\psi}-\ket{v}}.
\end{equation}
It measures how well the finite basis can represent $\ket{\psi}$ and does
not compare $\ket{\psi}$ with a chosen normalised eigenvector
$\ket{\psi_N^{\mathrm{eig}}}$ of $P\HBO P$.

That distinction matters in a degenerate or nearly degenerate ground
space, where a particular discrete eigenvector need not follow a
fixed continuum eigenvector. Even when a gap is available, one needs
a separate Galerkin eigenspace argument
\cite{BabuskaOsborn} to transfer an approximation rate to computed
eigenvectors. We make no such claim here. Within its stated scope,
the $N^{-5/6}$ truncation estimate is stronger than the generic
$N^{-1/2}$ estimate obtained by taking a square root of an
$N^{-1}$ energy error in a gap-based quadratic-form bound. The
momentum-tail method is also useful for the later dynamical analysis,
where the relevant approximation is the truncation of the exact
time-evolved state and no discrete-eigenvector comparison is required.
\end{remark}

\medskip\noindent\emph{What the eigenvalue equation adds.}
We began with only the $H^2$ regularity supplied by
Proposition~\ref{prop:F2}. The stronger decay
$t(K)\lesssim K^{-5/2}$ was then derived, rather than assumed, by
one use of the eigenvalue equation after estimating the potential
tail. This is the single elliptic bootstrap step used in the proof:
we use the kinetic operator's quadratic Fourier multiplier to
improve decay of the solution from decay of the right-hand side.
The tail estimate also gives Sobolev regularity below order $5/2$.
To see how the one-electron restrictions control the full Sobolev
norm, take a sufficiently large total-momentum threshold $\rho$.
If $\|\vec{\bm{k}}\|>\rho$, then at least one electron has
$\norm{\bm{k}_{\bm{p}_j}}>\rho/\sqrt{\eta}$, because
$\|\vec{\bm{k}}\|^2=\sum_j\norm{\bm{k}_{\bm{p}_j}}^2$. The same indicator argument as in
Eq.~\eqref{eq:red} therefore gives
\begin{equation}
 \norm{\ind{\|\vec{\bm{k}}\|>\rho}\ket{\psi}}^2
 \le\sum_{j=1}^{\eta}t_j(\rho/\sqrt{\eta})^2
 \le\eta\,t(\rho/\sqrt{\eta})^2
 =\mathcal O(\rho^{-5}),
\end{equation}
where the constant may depend on the fixed particle number.
We now group the total momenta into shells
$2^j<\|\vec{\bm{k}}\|\le2^{j+1}$, with $j$ an integer. For sufficiently large
$j$, the contribution of such a shell to the squared $H^s$ norm is
bounded by
\begin{equation}
 \sum_{2^j<\|\vec{\bm{k}}\|\le2^{j+1}}
 (1+\|\vec{\bm{k}}\|^2)^s\|\psi_{\vec{\bm{p}}}\|^2
 \le C_s2^{2sj}\norm{\ind{\|\vec{\bm{k}}\|>2^j}\ket{\psi}}^2
 \le C_s'2^{(2s-5)j}.
\end{equation}
As before, $\|\psi_{\vec{\bm{p}}}\|^2$ includes the spin sum.
The resulting geometric series converges for every $s<5/2$,
and the remaining bounded-momentum region contributes a finite
amount. Thus $\ket{\psi}\in H^s$ for every $s<5/2$. At $s=5/2$, this
argument only bounds each shell contribution by a constant and
does not make their sum finite, so it does not establish
$\ket{\psi}\in H^{5/2}$. None of
this uses a structural cusp expansion of the
Fournais--Hoffmann-Ostenhof--\O{}stergaard S\o{}rensen type
\cite{FHOS}. That additional structure is relevant only to the
proposed sharpness argument above.

\medskip\noindent\emph{Scope of the bound.}
The cell volume is fixed throughout this periodic analysis. The
lattice-sum constants $c_\Omega$ and $c_\Omega'$ and the resulting
energy constant depend on $\Omega$; we therefore make no claim of
uniformity in a thermodynamic limit with
$\Omega\propto\eta\to\infty$. The conditional comparison with an
isolated molecule is treated separately in Sec.~\ref{sec:free}.
Although the constants are polynomial in $\eta$ and
$\lambda_\zeta$, we have not optimised those polynomials.

Finally, the variational cancellation is specific to the energy of
a stationary state. A generic observable evaluated in a discrete
eigenstate does not have the same cancellation, so the energy
convergence rate alone does not give that rate for its expectation.
For time-dependent smooth observables, the later sections recover
an energy-rate coefficient for the error growing with time through
a separate duality argument. That mechanism uses regularity of a
backward-evolved test state in place of the ground-state
stationarity used here.

\section{Ground-state energy with quantum and clamped nuclei}
\label{app:nbo}
\label{sec:nonBO}

We now allow some or all nuclei to move as quantum particles, with
any remaining nuclei held fixed, and prove the same
$\mathcal O(N^{-1})$ ground-energy convergence rate as in
Appendix~\ref{app:bo}. The variational part of the argument is unchanged:
we project an exact ground state into the finite basis and control its
energy error by the discarded component. The new task is to keep track
of the masses when estimating that component. At a fixed momentum,
a heavy particle has less kinetic energy than an electron, so a bound
on kinetic energy gives weaker control of its high-momentum tail.
We make this dependence explicit while preserving the cutoff exponent.

Two features organise the proof. First, we measure regularity using the
mass-weighted kinetic operator, which is the kinetic energy appearing in
the eigenvalue equation. Second, we keep a separate tail bound for each
particle, rather than assuming all particle coordinates have the same
statistics or mass. This lets us treat electronic antisymmetry and the
exchange symmetry of each nuclear species within the same argument.
The final estimates require no spectral-gap assumption.

\subsection{Setting}
\label{sec:nbo-setting}

We retain the fixed periodic cell $\T=\R^3/L\Z^3$, the Coulomb kernel
$w$ of Eq.~\eqref{eq:w}, and the Fourier convention of
Eq.~\eqref{eq:pw}. Of the $L_{\mathrm{nuc}}$ nuclei, let the first
$L_{\mathrm q}$ be quantum particles and let the remaining nuclei be
clamped at positions $\bm{R}_\ell\in\T$, for
$L_{\mathrm q}<\ell\le L_{\mathrm{nuc}}$.
There are $\eta_{\mathrm{tot}}:=\eta+L_{\mathrm q}$ quantum particles. We label
them by $a\in\{1,\ldots,\eta_{\mathrm{tot}}\}$, with position
$\bm{x}_a\in\T$, mass $m_a$, and charge $q_a$. For an electron
($a\le\eta$), $m_a=1$ and $q_a=-1$. For quantum nucleus
$\ell\le L_{\mathrm q}$, the label is
$a=\eta+\ell$, its mass is $m_a=m_{\eta+\ell}\ge1$, and its charge is
$q_a=\zeta_\ell>0$. As before,
$\lambda_\zeta:=\sum_{\ell=1}^{L_{\mathrm{nuc}}}\zeta_\ell$ is the
total charge of both the quantum and clamped nuclei, with
$\zeta_\ell>0$ for every nucleus. We write $m_{\max}:=\max_a m_a$
for the largest quantum-particle mass. The fully quantum case has
$L_{\mathrm q}=L_{\mathrm{nuc}}$, the BO case has $L_{\mathrm q}=0$,
and intermediate choices give hybrid models.

Each particle may also have a finite internal-spin space
$\mathbb C^{d_a}$. We take $d_a=2$ for electrons and the appropriate
spin multiplicity for each nuclear species, or $d_a=1$ if no internal
spin is retained. Thus the complete internal space is
$\mathcal V:=\bigotimes_{a=1}^{\eta_{\mathrm{tot}}}\mathbb C^{d_a}$,
of dimension $\prod_a d_a$. On
$L^2(\T^{\eta_{\mathrm{tot}}};\mathcal V)$, the spin-independent
Hamiltonian is, retaining the notation $\Hnbo$ for this general setting,
\begin{equation}
\begin{aligned}
\Hnbo:=\sum_{j=1}^{\eta_{\mathrm{tot}}}
\left(-\frac{\nabla^2_j}{2m_j}\right)
&+\sum_{1\le i<j\le\eta_{\mathrm{tot}}}q_iq_j\,w(\bm{x}_i-\bm{x}_j)\\
&+\sum_{a=1}^{\eta_{\mathrm{tot}}}
  \sum_{\ell=L_{\mathrm q}+1}^{L_{\mathrm{nuc}}}
       q_a\zeta_\ell\,w(\bm{x}_a-\bm{R}_\ell).
\end{aligned}
\label{eq:Hnbo}
\end{equation}
The first potential sum couples two quantum coordinates; the second
couples a quantum particle to a clamped nucleus and is absent in the
fully quantum case. The charge products distinguish electron--nucleus
attraction from repulsion between two electrons or two nuclei. We use
$V_{\mathrm{tot}}=U+V+V_{\mathrm{nuc}}$ for the sum of these three
potential contributions. Here $V_{\mathrm{nuc}}$ includes the nuclear
repulsion terms involving at least one quantum nucleus. Repulsion
between two clamped nuclei is omitted as a common additive constant;
restoring it shifts the continuum and finite-basis ground energies
equally and leaves their difference unchanged.

For each particle we retain the same spatial Fourier cube $G$ of
Eq.~\eqref{eq:G}, with $N=n^3$ for odd $n$ and component cutoff
$K=\pi(n-1)/L$. The product basis is
$\ket{\vec{\bm{p}}}:=\bigotimes_a\ket{\bm{p}_a}$, where
$\vec{\bm{p}}=(\bm{p}_1,\ldots,\bm{p}_{\eta_{\mathrm{tot}}})$. Its position amplitude is
$\Phi_{\vec{\bm{p}}}(\bm{x}_1,\ldots,\bm{x}_{\eta_{\mathrm{tot}}})
:=\langle \bm{x}_1,\ldots,\bm{x}_{\eta_{\mathrm{tot}}}|\vec{\bm{p}}\rangle
=\prod_a\phi_{\bm{p}_a}(\bm{x}_a)$. Let $P=P_N$ project onto
the span of the configurations with $\bm{p}_a\in G$ for every particle,
and act as the identity on $\mathcal V$. The spin identity will be
implicit below. Before imposing exchange symmetry, the retained space
has dimension $N^{\eta_{\mathrm{tot}}}\prod_a d_a$.

When all nuclei are quantum particles, the matrix of $P\Hnbo P$
on this space agrees with the non-BO
Hamiltonian in Eqs.~(9)--(14) of
Ref.~\cite{SuBerryWiebeRubinBabbush2021}. To see the structure, the
kinetic term for particle $a$ is diagonal with entry
$\norm{\bm{k}_{\bm{p}_a}}^2/(2m_a)$. A Fourier mode of $w(\bm{x}_a-\bm{x}_b)$ transfers
opposite momenta to particles $a$ and $b$, conserving their sum.
Both output indices must still belong to $G$, because of the
projection. These are exactly the momentum-conservation and
basis-membership restrictions in the cited pair operators, with
$q_aq_b$ supplying their charge factors. The same calculation applies
to electron--electron, electron--nucleus, and nucleus--nucleus terms;
only the charges and kinetic masses change. A clamped-nucleus term
instead transfers momentum in one quantum register, with the phase
from the fixed position $\bm{R}_\ell$, as in the BO construction.

We choose a nonzero closed exchange-symmetry sector $\mathcal S$.
Electronic states are antisymmetric under simultaneous exchange of
position and spin. For each identical nuclear species we impose the
appropriate symmetry or antisymmetry on its quantum position--spin pairs;
different species need not be exchanged. The Hamiltonian and the
projector both preserve this sector, because particles of an identical
species have the same mass, charge, and spatial cutoff, and interact
with the same clamped nuclear configuration. In the
terminology used in Appendix~\ref{app:bo}, the sector reduces both
operators: each acts separately on the sector and its orthogonal
complement. The analytic extension to other such sectors discussed
there applies as well.

Let $E$ be the ground energy of $\Hnbo$ on $\mathcal S$ and let $\ket{\psi}$
be any corresponding normalised eigenfunction. Let $E_N$ be the
lowest eigenvalue of $P\Hnbo P$ on
$\mathrm{ran}\,P\cap\mathcal S$, for cutoffs where this finite sector
is nonzero, and set $\ket{\delta}:=(1-P)\ket{\psi}$. We use the momentum and spin basis
$\ket{\vec{\bm{p}},\boldsymbol\sigma}:=\ket{\vec{\bm{p}}}\otimes\ket{\boldsymbol\sigma}$,
where $\boldsymbol\sigma$ labels an orthonormal basis of $\mathcal V$.
The momentum amplitudes of the ground state are
\begin{equation}
 \psi_{\vec{\bm{p}},\boldsymbol\sigma}:=\langle\vec{\bm{p}},\boldsymbol\sigma|\psi\rangle,
 \qquad
 \ket{\psi}=\sum_{\vec{\bm{p}},\boldsymbol\sigma}
 \psi_{\vec{\bm{p}},\boldsymbol\sigma}\ket{\vec{\bm{p}},\boldsymbol\sigma}.
\end{equation}
The amplitude of $B\ket{\psi}$ is
$\langle\vec{\bm{p}},\boldsymbol\sigma|B|\psi\rangle$. When spin labels are
suppressed, $\psi_{\vec{\bm{p}}}$ denotes the vector of these amplitudes in
$\mathcal V$, with $\|\psi_{\vec{\bm{p}}}\|$ denoting its Euclidean norm. All norms and inner products include the internal-spin
sum. The componentwise estimates and Cauchy--Schwarz argument in
Appendix~\ref{app:bo} therefore apply without a factor from the
number of spin components.

The kinetic operator and its Fourier multiplier are
\begin{equation}
A(\vec{\bm{k}}):=\sum_{a=1}^{\eta_{\mathrm{tot}}}
\frac{\norm{\bm{k}_{\bm{p}_a}}^2}{2m_a},\qquad
\Ttot:=\sum_a\left(-\frac{\Delta_a}{2m_a}\right).
\label{eq:Asymbol}
\end{equation}
Thus $\Ttot\ket{\vec{\bm{p}},\boldsymbol\sigma}
=A(\vec{\bm{k}})\ket{\vec{\bm{p}},\boldsymbol\sigma}$, and the momentum amplitude of
$\Ttot\ket{\psi}$ is $A(\vec{\bm{k}})\psi_{\vec{\bm{p}},\boldsymbol\sigma}$.
We measure the available regularity by
\begin{equation}
\Theta:=\norm{(1+\Ttot)\ket{\psi}}
=\left(\sum_{\vec{\bm{p}},\boldsymbol\sigma}
 [1+A(\vec{\bm{k}})]^2|\psi_{\vec{\bm{p}},\boldsymbol\sigma}|^2\right)^{1/2}.
\label{eq:Theta}
\end{equation}
This quantity controls
both the state norm and the norm after applying kinetic energy;
it is comparable to the kinetic graph norm
$\norm{\ket{\psi}}+\norm{\Ttot\ket{\psi}}$. Using $A$ rather than the unweighted
total squared momentum keeps the particle masses visible in the
estimates. Proposition~\ref{prop:nbo-KR} will bound $\Theta$ in terms
of $E$ and the system parameters, so it is not an additional
assumption on the ground state.

When we sum the particlewise bounds, the sums
$S_r:=\sum_{a=1}^{\eta_{\mathrm{tot}}}m_a^r$, for $r=1,2,3$, will
collect the corresponding mass factors. In particular, $S_1S_2$
appears in the following energy estimate.

\begin{theorem}[Basis-set error with quantum and clamped nuclei]
\label{thm:nbo}
Fix $\eta_{\mathrm{tot}}$, $L$, the charges, the masses and any
clamped nuclear positions. There
exist $C, N_0$ --- with $C$ polynomial in $\eta_{\mathrm{tot}}$,
$\lambda_\zeta$, and $m_{\max}$, and $N_0$ polynomial in $m_{\max}$ and
$1 + |E|$ (the bootstrap absorption requires $K^2 \ge 4m_{\max}|E|$,
and we track $|E|$ explicitly rather than bounding it in the system
parameters) ---
such that for admissible $N \ge N_0$
(in the sense of Sec.~\ref{sec:setting}),
\begin{equation}
0 \;\le\; E_N - E \;\le\;
C C_0^2\, (1 + |E|)^3\, m_{\max}^4\, S_1 S_2\, K^{-3}
\;\le\; C'\, N^{-1},
\label{eq:nbo-main}
\end{equation}
using Proposition~\ref{prop:nbo-KR} below to eliminate the
state-dependent graph norm. Coarsely,
$E_N - E \le C''\, \eta_{\mathrm{tot}}^2\, m_{\max}^{7} (1+|E|)^3
K^{-3}$, the last factor $(1+|E|)$ beyond the squared graph
budget coming from the variational-sandwich constant $C_1$. No
spectral gap enters.
\end{theorem}

The squared graph budget mentioned in the theorem is $\Theta^2$,
the square of the kinetic-regularity quantity in Eq.~\eqref{eq:Theta}.

The mass factors in this theorem are upper bounds from the proof;
we do not optimise them. To establish the result, we first bound
$\Theta$ using the eigenvalue equation and the Hardy estimates
for quantum pairs and fixed centres. We then repeat the variational comparison in the
unweighted $H^1$ norm and estimate the discarded Fourier component
particle by particle. The potential-tail lemma and a final division
by $A(\vec{\bm{k}})$ give the same powers of $K$ as in the BO case, while
retaining the masses in their coefficients. The remarks at the end
explain which conclusions about nuclear resolution follow from
these bounds and which remain heuristic.

\subsection{Preliminaries revisited}

The kernel decomposition in Proposition~\ref{prop:F1} is unchanged:
$w$ consists of a local Coulomb singularity and a bounded periodic
remainder. What changes is the kinetic operator used to control a
derivative. For an interacting pair, we may apply Hardy's inequality
with either particle carrying the derivative. We choose the lighter
particle because converting that derivative into kinetic energy
costs a factor proportional to its mass. This choice avoids an
unnecessary nuclear mass factor whenever the pair contains an
electron. For an interaction with a clamped nucleus, only the
quantum particle carries a derivative, so its mass enters instead.

\begin{proposition}[Self-adjointness and mass-weighted regularity]
\label{prop:nbo-KR}
$\Hnbo$ is self-adjoint on $H^2(\T^{\eta_{\mathrm{tot}}};\mathcal V)$, bounded
below, with compact resolvent; the ground state exists on every
sector $\mathcal{S}$. Moreover every normalised eigenfunction with
eigenvalue $E$ satisfies
\begin{equation}
\Theta \;\le\; C_0\, m_{\max}\, (1 + |E|),
\label{eq:nbo-H2}
\end{equation}
with $C_0$ polynomial in $\eta_{\mathrm{tot}}, \lambda_\zeta$
(and $\Omega$-dependent).
\end{proposition}

\begin{proofx}
We begin with one pair $\{a,b\}$. Let
$\lambda=\lambda(a,b)\in\{a,b\}$ be a lighter member of the pair,
so $m_\lambda=\min\{m_a,m_b\}$; if the masses agree, either choice
works. Hold the other coordinates fixed and apply the
three-dimensional Hardy inequality in the relative variable
$\bm{x}_a-\bm{x}_b$, differentiating with respect to $\bm{x}_\lambda$.
Using the cutoff decomposition of Proposition~\ref{prop:F1}(ii)
and then integrating the fixed coordinates gives
\begin{equation}
 \norm{w(\bm{x}_a-\bm{x}_b)\ket{u}}
 \le C\big(\norm{\nabla_\lambda\ket{u}}+\norm{\ket{u}}\big).
\end{equation}
The constant here is independent of the masses. To express the
right side in terms of the kinetic operator, we use its positivity:
\begin{equation}
 \norm{\nabla_\lambda\ket{u}}^2
 =\langle u|-\Delta_\lambda|u\rangle
 \le2m_\lambda\langle u|\Ttot|u\rangle
 \le2m_\lambda\norm{\Ttot\ket{u}}\norm{\ket{u}}.
\end{equation}
The first inequality follows because
$-\Delta_\lambda=2m_\lambda T_\lambda$, where
$T_\lambda=-\Delta_\lambda/(2m_\lambda)$ is one nonnegative
summand of $\Ttot$. The second is Cauchy--Schwarz. Taking square
roots and applying Young's inequality to the product gives, for
any $\varepsilon>0$,
\begin{equation}
 \norm{w(\bm{x}_a-\bm{x}_b)\ket{u}}
 \le\varepsilon\norm{\Ttot\ket{u}}
       +C\left(\frac{m_\lambda}{\varepsilon}+1\right)\norm{\ket{u}}.
 \label{eq:nbo-KRpair}
\end{equation}
Thus the coefficient of the kinetic norm can be made arbitrarily
small. The compensating coefficient of the state norm carries
$m_\lambda$, which equals one for every electron-containing pair.
Among pairs of quantum particles, only nucleus--nucleus interactions
can introduce a large mass at this step.

For a clamped nucleus, the translated one-particle Hardy estimate gives
$\norm{w(\bm{x}_a-\bm{R}_\ell)\ket{u}}
\le C(\norm{\nabla_a\ket{u}}+\norm{\ket{u}})$,
uniformly in $\bm{R}_\ell$. The same interpolation with the kinetic energy
of particle $a$ therefore gives
\begin{equation}
 \norm{w(\bm{x}_a-\bm{R}_\ell)\ket{u}}
 \le\varepsilon\norm{\Ttot\ket{u}}
       +C\left(\frac{m_a}{\varepsilon}+1\right)\norm{\ket{u}}.
 \label{eq:nbo-KRclamped}
\end{equation}
This coefficient remains mass-independent for an electron, but can
carry a nuclear mass when a quantum nucleus interacts with a fixed one.

We now include the charge factors and sum over both kinds of interaction. Their
absolute weights obey
\begin{equation}
 \sum_{a<b}|q_aq_b|
 +\sum_a\sum_{\ell=L_{\mathrm q}+1}^{L_{\mathrm{nuc}}}|q_a|\zeta_\ell
 \le\tfrac12\left(\sum_a|q_a|
       +\sum_{\ell=L_{\mathrm q}+1}^{L_{\mathrm{nuc}}}\zeta_\ell\right)^2
 =\tfrac12(\eta+\lambda_\zeta)^2.
\end{equation}
Choosing the individual relative bound in
Eqs.~\eqref{eq:nbo-KRpair} and~\eqref{eq:nbo-KRclamped} small enough
in terms of this total weight
therefore yields
\begin{equation}
 \norm{V_{\mathrm{tot}}\ket{u}}
 \le\tfrac12\norm{\Ttot\ket{u}}+Cm_{\max}\norm{\ket{u}},
\end{equation}
where $C$ is polynomial in the particle and charge parameters and
independent of the masses. More generally, the factor $1/2$ may be
replaced by any positive relative bound, with a different constant.
The corresponding form estimate follows from the same Hardy
inequality and Cauchy--Schwarz. In particular, the potential
expectation can be bounded by an arbitrarily small multiple of
$\langle u|\Ttot|u\rangle$ plus a constant times $\norm{\ket{u}}^2$.
This also establishes a lower bound on the total energy.

To apply Kato--Rellich on the claimed domain, we must identify the
domain of $\Ttot$. Writing
$\|\vec{\bm{k}}\|^2:=\sum_a\norm{\bm{k}_{\bm{p}_a}}^2$, the assumptions
$1\le m_a\le m_{\max}$ give the two-sided comparison
\begin{equation}
 \frac{\|\vec{\bm{k}}\|^2}{2m_{\max}}\le A(\vec{\bm{k}})\le\frac{\|\vec{\bm{k}}\|^2}{2}.
\end{equation}
For fixed finite masses, the weights $1+A(\vec{\bm{k}})$ and
$1+\|\vec{\bm{k}}\|^2$ are therefore comparable, so the domain of $\Ttot$
is precisely the unweighted space
$H^2(\T^{\eta_{\mathrm{tot}}};\mathcal V)$.
Both inequalities are needed: the upper inequality alone would only
show that every $H^2$ state lies in the kinetic domain.
Kato--Rellich now gives self-adjointness of $\Hnbo$ on that same
domain. The relative bounds also make its graph norm comparable
to the $H^2$ norm. Since the configuration torus is compact, the
embedding of this domain into $L^2$ is compact, giving a compact
resolvent and hence a discrete spectrum. Restriction to a nonzero
closed sector preserved by these operators gives a ground-state
eigenfunction there as well. As in Appendix~\ref{app:bo}, all
estimates extend to the finite internal-spin space by summing
componentwise, without changing their constants.

Finally, for a normalised eigenfunction the equation
$\Ttot\ket{\psi}=(E-V_{\mathrm{tot}})\ket{\psi}$ gives
\begin{equation}
 \norm{\Ttot\ket{\psi}}
 \le |E|+\norm{V_{\mathrm{tot}}\ket{\psi}}
 \le |E|+\tfrac12\norm{\Ttot\ket{\psi}}+Cm_{\max}.
\end{equation}
Absorbing the half of the kinetic norm on the right yields
$\norm{\Ttot\ket{\psi}}\le2|E|+2Cm_{\max}$. Consequently,
\begin{equation}
 \Theta=\norm{(1+\Ttot)\ket{\psi}}
 \le1+\norm{\Ttot\ket{\psi}}
 \le1+2|E|+2Cm_{\max}
 \le C_0m_{\max}(1+|E|),
\end{equation}
which proves Eq.~\eqref{eq:nbo-H2} with $C_0$ independent of the
masses.
\end{proofx}

\subsection{Variational comparison and tail reduction}

We next reduce the energy error to Fourier tails. The cancellation
in Proposition~\ref{prop:sandwich} depends on the eigenvalue
equation and sector preservation, so it applies to the present
Hamiltonian as well. Indeed, if $P\ket{\psi}\ne0$, its normalisation is
an admissible trial state in $\mathcal S$, and
\begin{equation}
 0\le E_N-E\le
 \frac{\langle\psi|P(\Hnbo-E)P|\psi\rangle}{\norm{P\ket{\psi}}^2}
 =\frac{\langle\delta|\Hnbo-E|\delta\rangle}{\norm{P\ket{\psi}}^2}.
\end{equation}
The equality follows by expanding $P\ket{\psi}=\ket{\psi}-\ket{\delta}$ and using
$(\Hnbo-E)\ket{\psi}=0$ on both sides of the inner product. Thus the
quadratic cancellation survives when the nuclei are quantum
particles.

We continue to bound the numerator with the unweighted $H^1$ norm.
This choice introduces no additional explicit mass factor into the
comparison constant. For the kinetic term, $m_a\ge1$ gives
\begin{equation}
 \langle\delta|\Ttot|\delta\rangle
 =\sum_a\frac{\norm{\nabla_a\ket{\delta}}^2}{2m_a}
 \le\tfrac12\norm{\nabla\ket{\delta}}^2.
\end{equation}
For a pair potential, the Hardy estimate before conversion to
kinetic energy gives
\begin{equation}
 |\langle\delta|w(\bm{x}_a-\bm{x}_b)|\delta\rangle|
 \le C\big(\norm{\nabla_{\lambda(a,b)}\ket{\delta}}\norm{\ket{\delta}}
                 +\norm{\ket{\delta}}^2\big)
 \le C\norm{\ket{\delta}}_{H^1}^2.
\end{equation}
Here $\lambda(a,b)$ is the lighter particle, as above, although no
mass conversion is needed when the derivative is left in the
unweighted norm. The clamped-nucleus term $w(\bm{x}_a-\bm{R}_\ell)$ obeys
the same form bound with $\nabla_a$ in place of
$\nabla_{\lambda(a,b)}$. Summing both kinds of interaction with
their charge factors and adding the
term $|E|\norm{\ket{\delta}}^2$ gives
\begin{equation}
 0\le E_N-E\le2C_1\norm{\ket{\delta}}_{H^1}^2
 \qquad\text{once }\norm{P\ket{\psi}}^2\ge\tfrac12.
 \label{eq:nbo-sandwich}
\end{equation}
We may take $C_1$ to be a polynomial in the particle and charge
parameters times $1+|E|$. Its mass dependence enters through the
value of $E$, rather than through a separate mass factor in this
form comparison. This distinguishes the present estimate from the
mass-dependent conversion to the kinetic norm in
Eq.~\eqref{eq:nbo-KRpair}.

We now express the discarded norm as a sum of one-particle tails.
A configuration outside the product cube has at least one particle
outside its retained cube and hence outside the Euclidean ball of
radius $K$. Therefore
\begin{equation}
 \ind{\exists a:\bm{p}_a\notin G}
 \le\sum_{a=1}^{\eta_{\mathrm{tot}}}
                    \ind{\norm{\bm{k}_{\bm{p}_a}}>K}.
\end{equation}
For $a,m\in\{1,\ldots,\eta_{\mathrm{tot}}\}$, define
\begin{align}
 t_a(K)&:=\norm{\ind{\norm{\bm{k}_{\bm{p}_a}}>K}\ket{\psi}},\\
 s_{a,m}(K)&:=\norm{\ind{\norm{\bm{k}_{\bm{p}_a}}>K}
                         \norm{\bm{k}_{\bm{p}_m}}\ket{\psi}}.
\end{align}
Here the indicators and momentum weights act diagonally in the
momentum basis, and the norms include all spin components.
The index $a$ specifies which particle has high momentum, while
$m$ specifies the derivative used to measure the same tail.
Here $m$ is a particle index; its mass is $m_m$. The two indices
need not agree because the $H^1$ norm includes derivatives in
every particle coordinate. Multiplying the indicator inequality
by the Fourier weights and summing gives
\begin{equation}
 \norm{\ket{\delta}}^2\le\sum_a t_a(K)^2,\qquad
 \norm{\ket{\delta}}_{H^1}^2
 \le\sum_a\left[t_a(K)^2+\sum_m s_{a,m}(K)^2\right].
 \label{eq:nbo-red}
\end{equation}
This reduction uses no permutation symmetry of the Fourier
coefficients. We can therefore retain each particle's mass and
allow different exchange statistics for different species.

The initial tail estimates follow from $\Theta$, but their mass
dependence is now essential. On $\norm{\bm{k}_{\bm{p}_a}}>K$,
$A(\vec{\bm{k}})>K^2/(2m_a)$, so
\begin{equation}
 1\le\frac{2m_a}{K^2}A(\vec{\bm{k}}),\qquad
 \norm{\bm{k}_{\bm{p}_m}}
 \le(2m_m)^{1/2}A(\vec{\bm{k}})^{1/2}
 \le\frac{2(m_am_m)^{1/2}}{K}A(\vec{\bm{k}}).
\end{equation}
The second inequality uses the contribution of particle $m$ to
$A$ and then the lower bound supplied by the restricted particle
$a$. Multiplying by $\|\psi_{\vec{\bm{p}}}\|$ and taking norms yields
\begin{equation}
 t_a(K)\le\frac{2m_a}{K^2}\Theta,\qquad
 s_{a,m}(K)\le\frac{2(m_am_m)^{1/2}}{K}\Theta.
 \label{eq:nbo-apriori}
\end{equation}
At a fixed value of $\Theta$, the amplitude-tail estimate for a
nucleus of mass $m_\ell$ is weaker by a factor $m_\ell$ than for
an electron at the same cutoff. The derivative estimate records
both the restricted particle's mass and the mass associated with
the derivative. As in the BO proof, these initial estimates give
only $K^{-2}$ for the squared $H^1$ tail at large $K$. We next
use the Coulomb interaction and the eigenvalue equation to gain
the additional inverse power required for the energy theorem.
\subsection{Mass-weighted potential-tail estimate}

We now estimate the high-momentum part of the potential-applied state,
which is the input needed to improve the state tails through the
eigenvalue equation. For each restricted particle $a$, define
\begin{equation}
 \Lambda_a(K):=\norm{\ind{\norm{\bm{k}_{\bm{p}_a}}>K}
                         V_{\mathrm{tot}}\ket{\psi}},
\end{equation}
where $V_{\mathrm{tot}}=U+V+V_{\mathrm{nuc}}$ is the full potential
in Eq.~\eqref{eq:Hnbo}. Keeping the index $a$ will let us identify
which mass enters when a large output momentum is created in that
particle's coordinate.

\begin{lemma}[High-momentum content of $V_{\mathrm{tot}}\ket{\psi}$]
\label{lem:nbo-tail}
For $K \ge 1$ and every particle $a$,
\begin{equation}
\Lambda_a(K) \;\le\; C_\Lambda\, (m_a m_{\max})^{1/2}\, \Theta\,
K^{-1/2},
\label{eq:nbo-lambda}
\end{equation}
with $C_\Lambda$ polynomial in $\eta_{\mathrm{tot}}, \lambda_\zeta$
(and $\Omega$-dependent) and independent of the masses and of $N$.
\end{lemma}

\begin{proofx}
Fix $a$, and let $Q_a(K)$ be the Fourier projector onto
$\norm{\bm{k}_{\bm{p}_a}}>K$. We follow the decomposition used in
Lemma~\ref{lem:tail}, now retaining the mass in each estimate.
An interaction that does not involve $a$ cannot change its momentum.
For an interaction involving $a$, we distinguish small transfers,
which require an already large input momentum, from large transfers,
which we control through the Fourier decay of the Coulomb kernel.
We first bound the kernels without their charge factors and include
those factors when summing the interactions.

\emph{Pairs not involving the restricted particle.}
For $b,c\ne a$, write $w_{bc}:=w(\bm{x}_b-\bm{x}_c)$ and let
$\lambda=\lambda(b,c)$ denote the lighter member of the pair.
The operator $w_{bc}$ commutes with $Q_a(K)$, so the Hardy estimate
from Proposition~\ref{prop:nbo-KR}, before its interpolation step,
gives
\begin{align}
 \norm{Q_a(K)(w_{bc}\ket{\psi})}
 &=\norm{w_{bc}Q_a(K)\ket{\psi}}\\
 &\le C\big(\norm{\nabla_\lambda Q_a(K)\ket{\psi}}
                       +\norm{Q_a(K)\ket{\psi}}\big)\\
 &=C\big(s_{a,\lambda}(K)+t_a(K)\big).
\end{align}
Applying Eq.~\eqref{eq:nbo-apriori}, we obtain
\begin{equation}
 \norm{Q_a(K)(w_{bc}\ket{\psi})}
 \le C\left[\frac{(m_am_\lambda)^{1/2}}{K}
                   +\frac{m_a}{K^2}\right]\Theta
 \le C\frac{(m_am_{\max})^{1/2}}{K}\Theta.
\end{equation}
The two terms require different comparisons. For the derivative tail
we use $m_\lambda\le m_{\max}$; for the amplitude tail we use
$m_a\le(m_am_{\max})^{1/2}$ and $K^{-2}\le K^{-1}$ for $K\ge1$.
This contribution therefore decays as $K^{-1}$, faster than the
claimed $K^{-1/2}$.

A clamped-nucleus term $w(\bm{x}_b-\bm{R}_\ell)$ with $b\ne a$ also commutes
with $Q_a(K)$. Its one-particle Hardy estimate gives
$C(s_{a,b}(K)+t_a(K))$, and the same comparisons yield
$C(m_am_{\max})^{1/2}\Theta K^{-1}$.

\emph{Pairs involving the restricted particle: small transfers.}
For a pair $\{a,b\}$, split $w=w_<+w_>$ at Fourier radius $K/2$.
The kernel $w_<$ contains the modes with
$0<\norm{\bm{k}_{\bm{\nu}}}\le K/2$, while $w_>$ contains those above that
radius. Since $w_<(\bm{x}_a-\bm{x}_b)$ changes the momentum of particle $a$
by at most $K/2$, the triangle inequality gives the support identity
\begin{equation}
 Q_a(K)w_<(\bm{x}_a-\bm{x}_b)\ket{\psi}
 =Q_a(K)w_<(\bm{x}_a-\bm{x}_b)Q_a(K/2)\ket{\psi}.
\end{equation}
The first projector is a contraction in $L^2$. Combining the
three-dimensional lattice-sum estimate
$\norm{w_<}_{L^\infty}\le c_\Omega K$ from the BO proof with the
amplitude bound in Eq.~\eqref{eq:nbo-apriori} at cutoff $K/2$,
we obtain
\begin{align}
 \norm{Q_a(K)(w_<\ket{\psi})}
 &\le\norm{w_<}_{L^\infty}t_a(K/2)\\
 &\le c_\Omega K\,\frac{8m_a}{K^2}\Theta
 =8c_\Omega m_a\Theta K^{-1}.
\end{align}
Here the factor $m_a$ comes from controlling the large input
momentum by its kinetic energy. Again we may bound it by
$(m_am_{\max})^{1/2}$, and $K^{-1}$ by $K^{-1/2}$ for $K\ge1$.
For $w_<(\bm{x}_a-\bm{R}_\ell)$, the same support identity and bound apply:
translation by $\bm{R}_\ell$ changes only Fourier phases and does not
change the transfer support or the kernel's supremum norm.

\emph{Pairs involving the restricted particle: large transfers.}
For $w_>$, we may bound the full product norm and omit the output
projector. The Fourier coefficients of the kernel give
$\norm{w_>}_{L^2}\le(c_\Omega')^{1/2}K^{-1/2}$, as in the proof
of Lemma~\ref{lem:tail}. To bound this kernel acting on $\ket{\psi}$, we hold
all coordinates other than $\bm{x}_a$ fixed. The partner coordinate
$\bm{x}_b$ is then fixed as well, and translation of the periodic kernel
does not change its $L^2$ norm. Applying the three-dimensional
embedding $H^2(\T)\hookrightarrow L^\infty(\T)$ on this slice
and then integrating the fixed coordinates gives
\begin{equation}
 \norm{w_>(\bm{x}_a-\bm{x}_b)\ket{\psi}}
 \le C_S\norm{w_>}_{L^2}
             \norm{(1+\norm{\bm{k}_{\bm{p}_a}}^2)\ket{\psi}}.
\end{equation}
This is the same slice argument as Eq.~\eqref{eq:sliceSobolev}:
the square of the last norm is the integral, over the other
coordinates, of the squared $H^2$ norm in $\bm{x}_a$ alone. We never
require a supremum bound in the full many-particle configuration
space. For $w_>(\bm{x}_a-\bm{R}_\ell)$, we use the same slice with the
fixed centre $\bm{R}_\ell$ in place of $\bm{x}_b$. Translation again leaves
the kernel's $L^2$ norm unchanged, so its product with the state
has the same bound.

The remaining step converts this one-particle Sobolev weight to the
mass-weighted kinetic weight defining $\Theta$. Since
$\norm{\bm{k}_{\bm{p}_a}}^2\le2m_aA(\vec{\bm{k}})$ and $m_a\ge1$,
\begin{equation}
 1+\norm{\bm{k}_{\bm{p}_a}}^2\le1+2m_aA(\vec{\bm{k}})
 \le3m_a(1+A(\vec{\bm{k}})),\qquad
 \norm{(1+\norm{\bm{k}_{\bm{p}_a}}^2)\ket{\psi}}\le3m_a\Theta.
\end{equation}
Thus each large-transfer term contributes at most
$C m_a\Theta K^{-1/2}$. The factor $m_a$ is the cost of obtaining
two derivatives in the restricted particle's coordinate from a
kinetic-energy bound. All slice estimates hold componentwise in
internal spin and then sum in squared norm, with no spin
multiplicity factor.

Finally, the sum of the absolute interaction weights satisfies
\begin{equation}
 \sum_{b<c}|q_bq_c|
 +\sum_b\sum_{\ell=L_{\mathrm q}+1}^{L_{\mathrm{nuc}}}|q_b|\zeta_\ell
 \le\tfrac12\left(\sum_b|q_b|
       +\sum_{\ell=L_{\mathrm q}+1}^{L_{\mathrm{nuc}}}\zeta_\ell\right)^2
 =\tfrac12(\eta+\lambda_\zeta)^2.
\end{equation}
This includes the nucleus--nucleus products, whose total weight can
be of order $\lambda_\zeta^2$, including interactions between
quantum and clamped nuclei. We combine the three classes by the
triangle inequality and use $m_a\le(m_am_{\max})^{1/2}$ in the last
two classes. This proves Eq.~\eqref{eq:nbo-lambda}, with
$C_\Lambda$ independent of the masses and the cutoff. The
large-transfer class determines the slowest decay, $K^{-1/2}$;
the other two classes satisfy stronger cutoff bounds.
\end{proofx}

The proof of Lemma~\ref{lem:nbo-tail} uses only finiteness of
$\Theta$, not the eigenvalue equation. The same potential-tail
estimate is therefore available for other states with this
regularity. We now use stationarity to obtain the stronger tails
needed for the ground-energy bound.

\subsection{Bootstrap and energy bound}

We complete the proof by dividing the Fourier eigenvalue equation
by the kinetic multiplier. The lower bound on that multiplier
retains the mass of the particle whose momentum is restricted.
After deriving the improved tails, we will sum them and eliminate
$\Theta$ using Proposition~\ref{prop:nbo-KR}.

\begin{proofx}[Completion of the proof of Theorem~\ref{thm:nbo}]
For every momentum and spin configuration, the eigenvalue equation is
\begin{equation}
 A(\vec{\bm{k}})\psi_{\vec{\bm{p}},\boldsymbol\sigma}
 =E\psi_{\vec{\bm{p}},\boldsymbol\sigma}
 -\langle\vec{\bm{p}},\boldsymbol\sigma|V_{\mathrm{tot}}|\psi\rangle.
\end{equation}
On the region $\norm{\bm{k}_{\bm{p}_a}}>K$, we have
$A(\vec{\bm{k}})>K^2/(2m_a)$. Moreover,
$\norm{\bm{k}_{\bm{p}_m}}\le(2m_m)^{1/2}A(\vec{\bm{k}})^{1/2}$ for every
particle index $m$. These two facts give
\begin{equation}
 \frac1{A(\vec{\bm{k}})}\le\frac{2m_a}{K^2},\qquad
 \frac{\norm{\bm{k}_{\bm{p}_m}}}{A(\vec{\bm{k}})}
 \le\frac{2(m_am_m)^{1/2}}{K}.
 \label{eq:nbo-weights}
\end{equation}
In the second estimate, we first bound the ratio by
$(2m_m)^{1/2}A(\vec{\bm{k}})^{-1/2}$ and then use the lower bound on
$A$. This derivation covers both $m=a$ and $m\ne a$.
It shows explicitly the two inverse powers of $K$ gained for
the state tail and the one inverse power retained after inserting
a derivative.

Dividing the eigenvalue equation by $A$, restricting to the
chosen high-momentum region, and applying the triangle inequality,
we obtain
\begin{align}
 t_a(K)&\le\frac{2m_a}{K^2}
                  \big(|E|t_a(K)+\Lambda_a(K)\big),\\
 s_{a,m}(K)&\le\frac{2(m_am_m)^{1/2}}{K}
                  \big(|E|t_a(K)+\Lambda_a(K)\big).
\end{align}
Choose $K^2\ge4m_{\max}|E|$. Then
$2m_a|E|/K^2\le1/2$ for every particle, so moving the $t_a$
term in the first inequality to the left gives
\begin{equation}
 \left(1-\frac{2m_a|E|}{K^2}\right)t_a(K)
 \le\frac{2m_a}{K^2}\Lambda_a(K),\qquad
 t_a(K)\le\frac{4m_a}{K^2}\Lambda_a(K).
\end{equation}
The same cutoff condition implies
$|E|t_a(K)\le\Lambda_a(K)$. Thus the right side of the
derivative-tail inequality contains at most twice the potential
tail. Applying Lemma~\ref{lem:nbo-tail}, also with $K\ge1$,
we obtain
\begin{align}
 t_a(K)&\le\frac{4m_a\Lambda_a(K)}{K^2}
 \le4C_\Lambda m_a^{3/2}m_{\max}^{1/2}\Theta K^{-5/2},
 \label{eq:nbo-tfinal}\\
 s_{a,m}(K)&\le\frac{4(m_am_m)^{1/2}\Lambda_a(K)}{K}
 \le4C_\Lambda m_a(m_mm_{\max})^{1/2}\Theta K^{-3/2}.
 \label{eq:nbo-sfinal}
\end{align}
The powers of $K$ match the BO estimates. The masses record how
much weaker the kinetic multiplier is on the corresponding
particle coordinates.

To recover the full discarded $H^1$ norm, square these estimates
and sum as in Eq.~\eqref{eq:nbo-red}. The amplitude tails involve
$\sum_a m_a^3=S_3$, while the derivative tails involve
$\sum_a m_a^2\sum_m m_m=S_2S_1$. Hence
\begin{align}
 \norm{\ket{\delta}}_{H^1}^2
 &\le16C_\Lambda^2\Theta^2
 \left[m_{\max}\sum_a m_a^3 K^{-5}
       +m_{\max}\sum_a m_a^2\sum_m m_m K^{-3}\right]
 \\
 &\le32C_\Lambda^2m_{\max}^2S_1S_2\Theta^2K^{-3}.
 \label{eq:nbo-assembly}
\end{align}
For the last inequality, $S_3\le m_{\max}S_2$ and $K^{-5}\le K^{-3}$.
The first term is then bounded by
$m_{\max}^2S_1S_2K^{-3}$ because $S_1\ge1$; the second obeys
the same bound because $m_{\max}\ge1$. Combining the two terms
accounts for the factor $32$.

We must also check the norm of the trial state. The amplitude
estimate alone gives
\begin{equation}
 \norm{\ket{\delta}}^2
 \le16C_\Lambda^2m_{\max}S_3\Theta^2K^{-5}.
\end{equation}
Thus $\norm{P\ket{\psi}}^2=1-\norm{\ket{\delta}}^2\ge1/2$ whenever
$K^5\ge32C_\Lambda^2m_{\max}S_3\Theta^2$.
Using Eq.~\eqref{eq:nbo-H2}, it suffices to choose $K$ so that
\begin{equation}
 K\ge1,\qquad K^2\ge4m_{\max}|E|,\qquad
 K^5\ge32C_\Lambda^2C_0^2m_{\max}^3S_3(1+|E|)^2.
\end{equation}
Since $S_3\le\eta_{\mathrm{tot}}m_{\max}^3$, these conditions
admit a threshold bounded polynomially in $m_{\max}$ and
$1+|E|$, with the other system data fixed. The corresponding
admissible basis-size threshold $N_0$ is also polynomially
bounded, using $K=\pi(N^{1/3}-1)/L$ at fixed $L$.

With the trial-state condition verified, we combine
Eqs.~\eqref{eq:nbo-sandwich} and~\eqref{eq:nbo-assembly}:
\begin{equation}
 0\le E_N-E
 \le64C_1C_\Lambda^2m_{\max}^2S_1S_2\Theta^2K^{-3}.
\end{equation}
The graph-norm estimate contributes
$\Theta^2\le C_0^2m_{\max}^2(1+|E|)^2$, and the variational
constant $C_1$ contributes at most a further factor $1+|E|$,
times a polynomial in the particle and charge parameters. We
therefore obtain precisely the form of Eq.~\eqref{eq:nbo-main},
\begin{equation}
 0\le E_N-E
 \le CC_0^2(1+|E|)^3m_{\max}^4S_1S_2K^{-3}.
\end{equation}
The coarse estimate follows from
$S_1\le\eta_{\mathrm{tot}}m_{\max}$ and
$S_2\le\eta_{\mathrm{tot}}m_{\max}^2$, with $C_0$ absorbed into
the coarse constant. Finally, the admissible-cutoff conversion in
Sec.~\ref{sec:setting} gives $K^{-3}\le C_LN^{-1}$ for
sufficiently large admissible $N$. This completes the theorem
without comparing exact and discrete eigenvectors and without a
spectral-gap estimate.
\end{proofx}

\subsection{Remarks on quantum nuclei}

The theorem establishes a cutoff exponent and explicit upper bounds
on its coefficient. We now explain the role of the periodic cell,
what the mass factors do and do not imply, and how the same tail
argument can guide a choice of different cutoffs for different
species.

\medskip\noindent\emph{Why translation invariance causes no spectral
obstruction on the torus.}
This translation-invariance discussion concerns the fully quantum
case, with no clamped nuclei.
In free space, the centre of mass of an isolated translation-invariant
system can move freely. Even if the internal relative motion is
bound, this free motion prevents the internal ground state from
being a normalisable ground-state eigenfunction of the full
translation-invariant Hamiltonian. One therefore separates the
centre-of-mass motion, or resolves the operator at fixed total
momentum, when formulating the internal eigenvalue problem.
The bottom of the full spectrum can still be defined as an energy
infimum without such an eigenfunction.

On the torus, the whole configuration space is compact.
Proposition~\ref{prop:nbo-KR} gives a compact resolvent and a
discrete spectrum even though the pair interactions remain
translation invariant. Here translation invariance organises
states into sectors of fixed total momentum, the sum of their
particle momenta. The theorem applies to the bottom of the chosen
exchange sector, and the same proof applies after also fixing total
momentum whenever the resulting sector is nonzero. This restriction
is compatible with $P$, which is diagonal in the particle momenta
and therefore commutes with the total-momentum projection.
Clamped nuclei generally break translation invariance, so the
hybrid result does not use a total-momentum decomposition. Its
compact-resolvent and ground-energy arguments remain valid.

Nor do closely spaced molecular rotational and vibrational levels
create a denominator in the proof. We use an exact eigenfunction
as a variational trial state after projection, rather than identify
a discrete eigenvector by perturbation theory. Thus neither ground-state
degeneracy nor a small gap to nearby levels changes the error
argument. A separate eigenvector comparison would require its own
spectral assumptions, as explained in Remark~\ref{rem:trunc}.

\medskip\noindent\emph{Where the mass factors enter.}
Beyond the mass dependence of $E$ itself, the explicit mass factors
come from three conversions. First, a high momentum of particle
$a$ only implies the lower kinetic-energy bound
$A\ge K^2/(2m_a)$, which enters both
Eqs.~\eqref{eq:nbo-apriori} and~\eqref{eq:nbo-weights}.
Second, the large-transfer part of Lemma~\ref{lem:nbo-tail}
uses two derivatives in that particle's coordinate, and converting
them to the kinetic weight gives
$\norm{(1+\norm{\bm{k}_{\bm{p}_a}}^2)\ket{\psi}}\le3m_a\Theta$.
Third, eliminating $\Theta$ through
Proposition~\ref{prop:nbo-KR} can incur a large mass in the
Hardy estimate for a nucleus--nucleus interaction, including one
between a quantum nucleus and a clamped nucleus. These factors measure
how strongly the energy norm controls high momenta. They are not
measurements of the high-momentum amplitude actually present in
a particular molecular ground state.

This distinction motivates the expectation that the constants may
be much smaller than our bound suggests. More detailed high-momentum
information can come from coalescence cusps, the non-smooth local
behaviour of a Coulomb wavefunction when two charged particles
approach the same position. The relative-coordinate equation for
particles $a$ and $b$ contains their reduced mass
\begin{equation}
 \mu_{ab}:=\frac{m_am_b}{m_a+m_b}.
\end{equation}
For two electrons $\mu_{ab}=1/2$, and for an electron--nucleus
pair it is at most one, independently of the nuclear mass.
For two nuclei of equal mass $m$, however, it equals $m/2$.
Thus the mass entering a local cusp relation need not be the large
$m_{\max}$ used in a uniform energy-based estimate.

A bounded reduced mass alone does not prove that the Fourier-tail
coefficient is uniform in the nuclear masses. That coefficient also
involves the wavefunction near coincidence, whose size and mass
dependence must be controlled. Nuclear repulsion may make
nucleus--nucleus coincidence contributions small in molecular
states, but this is an additional property to establish, not an
estimate supplied by the present argument. The expectation that
actual error constants could be essentially mass-uniform is
therefore a physical and analytic motivation for a sharper result,
not a conclusion of Theorem~\ref{thm:nbo}. Proving such an
improvement would require structural cusp information of the type
discussed in Ref.~\cite{FHOS}, adapted to moving nuclei, together
with control of its coefficients. Our proof deliberately obtains
the $N^{-1}$ exponent without that additional input.

The cutoff threshold must be interpreted with the same care.
Our proof supplies sufficient conditions, including
$K^2\ge4m_{\max}|E|$, under which the high-momentum absorption
and trial-state normalisation work. These conditions give a
polynomially bounded choice of $N_0$ in the masses and
$1+|E|$. They do not prove that every system needs a cutoff growing
with precisely those powers before the asymptotic rate is visible.
They quantify the resolution certified by this argument; whether
a particular nuclear state needs that resolution depends on its
actual momentum distribution.

\medskip\noindent\emph{What the per-particle truncation estimate
certifies.}
Equation~\eqref{eq:nbo-tfinal} gives the explicit bound
\begin{equation}
 t_a(K)\le4C_\Lambda m_a^{3/2}m_{\max}^{1/2}\Theta K^{-5/2}.
\end{equation}
At fixed $\Theta$ and common cutoff, its coefficient is larger
for a nuclear coordinate than for an electronic one by the factor
$m_a^{3/2}$. The graph-norm estimate can eliminate $\Theta$, but
then introduces its own system and mass dependence. Keeping
$\Theta$ visible here separates the particle-specific factor from
the common regularity bound.

As in Remark~\ref{rem:trunc}, this is a statement about the Fourier
truncation of the exact eigenfunction. It does not bound the
difference from a selected eigenvector of the finite Hamiltonian.
Consequently, convergence of the computed energy at rate
$\mathcal O(K^{-3})$ does not by itself certify convergence of
nuclear-coordinate observables evaluated in that eigenvector.
Such a conclusion requires an observable or eigenspace analysis
in addition to the energy theorem. The per-particle estimate
identifies one approximation quantity available for that analysis;
it does not replace it.

\medskip\noindent\emph{Species-dependent cutoffs.}
The proof can also be used with different cutoffs for distinct
species, provided all particles of any one identical species
receive the same cutoff. This proviso preserves exchange symmetry:
exchanging two identical particles must leave the product Fourier
space unchanged. If their cutoffs differ, the product projector
need not preserve the chosen exchange sector, and the projected
state need not be an admissible trial state for our variational
argument.

To see how the estimates change, let $K_a$ be the cutoff assigned
to particle $a$, with $K_a=K_s$ for all particles of species $s$.
The tail proof is local to the restricted particle and therefore
applies with $K$ replaced by $K_a$. In particular, for
$K_a\ge1$ and $K_a^2\ge4m_a|E|$, the same calculation gives
\begin{equation}
 \norm{\ket{\delta}}_{H^1}^2
 \le16C_\Lambda^2\Theta^2m_{\max}
 \left[\sum_a m_a^3K_a^{-5}
              +S_1\sum_a m_a^2K_a^{-3}\right].
\end{equation}
Here $P$ and $\ket{\delta}=(1-P)\ket{\psi}$ use the new product cutoff.
Since $m_a\le m_{\max}$ and $K_a\ge1$, this implies the simpler
bound
\begin{equation}
 \norm{\ket{\delta}}_{H^1}^2
 \le16C_\Lambda^2\Theta^2m_{\max}(m_{\max}+S_1)
                  \sum_a m_a^2K_a^{-3}.
\end{equation}
The trial-state norm condition must also hold; a sufficient
condition is
$16C_\Lambda^2\Theta^2m_{\max}\sum_a m_a^3K_a^{-5}\le1/2$.
Subject to these conditions, the variational estimate therefore
reduces the allocation problem to the weighted sum
$\sum_a m_a^2K_a^{-3}$, with common prefactors independent of
the individual cutoffs.

For a simple continuous approximation to a fixed qubit budget,
let $n_s$ be the number of particles of species $s$, each of
mass $m_s$. At fixed cell size and large cutoff, the number of
bits per Cartesian momentum component grows as $\log_2 K_s$,
up to a common additive constant and integer rounding. Ignoring
rounding and assuming the cutoff thresholds are already met, the
allocation problem is therefore to minimise
$\sum_s n_sm_s^2K_s^{-3}$ while fixing
$\sum_s n_s\log K_s$; the common factor of three Cartesian
components does not affect the optimum. Differentiating with
respect to $\log K_s$ in a Lagrange-multiplier calculation gives
\begin{equation}
 -3n_sm_s^2K_s^{-3}+\alpha n_s=0,
\end{equation}
where $\alpha$ enforces the budget. Thus $m_s^2K_s^{-3}$ is
constant across the species and
$K_s\propto m_s^{2/3}$. Relative to an electron, this suggests
about $\tfrac23\log_2m_\ell$ extra bits per Cartesian component
of a nuclear register, roughly $7$--$12$ bits for typical nuclear
masses in electron-mass units. Integer rounding and any active
cutoff thresholds can change this allocation.

This prescription optimises the stated upper bound. It need not
optimise the error of a specific molecular state, since the bound
may substantially overestimate its nuclear tails. For comparison,
a local harmonic mode with effective mass $m_\ell$ and frequency
$\omega$ has a characteristic momentum of order
$(m_\ell\omega)^{1/2}$. Here the mode is a small-oscillation
coordinate near equilibrium, approximated by a one-dimensional
harmonic oscillator; its effective mass is the parameter in the
kinetic energy $p^2/(2m_\ell)$, where $p$ is the momentum conjugate
to that coordinate. At fixed $\omega$, this grows more slowly
with mass than $m_\ell^{2/3}$. Such physical momentum information
may justify a different allocation, but it requires information
about the state and its modes. The two considerations both
motivate additional nuclear momentum resolution in this setting;
only the former follows from the conservative bound just derived.

\section{Smooth observables under Galerkin evolution}
\label{app:dynamics}
\label{sec:dyn}

We now study an expectation value in a time-dependent state,
rather than the energy of a stationary state. Our goal is to show
that a smooth position observable has discretisation error of order
$N^{-5/3}+tN^{-1}$ when each particle is represented by $N$ plane
waves. There are two contributions to control. The exact evolved state
has a high-momentum component outside the retained basis, and
coupling to that omitted component changes the evolution inside
the basis. Smoothness makes the first contribution quadratic in
small Fourier tails. For the second, we estimate the inner products
that enter the expectation value, keeping more information than a
direct bound on the state-error norm would retain.

The central step is to move the discrete propagator onto a test
state obtained by applying the observable. We then evolve that
state backwards from the observation time. Its regularity lets us
pair a $K^{-5/2}$ discarded-state tail with a $K^{-1/2}$ potential
tail, giving the coefficient $K^{-3}$ in the error growing with
time. Here $K$ is the momentum cutoff, with $K=\Theta(N^{1/3})$
at fixed cell size. We first establish the regularity and tail
estimates needed during both exact evolution and this backward
discrete evolution.

\subsection{Setting and hypotheses}
\label{sec:dyn-setting}

We use the periodic Coulomb Hamiltonian $\Hnbo=\Ttot+V_{\mathrm{tot}}$
of Eq.~\eqref{eq:Hnbo} and the spin-inclusive exchange sector
$\mathcal S$ from Sec.~\ref{sec:nonBO}. The quantum particles may
include nuclei, and any remaining nuclei may be clamped. Thus the
same statement covers fully quantum, BO and hybrid nuclear treatments.
The fixed system data in the constants below include the charges and
positions of the clamped nuclei. Recall that $\Ttot$ is diagonal in the
Fourier basis, with multiplier
$A(\vec{\bm{k}})=\sum_a\norm{\bm{k}_{\bm{p}_a}}^2/(2m_a)$, and that $P=P_N$
retains the cube $\norm{\bm{k}_{\bm{p}_a}}_\infty\le K$ for every particle.
We use kets for states and momentum-basis amplitudes for their
coefficients. With $\vec{\bm{k}}$ denoting the full momentum configuration
and $\sigma$ the internal spin configuration, our convention is
\begin{equation}
 u_{\vec{\bm{k}},\sigma}:=\langle\vec{\bm{k}},\sigma|u\rangle,\qquad
 \ket{u}=\sum_{\vec{\bm{k}},\sigma}u_{\vec{\bm{k}},\sigma}\ket{\vec{\bm{k}},\sigma}.
\end{equation}
The spin label and its sum are omitted in a spinless setting.
For an operated state, its amplitude is written directly as
$\langle\vec{\bm{k}},\sigma|B|u\rangle$. By orthonormality of the
plane-wave basis, the kinetic graph norm can equivalently be written
\begin{equation}
 \Theta(\ket{u})
 :=\norm{(1+\Ttot)\ket{u}}
 =\left(\sum_{\vec{\bm{k}},\sigma}(1+A(\vec{\bm{k}}))^2
                              |u_{\vec{\bm{k}},\sigma}|^2\right)^{1/2}.
\end{equation}
An indicator such as $\ind{\norm{\bm{k}_{\bm{p}_a}}>K}$ acting on a ket
denotes the diagonal momentum projector that retains exactly those
basis states whose particle-$a$ momentum satisfies that condition.
Hats are retained for Fourier coefficients or Fourier transforms
of scalar functions. In particular, $\hat\chi(\bm{\nu})$ below uses
the Fourier-series convention of Eq.~\eqref{eq:fourier-sign};
these coefficients are distinct from the state amplitudes just defined.

The exact and finite-basis evolved states are
\begin{equation}
\ket{\psi(t)}:=e^{-i\Hnbo t}\ket{\psi_0},\qquad
\ket{\psi_N(t)}:=e^{-iH_Nt}P\ket{\psi_0},\qquad
H_N:=P\Hnbo P.
\label{eq:dyn-def}
\end{equation}
We regard $H_N$ as an operator on the full finite-dimensional space
$\mathrm{ran}\,P$, where it is self-adjoint and its evolution is
unitary. Its restriction preserves $\mathcal S$: permutations of
identical quantum particles preserve masses, charges, and cutoffs,
and each such particle interacts with the same clamped nuclear
configuration. These permutations therefore commute with both
$\Hnbo$ and $P$. Consequently,
$\ket{\psi_N(t)}\in\mathrm{ran}\,P\cap\mathcal S$ whenever
$\ket{\psi_0}\in\mathcal S$.

We distinguish the full state error from the part of the exact state
omitted by the Fourier projection:
\begin{equation}
\ket{e(t)}:=\ket{\psi(t)}-\ket{\psi_N(t)},\qquad
\ket{\delta(t)}:=(1-P)\ket{\psi(t)}.
\end{equation}
The remaining part of $\ket{e(t)}$ lies inside $\mathrm{ran}\,P$ and
will be represented by a Duhamel integral below.

\emph{Initial state.}
We assume $\norm{\ket{\psi_0}}=1$ and
$\ket{\psi_0}\in D(\Hnbo^2)\cap\mathcal S$.
The operator-domain condition means that $\ket{\psi_0}\in D(\Hnbo)$
and $\Hnbo\ket{\psi_0}\in D(\Hnbo)$, so the Hamiltonian may be applied
twice with an $L^2$ result. We record these two applications in
\begin{equation}
\mathcal D:=\norm{\ket{\psi_0}}+\norm{\Hnbo\ket{\psi_0}}
                         +\norm{\Hnbo^2\ket{\psi_0}}.
\label{eq:Ddef}
\end{equation}
Under exact evolution all three norms are conserved. We will use
them to obtain uniform kinetic regularity for both $\ket{\psi(t)}$ and
$\Hnbo\ket{\psi(t)}$; the latter replaces $E\ket{\psi}$ in the ground-state
tail argument.

\begin{remark}[Examples of initial states]
Finite linear combinations of eigenstates satisfy the operator-domain
assumption. Two other analytic ways to obtain admissible regularity
are imaginary-time smoothing $e^{-\tau\Hnbo}$, for $\tau>0$, and
the squared resolvent $(\Hnbo-z)^{-2}$, where $z$ lies outside the
spectrum. Applied to a nonzero vector in $\mathcal S$, these
operators preserve the sector; the output can then be normalised.
For imaginary-time smoothing, the multiplier
$\lambda^j e^{-\tau\lambda}$ is bounded on the spectrum for every
integer $j\ge0$, since $\Hnbo$ is bounded below. Thus the output
lies in $D(\Hnbo^j)$ for every $j$. For the squared resolvent,
$\lambda^2/(\lambda-z)^2$ is bounded, giving membership in
$D(\Hnbo^2)$. A single resolvent guarantees only $D(\Hnbo)$ in
general. These examples describe analytic regularity, rather than
additional formal state-preparation claims.

The domain of the squared Hamiltonian must not be confused with
fourth-order Sobolev regularity. A Coulomb eigenfunction lies in
$D(\Hnbo^j)$ for every $j$, but its cusp can produce a
 distributional singularity when four spatial derivatives are taken;
this already occurs for the hydrogen ground state. Conversely, a
smooth function that does not vanish at a Coulomb singularity can
be taken out of $H^2$ by multiplication with the potential, so
$H^4$ membership does not generally ensure membership in
$D(\Hnbo^2)$. We therefore do not replace the stated domain
assumption by an $H^4$ assumption. The proof below obtains the
needed momentum decay from the Hamiltonian equation itself.
\end{remark}

\emph{Observable.}
Choose a particle label $a\in\{1,\ldots,\eta_{\mathrm{tot}}\}$,
for example a quantum nuclear coordinate, and let
$\chi:\T\to[0,1]$ be a smooth approximation to the indicator of
a region $\mathcal R\subset\T$. Set $O:=\chi(\bm{x}_a)$ and, when
convenient, write $\chi_a$ for the same multiplication operator.
Then $\langle\psi(t)|O|\psi(t)\rangle$ is the smoothed probability of finding
particle $a$ in that region. The Galerkin state in
Eq.~\eqref{eq:dyn-def} is deliberately left unnormalised:
$\norm{\ket{\psi_N(t)}}=\norm{P\ket{\psi_0}}\le1$. Its observable quadratic
form is therefore not itself a probability unless this norm is one.
The theorem first compares these two quadratic forms;
Corollary~\ref{cor:norm} accounts for normalisation afterwards.

We require the following quantitative smoothness properties.

\smallskip\noindent
\textbf{Hypothesis (H$\chi$).} The Fourier coefficients satisfy
\begin{equation}
C_\chi:=\sum_{\bm{\nu}\in\Z^3}|\hat\chi(\bm{\nu})|<\infty,\qquad
\sum_{\norm{\bm{k}_{\bm{\nu}}}_\infty>\rho}|\hat\chi(\bm{\nu})|
\le C_{\chi,M}\rho^{-M}
\end{equation}
for every $M\ge1$ and $\rho\ge1$, and
\begin{equation}
D_\chi:=\norm{\nabla\chi}_{L^\infty}
              +\norm{\Delta\chi}_{L^\infty}<\infty.
\end{equation}
The first constant controls the total strength of the observable's
Fourier transfers; the second family of constants controls its
large transfers. The derivative bound will control the commutator
between $O$ and kinetic energy. For use in the subsequent grid
analysis, we also retain the finite weighted Fourier sum
\begin{equation}
D_\chi^\prime:=\sum_{\bm{\nu}\in\Z^3}\|\bm{\nu}\|^2|\hat\chi(\bm{\nu})|.
\label{eq:Dchi-prime}
\end{equation}
This second Fourier moment is an absolute coefficient sum weighted
by the squared lattice index. Its finiteness follows by summing the
rapid tail bound in (H$\chi$), for example with an exponent greater
than two. It is not an additional regularity hypothesis.

Convolving a region indicator with a smooth compactly supported
averaging kernel produces an example satisfying (H$\chi$).
We quantify the dependence on the averaging width in the final
remarks. For the stated powers of $N$, this width and the resulting
observable constants are held fixed as the basis grows.
For identical particles, a physically symmetric observable is
$\sum_{a\in\mathrm{species}}\chi(\bm{x}_a)$; its error follows by
summing the one-particle estimates. A single labelled multiplier
$\chi(\bm{x}_a)$ need not preserve $\mathcal S$. For this reason our
backward test states will evolve on the full retained space,
where all the regularity estimates remain valid.

\begin{theorem}[Observable error after periodic Coulomb evolution:
energy-rate secular bound]
\label{thm:dyn}
Assume (H$\chi$) and $\ket{\psi_0} \in D(\Hnbo^2) \cap \mathcal{S}$. There are constants $C$, $K_0$ depending only on the fixed
particle, mass, charge and cell data, such that for
all $K \ge K_0$, every $M$, and \emph{all} $t \ge 0$,
\begin{align}
\big| \langle\psi(t)|O|\psi(t)\rangle &- \langle\psi_N(t)|O|\psi_N(t)\rangle \big|
\\
&\le\;
C\, (1 + C_\chi + D_\chi)\, \Xi\, (\Xi + m_{\max})\,
\Big[ K^{-5} \;+\; t\, K^{-3} \Big]
+ C\, C_{\chi,M}\, \Xi\, K^{-M},
\label{eq:dyn-main}
\end{align}
where $\Xi := C_0\, m_{\max}\, \mathcal{D}$ is the uniform-in-time
regularity constant of Proposition~\ref{prop:dyn-reg}. Choosing any
fixed $M \ge 5$, as (H$\chi$) permits --- so that the $K^{-M}$
remainder is dominated by the $K^{-5}$ term --- this gives
\begin{equation}
\big| \Delta\langle O\rangle(t) \big|
\;=\; \mathcal{O}\!\big( N^{-5/3} + t\, N^{-1} \big) :
\label{eq:dyn-simple}
\end{equation}
the non-secular term is second order in the discarded tail, and the
secular coefficient of the \emph{bound} matches the
energy-error rate of
Theorems~\ref{thm:main} and~\ref{thm:nbo}. No spectral gap enters.
We do not prove that the secular coefficient is attained
(discussed after the proof).
\end{theorem}

The two powers of the cutoff come from different pairings. For the
time-independent contribution, the omitted state and the omitted
part of the observable-multiplied state each have norm of order
$K^{-5/2}$, so their inner product is of order $K^{-5}$.
For the accumulated error, the backward test state instead supplies
a potential tail of order $K^{-1/2}$, which pairs with the
$K^{-5/2}$ state tail. We prove both statements below, then check
the quadratic terms in the observable error. This last check is
essential: simply squaring a time-growing norm estimate would
introduce a $t^2$ term absent from the theorem.

The resulting $K^{-3}$ coefficient agrees with the energy-error
power, but the theorem asserts an upper bound, not an attained
rate of drift. The elementary state-norm estimate established here
is weaker; a sharper estimate for the retained-state error appears
later in Eq.~\eqref{eq:kinetic-retained-galerkin}. Our argument
does not require a spectral gap. The constants may depend on all
fixed masses, as in Appendix~\ref{app:nbo}; the explicit inverse
mass in the observable's kinetic commutator does not by itself
make the complete estimate uniform as masses vary.

To use the result as a comparison of probabilities, we must also
account for normalising the initial projection. Unitary discrete
evolution preserves its norm, so the correction depends only on
the probability discarded at time zero. The next corollary makes
this correction explicit. Its cutoff threshold depends on the
initial state, whereas the threshold in the theorem depends only
on the fixed system.

\begin{corollary}[Normalised Galerkin state]
\label{cor:norm}
For
\begin{equation}
K \;\ge\; K_{\mathrm{norm}}(\ket{\psi_0}) := \max\Big\{ K_0,\;
\big( 2 C_{\mathrm{tail}}\, m_{\max} S_3 \Xi^2 \big)^{1/5} \Big\} ,
\end{equation}
so that Theorem~\ref{thm:dyn} applies \emph{and}
$\norm{(1-P)\ket{\psi_0}}^2 \le \tfrac12$ by \eqref{eq:dyn-t} at $t=0$.
Unlike the system-only $K_0$, this threshold is polynomially bounded
in the initial-state regularity $\mathcal{D}$ (through $\Xi$,
scaling as $\mathcal{D}^{2/5}$), one has
$\norm{P\ket{\psi_0}}^2 \ge \tfrac12$, so the
normalised Galerkin state
$\ket{\widetilde\psi_N(t)} := e^{-iH_N t}\, P\ket{\psi_0} / \norm{P\ket{\psi_0}}$ is
well defined, and $\langle\widetilde\psi_N(t)|O|\widetilde\psi_N(t)\rangle$
is a genuine probability. For $0 \le O \le 1$,
\begin{align}
\big| \langle\widetilde\psi_N(t)|O|\widetilde\psi_N(t)\rangle
- \langle\psi_N(t)|O|\psi_N(t)\rangle \big|
&= \Big( \norm{P\ket{\psi_0}}^{-2} - 1 \Big)\,
\langle\psi_N(t)|O|\psi_N(t)\rangle
\\
&\le \norm{(1-P)\ket{\psi_0}}^2
\;\le\; C_{\mathrm{tail}}\, m_{\max}\, S_3\, \Xi^2\, K^{-5},
\end{align}
using $\langle\psi_N|O|\psi_N\rangle \le \norm{\ket{\psi_N}}^2 = \norm{P\ket{\psi_0}}^2$
in the middle step and \eqref{eq:dyn-t} at $t = 0$ in the last, whose
constant we here name $C_{\mathrm{tail}}$ (it is the same constant
that defines $K_{\mathrm{norm}}$ above).
Hence the bound \eqref{eq:dyn-main} holds verbatim, with an enlarged
constant, for
$\langle\psi(t)|O|\psi(t)\rangle -
\langle\widetilde\psi_N(t)|O|\widetilde\psi_N(t)\rangle$: the normalisation
correction is of the same order as the non-secular term.
\end{corollary}

\subsection{Propagated regularity and uniform-in-time tails}

We first establish bounds that remain valid throughout exact
evolution. The equation $\Hnbo=\Ttot+V_{\mathrm{tot}}$ lets us
control the kinetic graph norm---the norm of the state together
with its kinetic-energy image---by the conserved Hamiltonian norms.
Applying this comparison to both $\ket{\psi(t)}$ and $\Hnbo\ket{\psi(t)}$
provides the two inputs needed to repeat the Fourier-tail argument
for a state that is not an eigenstate.

\begin{proposition}[Uniform-in-time regularity]
\label{prop:dyn-reg}
For every $\ket{u} \in D(\Hnbo)$ the display in the proof of
Proposition~\ref{prop:nbo-KR} gives the general-state elliptic
estimate
\begin{equation}
\norm{\Ttot \ket{u}} \;\le\; 2 \norm{\Hnbo \ket{u}} + C m_{\max} \norm{\ket{u}} .
\label{eq:ell-gen}
\end{equation}
Consequently, with
$\Theta(t) := \norm{(1+\Ttot)\ket{\psi(t)}}$ and
$\Theta^{(1)}(t) := \norm{(1+\Ttot)\Hnbo\ket{\psi(t)}}$,
conservation of $\norm{\Hnbo^j \ket{\psi(t)}}$ yields, uniformly in $t$,
\begin{equation}
\Theta(t) + \Theta^{(1)}(t) \;\le\; \Xi := C_0\, m_{\max}\, \mathcal{D}.
\label{eq:Xi}
\end{equation}
\end{proposition}

\begin{proofx}
We use the relative potential estimate from
Proposition~\ref{prop:nbo-KR} in the form
\begin{equation}
 \norm{V_{\mathrm{tot}}\ket{u}}
 \le\tfrac12\norm{\Ttot \ket{u}}+Cm_{\max}\norm{\ket{u}}.
\end{equation}
Since $\Ttot \ket{u}=\Hnbo \ket{u}-V_{\mathrm{tot}}\ket{u}$, the triangle
inequality gives
\begin{equation}
 \norm{\Ttot \ket{u}}
 \le\norm{\Hnbo \ket{u}}+\tfrac12\norm{\Ttot \ket{u}}
                         +Cm_{\max}\norm{\ket{u}}.
\end{equation}
Move the half of the kinetic norm to the left to obtain
Eq.~\eqref{eq:ell-gen}, enlarging the fixed-system constant.
Under the Fourier transform, $1+\Ttot$ is multiplication by
$1+A$, so this also bounds
$\Theta(\ket{u})=\norm{(1+\Ttot)\ket{u}}$ by
$C(\norm{\Hnbo \ket{u}}+m_{\max}\norm{\ket{u}})$.

The spectral calculus for the self-adjoint Hamiltonian shows that
its unitary evolution commutes with its powers and preserves their
domains. Hence $\ket{\psi(t)}\in D(\Hnbo^2)$ and
\begin{equation}
 \norm{\Hnbo^j\ket{\psi(t)}}=\norm{\Hnbo^j\ket{\psi_0}},
 \qquad j=0,1,2.
\end{equation}
We may therefore apply the preceding comparison first to $\ket{\psi(t)}$
and then to $\Hnbo\ket{\psi(t)}\in D(\Hnbo)$. The results are
\begin{align}
 \Theta(t)&\le C\big(\norm{\Hnbo\ket{\psi_0}}
                             +m_{\max}\norm{\ket{\psi_0}}\big),\\
 \Theta^{(1)}(t)&\le C\big(\norm{\Hnbo^2\ket{\psi_0}}
                             +m_{\max}\norm{\Hnbo\ket{\psi_0}}\big).
\end{align}
Adding these inequalities and using $m_{\max}\ge1$ proves
Eq.~\eqref{eq:Xi}. This is where the second Hamiltonian application
in the initial-state hypothesis is needed.
\end{proofx}

We next recall which part of the ground-state tail proof applies
to general states. Lemma~\ref{lem:nbo-tail} bounds the potential
applied to a state using only its finite kinetic graph norm; the
eigenvalue equation is not used in that lemma. Thus for any
$\ket{\varphi}$ with $\Theta(\ket{\varphi})<\infty$ we have
\begin{equation}
\Lambda_a(K;\ket{\varphi}):=
\norm{\ind{\norm{\bm{k}_{\bm{p}_a}}>K}\,
                   V_{\mathrm{tot}}\ket{\varphi}}
\le C_\Lambda(m_am_{\max})^{1/2}\Theta(\ket{\varphi})K^{-1/2}.
\label{eq:lam-gen}
\end{equation}
Here the restriction measures the high-momentum part in particle
$a$, with the other coordinates and spins included in the norm.
The estimate includes both quantum-pair interactions and the
one-particle terms from clamped nuclei, as in Lemma~\ref{lem:nbo-tail}.
We will apply this estimate first to the exact state and later to
the backward discrete test state. The latter application is why
we have stated it without an exchange-sector restriction.

\begin{proposition}[Dynamical bootstrap]
\label{prop:dyn-boot}
For all $t$, every register $a$, and $K \ge\max\{1,4\pi/L\}$,
\begin{align}
\norm{\ind{\norm{\bm{k}_{\bm{p}_a}}>K}(1+\Ttot)\ket{\psi(t)}}
&\le C\, (m_a m_{\max})^{1/2}\, \Xi\, K^{-1/2},
\label{eq:dyn-wtail}\\
t_a(K; t) &\le C\, m_a^{3/2}\, m_{\max}^{1/2}\, \Xi\, K^{-5/2}.
\label{eq:dyn-t}
\end{align}
Here $t_a(K;t)$ is the register tail of Sec.~\ref{sec:nonBO}
evaluated on $\ket{\psi(t)}$.
\end{proposition}

\begin{proofx}
To improve the state tail, we isolate kinetic energy in the
Hamiltonian equation, for each momentum and spin configuration:
\begin{equation}
 A(\vec{\bm{k}})\psi_{\vec{\bm{k}},\sigma}(t)
 =\langle\vec{\bm{k}},\sigma|\Hnbo|\psi(t)\rangle
  -\langle\vec{\bm{k}},\sigma|V_{\mathrm{tot}}|\psi(t)\rangle.
\end{equation}
This identity replaces the eigenvalue equation used in
Appendix~\ref{app:nbo}. The Hamiltonian image
$\Hnbo\ket{\psi(t)}$ need not be a scalar multiple of
$\ket{\psi(t)}$, but its kinetic regularity is already controlled
by $\Theta^{(1)}(t)$. Taking the $\ell^2$ norm over amplitudes
with $\norm{\bm{k}_{\bm{p}_a}}>K$ and using the triangle inequality gives
\begin{equation}
 \norm{\ind{\norm{\bm{k}_{\bm{p}_a}}>K}\Ttot\ket{\psi(t)}}
 \le t_a(K;\Hnbo\ket{\psi(t)})+\Lambda_a(K;\ket{\psi(t)}).
\end{equation}
On this set $A\ge K^2/(2m_a)$, so the a priori estimate for the
first term and Eq.~\eqref{eq:lam-gen} for the second yield
\begin{equation}
 \norm{\ind{\norm{\bm{k}_{\bm{p}_a}}>K}\Ttot\ket{\psi(t)}}
 \le 2m_a\Theta^{(1)}(t)K^{-2}
       +C_\Lambda(m_am_{\max})^{1/2}\Theta(t)K^{-1/2}.
\end{equation}
For $K\ge1$ and $m_a\le m_{\max}$, the first coefficient is
bounded by the same mass and cutoff dependence as the second,
up to a numerical constant. The unweighted tail also satisfies
$t_a(K;t)\le2m_a\Theta(t)K^{-2}$. Adding it accounts for the
$1$ in $1+A$, and Eq.~\eqref{eq:Xi} proves
Eq.~\eqref{eq:dyn-wtail}.

Finally, on the same restricted set we may divide by the kinetic
weight:
\begin{equation}
 t_a(K;t)\le\frac{2m_a}{K^2}
                \norm{\ind{\norm{\bm{k}_{\bm{p}_a}}>K}\Ttot\ket{\psi(t)}}.
\end{equation}
Substitution gives Eq.~\eqref{eq:dyn-t}. Thus the potential-tail
power $K^{-1/2}$ gains two further inverse powers of momentum
when converted to a state tail. No eigenvalue-dependent absorption
or separate gradient-tail estimate is needed, and the cutoff
threshold is independent of the initial state's regularity budget.
\end{proofx}

We have obtained the same per-register tail powers as for an
eigenstate, uniformly in time, using the conserved Hamiltonian
content of the initial state in place of an eigenvalue. The displayed
mass factors are analytic refinements; coarser powers of
$m_{\max}$ suffice for the formal uniform bounds because
$m_a\le m_{\max}$. We do not identify these smaller prefactors
with independently checked Lean constants.

\subsection{Duhamel decomposition and the norm-level error}

We now separate the part of the error outside the basis from the
error that develops inside it. The latter starts at zero because
both retained states initially equal $P\ket{\psi_0}$. Its driving term
comes from the potential coupling the omitted exact state back
into the basis. Duhamel's formula expresses this driven evolution
as an integral of the source propagated by the discrete Hamiltonian.

\begin{proposition}[Norm-level error]
\label{prop:duhamel}
Write $\ket{e(t)} = \ket{\delta(t)} + \ket{e_P(t)}$ with
$\ket{e_P(t)} := P\ket{\psi(t)} - \ket{\psi_N(t)}$. Then
\begin{equation}
\ket{e_P(t)} = -i \int_0^t e^{-i H_N (t-s)}\, P\, V_{\mathrm{tot}}\, (1-P)\, \ket{\psi(s)}\, ds,
\label{eq:duhamel}
\end{equation}
and for $K \ge 1$,
\begin{align}
\norm{P V_{\mathrm{tot}} (1-P) \ket{\psi(s)}}
&\le C_{\mathrm{sec}}\, m_{\max}\, S_2^{1/2}\, \Xi\, K^{-3/2}
\quad \text{uniformly in } s,
\label{eq:secular}\\
\norm{\ket{e(t)}}
&\le C \Xi \big( m_{\max}^{1/2} S_3^{1/2} K^{-5/2}
+ m_{\max} S_2^{1/2}\, t\, K^{-3/2} \big),
\label{eq:normlevel}
\end{align}
with $C_{\mathrm{sec}}, C$ depending only on the fixed particle,
charge, mass and cell data. We emphasise that
\eqref{eq:secular}--\eqref{eq:normlevel} are \emph{norm-level}
statements; they are not the final input for the secular part of the
observable estimate, which is obtained at the better rate $K^{-3}$
by duality in Sec.~\ref{sec:dual}. They are, however, used there for
the cross term and for the regularity of $\ket{e_P(t)}$.
\end{proposition}

\begin{proofx}
The initial-state assumption makes $t\mapsto\ket{\psi(t)}$ continuously
differentiable in $D(\Hnbo)$, so we may differentiate the
projected state in the interaction representation:
\begin{align}
 \frac{d}{dt}\big(e^{iH_Nt}P\ket{\psi(t)}\big)
 &=i e^{iH_Nt}(H_NP-P\Hnbo)\ket{\psi(t)}\\
 &=-i e^{iH_Nt}P\Hnbo(1-P)\ket{\psi(t)}.
\end{align}
Both $P$ and $\Ttot$ are Fourier multipliers, so
$[\Ttot,P]=0$ and $P\Ttot(1-P)=0$. The source therefore
contains only the potential:
$P\Hnbo(1-P)=PV_{\mathrm{tot}}(1-P)$. Integrating from zero
to $t$ and multiplying by $e^{-iH_Nt}$ proves
Eq.~\eqref{eq:duhamel}. Unitarity on $\mathrm{ran}\,P$ then gives
\begin{equation}
 \norm{\ket{e_P(t)}}\le\int_0^t
                 \norm{PV_{\mathrm{tot}}\ket{\delta(s)}}\,ds.
\end{equation}

To bound this source we interpolate between the plain and
kinetically weighted tails. In physical space write
$\mathcal E:=1+\Ttot$, whose Fourier multiplier is $1+A$.
Since a mode excluded by the product cutoff is excluded in at
least one particle coordinate, the squared norm of its restriction
is at most the sum of the squared per-register restricted norms.
The Euclidean restrictions in Proposition~\ref{prop:dyn-boot}
also bound the cubical restrictions defining $P$. Consequently,
with $S_r=\sum_bm_b^r$ as before,
\begin{equation}
 \norm{\ket{\delta(s)}}\le C\Xi m_{\max}^{1/2}S_3^{1/2}K^{-5/2},
 \qquad
 \norm{\mathcal E\ket{\delta(s)}}
       \le C\Xi m_{\max}^{1/2}S_1^{1/2}K^{-1/2}.
\end{equation}
Positivity of $\mathcal E$ and Cauchy--Schwarz imply
\begin{equation}
 \norm{\mathcal E^{1/2}\ket{\delta}}^2
 =\langle\delta|\mathcal E|\delta\rangle
 \le\norm{\ket{\delta}}\,\norm{\mathcal E\ket{\delta}}.
 \label{eq:secular-interpolation}
\end{equation}
Thus the half-weighted norm is of order $K^{-3/2}$: the
product on the right has power $K^{-3}$ and we take its square
root. The Coulomb Hardy bound gives
$\norm{V_{\mathrm{tot}}\ket{z}}\le C_{\rm sys}
\norm{\mathcal E^{1/2}\ket{z}}$. Applying it to $\ket{z}=\ket{\delta(s)}$ and
using $\norm{P}\le1$ proves
$\norm{PV_{\mathrm{tot}}\ket{\delta(s)}}
\le C_{\rm sys}\Xi K^{-3/2}$, uniformly in $s$.
At fixed masses this has the form in Eq.~\eqref{eq:secular};
the constant there is allowed to depend on the masses as well
as the displayed factors.

The tail proposition directly supplies these estimates for
$K\ge\max\{1,4\pi/L\}$. To cover the remaining bounded interval
$1\le K<\max\{1,4\pi/L\}$ in the present statement, use
$\norm{\ket{\delta}}\le\norm{\ket{\psi}}$ and
$\norm{\mathcal E\ket{\delta}}\le\Theta(\ket{\psi})$, then enlarge
the constants by powers of this fixed system threshold.
This preserves all asserted dependencies. Integrating the source
bound and adding the bound on $\ket{\delta(t)}$ proves
Eq.~\eqref{eq:normlevel}.
\end{proofx}

\subsection{Smoothing makes the never-projected tail second order}
\label{sec:chitail}

We next turn the tail estimates into an observable estimate.
The never-projected tail is $\ket{\delta(t)}=(1-P)\ket{\psi(t)}$, the part
of the exact state that lies outside the finite basis at the
observation time. It enters the difference of quadratic forms
through
\begin{equation}
\Delta\langle O\rangle(t):=\langle\psi(t)|O|\psi(t)\rangle
                  -\langle\psi_N(t)|O|\psi_N(t)\rangle
=2\mathrm{Re}\langle e(t)|O|\psi(t)\rangle-\langle e(t)|O|e(t)\rangle.
\label{eq:obs-split}
\end{equation}
This identity follows by substituting $\ket{\psi_N}=\ket{\psi}-\ket{e}$ and
using self-adjointness of $O$. The apparently first-order term
involving $\ket{\delta}$ has an additional small factor: $\ket{\delta}$ is
orthogonal to every retained vector. We may therefore discard
the retained part of $O\ket{\psi}$ inside this pairing:
\begin{align}
 \langle\delta(t)|O|\psi(t)\rangle
 &=\langle\delta(t)|(1-P)\chi_a|\psi(t)\rangle,\\
 |\langle\delta|O|\psi\rangle|
 &\le\norm{\ket{\delta}}\,\norm{(1-P)(\chi_a\ket{\psi})}.
 \label{eq:support-pairing}
\end{align}
To obtain a product of two small tails, it remains to show that
multiplying by the smooth observable does not create a large
high-momentum component. The following lemma gives precisely that
estimate.

\begin{lemma}[Tail of the multiplied state]
\label{lem:chitail}
Under (H$\chi$), for every $M$ and $K \ge 2$,
\begin{equation}
\norm{ (1-P)\big( \chi(\bm{x}_a)\, \ket{\psi(t)} \big) }
\;\le\; (1 + C_\chi) \sum_b t_b(K/2;\, t)
\;+\; C_{\chi,M}\, K^{-M} .
\end{equation}
\end{lemma}

\begin{proofx}
We bound the omitted part by summing its restrictions outside the
cutoff cube in each coordinate. If $b\ne a$, the observable
acts on a different particle from the restriction, so the two
operations commute. Since $\norm{\chi}_\infty\le1$, this
contribution is bounded by the original tail $t_b(K;t)$, and
hence by $t_b(K/2;t)$.

For the observed coordinate $a$, multiplication shifts momenta.
Expand $\chi_a=\sum_{\bm{\nu}}\hat\chi(\bm{\nu})e^{i\bm{k}_{\bm{\nu}}\cdot \bm{x}_a}$.
By the Fourier convention in Eq.~\eqref{eq:fourier-sign}, an
input momentum $\bm{k}_{\bm{q}}$ becomes $\bm{k}_{\bm{q}}-\bm{k}_{\bm{\nu}}$. Split the shifts
into those with $\norm{\bm{k}_{\bm{\nu}}}_\infty\le K/2$ and those with
$\norm{\bm{k}_{\bm{\nu}}}_\infty>K/2$. For the first group, an output
outside the cube forces the input outside the cube of radius
$K/2$, by the triangle inequality. A Fourier shift preserves
the norm, so this group contributes at most
\begin{equation}
 \sum_{\norm{\bm{k}_{\bm{\nu}}}_\infty\le K/2}|\hat\chi(\bm{\nu})|
                         t_a(K/2;t)
 \le C_\chi t_a(K/2;t).
\end{equation}
For the second group we use $\norm{\ket{\psi(t)}}=1$, rather than
any input-tail bound. Its contribution is at most
\begin{equation}
 \sum_{\norm{\bm{k}_{\bm{\nu}}}_\infty>K/2}|\hat\chi(\bm{\nu})|
 \le C_{\chi,M}(K/2)^{-M}.
\end{equation}
Here $K\ge2$ ensures that the tail hypothesis applies.
The constants in (H$\chi$) are chosen upper bounds and may be
enlarged for each $M$; we absorb $2^M$ into $C_{\chi,M}$,
as in the lemma statement. Summing all particle contributions
proves the claim. Absolute summability justifies the Fourier
expansion and the triangle inequalities in $L^2$.
\end{proofx}

For the final theorem we may take the fixed threshold $K_0$ large
enough that $K/2\ge\max\{1,4\pi/L\}$ and $K\ge2$.
The dynamical tail bound then shows that both factors in
Eq.~\eqref{eq:support-pairing} have order $K^{-5/2}$, apart
from the arbitrarily rapid Fourier remainder of $\chi$.
This is the origin of the $K^{-5}$ term. We still have to control
the error inside the basis, for which the next pairing argument
uses a different second factor.

\subsection{Duality: regularity of the discrete backward flow and
the energy-rate secular bound}
\label{sec:dual}

We now control the part of the error that has accumulated inside
the retained basis. The direct estimate
$|\langle e_P(t)|\chi_a|v\rangle|\le\norm{\ket{e_P(t)}}\norm{\chi_a \ket{v}}$
would use only the rate $tK^{-3/2}$ from
Proposition~\ref{prop:duhamel}. To improve it, we keep the
Duhamel source inside the inner product and move the finite-dimensional
propagator onto the test vector. The resulting vector is the
observable applied to $\ket{v}$, projected into the finite basis and
evolved backwards. We first prove that its kinetic regularity stays
bounded independently of the cutoff and the duration of that
backward evolution.

This is the role of duality here: rather than estimate the error
against every possible unit vector, we use the additional
regularity of the particular vectors that test the observable.
It is analogous to Aubin--Nitsche duality in finite-element error
analysis, where regularity of a second equation gives a better
bound on a chosen functional of the error. The argument below
spells out the corresponding operation for the discrete quantum
evolution. To prepare the test vector, we first form the smeared
state $\chi(\bm{x}_a)\ket{v}$, the product with the smooth observable, and
then project it into the retained basis.

\begin{lemma}[$H_N$-regularity of smeared states]
\label{lem:smear}
For every $\ket{v} \in L^2$ with $\Theta(\ket{v}) := \norm{(1+\Ttot)\ket{v}} < \infty$,
\begin{equation}
\norm{ H_N\, P\big( \chi(\bm{x}_a)\, \ket{v} \big) }
\;\le\; C\, (1 + D_\chi)\, \big( \Theta(\ket{v}) + m_{\max} \norm{\ket{v}} \big),
\label{eq:smear}
\end{equation}
with $C$ depending only on the fixed particle, charge, mass and cell
data. Consequently, for the backward flow
\begin{equation}
\ket{\varphi_v(s)} := e^{\,i H_N (t-s)}\, P\big( \chi(\bm{x}_a)\, \ket{v} \big),
\qquad 0 \le s \le t,
\label{eq:testflow}
\end{equation}
the in-$\mathrm{ran}\,P$ elliptic estimate
$\Theta(\ket{\varphi}) \le C(\norm{H_N \ket{\varphi}} + m_{\max} \norm{\ket{\varphi}})$
(valid for all $\ket{\varphi} \in \mathrm{ran}\,P$ by the argument of
\eqref{eq:ell-gen}, using $[\Ttot, P] = 0$ and $\norm{P} \le 1$)
together with conservation of $\norm{H_N \ket{\varphi_v(s)}}$ gives,
uniformly in $0 \le s \le t$,
\begin{equation}
\Theta\big( \ket{\varphi_v(s)} \big)
\;\le\; \Xi_\chi(\ket{v}) := C\, m_{\max}\, (1 + D_\chi)\,
\big( \Theta(\ket{v}) + m_{\max} \norm{\ket{v}} \big).
\label{eq:testreg}
\end{equation}
\end{lemma}

\begin{proofx}
There are two obligations: control the discrete Hamiltonian applied
to the initial test vector $P(\chi_a \ket{v})$, and convert that control
back into kinetic regularity after evolution. We begin by accounting
for the effect of projection:
\begin{equation}
 H_NP(\chi_a \ket{v})
 =P\Hnbo(\chi_a \ket{v})-PV_{\mathrm{tot}}(1-P)(\chi_a \ket{v}).
\end{equation}
Indeed, insert $P=1-(1-P)$ immediately to the right of $\Hnbo$.
The kinetic part of the correction vanishes because
$P\Ttot(1-P)=0$; both operators are diagonal in the Fourier basis.
Thus only a potential correction remains.

To estimate the first term, multiplication by $\chi_a$ commutes
with every potential term, including the clamped-nucleus terms,
but we must keep its commutator with kinetic energy:
\begin{equation}
 \Hnbo(\chi_a \ket{v})
 =\chi_a\Ttot \ket{v}+\chi_aV_{\mathrm{tot}}\ket{v}
                       +[\Ttot,\chi_a]\ket{v}.
\end{equation}
The first term has norm at most $\Theta(\ket{v})$, and the relative
potential bound from Proposition~\ref{prop:nbo-KR} gives
\begin{equation}
 \norm{\chi_aV_{\mathrm{tot}}\ket{v}}
 \le\norm{V_{\mathrm{tot}}\ket{v}}
 \le\tfrac12\Theta(\ket{v})+Cm_{\max}\norm{\ket{v}}.
\end{equation}
For the commutator, only differentiation in coordinate $\bm{x}_a$
acts on $\chi$. Expanding the Laplacian of a product gives
\begin{equation}
 [\Ttot,\chi(\bm{x}_a)]
 =-\frac1{2m_a}(\Delta\chi)(\bm{x}_a)
    -\frac1{m_a}(\nabla\chi)(\bm{x}_a)\cdot\nabla_a.
\end{equation}
The position-space wavefunction $v(\vec{\bm{x}},\sigma)=\langle \vec{\bm{x}},\sigma|v\rangle$
has a gradient controlled by the kinetic graph norm of $\ket{v}$:
\begin{equation}
 \norm{\nabla_a v}_{L^2}^2
 \le2m_a\langle v|\Ttot|v\rangle
 \le2m_a\Theta(\ket{v})\norm{\ket{v}}.
\end{equation}
It follows that
\begin{equation}
 \norm{[\Ttot,\chi_a]\ket{v}}
 \le\frac{D_\chi}{2m_a}\norm{\ket{v}}
       +\frac{D_\chi}{m_a}(2m_a)^{1/2}
                    \big(\Theta(\ket{v})\norm{\ket{v}}\big)^{1/2}
 \le CD_\chi\big(\Theta(\ket{v})+\norm{\ket{v}}\big).
\end{equation}
The last step uses $m_a\ge1$ and
$2\sqrt{xy}\le x+y$. In particular, these explicit commutator
coefficients decrease as the mass of the observed particle grows.

For the projection correction, apply the same relative potential
bound to $(1-P)(\chi_a \ket{v})$:
\begin{equation}
 \norm{V_{\mathrm{tot}}(1-P)(\chi_a \ket{v})}
 \le\tfrac12\norm{\Ttot(1-P)(\chi_a \ket{v})}
       +Cm_{\max}\norm{(1-P)(\chi_a \ket{v})}.
\end{equation}
Because $1-P$ commutes with $\Ttot$ and is a contraction, the
first norm is at most
\begin{equation}
 \norm{\Ttot(\chi_a \ket{v})}
 \le\norm{\Ttot \ket{v}}+\norm{[\Ttot,\chi_a]\ket{v}}.
\end{equation}
The other norm is at most $\norm{\ket{v}}$. All terms have now been
bounded by the right side of Eq.~\eqref{eq:smear}, which proves
that estimate. These product calculations are legitimate for
$\ket{v}\in D(\Ttot)=H^2$, since the bounded derivatives of $\chi$
through order two preserve this domain.

Next let $\ket{z}\in\mathrm{ran}\,P$. Within this space,
$H_N \ket{z}=\Ttot \ket{z}+PV_{\mathrm{tot}}\ket{z}$, so
\begin{equation}
 \norm{\Ttot \ket{z}}
 \le\norm{H_N \ket{z}}+\norm{V_{\mathrm{tot}}\ket{z}}
 \le\norm{H_N \ket{z}}+\tfrac12\norm{\Ttot \ket{z}}
                         +Cm_{\max}\norm{\ket{z}}.
\end{equation}
Absorbing the kinetic term gives the cutoff-independent comparison
\begin{equation}
 \Theta(\ket{z})\le C\big(\norm{H_N \ket{z}}+m_{\max}\norm{\ket{z}}\big).
\end{equation}
Apply this to $\ket{z}=\ket{\varphi_v(s)}$. The discrete unitary conserves
both $\norm{\ket{\varphi_v(s)}}$ and $\norm{H_N\ket{\varphi_v(s)}}$:
\begin{equation}
 \norm{\ket{\varphi_v(s)}}=\norm{P(\chi_a \ket{v})}\le\norm{\ket{v}},\qquad
 \norm{H_N\ket{\varphi_v(s)}}=\norm{H_NP(\chi_a \ket{v})}.
\end{equation}
Combining these identities with Eq.~\eqref{eq:smear} proves the
uniform bound in Eq.~\eqref{eq:testreg}, allowing the stated
extra factor $m_{\max}\ge1$ in $\Xi_\chi(\ket{v})$. Neither the
comparison nor the unitary conservation requires the test vector
to remain in an exchange-symmetry sector.
\end{proofx}

We can now apply the potential-tail estimate to the backward test
state. In summing over particles, we use
$S_{1/2}:=\sum_b m_b^{1/2}$, extending the mass-sum notation of
Appendix~\ref{app:nbo} to this exponent.

\begin{proposition}[Secular pairing bound]
\label{prop:duality}
For $\ket{v}$ with $\Theta(\ket{v})<\infty$, let $\ket{\varphi_v}$ be the backward
flow in \eqref{eq:testflow}. For $K\ge\max\{1,4\pi/L\}$,
\begin{equation}
\big| \langle e_P(t)|\chi_a|v\rangle \big|
\;\le\; t \cdot
\Big[ \sup_s \norm{\ket{\delta(s)}} \Big] \cdot
C_\Lambda\, m_{\max}^{1/2} S_{1/2}\, \Xi_\chi(\ket{v})\, K^{-1/2}
\;\le\; C\, \Xi\, \Xi_\chi(\ket{v})\, t\, K^{-3},
\label{eq:dualbound}
\end{equation}
with $C$ depending only on the fixed particle, charge, mass and cell data.
\end{proposition}

\begin{proofx}
Since $\ket{e_P(t)}$ belongs to $\mathrm{ran}\,P$, its inner product
with $\chi_a \ket{v}$ equals its inner product with $P(\chi_a \ket{v})$.
Insert the Duhamel representation \eqref{eq:duhamel} and move
$e^{-iH_N(t-s)}$ to the second entry of the inner product as its
adjoint. This produces precisely the backward state
$\ket{\varphi_v(s)}=e^{iH_N(t-s)}P(\chi_a \ket{v})$. Symmetry of the real
potential then moves it onto that state. With the convention
that the inner product is linear in its second entry, the result is
\begin{equation}
 \langle e_P(t)|\chi_a|v\rangle
 =i\int_0^t\langle\delta(s)|V_{\mathrm{tot}}|\varphi_v(s)\rangle\,ds
 =i\int_0^t\langle\delta(s)|(1-P)V_{\mathrm{tot}}|\varphi_v(s)\rangle\,ds.
\end{equation}
The second equality uses $\ket{\delta(s)}\perp\mathrm{ran}\,P$.
All potential applications are well defined because both states
belong to $H^2$; no exchange symmetry of the test state is needed.
Taking absolute values gives
\begin{equation}
 |\langle e_P(t)|\chi_a|v\rangle|
 \le\int_0^t\norm{\ket{\delta(s)}}
              \norm{(1-P)V_{\mathrm{tot}}\ket{\varphi_v(s)}}\,ds.
\end{equation}
The important change is that the potential now acts on the regular
backward state, while the first factor retains the faster decay of
the plain discarded-state tail.

For the remaining potential norm, the union bound over the omitted
particle indices gives
\begin{equation}
 \norm{(1-P)V_{\mathrm{tot}}\ket{\varphi_v(s)}}
 \le\left(\sum_b\Lambda_b(K;\ket{\varphi_v(s)})^2\right)^{1/2}
 \le\sum_b\Lambda_b(K;\ket{\varphi_v(s)}).
\end{equation}
Apply Eq.~\eqref{eq:lam-gen} to each term and then use
Eq.~\eqref{eq:testreg}:
\begin{equation}
 \sum_b\Lambda_b(K;\ket{\varphi_v(s)})
 \le C_\Lambda m_{\max}^{1/2}S_{1/2}
                         \Xi_\chi(\ket{v})K^{-1/2}.
\end{equation}
Meanwhile Eq.~\eqref{eq:dyn-t} gives
$\sup_s\norm{\ket{\delta(s)}}
\le Cm_{\max}^{1/2}S_3^{1/2}\Xi K^{-5/2}$.
Both bounds are uniform over the integration interval. Their
product has cutoff dependence $K^{-5/2}K^{-1/2}=K^{-3}$,
and integration contributes only the interval length $t$.
This proves Eq.~\eqref{eq:dualbound} with the stated fixed-system
constant.
\end{proofx}

\begin{proofx}[Proof of Theorem~\ref{thm:dyn}]
We now combine the two pairing estimates and check that every
quadratic term has the same permitted dependence on time. Choose
$K_0$ large enough for the preceding estimates, including the
tail bound at $K/2$. This choice depends only on the fixed system.
Expanding Eq.~\eqref{eq:obs-split} with $\ket{e}=\ket{\delta}+\ket{e_P}$ gives
\begin{align}
 \Delta\langle O\rangle(t)
 ={}&2\mathrm{Re}\langle\delta|O|\psi\rangle
       +2\mathrm{Re}\langle e_P|O|\psi\rangle
       -\langle\delta|O|\delta\rangle\\
    &-2\mathrm{Re}\langle\delta|O|e_P\rangle
       -\langle e_P|O|e_P\rangle,
\end{align}
where every vector in this display is evaluated at time $t$.
We estimate the five terms in their displayed order.

For the first term, orthogonality in
Eq.~\eqref{eq:support-pairing}, followed by
Lemma~\ref{lem:chitail} and Eq.~\eqref{eq:dyn-t}, gives
\begin{align}
 |\langle\delta|O|\psi\rangle|
 &\le C\Xi K^{-5/2}
       \big((1+C_\chi)\Xi K^{-5/2}+C_{\chi,M}K^{-M}\big)\\
 &\le C(1+C_\chi)\Xi^2K^{-5}
                         +CC_{\chi,M}\Xi K^{-M}.
\end{align}
In the last step we used $K\ge1$ to drop the extra
$K^{-5/2}$ in the Fourier remainder. Fixed particle and mass
factors have been included in $C$.

For the second term we use Proposition~\ref{prop:duality}
with the terminal vector $\ket{v}=\ket{\psi(t)}$. It has norm one and
$\Theta(\ket{v})\le\Xi$, so Eq.~\eqref{eq:testreg} gives
$\Xi_\chi(\ket{\psi(t)})\le Cm_{\max}(1+D_\chi)(\Xi+m_{\max})$.
Therefore
\begin{equation}
 |\langle e_P|O|\psi\rangle|
 \le C(1+D_\chi)\Xi(\Xi+m_{\max})tK^{-3}.
\end{equation}
The backward pairing, rather than the direct norm estimate,
supplies the required cutoff power.

For the third term, $0\le O\le1$ immediately implies
\begin{equation}
 |\langle\delta|O|\delta\rangle|\le\norm{\ket{\delta}}^2
                                  \le C\Xi^2K^{-5}.
\end{equation}
For the fourth, the elementary retained-state norm bound is already
sufficient. The Duhamel source estimate yields
$\norm{\ket{e_P(t)}}\le C\Xi tK^{-3/2}$, so
\begin{equation}
 |\langle\delta|O|e_P\rangle|
 \le\norm{\ket{\delta}}\norm{\ket{e_P}}
 \le C\Xi^2tK^{-4}\le C\Xi^2tK^{-3}.
\end{equation}

The fifth term requires more care. Squaring the preceding bound
on $\norm{\ket{e_P}}$ would introduce $t^2$, so instead we apply the
pairing proposition again, this time with $\ket{v}=\ket{e_P(t)}$. Although
$\ket{e_P(t)}$ is itself an error, it has a kinetic graph norm bounded
uniformly in time. To see this, use the discrete elliptic comparison
proved in Lemma~\ref{lem:smear} and conservation under $H_N$:
\begin{equation}
\Theta(\ket{e_P(t)})\le\Theta(P\ket{\psi(t)})+\Theta(\ket{\psi_N(t)})
\le\Xi+C\big(\norm{H_NP\ket{\psi_0}}+m_{\max}\big)
\le C\Xi.
\label{eq:etareg}
\end{equation}
We spell out the last two steps because they ensure that no hidden
time or cutoff factor enters. First, $P$ commutes with the kinetic
weight, so $\Theta(P\ket{\psi(t)})\le\Theta(\ket{\psi(t)})\le\Xi$.
Second, discrete evolution conserves both
$\norm{\ket{\psi_N(t)}}=\norm{P\ket{\psi_0}}\le1$ and
$\norm{H_N\ket{\psi_N(t)}}=\norm{H_NP\ket{\psi_0}}$. The initial
Hamiltonian norm is bounded by
\begin{align}
 \norm{H_NP\ket{\psi_0}}
 &\le\norm{\Hnbo\ket{\psi_0}}
                   +\norm{V_{\mathrm{tot}}(1-P)\ket{\psi_0}}\\
 &\le\mathcal D+\tfrac12\norm{\Ttot(1-P)\ket{\psi_0}}
                        +Cm_{\max}\norm{(1-P)\ket{\psi_0}}\\
 &\le\mathcal D+\tfrac12\Theta(\ket{\psi_0})+Cm_{\max}
 \le C\Xi.
\end{align}
The first line uses
$H_NP\ket{\psi_0}=P\Hnbo\ket{\psi_0}-PV_{\mathrm{tot}}(1-P)\ket{\psi_0}$;
the next lines use the relative potential bound and commutation
of the kinetic operator with $P$.

We also have the elementary uniform bound
$\norm{\ket{e_P(t)}}\le\norm{P\ket{\psi(t)}}+\norm{\ket{\psi_N(t)}}\le2$.
Combining it with Eq.~\eqref{eq:etareg} shows that
\begin{equation}
 \Xi_\chi(\ket{e_P(t)})
 \le Cm_{\max}(1+D_\chi)(\Xi+m_{\max}).
\end{equation}
Proposition~\ref{prop:duality} now gives
\begin{equation}
 |\langle e_P|O|e_P\rangle|
 \le C(1+D_\chi)\Xi(\Xi+m_{\max})tK^{-3}.
\end{equation}
The quadratic observable term is thus also controlled by a
single time integral, with a uniformly regular terminal test vector.

Adding the five estimates, and using
$\Xi^2\le\Xi(\Xi+m_{\max})$, proves
Eq.~\eqref{eq:dyn-main}. No step restricts $t$ to a small-time
interval. Both quadratic forms lie in $[0,1]$, so their difference
is also bounded by one; at sufficiently long times this trivial
bound can be stronger than the stated time-growing estimate.
Finally, for fixed $M\ge5$, the Fourier remainder is of order
$K^{-5}$ or smaller, and $K=\Theta(N^{1/3})$ gives
Eq.~\eqref{eq:dyn-simple}.
\end{proofx}

\subsection{Remarks on the dynamical result}

\medskip\noindent\emph{What the two terms measure.}
We can now distinguish the two mechanisms in the theorem.
The $N^{-5/3}$ term comes from the omitted component at the
observation time. Smooth multiplication preserves enough momentum
decay that its contribution is a product of two small tails.
This resembles the quadratic protection of the variational energy
error, but the reason here is the support identity
\eqref{eq:support-pairing}, rather than stationarity of the energy.

The time-dependent contribution comes from
$PV_{\mathrm{tot}}(1-P)\ket{\psi(s)}$: the potential transfers amplitude
from omitted momenta back into retained momenta. Duhamel's formula
integrates that source over the evolution. The pairing proof bounds
its effect on the observable by $tK^{-3}$, using regularity of the
backward test state. This explains why the coefficient of the
bound has the same cutoff power as the energy error. It does not
show that an observable actually drifts at this rate, nor does
it establish a small operator norm for a difference between the
full and projected Hamiltonians. No lower bound showing attainment
of the secular coefficient is proved here. The explicit observed-mass
factors in the commutator are favourable, but other constants and
$\Xi$ retain their stated dependence on the fixed masses.

\medskip\noindent\emph{What the norm estimate does and does not say.}
The elementary bound in Proposition~\ref{prop:duhamel} controls
the source in norm and then integrates its norm:
\begin{equation}
 \norm{PV_{\mathrm{tot}}(1-P)\ket{\psi(s)}}
       \le C\Xi K^{-3/2},\qquad
 \norm{\ket{e_P(t)}}\le C\Xi tK^{-3/2}.
\end{equation}
The observable argument retains the inner product during that
integration. Its gain follows from a definite identity: moving the
propagator and the potential to the test side leaves the discarded
state paired with a potential tail of the backward state. The
two available tail estimates then multiply. Conservation of the
discrete Hamiltonian norm and the discrete elliptic comparison
provide the test-state regularity without any spectral separation.

The norm estimate above is only the elementary estimate used in
this appendix. Later, Eq.~\eqref{eq:kinetic-retained-galerkin}
establishes the sharper retained-state comparison
\begin{equation}
 \norm{P\ket{\psi(t)}-\ket{\psi_N(t)}}
 \le C\mathcal D K^{-3}\big[1+t+\log(2+K)\big]
\end{equation}
by a spectral argument. Thus the $tK^{-3/2}$ estimate should
not be interpreted as an attained asymptotic rate or as an
obstruction to better norm control under the present hypotheses.
The full state error also contains the orthogonal discarded
component, whose available norm bound is $K^{-5/2}$.
Our direct observable proof has its own advantage: it combines
the $K^{-5}$ time-independent term with a $tK^{-3}$ term
without adding the logarithm from that later norm comparison.

One may picture the potential as returning omitted amplitude
through momenta near the cutoff, but the present proof establishes
neither localisation of the error in such a shell nor saturation
of the norm estimate there. Likewise, more detailed pointwise or
Besov-type Fourier information---control of coefficients within
momentum scales beyond the weighted $L^2$ estimates used here---
would require additional structural input and a time-propagation
argument, for example building on the cusp analysis of
Ref.~\cite{FHOS}. None of that is needed for
Eq.~\eqref{eq:dyn-main}. The unresolved attainment question
relevant to this theorem is whether its secular observable
coefficient $K^{-3}$ can be realised by an example.

\medskip\noindent\emph{What smoothing buys, and its cost.}
To relate the hypothesis to a spatial measurement, fix a region
$\mathcal R\subset\T$ with $C^\infty$ boundary. We smooth its
indicator by local averaging:
\begin{equation}
 \chi=\mathbf1_{\mathcal R}*\phi_\ell,\qquad
 \phi_\ell(\bm{x})=\ell^{-3}\phi(\bm{x}/\ell),
 \quad \phi\in C_c^\infty,\quad \phi\ge0,\quad\int\phi=1,
\end{equation}
with the kernel periodised on the cell. The parameter $\ell>0$
is the averaging width; the unit integral and nonnegativity ensure
$0\le\chi\le1$. Differentiation falls on the smooth kernel, so
\begin{equation}
 D_\chi\le\norm{\nabla\phi_\ell}_{L^1}
              +\norm{\Delta\phi_\ell}_{L^1}
 \le C(\ell^{-1}+\ell^{-2})=\mathcal O(\ell^{-2})
\end{equation}
as $\ell\downarrow0$. This is the cost of sharpening the
observable in the backward-regularity argument.

The Fourier constants also grow as the averaging width shrinks.
For nonzero $\bm{k}_{\bm{\nu}}$, the divergence theorem converts the Fourier
integral over the region into a boundary integral and gives
\begin{equation}
 |\widehat{\mathbf1_{\mathcal R}}(\bm{k}_{\bm{\nu}})|
 \le C|\partial\mathcal R|\norm{\bm{k}_{\bm{\nu}}}^{-1}.
\end{equation}
If the boundary has everywhere nonvanishing Gaussian curvature,
the product of its two principal curvatures, stationary phase
improves this to $\mathcal O(\norm{\bm{k}_{\bm{\nu}}}^{-2})$.
The principal curvatures are the largest and smallest signed
curvatures obtained by cutting the surface with planes containing
its normal at the point.
In this use of stationary phase, oscillations cancel the boundary
integral except near points whose normals are parallel to the
momentum; nonzero curvature makes those stationary points
nondegenerate and supplies an additional inverse power.
Strict convexity alone does not guarantee this uniform improvement:
a strictly convex surface can have zero-curvature points, for
example with quartic contact, at which the decay estimate weakens.
At such a point, the departure from the tangent plane can begin
at fourth order in a tangent direction, rather than at the
quadratic order measured by curvature.

Smoothness of the averaging kernel gives
$|\hat\phi(\ell \bm{k})|\le C_q(1+\ell\norm{\bm{k}})^{-q}$
for every fixed $q$. Since the Fourier transform of a convolution
is the product of the transforms, we multiply this bound by the
region-indicator bound and sum over lattice shells. In three
dimensions a shell at radius $r$ has order $r^2$ lattice points
per unit radial width. The general $r^{-1}$ coefficient bound
therefore leaves an integrand of order
$r(1+\ell r)^{-q}$. Its integral is of order $\ell^{-2}$
for $q>2$. With the curvature improvement the corresponding
integrand is $(1+\ell r)^{-q}$, of integral order $\ell^{-1}$.
The finitely many small modes are absorbed into the constants.
Consequently,
\begin{align}
 C_\chi&=\mathcal O(\ell^{-2})
 \quad\big(\mathcal O(\ell^{-1})\text{ for nonvanishing
 Gaussian curvature}\big),\\
 \sum_{\norm{\bm{k}_{\bm{\nu}}}_\infty>\rho}|\hat\chi(\bm{\nu})|
 &\le C_M'\ell^{-(M+2)}\rho^{-M},\qquad \rho\ge1.
\end{align}
For the second line, choose kernel decay exponent $q=M+2$;
the general shell estimate is bounded by
$C\ell^{-(M+2)}\int_\rho^\infty r^{-M-1}\,dr$,
with changes of norm and lattice scale absorbed in $C_M'$.
This calculation explains the two extra inverse powers of $\ell$
in $C_{\chi,M}$ beyond a simple $\ell^{-M}$ count.

The constants therefore deteriorate polynomially as the boundary
is sharpened. For a fixed smoothing width they are independent
of $N$, and the powers of $N$ in the theorem are unchanged.
If the width is instead allowed to shrink as $N$ grows, this
dependence must be inserted into the bound; the fixed-observable
rates alone do not describe that joint limit. The two uses of
smoothness are distinct: the Fourier sums control the omitted
part of $\chi_a\ket{\psi}$ in Lemma~\ref{lem:chitail}, and the
derivatives control the backward test state in
Lemma~\ref{lem:smear}.

For a sharp indicator of a nontrivial region, neither argument
is available in this form. Absolute summability of its Fourier
coefficients would give a continuous representative, which a
region indicator with a genuine boundary does not have.
This would be a continuous function agreeing with the indicator
except on a set of measure zero.
The preceding pointwise bounds give truncated absolute sums
bounded above by $\mathcal O(R^2)$ at momentum radius $R$
in general, or by $\mathcal O(R)$ under the curvature assumption;
these are upper bounds, not matching growth laws. Moreover,
multiplication by a sharp indicator need not preserve even the
$H^1$ form domain, let alone the $H^2$ domain used in the
backward-regularity lemma. Special cancellation, such as a
vanishing boundary trace, can help with the form domain but does
not provide the general product estimate used here.
Here the boundary trace is the boundary value of the state,
extended from ordinary restriction of smooth functions to the
$H^1$ functions under consideration; vanishing means that this
boundary value is zero.

For such a bounded but nonsmooth observable, the elementary
state-norm route still gives a first-order discarded-tail error
and an accumulated error of order
$K^{-5/2}+tK^{-3/2}$. This describes what that elementary
route proves, not the best available conclusion in every setting:
the later retained-state comparison above gives additional control,
and special geometric or state-dependent information may allow
sharper boundary estimates. In particular, failure of the present
smoothness hypotheses is not a lower bound on sharp-observable
accuracy.

There is a limited geometric simplification in the BO case.
Its electron--nucleus cusp locations are fixed, so a one-particle
region boundary may be placed away from them; in suitable
restricted settings, boundary oscillation estimates could then
help control a sharp observable. Electron--electron coincidence
sets $\bm{x}_i=\bm{x}_j$, however, vary with the other electronic coordinates
and intersect generic one-particle boundaries even in BO.
An argument based only on avoiding the fixed nuclei therefore
applies directly to one-electron systems, or requires separate
control of the electron--electron contributions. In non-BO
coordinates the electron--nucleus coincidences also depend on
other particle positions, so avoiding a fixed set of nuclear
locations no longer resolves that issue. Smooth observables
provide the uniform sufficient condition used in this appendix;
we do not claim they are the only possible formulation.

\medskip\noindent\emph{BO and hybrid nuclear treatments.}
The BO Hamiltonian $H_{\mathrm{per}}$ from Sec.~\ref{sec:setting}
is the special case in which only the electrons are quantum
particles, with mass one, and all nuclei are clamped. In a hybrid
model, the quantum nuclei also contribute to $\Ttot$ and to the
quantum-pair potential, while the clamped nuclei contribute
$q_a\zeta_\ell w(\bm{x}_a-\bm{R}_\ell)$ for every quantum particle $a$.
The latter terms obey the one-particle Hardy and Fourier-tail
estimates included in Appendix~\ref{app:nbo}. They commute with
the observable, while $\Ttot$ still commutes with the Fourier
projection. Consequently the same Duhamel identity, backward
regularity estimate and duality argument apply to all three
treatments. The rates and initial-state domain condition are
unchanged; the system constants include all nuclear charges.
The possible geometric treatment of sharp boundaries remains
subject to the preceding qualifications about electronic
coincidence sets; it is not needed for the smoothed result.

\section{Spatial-grid discretisation and aliasing}
\label{app:grid}
\label{sec:grid}

We now replace the projected potential by a potential that is
diagonal on a spatial grid. This makes its action a multiplication
of each grid amplitude by a sampled potential value. The resulting
finite operator is not the plane-wave Galerkin operator, even
though the grid and plane-wave bases span the same space. The
difference comes from aliasing: on a finite grid, Fourier modes
whose indices differ by a multiple of the grid size have identical
sample values. A momentum transfer can therefore leave one edge
of the register and re-enter at the opposite edge.

Our goal is to control these additional matrix elements well
enough to retain the $N^{-1}$ energy-error bound and the
$N^{-5/3}+tN^{-1}$ smooth-observable bound. We choose the
potential's transfer bandwidth together with the register size so
that every such wrap involves high momentum in at least one
interacting coordinate. The tails proved in the preceding
appendices can then make these matrix elements small when paired
with the states of interest. Smallness of these pairings, rather
than smallness of the operator on every register vector, is the
property the proof uses.

We use effective resolution $K$, register cutoff $2K$, and
interaction transfers bounded by $2K$. If $N=n^3$ with $n$ odd,
the register has $n'=2n-1$ points in each direction and
$N'=(2n-1)^3=\Theta(N)$ spatial modes per particle. We call
this the padded parametrisation. The factor of two separates the
effective resolution from the register edge; it does not change
the asymptotic powers of the basis size. After deriving the
estimates, we explain the equivalent convention that starts with
a fixed register and chooses its kernel bandwidth accordingly.

\subsection{Grid basis and the fold identity}
\label{sec:grid-fold}

We first derive the matrix represented by grid multiplication.
For this derivation let $n'$ be any admissible odd grid size and set
\begin{equation}
 G':=[-(n'-1)/2,(n'-1)/2]^3\cap\Z^3,
 \qquad K':=\pi(n'-1)/L.
\end{equation}
The projector $\Pp$ retains these indices in every particle
coordinate and acts as the identity on spin. The spatial grid
points for one particle are
$\bm{x}_{\bm{j}}:=\bm{j}L/n'$, with $\bm{j}\in\{0,\ldots,n'-1\}^3$.
The discrete Fourier transform rotates the plane-wave basis
$\{\ket{\bm{p}}:\bm{p}\in G'\}$ into an orthonormal grid basis of the
same $n'^3$-dimensional one-particle space. Thus choosing a grid
basis alone does not approximate an operator further. The new
approximation enters when we define the potential by its sampled
values, a procedure called collocation.

For a trigonometric polynomial $f$, meaning a function with
finitely many Fourier modes, let $\mathrm{Diag}[f]$ be the
operator with entries $f(\bm{x}_{\bm{j}_a}-\bm{x}_{\bm{j}_b})$ for a pair interaction,
or $f(\bm{x}_{\bm{j}_a})$ for a one-particle interaction, in this grid
basis. We compute its matrix in the plane-wave basis to compare
it with Galerkin projection. The following fold identity collects
all Fourier coefficients that the sampled grid cannot distinguish.

\begin{lemma}[Fold identity]
\label{lem:fold}
In the momentum basis of $\mathrm{ran}\,\Pp$, use the multiplier
convention of Eq.~\eqref{eq:fourier-sign}. For the matrix element
$\langle \bm{p}_a,\bm{p}_b|\mathrm{Diag}[f]|\bm{q}_a,\bm{q}_b\rangle$, with $\vec{\bm{q}}$ the input and
$\vec{\bm{p}}$ the output, put $\bm{d}_a:=\bm{q}_a-\bm{p}_a$ and $\bm{d}_b:=\bm{q}_b-\bm{p}_b$.
All other spatial registers and all spin indices are diagonal. Then the
matrix element is
\begin{equation}
\sum_{\substack{\bm{\ell}_a, \bm{\ell}_b \in \Z^3 \\
\bm{d}_a + n' \bm{\ell}_a \,=\, -( \bm{d}_b + n' \bm{\ell}_b )}}
\hat f\big( \bm{d}_a + n' \bm{\ell}_a \big),
\label{eq:fold}
\end{equation}
the fold indices $\bm{\ell}_a, \bm{\ell}_b$ (integer triples, not to be
confused with particle masses) recording the aliasing of each
register separately, and the sum finite since $f$ is band-limited.
Equivalently, the element vanishes unless
$\bm{p}_a + \bm{p}_b \equiv \bm{q}_a + \bm{q}_b \ (\mathrm{mod}\ n')$: the grid pair
operator conserves total momentum only \emph{modulo} $n'$, whereas
the Galerkin operator conserves it exactly. The Galerkin part is
precisely the term $(\bm{\ell}_a, \bm{\ell}_b) = (0,0)$, which exists only
when $\bm{p}_a + \bm{p}_b = \bm{q}_a + \bm{q}_b$ exactly; a term with $\bm{\ell}_a = 0$ but
$\bm{\ell}_b \ne 0$ is a wrap, not a Galerkin element. \emph{All} terms
with $(\bm{\ell}_a, \bm{\ell}_b) \ne (0,0)$ --- including those with
$\bm{\ell}_a + \bm{\ell}_b \ne 0$, which violate exact total-momentum
conservation while preserving it modulo $n'$ --- constitute the wrap
contribution. For the single-register version
$\mathrm{Diag}[f(\bm{x}_a)]$ no total-momentum condition arises and the
element is $\sum_{\bm{\ell}_a} \hat f( \bm{d}_a + n' \bm{\ell}_a )$. This covers
in particular the clamped-nucleus external terms of the BO setting,
$f(\bm{x}) = -\sum_\ell \zeta_\ell\, w_{\le 2K}(\bm{x} - \bm{R}_\ell)$ with the
nuclear positions $\bm{R}_\ell \in \T$ \emph{arbitrary} (not grid
points): the entries $f(\bm{x}_{\bm{j}})$ are evaluated at the electron grid
points, and $\hat f(\bm{\nu}) = -\sum_\ell \zeta_\ell\,
\hat w(\bm{\nu})\,e^{-i\bm{k}_{\bm{\nu}}\cdot \bm{R}_\ell}$ has the same support and is
bounded in magnitude by $\lambda_\zeta\hat w(\bm{\nu})$. Thus the
coefficient estimates apply with the nuclear charge weights included.
\end{lemma}

\begin{proofx}
Consider one Fourier mode $\hat f(\bm{\nu})e^{i\bm{k}_{\bm{\nu}}\cdot(\bm{x}_a-\bm{x}_b)}$.
With the negative-exponent plane-wave convention of
Eq.~\eqref{eq:fourier-sign}, its matrix element contains the
two discrete averages
\begin{equation}
 \frac1{n'^3}\sum_{\bm{j}_a}
 e^{i\bm{k}_{\bm{\nu}-\bm{d}_a}\cdot \bm{x}_{\bm{j}_a}},\qquad
 \frac1{n'^3}\sum_{\bm{j}_b}
 e^{i\bm{k}_{-\bm{\nu}-\bm{d}_b}\cdot \bm{x}_{\bm{j}_b}}.
\end{equation}
For any integer triple $\bm{\mu}$, the finite geometric sums in the
three directions give
\begin{equation}
 \frac1{n'^3}\sum_{\bm{j}} e^{i\bm{k}_{\bm{\mu}}\cdot \bm{x}_{\bm{j}}}
 =\mathbf1[\bm{\mu}\in n'\Z^3].
\end{equation}
The mode therefore contributes precisely when
$\bm{\nu}\equiv \bm{d}_a$ and $-\bm{\nu}\equiv \bm{d}_b$ modulo $n'$,
component by component. Writing the two congruences as
\begin{equation}
 \bm{\nu}=\bm{d}_a+n'\bm{\ell}_a=-(\bm{d}_b+n'\bm{\ell}_b)
\end{equation}
and summing their coefficients gives Eq.~\eqref{eq:fold}.
Compatibility requires $\bm{d}_a+\bm{d}_b\equiv0$, or equivalently
conservation of the sum of the two particle momenta modulo the
register size. The other coordinates and spin indices are
unchanged by the multiplication and contribute Kronecker deltas.

The ordinary Galerkin matrix retains only transfers that connect
the input and output indices without reducing either index modulo
$n'$. These are exactly the terms with both fold indices zero.
If either fold index is nonzero, the contribution is a wrap even
when the other register does not fold. In particular, exact
total-momentum conservation is not a sufficient way to classify
all possible grid matrix elements: the congruence also admits
terms with nonzero total fold index. For a one-particle potential
there is just the first grid average, which gives the single sum
in the statement. Translating a kernel by a clamped nuclear
position multiplies each original Fourier coefficient by its
phase; it does not change either the congruence or the coefficient
magnitude bound.
\end{proofx}

This calculation identifies two issues we must address. First,
sampling the bare Coulomb kernel at coincidence is not defined;
we instead use a finite Fourier series, which has a finite value
at every grid separation. Second, limiting the Fourier bandwidth
does not eliminate wraps. Even a low-frequency mode can wrap
between indices near opposite faces of the register. We must
therefore control the momentum support of the states coupled by
a wrap, as well as the size of its coefficient. The next
construction arranges that these two controls complement one
another.

\subsection{The truncated-kernel method and its structure}
\label{sec:grid-def}

We choose the register and the kernel together so that a large
Coulomb coefficient cannot wrap a low-momentum state directly
to another low-momentum state. Let $n$ be odd, set $n':=2n-1$,
and define $K:=\pi(n-1)/L$. The register cutoff is then
$K'=2K$. In cutoff estimates we take $n\ge3$, so $K>0$.
The largest possible transfer between two register indices is
$4K$, but we retain kernel transfers only up to $2K$:
\begin{equation}
w_{\le 2K}(\bm{x}):=\frac{4\pi}{\Omega}
\sum_{0<\norm{\bm{k}_{\bm{\nu}}}_\infty\le2K}
\frac{e^{i\bm{k}_{\bm{\nu}}\cdot \bm{x}}}{\norm{\bm{k}_{\bm{\nu}}}^2},
\qquad
\norm{\bm{\nu}}_\infty\le s:=n-1.
\label{eq:wtrunc}
\end{equation}
Thus the kernel bandwidth equals the register cutoff and is half
the width of its difference set. This is a truncation of momentum
transfer, not a cutoff in the distance between particles. The
function is evaluated at every grid separation, so distant pairs
still interact. In position space the finite Fourier sum regularises
the Coulomb singularity: it is finite at coincidence, where the
sum of the positive coefficients grows as $K$.
The finite sums converge to the periodic Coulomb kernel in
$L^2$: square summability of the omitted coefficients gives
$\|w-w_{\le2K}\|_{L^2}\le CK^{-1/2}$ at fixed cell size.
This norm convergence is the kernel comparison used in the
estimates below. Sharp cubical truncation need not converge
pointwise at every noncoincident separation.

In this appendix $H$ denotes the continuum Hamiltonian being
approximated: $\Hnbo$ in the fully quantum setting, or
$H_{\mathrm{per}}$ in the BO setting. We allow the notation
for clamped nuclei as well as quantum particles so that the same
construction covers both. Define the grid Hamiltonian on
$\mathrm{ran}\,\Pp$ by
\begin{equation}
\Hgrid:=\Ttot\big|_{\mathrm{ran}\Pp}
+\sum_{a<b}q_aq_b\,\mathrm{Diag}[w_{\le2K}]_{ab}
+\sum_{a,\ell}q_a\zeta_\ell\,
          \mathrm{Diag}[w_{\le2K}(\,\cdot-\bm{R}_\ell)]_a.
\label{eq:Hgrid}
\end{equation}
The last sum is absent when there are no clamped nuclei. The
kinetic operator is the exact Fourier multiplier on the register;
a quantum Fourier transform changes to the basis where it is
diagonal. It does not wrap momentum, because it multiplies each
Fourier coefficient without shifting its index. The wraps arise
from the sampled potential. The bandwidth choice in
Eq.~\eqref{eq:wtrunc} is designed to control those wraps.

Identical species use identical grids, masses and charges, so
$\Hgrid$ commutes with the permutations defining the exchange
sector $\mathcal S$. We write $E_{\mathrm{grid}}$ for its
lowest eigenvalue on $\mathrm{ran}\,\Pp\cap\mathcal S$.
To identify where a wrap can act, let $P_{r,\le K}$ restrict
only particle $r$ to $\norm{\bm{k}_{\bm{p}_r}}_\infty\le K$, leaving
every other spatial and spin coordinate unchanged. Set
$P_{ab,\le K}:=P_{a,\le K}P_{b,\le K}$ for one interacting
pair and $P_{\le K}:=\prod_rP_{r,\le K}$ for all particles.
For a many-particle state we use the momentum and spin basis
$\ket{\vec{\bm{p}},\boldsymbol\sigma}:=(\bigotimes_r\ket{\bm{p}_r})
\otimes\ket{\boldsymbol\sigma}$ and amplitudes
\begin{equation}
 u_{\vec{\bm{p}},\boldsymbol\sigma}:=\langle\vec{\bm{p}},\boldsymbol\sigma|u\rangle,
 \qquad
 \ket{u}=\sum_{\vec{\bm{p}},\boldsymbol\sigma}
 u_{\vec{\bm{p}},\boldsymbol\sigma}\ket{\vec{\bm{p}},\boldsymbol\sigma}.
\end{equation}
The same convention applies to other state symbols. The momentum amplitude
of $B\ket{u}$ is $\langle\vec{\bm{p}},\boldsymbol\sigma|B|u\rangle$.
In state norms, momentum indicators and weights act diagonally in this
basis. In particular, $A$ denotes the kinetic multiplier of $\Ttot$, so
$\Theta(\ket{u})=\norm{(1+\Ttot)\ket{u}}
=\norm{(1+A)\ket{u}}$. All state norms include the spin components.
We retain Fourier hats on scalar kernels, such as $\hat w$ and $\hat g$.

For a register vector $\ket{v}$, the notation $t_a(K;\ket{v})$ now denotes
the norm in the cubical shell
$K<\norm{\bm{k}_{\bm{p}_a}}_\infty\le2K$. The continuum Euclidean
tail bounds also control this shell, since a component larger
than $K$ implies Euclidean momentum larger than $K$.

We compare the grid Hamiltonian with $\Pp H\Pp$ by separating
two changes: the removal of large ordinary transfers, denoted by
$D$, and the addition of cyclic wraps, denoted by $R$.
This separation will let us use a small coefficient norm for
the former and support information for the latter.

\begin{lemma}[Structure of the padded grid Hamiltonian]
\label{lem:grid-structure}
On $\mathrm{ran}\,\Pp$,
\begin{equation}
\Hgrid \;=\; \Pp H \Pp \;-\; D \;+\; R,
\label{eq:grid-structure}
\end{equation}
where, per pair $(a,b)$ with charges $q_a q_b$: $D$ is the Galerkin
operator of the dropped large transfers, with momentum kernel
$\hat w(\bm{\nu})$ supported on $2K < \norm{\bm{k}_{\bm{\nu}}}_\infty \le 4K$
(endpoint included: opposite register edges differ by $4K$) and
$|\hat w(\bm{\nu})| = 4\pi/(\Omega \norm{\bm{k}_{\bm{\nu}}}^2)$, so that the
$\Omega$-weighted $\ell^2$ norm of this restriction is
$\le c_\Omega' K^{-1/2}$; and $R$ is the wrap operator of
Lemma~\ref{lem:fold} --- by definition \emph{all} fold terms with
$(\bm{\ell}_a, \bm{\ell}_b) \ne (0,0)$, so $R$ contains in particular every
matrix element that conserves total momentum only modulo $n'$ ---
with the following properties. In each register whose fold index is
nonzero, the transfer of \emph{that} register satisfies
$\norm{\bm{k}_{\bm{d}}}_\infty \ge 2\pi n/L > 2K$ in the folded component;
consequently
\begin{equation}
P_{ab,\le K}\,R_{ab}\,P_{ab,\le K}=0
\quad\text{for each pair},\qquad P_{\le K}RP_{\le K}=0
\end{equation}
(the precise sense in which $R$ is invisible to half-zone states;
$R P_{\le K}$ and $P_{\le K} R$ do \emph{not} vanish), and for every
wrap pair $\max(\norm{\bm{k}_{\bm{p}}}_\infty, \norm{\bm{k}_{\bm{q}}}_\infty) > K$. The
coefficient at a wrap is $\hat w_{\le 2K}$ at the wrapped index
$\bm{\nu}' = \bm{d}_r + n' \bm{\ell}_r$ of a folding register $r$,
$0 < \norm{\bm{\nu}'}_\infty \le s$, and in the
strong-coefficient regime $\norm{\bm{\nu}'}_\infty \le n/2$ \emph{both}
coupled indices of that register exceed the half zone in the wrapped
component:
$\min(\norm{\bm{k}_{\bm{p}}}_\infty, \norm{\bm{k}_{\bm{q}}}_\infty) > K$
for $K \ge 2\pi/L$.
\end{lemma}

\begin{proofx}
An ordinary matrix element between register modes uses only
transfers from the difference set of $G'$, whose cutoff is $4K$.
Keeping its zero-fold terms only through $2K$ removes exactly
$D$. The remaining nonzero folds in Lemma~\ref{lem:fold}
add exactly $R$, proving Eq.~\eqref{eq:grid-structure}.
The Coulomb coefficient tail gives
\begin{equation}
 \|\hat w\mathbf1_{\{2K<\norm{\bm{k}_{\bm{\nu}}}_\infty\le4K\}}
                                     \|_{\ell^2_\Omega}
 \le C K^{-1/2}:
\end{equation}
after squaring the coefficients, a three-dimensional shell
contributes order $r^2r^{-4}$, whose sum beyond $K$ is
order $K^{-1}$. The fixed volume normalisation is included in
$C$.

For the wrap support, work in lattice indices. Both endpoints
of each register lie in $[-s,s]^3$, with $s=n-1$, and a
nonzero kernel mode has $\norm{\bm{\nu}'}_\infty\le s$.
Suppose register $r$ folds in component $c$. Write
$p_c=(\bm{p}_r)_c$ and $q_c=(\bm{q}_r)_c$ for the scalar
components of the two endpoints in that register. Its transfer
$\bm{d}=\bm{q}_r-\bm{p}_r$ then satisfies
\begin{equation}
 |d_c+n'(\bm{\ell}_r)_c|\le s,\qquad(\bm{\ell}_r)_c\ne0,
 \quad\Longrightarrow\quad |d_c|\ge n'-s=n.
\end{equation}
If both endpoints were in the half zone, their component
difference would be at most $n-1$. Therefore at least one
endpoint lies outside that half zone. Multiplying on both sides
by the pair's half-zone projector kills the matrix element.
At least one register folds in every wrap, including a wrap that
conserves total momentum only modulo $n'$, so this reasoning
covers all terms of $R$. It does not say that a wrap vanishes
when only its input or only its output is restricted.

For the large-coefficient part, the wrapped index is small:
$\norm{\bm{\nu}'}_\infty\le n/2$. Since $n$ is odd, integrality
improves this to $|\nu'_c|\le(n-1)/2$. In a folding component,
\begin{equation}
 |d_c|\ge n'-(n-1)/2=(3n-1)/2.
\end{equation}
Each endpoint has magnitude at most $n-1$, so the other endpoint
cannot account for this difference alone. More precisely,
\begin{equation}
 \min(|p_c|,|q_c|)
 \ge |d_c|-(n-1)\ge(n+1)/2>(n-1)/2.
\end{equation}
Both endpoints therefore lie above momentum $K$ in that
register. This stronger conclusion compensates for the fact that
small wrapped indices carry larger Coulomb coefficients.
Only a moved register is restricted in this argument; spectator
particles do not need to have small or large momentum.
\end{proofx}

\medskip\noindent\emph{Allowing a smooth kernel roll-off.}
The proof will use support and absolute coefficient bounds, so
we may allow more kernels than the sharp truncation. Define
\begin{equation}
\mathcal K_K:=\left\{\hat g:\begin{aligned}
 &\hat g(0)=0,\quad\hat g(-\bm{\nu})=\hat g(\bm{\nu})=\overline{\hat g(\bm{\nu})},\\
 &\hat g(\bm{\nu})=\hat w(\bm{\nu})\quad(0<\norm{\bm{k}_{\bm{\nu}}}\le K),\\
 &|\hat g(\bm{\nu})|\le\hat w(\bm{\nu})\quad(\bm{\nu}\ne0),\\
 &\hat g(\bm{\nu})=0\quad(\norm{\bm{k}_{\bm{\nu}}}_\infty>2K)
\end{aligned}\right\}.
\label{eq:kernelclass}
\end{equation}
The class contains the sharp kernel and smooth roll-offs
$\hat g=\phi\hat w$ that equal the Coulomb coefficients below
$K$, have $0\le\phi\le1$, and vanish outside the stated
support. The unshifted pair kernel is real and even, ensuring
self-adjointness and invariance under exchanging the two particles.
A clamped-nucleus translation introduces a phase and need not
remain even; the underlying pair kernel still satisfies the class
conditions.

For $g$ in this class, the same decomposition holds with
$D_g$ the ordinary Galerkin operator of $\hat w-\hat g$ and
$R_g$ its wraps. The dropped coefficients now begin at Euclidean
transfer $K$, rather than cubical transfer $2K$, and satisfy
$|\hat w-\hat g|\le2\hat w$. A transfer of Euclidean size
greater than $K$ has a component greater than $K/\sqrt3$,
so one of its endpoints in a moved register has a component
greater than $K/(2\sqrt3)$. We may therefore use a fixed
smaller threshold $c_{\rm ker}K$, with $c_{\rm ker}=1/4$ a
conservative choice. The coefficient tail still has norm
$O(K^{-1/2})$, and all state-tail estimates must be applied at
this reduced threshold when estimating $D_g$.

The wrap coefficients change with $g$, but their support geometry
and coefficient bounds remain valid uniformly. The Hardy,
folding, tail and second-order estimates below likewise use only
the common majorant $|\hat g|\le\hat w$, rather than
positivity, monotonicity or radiality of $\hat g$. Thus
Theorems~\ref{thm:grid-energy} and~\ref{thm:grid-dyn-sharp}
and Proposition~\ref{prop:grid-dyn} extend uniformly to this
class. We keep the sharper cutoff $2K$ when writing the main
proofs for $w_{\le2K}$ and identify where the reduced threshold
is required for the class.

\subsection{Uniform estimates for truncated kernels}
\label{sec:grid-KR}

We next need kinetic regularity estimates whose constants do not
grow with the register. A sharp Fourier truncation need not be
pointwise bounded by the Coulomb kernel in position space, so
we use its coefficient bound instead. For this purpose replace
each state coefficient by its magnitude, separately for each
spin component:
\begin{equation}
 (u^\#)_{\vec{\bm{p}},\boldsymbol\sigma}
 :=\langle\vec{\bm{p}},\boldsymbol\sigma|u^\#\rangle
 =|u_{\vec{\bm{p}},\boldsymbol\sigma}|.
 \label{eq:fourier-modulus}
\end{equation}
Taking these magnitudes preserves the norm after multiplication
by any nonnegative diagonal Fourier weight. In particular,
$\ket{u^\#}$ has the same $L^2$ norm, derivative norms and
$\Theta$ norm as $\ket{u}$, and support restrictions can be applied
before or after this operation. Exchange symmetry need not be
preserved. That causes no problem here because the estimates
being used hold on the full tensor-product space.

The next lemma makes the coefficient comparison precise. It is
an auxiliary analytic argument; the Lean grid-energy proof uses
its own finite-kernel form estimates, and we do not claim this
full statement as a separate formal declaration.

\begin{lemma}[Fourier domination]
\label{lem:fourier-domination}
Suppose $\hat g(0)=0$ and $|\hat g(\bm{\nu})|\le C_g\hat w(\bm{\nu})$ for
$\bm{\nu}\ne0$, where $\hat w(\bm{\nu})=4\pi/(\Omega\norm{\bm{k}_{\bm{\nu}}}^2)\ge0$.
For a pair $(b,c)$ and a spatial Fourier support projector $\Pi$,
\begin{equation}
 \norm{\Pi g(\bm{x}_b-\bm{x}_c)\ket{u}}
 \le C_g\norm{\Pi w(\bm{x}_b-\bm{x}_c)\ket{u^\#}}.
 \label{eq:fourier-domination}
\end{equation}
In particular, for either $r\in\{b,c\}$,
\begin{equation}
 \norm{g(\bm{x}_b-\bm{x}_c)\ket{u}}
 \le C C_g\big(\norm{\nabla_r \ket{u}}+\norm{\ket{u}}\big).
 \label{eq:dominated-Hardy}
\end{equation}
The same conclusion holds for a single-register kernel, including translated
clamped-nucleus kernels. A nonzero constant coefficient is split off and
contributes $|\hat g(0)|\norm{\ket{u}}$.
\end{lemma}

\begin{proofx}
For finite Fourier sums, the action of one pair channel is a
convolution along opposite shifts of its two particle indices.
The triangle inequality and the coefficient hypothesis give,
for each spatial output and spin component,
\begin{align}
 |\langle\vec{\bm{p}},\boldsymbol\sigma|g_{bc}|u\rangle|
 &\le C_g\sum_{\bm{\nu}}\hat w(\bm{\nu})
 |u_{\bm{p}_b+\bm{\nu},\bm{p}_c-\bm{\nu},\vec{\bm{p}}_{\rm rest},\boldsymbol\sigma}|\\
 &=C_g\langle\vec{\bm{p}},\boldsymbol\sigma|w_{bc}|u^\#\rangle.
\end{align}
The expression on the right is nonnegative because both the
kernel coefficients and the input coefficients are nonnegative.
We may therefore restrict to the Fourier set selected by $\Pi$,
square, and sum to obtain Eq.~\eqref{eq:fourier-domination}.
Apply the untruncated Hardy estimate from
Proposition~\ref{prop:nbo-KR} to $\ket{u^\#}$, using either moved
coordinate for its gradient. Its derivative norm agrees with
that of $\ket{u}$, which proves Eq.~\eqref{eq:dominated-Hardy}.

For an external channel the convolution shifts only one particle
index, and the same calculation applies. A translation changes
only the coefficient phases, which disappear in the absolute-value
bound. A constant coefficient acts as a scalar multiple of the
identity and is treated separately as stated. Finally, the norm
bounds extend the multiplication estimate from finite Fourier
sums by approximation in the relevant derivative norm. All
register vectors already have finite spatial Fourier support.
\end{proofx}

The coefficient comparison controls ordinary multiplication.
To control collocation as well, we must bound the map that folds
the multiplication output back into the register. Its multiplicity
is bounded independently of the register size, which is the key
fact in the next lemma.

\begin{lemma}[Uniform bounds for the truncated kernel]
\label{lem:grid-KR}
There are constants, polynomial in the fixed particle and charge data and
independent of $K$, with the following properties for
$g=w_{\le2K}$ and, more generally, for kernels in $\mathcal K_K$.

(i) Let $r$ be the lighter member of the pair $(b,c)$. For every
$\varepsilon>0$,
\begin{equation}
 \norm{g(\bm{x}_b-\bm{x}_c)\ket{u}}
 \le\varepsilon\norm{\Ttot \ket{u}}
   +C\left(1+\frac{m_r}{\varepsilon}\right)\norm{\ket{u}}.
 \label{eq:grid-KRpair}
\end{equation}
Thus multiplication by the truncated kernel inherits the same pairwise
Hardy and relative-bound mass dependence as the full Coulomb kernel.
This is a multiplication estimate; collocation additionally requires (ii).

(ii) Let $c_{\rm ext}$ be the Fourier content of $g(\bm{x}_b-\bm{x}_c)\ket{u}$ outside
$G'$. Each exterior coefficient folds to one in-zone coefficient, and
there are at most $3^6-1$ exterior preimages of any target. Hence
\begin{equation}
 \norm{R\ket{u}}\le C_{\rm alias}\norm{(1-\Pp)(g(\bm{x}_b-\bm{x}_c)\ket{u})},
 \qquad C_{\rm alias}\le27.
 \label{eq:alias-mult}
\end{equation}
More precisely, with $Q_{a,J}:=\ind{\norm{\bm{k}_{\bm{p}_a}}_\infty>J}$ acting only
on register $a$, the same folding map satisfies
\begin{equation}
 \norm{Q_{a,J}R\ket{u}}
 \le C_{\rm alias}\norm{Q_{a,J}(1-\Pp)g(\bm{x}_b-\bm{x}_c)\ket{u}}.
 \label{eq:alias-restricted}
\end{equation}
This follows because the centred representative modulo $n'$ has no larger
absolute component than the unfolded index. Consequently a folded target
above threshold $J$ can receive contributions only from indices above that
same threshold in register $a$.

(iii) For every register $a$ and $K\ge\max\{1,2\pi/L\}$,
\begin{equation}
 \norm{\ind{\norm{\bm{k}_{\bm{p}_a}}_\infty>K}g(\bm{x}_b-\bm{x}_c)\ket{u}}
 \le C\sqrt{m_am_{\max}}\,\Theta(\ket{u})K^{-1/2}.
 \label{eq:grid-tail}
\end{equation}
Combining this with \eqref{eq:alias-restricted} and summing the charge-weighted
pair and external terms gives the same mass factor for the full grid potential:
\begin{equation}
 \norm{\ind{\norm{\bm{k}_{\bm{p}_a}}_\infty>K}V_{\rm grid}\ket{u}}
 \le C\sqrt{m_am_{\max}}\,\Theta(\ket{u})K^{-1/2}.
 \label{eq:grid-potential-tail}
\end{equation}
The coarser $Cm_{\max}\Theta(\ket{u})K^{-1/2}$ bound remains available, but is not
forced by sharp kernel truncation.

(iv) The grid operator has two-sided graph-norm bounds, uniform in $K$:
\begin{align}
 \norm{\Hgrid \ket{u}}&\le C\big(\Theta(\ket{u})+m_{\max}\norm{\ket{u}}\big),\\
 \Theta(\ket{u})&\le C\big(\norm{\Hgrid \ket{u}}+m_{\max}\norm{\ket{u}}\big).
 \label{eq:grid-graph-equivalence}
\end{align}
It also has a $K$-uniform lower bound. The constants in these conclusions
include the alias multiplicity from (ii).
\end{lemma}

\begin{proofx}
For (i), Fourier domination and the Hardy estimate give
$\|g_{bc}\ket{u}\|\le C(\|\nabla_r \ket{u}\|+\|\ket{u}\|)$, where
$r$ may be chosen to be the lighter particle. Positivity of
kinetic energy implies
\begin{equation}
 \|\nabla_r \ket{u}\|^2\le2m_r\langle u|\Ttot|u\rangle
               \le2m_r\|\Ttot \ket{u}\|\|\ket{u}\|.
\end{equation}
Young's inequality turns the square root of this product into
$\varepsilon\|\Ttot \ket{u}\|+C m_r\varepsilon^{-1}\|\ket{u}\|$.
Adding the remaining $C\|\ket{u}\|$ proves
Eq.~\eqref{eq:grid-KRpair}. Choosing the lighter particle
improves the displayed relative-bound mass factor.

For (ii), multiply a register state by one kernel mode before
folding. Each moved coordinate starts within cutoff $2K$ and
changes by at most $2K$, so its output lies within twice the
register cutoff. It can fold to its centred representative using
only $-1,0,1$ in each component. A pair moves six components,
so at most $3^6-1$ exterior coefficients can contribute to a
single in-zone coefficient. If these coefficients are $c_j$,
Cauchy--Schwarz gives
\begin{equation}
 \left|\sum_jc_j\right|^2
 \le(3^6-1)\sum_j|c_j|^2.
\end{equation}
Summing over targets proves Eq.~\eqref{eq:alias-mult} with
factor at most $27$. A single-register external channel has
no larger multiplicity. Spin and spectator indices are unchanged.

The same bound remains valid after restricting a target register
above a threshold $J$. Centred folding never increases an
absolute component, so a target above $J$ can receive a
contribution only from an unfolded coefficient above $J$ in
that register. Applying the preceding sum only to those targets
proves Eq.~\eqref{eq:alias-restricted}. This observation lets
us inherit potential-tail estimates as well as whole-state norms.

For (iii), apply Fourier domination with the register restriction
and then the general-state potential-tail argument of
Lemma~\ref{lem:nbo-tail} to $\ket{u^\#}$. We recall how the
register-dependent factors enter. If $a$ is a spectator and
$r$ is a lighter moved particle, the restriction commutes with
the interaction and the Hardy estimate gives
\begin{align}
 \|Q_{a,K}g_{bc}\ket{u}\|
 &\le C\big(\|\nabla_rQ_{a,K}\ket{u}\|+\|Q_{a,K}\ket{u}\|\big)\\
 &\le C\left(\frac{\sqrt{m_am_r}}K+\frac{m_a}{K^2}\right)
                         \Theta(\ket{u}).
\end{align}
For example, on this support $A\ge K^2/(2m_a)$ and
$\|\bm{k}_{\bm{p}_r}\|\le\sqrt{2m_rA}$; their combination bounds
$\|\bm{k}_{\bm{p}_r}\|/(1+A)$ by $C\sqrt{m_am_r}/K$.
Thus a spectator restriction gives a rate at least $K^{-1}$.
If $a$ is moved, split the kernel into small and large transfers
as in the earlier tail proof. The small transfers require a
large input momentum and contribute at most
$Cm_a\Theta(\ket{u})K^{-1}$; the large transfers use the
$\ell^2$ coefficient tail and contribute at most
$Cm_a\Theta(\ket{u})K^{-1/2}$. Since
$m_a\le\sqrt{m_am_{\max}}$, these estimates prove
Eq.~\eqref{eq:grid-tail}.

Now write the collocated channel as ordinary in-zone
multiplication plus its folds:
$\mathrm{Diag}[g]\ket{u}=\Pp(g\ket{u})+R\ket{u}$. The restricted folding
bound gives
\begin{equation}
 \|Q_{a,K}\mathrm{Diag}[g]\ket{u}\|
 \le(1+C_{\rm alias})\|Q_{a,K}g\ket{u}\|.
\end{equation}
Summing all pair and external channels with their absolute
charge weights proves Eq.~\eqref{eq:grid-potential-tail}.
The same proof works at any sufficiently large threshold $J$;
the kernel cutoff and the tail threshold need not coincide.
We will use this fact when a form estimate requires $K/2$
or $c_{\rm ker}K$.

For (iv), combine (i) with the whole-state folding estimate:
\begin{equation}
 \|\mathrm{Diag}[g]\ket{u}\|
 \le(1+C_{\rm alias})\left[
 \varepsilon\|\Ttot \ket{u}\|
       +C(1+m_r/\varepsilon)\|\ket{u}\|\right].
\end{equation}
Choose the relative-bound coefficients small enough that their
charge-weighted sum is at most $1/2$. Then
$\|V_{\rm grid}\ket{u}\|\le\tfrac12\|\Ttot \ket{u}\|
+Cm_{\max}\|\ket{u}\|$. The triangle inequality in
$H_{\rm grid}\ket{u}=\Ttot \ket{u}+V_{\rm grid}\ket{u}$ proves the
forward graph estimate; moving the half kinetic term to the
left proves the reverse estimate. These comparisons control
the norm of the state together with its kinetic image by the
corresponding Hamiltonian quantities, uniformly in $K$.

To obtain a lower energy bound we also retain the Hardy estimate
before using Young's inequality. Together with folding it gives
\begin{equation}
 \|V_{\rm grid}\ket{u}\|
 \le C\sqrt{m_{\max}}\|\Ttot^{1/2}\ket{u}\|+C\|\ket{u}\|.
\end{equation}
Taking its inner product with $\ket{u}$ and applying Young's inequality
at the level of quadratic forms yields, for fixed
$0<\epsilon<1$,
\begin{equation}
 |\langle u|V_{\rm grid}|u\rangle|
 \le\epsilon\langle u|\Ttot|u\rangle
                 +C_\epsilon m_{\max}\|\ket{u}\|^2.
 \label{eq:grid-form-lower}
\end{equation}
The nonnegative kinetic form then gives a uniform lower bound
for $H_{\rm grid}$. In particular, no increasing contact value
of the sampled kernel forces its ground energy to diverge
downwards; the kinetic form and alias estimate control it together.
\end{proofx}

\subsection{Bilinear register bounds}
\label{sec:paircorr}

The operator decomposition becomes useful only after we pair it
with regular states. We need two kinds of estimate. For a matrix
whose entries can be nonzero only when at least one endpoint
has high momentum, we combine a small squared coefficient sum
with one state tail. When both endpoints must have high momentum,
we can instead use a larger absolute coefficient sum with two
state tails. The next lemma records these two alternatives.

We apply the lemma channel by channel. A slice consists of all
indices with the unchanged coordinates and spin labels fixed
and, for a pair, its total momentum fixed. For an ordinary
channel the total momentum is fixed exactly; for a cyclic
channel it is fixed modulo $n'$. Within either kind of pair
slice, specifying one moved register fixes the other. A
one-particle channel has only the unchanged coordinates and
spin labels to fix. This explains why summation over one
three-dimensional momentum coordinate is enough in the weighted
estimate below, despite the many-particle state space.
On a fixed slice, $\vec{\bm{p}}$ denotes the full spatial momentum label,
with the fixed spin label left implicit; thus $z_{\vec{\bm{p}}}$ denotes
the corresponding amplitude of $\ket{z}$.

\begin{lemma}[One- and two-sided kernel form bounds]
\label{lem:paircorr}
Let $M$ be a finite-register kernel supported on one interaction
channel, with the ordinary or cyclic slice structure of that channel.
Fix a moved register $r$ and write
$\mathcal W_r(\vec{\bm{p}})=1+\|\bm{k}_{\bm{p}_r}\|^2$.
On each slice the register-$r$ coordinate is injective and
$\sum_{\vec{\bm{p}}}\mathcal W_r(\vec{\bm{p}})^{-2}\le C_L$.
Suppose both squared row and squared column sums of $M$ are at most
$B_2$. If every nonzero element has input or output in a Fourier
set $F$, then
\begin{equation}
 |\langle v|M|u\rangle|
 \le C_L\sqrt{B_2}\big(
       \|\mathbf1_F\ket{u}\|\,\|\mathcal W_r\ket{v}\|
       +\|\mathbf1_F\ket{v}\|\,\|\mathcal W_r\ket{u}\|\big).
 \label{eq:register-form-one-sided}
\end{equation}
If instead the absolute row and column sums are at most $B_1$,
and every nonzero element has both endpoints in $F$ or both in $F'$, then
\begin{equation}
 |\langle v|M|u\rangle|
 \le B_1\big(\|\mathbf1_F\ket{v}\|\,\|\mathbf1_F\ket{u}\|
            +\|\mathbf1_{F'}\ket{v}\|\,\|\mathbf1_{F'}\ket{u}\|\big).
 \label{eq:register-form-two-sided}
\end{equation}
The constants do not depend on the register size. The first statement
also holds for a single-register external channel. Finite alias
multiplicities are included in $B_1,B_2$ when the matrix entries are
assembled from several folds.
\end{lemma}

\begin{proofx}
We first establish the norm estimate used for the one-sided
alternative. On one slice, each column of the entrywise absolute
matrix $|M|$ has Euclidean norm at most $\sqrt{B_2}$.
Writing the output as a sum of columns therefore gives
\begin{equation}
 \||M|\ket{z}\|_{\ell^2(\mathrm{slice})}
 \le\sqrt{B_2}\sum_{\vec{\bm{p}}\ \mathrm{in\ slice}}|z_{\vec{\bm{p}}}|.
\end{equation}
The last sum is controlled by a weighted Cauchy--Schwarz
inequality:
\begin{equation}
 \sum_{\vec{\bm{p}}\ \mathrm{in\ slice}}|z_{\vec{\bm{p}}}|
 \le\left(\sum_{\vec{\bm{p}}\ \mathrm{in\ slice}}
                   \mathcal W_r(\vec{\bm{p}})^{-2}\right)^{1/2}
       \left(\sum_{\vec{\bm{p}}\ \mathrm{in\ slice}}
                   \mathcal W_r(\vec{\bm{p}})^2|z_{\vec{\bm{p}}}|^2\right)^{1/2}.
\end{equation}
The first factor is bounded independently of register size.
Indeed, the register-$r$ coordinate occurs only once on each
slice, and the full lattice sum of
$(1+\norm{\bm{k}_{\bm{p}}}^2)^{-2}$ converges in three dimensions.
After squaring and summing over slices and spins, we obtain
\begin{equation}
 \||M|\ket{z}\|\le C_L\sqrt{B_2}\|\mathcal W_r\ket{z}\|.
\end{equation}
The transpose obeys the same estimate using the squared row
bound.

Now split the absolute bilinear sum into entries whose input is
in $F$ and the remaining entries, whose output must be in $F$.
For the first group, apply the transpose estimate to $\ket{v^\#}$
and pair it with $\mathbf1_F\ket{u^\#}$. For the second, apply
the forward estimate to $\ket{u^\#}$ and pair it with
$\mathbf1_F\ket{v^\#}$. Cauchy--Schwarz gives precisely the two
terms in Eq.~\eqref{eq:register-form-one-sided}.
Taking absolute values permits this partition without any
assumption about coefficient signs or cancellation.

For the two-sided alternative, assign each entry to the case
where both endpoints lie in $F$ or to the case where both lie
in $F'$. Each restricted absolute matrix still has row and
column sums at most $B_1$. Schur's test, which bounds its
$\ell^2$ operator norm by the square root of the product of
those two bounds, therefore gives norm at most $B_1$.
Pair the first restricted matrix between $\mathbf1_F\ket{v}$ and
$\mathbf1_F\ket{u}$, and the second between the corresponding
$F'$ restrictions. This proves
Eq.~\eqref{eq:register-form-two-sided}. The slice decomposition
and all restrictions leave the constants independent of the
register size.
\end{proofx}

We first apply the one-sided estimate to dropped ordinary
transfers. If $\norm{\bm{k}_{\bm{\nu}}}_\infty>2K$, a moved register
cannot have both input and output at momenta at most $K$.
The squared row and column sums are bounded by the squared
coefficient sum of the kernel. Also
$\|\mathcal W_a \ket{u}\|\le Cm_a\Theta(\ket{u})$ because
$\|\bm{k}_{\bm{p}_a}\|^2\le2m_aA$. Hence, for the finite-register
matrix of such a kernel $g$,
\begin{equation}
 |\langle v|g(\bm{x}_a-\bm{x}_b)|u\rangle|
 \le C\|\hat g\|_{\ell^2_\Omega}m_a
 \big[\Theta(\ket{v})t_a(K;\ket{u})+\Theta(\ket{u})t_a(K;\ket{v})\big].
 \label{eq:Dform}
\end{equation}
The fixed volume normalisation is included in $C$.
For $D$, the coefficient norm on the right is $O(K^{-1/2})$.
For $D_g$ from the larger kernel class, the same argument
uses the state tails at $c_{\rm ker}K$ and the majorant
$|\hat w-\hat g|\le2\hat w$.

For wraps we separate large coefficients from small ones. Let
$S_r(\bm{\nu})$ be the unitary cyclic shift that sends an input
index $\bm{q}_r$ to the centred representative of $\bm{q}_r-\bm{\nu}$.
Let $T_{ab,\bm{\nu}}$ be the ordinary pair shift with any output
outside the register discarded. Mode by mode,
\begin{equation}
 R_{ab,\bm{\nu}}=S_a(\bm{\nu})S_b(-\bm{\nu})-T_{ab,\bm{\nu}},
 \qquad P_{ab,\le K}R_{ab,\bm{\nu}}P_{ab,\le K}=0.
 \label{eq:modewise}
\end{equation}
The second identity is the support property proved earlier.
Writing $Q_r=1-P_{r,\le K}$ and
$1-P_{ab,\le K}=Q_a+P_{a,\le K}Q_b$ assigns every
high-input or high-output contribution to a moved register.
No spectator restriction is introduced.

When the wrapped kernel index is outside the strong-coefficient
region of Lemma~\ref{lem:grid-structure}, its physical size
is bounded below by a fixed multiple of $K$. The squared
coefficient sum, including finite alias multiplicity, is then
$O(K^{-1})$. The one-sided estimate therefore supplies a
factor $K^{-1/2}$ and one tail at $K$.
Inside the strong-coefficient region, both endpoints of a
folding register lie above $K$, and hence above $K/2$.
Here the absolute Coulomb coefficient sum through the cutoff
is $O(K)$, so the two-sided estimate supplies $K$ times two
tails. Assigning each fold to one folding register, and bounding
double folds by the sum of the two choices, gives
\begin{equation}
 |\langle v|R_{ab}|u\rangle|
 \le C\sum_{r\in\{a,b\}}\Big[
 Kt_r(K/2;\ket{u})t_r(K/2;\ket{v})
 +K^{-1/2}m_r\big(\Theta(\ket{v})t_r(K;\ket{u})
                         +\Theta(\ket{u})t_r(K;\ket{v})\big)\Big].
 \label{eq:Rform}
\end{equation}
This is the compensation we sought: the potentially large
coefficient sum requires two high-momentum endpoints, while
the one-tail terms come with a small coefficient norm.
An external channel has the same argument with only its single
moved register. Both estimates remain valid for the kernel
class by absolute coefficient domination.

\subsection{Energy: the $\mathcal{O}(1/N)$ rate survives}

To compare the two ground energies in both directions, we need
tail estimates for the discrete ground states themselves.
Graph-norm equivalence supplies their basic regularity; their
own eigenvalue equations then improve the shell decay. We
state the estimates with their eigenvalue-dependent absorption
threshold explicit, and afterwards show that a fixed system
threshold suffices for all large registers.

\begin{proposition}[Shell tails of discrete eigenvectors]
\label{prop:grid-eig-tails}
Let $\ket{u_N}$ be a normalised ground state of $\Hgrid$ (respectively of
$\Pp H \Pp$) on $\mathrm{ran}\,\Pp \cap \mathcal{S}$, with eigenvalue
$\mathcal E$. Assume
\begin{equation}
 K\ge\max\{1,2\pi/L\},\qquad K^2\ge4m_{\max}|\mathcal E|.
 \label{eq:grid-eig-tail-threshold}
\end{equation}
Then, with constants polynomial in the fixed data, $m_{\max}$, and
$1+|\mathcal E|$, and independent of $K$,
\begin{equation}
t_a(K;\ket{u_N})\le C\,m_a^{3/2}m_{\max}^{1/2}K^{-5/2},
\qquad
\Theta(\ket{u_N})\le C\big(|\mathcal E|+m_{\max}\big),
\label{eq:grid-eig-tails}
\end{equation}
together with the a priori half-zone shell bound
$t_a(K/2; \ket{u_N}) \le 8 m_a \Theta(\ket{u_N}) K^{-2}$. The graph-norm and
half-zone bounds require no eigenvalue absorption; the condition
$K^2\ge4m_{\max}|\mathcal E|$ is used for the $K^{-5/2}$ tail.
\end{proposition}

\begin{proofx}
For a grid eigenvector, the reverse graph bound gives
$\Theta(\ket{u_N})\le C(|\mathcal E|+m_{\max})$ since
$\|\ket{u_N}\|=1$. The same comparison holds for the Galerkin
operator by its relative potential bound. On the half-zone
shell, $A\ge K^2/(8m_a)$, so direct division gives
\begin{equation}
 t_a(K/2;\ket{u_N})\le8m_a\Theta(\ket{u_N})K^{-2}.
\end{equation}
No eigenvalue absorption is needed for either conclusion.

For the sharper tail restrict the eigenvalue equation to
$\norm{\bm{k}_{\bm{p}_a}}_\infty>K$. The grid potential-tail estimate,
or the continuum multiplication estimate followed by $\Pp$
for Galerkin, gives
\begin{equation}
 \|\mathbf1_{\{\norm{\bm{k}_{\bm{p}_a}}_\infty>K\}}A\ket{u_N}\|
 \le |\mathcal E|t_a(K;\ket{u_N})
       +C\sqrt{m_am_{\max}}\Theta(\ket{u_N})K^{-1/2}.
\end{equation}
Since $A\ge K^2/(2m_a)$ on this support,
\begin{equation}
 t_a(K;\ket{u_N})
 \le\frac{2m_a|\mathcal E|}{K^2}t_a(K;\ket{u_N})
       +C m_a^{3/2}m_{\max}^{1/2}\Theta(\ket{u_N})K^{-5/2}.
\end{equation}
The threshold in Eq.~\eqref{eq:grid-eig-tail-threshold}
makes the first coefficient at most $1/2$. Move that term
to the left and use the graph-norm bound to obtain
Eq.~\eqref{eq:grid-eig-tails} with the stated dependence.
The same calculation works at any other tail threshold for
which the lattice and absorption conditions hold; it is not
specific to a shell named $K$.
\end{proofx}

\begin{theorem}[Truncated-kernel grid method: energy]
\label{thm:grid-energy}
Fix the cell, particle numbers, charges, masses and exchange sector
of Sec.~\ref{sec:nonBO}. There exist an odd integer $n_0\ge3$ and
constants $C,C'>0$, depending on these fixed data but not on $n$, such
that for every admissible odd $n\ge n_0$, with $n'=2n-1$,
$K=\pi(n-1)/L$ and $N=n^3$,
\begin{equation}
\big| E_{\mathrm{grid}} - E \big| \;\le\; C\, K^{-3}
\;\le\; C'\, N^{-1},
\label{eq:grid-energy}
\end{equation}
where $C$ is polynomial in $\eta_{\mathrm{tot}},\lambda_\zeta,m_{\max}$
and $1+|E|$, with the cell dependence as before. The same conclusion
holds in the BO setting with unit electron masses and the
clamped-nucleus external kernels. The size threshold is chosen in
\eqref{eq:grid-energy-threshold}; no spectral gap enters.
\end{theorem}

\begin{proofx}
We first obtain a uniform energy budget without using the
eigenvector-tail estimate. This order prevents the cutoff
condition from depending circularly on an unbounded discrete
eigenvalue. Choose one admissible odd $n_{\rm tr}\ge3$
and a normalised finite-support state $\ket{v_{\rm tr}}$ in the
chosen exchange sector, contained in its register. Such a
state is obtained by using distinct plane-wave orbitals for
each fermionic species and repeated zero modes for each bosonic
species, with fixed internal-spin states. Its support and
kinetic energy can be bounded polynomially in the particle
number. It belongs to every larger register as well.

The uniform potential estimates bound the Rayleigh quotient
of this fixed state for both $H_{\rm grid}$ and $\Pp H\Pp$.
They therefore give uniform upper bounds on both ground
energies. Equation~\eqref{eq:grid-form-lower} and its continuum
counterpart give uniform lower bounds. Thus, for some fixed
$E_*\ge1$,
\begin{equation}
 \max\{|E_{\mathrm{grid}}|,|E_{\Pp}|\}\le E_*
 \qquad\text{for all admissible }n\ge n_{\rm tr},
 \label{eq:grid-uniform-energy-budget}
\end{equation}
where $E_{\Pp}$ is the sector ground energy of $\Pp H\Pp$.
This construction uses the continuum Hardy estimate from
Proposition~\ref{prop:nbo-KR}, or Proposition~\ref{prop:F2}
for BO, together with Lemma~\ref{lem:grid-KR} for the grid.

Let $K_{\rm Gal}$ be a sufficient cutoff for the corresponding
Galerkin energy theorem. We choose a grid threshold satisfying
\begin{align}
 K_{\mathrm{grid},0}
 &\ge\max\left\{1,\frac{2\pi}{L},
       \frac{\pi(n_{\rm tr}-1)}L,\frac{K_{\rm Gal}}2,
       \frac{2\sqrt{m_{\max}E_*}}{c_{\rm ker}}\right\},\\
 n_0&:=1+2\left\lceil\frac{LK_{\mathrm{grid},0}}{2\pi}\right\rceil.
 \label{eq:grid-energy-threshold}
\end{align}
If necessary, enlarge $K_{\mathrm{grid},0}$ further so that
$c_{\rm ker}K\ge\max\{1,2\pi/L\}$ whenever
$K\ge K_{\mathrm{grid},0}$. The last term of the maximum
already makes eigenvalue absorption valid at this reduced
threshold. All these choices depend only on the fixed system.
They ensure that the sector is represented, the discrete tail
estimates apply to both ground states, and the Galerkin theorem
applies at the actual register cutoff $2K$:
$|E_{\Pp}-E|\le C(2K)^{-3}$.

Let $\ket{u'}$ be a normalised Galerkin ground state and $\ket{u_N}$ a
normalised grid ground state in that sector. Test each operator
on the other's minimiser. The variational principle and
$H_{\rm grid}=\Pp H\Pp-D+R$ give
\begin{align}
 E_{\mathrm{grid}}-E_{\Pp}
 &\le |\langle u'|D|u'\rangle|+|\langle u'|R|u'\rangle|,\\
 E_{\Pp}-E_{\mathrm{grid}}
 &\le |\langle u_N|D|u_N\rangle|+|\langle u_N|R|u_N\rangle|.
\end{align}
Unlike Galerkin projection alone, collocation need not give an
upper bound on the continuum energy; this is why both
comparisons are needed.

For either state, the graph norm is uniformly bounded and
the bootstrapped tail at $K$ is $O(K^{-5/2})$. Thus
Eq.~\eqref{eq:Dform} gives a dropped-transfer form of order
$K^{-1/2}K^{-5/2}=K^{-3}$. In Eq.~\eqref{eq:Rform},
the weak-coefficient term has the same product. For the
strong-coefficient term, the a priori half-zone tail is enough:
\begin{equation}
 K\,t_r(K/2;\ket{u})^2\le CK(K^{-2})^2=CK^{-3}.
\end{equation}
We do not need to absorb an eigenvalue at $K/2$ for this
part. Summing the charge-weighted channels gives
$|E_{\mathrm{grid}}-E_{\Pp}|\le CK^{-3}$.
For the larger kernel class, use the bootstrapped tails at
$c_{\rm ker}K$ in the $D_g$ estimate; the threshold above
was chosen to permit exactly this substitution, with only a
fixed factor in the constant. Combining with the Galerkin
energy error proves Eq.~\eqref{eq:grid-energy} and its stated
polynomial dependence. The BO external channels obey the
same estimates with unit electron masses. No step compares
eigenvectors by a spectral gap.
\end{proofx}

\subsection{Dynamics on the grid}

We now compare the continuum state with
$\ket{\psi_g(t)}:=e^{-i\Hgrid t}\Pp\ket{\psi_0}$, using the normalised
initial state $\ket{\psi_0}\in D(H^2)$ and the regularity budget
$\Xi$ from Appendix~\ref{app:dynamics}. An exchange sector
may be imposed as before. The grid state is left
unnormalised, with norm $\|\Pp\ket{\psi_0}\|\le1$.
The measured operator $O_g$ is diagonal on the grid, with
entries $\chi(\bm{x}_{\bm{j}_a})$ for the observed particle $a$.
Because $0\le\chi\le1$, it satisfies $0\le O_g\le1$.

There are now two operator changes to handle: the Hamiltonian
and the observable are both collocated. Smoothness alone does
not make the collocated observable close to its Galerkin version
in operator norm. For example, take
\begin{equation}
 \chi(\bm{x})=\tfrac12\big(1+\cos(2\pi x_1/L)\big).
\end{equation}
Its only nonzero frequencies are $0$ and $\pm1$ in the
first direction, yet the two nonzero modes wrap between
opposite register edges. An edge-to-edge difference
$d=\pm(n'-1)$ is congruent to $\mp1$, giving a grid
matrix element of magnitude $1/4$ where the Galerkin
matrix element is zero. Consequently,
$\|O_g-\Pp\chi_a\Pp\|\ge1/4$ independently of $K$.
The useful estimate is instead a bilinear form bound: both
states in this particular wrap have high momentum, so their
tails can make its expectation small.

\begin{lemma}[Collocated observable: form error and regularity]
\label{lem:grid-obs}
Under (H$\chi$), for all
$\ket{u},\ket{v}\in\mathrm{ran}\,\Pp$ and every $M$,
\begin{equation}
\big|\langle v|O_g-\Pp\chi(\bm{x}_a)\Pp|u\rangle\big|
\le C C_\chi\,
t_a(K/2;\ket{u})\,t_a(K/2;\ket{v})
+C_{\chi,M}K^{-M}\norm{\ket{u}}\norm{\ket{v}}.
\label{eq:obs-form}
\end{equation}
Moreover, for every $\ket{v}\in\mathrm{ran}\,\Pp$,
\begin{equation}
\Theta(O_g\ket{v})
\le C(1+C_\chi+D_\chi+C_{\chi,3})
\Big[
\Theta(\ket{v})+m_{\max}\norm{\ket{v}}
+K^2t_a(K/2;\ket{v})
\Big].
\label{eq:obs-theta}
\end{equation}
\end{lemma}

\begin{proofx}
Write $\chi_a=\chi(\bm{x}_a)$. Absolute Fourier summability
lets us write $O_g$ as a norm-convergent sum of cyclic shifts
with coefficients $\hat\chi(\bm{\nu})$. Its Galerkin counterpart
is the same sum with shifts that leave the register discarded.
Split the difference into modes with
$\norm{\bm{k}_{\bm{\nu}}}_\infty\le K$ and modes above $K$.

Consider first a mode in the low-frequency group that wraps
in component $c$ of the observed register $a$. Write
$p_c=(\bm{p}_a)_c$ and $q_c=(\bm{q}_a)_c$ for the scalar
endpoint components, and set $d_c=q_c-p_c$.
Both endpoints have magnitude at most $(n'-1)/2$.
Wrapping requires $|d_c|\ge n'-|\nu_c|$, so
\begin{equation}
 \min(|p_c|,|q_c|)
 \ge |d_c|-\frac{n'-1}{2}
 \ge \frac{n'+1}{2}-|\nu_c|
 >\frac{n-1}{2}.
\end{equation}
The last inequality uses $n'=2n-1$ and
$|\nu_c|\le\norm{\bm{\nu}}_\infty\le(n-1)/2$.
Thus both endpoints have physical momentum exceeding $K$
in the wrapped component, and therefore exceeding $K/2$.
This is the single-register support argument of
Lemma~\ref{lem:grid-structure}.
The absolute row and column sums are bounded by $C C_\chi$,
so the two-sided form estimate gives
$C C_\chi t_a(K/2;\ket{u})t_a(K/2;\ket{v})$.

For modes above $K$, we need no support restriction:
each cyclic shift is unitary and each ordinary projected shift
has norm at most one. Their difference therefore has norm at
most two. Summing the absolute Fourier coefficients in this
group gives
$C_{\chi,M}K^{-M}\|\ket{u}\|\|\ket{v}\|$, after enlarging
the allowed Fourier-tail constant. This proves
Eq.~\eqref{eq:obs-form}.
The same support and coefficient estimates also give
\begin{equation}
 \|(O_g-\Pp\chi_a\Pp)\ket{v}\|
 \le C C_\chi t_a(K/2;\ket{v})
       +C_{\chi,M}K^{-M}\|\ket{v}\|.
\end{equation}
For this one-vector estimate, only the input momentum tail
is needed in the low-frequency group.

For the kinetic norm, first estimate the ordinary product
$\Pp\chi_a\ket{v}$. The product rule and the commutator calculation
in Lemma~\ref{lem:smear}, with $\Pp$ in place of $P$, give
\begin{equation}
 \Theta(\Pp\chi_a\ket{v})
 \le C(1+D_\chi)
       \big(\Theta(\ket{v})+m_{\max}\|\ket{v}\|\big).
\end{equation}
For the wrap part, every output remains in the finite register.
On that register, $1+A\le C(1+K^2)$, with the fixed particle
and mass data absorbed into $C$. Its kinetic norm is therefore
at most $C K^2$ times its norm for $K\ge1$.
Applying the preceding one-vector estimate with $M=3$ gives
\begin{equation}
 \Theta((O_g-\Pp\chi_a\Pp)\ket{v})
 \le C C_\chi K^2t_a(K/2;\ket{v})
       +C C_{\chi,3}K^{-1}\|\ket{v}\|.
\end{equation}
Since $\Pp\ket{v}=\ket{v}$, we have
$O_g\ket{v}=\Pp\chi_a\ket{v}
 +(O_g-\Pp\chi_a\Pp)\ket{v}$.
Combining the two bounds proves Eq.~\eqref{eq:obs-theta}.

Small admissible cutoffs, when relevant, can be included
by enlarging the fixed-system constants. In the dynamical
theorems we choose $K_0$ above all the required lattice scales.
\end{proofx}

The direct cyclic-shift estimate used in the formalisation instead
involves the second Fourier moment $D_\chi'$ from
Eq.~\eqref{eq:Dchi-prime}. This introduces no further independent
constant here: integrating the tail bound in (H$\chi$) with $M=3$
gives
\begin{equation}
 D_\chi'\le 3\left(\frac{L}{2\pi}\right)^2
                         (C_\chi+2C_{\chi,3}).
\end{equation}
Thus we absorb this moment into the observable constants already
listed in Eq.~\eqref{eq:obs-theta}.

The additional term $K^2t_a(K/2;\ket{v})$ in this estimate will
not grow with $K$ for our test vectors. We first show this
using only first-order graph norms. This gives a complete
observable bound, and identifies the one shell estimate that
we subsequently improve.

\begin{lemma}[Regularity of the grid state]
\label{lem:grid-traj}
First, the initial graph norm is uniformly bounded:
\begin{equation}
\norm{ \Hgrid\, \Pp \ket{\psi_0} }
\;\le\; \norm{ \Ttot \Pp \ket{\psi_0} }
+ \norm{ V_{\mathrm{grid}}\, \Pp \ket{\psi_0} }
\;\le\; C \big( \norm{\Ttot \ket{\psi_0}} + m_{\max} \norm{\ket{\psi_0}}
\big) \;\le\; C \Xi,
\label{eq:grid-init}
\end{equation}
by $[\Ttot, \Pp] = 0$ and the \emph{forward} graph-norm bound of
Lemma~\ref{lem:grid-KR}(iv) applied to $\Pp\ket{\psi_0}$
(with $\Theta(\Pp\ket{\psi_0}) \le \Theta(\ket{\psi_0})$), all constants
independent of $K$. Consequently, uniformly in $t$:
$\Theta(\ket{\psi_g(t)}) \le C \Xi$, by conservation of
$\norm{\Hgrid \ket{\psi_g(t)}} = \norm{\Hgrid \Pp \ket{\psi_0}}$ and the
\emph{elliptic} direction of Lemma~\ref{lem:grid-KR}(iv); and
$t_a(K/2;\, \ket{\psi_g(t)}) \le C m_a \Xi\, K^{-2}$, by one crude
division of $\Ttot\ket{\psi_g}=\Hgrid\ket{\psi_g}
-V_{\mathrm{grid}}\ket{\psi_g}$ by the kinetic symbol, the
potential term bounded by Lemma~\ref{lem:grid-KR}(iii). Likewise
$\Theta(\Pp \ket{\psi(t)}) \le \Xi$ and
$t_a(K/2; \Pp\ket{\psi(t)}) \le C \Xi K^{-5/2}$ by the dynamical
bootstrap of Proposition~\ref{prop:dyn-boot}, which holds at every
threshold $\lambda \ge 1$ (its proof nowhere uses the value of the
threshold) and is applied here at $\lambda = K/2$. Hence, by
\eqref{eq:obs-theta} with $K^2 t_a(K/2; \ket{v}) \le C m_{\max} \Xi$ for
both states,
$\Theta(O_g \ket{v}) \le C(1 + C_\chi + D_\chi + C_{\chi,3})\,
m_{\max}\, \Xi$ for $\ket{v} \in \{\Pp\ket{\psi(t)},\, \ket{\psi_g(t)}\}$, and the
backward flows
\begin{equation}
\ket{\varphi_v(s)} := e^{\, i \Hgrid (t-s)}\, O_g\, \ket{v},
\qquad \ket{v} \in \{ \Pp \ket{\psi(t)},\ \ket{\psi_g(t)} \},
\end{equation}
satisfy, by conservation of $\norm{\Hgrid \ket{\varphi_v(s)}}
= \norm{\Hgrid O_g \ket{v}}$ --- itself $\le C \Xi_\chi'$ by the
\emph{forward} graph-norm bound of Lemma~\ref{lem:grid-KR}(iv)
applied to $O_g \ket{v}$, whose $\Theta$-norm was just estimated --- and
then the \emph{elliptic} direction,
$\Theta(\ket{\varphi_v(s)}) \le \Xi_\chi' :=
C m_{\max} (1 + C_\chi + D_\chi + C_{\chi,3})\, \Xi$ and the a
priori shell bound
$t_a(K; \ket{\varphi_v(s)}) \le 2 m_a \Xi_\chi' K^{-2}$, uniformly in
$0 \le s \le t$.
\end{lemma}

\begin{proofx}
We give the chain of estimates explicitly to distinguish
preparing a test vector from propagating it. Since $\Pp$
commutes with kinetic energy and is a contraction, the forward
graph comparison bounds $\|\Hgrid\Pp\ket{\psi_0}\|$ by
$C\Xi$, as in Eq.~\eqref{eq:grid-init}. Unitary grid
evolution conserves this Hamiltonian norm and the state norm.
The reverse graph comparison therefore gives
$\Theta(\ket{\psi_g(t)})\le C\Xi$, uniformly in time.
On a half-zone shell $A\ge K^2/(8m_a)$, so direct
kinetic division also gives
\begin{equation}
 t_a(K/2;\ket{\psi_g(t)})
 \le8m_a\Theta(\ket{\psi_g(t)})K^{-2}
 \le C m_a\Xi K^{-2}.
\end{equation}
This a priori bound needs only the graph norm, not a discrete
eigenvalue equation.

For the projected exact state, kinetic commutation gives
$\Theta(\Pp\ket{\psi(t)})\le\Xi$, and the continuum bootstrap
gives its sharper $K^{-5/2}$ half-zone tail. We take $K_0$
large enough that $K/2\ge\max\{1,4\pi/L\}$ when
applying that bootstrap. Equivalently, the bounded interval
of smaller positive thresholds can be absorbed into its
system constants. Thus the threshold phrasing in the statement
does not require a new initial-state condition.

For either terminal state $\ket{v}$, the quantity
$K^2t_a(K/2;\ket{v})$ is now uniformly bounded. Equation~\eqref{eq:obs-theta}
therefore controls $\Theta(O_g\ket{v})$. To carry this control
backwards, use the forward graph estimate on $O_g\ket{v}$ to bound
$\|\Hgrid O_g\ket{v}\|$, conserve that norm under
$e^{i\Hgrid(t-s)}$, and finally use the reverse graph
estimate. These are precisely the operations producing
$\Xi_\chi'$ in the statement. A final division by
$A\ge K^2/(2m_a)$ gives the $K^{-2}$ tail of the
backward state. All estimates are on the full register:
the labelled observable need not preserve the original
exchange sector.
\end{proofx}

\begin{proposition}[Truncated-kernel grid method: observable after evolution]
\label{prop:grid-dyn}
With the hypotheses above, for all $t \ge 0$ and $K \ge K_0$,
\begin{equation}
\big| \langle\psi(t)|\chi(\bm{x}_a)|\psi(t)\rangle
- \langle\psi_g(t)|O_g|\psi_g(t)\rangle \big|
\;\le\; \min\Big\{ 1,\;
C \big[ K^{-5} + t\, K^{-5/2} \big] \Big\}
\;=\; \mathcal{O}\big( N^{-5/3} + t\, N^{-5/6} \big),
\label{eq:grid-dyn}
\end{equation}
where $K_0$ depends only on the fixed system and kernel-class data.
The error constant $C$ depends additionally on the regularity
budget and $1+C_\chi+D_\chi+C_{\chi,3}+C_{\chi,5}$; the
normalised-state transfer of Corollary~\ref{cor:norm} applies
verbatim at cutoff $2K$.
\end{proposition}

\begin{proofx}
We separate the error by which operation changes, so that each
piece uses the appropriate estimate. In this proof set
$\ket{u(t)}:=\Pp\ket{\psi(t)}$ and $\ket{e_{\Pp}(t)}:=\ket{u(t)}-\ket{\psi_g(t)}$.
Then
\begin{align}
 \Delta\langle O\rangle
 ={}&\underbrace{\big[\langle\psi|\chi_a|\psi\rangle
                         -\langle u|\chi_a|u\rangle\big]}_{\mathrm{(I)}}
 +\underbrace{\big[\langle u|\chi_a|u\rangle
                         -\langle u|O_g|u\rangle\big]}_{\mathrm{(II)}}\\
 &+\underbrace{\big[\langle u|O_g|u\rangle
                   -\langle\psi_g|O_g|\psi_g\rangle\big]}_{\mathrm{(III)}}.
\end{align}
The first term is the continuum projection error, the second
is observable collocation on the projected exact state, and
the third is the difference between the two retained
states measured with the same grid observable.

For (I), let $\ket{\delta'}=(1-\Pp)\ket{\psi(t)}$. Expanding the
quadratic forms gives
\begin{equation}
 \mathrm{(I)}=2\mathrm{Re}\langle\delta'|\chi_a|\psi\rangle
                              -\langle\delta'|\chi_a|\delta'\rangle.
\end{equation}
The support pairing from Eq.~\eqref{eq:support-pairing},
Lemma~\ref{lem:chitail} at cutoff $2K$, and the continuum
tails yield
\begin{equation}
 |\mathrm{(I)}|\le C(1+C_\chi)\Xi^2K^{-5}
                             +C C_{\chi,M}\Xi K^{-M}.
\end{equation}
For (II), use Eq.~\eqref{eq:obs-form} with both vectors
equal to $\ket{u(t)}$. Its half-zone tail is $O(\Xi K^{-5/2})$,
so
\begin{equation}
 |\mathrm{(II)}|\le C C_\chi\Xi^2K^{-5}
                                      +C_{\chi,M}K^{-M}.
\end{equation}
Fixing $M=5$ absorbs these rapid Fourier remainders into
the time-independent $K^{-5}$ term.

For (III), use the exact identity
\begin{equation}
 \mathrm{(III)}
 =\langle e_{\Pp}|O_g|u\rangle+\langle\psi_g|O_g|e_{\Pp}\rangle.
\end{equation}
Self-adjointness makes the second summand the complex
conjugate of $\langle e_{\Pp}|O_g|\psi_g\rangle$. Thus we need
only pairings of $\ket{e_{\Pp}}$ against $O_g\ket{v}$ with
$\ket{v}\in\{\ket{u(t)},\ket{\psi_g(t)}\}$. There is no separate squared
error norm. This matters because a direct estimate of the
new grid sources gives only an $O(K^{-1/2})$ norm bound
at this stage, whose time integral would produce an
unhelpful $t^2K^{-1}$ contribution if squared.

The two retained states agree initially. Differentiating
$e^{i\Hgrid t}\Pp\ket{\psi(t)}$, using kinetic commutation
and Eq.~\eqref{eq:grid-structure}, gives
\begin{equation}
 \ket{e_{\Pp}(t)}=-i\int_0^t e^{-i\Hgrid(t-s)}
 \big[\Pp V_{\mathrm{tot}}(1-\Pp)\ket{\psi(s)}
                     +(D-R)\Pp\ket{\psi(s)}\big]\,ds.
\end{equation}
The first source is the same omitted-state coupling as in
Galerkin evolution. The other two are the dropped-transfer
and wrap changes to the retained Hamiltonian. Move the
propagator to the test side of the inner product and set
$\ket{\varphi_v(s)}=e^{i\Hgrid(t-s)}O_g\ket{v}$. We obtain
\begin{equation}
 |\langle e_{\Pp}(t)|O_g|v\rangle|
 \le\int_0^t
 \left|\langle\delta'(s)|V_{\mathrm{tot}}\Pp|\varphi_v(s)\rangle
       +\langle u(s)|D-R|\varphi_v(s)\rangle\right|\,ds.
\end{equation}
Lemma~\ref{lem:grid-traj} supplies a time-independent
graph bound $\Theta(\ket{\varphi_v(s)})\le\Xi_\chi'$
and a shell tail of order $K^{-2}$.

For the Galerkin source, symmetry of the potential and
$\ket{\delta'}\perp\mathrm{ran}\,\Pp$ give
\begin{equation}
 \langle\delta'|V_{\mathrm{tot}}\Pp|\varphi_v\rangle
 =\langle\delta'|(1-\Pp)V_{\mathrm{tot}}|\varphi_v\rangle.
\end{equation}
Its magnitude is at most
$C\Xi\Xi_\chi'K^{-5/2}K^{-1/2}=C\Xi\Xi_\chi'K^{-3}$
by the general-state potential tail.
For $D$, Eq.~\eqref{eq:Dform} pairs the exact-state tail
$K^{-5/2}$ with the test graph norm, and the test tail
$K^{-2}$ with the exact-state graph norm. Consequently,
\begin{equation}
 |\langle u(s)|D|\varphi_v(s)\rangle|
 \le C\Xi\Xi_\chi'K^{-1/2}
                              (K^{-5/2}+K^{-2})
 \le C\Xi\Xi_\chi'K^{-5/2}.
\end{equation}
For $R$, Eq.~\eqref{eq:Rform} gives a strong-coefficient
part of order $K K^{-5/2}K^{-2}=K^{-7/2}$ and weak-coefficient
parts of orders $K^{-3}$ and $K^{-5/2}$. The last is again
the limiting term. All these estimates are uniform in $s$;
for the kernel class, substituting the fixed threshold
$c_{\rm ker}K$ leaves the powers unchanged.

Integrating over an interval of length $t$ now bounds (III)
by $C tK^{-5/2}$. Together with (I) and (II) this gives
the stated cutoff powers. Both quadratic forms lie in
$[0,1]$, which supplies the minimum with one, and
$K=\Theta(N^{1/3})$ gives the displayed basis-size rate.
Normalising the projected initial state changes the answer
by its discarded probability at cutoff $2K$, of order
$K^{-5}$, exactly as in Corollary~\ref{cor:norm}.
\end{proofx}

\medskip\noindent\emph{Dependence on the observable tail.}
The constant $C_{\chi,5}$ controls the high-frequency observable
remainder in the time-independent $K^{-5}$ term.
Hypothesis (H$\chi$) supplies its finiteness. This dependence cannot
in general be removed by another state-tail estimate: a Fourier
mode congruent to zero modulo the grid size aliases to a
constant, even on a state with only zero momentum. The sharper
theorem below includes $C_{\chi,5}$ explicitly.

The limiting pairing has now been identified: a weak-coefficient
wrap or dropped transfer meets a backward-state shell known
only to be $O(K^{-2})$. Improving that shell to $O(K^{-5/2})$
would make every source pairing $O(K^{-3})$. We establish
this improvement next without imposing an additional qualitative
smoothness condition on $\chi$.

\subsection{Second-order grid regularity and the energy-rate secular bound}
\label{sec:grid-upgrade}

The preceding observable proof loses half a power of $K$ because
its backward test states have only the a priori $K^{-2}$ shell
tail. We now obtain a $K^{-5/2}$ tail for those same states.
The extra input is a bound on two applications of kinetic energy.
This bound need not be uniform as the register grows: growth of
order $K^{3/2}$ is sufficient, because division by the squared
kinetic symbol contributes $K^{-4}$. Their product is the
required $K^{-5/2}$ tail.

We will keep the cyclic momentum shifts in the calculation from
the outset. This lets us control the actual grid operator without
replacing it by a continuum product and then differentiating a
separate wrap correction. For the observable constants define
\begin{equation}
 D_\chi^{(4)}:=\sum_{2\le|\alpha|\le4}
       \|\partial^\alpha\chi\|_\infty,\qquad
 \mathfrak M_{\chi,4}:=\sum_{\bm{\nu}}(1+\|\bm{k}_{\bm{\nu}}\|^4)|\hat\chi(\bm{\nu})|.
 \label{eq:D4}
\end{equation}
Here $\alpha=(\alpha_1,\alpha_2,\alpha_3)$ is a derivative
multi-index, $|\alpha|=\alpha_1+\alpha_2+\alpha_3$, and
$\partial^\alpha=\partial_1^{\alpha_1}
\partial_2^{\alpha_2}\partial_3^{\alpha_3}$. Thus the first
sum includes every second derivative, including mixed derivatives,
as well as derivatives of orders three and four. The second sum
is an absolute Fourier moment, weighted by four powers of
momentum. It controls the derivative norms in the first sum,
but those derivative norms alone do not bound the absolute
Fourier sum.

Both quantities are finite under (H$\chi$). To see the point
for the Fourier moment, group modes into shells with radii
$2^j$ to $2^{j+1}$. Weighting such a shell by $\norm{\bm{k}}^4$
costs at most $2^{4(j+1)}$, while the absolute coefficient tail
decays as $2^{-Mj}$ for any chosen $M>4$. The resulting
geometric series converges. We have therefore introduced a
quantitative constant, not an additional qualitative smoothness
assumption. A dependence of $\mathfrak M_{\chi,4}$ on the
smoothing width must nevertheless be estimated from the Fourier
sum itself; it cannot be inferred from a derivative bound alone.

We use $X(\ket{u}):=\|A^2\ket{u}\|$ for the norm after two kinetic
applications, together with $\Theta(\ket{u})=\|(1+A)\ket{u}\|$.
All vectors here belong to the finite register, so both norms
exist. The task is to bound their dependence on $K$.
Constants in this subsection may depend on
$C_\chi,D_\chi,D_\chi^{(4)},\mathfrak M_{\chi,4},C_{\chi,5}$
and the fixed system data, but not on the cutoff or time.

\begin{lemma}[Second-order grid estimates]
\label{lem:grid-KR2}
Set $\mathcal E=1+A$. For $K\ge\max\{1,4\pi/L\}$ there is
$c_2(K)\le C(1+K^{3/2})$, depending otherwise only on the fixed
system and kernel-majorant constants, such that
\begin{equation}
 \|\mathcal E V_{\rm grid}\ket{u}\|
 \le\tfrac14X(\ket{u})+c_2(K)\Theta(\ket{u}).
 \label{eq:wKR}
\end{equation}
Consequently, for every register vector,
\begin{align}
 X(\ket{u})&\le\tfrac43\big[\Theta(H_{\rm grid}\ket{u})+c_2(K)\Theta(\ket{u})\big],
       \label{eq:ell2}\\
 \Theta(H_{\rm grid}\ket{u})&\le\tfrac54X(\ket{u})+(1+c_2(K))\Theta(\ket{u}).
       \label{eq:grid-second-forward}
\end{align}
Let $\mathcal U$ be any unitary on the register commuting with $H_{\rm grid}$.
Choose fixed first-order constants $c_f,c_0,c_h$ such that
$\Theta(\mathcal U\ket{z})\le c_f\Theta(\ket{z})+c_0\|\ket{z}\|$ and
$\|H_{\rm grid}\ket{z}\|\le c_h\Theta(\ket{z})$.
Then
\begin{equation}
 \begin{split}
 X(\mathcal U\ket{w})&\le f_K(X(\ket{w}),\Theta(\ket{w}),\|\ket{w}\|),\\
 f_K(x,y,z)&:=\tfrac43\left[
 \tfrac54c_f x+\big(c_f(1+2c_2(K))+c_0c_h\big)y
                  +c_0c_2(K)z\right].
 \end{split}
 \label{eq:grid-second-flow}
\end{equation}
In particular these constants are independent of time for
$\mathcal U=e^{\pm iH_{\rm grid}\tau}$.
\end{lemma}

\begin{proofx}
We first estimate the potential with one kinetic weight on its
output. Work on one interaction channel and let $\kappa(\bm{\tau})$
be its coefficient at integer transfer $\bm{\tau}$. The Coulomb
majorant gives
\begin{equation}
 |\kappa(\bm{\tau})|\,\|\bm{\tau}\|^2\le\gamma\qquad(\bm{\tau}\ne0),
\end{equation}
with $\gamma$ independent of $K$; fixed charges and the conversion
between lattice indices and physical momenta are included in
$\gamma$. The transfer support has radius of order $K$ in
lattice units at fixed cell size.

Let $\vec{\bm{p}}$ be an input index and $\vec{\bm{q}}$ the output after the cyclic
transfer. Before folding, the moved momenta change by opposite
copies of $\bm{k}_{\bm{\tau}}$ for a pair, or by one copy for an external
channel. The inequality $\|\bm{x}+\bm{y}\|^2\le2\|\bm{x}\|^2+2\|\bm{y}\|^2$ bounds
their kinetic energies. Choosing the centred representative
modulo $n'$ can only reduce each absolute component, so the
same comparison holds after folding:
\begin{equation}
 \mathcal E(\vec{\bm{q}})\le2\mathcal E(\vec{\bm{p}})+C_L\|\bm{\tau}\|^2,
 \qquad \mathcal E=1+A.
\end{equation}
Taking coefficientwise absolute values now separates the weighted
potential into two positive estimates. In the first, the factor
$\mathcal E(\vec{\bm{p}})$ remains on the input state and the channel
coefficients are bounded by a Coulomb kernel. In the second,
the transfer weight cancels the inverse-square coefficient and
leaves a bounded coefficient of size at most $C\gamma$.

For the first estimate apply the relative bound of
Lemma~\ref{lem:grid-KR} to the register vector with coefficients
$\mathcal E(\vec{\bm{p}})|u_{\vec{\bm{p}},\boldsymbol\sigma}|$,
separately for each spin component.
This is legitimate because the estimates hold on the full
register space, whether or not taking absolute values preserves
exchange symmetry. The bound has the form
\begin{equation}
 C\varepsilon\|A\mathcal E\ket{u}\|
       +C_\varepsilon\|\mathcal E\ket{u}\|
 \le C\varepsilon X(\ket{u})+C_\varepsilon'\Theta(\ket{u}),
\end{equation}
where $\|A\mathcal E\ket{u}\|\le X(\ket{u})+\Theta(\ket{u})$.
Choose $\varepsilon$ small enough that, after summing the
finitely many charge-weighted channels, the coefficient of $X(\ket{u})$
is at most $1/4$.

For the second estimate there are $O((1+K)^3)$ possible transfer
indices and each coefficient is bounded independently of $K$.
At a fixed input a transfer determines the output, with only a
fixed alias multiplicity. Hence the squared column sums of this
channel matrix are bounded by $C\gamma^2(1+K)^3$.
The weighted slice estimate used in Lemma~\ref{lem:paircorr}
then gives
\begin{equation}
 \|\text{second contribution}\|
 \le C(1+K^{3/2})\|\mathcal W_r\ket{u}\|
 \le C(1+K^{3/2})\Theta(\ket{u}),
\end{equation}
where $r$ is a moved register and its fixed mass is included in
$C$. Translated external kernels obey the same absolute
coefficient estimates. Combining both contributions proves
Eq.~\eqref{eq:wKR}. The power $K^{3/2}$ is the square root
of the number of allowed three-dimensional transfers after the
kinetic weight has cancelled their coefficient decay.

Next use $\Ttot\ket{u}=H_{\rm grid}\ket{u}
-V_{\rm grid}\ket{u}$ and $A\le\mathcal E$ to obtain
\begin{equation}
 X(\ket{u})\le\Theta(H_{\rm grid}\ket{u})
                   +\tfrac14X(\ket{u})+c_2(K)\Theta(\ket{u}).
\end{equation}
Absorbing the $X(\ket{u})/4$ term proves Eq.~\eqref{eq:ell2}.
For the reverse direction,
\begin{equation}
 \Theta(H_{\rm grid}\ket{u})
 \le\|\mathcal E A\ket{u}\|
             +\|\mathcal E V_{\rm grid}\ket{u}\|
 \le\tfrac54X(\ket{u})+(1+c_2(K))\Theta(\ket{u}),
\end{equation}
which is Eq.~\eqref{eq:grid-second-forward}.

Finally let $\mathcal U$ be as in the statement. The first-order
graph comparison and unitarity give constants $c_f,c_0,c_h$
independent of the duration of the evolution. Apply
Eq.~\eqref{eq:ell2} to $\mathcal U\ket{w}$ and commute the
Hamiltonian through $\mathcal U$:
\begin{align}
 X(\mathcal U\ket{w})
 &\le\tfrac43\big[
      \Theta(\mathcal U H_{\rm grid}\ket{w})
                     +c_2(K)\Theta(\mathcal U\ket{w})\big]\\
 &\le\tfrac43\big[
      c_f\Theta(H_{\rm grid}\ket{w})+c_0\|H_{\rm grid}\ket{w}\|
      +c_2(K)(c_f\Theta(\ket{w})+c_0\|\ket{w}\|)\big].
\end{align}
Substitute Eq.~\eqref{eq:grid-second-forward} for the first
Hamiltonian term and $\|H_{\rm grid}\ket{w}\|\le c_h\Theta(\ket{w})$
for the second. Collecting the coefficients gives precisely
$f_K$ in Eq.~\eqref{eq:grid-second-flow}. This estimate
propagates the finite-register kinetic norm from its initial
value; it does not assume a cutoff-independent second
Hamiltonian graph norm for the initial test vector.
\end{proofx}

We next supply the initial kinetic norm required by the flow
estimate. Although the exact Coulomb state need not have four
square-integrable spatial derivatives, its finite Fourier
projection does. The propagated tail bound quantifies how
that projected norm grows with the cutoff.

\begin{lemma}[Second moments of the projected and grid states]
\label{lem:grid-data2}
For $\ket{\psi_0}\in D(H^2)$,
\begin{equation}
 X(P'\ket{\psi(t)})\le C\Xi K^{3/2}
 \quad\text{uniformly in }t.
 \label{eq:data2}
\end{equation}
The grid state evolved from the projected initial state also satisfies
$X(\ket{\psi_g(t)})\le C\Xi K^{3/2}$ uniformly in time.
\end{lemma}

\begin{proofx}
Group Fourier modes according to the largest Euclidean momentum
among the particle registers. Above a fixed low-momentum scale,
use dyadic shells
$\lambda<\max_r\norm{\bm{k}_{\bm{p}_r}}\le2\lambda$.
On such a shell $A\le C\lambda^2$. Also at least one
register has momentum above $\lambda$, so the union bound
and Eq.~\eqref{eq:dyn-wtail} give
\begin{equation}
 \|(1+A)\mathbf1_{\rm shell}\ket{\psi(t)}\|
 \le C\Xi\lambda^{-1/2}.
\end{equation}
Choose the low-momentum scale above the fixed lattice threshold
for that estimate. Since $A^2\le C\lambda^2(1+A)$ on
the shell,
\begin{equation}
 \|A^2\mathbf1_{\rm shell}\ket{\psi(t)}\|
 \le C\Xi\lambda^{3/2}.
\end{equation}
The projected state has maximal Euclidean register momentum
at most $2\sqrt3K$. Its shells are orthogonal, so the
squared norms add. The dyadic sum of $\lambda^3$ is
bounded by a constant times its largest term, of order
$K^3$. Taking the square root and including the bounded
low-momentum block proves Eq.~\eqref{eq:data2}, uniformly
in time.

Now take $\ket{w}=\Pp\ket{\psi_0}$. The result just proved gives
$X(\ket{w})\le C\Xi K^{3/2}$, while kinetic commutation gives
$\Theta(\ket{w})+\|\ket{w}\|\le C\Xi$. Insert these bounds into
Eq.~\eqref{eq:grid-second-flow} with
$\mathcal U=e^{-iH_{\rm grid}t}$. Every term in its affine
bound is at most $C\Xi K^{3/2}$ because
$c_2(K)=O(K^{3/2})$ for $K\ge1$. This proves the grid
state estimate without assuming that a second grid
Hamiltonian application to the projected initial states is
uniformly bounded as $K$ grows.
\end{proofx}

It remains to apply the observable and then evolve backwards.
The fourth Fourier moment in Eq.~\eqref{eq:D4} pays for
the kinetic weight added by each observable transfer, including
transfers that wrap.

\begin{lemma}[Second moment of the smeared states]
\label{lem:grid-obs2}
For every $\ket{v}\in\operatorname{ran}P'$,
\begin{equation}
 X(O_g\ket{v})\le C C_\chi\big(X(\ket{v})+\Theta(\ket{v})\big)
                   +C\mathfrak M_{\chi,4}\|\ket{v}\|.
 \label{eq:grid-obs2-moment}
\end{equation}
For $\ket{v}=P'\ket{\psi(t)}$ or $\ket{v}=\ket{\psi_g(t)}$, this and the direct flow estimate give
\begin{equation}
 X(O_g\ket{v})\le C\Xi K^{3/2},\qquad
 X(e^{iH_{\rm grid}\tau}O_g\ket{v})\le C\Xi K^{3/2}
 \quad\text{for every real }\tau.
 \label{eq:obs2}
\end{equation}
\end{lemma}

\begin{proofx}
Write $O_g$ as its absolutely convergent Fourier sum of
unitary cyclic shifts in the observed register. If mode $\bm{\nu}$
sends input $\vec{\bm{p}}$ to output $\vec{\bm{q}}$, the same centred-index
comparison as in Lemma~\ref{lem:grid-KR2} gives
\begin{equation}
 \mathcal E(\vec{\bm{q}})^2
 \le C\mathcal E(\vec{\bm{p}})^2+C\|\bm{k}_{\bm{\nu}}\|^4.
\end{equation}
This remains valid for a mode larger than the register:
centred reduction can only decrease the magnitude relative
to the unfolded output. For the unitary shift $S_a(\bm{\nu})$,
we may therefore estimate
\begin{equation}
 X(S_a(\bm{\nu})\ket{v})
 \le\|\mathcal E^2S_a(\bm{\nu})\ket{v}\|
 \le C\|\mathcal E^2\ket{v}\|
                         +C\|\bm{k}_{\bm{\nu}}\|^4\|\ket{v}\|.
\end{equation}
Sum this inequality with weights $|\hat\chi(\bm{\nu})|$.
The identity $\mathcal E^2=1+2A+A^2$ implies
$\|\mathcal E^2\ket{v}\|\le X(\ket{v})+2\Theta(\ket{v})$,
so the sum is Eq.~\eqref{eq:grid-obs2-moment}.
Absolute convergence of the weighted sum follows from
$\mathfrak M_{\chi,4}<\infty$.

For either $\ket{v}=\Pp\ket{\psi(t)}$ or $\ket{v}=\ket{\psi_g(t)}$,
Lemma~\ref{lem:grid-data2} bounds $X(\ket{v})$ by
$C\Xi K^{3/2}$ and the first-order estimates bound
$\Theta(\ket{v})$ and $\|\ket{v}\|$ by $C\Xi$.
This proves the first inequality of Eq.~\eqref{eq:obs2}.
To propagate it, we also need the lower-order norms of
$O_g\ket{v}$. The observable norm is at most one, and
Eq.~\eqref{eq:obs-theta} gives
$\Theta(O_g\ket{v})\le C\Theta(\ket{v})$: its extra term satisfies
$K^2t_a(K/2;\ket{v})\le8m_a\Theta(\ket{v})$ for every register
vector. Apply Eq.~\eqref{eq:grid-second-flow} to
$\ket{w}=O_g\ket{v}$ and $\mathcal U=e^{iH_{\rm grid}\tau}$.
The same affine-budget calculation as above yields the
second inequality of Eq.~\eqref{eq:obs2}, uniformly for
every real $\tau$.
\end{proofx}

We have now controlled the terminal test vectors and their
backward evolution with the same $K^{3/2}$ kinetic budget.
The following theorem uses this budget to recover the
Galerkin observable rate. As in Appendix~\ref{app:dynamics},
the energy-rate coefficient describes an upper bound; it
does not assert that an observable's actual drift attains
that power.

\begin{theorem}[Truncated-kernel grid method: energy-rate secular bound]
\label{thm:grid-dyn-sharp}
Let $\ket{\psi_0}\in D(H^2)$ be normalised and let $\ket{\psi_g(t)}=e^{-iH_{\rm grid}t}\Pp\ket{\psi_0}$. Under (H$\chi$), with the Fourier moment in \eqref{eq:D4} included in the constant, for all
$t \ge 0$ and $K \ge K_0$,
\begin{equation}
\big| \langle\psi(t)|\chi(\bm{x}_a)|\psi(t)\rangle
- \langle\psi_g(t)|O_g|\psi_g(t)\rangle \big|
\;\le\; \min\Big\{ 1,\; C \big[ K^{-5} + t\, K^{-3} \big] \Big\}
\;=\; \mathcal{O}\big( N^{-5/3} + t\, N^{-1} \big):
\label{eq:grid-dyn-sharp}
\end{equation}
the truncated-kernel grid method attains the same energy-rate secular
coefficient as the Galerkin method
(Theorem~\ref{thm:dyn}), gap-free.
The constant includes the system and state data and
$C_\chi,D_\chi,D_\chi^{(4)},\mathfrak M_{\chi,4},C_{\chi,5}$.
The Hessian terms $2\le|\alpha|\le4$ are part of $D_\chi^{(4)}$;
(H$\chi$) supplies all these quantities.
\end{theorem}
The order-two terms in $D_\chi^{(4)}$ constitute the Hessian,
the matrix of second derivatives; the same sum also contains
the order-three and order-four derivatives.

\begin{proofx}
Use the same three-term decomposition as in the proof of
Proposition~\ref{prop:grid-dyn}. The projection term (I)
and the observable-collocation term (II) are already
$O(K^{-5})$, including their $C_{\chi,5}$ remainders.
The retained-state difference (III) is still the sum
of two linear pairings against backward states
\begin{equation}
 \ket{\varphi_v(s)}=e^{iH_{\rm grid}(t-s)}O_g\ket{v},
 \qquad \ket{v}\in\{\Pp\ket{\psi(t)},\ket{\psi_g(t)}\}.
\end{equation}
Only the estimates inside their time integrals need to
change.

Equations~\eqref{eq:grid-second-flow} and~\eqref{eq:obs2}
give $X(\ket{\varphi_v(s)})\le C\Xi K^{3/2}$ uniformly in
$s$. On the shell of register $r$ above $K$, the squared
kinetic symbol satisfies $A^2\ge(K^2/(2m_r))^2$.
Dividing by it gives
\begin{equation}
 t_r(K;\ket{\varphi_v(s)})
 \le\left(\frac{2m_r}{K^2}\right)^2X(\ket{\varphi_v(s)})
 \le C m_r^2\Xi K^{-5/2}.
\end{equation}
The same argument at $K/2$ gives
\begin{equation}
 t_r(K/2;\ket{\varphi_v(s)})
 \le\left(\frac{8m_r}{K^2}\right)^2X(\ket{\varphi_v(s)})
 \le C m_r^2\Xi K^{-5/2}.
\end{equation}
Thus both shell thresholds used by the wrap form have the
improved power. A fixed smaller threshold $c_{\rm ker}K$
is handled in the same way for the larger kernel class.

The Galerkin source keeps its previous estimate
$K^{-5/2}K^{-1/2}=K^{-3}$. For the dropped-transfer
and weak-wrap terms, both the exact state and the backward
state now have $K^{-5/2}$ tails, while their first-order
graph norms remain uniformly bounded. Equations~\eqref{eq:Dform}
and~\eqref{eq:Rform} therefore give
\begin{equation}
 K^{-1/2}\big[
 \Theta(\ket{\varphi_v})t_r(K;\Pp\ket{\psi})
       +\Theta(\Pp\ket{\psi})t_r(K;\ket{\varphi_v})\big]
 \le CK^{-3}.
\end{equation}
The strong-wrap term is even smaller:
\begin{equation}
 K\,t_r(K/2;\Pp\ket{\psi})t_r(K/2;\ket{\varphi_v})
 \le CK\,K^{-5/2}K^{-5/2}=CK^{-4}.
\end{equation}
All constants here include the fixed system, initial-state
and observable budgets specified in the theorem.
Integrating gives $|\mathrm{(III)}|\le CtK^{-3}$.
Adding (I) and (II), using the trivial bound by one, and
converting $K$ to $N^{1/3}$ proves
Eq.~\eqref{eq:grid-dyn-sharp} for all $t\ge0$.
\end{proofx}

\medskip\noindent\emph{Why the truncated kernel supplies the improvement.}
The second-order estimate succeeds because the kinetic
transfer weight cancels the inverse-square Coulomb coefficient
on a finite set of $O(K^3)$ transfers. Its squared column
sum then grows as $K^3$, giving norm growth only $K^{3/2}$.
Two kinetic divisions turn that growth into a $K^{-5/2}$
backward tail. This is a finite-register argument; it does
not assume that multiplying a continuum Coulomb state by a
generic smooth observable preserves $D(H^2)$.

Hypothesis (H$\chi$) already gives all the observable
moments used in this improvement. The sharper theorem
therefore has no additional qualitative smoothness assumption,
although its quantitative constant involves the fourth
Fourier moment and higher derivatives. The preliminary
proposition remains useful for seeing which first-order
estimates suffice and where the half-power loss occurs.
Neither bound proves attainment of its secular coefficient.
An extension to a weaker observable class would have to
state the Fourier-tail and derivative bounds used in that
argument explicitly; selected $C^2$ bounds alone do not
supply (H$\chi$).

\medskip\noindent\emph{Choosing bandwidth on a fixed register.}
The padded parametrisation describes a single family:
a register of cutoff $K_{\rm reg}$ with a kernel whose
transfer cutoff is also $K_{\rm reg}$. The ordinary
Galerkin difference set reaches $2K_{\rm reg}$, so the
method drops its upper half, and wraps cannot connect two
states both below $K_{\rm reg}/2$. In our notation
$K_{\rm reg}=2K$.

Equivalently, start with an existing odd grid of $n$ points
per direction and truncate its kernel at
$\pi(n-1)/L$. The estimates then use effective resolution
half that register cutoff. No register enlargement is needed
to define this version. At a fixed effective resolution,
however, the padded description uses asymptotically eight
times as many spatial modes as an $n^3$ register. These
are two ways of accounting for the same fixed factor, not
a claim that enlarging a register has no resource cost.
The kernel bandwidth is the substantive choice.

Using the entire Galerkin difference set as a collocated
kernel makes a different operator. Its additional large
transfer coefficients fold onto bulk cyclic transfers with
coefficients bounded by $CK^{-2}$. They need not carry
the high-endpoint support that protected the truncated
method. The next proposition quantifies this alternative;
changing the sampled kernel values can remove these
additional coefficients without changing the register.

\begin{proposition}[Full-kernel collocation: energy bound]
\label{prop:fullkernel}
Let $H_{\rm full}$ be collocation on the register of cutoff $2K$ of
the maximal band-limited Coulomb kernel, with transfers up to $4K$,
including translated clamped-nucleus terms when present. For sufficiently
large admissible registers, with a threshold depending only on the fixed
system data,
\begin{equation}
 |E_{\rm full}-E|\le CK^{-3/2}
       \le C'N_{\rm reg}^{-1/2}.
 \label{eq:fullkernel}
\end{equation}
The bound improves to $C(1+C_{\rm corr})K^{-2}$ if there exist
normalised ground states $\ket{u_{\rm gal}}$ of $P'HP'$ and $\ket{u_{\rm full}}$
of $H_{\rm full}$, in the sector under consideration, for which
\begin{align}
 \sum_{\bm{d}\ne0}|\langle u|S_a(\bm{d})S_b(-\bm{d})|u\rangle|&\le C_{\rm corr}
       &&\text{for each quantum pair},\\
 \sum_{\bm{d}\ne0}|\langle u|S_a(\bm{d})|u\rangle|&\le C_{\rm corr}
       &&\text{for each particle with a clamped interaction}.
 \label{eq:fullkernel-correlation}
\end{align}
These conditions are required for both selected ground states, uniformly
in $K$. They include one-particle density correlations; the quantum-pair
condition alone would be vacuous in a one-electron clamped-nucleus model.
No matching lower bound is asserted.
\end{proposition}
For a one-register shift, the displayed overlap is a Fourier
coefficient of that particle's position probability distribution
on the grid. Indeed, the cyclic momentum shift is diagonal in
the grid basis, so its expectation averages its phase over that
distribution.

\begin{proofx}
We compare the full-difference-set method with the truncated
method already analysed. Its extra Fourier modes are precisely
those with $2K<\|\bm{k}_{\bm{\nu}}\|_\infty\le4K$, applied as
cyclic shifts. Thus
\begin{equation}
 H_{\rm full}=\Pp H\Pp+\Delta_K+V_{\rm alias},
 \qquad\Delta_K:=H_{\rm grid}-\Pp H\Pp=R-D.
\end{equation}
The operator $V_{\rm alias}$ collects these additional
cyclic coefficients. Reduce a transfer to its centred
representative $\bm{d}$ modulo the register size. There are
only a bounded number of original indices congruent to
$\bm{d}$, and every additional index has momentum greater
than $2K$. Each extra coefficient is therefore at most
$CK^{-2}$ in magnitude. For a translated nuclear kernel,
the phase is attached to each original index before this
sum is taken; substituting the phase of $\bm{d}$ would give a
different operator.

Before estimating its energy contribution, we need uniform
regularity for the full-kernel ground state. For nonzero
centred $\bm{d}$, the complete folded coefficient is bounded by
a fixed multiple of the Coulomb coefficient at $\bm{d}$.
Indeed the unaliased coefficient has that value, and the
bounded number of additional $O(K^{-2})$ terms is no
larger than a fixed multiple of $\hat w(\bm{d})$ within the
register. Any zero-transfer contribution is a bounded
constant and is split off. The resulting centred kernel
therefore obeys the same coefficient-majorant and finite-fold
estimates used in Lemma~\ref{lem:grid-KR}.

Consequently the full operator has uniform graph and form
lower bounds. A fixed finite-support sector trial state,
as in the proof of Theorem~\ref{thm:grid-energy}, supplies
an upper energy bound without using a tail bootstrap.
We may then choose a common fixed threshold for the
eigenvalue absorptions. Both selected ground states
$\ket{u_{\rm full}}$ and $\ket{u_{\rm gal}}$ have uniformly bounded
graph norms and $J^{-5/2}$ tails at every sufficiently
large threshold $J$ within the register. The estimates
for $D$ and $R$ therefore give
$|\langle u|\Delta_K|u\rangle|\le CK^{-3}$ for either
state, just as in the earlier two-sided energy comparison.

We now estimate the extra cyclic terms. A nonzero centred
displacement must move at least one of the endpoints
outside a small interval. The exact support statement
uses a strict projector in a moved register $r$:
\begin{equation}
 P_{r,<\|\bm{k}_{\bm{d}}\|_\infty/2}\,S_a(\bm{d})S_b(-\bm{d})\,
 P_{r,<\|\bm{k}_{\bm{d}}\|_\infty/2}=0.
 \label{eq:cyclic-strict-support}
\end{equation}
To check it, choose a component with largest $|d_c|$.
Because $\bm{d}$ is centred, $|d_c|$ is the shortest cyclic
displacement in that component. Two indices both strictly
inside the interval of radius $|d_c|/2$ cannot have this
displacement, whether or not the shift crosses a register
edge. Strictness matters: when $d_c$ is even, the two
endpoints $-d_c/2$ and $d_c/2$ would both belong to the
closed interval. The single-register shift has the same
support identity with $r=a$.

Split a normalised state into its low part and its
complement in Eq.~\eqref{eq:cyclic-strict-support}.
The low--low pairing vanishes, and unitarity bounds each
remaining pairing by a high-momentum norm. Using the
smaller threshold $\|\bm{k}_{\bm{d}}\|_\infty/4$ avoids endpoint
issues and gives
\begin{equation}
 |\langle u|S_a(\bm{d})S_b(-\bm{d})|u\rangle|
 \le C\min\{1,\|\bm{k}_{\bm{d}}\|^{-5/2}\}.
\end{equation}
The constant includes the bounded small-threshold range;
the same inequality holds for an external one-register
correlation. In three dimensions, summing this decay over
centred transfer shells through radius $O(K)$ gives
\begin{equation}
 \sum_{\bm{d}}|\langle u|S_j(\bm{d})|u\rangle|
 \le C(1+K^{1/2}),
\end{equation}
where $S_j$ denotes the cyclic shift of the channel under
consideration. The exponent follows from the shell sum
$\int_1^{O(K)}r^2r^{-5/2}\,dr=O(K^{1/2})$.

Multiply by the $CK^{-2}$ coefficient bound and sum
the finitely many charge-weighted channels:
\begin{equation}
 |\langle u|V_{\rm alias}|u\rangle|
 \le CK^{-2}\sum_j\sum_{\bm{d}}
              |\langle u|S_j(\bm{d})|u\rangle|
 \le CK^{-3/2}.
\end{equation}
Under Eq.~\eqref{eq:fullkernel-correlation}, replace the
shell sum by the uniform budget $1+C_{\rm corr}$; the
$1$ includes the zero displacement. The resulting bound
is $C(1+C_{\rm corr})K^{-2}$.
Apply these estimates to both selected ground states
in the two directions of the variational comparison.
Together with the $O(K^{-3})$ form of $\Delta_K$ and
the Galerkin energy theorem at cutoff $2K$, this proves
both claims. Since the actual register has
$N_{\rm reg}=\Theta(K^3)$ spatial modes per particle,
the first rate is $N_{\rm reg}^{-1/2}$.
\end{proofx}

\medskip\noindent\emph{Scope of the full-kernel comparison.}
The proposition concerns a specified maximal finite Fourier
kernel, not raw sampling of the singular Coulomb function.
It provides upper bounds, with no matching lower bound
asserted. The improved rate requires the absolute
correlation sums for both chosen ground states. It also
requires the one-particle correlations for external
channels: a quantum-pair condition alone says nothing
about a one-electron problem with clamped nuclei. These
are additional assumptions, not consequences of the
truncated-kernel energy theorem.

\medskip\noindent\emph{Finite-difference kinetics: a heuristic comparison.}
The preceding theorems use exact kinetic energy on the
Fourier register. A $2p$-th-order finite-difference stencil
instead changes its symbol by
$O(h^{2p}\|\bm{k}\|^{2p+2}/m)$, with grid spacing $h=L/n'$.
Such a stencil approximates the kinetic differential operator
by a weighted sum of state values at finitely many neighbouring
grid points; its symbol is the multiplier it applies to each
Fourier mode.
Testing this change on the exact ground state $\ket{\psi}$ gives useful
guidance, but only one direction of an energy comparison.
The kinetic graph norm bounds its fourth momentum moment
uniformly:
\begin{equation}
 \sum_{\mathrm{ran}\Pp}\|\bm{k}\|^4|\psi_{\vec{\bm{p}},\boldsymbol\sigma}|^2
 \le(2m_{\max}\Theta(\ket{\psi}))^2
\end{equation}
where the sum runs over the retained momentum and spin labels and
$\bm{k}=\bm{k}_{\bm{p}_a}$ is the momentum of a chosen register. Fixed particle factors
are included if registers are summed. The improved tail bound,
applied on dyadic shells, also gives
$\sum_{\mathrm{ran}\Pp}\|\bm{k}\|^6|\psi_{\vec{\bm{p}},\boldsymbol\sigma}|^2\le CK$.
Thus a second-order stencil suggests a contribution
$O(h^2)=O(N^{-2/3})$, whereas a fourth-order stencil
suggests $O(h^4K)=O(N^{-1})$.
For higher orders, the corresponding shell estimate is
$O(K^{2p-3})$ for the $(2p+2)$-th moment, giving the
same $O(K^{-3})$ cutoff power when multiplied by
$h^{2p}$.

These are heuristic energy comparisons. A theorem would
also require the stencil symbol to remain uniformly
comparable to the exact kinetic symbol in both directions,
a tail bootstrap for its own discrete ground state, and
the reverse variational comparison. We have not established
those ingredients here. Exact kinetics via the quantum
Fourier transform is covered by our results; the calculation
suggests considering stencils of order at least four if
a finite-difference implementation is chosen, subject to
those additional checks.

\subsection{Implementation and additional errors}
\label{sec:grid-implementation}

The kernel bandwidth specifies the function represented by
the diagonal potential operation. It does not require
filtering the evolving state into the effective half zone:
all register amplitudes remain part of the simulation and
the measurement. One may represent the kernel by a table
of values at grid separations, or evaluate an approximation
arithmetically. In the latter case the approximation must
be controlled uniformly to obtain an operator-norm error.
The present discretisation results do not provide a
compiled reversible circuit or a gate-count theorem for
that arithmetic evaluation.

\begin{remark}[Implementation errors]
Suppose the implemented diagonal potential differs from
the target by at most $\epsilon_V$ in operator norm.
For diagonal matrices this is the largest absolute error
in a full potential entry. The variational principle
then bounds every eigenvalue shift by $\epsilon_V$.
Duhamel's formula bounds the difference of the two
unitary states by $t\epsilon_V$ for initial norm
at most one, and hence the induced expectation difference
by $2t\epsilon_V$ when $\|O\|\le1$.
When starting with an error per pair or external kernel,
the interaction charge weights and the sum over channels
must be included in $\epsilon_V$.
Errors in the kinetic phase, time-propagation algorithm
and state preparation are additional to these errors.
These implementation observations are not separate formal
endpoints.
\end{remark}

The transfer implementation must also match the finite
operator being analysed. Modular momentum shifts implement
the collocation matrix of Lemma~\ref{lem:fold}; rejecting
a transfer when it leaves the register implements
Galerkin projection instead. Changing between these rules
changes the Hamiltonian, not merely its representation.
Finally, phase estimation returns an eigenvalue of the
chosen finite Hamiltonian. To obtain its ground energy
with appreciable probability, the prepared state must
have sufficient overlap with the corresponding eigenspace.
The discretisation error theorem does not supply that
state-preparation success bound.

\section{Free-boundary systems}
\label{app:free}
\label{sec:free}

We seek error bounds for an isolated Coulomb system simulated in a finite
periodic cell and then on a plane-wave grid. These are two distinct
approximations. We first arrange that the cell interaction agrees exactly
with the free-space interaction wherever the particles are retained.
Localisation or leakage estimates then control the spatial truncation.
We next apply the preceding grid analysis in physical momentum units,
checking that its constants remain bounded as the cell grows. This
uniformity is essential: increasing the cell size should reduce the
boundary error without spoiling the grid error bound.

We treat stationary energies and time evolution separately. For energies
we fix the nuclei and compare the bottoms of the Hamiltonians in a
specified scalar exchange sector. For dynamics the nuclei may be fixed
or quantum particles. In the latter case a normalisable molecular
wavepacket may have a moving centre of mass; our comparison retains
that motion. Write $d=\eta_{\rm tot}$ for the number of moving particles,
with fixed masses $m_j\ge1$ and charges $q_j$. Optional clamped nuclei
have charges $\zeta_\ell$ and positions $\bm{R}_\ell\in B_{R_0}(0)$. The
Hamiltonian on $L^2(\mathbb R^{3d})$ is
\begin{equation}
 H_\infty=T+V_\infty,\qquad
 T=-\sum_{j=1}^d\frac{\Delta_j}{2m_j},\qquad
 V_\infty=\sum_{i<j}\frac{q_iq_j}{\|\bm{x}_i-\bm{x}_j\|}
       +\sum_{j,\ell}\frac{q_j\zeta_\ell}{\|\bm{x}_j-\bm{R}_\ell\|}.
 \label{eq:free-H}
\end{equation}
We omit the repulsion between clamped nuclei, which is a common scalar
and can be added to every energy. A scalar exchange sector specifies the
species labels and the signs under exchange of identical particles,
without internal spin coordinates. Proposition~\ref{prop:free} compares
ground energies within that specified sector. It does not identify the
physical spin-inclusive ground energy with a minimum over scalar sectors.
For smooth observables we first prove the dynamical comparison for scalar
states, then include finite internal spin in
Theorem~\ref{thm:free-dyn-spin}. The kinetic-energy comparison in
Proposition~\ref{prop:free-kinetic-log} includes its own spin extension.
The precise formal scope is described in Sec.~\ref{sec:exactdisc}.

The proofs have two different spatial inputs. The energy comparison uses
exponentially localised ground states and a variational argument. The
dynamical comparison uses the amount of wavefunction and gradient in the
cutoff region along a supplied free-space solution. The later domain
lemmas construct cutoffs compatible with Coulomb singularities; they
verify regularity of the cut initial state, but do not assert that the
required leakage is small.

We write scalar states as $\ket{u}$ and their position-space wavefunctions
as $u(\vec{\bm{x}})=\langle \vec{\bm{x}}|u\rangle$. During time evolution this convention is
$u(s,\vec{\bm{x}})=\langle \vec{\bm{x}}|u(s)\rangle$. Unbracketed functions in spatial
derivatives, integrals and pointwise quotients denote these wavefunctions;
their norms are the corresponding $L^2$ norms. We retain hats for Fourier
coefficients or transforms of scalar interaction kernels and observables,
and write state coefficients as momentum-basis amplitudes.

\subsection{Range truncation, zero mode and physical units}

We construct the periodic interaction by switching off the Coulomb
kernel before it can encounter another periodic image. Fix a transition
width $\rho>0$, and choose a smooth radial profile
$0\le\theta_{L,\rho}\le1$ that equals one for $r\le L/2-\rho$ and
zero for $r\ge L/2$. A monotone profile suffices. More generally, we
allow profiles with a common upper bound on their total variation
$V_\theta=\int_0^{L/2}|\theta'_{L,\rho}(r)|\,dr$, the total amount
by which the profile changes along the radial interval. This bound
will control its nonzero Fourier coefficients.
Write $\|\bm{x}\|_{\rm ni}:=\min_{\bm{\ell}\in\mathbb Z^3}\|\bm{x}+L\bm{\ell}\|_2$
for the Euclidean distance to the nearest periodic image of the origin.
Define the range-truncated kernel~\cite{IsmailBeigi2006} by
\begin{equation}
 g_L(\bm{x})=\frac{\theta_{L,\rho}(\|\bm{x}\|_{\rm ni})}{\|\bm{x}\|_{\rm ni}},\qquad
 |\widehat g_L(0)|\le\frac{\pi}{2L},\qquad
 |\widehat g_L(\bm{\nu})|\le(1+V_\theta)
                 \frac{4\pi}{L^3\|\bm{k}_{\bm{\nu}}\|^2}\quad(\bm{\nu}\ne0).
 \label{eq:rangecut}
\end{equation}
The kernel is singular where two positions coincide, but assigning it a
value on that measure-zero set does not change the multiplication
operator. For a monotone transition, $V_\theta=1$. To verify the
coefficient estimates and their dependence on $L$, for $k=\|\bm{k}_{\bm{\nu}}\|>0$ the radial Fourier integral is
\begin{equation}
 \widehat g_L(\bm{\nu})=\frac{4\pi}{L^3k}
                  \int_0^{L/2}\theta_{L,\rho}(r)\sin(kr)\,dr.
\end{equation}
Integration by parts exposes the contribution from the Coulomb
singularity at the origin:
\begin{equation}
 \int_0^{L/2}\theta_{L,\rho}(r)\sin(kr)\,dr
 =\frac1k+\frac1k\int_0^{L/2}\theta'_{L,\rho}(r)\cos(kr)\,dr.
\end{equation}
The boundary values are $\theta_{L,\rho}(0)=1$ and
$\theta_{L,\rho}(L/2)=0$, so the absolute value is at most
$(1+V_\theta)/k$.
At zero momentum the Fourier coefficient is
$4\pi L^{-3}\int_0^{L/2}r\theta_{L,\rho}(r)\,dr\le\pi/(2L)$.
Thus fixed width is accompanied by a specified profile, not by an
unrestricted family of oscillatory cutoffs.

The zero Fourier mode is the spatial average of the interaction. It is
a constant multiplier, but it must be retained when we compare absolute
energies with free space. Throughout this appendix the physical
finite-volume Hamiltonian $H_L$ \emph{includes} this mode. We write $H_L^{(0)}$ for the operator
with that mode omitted. The two conventions are related by
\begin{equation}
 \begin{split}
 C_{\rm zm}(L)&=\widehat g_L(0)
   \left(\sum_{i<j}q_iq_j+\sum_{j,\ell}q_j\zeta_\ell\right),\\
 H_L&=H_L^{(0)}+C_{\rm zm}(L)I,\qquad
 E_L=E_L^{(0)}+C_{\rm zm}(L),\qquad
 E_{N,L}=E_{N,L}^{(0)}+C_{\rm zm}(L).
 \end{split}
 \label{eq:free-zero-mode}
\end{equation}
The scalar is represented exactly by both discretisations. With
$Q_{\rm ch}=\sum_{i<j}|q_iq_j|+\sum_{j,\ell}|q_j\zeta_\ell|$,
$|C_{\rm zm}(L)|\le Q_{\rm ch}\pi/(2L)\le Q_{\rm ch}/4$ for $L\ge2\pi$.
Dropping this shift from the energy comparison would introduce an
$O(L^{-1})$ error. In dynamics it gives only the phase
$e^{-itH_L}=e^{-itC_{\rm zm}(L)}e^{-itH_L^{(0)}}$.

We use physical coordinates and the Fourier convention of
Eq.~\eqref{eq:fourier-sign}. Consequently the sampled observable is
$O_g=f(\bm{x}_{\bm{j}_a})$ at physical grid points. In the unit-torus synthesis
convention of the Lean library the physical frame is reflected: a nucleus
at $\bm{R}_\ell$ has unit-torus coordinate $-\bm{R}_\ell/L$, and the observable is
\texttt{chiRefl(cellObs)}. These changes of coordinates do not alter the
physical Hamiltonian or expectation values.

For the grid we distinguish momentum spacing, physical resolution and
register size. Write
\begin{equation}
 g=\frac{2\pi}{L},\qquad K=gm,\qquad
 m\in\mathbb N,\qquad N=(4m+1)^3.
 \label{eq:free-grid-count}
\end{equation}
Here $g$ is the spacing between adjacent physical momenta, while $K$ is
the physical cutoff used in the error estimates. The padded register
contains indices from $-2m$ to $2m$ in each spatial direction, so $N$
counts its actual spatial modes per particle. This differs from the effective basis count in
Appendix~\ref{app:grid} by only a fixed factor. Unless stated otherwise,
the grid interaction uses the sharp transfer truncation
$\widehat g_L\mathbf1_{\|\bm{k}_{\bm{\nu}}\|_\infty\le2K}$. The same operator estimates
hold for the real, even kernel class of Eq.~\eqref{eq:kernelclass}, with
$\widehat g_L$ as reference and its Coulomb coefficient majorant as the
bound. Agreement at low transfers and the register bandwidth remain
required; arbitrary sampling of the singular function is not included.

To use the earlier error bounds while varying $L$, we need estimates
at a fixed \emph{physical} momentum threshold. At fixed threshold the
number of Fourier modes grows like $L^3$, while each coefficient carries
a volume factor. The next lemma verifies the cancellations needed in
both the interaction estimates and the three-dimensional Sobolev bound.
The latter is applied to one particle coordinate with the others held
fixed, which is why we call it a slice bound.

\begin{lemma}[Scaling uniformity]
\label{lem:scaling}
Fix $L_*\ge2\pi$ and a bound on $V_\theta$. Uniformly for $L\ge L_*$,
$J\ge2\pi/L_*$ and $K\ge2\pi/L_*$,
\begin{align}
 \sum_{0<\|\bm{k}_{\bm{\nu}}\|\le J}|\widehat g_L(\bm{\nu})|&\le CJ,&
 \left(L^3\sum_{\|\bm{k}_{\bm{\nu}}\|>J}|\widehat g_L(\bm{\nu})|^2\right)^{1/2}
       &\le CJ^{-1/2},\label{eq:free-coeff-uniform}\\
 \left(L^3\sum_{0<\|\bm{k}_{\bm{\nu}}\|\le2K}\|\bm{k}_{\bm{\nu}}\|^{2j}
                  |\widehat g_L(\bm{\nu})|^2\right)^{1/2}
       &\le CK^{j-1/2},&&j=1,2.\label{eq:free-moments-uniform}
\end{align}
The physical $H^2(\T_L)\to L^\infty(\T_L)$ slice bound has an
$L$-independent constant.
\end{lemma}
\begin{proofx}
We first convert each physical threshold to an integer-lattice radius.
Equation~\eqref{eq:rangecut} and $\bm{k}_{\bm{\nu}}=(2\pi/L)\bm{\nu}$ give
$|\widehat g_L(\bm{\nu})|\le C/(L\|\bm{\nu}\|^2)$ for $\bm{\nu}\ne0$.
For $R=JL/(2\pi)$, the assumed lower bounds on $J$ and $L$ ensure
$R\ge1$. Summing over three-dimensional lattice shells gives
\begin{equation}
 \sum_{0<\|\bm{\nu}\|\le R}\|\bm{\nu}\|^{-2}\le CR,\qquad
 \sum_{\|\bm{\nu}\|>R}\|\bm{\nu}\|^{-4}\le C/R,\qquad
 \#\{\bm{\nu}:\|\bm{\nu}\|\le R\}\le CR^3.
\end{equation}
Indeed, a shell of radius comparable to $r$ has $O(r^2)$ lattice
points. Thus the first sum grows like $\int_1^Rdr$, the second
decays like $\int_R^\infty r^{-2}dr$, and the count grows like
$\int_0^R r^2dr$. The low-transfer sum is consequently at most
$CR/L=C'J$. For the squared tail we obtain
\begin{equation}
 L^3\sum_{\|\bm{k}_{\bm{\nu}}\|>J}|\widehat g_L(\bm{\nu})|^2
 \le CL\sum_{\|\bm{\nu}\|>R}\|\bm{\nu}\|^{-4}
 \le CL/R=C'/J.
\end{equation}
Taking a square root proves both estimates in
Eq.~\eqref{eq:free-coeff-uniform}.

For the moments, set $R_K=2K/g$. With $j=1,2$, the same shell
count gives $\sum_{0<\|\bm{\nu}\|\le R_K}\|\bm{\nu}\|^{2j-4}
\le C R_K^{2j-1}$. Hence
\begin{equation}
 L^3\sum_{0<\|\bm{k}_{\bm{\nu}}\|\le2K}\|\bm{k}_{\bm{\nu}}\|^{2j}
                   |\widehat g_L(\bm{\nu})|^2
 \le CLg^{2j}R_K^{2j-1}
 =C(Lg)(2K)^{2j-1}\le C'K^{2j-1}.
\end{equation}
The identity $Lg=2\pi$ removes the cell size. Taking square roots
proves Eq.~\eqref{eq:free-moments-uniform}.

We finally check the normalisation in the slice embedding. To see the
required power of the momentum spacing, group integer vectors by
$n=\|\bm{\nu}\|_\infty$. For $n\ge1$ this shell contains
$(2n+1)^3-(2n-1)^3\le26n^2$ points. Since $g\le1$, put
$R=1/g\ge1$ and split the sum at $R$. On the inner shells we use
$\sum_{1\le n\le R}n^2\le R^3$; on the outer shells we use
$\sum_{n>R}n^{-2}\le2/R$. Thus
\begin{equation}
 \sum_{\bm{\nu}\ne0}(1+g^2\|\bm{\nu}\|^2)^{-2}
 \le26\left(\sum_{1\le n\le R}n^2
                +g^{-4}\sum_{n>R}n^{-2}\right)
 \le78g^{-3}.
\end{equation}
Adding the zero vector and multiplying by the physical volume
normalisation gives
\begin{equation}
 \sum_{\bm{\nu}\in\Z^3}(1+g^2\|\bm{\nu}\|^2)^{-2}\le1+78g^{-3},\qquad
 L^{-3}\sum_{\bm{\nu}}(1+\|\bm{k}_{\bm{\nu}}\|^2)^{-2}
       \le L_*^{-3}+\frac{78}{(2\pi)^3}.
 \label{eq:free-sobolev-uniform}
\end{equation}
For the orthonormal one-particle plane-wave basis, write
$v_{\bm{k}_{\bm{\nu}}}:=\langle \bm{k}_{\bm{\nu}}|v\rangle$, so that
$\ket{v}=\sum_{\bm{\nu}} v_{\bm{k}_{\bm{\nu}}}\ket{\bm{k}_{\bm{\nu}}}$ and
$v(\bm{x})=L^{-3/2}\sum_{\bm{\nu}} v_{\bm{k}_{\bm{\nu}}}e^{-i\bm{k}_{\bm{\nu}}\cdot \bm{x}}$.
Weighted Cauchy--Schwarz gives
\begin{equation}
 |v(\bm{x})|\le
 \left[L^{-3}\sum_{\bm{\nu}}(1+\|\bm{k}_{\bm{\nu}}\|^2)^{-2}\right]^{1/2}
 \left[\sum_{\bm{\nu}}(1+\|\bm{k}_{\bm{\nu}}\|^2)^2|v_{\bm{k}_{\bm{\nu}}}|^2\right]^{1/2}.
\end{equation}
The second factor is the physical $H^2$ norm; the first is uniformly
bounded by Eq.~\eqref{eq:free-sobolev-uniform}. This proves the slice
bound. Keeping the $g^{-3}$ mode count is essential: its factor of
$L^3$ cancels the Fourier volume normalisation. A coarser $g^{-4}$
bound would leave an artificial dependence on $L$.
\end{proofx}

\paragraph{Uniform operator bounds.}
We next check that the operator estimates, not just the Fourier
coefficient sums, are uniform in the cell size. In this paragraph
$w(\vec{\bm{k}})=1+\sum_j\|\bm{k}_j\|^2/m_j=1+2A(\vec{\bm{k}})$ denotes the
physical kinetic weight; it is not the periodic interaction kernel
denoted by $w$ earlier in the paper. We write $W:=1+2T$ for the
operator diagonal in the momentum basis with multiplier $w(\vec{\bm{k}})$.
Applying $W$ measures two spatial derivatives in physical units. Let
$\mathcal U_g(t)$ be the grid evolution with the zero mode omitted.
The physical-unit grid estimates underlying
Theorem~\ref{thm:grid-dyn-sharp} give constants $a_0,c_2$ independent of
$L\ge2\pi$ such that
\begin{align}
 \|V_{\rm grid}^{(0)}\ket{v}\|&\le a_0\|\ket{v}\|+\tfrac18\|W\ket{v}\|,\\
 \|WV_{\rm grid}^{(0)}\ket{v}\|&\le\tfrac14\|W^2\ket{v}\|
                 +c_2K^{3/2}\|W\ket{v}\|,\label{eq:free-physical-relative}\\
 \|W\mathcal U_g(t)\ket{v}\|&\le\tfrac53\|W\ket{v}\|
                 +\tfrac43(1+4a_0)\|\ket{v}\|.
\end{align}
Here $m\ge8$, $K\ge8$ suffice for the dynamical application.
For the first estimate, we split momentum transfers at a fixed physical
threshold. Lemma~\ref{lem:scaling} controls the absolute sum of the
low transfers and the squared sum of the high transfers. The slice
bound in Eq.~\eqref{eq:free-sobolev-uniform} then converts the latter
coefficient estimate into a relative operator bound. Increasing the
fixed threshold makes the coefficient of $\|W\ket{v}\|$ as small as $1/8$;
the remaining constant is absorbed into $a_0$.

For the second estimate, moving the kinetic weight through a cyclic
transfer gives $w(\vec{\bm{q}})\le2w(\vec{\bm{p}})+C\|\bm{k}_{\bm{\nu}}\|^2$. The first term is handled
by the relative bound. In the second, the factor $\|\bm{k}_{\bm{\nu}}\|^2$ cancels
the Coulomb decay, leaving coefficients of size $O(L^{-3})$.
There are $O((LK)^3)$ transfers on each three-dimensional slice,
so their coefficient $\ell^2$ norm is $O(L^{-3/2}K^{3/2})$.
Weighted Cauchy--Schwarz for the input slice contributes $O(L^{3/2})$,
by the Sobolev sum just proved. The volume factors cancel, leaving
$K^{3/2}$. Equivalently, the transfer-count factor in physical units
is bounded by $[g(4m+1)]^{3/2}\le(5K)^{3/2}$.

The final estimate controls the kinetic weight throughout the grid
evolution. To see why its constant is independent of time, shift the
zero-mode-omitted Hamiltonian by $1/2$ and write it as
$G=W/2+V_{\rm grid}^{(0)}$. The first relative bound gives
\begin{equation}
 \|G\ket{v}\|\le\tfrac58\|W\ket{v}\|+a_0\|\ket{v}\|,\qquad
 \|W\ket{v}\|\le\tfrac83\|G\ket{v}\|+\tfrac83a_0\|\ket{v}\|.
\end{equation}
The flow conserves both $\|G\ket{v}\|$ and $\|\ket{v}\|$, since $G$ differs
from its generator by a scalar. Combining the two inequalities
gives $\|W\mathcal U_g(t)\ket{v}\|\le\tfrac53\|W\ket{v}\|
+\tfrac{16}{3}a_0\|\ket{v}\|$, which in particular implies the displayed
bound in Eq.~\eqref{eq:free-physical-relative}.

The remaining input is the second kinetic weight of the projected
continuum state. We use the momentum-shell argument of
Lemma~\ref{lem:grid-data2}, with shells defined in physical units.
If $\ket{u}$ has a uniformly bounded second Hamiltonian graph norm, its
weighted tail above a physical radius $J\ge1$ is $O(J^{-1/2})$.
On a shell $J<\max_j\|\bm{k}_j\|\le2J$, an additional kinetic weight
costs $O(J^2)$, so the squared second-weight norm on that shell
is $O(J^3)$. Orthogonality lets us sum these squared norms over
dyadic radii, giving $\sum_{2^j\lesssim K}2^{3j}=O(K^3)$.
The resulting norm is $O(K^{3/2})$, with a bounded contribution
from momenta below one. Increasing $L$ adds modes inside each
physical shell, but does not add shells between one and $K$.
This is why a count in integer Fourier indices must retain the
factors of $g=2\pi/L$ when converted to kinetic weights.
These are the uniform versions of the relative, moment and flow bounds
used in Appendix~\ref{app:grid}. The alias multiplicity is independent
of $L$. Thus the $g_L$ grid energy and smooth-observable rates have
$L$-independent constants when their energy, state and observable
budgets are uniformly bounded. Coefficient scaling alone would not
justify that conclusion.

\subsection{Localised ground energy and basis size}

We now compare ground energies in a fixed scalar exchange sector, with
the nuclei clamped. The variational principle gives the two directions
of comparison from two different trial states: a localised free-space
ground state gives an upper bound on the cell energy, whereas a
localised cell ground state gives the reverse bound. The first direction
uses only free-space localisation; the second needs a separate
localisation hypothesis for the cell ground state.

We first choose a region on which the two potentials coincide. Set
\begin{equation}
 R_L=\frac L4-R_0-\rho,\qquad
 \mathcal B_L=\{\vec{\bm{x}}:\ \|\bm{x}_j\|<R_L\text{ for every moving particle }j\}.
 \label{eq:free-retained-region}
\end{equation}
On this product region, $\|\bm{x}_j-\bm{R}_\ell\|\le R_L+R_0<L/2-\rho$ and
$\|\bm{x}_i-\bm{x}_j\|\le2R_L<L/2-\rho$. Therefore the physical $g_L$ interaction
agrees identically with $V_\infty$ there.

To turn a state into a trial state supported in this region, we use a
smooth partition $\chi_L^2+\xi_L^2=1$. The first multiplier retains
the interior and the second accounts for what is removed. The IMS
localisation identity expresses the original kinetic form as the sum
of the forms of these two pieces, minus a derivative cost. We therefore
need to control the derivatives of \emph{both} multipliers.
Choose a fixed smooth transition
$\vartheta:\mathbb R\to[0,1]$, equal to zero on $(-\infty,0]$ and
one on $[1,\infty)$, with $\|\vartheta'\|_\infty\le B$.
For $w_0>0$, $w_0<R_L$, put $R_1=R_L-w_0$ and
\begin{equation}
 b_j(\vec{\bm{x}})=1-\vartheta\!\left(\frac{\|\bm{x}_j\|^2-R_1^2}{R_L^2-R_1^2}\right),\quad
 a_L(\vec{\bm{x}})=\frac\pi2\left(1-\prod_j b_j(\vec{\bm{x}})\right),\quad
 \chi_L=\cos a_L,\quad \xi_L=\sin a_L.
 \label{eq:free-IMS-pair}
\end{equation}
Then $\chi_L^2+\xi_L^2=1$, $\chi_L=1$ on the product region of radius
$R_1$, and $\chi_L=0$ when any particle lies outside the radius $R_L$.
Both multipliers preserve the scalar exchange sector. On the transition
region $\|\nabla b_j\|\le2B/w_0$, so
\begin{equation}
 \sum_j\bigl(\|\nabla_j\chi_L\|^2+\|\nabla_j\xi_L\|^2\bigr)
       =\sum_j\|\nabla_j a_L\|^2\le d\pi^2B^2/w_0^2.
 \label{eq:free-IMS-bound}
\end{equation}
The common phase $a_L$ is what makes the derivative cost explicit:
$\nabla\chi_L=-\sin(a_L)\nabla a_L$ and
$\nabla\xi_L=\cos(a_L)\nabla a_L$, so their squared derivatives
add to $\|\nabla a_L\|^2$. This avoids having to estimate derivatives
of a complementary multiplier introduced only as a square root.

We will apply the localisation identity to smooth trial states whose
energies approach the ground energy. The next lemma supplies such
states while retaining an upper bound on their exterior probability.
It uses approximation in the energy form, so a Coulomb eigenfunction
need not be pointwise smooth at a collision.

\begin{lemma}[Smooth trial states from localised ground states]
\label{lem:free-form-approximation}
Let a normalised free-space scalar-sector ground state $\ket{\psi_\infty}$
satisfy $H_\infty\ket{\psi_\infty}=E_\infty\ket{\psi_\infty}$ and
\begin{equation}
 \sum_j\int e^{2\alpha\|\bm{x}_j\|}|\psi_\infty(\vec{\bm{x}})|^2\,d\vec{\bm{x}}\le A_\infty,
 \qquad \alpha>0.
 \label{eq:free-agmon-data}
\end{equation}
For each fixed $r>0$ it admits normalised, compactly supported $C^2$
sector-preserving trial states with energies tending to $E_\infty$
and mass outside $\{\|\bm{x}_j\|<r\ \forall j\}$ at most
$2\max\{A_\infty,1\}e^{-2\alpha r}$.
Likewise, a normalised $H^2$ torus ground eigenvector has smooth
sector-preserving near-minimisers whose mass on any fixed measurable
outer region is at most twice any given positive bound for its own mass.
\end{lemma}
\begin{proofx}
We first approximate the free-space ground state without losing its
exchange symmetry. The weak Hamiltonian relation and elliptic regularity
give $\ket{\psi_\infty}\in H^1$. Here the Hamiltonian equation is understood
by testing against smooth compactly supported functions and transferring
spatial derivatives to those test functions by integration by parts.
Convolve it with a product of identical
one-particle mollifiers, which smooth the state on a small length scale,
and then multiply by the same smooth large-radius cutoff in each
particle coordinate. Both operations commute with particle exchanges.
By taking the smoothing scale to zero and the cutoff radius to infinity
along a diagonal sequence, we obtain compactly supported smooth states
converging to $\ket{\psi_\infty}$ in $H^1$ in the same sector.

The kinetic form is continuous in $H^1$. Hardy's inequality gives the
same continuity for every Coulomb form term, including the translated
nuclear terms. For example, applying Cauchy--Schwarz to a difference of
potential expectations reduces it to norms of the approximating states
and of their Coulomb quotients, and Hardy bounds the latter by their
$H^1$ norms. After normalising the approximants, their energies
therefore tend to $E_\infty$.

We also need convergence of the mass in a fixed exterior region. If
$A_r=\{\vec{\bm{x}}:\exists j,\ \|\bm{x}_j\|\ge r\}$, then multiplication by
$\mathbf1_{A_r}$ is an $L^2$ contraction. Thus $L^2$ convergence
implies convergence of $\|\mathbf1_{A_r}\ket{\psi}\|^2$. For the exact
ground state, the union bound and the exponential moment give
\begin{equation}
 \int_{A_r}|\psi_\infty|^2
 \le\sum_j\int_{\|\bm{x}_j\|\ge r}|\psi_\infty|^2
 \le e^{-2\alpha r}\sum_j\int e^{2\alpha\|\bm{x}_j\|}|\psi_\infty|^2
 \le A_\infty e^{-2\alpha r}.
\end{equation}
For each fixed $r$, sufficiently accurate normalised approximants
therefore satisfy the stated bound with the factor-two slack.

On the torus we instead truncate the Fourier series in a common cube
for every particle. These smooth approximants preserve the sector and
converge in $H^2$. Consequently their energies and their masses on
any fixed measurable exterior set converge. Normalisation followed
by the same positive slack gives the torus assertion. Neither
construction assumes $H^4$ regularity of a Coulomb eigenfunction.
\end{proofx}

\begin{proposition}[Free-boundary energy under localisation]
\label{prop:free}
Fix a clamped-nuclei system, a scalar exchange sector, $\rho,w_0>0$,
and a normalised free-space ground eigenfunction satisfying
Eq.~\eqref{eq:free-agmon-data}. Assume $E_\infty$ is the bottom of the
Coulomb form on that sector. Choose the profiles above with a fixed
bound on $V_\theta$. There exist $L_0,K_0,C,C_+>0$, independent of
$L,m$, such that for $L\ge L_0$, $m\ge d+8$ and $K=2\pi m/L\ge K_0$,
\begin{equation}
 E_{N,L}\le E_\infty+CK^{-3}+C_+e^{-\alpha L/2}.
 \label{eq:free-energy-upper}
\end{equation}
This upper bound does not require the finite-volume localisation
assumption below.

For the two-sided comparison, assume that for each sufficiently large
$L$ there is a normalised $H^2$ ground eigenvector $\ket{u_L}$ of the completed
range-cut torus operator, in the same sector, and that
\begin{equation}
 \text{(A):}\qquad
 \sum_j\int_{\T_L^d}
   e^{2\beta(\|\bm{x}_j\|-R_{L/2})}|u_L(\vec{\bm{x}})|^2\,d\vec{\bm{x}}\le A_{\rm tor},
 \qquad \beta>0,
 \label{eq:free-localisation}
\end{equation}
with fixed $\beta,A_{\rm tor}$; $\bm{x}_j$ are centred cell representatives.
Then, after increasing $L_0,K_0$ if necessary,
\begin{equation}
 |E_{N,L}-E_\infty|\le CK^{-3}+C'e^{-\gamma L},
 \qquad \gamma=\frac12\min\{\alpha,\beta/2\}>0.
 \label{eq:free-energy}
\end{equation}
The constants are uniform for the fixed molecule, sector, profiles and
localisation data. For every $0<\epsilon\le1/2$ a sufficient actual grid
size per particle is
\begin{equation}
 N=(4m+1)^3=O\!\left(\epsilon^{-1}\log^3(1/\epsilon)\right).
 \label{eq:free-complexity}
\end{equation}
\end{proposition}
The torus operator in assumption (A) is the full physical operator
$H_L$ on the specified sector, with domain $H^2$, before any Fourier
projection. In particular, the assumed $\ket{u_L}$ is a cell eigenstate,
not an eigenvector of a finite grid matrix. The zero mode is included
as in Eq.~\eqref{eq:free-zero-mode}.
\begin{proofx}
We first prove a localisation estimate that can be used in either
direction. Let $q$ be the quadratic form in the starting space and
$E$ its bottom in the chosen sector. Take a normalised trial state
$\ket{v}$ with $q(\ket{v})\le E+\delta$, and suppose that its mass outside the
product region of radius $R_L-w_0$ is at most $\tau<1$.
The IMS identity for Eq.~\eqref{eq:free-IMS-pair} reads
\begin{equation}
 q(\ket{v})=q(\chi_L\ket{v})+q(\xi_L\ket{v})
 -\sum_j\frac1{2m_j}\int
   (\|\nabla_j\chi_L\|^2+\|\nabla_j\xi_L\|^2)|v|^2.
\end{equation}
Both cutoffs preserve the sector, so
$q(\xi_L\ket{v})\ge E\|\xi_L\ket{v}\|^2$. Their derivatives vanish on the
inner configuration region, and Eq.~\eqref{eq:free-IMS-bound} bounds
the last term by $d\pi^2B^2\tau/(2w_0^2)$, using $m_j\ge1$.
Also $\|\chi_L\ket{v}\|^2\ge1-\tau$. Subtracting
$E\|\chi_L\ket{v}\|^2$ before dividing therefore gives
\begin{equation}
 \frac{q(\chi_L\ket{v})}{\|\chi_L\ket{v}\|^2}
 \le E+\frac{\delta+d\pi^2B^2\tau/(2w_0^2)}{1-\tau}.
\end{equation}
This estimate makes the small quantity explicit: only the mass in the
transition and exterior region multiplies the derivative cost. The
retained piece can then be transplanted isometrically to the other
space. Its kinetic form is unchanged because it is supported away
from the cell boundary, and its physical potential is unchanged by
Eq.~\eqref{eq:free-retained-region}.

For the upper bound on $E_L$, start with the free-space trial states
from Lemma~\ref{lem:free-form-approximation}, taking $r=R_L-w_0$.
Their exterior mass satisfies
$\tau\le2\max\{A_\infty,1\}e^{-2\alpha(R_L-w_0)}$.
Let their energy excess $\delta$ tend to zero. When $L$ is large
enough that $\tau\le1/2$, the denominator above is bounded below
by $1/2$. Since $R_L=L/4-R_0-\rho$, we obtain
\begin{equation}
 E_L\le E_\infty+C_+e^{-\alpha L/2},
\end{equation}
with the fixed factors involving $R_0,\rho,w_0$ absorbed into $C_+$.
The existence of an Agmon bound is an explicit input, rather than a
consequence proved here from an ionisation gap~\cite{Agmon1982}.
The Agmon bound used here is the exponentially weighted probability
estimate in Eq.~\eqref{eq:free-agmon-data}; an ionisation gap is a
positive separation between the bound-state energy and the threshold
energy for escape of a particle to infinity.

For the lower bound on $E_L$, we must construct a free-space trial
state from the cell ground state. This is where the independent
assumption (A) enters. If any particle satisfies
$\|\bm{x}_j\|\ge R_L-w_0$, its exponential weight in (A) is at least
$e^{2\beta(R_L-w_0-R_{L/2})}$. Summing over these events gives
\begin{equation}
 \int_{\{\exists j:\ \|\bm{x}_j\|\ge R_L-w_0\}}|u_L|^2
 \le A_{\rm tor}e^{-2\beta(R_L-w_0-R_{L/2})}
 =A_{\rm tor}e^{-2\beta(L/8-w_0)}.
 \label{eq:free-product-tail}
\end{equation}
Smooth torus near-minimisers inherit twice this positive tail bound.
The same IMS argument and transplantation back to free space give
$E_\infty\le E_L+C_-e^{-\beta L/4}$. Thus the continuous-volume error
has the exponent in Eq.~\eqref{eq:free-energy}.
The alternative weighted-distance condition
$\|e^{\beta\operatorname{dist}(\cdot,\mathcal B_{L/2})}\ket{u_L}\|\le C_A$
implies (A) with $A_{\rm tor}=dC_A^2$, since
$\operatorname{dist}(\vec{\bm{x}},\mathcal B_{L/2})\ge \|\bm{x}_j\|-R_{L/2}$ for each $j$.
Consequently the product geometry has not been replaced by a smaller
configuration-space ball.

We can now choose a cell threshold that validates the two denominator
bounds used above. It is enough that the exterior masses of the smooth
trial states be at most $1/2$. Solving the two exponential inequalities
gives the following sufficient choice, including the retained-region
and transition-width requirements:
\begin{equation}
 \begin{split}
 L_+&=\max\!\left\{2\pi,
 4\left(R_0+\rho+w_0+
       \frac{\log(4\max\{A_\infty,1\})}{2\alpha}\right)\right\},\\
 L_0&=\max\!\left\{L_+,\,
 4\left(R_0+\rho+2w_0+
       \frac{\log(4\max\{A_{\rm tor},1\})}{\beta}\right)\right\}.
 \end{split}
 \label{eq:free-energy-L0}
\end{equation}
For the upper bound alone $L_+$ suffices.

It remains to make the grid-energy constant uniform in $L$.
The physical-unit version of Theorem~\ref{thm:grid-energy} gives
\begin{equation}
 |E_{N,L}^{(0)}-E_L^{(0)}|\le C(E_b)K^{-3}
 \quad\text{for }m\ge d+8,\ K\ge K_0(E_b),\quad
 E_b\ge |E_L^{(0)}|+1,
 \label{eq:free-uniform-grid-energy}
\end{equation}
where $C(E_b),K_0(E_b)$ do not depend on $L\ge2\pi$.
To use this estimate we must choose $E_b$ before invoking the grid
comparison. The uniform Coulomb form bound supplies a fixed constant
$B_{\rm rel}$ for which
$E_L^{(0)}\ge-(B_{\rm rel}^2+1/4)$.
Together with the upper comparison and Eq.~\eqref{eq:free-zero-mode},
this permits the noncircular common energy budget
\begin{equation}
 E_b=1+\max\{|E_\infty|+C_++Q_{\rm ch}/4,
                         B_{\rm rel}^2+1/4\}
\end{equation}
for the upper bound, without using (A). For the two-sided bound use
$E_b=1+|E_\infty|+C'+Q_{\rm ch}/4$.
Equation~\eqref{eq:free-uniform-grid-energy} now completes both grid
comparisons, since the scalar shifts cancel between grid and torus.

Finally, to obtain accuracy $\epsilon$, we allocate at most
$\epsilon/2$ to each term in Eq.~\eqref{eq:free-energy}. The
exponential spatial term asks for $L=O(\log(1/\epsilon))$, while
the grid term asks for $K=O(\epsilon^{-1/3})$. An explicit admissible
choice is
\begin{equation}
 L=L_0+\gamma^{-1}\bigl[\log(2C'+2)+\log(1/\epsilon)\bigr],\qquad
 K_* =\max\{K_0,((2C+2)/\epsilon)^{1/3}\},
\end{equation}
and $m=\max\{d+8,\lceil LK_*/(2\pi)\rceil\}$.
Both error terms are at most $\epsilon/2$ and
$m=O(\epsilon^{-1/3}\log(1/\epsilon))$, proving
Eq.~\eqref{eq:free-complexity}, including admissible-size rounding.
\end{proofx}

The hypotheses on free-space decay and finite-volume localisation are
not proved by this comparison. Nor is a normalised torus ground
eigenvector constructed in the free-boundary endpoint: it is supplied
in its lower-bound hypothesis. These distinctions remain even though
the approximation, energy comparison and complexity statements are
formalised. A stationary isolated-system theorem with all particles
quantum would require an internal-energy formulation after treating the
centre of mass separately; no such ground-eigenfunction statement is
asserted here. The dynamical result below instead keeps the full motion.

\subsection{Uniform free-boundary dynamics}

We compare a supplied exact free-space evolution
with the cell evolution from its cut initial state, and then
with the grid evolution from the projected cell initial state. The
spatial comparison will use the small amount of wavefunction outside
the inner region and the small gradient on the cutoff shell. Before
stating it, we check that a fixed physical observable has uniform
Fourier bounds as the cell grows.

Let $f\in C_c^\infty(\mathbb R^3;\mathbb R)$, $0\le f\le1$, be a fixed
one-particle observable supported in $B_{R_f}(0)$. Its cell version
$\chi_L(\bm{x}_a)=f(\bm{x}_a)$ is continuous on the torus when $L>2R_f$.
Here $\chi_L$ denotes the one-particle observable; the configuration
localisation multiplier used in the energy proof is no longer in use.
In the Fourier-transform convention
$\widehat f(\bm{\xi})=\int f(\bm{x})e^{-2\pi i \bm{x}\cdot\bm{\xi}}\,d\bm{x}$, its multiplier
coefficients satisfy $\widehat\chi_L(\bm{\nu})=L^{-3}\widehat f(\bm{\nu}/L)$.
For $B_j(f)=\sup_{\bm{\xi}}(1+\|\bm{\xi}\|)^j|\widehat f(\bm{\xi})|$,
\begin{align}
 C_\chi,\ \|O\|&\le27B_6(f),&
 D_2&\le27(2\pi)^2B_8(f),\\
 D_4&\le27(2\pi)^4B_{10}(f),&
 C_{\chi,j}&\le27(2\pi)^jB_{j+6}(f).
 \label{eq:free-observable-uniform}
\end{align}
Here $D_r=\sum_{\bm{\nu}}\|\bm{k}_{\bm{\nu}}\|^r|\widehat\chi_L(\bm{\nu})|$.
We verify uniformity by measuring decay in the physical variable
$\bm{\nu}/L$. Put $h=1/L\le1$. Since
$\prod_{c=1}^3(1+h|\nu_c|)\le(1+h\|\bm{\nu}\|)^3$, comparison with
three one-dimensional sums gives
\begin{equation}
 \sum_{\bm{\nu}\in\mathbb Z^3}L^{-3}(1+\|\bm{\nu}\|/L)^{-6}
 \le\left(\sum_{n\in\mathbb Z}\frac{h}{(1+h|n|)^2}\right)^3
 \le(h+2)^3\le27.
\end{equation}
For the $r$th physical moment we also use
$\|\bm{k}_{\bm{\nu}}\|^r=(2\pi)^r\|\bm{\nu}/L\|^r$ and absorb this power into
$B_{r+6}(f)$. Restricting the same sum to $\|\bm{k}_{\bm{\nu}}\|>J$ then gives
the required $J^{-j}$ observable-tail bound from its $j$th moment.
This proves Eq.~\eqref{eq:free-observable-uniform}, including the
constants in (H$\chi$). The derivative estimates used in the grid
proof are also in physical units: an index moment alone may grow
with $L$, whereas the corresponding factor of $g$ restores the
physical moment bounded here.

There is one further regularity issue. A smooth spatial cutoff does
not automatically preserve $D(H_\infty^2)$ for a Coulomb Hamiltonian,
because a second Hamiltonian application differentiates the potential.
We therefore state a uniform squared-domain condition explicitly in
the next proposition. The later cutoff lemmas verify it by controlling
the singular commutators for two concrete families of cutoffs.

\begin{proposition}[Free-boundary dynamics under leakage and cutoff regularity]
\label{prop:free-dyn}
Let $\ket{\psi_\infty}$ be a strong solution of the free-space system in
Eq.~\eqref{eq:free-H}, including any clamped nuclei, with initial state
$\ket{\psi_0}\in D(H_\infty^2)$. The solution obeys
$i\partial_s\ket{\psi_\infty}=H_\infty\ket{\psi_\infty}$.
In the weak formulation the curve, its $L^2$ Hamiltonian image and the
weak spatial gradient are supplied.

Let $X_L$ be a real $C^2$ configuration cutoff with $0\le X_L\le1$,
compactly supported in the retained region and equal to one on a
measurable inner region $\mathcal B_L^{\rm one}$. Choose a measurable
shell $S_L$ in the complement of this inner region. All first and second
coordinate derivatives of $X_L$ vanish off $S_L$, and have uniform bounds
\begin{equation}
 \sup_{L,j,c}\|\partial_{j,c}X_L\|_\infty\le D_1,\qquad
 \sup_{L,j,c}\|\partial_{j,c}^2X_L\|_\infty\le D_2^{\rm cut}.
\end{equation}
If a sector is imposed, choose the cutoff to preserve it.
For $0\le s\le t$ assume
\begin{equation}
 \text{(H$_{\rm leak}$):}\qquad
 \|\mathbf1_{(\mathcal B_L^{\rm one})^c}\ket{\psi_\infty(s)}\|\le\epsilon_{\rm leak},
 \quad
 \|\mathbf1_{S_L}\nabla\psi_\infty(s,\cdot)\|\le\epsilon_{\rm leak}\le1.
 \label{eq:free-leakage}
\end{equation}
Let $\mathcal T_L$ denote transplantation into the centred cell and put
$\ket{\varphi_{0,L}}=\mathcal T_L(X_L\ket{\psi_0})$. Also assume
\begin{equation}
 \text{(H$_{\rm dom}$):}\qquad
 \ket{\varphi_{0,L}}\in D(H_L^2),\qquad
 \sup_{L\ge L_0}\sum_{j=0}^2\|H_L^j\ket{\varphi_{0,L}}\|\le D_*.
 \label{eq:free-domain-hyp}
\end{equation}
For a fixed observable $f$ as above, all $L\ge L_0\ge2\pi$ with
$L>2R_f$, $m\ge8$ and $K=2\pi m/L\ge8$, the truncated-kernel grid
state evolved from $P'\ket{\varphi_{0,L}}$ obeys
\begin{equation}
 \left|\langle\psi_\infty(t)|f(\bm{x}_a)|\psi_\infty(t)\rangle
       -\langle\psi_g(t)|O_g|\psi_g(t)\rangle\right|
 \le C(K^{-5}+tK^{-3})+C(1+t)\epsilon_{\rm leak}.
 \label{eq:free-dyn}
\end{equation}
The constant depends only on the fixed system, a uniform upper bound
on $V_\theta$, the observable $f$, the cutoff bounds
$D_1,D_2^{\rm cut}$, the domain bound $D_*$ and $\|\ket{\psi_0}\|$,
not on $L,K,t$ or $\epsilon_{\rm leak}$. In particular, the
interaction profile may vary with $L$ while these bounds remain fixed.
The same conclusion holds for the admissible
reference-kernel class described after Eq.~\eqref{eq:free-grid-count}.
The scalar formulation here extends to finite internal spin as in
Theorem~\ref{thm:free-dyn-spin}, with Hilbert norms and spin-summed leakage.
Corollaries~\ref{cor:free-dyn-nbo} and~\ref{cor:free-dyn-bo} supply
(H$_{\rm dom}$) for their specified cutoffs. No free-space propagator
construction is assumed to be part of this error proof: it applies to
the specified strong solution.
\end{proposition}
\begin{proofx}
We first compare free space with the exact cell evolution. The norm of
the supplied strong solution is conserved: reality of the Hamiltonian
form gives
\begin{equation}
 \frac{d}{ds}\|\ket{\psi_\infty(s)}\|^2
 =2\operatorname{Re}\langle\psi_\infty(s)|-iH_\infty|\psi_\infty(s)\rangle=0.
\end{equation}
Write $\ket{\varphi(s)}=\mathcal T_L(X_L\ket{\psi_\infty(s)})$ and
$\ket{u_L(s)}=e^{-isH_L}\ket{\varphi_{0,L}}$. The first vector is obtained by
cutting the free solution at each time; the second evolves the cut
initial state with the cell Hamiltonian. They agree at time zero.
On the support of $X_L$ the two physical potentials coincide, so
the only source in their evolution difference comes from derivatives
of the cutoff:
\begin{equation}
 i\partial_s\ket{\varphi}=H_L\ket{\varphi}+
       \mathcal T_L([X_L,T]\ket{\psi_\infty(s)}),\qquad
 [X_L,T]=\sum_j\frac1{m_j}
  \left(\nabla_jX_L\cdot\nabla_j+\tfrac12\Delta_jX_L\right).
 \label{eq:free-cutoff-duhamel}
\end{equation}
The first term uses the gradient leakage on $S_L$. The second uses
wavefunction leakage there, since $S_L$ lies outside the inner region.
The derivative bounds and $m_j\ge1$ therefore give
$\|[X_L,T]\ket{\psi_\infty(s)}\|\le C_{\rm cut}\epsilon_{\rm leak}$,
where $C_{\rm cut}=3d(D_1+D_2^{\rm cut}/2)$ is sufficient.
Integrating the source with the unitary cell propagator yields
\begin{equation}
 \ket{\varphi(t)}-\ket{u_L(t)}
 =-i\int_0^t e^{-i(t-s)H_L}
       \mathcal T_L([X_L,T]\ket{\psi_\infty(s)})\,ds,
 \qquad
 \|\ket{\varphi(t)}-\ket{u_L(t)}\|\le C_{\rm cut}t\epsilon_{\rm leak}.
\end{equation}
This is Duhamel's comparison formula: the state difference is the
accumulated cutoff source. Using $H_L^{(0)}$ instead would add the
common phase $e^{-itC_{\rm zm}(L)}$, which cancels from expectations.

We next account for measuring the cut state rather than the free state.
Although $f$ has one-particle support, another particle may be outside
the inner region; we therefore cannot assume $fX_L=f$. Instead,
the multiplication operators commute, $0\le f\le1$, and $X_L=1$
on $\mathcal B_L^{\rm one}$. Consequently
\begin{equation}
 |\langle\psi_\infty|f|\psi_\infty\rangle
       -\langle\varphi|f|\varphi\rangle|
 =|\langle\psi_\infty|f(1-X_L^2)|\psi_\infty\rangle|
 \le\|\mathbf1_{(\mathcal B_L^{\rm one})^c}\ket{\psi_\infty}\|^2
 \le\epsilon_{\rm leak}^2.
\end{equation}
Both $\|\ket{\varphi(t)}\|$ and $\|\ket{u_L(t)}\|$ are at most $\|\ket{\psi_0}\|$.
The usual two-term expansion of a bounded-observable expectation
therefore bounds the cost of replacing $\ket{\varphi(t)}$ by $\ket{u_L(t)}$
by $2\|\ket{\psi_0}\|C_{\rm cut}t\epsilon_{\rm leak}$.

It remains to compare the exact cell evolution with the grid evolution.
The reference-kernel version of Theorem~\ref{thm:grid-dyn-sharp}
applies with the uniform operator bounds of
Eq.~\eqref{eq:free-physical-relative}, the observable bounds of
Eq.~\eqref{eq:free-observable-uniform}, and the supplied budget
(H$_{\rm dom}$). It gives $C(K^{-5}+tK^{-3})$ with one constant
for all admissible $L$. The initial state need not be normalised:
the comparison is quadratic in the state, so rescaling the state
rescales its error and the corresponding graph-norm and leakage bounds consistently.
We use the actual projected initial state $P'\ket{\varphi_{0,L}}$ without
renormalisation. Adding the two spatial costs and using
$\epsilon_{\rm leak}^2\le\epsilon_{\rm leak}$ proves
Eq.~\eqref{eq:free-dyn}.
\end{proofx}

\subsection{Kinetic energy on the isolated-system grid}
\label{sec:free-kinetic-log}

We now extend the logarithmic kinetic-energy bound to the isolated
system. There are two points to establish. First, the cell-to-grid
estimate of Theorem~\ref{thm:kinetic-log} must have a constant uniform
in the cell size when we use the range-cut reference interaction.
Second, the spatial cutoff and free-space-to-cell comparison must
control an unbounded observable. The existing leakage assumptions
suffice for the second task because the states being compared have
bounded kinetic graph norms. We do not need a small kinetic graph
norm for their difference.

We retain the free-space Hamiltonian of Eq.~\eqref{eq:free-H}, with
fixed masses $m_j\ge1$, charges, and any clamped nuclei in
$B_{R_0}(0)$. Let $\ket{\psi_\infty(s)}$ be the strong solution with
$\ket{\psi_0}\in D(H_\infty^2)$, and write
\begin{equation}
 \mathcal D_\infty
 :=\norm{\ket{\psi_0}}+\norm{H_\infty\ket{\psi_0}}
                    +\norm{H_\infty^2\ket{\psi_0}}.
 \label{eq:free-kinetic-budget}
\end{equation}
The solution may be supplied directly, as in
Proposition~\ref{prop:free-dyn}. The reference cell interaction is
the range-cut $g_L$ of Eq.~\eqref{eq:rangecut}, with fixed $\rho$,
$0\le\theta_{L,\rho}\le1$, and a uniform bound on $V_\theta$.
The physical cell Hamiltonian $H_L$ includes its zero mode.

Choose a real $C^2$ cutoff $0\le X_L\le1$, compactly supported in
the retained region of Eq.~\eqref{eq:free-retained-region} and equal
to one on $\mathcal B_L^{\rm one}$. Its first and pure second
coordinate derivatives have the uniform bounds $D_1,D_2^{\rm cut}$
of Proposition~\ref{prop:free-dyn}, and vanish off a shell
$S_L\subset(\mathcal B_L^{\rm one})^c$. Preserve any imposed
exchange sector. We assume precisely the leakage conditions
Eq.~\eqref{eq:free-leakage} throughout the comparison interval and
the cutoff-domain condition Eq.~\eqref{eq:free-domain-hyp}, with
\begin{equation}
 \ket{\varphi_{0,L}}:=\mathcal T_L(X_L\ket{\psi_0}),\qquad
 \sup_{L\ge L_0}\sum_{j=0}^2\norm{H_L^j\ket{\varphi_{0,L}}}\le D_*.
\end{equation}
No smooth position observable, compact support of an observable, or
condition (H$\chi$) is imposed in this kinetic-energy comparison.

For clarity, the allowed grid kernels are specified relative to
$g_L$, rather than to the zero-mean periodic Coulomb kernel.
Let $\widetilde g_{L,K}$ have real, even Fourier coefficients, preserve
$\widehat g_L(0)$, agree with $\widehat g_L(\bm{\nu})$ for
$0<\|\bm{k}_{\bm{\nu}}\|\le K$, and vanish for $\|\bm{k}_{\bm{\nu}}\|_\infty>2K$. Require
\begin{equation}
 |\widehat{\widetilde g}_{L,K}(\bm{\nu})|
 \le\frac{4\pi(1+V_\theta)}{L^3\|\bm{k}_{\bm{\nu}}\|^2},
 \qquad \bm{\nu}\ne0.
 \label{eq:free-kinetic-kernel}
\end{equation}
The sharp transfer truncation and the permitted roll-offs satisfy
these conditions. Translated nuclear terms include their Fourier
phases. Use the padded register of Eq.~\eqref{eq:free-grid-count},
with $K=2\pi m/L$, register cutoff $2K$, and projector $\Pp$.
The resulting grid Hamiltonian is denoted $\Hgrid$; its kinetic term
is the exact Fourier multiplier. We compare with
$\ket{\psi_g(s)}=e^{-i\Hgrid s}\Pp\ket{\varphi_{0,L}}$, without renormalising
either the spatial cutoff or its momentum projection.

\begin{proposition}[Logarithmic kinetic-energy error for the isolated grid]
\label{prop:free-kinetic-log}
Under the preceding system, kernel, cutoff, leakage and domain
hypotheses, there are constants $C,L_0,K_0$, independent of $L,m,t$,
such that for $L\ge L_0$, $m\ge8$, $K=2\pi m/L\ge K_0$ and
every particle $j$,
\begin{align}
 &\left|\langle\psi_\infty(t)|T_j|\psi_\infty(t)\rangle
             -\langle\psi_g(t)|T_j|\psi_g(t)\rangle\right|\\
 &\quad\le C D_*^2 K^{-3}\bigl[1+t+\log(2+K)\bigr]
       +C(\mathcal D_\infty+D_*)(1+t)\epsilon_{\rm leak}.
 \label{eq:free-kinetic-log}
\end{align}
The estimate holds for every $t\ge0$ for which
Eq.~\eqref{eq:free-leakage} holds throughout $[0,t]$.
Constants and thresholds depend on the fixed physical system,
kernel-profile and cutoff bounds, but not on the initial state, $L,K,t$
or $\epsilon_{\rm leak}$. The same statement holds in finite
internal-spin Hilbert space, with the spin-summed leakage conditions
and no spin-dimension factor.
\end{proposition}

\begin{proofx}
We use the same free-space--cell--grid comparison as for the bounded
observable, but must now control the unbounded multiplier
$T_j=-\Delta_j/(2m_j)$. We insert the exact cell state
$\ket{u_L(t)}:=e^{-iH_Lt}\ket{\varphi_{0,L}}$ between the free and grid
expectations. We first control the spatial localisation error, then
verify that the spectral discretisation proof applies uniformly to
this family of cell operators.

\emph{Spatial comparison.}
The useful replacement for a bounded-observable estimate is a pairing
with the individual kinetic images. For $\ket{u},\ket{v}\in D(T_j)$ in the same
Hilbert space, self-adjointness gives
\begin{align}
 \langle u|T_j|u\rangle-\langle v|T_j|v\rangle
 &=(\langle u|-\langle v|)T_j\ket{u}+\langle v|T_j(\ket{u}-\ket{v}),\\
 \left|\langle u|T_j|u\rangle-\langle v|T_j|v\rangle\right|
 &\le\norm{\ket{u}-\ket{v}}\bigl(\norm{T_j\ket{u}}+\norm{T_j\ket{v}}\bigr).
 \label{eq:free-kinetic-pairing}
\end{align}
Unlike an operator-norm bound for $T_j$, this estimate uses its action
only on the two states being compared. We will bound those kinetic
norms globally and use the leakage assumptions only to make their
state-norm difference small.

We first obtain the two individual kinetic bounds needed in this pairing.
For Coulomb Hamiltonians, the graph norm $\|\ket{u}\|+\|H_\infty \ket{u}\|$
controls $\|T\ket{u}\|$ and $\|\nabla u\|$; this follows from the
relative potential bound and absorption, as in the regularity lemmas
below. Conservation of the Hamiltonian graph norm therefore gives
\begin{equation}
 \norm{\ket{\psi_\infty(s)}}+\norm{\nabla\psi_\infty(s,\cdot)}
       +\norm{T\ket{\psi_\infty(s)}}
 \le C\mathcal D_\infty.
 \label{eq:free-kinetic-firstgraph}
\end{equation}
This remains valid for a supplied strong solution. To see the
conservation directly, let
$F_M=\ind{|H_\infty|\le M}$ be a bounded spectral filter.
The vector $H_\infty F_M\ket{\psi_\infty(s)}$ satisfies the same
Schr\"odinger equation, with a bounded Hamiltonian on its spectral
support. Differentiating its squared norm and using self-adjointness
shows that $\norm{H_\infty F_M\ket{\psi_\infty(s)}}$ is constant.
Letting $M$ increase gives conservation of
$\norm{H_\infty\ket{\psi_\infty(s)}}$. The Coulomb graph bound then supplies
Eq.~\eqref{eq:free-kinetic-firstgraph}; no regularity assumption beyond
the stated initial domain has been introduced.

At a fixed time write $\ket{u}=\ket{\psi_\infty(t)}$. The product rule gives
\begin{equation}
 T_jX_L=X_LT_j-\frac1{m_j}\nabla_jX_L\cdot\nabla_j
                       -\frac1{2m_j}\Delta_jX_L.
 \label{eq:free-kinetic-product}
\end{equation}
The uniform first and second derivative bounds on $X_L$, together
with Eq.~\eqref{eq:free-kinetic-firstgraph}, therefore give
$\norm{T_j\ket{u}}+\norm{T_j(X_L\ket{u})}\le C\mathcal D_\infty$.
Since $X_L=1$ on the inner region,
$\norm{\ket{u}-X_L\ket{u}}\le\epsilon_{\rm leak}$.
Applying Eq.~\eqref{eq:free-kinetic-pairing} in free space shows that
cutting off the state changes its kinetic expectation by at most
$C\mathcal D_\infty\epsilon_{\rm leak}$. In particular, we do not
require a small gradient throughout the exterior region: the only
small exterior quantity used in this pairing is the wavefunction norm.

Next let $\ket{\varphi(s)}=\mathcal T_L(X_L\ket{\psi_\infty(s)})$.
Compact support inside the cell makes transplantation preserve the
kinetic expectation and the kinetic graph norm. Equation
\eqref{eq:free-cutoff-duhamel} and the shell leakage give
\begin{equation}
 \norm{\ket{\varphi(t)}-\ket{u_L(t)}}
 \le\int_0^t\norm{[X_L,T]\ket{\psi_\infty(s)}}\,ds
 \le C t\epsilon_{\rm leak}.
\end{equation}
Here the term containing $\nabla X_L$ uses the gradient leakage
on $S_L$, and the term containing $\Delta X_L$ uses wavefunction
leakage there. The product rule above bounds
$\norm{T_j\ket{\varphi(t)}}$ by $C\mathcal D_\infty$.
The cell graph equivalence and conservation of its Hamiltonian norm
give $\norm{T_j\ket{u_L(t)}}\le CD_*$, uniformly in $L$.
We may therefore apply Eq.~\eqref{eq:free-kinetic-pairing} on the
torus. Adding the preceding cutoff cost yields
\begin{equation}
 \left|\langle\psi_\infty(t)|T_j|\psi_\infty(t)\rangle
                    -\langle u_L(t)|T_j|u_L(t)\rangle\right|
 \le C(\mathcal D_\infty+D_*)(1+t)\epsilon_{\rm leak}.
 \label{eq:free-kinetic-cell}
\end{equation}
This establishes the free-space comparison using exactly the existing
wavefunction and shell-gradient leakage assumptions.

\emph{Uniform cell-to-grid inputs.}
The spatial estimate is complete. We now verify that every input to
the logarithmic cell-to-grid proof has a constant independent of $L$.
It is not enough
to apply a fixed-cell theorem and leave its constant dependent on $L$.
The range-cut coefficients satisfy the same Coulomb majorant as
Eq.~\eqref{eq:free-kinetic-kernel}. Lemma~\ref{lem:scaling} gives,
at every physical threshold $J\ge1$,
\begin{align}
 \sum_{0<\|\bm{k}_{\bm{\nu}}\|\le J}|\widehat g_L(\bm{\nu})|&\le CJ,\\
 \left(L^3\sum_{\|\bm{k}_{\bm{\nu}}\|>J}|\widehat g_L(\bm{\nu})|^2\right)^{1/2}
       &\le CJ^{-1/2},&
 L^{-3}\sum_{\bm{\nu}}(1+\|\bm{k}_{\bm{\nu}}\|^2)^{-2}&\le C.
 \label{eq:free-kinetic-volume-inputs}
\end{align}
The same bounds hold for every coefficient array dominated by the
majorant. In particular, $\widehat g_L-\widehat{\widetilde g}_{L,K}$
vanishes below $K$, including at zero, and is bounded by twice that
majorant. Its coefficient $\ell^2$ norm is
$O(L^{-3/2}K^{-1/2})$. On each three-dimensional interaction slice,
weighted Cauchy--Schwarz bounds the state coefficient $\ell^1$ norm
by $O(L^{3/2})$ times its physical kinetic graph norm. Their product is
$O(L^{-3/2}K^{-1/2})\,O(L^{3/2})=O(K^{-1/2})$.
The dropped-transfer norm and bilinear bounds consequently
retain their $K^{-1/2}$ prefactor with no volume dependence. The same
calculation applies to a one-register nuclear term; its translation
phase does not change a coefficient magnitude.

The wrap terms, which arise when a Fourier transfer is reduced modulo
the register width, require the same volume cancellation and a support
argument. Their estimates are uniform for two related reasons. The number of
unfolded coefficients mapped to any retained coefficient is bounded
independently of $L$ and the register size. Moreover, taking a centred
representative cannot increase a momentum component, so a folded output
above a physical threshold must come from an unfolded coefficient above
the same threshold. Thus both ordinary and threshold-restricted folding
preserve the tail bounds. For the wrap part with two high-momentum
endpoints, the absolute coefficient sum in
Eq.~\eqref{eq:free-kinetic-volume-inputs} is $O(K)$, which is the
uniform Schur bound needed in Eq.~\eqref{eq:Rform}.

These coefficient estimates and the physical relative bounds
Eq.~\eqref{eq:free-physical-relative} give a common fixed shift $c$
such that
\begin{equation}
 \mathsf H_L:=H_L+c\ge1,\qquad
 \mathsf H_g:=\Hgrid+c\ge1.
\end{equation}
Writing $\mathcal E:=1+T$ in the remainder of this uniformity
calculation, we have
\begin{equation}
 \norm{\mathsf H_L\ket{u}}\asymp\norm{\mathcal E \ket{u}},\qquad
 \norm{\mathsf H_g\ket{v}}\asymp\norm{\mathcal E \ket{v}},
 \label{eq:free-kinetic-graphs}
\end{equation}
with constants independent of $L,K$. The zero-mode contribution is
$O(L^{-1})$ and therefore uniformly bounded for $L\ge L_0$.
It is represented equally by $H_L$ and $\Hgrid$, so it cancels from
the consistency residual
$\mathcal C_K:=\Hgrid\Pp-\Pp H_L$.
Omitting this common scalar consistently would change the two
evolutions only by scalar phases and would leave their kinetic
expectations unchanged.

\emph{Momentum tails throughout evolution.}
We next turn the coefficient bounds into the regularity estimates used
by the spectral comparison. Let
$Q_{r,J}=\ind{\|\bm{k}_{\bm{p}_r}\|_\infty>J}$ project onto Fourier
configurations in which particle register $r$ has a momentum
component larger than $J$. Here $\bm{p}_r\in\mathbb Z^3$ is the Fourier
index in that register, and $\bm{k}_{\bm{p}_r}=g\bm{p}_r$ is its physical momentum.
Coefficient domination and the physical slice estimate give
\begin{equation}
 \norm{Q_{r,J}V_{{\rm tot},L}\ket{u}}
 \le CJ^{-1/2}\norm{\mathcal E \ket{u}},
\end{equation}
and the same bound for the grid potential. Here
$V_{{\rm tot},L}=H_L-T$ and the grid potential is
$\Hgrid-T|_{\ran\Pp}$.
For an initial state in $D(H_L^2)$, the dropped, folded and exterior
terms can be paired with an arbitrary unit grid vector. Its first
kinetic graph norm is at most $CK^2$ because the register is finite,
while the initial state has an $O(K^{-5/2})$ momentum tail. The form bounds
therefore give, for the initial state used here,
$\norm{\mathcal C_K\ket{\varphi_{0,L}}}\le CD_*K^{-1/2}$.
This residual estimate depends on the initial state and is required
for the grid state; Lemma~\ref{lem:kinetic-strong} gives the same
argument in the periodic notation.
For the exact cell state we use
$T\ket{u_L}=H_L\ket{u_L}-V_{{\rm tot},L}\ket{u_L}$:
$H_L\ket{u_L}$ has a uniformly bounded first graph norm by the initial
second graph budget $D_*$, so its momentum tail is $O(D_*J^{-2})$.
The potential term has the larger $O(D_*J^{-1/2})$ tail.

On the grid, we must not assume a uniform second discrete graph norm
for the projected initial state. Instead we split
\begin{equation}
 \Hgrid\ket{\psi_g(s)}
 =e^{-i\Hgrid s}\Pp H_L\ket{\varphi_{0,L}}
       +e^{-i\Hgrid s}\mathcal C_K\ket{\varphi_{0,L}}.
\end{equation}
The first term has first graph norm at most $CD_*$, by
Eq.~\eqref{eq:free-kinetic-graphs} and conservation. The second has
norm at most $CD_*K^{-1/2}$ by the preceding residual estimate. Restricting
$T\ket{\psi_g}=\Hgrid\ket{\psi_g}-V_{\rm grid}\ket{\psi_g}$ to $Q_{r,J}$ therefore
gives a bound
$CD_*(J^{-2}+J^{-1/2}+K^{-1/2})$ for the weighted tail.
For $1\le J\le K$ this is at most $CD_*J^{-1/2}$. We have obtained
\begin{equation}
 \norm{Q_{r,J}\mathcal E \ket{u_L(s)}}\le CD_*J^{-1/2},\qquad
 \norm{Q_{r,J}\mathcal E\ket{\psi_g(s)}}\le CD_*J^{-1/2},
 \label{eq:free-kinetic-tails}
\end{equation}
with the first estimate valid for all $J\ge1$ and the second for
$1\le J\le K$. The final finite grid shell is controlled at $J=K$.
On the range of $Q_{r,J}$, the kinetic weight is at least a constant
times $J^2$. Thus the continuum estimate also gives
$\|Q_{r,J}\ket{u_L(s)}\|\le CD_*J^{-5/2}$ by division by this weight.
The distinction matters: Eq.~\eqref{eq:free-kinetic-tails} controls
a weighted tail, whereas this last bound controls the state norm
of the discarded component. These are the estimates for the evolved states used
in Theorem~\ref{thm:kinetic-log}.

\emph{Comparison below a physical energy cutoff.}
We now match the two available Hamiltonian derivatives in the initial
state to the inverse Hamiltonian weights in the consistency estimate.
Introduce spectral projectors
$S=\ind{\mathsf H_L\le\Lambda}$ and
$S_g=\ind{\mathsf H_g\le2\Lambda}$, with $\Lambda=aK^2$.
They select low Hamiltonian energies for the proof; they are not
additional operations in the simulation. The grid threshold is twice
the cell threshold so that the complementary grid energies are
separated from the retained cell energies.

We choose $a$ small enough that a low Hamiltonian energy cannot hide
a large kinetic tail. For a fixed-fraction momentum projector
$Q=Q_{r,\theta K}$, with fixed $\theta>0$, the kinetic energy on
its range is at least $bK^2$ for a fixed $b>0$. Put
$B=\mathsf H_L^{-1}S$. Graph equivalence gives
$\|\mathcal E B\|\le C$, so the potential-tail estimate gives
$\|QV_{{\rm tot},L}B\|\le CK^{-1/2}$. Since $Q$ commutes with
$T$, we have
\begin{equation}
 bK^2\|QB\|\le\|QTB\|
 \le(\Lambda+c)\|QB\|+CK^{-1/2}.
\end{equation}
In the last step we used $H_LB=BH_LS$; we did not commute $Q$
with $H_L$. The grid calculation has $2\Lambda+c$ in place of
$\Lambda+c$. Choose $a$ so that $2a<b/2$ and increase $K_0$
until $c\le bK^2/4$. Absorbing these terms proves
$\|Q\mathsf H_L^{-1}S\|+\|Q\mathsf H_g^{-1}S_g\|
\le CK^{-5/2}$. This choice uses only physical scales and is
uniform in $L$.
Testing the residual against arbitrary unit grid vectors first gives
$\norm{\mathcal C_K\mathsf H_L^{-1}S}\le CK^{-1/2}$.
Restricting and inverse-weighting the test vector as well gives
\begin{equation}
 \norm{S_g\mathsf H_g^{-1}\mathcal C_K
                     \mathsf H_L^{-1}S}\le CK^{-3}.
 \label{eq:free-kinetic-low-consistency}
\end{equation}
The dropped and weak-wrap pairings combine $K^{-1/2}$ with
$K^{-5/2}$; the two-high-endpoint wrap costs $K(K^{-5/2})^2$.
The exterior term pairs the potential tail with the omitted state
tail, again giving $K^{-3}$. Low-transfer agreement is with $g_L$
itself throughout this argument. The proof uses its coefficient
majorant, not an equality between $g_L$ and the periodic Coulomb
Fourier kernel.

To apply this consistency estimate to evolution, define the two full
register vectors
\begin{equation}
 \ket{x(s)}=\Pp e^{-iH_Ls}S\ket{\varphi_{0,L}},\qquad
 \ket{y(s)}=e^{-i\Hgrid s}\Pp S\ket{\varphi_{0,L}},
\end{equation}
which have the same initial vector. Their low grid-energy difference
satisfies
\begin{equation}
 S_g(\ket{x(t)}-\ket{y(t)})
 =i\int_0^t e^{-i\Hgrid(t-s)}S_g\mathcal C_K
                      e^{-iH_Ls}S\ket{\varphi_{0,L}}\,ds.
\end{equation}
The two inverse weights in Eq.~\eqref{eq:free-kinetic-low-consistency}
match the initial regularity through
$S\ket{\varphi_{0,L}}=\mathsf H_L^{-2}
(S\mathsf H_L^2\ket{\varphi_{0,L}})$, where the vector in parentheses
has norm at most $CD_*$. Lemma~\ref{lem:kinetic-dyadic} therefore
gives
\begin{equation}
 \|S_g(\ket{x(t)}-\ket{y(t)})\|
 \le CD_*K^{-3}[t+\log(2+K)].
\end{equation}
More explicitly, we divide the positive Hamiltonian energies into
bands $[2^j,2^{j+1})$. On neighbouring bands the output/input energy
ratio is bounded, so direct integration gives the term proportional
to $t$. On separated bands the relative oscillation supplies the
inverse energy separation, and summing these contributions gives
the logarithm. There are $O(\log(2+K))$ bands below the physical
ceiling $2\Lambda=O(K^2)$, regardless of how many modes or
eigenvalues lie within a band. Thus the dyadic estimate introduces
no factor depending on the cell volume or a spectral gap.

\emph{The complementary grid energies.}
The separated-spectrum estimate of Lemma~\ref{lem:kinetic-leakage}
uses the gap between energies at most $\Lambda$ and energies greater
than $2\Lambda$. Together with the one-sided residual bound
$\|\mathcal C_K\mathsf H_L^{-1}S\|\le CK^{-1/2}$, it gives
$\norm{(1-S_g)(\ket{x(s)}-\ket{y(s)})}\le CD_*K^{-5/2}$.
This is a comparison of the high-energy parts of the full register
vectors, not an assumption about the momentum tails of $S_g\ket{x}$ or
$S_g\ket{y}$.
The initial state $S\ket{\varphi_{0,L}}$ still has a second graph norm bounded by
a fixed multiple of $D_*$. We can therefore apply
Eq.~\eqref{eq:free-kinetic-tails} to both evolved states and repeat
the shell calculation of Lemma~\ref{lem:kinetic-high}. On a shell
of physical momentum scale $J$, the additional kinetic multiplier
contributes at most $CJ^2$, so its squared contribution to
$\norm{\mathcal E T_j\ket{w}}^2$ is at most $CD_*^2J^3$.
For clarity, these are momentum shells, distinct from the energy
bands used in Duhamel's formula. Squared norms of disjoint momentum
shells add because $\mathcal E T_j$ is diagonal in momentum.
The low shell is controlled by the first graph norm, and the
remaining shells give $\sum_{2^j\le K}(2^j)^3\le CK^3$.
The top shell up to the register cutoff $2K$ is controlled by the
tail at a threshold between $K/2$ and $K$, so we do not require
a grid tail estimate beyond its stated range. Taking square roots gives
\begin{equation}
 \norm{\mathsf H_gT_j\ket{w(s)}}
 \le C\norm{\mathcal E T_j\ket{w(s)}}\le CD_*K^{3/2},
 \qquad \ket{w}\in\{\ket{x},\ket{y}\}.
\end{equation}
All shells begin at a fixed physical threshold, such as one. The
increasing number of lower-momentum modes in a growing cell is
controlled by the first graph norm, so neither this sum nor graph
equivalence introduces an additional factor of $L$.

We now pair $\ket{x}-\ket{y}$ directly with $T_j\ket{x}$ and $T_j\ket{y}$, as in
Eq.~\eqref{eq:kinetic-grid-pairing}. On the high grid-energy space,
\begin{equation}
 \norm{(1-S_g)T_j\ket{w}}\le(2\Lambda)^{-1}\norm{\mathsf H_gT_j\ket{w}}
 \le CD_*K^{-1/2}.
\end{equation}
Together with the $CD_*K^{-5/2}$ state difference, this gives an
$O(D_*^2K^{-3})$ contribution. The low-energy part supplies the
stated logarithmic dynamical term. This argument uses the momentum
tails of the full register states and does not assume that
$S_g$ preserves those tails or commutes with $T_j$.

\emph{Restoring the omitted components.}
We finally account for the two projections used in the proof. Since
$P'$ commutes with $T_j$, the error from projecting the exact filtered
state has no cross terms. It is bounded by the product of its
ordinary and weighted omitted tails,
$CD_*K^{-5/2}\,CD_*K^{-1/2}=CD_*^2K^{-3}$.
For the initial energy filter, the second Hamiltonian budget gives
\begin{equation}
 \|(1-S)\ket{\varphi_{0,L}}\|
 \le\Lambda^{-2}\|\mathsf H_L^2\ket{\varphi_{0,L}}\|
 \le CD_*K^{-4}.
\end{equation}
Both the filtered and unfiltered evolutions have first graph norms
at most $CD_*$. Unitarity and Eq.~\eqref{eq:free-kinetic-pairing}
therefore show that removing this initial spectral filter costs
$CD_*^2K^{-4}$ in each expectation.
The assembly in Theorem~\ref{thm:kinetic-log} consequently gives
\begin{equation}
 \left|\langle u_L(t)|T_j|u_L(t)\rangle
                    -\langle\psi_g(t)|T_j|\psi_g(t)\rangle\right|
 \le CD_*^2K^{-3}[1+t+\log(2+K)]
 \label{eq:free-kinetic-grid}
\end{equation}
with uniform constants. Adding Eq.~\eqref{eq:free-kinetic-cell}
proves Eq.~\eqref{eq:free-kinetic-log}. All steps use Hilbert norms
and spatial operators acting as the identity on spin. They therefore
hold in the finite Hilbert direct sum under spin-summed leakage, with
no spin-dimension factor.
\end{proofx}

The cutoffs in Corollaries~\ref{cor:free-dyn-nbo} and
\ref{cor:free-dyn-bo} give $D_*\le C\mathcal D_\infty$.
The explicit BO product cutoff retains its equal-moving-mass
restriction. For a general admissible cutoff, including unequal
moving masses with fixed centres, the condition
Eq.~\eqref{eq:free-domain-hyp} remains unless it is supplied by a
compatible domain-transfer construction.
The different powers of the graph-norm bounds in
Eq.~\eqref{eq:free-kinetic-log} are intentional:
$\epsilon_{\rm leak}$ bounds absolute state norms, so the spatial
comparison is linear in the graph budget and in the leakage norm.

For basis counting, Eq.~\eqref{eq:free-grid-count} gives
$N=(4m+1)^3=\Theta((LK)^3)$ for sufficiently large admissible
cutoffs. Thus, for fixed graph-norm bounds, the kinetic bound reads
\begin{equation}
 \mathcal O\!\left(
 \frac{L^3}{N}
 \left[1+t+\log\left(2+\frac{N^{1/3}}{L}\right)\right]
 +(1+t)\epsilon_{\rm leak}\right).
 \label{eq:free-kinetic-N}
\end{equation}
At fixed cell size and fixed time, the basis term is
$\mathcal O(\log N/N)$. The factor $L^3$ must be retained when the
cell also grows, and the leakage bound remains a separate input.

\begin{corollary}[Isolated-system finite-window stopping estimate]
\label{cor:free-stopping-log}
Under Proposition~\ref{prop:free-kinetic-log}, assume the leakage
bounds throughout $[0,\tau]$, where $\tau>0$, and fix a reference
speed $v_j>0$. Compare the signed secant
$[\langle T_j\rangle(\tau)-\langle T_j\rangle(0)]/(v_j\tau)$
for the exact free-space and grid evolutions. Its error satisfies
\begin{align}
 |\Delta\widehat S_\tau|
 &\le \frac{CD_*^2}{v_j}K^{-3}
       \left[1+\frac{1+\log(2+K)}{\tau}\right]
       +\frac{C(\mathcal D_\infty+D_*)}{v_j}
                      (1+\tau^{-1})\epsilon_{\rm leak}.
 \label{eq:free-stopping-log}
\end{align}
\end{corollary}
\begin{proofx}
Apply Proposition~\ref{prop:free-kinetic-log} at $t=0$ and
$t=\tau$. Adding these bounds and dividing by $v_j\tau$ gives a
basis contribution proportional to
$K^{-3}[1+2(1+\log(2+K))/\tau]$ and a leakage contribution
proportional to $(1+2/\tau)\epsilon_{\rm leak}$.
Absorbing the factors of two into the constant proves
Eq.~\eqref{eq:free-stopping-log}. This controls discretisation
of the stated secant; the modelling bias of a finite-window slope
is a separate issue.
\end{proofx}

\subsection{The cutoff-domain condition with all particles quantum}

The dynamical estimates are now established under leakage and
(H$_{\rm dom}$). We next discharge the domain condition for an explicit
cutoff when all particles are quantum. This requires more than a bound
on cutoff derivatives: applying the Hamiltonian twice differentiates
the Coulomb potential. We will choose the cutoff so that the dangerous
coefficient vanishes at a pair collision.

The following construction derives (H$_{\rm dom}$) from the initial
Hamiltonian graph norm in the non-Born--Oppenheimer case. Fixed Coulomb
centres are treated separately in Lemma~\ref{lem:free-bo-domain}, where
nuclear flatness is required in addition to pair cancellation. Neither
argument asserts domain preservation for arbitrary smooth cutoffs.

\begin{lemma}[Free-space regularity from two Hamiltonian relations]
\label{lem:free-regularity-bridge}
Suppose there are no clamped nuclei and $\ket{u},\ket{h},\ket{k}\in L^2(\R^{3d})$
satisfy $H_\infty \ket{u}=\ket{h}$, $H_\infty \ket{h}=\ket{k}$ in the weak Hamiltonian sense,
including $L^2$ potential images. Then $\ket{u},\ket{h}\in H^2$, the functions
$u/\|\bm{x}_i-\bm{x}_j\|$ and $\partial_a u/\|\bm{x}_i-\bm{x}_j\|$ belong to $L^2$, and
\begin{equation}
 \begin{split}
 \mathcal N_\infty:={}&\|\ket{u}\|+\|\ket{h}\|+\|\ket{k}\|
  +\sum_a\bigl(\|\partial_a u\|+\|\partial_a h\|\bigr)
  +\sum_{a,b}\|\partial_a\partial_b u\|
  +\sum_{i<j}\|u/\|\bm{x}_i-\bm{x}_j\|\|\\
 &\le C_{\rm sys}(\|\ket{u}\|+\|\ket{h}\|+\|\ket{k}\|).
 \end{split}
 \label{eq:free-regularity-bridge}
\end{equation}
Here $a,b$ are coordinate indices. There is no assumption of $H^4$
regularity and no $L$ in the constant.
\end{lemma}
\begin{proofx}
We first obtain derivatives from the weak kinetic equation, then bound
the potential by those derivatives. This order avoids assuming the
regularity that the estimate is meant to establish. If $T\ket{f}=\ket{z}$ with
$\ket{f},\ket{z}\in L^2$, mollification and integration by parts give
\begin{equation}
 \sum_a m_{j(a)}^{-1}\|\partial_a f\|^2\le2\|\ket{f}\|\|\ket{z}\|,
 \qquad
 \sum_{a,b}m_{j(a)}^{-1}m_{j(b)}^{-1}
     \|\partial_a\partial_b f\|^2=4\|\ket{z}\|^2.
\end{equation}
Here $j(a)$ is the particle to which coordinate $a$ belongs.
Equivalently, the identities follow from the kinetic Fourier
multiplier; mollification justifies them for the weak equation.
Apply this first to $\ket{f}=\ket{u}$, $\ket{z}=\ket{h}-V_\infty \ket{u}$. The assumed $L^2$
potential image makes $\ket{z}$ an $L^2$ vector before any estimate
on its size is used. The same reasoning applies to $\ket{f}=\ket{h}$.

We next control the Coulomb quotients. For a pair separation,
the regularised vector field
$\bm{F}_\delta=(\bm{x}_i-\bm{x}_j)/(\|\bm{x}_i-\bm{x}_j\|^2+\delta^2)$ obeys
$\operatorname{div}_i \bm{F}_\delta\ge
(\|\bm{x}_i-\bm{x}_j\|^2+\delta^2)^{-1}\ge\|\bm{F}_\delta\|^2$.
Integrating the divergence against $|f|^2$ and applying
Cauchy--Schwarz gives
\begin{equation}
 \int\frac{|f|^2}{\|\bm{x}_i-\bm{x}_j\|^2+\delta^2}
 \le 2\left(\int\frac{|f|^2}{\|\bm{x}_i-\bm{x}_j\|^2+\delta^2}\right)^{1/2}
       \left(\sum_{c=1}^3\|\partial_{i,c}f\|^2\right)^{1/2}.
\end{equation}
Approximation extends this inequality to $H^1$ functions. Squaring
and taking $\delta\downarrow0$ by monotone convergence yields
\begin{equation}
 \|f/\|\bm{x}_i-\bm{x}_j\|\|^2\le4\sum_{c=1}^3\|\partial_{i,c}f\|^2.
 \label{eq:free-Hardy}
\end{equation}

Let $M=\max_j m_j$ and $Q=\sum_{i<j}|q_iq_j|$ for this estimate.
The triangle inequality, Hardy's bound and the kinetic gradient
estimate above imply
$\|V_\infty \ket{f}\|^2\le8MQ^2\|\ket{f}\|\|\ket{z}\|$. In the relation
$\ket{z}=H_\infty \ket{f}-V_\infty \ket{f}$, use
$\sqrt{8MQ^2\|\ket{f}\|\|\ket{z}\|}\le\tfrac12\|\ket{z}\|+4MQ^2\|\ket{f}\|$
and absorb the first term. We obtain
\begin{equation}
 \|\ket{z}\|\le2\|H_\infty \ket{f}\|+8MQ^2\|\ket{f}\|.
\end{equation}
Applying this to $\ket{u}$ and $\ket{h}$ bounds their kinetic images, and hence
their first derivatives and the second derivatives of $\ket{u}$, by
$\|\ket{u}\|+\|\ket{h}\|+\|\ket{k}\|$. Apply Eq.~\eqref{eq:free-Hardy} once
more to $\partial_a u\in H^1$ to obtain
$(\partial_a u)/\|\bm{x}_i-\bm{x}_j\|\in L^2$. We do not differentiate
$u/\|\bm{x}_i-\bm{x}_j\|$, which would introduce a stronger singularity.
Summing the finitely many derivative and quotient estimates proves
Eq.~\eqref{eq:free-regularity-bridge}. For example, the formal proof
permits $C_{\rm sys}=1+2M(3+8MQ^2)(6d+9d^2+\binom d2)$.
The calculation has taken place entirely in free space, so no cell
size occurs in this constant.
\end{proofx}

We can now design the cutoff. The kinetic commutator contains
$m_j^{-1}\nabla_jX_L$, so a radius weighted by $m_j$ will cancel
these inverse masses when differentiated. This makes the difference
between the two particle coefficients proportional to $\bm{x}_i-\bm{x}_j$,
which is the factor needed at a pair collision.

\begin{lemma}[Mass-weighted cutoff and uniform squared-domain transfer]
\label{lem:free-nbo-domain}
Suppose all particles are quantum. Fix a smooth $F:\R\to[0,1]$
that is one on $(-\infty,1]$ and zero on $[2,\infty)$, and set
\begin{equation}
 s=L/8,\qquad Q_m(\vec{\bm{x}})=\sum_j m_j\|\bm{x}_j\|^2,\qquad
 X_L(\vec{\bm{x}})=F(Q_m(\vec{\bm{x}})/s^2).
 \label{eq:free-mass-cutoff}
\end{equation}
For $L\ge8$ the derivatives of $X_L$ through order four have uniform
bounds depending only on the masses and $F$. For sufficiently large $L$
its support is inside the retained region. For every
$\ket{u}\in D(H_\infty^2)$,
\begin{equation}
 \mathcal T_L(X_L\ket{u})\in D(H_L^2),\qquad
 \sup_{L\ge L_0}\sum_{j=0}^2\|H_L^j\mathcal T_L(X_L\ket{u})\|
 \le C_{\rm sys,F}\sum_{j=0}^2\|H_\infty^j\ket{u}\|.
 \label{eq:free-nbo-domain}
\end{equation}
\end{lemma}
\begin{proofx}
We must control two Hamiltonian applications after multiplication by
$X_L$. A bound on the derivatives of $X_L$ alone does not establish
this: commuting a derivative through a Coulomb potential produces an
inverse-square singularity. We chose the mass-weighted radius $Q_m$
so that the coefficient of that singularity vanishes linearly at each
pair collision. We first verify this cancellation, then use it to
justify the operator-domain claim.

On the transition region of $F(Q_m/s^2)$, every $\|\bm{x}_j\|$ is $O(s)$.
The chain rule therefore gives
$\|\partial^\alpha X_L\|_\infty\le C_\alpha s^{-|\alpha|}$ for
$1\le|\alpha|\le4$. Since $s=L/8\ge1$, these bounds are uniform
in $L$. Also $Q_m\le2s^2$ on the support and $m_j\ge1$, so
$\|\bm{x}_j\|\le\sqrt2s$ for every particle. For sufficiently large $L$,
this support is strictly inside the retained region.

The inverse masses in the kinetic energy explain the weights in
$Q_m$. Differentiating gives
$m_j^{-1}\nabla_jX_L=2F'(Q_m/s^2)\bm{x}_j/s^2$, and thus
\begin{equation}
 \frac{\nabla_iX_L}{m_i}-\frac{\nabla_jX_L}{m_j}
   =\frac{2F'(Q_m/s^2)}{s^2}(\bm{x}_i-\bm{x}_j).
 \label{eq:free-mass-cancellation}
\end{equation}
The difference vanishes at $\bm{x}_i=\bm{x}_j$ even when the two masses differ.
It therefore treats electron--nucleus and nucleus--nucleus collisions
as well as electron--electron collisions when all nuclei are quantum.

Put $B_L=[T,X_L]$. Explicitly,
\begin{equation}
 B_L=-\sum_j\frac1{m_j}
       \left(\nabla_jX_L\cdot\nabla_j
                       +\tfrac12\Delta_jX_L\right).
\end{equation}
This is a first-order differential operator. Commuting it with $T$
produces an operator $[T,B_L]$ of order at most two: its coefficients
contain derivatives of $X_L$ of orders two through four and fixed
inverse masses. These coefficients are uniformly bounded. For the
pair multiplier $V_{ij}=q_iq_j/\|\bm{x}_i-\bm{x}_j\|$, the other commutator is
\begin{equation}
 [V_{ij},B_L]
 =\left(\frac{\nabla_iX_L}{m_i}-\frac{\nabla_jX_L}{m_j}\right)
       \cdot\nabla_i V_{ij}
 =-\frac{2q_iq_jF'(Q_m/s^2)}{s^2\|\bm{x}_i-\bm{x}_j\|}.
\end{equation}
Equation~\eqref{eq:free-mass-cancellation} has reduced the apparent
inverse-square singularity to a Coulomb singularity. The Hardy
estimate in Lemma~\ref{lem:free-regularity-bridge} controls its
action on $\ket{u}$ in $L^2$.

To see the complete second image, commute one Hamiltonian application
at a time. We have $H_\infty X_L=X_LH_\infty+B_L$ and
$H_\infty B_L=B_LH_\infty+[T,B_L]+[V_\infty,B_L]$.
Combining these identities gives
\begin{equation}
 H_\infty^2(X_L\ket{u})=X_LH_\infty^2\ket{u}+2B_LH_\infty \ket{u}
                   +[T,B_L]\ket{u}+[V_\infty,B_L]\ket{u}
 \label{eq:free-second-commutator}
\end{equation}
in the weak sense. Each term on the right has a uniform $L^2$ bound
by $C\mathcal N_\infty$: the first uses $H_\infty^2\ket{u}\in L^2$;
the second uses the first derivatives of $H_\infty \ket{u}$; the third
uses the derivatives of $\ket{u}$ through order two; and the last uses
the Coulomb quotients of $\ket{u}$. These are precisely the quantities
supplied by Lemma~\ref{lem:free-regularity-bridge}.

We also need to show that this weak calculation identifies an actual
operator image. Replace each $r^{-1}$ by
$(r^2+\delta^2)^{-1/2}$, integrate by parts, and then take
$\delta\downarrow0$. For a pair separation $r=\|\bm{x}_i-\bm{x}_j\|$, the regularised commutator
contains the factor
\begin{equation}
 \left|\frac{2q_iq_jF'(Q_m/s^2)}{s^2}
             \frac{r^2}{(r^2+\delta^2)^{3/2}}\,u\right|
 \le C\frac{|u|}{r}.
\end{equation}
The right-hand side is in $L^2$ and is independent of $\delta$.
The other potential products are dominated by $C|u|/r$ or
$C|\partial_a u|/r$, which are also in $L^2$ by the regularity
bridge. Dominated convergence therefore justifies the limits in
the weak identities. We combine the two particle derivatives
before taking this limit, so that the cancellation is present
in the dominating function. The resulting identity
for $H_\infty B_L\ket{u}$ has an $L^2$ right-hand side. Moreover,
$B_L\ket{u}\in H^1$ because $\ket{u}\in H^2$ and the cutoff derivatives are
bounded. Hardy therefore gives $V_\infty B_L\ket{u}\in L^2$; subtracting
it from the weak Hamiltonian relation gives $TB_L\ket{u}\in L^2$.
Weak kinetic regularity now yields
$B_L\ket{u}\in H^2=D(H_\infty)$. Since $H_\infty \ket{u}\in H^2$,
we also have $X_LH_\infty \ket{u}\in H^2$. Thus
$H_\infty(X_L\ket{u})=X_LH_\infty \ket{u}+B_L\ket{u}$ lies in $D(H_\infty)$,
which proves $X_L\ket{u}\in D(H_\infty^2)$. No $H^4$ regularity of
the Coulomb state is needed.

Finally, we transfer this conclusion to the cell. Both $X_L\ket{u}$ and
its first Hamiltonian image are supported in the retained region,
where the free and physical cell potentials agree. Transplanting
these two weak Hamiltonian identities successively therefore gives
the first and second cell images, with the same $L^2$ norms.
In the convention omitting the zero mode, we subtract
$C_{\rm zm}(L)$ at each application; its magnitude is uniformly
bounded. One explicit bound on the three images for $H_L^{(0)}$ is
$\mathcal N_\infty$, $\Lambda\mathcal N_\infty$ and
$\Lambda^2\mathcal N_\infty$, where
\begin{equation}
 \Lambda=1+c_0+6(1+3d)^2D+C_{\rm can}\sum_{i<j}|q_iq_j|.
\end{equation}
Here $D$ bounds the cutoff derivatives, $C_{\rm can}$ bounds the
coefficient of $\bm{x}_i-\bm{x}_j$ in
Eq.~\eqref{eq:free-mass-cancellation}, and
$c_0\ge|C_{\rm zm}(L)|$. All are independent of $L$.
Returning to $H_L$ through Eq.~\eqref{eq:free-zero-mode} and
using the regularity bound for $\mathcal N_\infty$ proves
Eq.~\eqref{eq:free-nbo-domain}.
\end{proofx}

\begin{corollary}[Non-Born--Oppenheimer free-boundary dynamics]
\label{cor:free-dyn-nbo}
In the system with no clamped nuclei, let $\ket{\psi_\infty(s)}$ be the supplied
strong solution, $\ket{\psi_0}\in D(H_\infty^2)$, and fix $f$ as above.
Choose $X_L$ from Eq.~\eqref{eq:free-mass-cutoff} and take
\begin{equation}
 \mathcal B_L^{\rm one}=\{Q_m<(L/8)^2\},\qquad
 S_L=\{(L/8)^2\le Q_m\le2(L/8)^2\}.
 \label{eq:free-nbo-shell}
\end{equation}
For any fixed $\rho>0$ a sufficient threshold is
\begin{equation}
 L_0=\max\{8,2\pi,2R_f+1,16\rho\}.
 \label{eq:free-nbo-L0}
\end{equation}
There is one constant $C$ for every $L\ge L_0$, $m\ge8$, $K\ge8$,
and $t\ge0$ satisfying (H$_{\rm leak}$), for which
Eq.~\eqref{eq:free-dyn} holds. Its state dependence is only through
$\mathcal D_\infty=\|\ket{\psi_0}\|+\|H_\infty\ket{\psi_0}\|+\|H_\infty^2\ket{\psi_0}\|$.
The constant is uniform over admissible interaction profiles whose
total variations have a common fixed upper bound.
No independent cutoff-domain, weak-derivative or Hardy-quotient
assumption on the initial state is required.
\end{corollary}
\begin{proofx}
The chosen $L_0$ ensures $\sqrt2L/8\le3L/16\le L/4-\rho$,
so the support lies in the retained region (take $R_0=0$).
The shell is disjoint from the open inner region, and all first and
second cutoff derivatives vanish off it. The fixed derivative bounds
and Lemma~\ref{lem:free-nbo-domain} verify the hypotheses of
Proposition~\ref{prop:free-dyn}, uniformly in $L$.
\end{proofx}

To interpret the leakage assumption for this cutoff, let
$M=\sum_j m_j$ be the total mass and
$\bm{X}_{\rm cm}=M^{-1}\sum_j m_j \bm{x}_j$ the centre-of-mass position.
The identity
$Q_m=M\|\bm{X}_{\rm cm}\|^2+\sum_j m_j\|\bm{x}_j-\bm{X}_{\rm cm}\|^2$ shows that the
leakage hypotheses for this cutoff control centre-of-mass spreading
and internal separation together. No bound on leakage for arbitrary
initial states is inferred from $D(H_\infty^2)$ membership. Boundary occupancy of
the simulated state is not a certificate of the exact local
$H^1$ leakage.

\subsection{Cutoff-domain preservation with clamped nuclei}

With clamped nuclei, cancelling the quantum-pair singularities is only
part of the domain argument. A nuclear potential depends on one moving
coordinate and a fixed position, so there is no second moving-particle
derivative with which to combine its commutator. We control this term
by making the cutoff constant in that particle coordinate near each
nucleus. This is an additional condition on the cutoff, not additional
regularity of the initial state. We first extend the regularity
bridge and then choose a cutoff that is flat in each electron coordinate
near every nucleus.

\begin{lemma}[Free-space regularity with clamped nuclei]
\label{lem:free-nuclear-regularity}
For the Hamiltonian in Eq.~\eqref{eq:free-H}, including the fixed
nuclei, let $H_\infty \ket{u}=\ket{h}$ and $H_\infty \ket{h}=\ket{k}$ be the two weak
Hamiltonian relations of Lemma~\ref{lem:free-regularity-bridge}.
Then $\ket{u},\ket{h}\in H^2$, all pair quotients in that lemma belong to $L^2$,
and so do
\begin{equation}
 \frac{u}{\|\bm{x}_j-\bm{R}_\ell\|},\qquad
 \frac{\partial_a u}{\|\bm{x}_j-\bm{R}_\ell\|}
 \quad\text{for every }j,\ell,a.
\end{equation}
The budget $\mathcal N_\infty$ of Eq.~\eqref{eq:free-regularity-bridge}
satisfies
\begin{equation}
 \mathcal N_\infty\le C_{\rm reg}^{\rm nuc}
       (\|\ket{u}\|+\|\ket{h}\|+\|\ket{k}\|),\qquad
 C_{\rm reg}^{\rm nuc}
 =1+2M(3+8M\mathcal Q^2)\left(6d+9d^2+\binom d2\right),
 \label{eq:free-nuclear-regularity}
\end{equation}
where $M=\max_j m_j$ and
\begin{equation}
 \mathcal Q=\sum_{i<j}|q_iq_j|+\sum_{j,\ell}|q_j\zeta_\ell|.
\end{equation}
The constant is independent of $L$. The nuclear quotient witnesses are
needed for the weak identities below, but their norms need not be added
to $\mathcal N_\infty$.
\end{lemma}
Here the witnesses are the square-integrable functions representing the
displayed nuclear quotients, including those of the weak derivatives.
\begin{proofx}
We repeat the two-stage argument of
Lemma~\ref{lem:free-regularity-bridge}. The supplied potential images
first make the weak kinetic right-hand sides square integrable, so
$\ket{u},\ket{h}\in H^2$. We then estimate those images, now including nuclear
terms. For a fixed centre,
the regularised vector field
$(\bm{x}_j-\bm{R}_\ell)/(\|\bm{x}_j-\bm{R}_\ell\|^2+\delta^2)$ gives the translated Hardy bound
\begin{equation}
 \|f/\|\bm{x}_j-\bm{R}_\ell\|\|^2
 \le4\sum_{c=1}^3\|\partial_{j,c}f\|^2.
 \label{eq:free-Hardy-nucleus}
\end{equation}
Applying Eq.~\eqref{eq:free-Hardy-nucleus} to $\ket{f}$ controls its nuclear
Coulomb quotient. Once $\ket{u}\in H^2$ is known, applying it to each
$\partial_a u\in H^1$ supplies the derivative quotients needed
when passing to weak commutator identities. Translation of the centre
does not change the constant. With $\ket{z}=H_\infty \ket{f}-V_\infty \ket{f}$, the pair and nuclear Hardy
bounds give
\begin{equation}
 \|V_\infty \ket{f}\|^2\le8M\mathcal Q^2\|\ket{f}\|\|\ket{z}\|,
 \qquad
 \|\ket{z}\|\le2\|H_\infty \ket{f}\|+8M\mathcal Q^2\|\ket{f}\|.
\end{equation}
Apply the latter estimate to $\ket{u}$ and $\ket{h}$ and sum the derivative and
pair-quotient bounds exactly as in
Lemma~\ref{lem:free-regularity-bridge}. This gives
Eq.~\eqref{eq:free-nuclear-regularity}. No symmetry assumption is used
by the regularity argument.
\end{proofx}

\begin{lemma}[Nuclear flatness and squared-domain transfer]
\label{lem:free-bo-domain}
Let $X_L$ be a family of real $C^4$ cutoffs with $0\le X_L\le1$,
compactly supported in the retained region. Suppose its coordinate
derivatives through order four are bounded by $D$, uniformly in $L$,
and that
\begin{equation}
 \left\|\frac{\nabla_iX_L}{m_i}-\frac{\nabla_jX_L}{m_j}\right\|
 \le C_{\rm can}\|\bm{x}_i-\bm{x}_j\|.
 \label{eq:free-general-cancellation}
\end{equation}
For a fixed $d_0>0$, also require coordinate-wise nuclear flatness:
\begin{equation}
 \nabla_jX_L(\vec{\bm{x}})=0
 \quad\text{whenever }\|\bm{x}_j-\bm{R}_\ell\|<d_0,
 \quad\text{for every }j,\ell.
 \label{eq:free-nuclear-flatness}
\end{equation}
This condition holds irrespective of the positions of the other
particles. Then for every $\ket{u}\in D(H_\infty^2)$,
\begin{equation}
 \mathcal T_L(X_L\ket{u})\in D(H_L^2),\qquad
 \sup_{L\ge L_0}\sum_{j=0}^2\|H_L^j\mathcal T_L(X_L\ket{u})\|
 \le C_{\rm sys,cut}\sum_{j=0}^2\|H_\infty^j\ket{u}\|.
 \label{eq:free-bo-domain}
\end{equation}
Here $C_{\rm sys,cut}$ depends on the fixed system, $D,C_{\rm can}$,
$d_0$ and the uniform zero-mode bound, but not on $L$ or $\ket{u}$.
\end{lemma}
\begin{proofx}
We use the same second-commutator identity as in the fully quantum
case and check the new fixed-centre term separately. Put $B_L=[T,X_L]$.
Its explicit first-order expression was given in the proof of
Lemma~\ref{lem:free-nbo-domain}. The quantum-pair terms of $[V_\infty,B_L]$ have a Coulomb majorant
by Eq.~\eqref{eq:free-general-cancellation}. The additional nuclear
multiplier is
\begin{equation}
 F_{{\rm nuc},L}(\vec{\bm{x}})
 =-\sum_{j,\ell}\frac{q_j\zeta_\ell}{m_j}
     \frac{\nabla_jX_L(\vec{\bm{x}})\cdot(\bm{x}_j-\bm{R}_\ell)}{\|\bm{x}_j-\bm{R}_\ell\|^3}.
 \label{eq:free-nuclear-commutator}
\end{equation}
The apparent singularity in each summand is proportional to
$\|\bm{x}_j-\bm{R}_\ell\|^{-2}$. Coordinate-wise nuclear flatness makes that
summand zero throughout its nuclear neighbourhood, regardless of where
the other particles are. Outside that neighbourhood,
$\|\bm{x}_j-\bm{R}_\ell\|^{-2}\le d_0^{-2}$, so the coordinate derivative bound
and $m_j\ge1$ give the sufficient estimate
\begin{equation}
 \|F_{{\rm nuc},L}\|_\infty
 \le 3dD\,d_0^{-2}\sum_{j,\ell}|q_j\zeta_\ell|.
 \label{eq:free-nuclear-commutator-bound}
\end{equation}
Thus the fixed-centre contribution is a bounded multiplier, rather
than an uncontrolled inverse-square singularity.

Lemma~\ref{lem:free-nuclear-regularity} supplies the first and second
weak derivatives and both kinds of Hardy quotient. Establish the
second-commutator identity weakly with regularised Coulomb kernels.
In the nuclear terms, flatness gives the bound
Eq.~\eqref{eq:free-nuclear-commutator-bound} uniformly in the
regularisation parameter; the remaining integrations by parts are
controlled by the nuclear Hardy quotients. Dominated convergence
therefore yields the $L^2$ identity
Eq.~\eqref{eq:free-second-commutator}, now with all nuclear terms
included. To obtain domain membership, recall that $B_L\ket{u}\in H^1$ and that
the weak commutator calculation gives $H_\infty B_L\ket{u}\in L^2$.
Pair and nuclear Hardy estimates give $V_\infty B_L\ket{u}\in L^2$;
subtracting leaves $TB_L\ket{u}\in L^2$, and weak kinetic regularity
gives $B_L\ket{u}\in H^2=D(H_\infty)$. Thus the first image
$X_LH_\infty \ket{u}+B_L\ket{u}$ belongs to $D(H_\infty)$, proving
$X_L\ket{u}\in D(H_\infty^2)$ as an operator-domain statement. Transplant the cut state and its
first Hamiltonian image successively; both have support in the region
of exact potential agreement.

For the zero-mode-omitted Hamiltonian $H_L^{(0)}$, the resulting images
can be bounded by $\mathcal N_\infty$, $\Lambda_{\rm nuc}\mathcal N_\infty$
and $\Lambda_{\rm nuc}^2\mathcal N_\infty$, where
\begin{equation}
 \Lambda_{\rm nuc}
 =1+c_0+6(1+3d)^2D
   +C_{\rm can}\sum_{i<j}|q_iq_j|
   +\frac{3dD}{d_0^2}\sum_{j,\ell}|q_j\zeta_\ell|,
 \qquad c_0\ge|C_{\rm zm}(L)|.
 \label{eq:free-nuclear-domain-constant}
\end{equation}
Equation~\eqref{eq:free-zero-mode} converts these to the physical
$H_L$ by a bounded scalar shift. Together with
Eq.~\eqref{eq:free-nuclear-regularity}, this proves
Eq.~\eqref{eq:free-bo-domain}. The nuclear quotient norms do not
enter the final budget separately because the multiplier
$F_{{\rm nuc},L}$ is bounded.
\end{proofx}

For equal moving masses, a product of identical one-particle bumps
satisfies both cutoff requirements. Its derivative in coordinate $j$
vanishes near every nucleus whenever that one-particle bump is flat
there, independently of all other coordinates. Using the same bump
also makes the difference of the $i$ and $j$ derivatives vanish when
$\bm{x}_i=\bm{x}_j$; a bounded derivative then makes this difference proportional
to $\|\bm{x}_i-\bm{x}_j\|$. The next corollary gives an explicit scale and threshold.

\begin{corollary}[Born--Oppenheimer free-boundary dynamics]
\label{cor:free-dyn-bo}
Let all moving particles have the same mass $m_e\ge1$, and let the
clamped nuclei lie in $B_{R_0}(0)$. Fix $d_0,\rho>0$ and a smooth
one-particle bump $q:\R^3\to[0,1]$, equal to one on $B_1(0)$ and zero
for $\|\bm{x}\|\ge2$. Use the same bump for every moving particle and set
\begin{equation}
 s=L/16,\qquad q_s(\bm{x})=q(\bm{x}/s),\qquad
 X_L^{\rm BO}(\vec{\bm{x}})=\prod_{j=1}^d q_s(\bm{x}_j).
 \label{eq:free-bo-cutoff}
\end{equation}
Take
\begin{equation}
 \begin{split}
 \mathcal B_L^{\rm one}&=\{\vec{\bm{x}}:\ \|\bm{x}_j\|<L/16\ \text{for every }j\},\\
 S_L&=\{\vec{\bm{x}}:\ \exists j,\ \|\bm{x}_j\|\ge L/16;\
                    \|\bm{x}_i\|\le L/8\ \text{for every }i\},\\
 L_0^{\rm BO}&=\max\{16,\ 2R_f+1,\ 16(R_0+\rho+d_0)\}.
 \end{split}
 \label{eq:free-bo-shell-L0}
\end{equation}
For the fixed system, kernel family and observable of
Proposition~\ref{prop:free-dyn}, there is one constant $C$ for all
$L\ge L_0^{\rm BO}$, $m\ge8$, $K=2\pi m/L\ge8$ and $t\ge0$ such that
Eq.~\eqref{eq:free-dyn} holds whenever the supplied strong solution
has $\ket{\psi_0}\in D(H_\infty^2)$ and satisfies
Eq.~\eqref{eq:free-leakage} on these regions. Its state dependence is
only through
$\mathcal D_\infty=\|\ket{\psi_0}\|+\|H_\infty\ket{\psi_0}\|+\|H_\infty^2\ket{\psi_0}\|$.
The constant is uniform over admissible interaction profiles whose
total variations have a common fixed upper bound.
No independent (H$_{\rm dom}$) or Hardy-quotient assumption is required.
\end{corollary}
\begin{proofx}
Since $s\ge1$, derivatives through order four are bounded uniformly
by the corresponding product-bump constants. For $i\ne j$,
\begin{equation}
 \nabla_iX_L^{\rm BO}-\nabla_jX_L^{\rm BO}
 =\prod_{k\ne i,j}q_s(\bm{x}_k)
       \big[q_s(\bm{x}_j)\nabla q_s(\bm{x}_i)-q_s(\bm{x}_i)\nabla q_s(\bm{x}_j)\big].
\end{equation}
To make the cancellation explicit, for $\bm{x}=\bm{x}_i$ and $\bm{y}=\bm{x}_j$ write
\begin{equation}
 q_s(\bm{y})\nabla q_s(\bm{x})-q_s(\bm{x})\nabla q_s(\bm{y})
 =q_s(\bm{y})[\nabla q_s(\bm{x})-\nabla q_s(\bm{y})]
   +[q_s(\bm{y})-q_s(\bm{x})]\nabla q_s(\bm{y}).
\end{equation}
The first difference is bounded by $s^{-2}\|D^2q\|_\infty\|\bm{x}-\bm{y}\|$;
the second term is bounded by
$s^{-2}\|\nabla q\|_\infty^2\|\bm{x}-\bm{y}\|$. The remaining product factors
are at most one, and division by the common mass gives
\begin{equation}
 \left\|\frac{\nabla_iX_L^{\rm BO}}{m_e}
            -\frac{\nabla_jX_L^{\rm BO}}{m_e}\right\|
 \le\frac{\|D^2q\|_\infty+\|\nabla q\|_\infty^2}{m_es^2}
             \|\bm{x}_i-\bm{x}_j\|.
 \label{eq:free-bo-cancellation}
\end{equation}
If $\|\bm{x}_j-\bm{R}_\ell\|<d_0$, then
$\|\bm{x}_j\|<R_0+d_0\le L/16$, so $\nabla q_s(\bm{x}_j)=0$ irrespective of all
other coordinates. This verifies nuclear flatness. The support has
$\|\bm{x}_j\|\le L/8$, and the threshold ensures
$L/8<R_L=L/4-R_0-\rho$ and $L>2R_f$. The stated shell contains every
nonzero first and second cutoff derivative and is disjoint from the
open inner region. The common product cutoff also preserves exchanges
of identical particles. Lemma~\ref{lem:free-bo-domain} supplies
(H$_{\rm dom}$) with a uniform bound, so
Proposition~\ref{prop:free-dyn} applies.
\end{proofx}

The equal-mass restriction concerns this explicit product construction.
For unequal moving masses with fixed centres,
Lemma~\ref{lem:free-bo-domain} still applies to any cutoff satisfying
both mass-weighted cancellation and nuclear flatness, but the present
construction does not supply such a cutoff. For arbitrary smooth
cutoffs, Proposition~\ref{prop:free-dyn} retains (H$_{\rm dom}$).

\subsection{Free-boundary dynamics with internal spin}

We finally include finite internal spin without allowing the error
constant to grow with the number of spin components. Since the
Hamiltonian and observable are spin independent, each component obeys
the same spatial equation. The essential point is to combine their
norms in quadrature before estimating leakage or dependence on the initial state.

Let $\mathcal J$ be a finite set of spin configurations. With
$\{\ket{\sigma}:\sigma\in\mathcal J\}$ the orthonormal spin basis,
write $\ket{\Psi}=\sum_\sigma\ket{\Psi_\sigma}\otimes\ket{\sigma}$,
where $\ket{\Psi_\sigma}$ is the spatial component at spin label $\sigma$.
Then
\begin{equation}
 \mathcal H_{\rm spin}
 =\bigoplus_{\sigma\in\mathcal J}L^2(\R^{3d})
 \cong L^2(\R^{3d})\otimes\mathbb C^{|\mathcal J|},\qquad
 \|\ket{\Psi}\|^2=\sum_\sigma\|\ket{\Psi_\sigma}\|^2.
\end{equation}
The Hamiltonians, cutoffs, spatial projectors and measured observable
act as the identity on spin. In this subsection their action on the
direct sum is understood. In particular the initial budget is the
Hilbert-space quantity
\begin{equation}
 \mathcal D_\infty^{\rm spin}
 =\|\ket{\Psi_0}\|+\|H_\infty\ket{\Psi_0}\|+\|H_\infty^2\ket{\Psi_0}\|,
 \label{eq:free-spin-budget}
\end{equation}
not the sum of the component budgets.

\begin{theorem}[Spin-inclusive free-boundary dynamics]
\label{thm:free-dyn-spin}
Use either the fully quantum system and cutoff of
Corollary~\ref{cor:free-dyn-nbo}, or the clamped-nuclei and cutoff of Corollary~\ref{cor:free-dyn-bo}. Let $\ket{\Psi_\infty}$
be a supplied strong solution in $\mathcal H_{\rm spin}$, with
$\ket{\Psi_0}\in D(H_\infty^2)$. At every $s\in[0,t]$ assume the
spin-summed leakage conditions
\begin{equation}
 \sum_\sigma
   \|\mathbf1_{(\mathcal B_L^{\rm one})^c}\ket{\Psi_{\infty,\sigma}(s)}\|^2
       \le\epsilon_{\rm leak}^2,\qquad
 \sum_\sigma
   \|\mathbf1_{S_L}\nabla\Psi_{\infty,\sigma}(s,\cdot)\|^2
       \le\epsilon_{\rm leak}^2,\quad
 0\le\epsilon_{\rm leak}\le1.
 \label{eq:free-spin-leakage}
\end{equation}
Evolve every component of $P'\mathcal T_L(X_L\ket{\Psi_0})$ with the same
grid Hamiltonian. With the cell and cutoff thresholds of the respective
corollary,
\begin{equation}
 \left|\sum_\sigma
   \langle\Psi_{\infty,\sigma}(t)|f(\bm{x}_a)|\Psi_{\infty,\sigma}(t)\rangle
       -\sum_\sigma
   \langle\Psi_{g,\sigma}(t)|O_g|\Psi_{g,\sigma}(t)\rangle\right|
 \le C(K^{-5}+tK^{-3})+C(1+t)\epsilon_{\rm leak}.
 \label{eq:free-dyn-spin}
\end{equation}
The constant depends on the fixed system, a uniform upper bound on
$V_\theta$, the observable, the cutoff bounds and
$\mathcal D_\infty^{\rm spin}$, but not on $L,K,t$,
$\epsilon_{\rm leak}$ or $|\mathcal J|$. No normalisation
or antisymmetry assumption is used. The estimate therefore applies,
in particular, on physical position--spin exchange sectors.
\end{theorem}
\begin{proofx}
We carry out the spatial comparison in the full spin Hilbert space,
whose squared norm is the sum of the component squared norms. This
lets us use the stated spin-summed leakage at each time directly.
The same cutoff acts on every component.
With $C_{\rm cut}=3d(D_1+D_2^{\rm cut}/2)$, the derivative and shell
bounds give at each time
\begin{equation}
 \left(\sum_\sigma\|[X_L,T]\ket{\Psi_{\infty,\sigma}(s)}\|^2\right)^{1/2}
 \le C_{\rm cut}\epsilon_{\rm leak}.
 \label{eq:free-spin-commutator}
\end{equation}
This follows by the triangle inequality in the direct sum, using the
two sums in Eq.~\eqref{eq:free-spin-leakage}. Duhamel then gives a
norm error $C_{\rm cut}t\epsilon_{\rm leak}$, while the observable
cutoff error is at most $\epsilon_{\rm leak}^2$. Conservation of the
Hilbert norm bounds the remaining observable difference by
$2\|\ket{\Psi_0}\|C_{\rm cut}t\epsilon_{\rm leak}$. It is important to perform
this step in the direct sum: summing separately chosen componentwise
time suprema would require a stronger leakage hypothesis.

For the grid comparison, the scalar error is quadratic in the initial
state. We therefore sum bounds quadratic in the initial graph norms, rather than summing
separately chosen scalar error constants. Set
\begin{equation}
 d_\sigma=\|\ket{\Psi_{0,\sigma}}\|
           +\|H_\infty\ket{\Psi_{0,\sigma}}\|
           +\|H_\infty^2\ket{\Psi_{0,\sigma}}\|.
\end{equation}
The regularity and domain lemmas provide common constants
$C_{\rm reg},\Lambda$ such that the three transplanted image norms
are bounded by $N_\sigma,\Lambda N_\sigma,\Lambda^2N_\sigma$, where
$N_\sigma=C_{\rm reg}d_\sigma$; incorporate the bounded zero-mode
shift into $\Lambda$. For $N_\sigma>0$, divide the transplanted initial state by $N_\sigma$.
Its three image norms are then bounded by the same fixed budget
$(1,\Lambda,\Lambda^2)$ for every component. Apply the scalar
grid estimate at that budget and multiply the quadratic error by
$N_\sigma^2$.
The case $N_\sigma=0$ contributes zero. By the triangle inequality
in the finite-dimensional $\ell^2$ space,
\begin{equation}
 \sum_\sigma N_\sigma^2
 \le C_{\rm reg}^2
    \left(\|\ket{\Psi_0}\|+\|H_\infty\ket{\Psi_0}\|
                     +\|H_\infty^2\ket{\Psi_0}\|\right)^2.
 \label{eq:free-spin-data-sum}
\end{equation}
The inequality follows by applying the $\ell^2$ triangle inequality
to the three sequences of component norms. Each sequence has
$\ell^2$ norm equal to the corresponding full Hilbert norm, so no
factor $|\mathcal J|$ is introduced. Thus the sum of the quadratic
grid errors is bounded by
$C(K^{-5}+tK^{-3})$ with no spin-multiplicity factor. Adding the
free-space step and using
$\epsilon_{\rm leak}^2\le\epsilon_{\rm leak}$ proves
Eq.~\eqref{eq:free-dyn-spin}.
\end{proofx}

The spin extension concerns dynamics. Proposition~\ref{prop:free}
remains a ground-energy comparison in the specified scalar exchange
sector; no identification of the physical spin-inclusive free-space
ground energy with a minimum over such sectors is used or formalised
here. The exact free-space solution and the stated localisation or
leakage properties remain inputs. The torus and grid propagators used
in the comparison are constructed, but the canonical free-space
propagator is not constructed by these endpoints.

\section{Kinetic-energy bounds and finite-window estimators}
\label{app:kinetic}
\label{sec:stopping}

We seek bounds on the error in the kinetic energy of a selected
particle, which we call the projectile and label by $p$. Its kinetic
energy $T_p$ multiplies each momentum amplitude by
$\norm{\bm{k}_p}^2/(2m_p)$. This symbol grows with momentum, so the
bounded smooth-observable estimates of Appendix~\ref{app:dynamics}
do not apply directly. We first control the error using the available
state-norm and kinetic-regularity estimates. We then improve the
time dependence by temporarily replacing the kinetic symbol by a
capped symbol in the proof and estimating the contribution removed
by that replacement. The final bounds concern the original $T_p$.

Theorem~\ref{thm:kinetic-log}, proved in Appendix~\ref{app:source},
gives the stronger logarithmic cutoff dependence. The direct
estimates below have simpler proofs, while the capped estimates
provide a complementary error term growing as $t^{2/3}$. We also
translate both estimates into errors for finite-window
stopping-power estimators, following the application in
Ref.~\cite{Rubin2024Stopping}.

Throughout this appendix we use the standing hypotheses of
Appendix~\ref{app:dynamics}, including
$\norm{\ket{\psi_0}}=1$ and
$\ket{\psi_0}\in D(\Hnbo^2)\cap\mathcal S$.
The domain condition means that the Hamiltonian can be applied twice
to the initial state with a square-integrable result. We compare
\begin{equation}
\ket{\psi(t)}=e^{-i\Hnbo t}\ket{\psi_0},\qquad
\ket{\psi_N(t)}=e^{-iH_Nt}P\ket{\psi_0},\qquad
\ket{\psi_g(t)}=e^{-i\Hgrid t}\Pp\ket{\psi_0}.
\end{equation}
The discrete states are not renormalised. Here $H_N=P\Hnbo P$
acts on the range of $P$, which retains the one-particle momentum
cubes of radius $K$; the grid projection $\Pp$ retains cubes of
radius $2K$. Let $\ket{\vec{\bm{k}}}$ denote the orthonormal many-particle
plane-wave basis, with any spin labels included implicitly. We write
the momentum amplitudes of a state as
\begin{equation}
u_{\vec{\bm{k}}}:=\langle\vec{\bm{k}}|u\rangle,\qquad
\ket{u}=\sum_{\vec{\bm{k}}}u_{\vec{\bm{k}}}\ket{\vec{\bm{k}}}.
\end{equation}
The amplitude of $B\ket{u}$ is $\langle\vec{\bm{k}}|B|u\rangle$.
Time and discretisation labels carry over in the same way; for example,
$\psi_{N,\vec{\bm{k}}}(s):=\langle\vec{\bm{k}}|\psi_N(s)\rangle$.
We use $A$ for the momentum-space multiplier representing the total
kinetic energy, so $\Ttot\ket{\vec{\bm{k}}}=A(\vec{\bm{k}})\ket{\vec{\bm{k}}}$.
Its symbol and the kinetic graph norm are
\begin{align}
A(\vec{\bm{k}})&=\sum_a\frac{\norm{\bm{k}_{\bm{p}_a}}^2}{2m_a},\\
\Theta(\ket{u})&:=\norm{(1+A)\ket{u}}
=\left(\sum_{\vec{\bm{k}}}[1+A(\vec{\bm{k}})]^2|u_{\vec{\bm{k}}}|^2\right)^{1/2}.
\end{align}
Thus $\Theta$ controls both the state and its image under the total
kinetic energy. The constant $\Xi$ is the bound on kinetic graph norms in
Eq.~\eqref{eq:Xi}. Unless stated otherwise, constants denoted by
$C$ may depend on the fixed physical parameters but not on $K$,
$t$, the initial state beyond its displayed factor $\Xi$, or the
cap scale. We enlarge the fixed threshold $K_0$ when necessary.
At fixed cell size, the one-particle basis size obeys
$N=\Theta(K^3)$ for either discretisation.

The grid statements hold uniformly for the kernel class
$\mathcal K_K$ of Eq.~\eqref{eq:kernelclass}. For its sharp
truncation, the dropped-transfer bounds use tails at $K$; for a
general kernel they use tails at $c_{\rm ker}K$, with the fixed
$c_{\rm ker}>0$ specified there. Replacing $K$ by this fixed
fraction changes constants but none of the powers below. We use
$D$ for the appropriate dropped-transfer operator, including
$D_g$ for a general kernel, and $R$ for the corresponding wrap
operator. These are the two potential errors in
Eq.~\eqref{eq:grid-structure}.

\subsection{Direct kinetic-energy estimates}

We begin with a property special to Fourier multipliers. They do
not mix retained and discarded Fourier modes, so projecting a state
removes precisely the observable contribution carried by its tail.
For kinetic energy this contribution is non-negative.

\begin{lemma}[Multiplier basics]
\label{lem:mult-basic}
Let $O$ be a Fourier multiplier with symbol $\mu(\bm{k}_p)$ satisfying
$0\le\mu(\bm{k}_p)\le c_\mu A(\vec{\bm{k}})$. This includes $T_p$ and its
non-negative kinetic cap. Then $O$ commutes with $P$ and $\Ttot$.
For $\ket{\delta(t)}:=(1-P)\ket{\psi(t)}$,
\begin{align}
\langle\psi(t)|O|\psi(t)\rangle
-\langle\psi(t)|POP|\psi(t)\rangle
&=\langle\delta(t)|O|\delta(t)\rangle,\\
0\le\langle\delta(t)|O|\delta(t)\rangle
&\le c_\mu\norm{\ket{\delta(t)}}
       \norm{(1-P)(1+A)\ket{\psi(t)}}
 \le Cc_\mu\Xi^2K^{-3}.
\label{eq:mult-I}
\end{align}
Here and below, momentum-space weights and tail projections act
diagonally in the plane-wave basis. Moreover, for every state with
finite kinetic graph norm,
\begin{equation}
\norm{O\ket{v}}\le c_\mu\Theta(\ket{v}).
\end{equation}
In particular,
\begin{equation}
\norm{T_pP\ket{\psi(t)}}\le\Xi,\qquad
\norm{T_p\ket{\psi_N(t)}}\le C\Xi,\qquad
\norm{T_p\ket{\psi_g(t)}}\le C\Xi.
\end{equation}
\end{lemma}

\begin{proofx}
Because $P$ and $O$ are diagonal in the same basis,
$PO(1-P)=(1-P)OP=0$. Expanding
$\ket{\psi}=P\ket{\psi}+\ket{\delta}$ therefore leaves no
cross terms. Positivity gives the lower bound, and the pointwise
symbol inequality followed by Cauchy--Schwarz gives the upper bound
in Eq.~\eqref{eq:mult-I}. Equations~\eqref{eq:dyn-t}
and~\eqref{eq:dyn-wtail}, summed over the finitely many registers,
bound its two factors by $C\Xi K^{-5/2}$ and
$C\Xi K^{-1/2}$, respectively. Thus projection undercounts the
kinetic energy by at most $C\Xi^2K^{-3}$, without any smoothing
assumption on the observable.

The same symbol inequality gives
$\norm{O\ket{v}}\le c_\mu\norm{A\ket{v}}
\le c_\mu\Theta(\ket{v})$. The three state bounds follow
from the exact graph bound, Eq.~\eqref{eq:etareg}, and
Lemma~\ref{lem:grid-traj}. These estimates control $O$ applied to
the particular states; they do not assert a cutoff-independent
operator norm for $T_p$.
\end{proofx}

\begin{theorem}[Kinetic-energy observable, Galerkin evolution]
\label{thm:mult-gal}
For the evolution in Eq.~\eqref{eq:dyn-def}, all $t\ge0$ and
$K\ge K_0$ satisfy
\begin{equation}
\left|\langle\psi(t)|T_p|\psi(t)\rangle
-\langle\psi_N(t)|T_p|\psi_N(t)\rangle\right|
\le C\Xi^2\bigl[K^{-3}+tK^{-3/2}\bigr]
=\mathcal O\bigl(N^{-1}+tN^{-1/2}\bigr).
\label{eq:mult-gal}
\end{equation}
No observable-smoothing hypothesis or cutoff-dependent observable
norm enters the constant.
\end{theorem}

\begin{proofx}
We split the error into projection of the exact state and evolution
inside the retained space. The first contribution is
Eq.~\eqref{eq:mult-I}. For the second, set
$\ket{e_P(t)}:=P\ket{\psi(t)}-\ket{\psi_N(t)}$ and use the
exact identity
\begin{align}
&\langle\psi(t)|PT_pP|\psi(t)\rangle
-\langle\psi_N(t)|T_p|\psi_N(t)\rangle\\
&\hspace{1cm}=
\langle e_P(t)|T_pP|\psi(t)\rangle
+\langle\psi_N(t)|T_p|e_P(t)\rangle.
\end{align}
We have introduced the error only once in each term, so no
quadratic error term needs a separate estimate. The Duhamel
identity~\eqref{eq:duhamel} and source bound~\eqref{eq:secular}
give
$\norm{\ket{e_P(t)}}\le C\Xi tK^{-3/2}$.
Using self-adjointness of $T_p$ in the second pairing, followed by
Cauchy--Schwarz and Lemma~\ref{lem:mult-basic}, gives
\begin{equation}
\left|\langle e_P|T_pP|\psi\rangle\right|
+\left|\langle\psi_N|T_p|e_P\rangle\right|
\le\norm{\ket{e_P}}
\left(\norm{T_pP\ket{\psi}}+\norm{T_p\ket{\psi_N}}\right)
\le C\Xi^2tK^{-3/2}.
\end{equation}
Adding the projection contribution proves the claim. Later we
will improve this time-dependent term by retaining more information
about the source inside the pairing.
\end{proofx}

\begin{proposition}[Kinetic-energy observable, truncated-kernel grid evolution]
\label{prop:mult-grid}
For the grid evolution of Sec.~\ref{sec:grid}, with any kernel in
$\mathcal K_K$, all $t\ge0$ and $K\ge K_0$ satisfy
\begin{equation}
\left|\langle\psi(t)|T_p|\psi(t)\rangle
-\langle\psi_g(t)|T_p|\psi_g(t)\rangle\right|
\le C\Xi^2\bigl[K^{-3}+tK^{-1}\bigr]
=\mathcal O\bigl(N^{-1}+tN^{-1/3}\bigr).
\label{eq:mult-grid}
\end{equation}
The kinetic multiplier is represented exactly on the grid
register. The additional error compared with
Eq.~\eqref{eq:mult-gal} comes from the dropped-transfer and wrap
sources in the dynamics.
\end{proposition}

\begin{proofx}
Projection again costs at most $C\Xi^2K^{-3}$, now using the
cutoff $2K$. In Fourier coordinates, obtained by the quantum Fourier transform,
the register observable is exactly the diagonal multiplier
$\Pp T_p\Pp$. Consequently,
there is no additional error from sampling the observable in
position space, unlike term (II) in the smooth position-observable
comparison of Appendix~\ref{app:grid}.

For the retained-space error
$\ket{e_{\Pp}(t)}:=\Pp\ket{\psi(t)}-\ket{\psi_g(t)}$,
the same linear identity as above produces two pairings. For each
endpoint $\ket{v}\in\{\Pp\ket{\psi(t)},\ket{\psi_g(t)}\}$,
we move its kinetic image backwards through the grid evolution:
\begin{equation}
\ket{\varphi_v(s)}:=e^{i\Hgrid(t-s)}T_p\ket{v},
\qquad 0\le s\le t.
\end{equation}
The Duhamel formula then pairs each source at time $s$ with
$\ket{\varphi_v(s)}$. We need its graph norm as well as its
ordinary norm. Writing $X(\ket{v}):=\norm{A^2\ket{v}}$,
Lemma~\ref{lem:grid-data2} gives
\begin{equation}
\Theta(T_p\ket{v})
\le\norm{(1+A)A\ket{v}}
\le X(\ket{v})+\Theta(\ket{v})
\le C\Xi K^{3/2}.
\end{equation}
The forward and inverse graph comparisons of
Lemma~\ref{lem:grid-KR}(iv), together with conservation of the
Hamiltonian norm, propagate this estimate uniformly in $s$.
Unitarity and Lemma~\ref{lem:mult-basic} give the sharper ordinary
norm bound. Hence
\begin{equation}
\Theta(\ket{\varphi_v(s)})\le C\Xi K^{3/2},\qquad
\norm{\ket{\varphi_v(s)}}\le C\Xi,
\qquad t_r(K;\ket{\varphi_v(s)})\le C\Xi K^{-1/2}.
\end{equation}
Here $t_r(\lambda;\ket{u})$ denotes the Fourier tail in register
$r$ beyond the stated momentum threshold. The last inequality
uses the kinetic weight $A\ge K^2/(2m_r)$ on that tail. The same
power holds at a fixed fraction of $K$.

We now estimate the three Duhamel sources. Let
$\ket{u(s)}:=\Pp\ket{\psi(s)}$ and
$\ket{\delta'(s)}:=(1-\Pp)\ket{\psi(s)}$.
For the source already present in Galerkin evolution,
Eq.~\eqref{eq:secular} and the ordinary norm bound give
\begin{equation}
\left|\langle\varphi_v(s)|\Pp V_{\mathrm{tot}}|\delta'(s)\rangle\right|
\le C\Xi\,C\Xi K^{-3/2}.
\end{equation}
For the dropped transfers, Eq.~\eqref{eq:Dform} combines the
exact-state tail $t_r(K;\ket{u})\le C\Xi K^{-5/2}$ with the
backward-state tail:
\begin{align}
\left|\langle\varphi_v|D|u\rangle\right|
&\le CK^{-1/2}\sum_r
\left[\Theta(\ket{\varphi_v})t_r(K;\ket{u})
+\Theta(\ket{u})t_r(K;\ket{\varphi_v})\right]\\
&\le C\Xi^2K^{-1/2}
\left[K^{3/2}K^{-5/2}+K^{-1/2}\right]
\le C\Xi^2K^{-1}.
\end{align}
For general kernels the $D$ estimate uses $c_{\rm ker}K$ as
explained above. The weak-coefficient part of the wrap has the
same bound by Eq.~\eqref{eq:Rform}. Its strong-coefficient part
requires both endpoints to have high momentum and is smaller:
\begin{equation}
CK\sum_r t_r(K/2;\ket{u})t_r(K/2;\ket{\varphi_v})
\le C\Xi^2K\,K^{-5/2}K^{-1/2}
=C\Xi^2K^{-2}.
\end{equation}
Thus each retained-space pairing accumulates at most
$C\Xi^2tK^{-1}$. Adding the projection error completes the proof.
\end{proofx}

We translate an energy error into a stopping-power error by comparing
the same finite-window functional for the exact and discrete
evolutions. This comparison does not estimate any separate bias
between a finite-window slope and an instantaneous energy derivative.

\begin{corollary}[Stopping-power estimation]
\label{cor:stopping}
Let $T>0$ be the window length and $v_p>0$ the prescribed projectile
speed. Define the signed secant estimator for either evolution by
\begin{equation}
\widehat S:=
\frac{\langle T_p\rangle(T)-\langle T_p\rangle(0)}{v_pT},
\end{equation}
where $\langle T_p\rangle(t)$ is the kinetic quadratic form of the
corresponding state at that time. If $\Delta\widehat S$ denotes the discrete estimator
minus the exact estimator, then
\begin{equation}
|\Delta\widehat S|\le\frac{C\Xi^2}{v_p}
\begin{cases}
K^{-3/2}+2K^{-3}/T,&\text{Galerkin},\\
K^{-1}+2K^{-3}/T,&\text{grid}.
\end{cases}
\label{eq:stopping}
\end{equation}
In particular, at fixed $T$, $v_p$, and cell size, these errors are
$\mathcal O(N^{-1/2})$ and $\mathcal O(N^{-1/3})$, respectively.
\end{corollary}

\begin{proofx}
Let $\varepsilon(t)$ denote the difference between the discrete
and exact kinetic energies. Subtracting the two secants gives
\begin{equation}
|\Delta\widehat S|
=\frac{|\varepsilon(T)-\varepsilon(0)|}{v_pT}
\le\frac{|\varepsilon(T)|+|\varepsilon(0)|}{v_pT}.
\end{equation}
Substitute Eq.~\eqref{eq:mult-gal} or Eq.~\eqref{eq:mult-grid}.
The time-dependent term at $T$ loses its factor $T$, and the two
endpoint projection errors give $2K^{-3}/T$. Reversing the sign to
use a positive energy-loss convention leaves this absolute-error
bound unchanged.
\end{proofx}

\begin{remark}[Least-squares fits]
For a fixed nondegenerate sampling pattern rescaled to a window of
length $T$, a least-squares slope is a linear combination of the
sampled energies. Here the pattern specifies fixed relative sampling
times, with at least two distinct times so that the fitted slope is
well defined. The sum of the absolute values of its coefficients
is $C_{\rm fit}/T$, with $C_{\rm fit}$ depending only on that
pattern. A uniform perturbation of the sampled energies by at most
$\epsilon$ therefore changes the fitted slope by at most
$C_{\rm fit}\epsilon/T$, or the stopping estimator by at most
$C_{\rm fit}\epsilon/(v_pT)$. Applying the energy bounds uniformly
on the window gives analogous rates, up to the sampling-dependent
constant. This fitting observation is not a separate formal endpoint.
\end{remark}

\subsection{Momentum tails during discrete evolution}

To improve the time dependence, we will cap the kinetic symbol at a
momentum scale $\kappa$ below the basis cutoff. We must then control
the momentum contribution above $\kappa$ for both the exact and
discrete states. The exact-state estimates already hold at every
threshold. We next establish the corresponding discrete estimates;
a bound only at the register edge would not suffice for a variable
cap.

\begin{lemma}[Momentum tails, Galerkin evolution]
\label{lem:disc-tails}
Let $\ket{\psi_N(s)}=e^{-iH_Ns}P\ket{\psi_0}$ with
$\ket{\psi_0}\in D(\Hnbo^2)\cap\mathcal S$ and the standing
normalisation. For every $s\in\mathbb R$, register $a$, and
$1\le\lambda<\sqrt3K$,
\begin{align}
\norm{\ind{\norm{\bm{k}_{\bm{p}_a}}>\lambda}\ket{\psi_N(s)}}
&\le C m_a^{3/2}m_{\max}^{1/2}\Xi\lambda^{-5/2},\\
\norm{\ind{\norm{\bm{k}_{\bm{p}_a}}>\lambda}(1+A)\ket{\psi_N(s)}}
&\le C(m_am_{\max})^{1/2}\Xi\lambda^{-1/2}.
\label{eq:disc-tails}
\end{align}
These Euclidean tails vanish for $\lambda\ge\sqrt3K$. With an
explicit $\infty$-norm restriction, the same estimates hold and
vanish for $\lambda\ge K$. Extending the lower-threshold estimates
to the full Euclidean range costs at most a factor six in the
coarser uniform bound.
\end{lemma}

\begin{proofx}
The exact-state bootstrap isolates $A\ket{\psi}$ in the
Hamiltonian equation and bounds the tail of
$\Hnbo\ket{\psi}$ using its graph norm. We cannot
simply replace $\Hnbo$ by $H_N$ and assume the same graph bound:
projection of the initial state introduces an additional source.
Indeed, with $\ket{\delta_0}:=(1-P)\ket{\psi_0}$,
\begin{equation}
H_NP\ket{\psi_0}
=P\Hnbo\ket{\psi_0}-PV_{\mathrm{tot}}\ket{\delta_0},
\end{equation}
because $P\Ttot(1-P)=0$. The second term is small in norm, but
that does not give a uniform graph norm after multiplication by
kinetic energy. We therefore propagate the two terms separately:
\begin{align}
H_N\ket{\psi_N(s)}&=\ket{\chi_N(s)}-\ket{\rho_N(s)},\\
\ket{\chi_N(s)}&:=e^{-iH_Ns}P\Hnbo\ket{\psi_0},\qquad
\ket{\rho_N(s)}:=e^{-iH_Ns}PV_{\mathrm{tot}}\ket{\delta_0}.
\end{align}

The first evolved state has a uniform graph bound because
$\Hnbo\ket{\psi_0}$ itself lies in the Hamiltonian domain.
More explicitly, the forward graph estimate and the fact that
Fourier projection decreases $\Theta$ give
\begin{align}
\norm{H_N\ket{\chi_N(s)}}
&=\norm{H_NP\Hnbo\ket{\psi_0}}\\
&\le C\bigl[\Theta(P\Hnbo\ket{\psi_0})
              +m_{\max}\norm{P\Hnbo\ket{\psi_0}}\bigr]
\le C\Xi.
\end{align}
To recover the kinetic graph norm from this Hamiltonian bound on
$\operatorname{ran}P$, use
$\Ttot\ket{u}=H_N\ket{u}-PV_{\mathrm{tot}}\ket{u}$ and the
relative potential bound. Absorbing its term proportional to
$\tfrac12\Theta(\ket{u})$ yields
$\Theta(\ket{u})\le C[\norm{H_N\ket{u}}+
 m_{\max}\norm{\ket{u}}]$. Hamiltonian-norm conservation and
unitarity therefore imply
$\Theta(\ket{\chi_N(s)})\le C\Xi$, and hence
\begin{equation}
t_a(\lambda;\ket{\chi_N(s)})\le C m_a\Xi\lambda^{-2}.
\end{equation}
For the second evolved state we need only norm conservation and
Eq.~\eqref{eq:secular} at time zero:
\begin{equation}
\norm{\ket{\rho_N(s)}}
=\norm{PV_{\mathrm{tot}}\ket{\delta_0}}
\le C m_{\max}S_2^{1/2}\Xi K^{-3/2}.
\end{equation}
We make no support claim about this evolved remainder. Its small
norm already controls every one of its tails. For
$1\le\lambda\le\sqrt3K$, the two estimates imply
\begin{equation}
t_a(\lambda;H_N\ket{\psi_N(s)})
\le C\Xi(\lambda^{-2}+K^{-3/2})
\le C\Xi\lambda^{-1/2}.
\end{equation}

We now return to the finite-basis Hamiltonian equation,
\begin{equation}
A(\vec{\bm{k}})\psi_{N,\vec{\bm{k}}}(s)
=\langle\vec{\bm{k}}|H_N|\psi_N(s)\rangle
 -\langle\vec{\bm{k}}|PV_{\mathrm{tot}}|\psi_N(s)\rangle.
\end{equation}
Projection cannot enlarge a Fourier tail. The general potential
estimate~\eqref{eq:lam-gen} and
$\Theta(\ket{\psi_N(s)})\le C\Xi$ therefore control the
second term by $C(m_am_{\max})^{1/2}\Xi\lambda^{-1/2}$.
Adding the two contributions bounds the tail of $A\ket{\psi_N}$;
the a priori tail
$t_a(\lambda;\ket{\psi_N})\le2m_a\Theta(\ket{\psi_N})
\lambda^{-2}$ supplies the $1$ in $1+A$. Finally,
\begin{equation}
\norm{\ind{\norm{\bm{k}_{\bm{p}_a}}>\lambda}\ket{\psi_N}}
\le\frac{2m_a}{\lambda^2}
\norm{\ind{\norm{\bm{k}_{\bm{p}_a}}>\lambda}A\ket{\psi_N}},
\end{equation}
which gives the extra two powers of $\lambda$ in the plain tail.

The support endpoints follow from the geometry of the retained
cube, whose largest Euclidean momentum is $\sqrt3K$. Equivalently,
one can extend a bound known only up to $K$ by monotonicity:
for $K\le\lambda\le\sqrt3K$ the plain-tail constant increases
by at most $3^{5/4}<6$, and the weighted-tail constant by at most
$3^{1/4}<6$. For an $\infty$-norm tail the cube endpoint is $K$.
\end{proofx}

\begin{lemma}[Momentum tails, grid evolution]
\label{lem:grid-tails}
Let $\ket{\psi_g(s)}=e^{-i\Hgrid s}\Pp\ket{\psi_0}$ with
$\ket{\psi_0}\in D(\Hnbo^2)\cap\mathcal S$ and the standing
normalisation. For every $s\in\mathbb R$, register $a$, and
$1\le\lambda\le K$,
\begin{align}
\norm{\ind{\norm{\bm{k}_{\bm{p}_a}}>\lambda}(1+A)\ket{\psi_g(s)}}
&\le C(m_am_{\max})^{1/2}\Xi\lambda^{-1/2},\\
\norm{\ind{\norm{\bm{k}_{\bm{p}_a}}>\lambda}\ket{\psi_g(s)}}
&\le C m_a^{3/2}m_{\max}^{1/2}\Xi\lambda^{-5/2}.
\label{eq:grid-tails}
\end{align}
The constants are uniform over the admitted grid kernels.
\end{lemma}

\begin{proofx}
We again isolate the kinetic term, now on the grid register:
\begin{equation}
A(\vec{\bm{k}})\psi_{g,\vec{\bm{k}}}(s)
=\langle\vec{\bm{k}}|\Hgrid|\psi_g(s)\rangle
 -\langle\vec{\bm{k}}|V_{\mathrm{grid}}|\psi_g(s)\rangle,
\qquad V_{\mathrm{grid}}:=\Hgrid-\Ttot.
\end{equation}
We first control the potential term. In each interaction channel,
$V_{\mathrm{grid}}$ consists of ordinary multiplication by the
retained kernel, projected to the register, together with its wrap
term. The ordinary multiplication inherits
Eq.~\eqref{eq:lam-gen} from Fourier-coefficient domination.
The alias bound~\eqref{eq:alias-mult} controls the wrap by the
part of that product lying outside the register. Its potential-tail
bound is $C\Theta(\ket{v})K^{-1/2}$ for a register state
$\ket{v}$. Applying these bounds in each interaction channel and
summing gives, for $1\le\lambda\le K$,
\begin{equation}
\norm{\ind{\norm{\bm{k}_{\bm{p}_a}}>\lambda}
       V_{\mathrm{grid}}\ket{v}}
\le C\Theta(\ket{v})(\lambda^{-1/2}+K^{-1/2})
\le C\Theta(\ket{v})\lambda^{-1/2}.
\end{equation}
This uses the three-dimensional transfer coordinate of each
channel; it does not require an $\ell^1$ bound on the full
many-particle Fourier array. For $\ket{v}=\ket{\psi_g(s)}$,
Lemma~\ref{lem:grid-traj} bounds $\Theta(\ket{v})$ by $C\Xi$.

For the Hamiltonian term we separate an evolution with bounded graph norm
from the evolution of a small initial residual. Define, in this proof,
\begin{equation}
\Delta:=\Hgrid-\Pp\Hnbo\Pp=R-D,
\qquad \ket{\delta'_0}:=(1-\Pp)\ket{\psi_0}.
\end{equation}
Commutation of the kinetic operator with $\Pp$ gives
\begin{align}
\Hgrid\Pp\ket{\psi_0}
 &=\Pp\Hnbo\ket{\psi_0}+\ket{r_g},\\
\ket{r_g}&:=\Delta\Pp\ket{\psi_0}
                -\Pp V_{\mathrm{tot}}\ket{\delta'_0}.
\end{align}
The second term has norm at most $C\Xi K^{-3/2}$ by
Eq.~\eqref{eq:secular}. To bound the first, set
$\ket{u_0}:=\Pp\ket{\psi_0}$ and
$\ket{z}:=\Delta\ket{u_0}$. The bilinear bounds
\eqref{eq:Dform} and~\eqref{eq:Rform} apply to
$\langle z|\Delta|u_0\rangle=\norm{\ket{z}}^2$.
For $\ket{u_0}$, the relevant tails at fixed fractions of $K$
are at most $C\Xi K^{-5/2}$ and its graph norm is at most $C\Xi$.
For $\ket{z}$, which lies in the finite register, the elementary
bounds are
\begin{equation}
t_r(cK;\ket{z})\le\norm{\ket{z}},\qquad
\Theta(\ket{z})\le CK^2\norm{\ket{z}}
\end{equation}
for any fixed $c>0$ used in those form estimates. Substitution gives
\begin{align}
\norm{\ket{z}}^2
&\le C\Xi\norm{\ket{z}}
 \left[K^{-1/2}\bigl(K^2K^{-5/2}+1\bigr)
                    +KK^{-5/2}\right]\\
&\le C\Xi K^{-1/2}\norm{\ket{z}}.
\end{align}
Dividing by $\norm{\ket{z}}$ when it is nonzero proves
$\norm{\ket{r_g}}\le C\Xi K^{-1/2}$; the zero case is immediate.
This argument covers the full kernel class, including kernels
whose dropped transfers begin above $K$ rather than $2K$.

Now let
$\ket{\chi_g(s)}:=e^{-i\Hgrid s}\Pp\Hnbo\ket{\psi_0}$.
The forward graph estimate, Hamiltonian-norm conservation, and its
inverse comparison show that
\begin{equation}
\Theta(\ket{\chi_g(s)})\le C\Xi,
\qquad t_a(\lambda;\ket{\chi_g(s)})
\le C m_a\Xi\lambda^{-2}.
\end{equation}
Since
$\Hgrid\ket{\psi_g(s)}=\ket{\chi_g(s)}+
 e^{-i\Hgrid s}\ket{r_g}$, the residual contributes at most
$C\Xi K^{-1/2}$ to any tail. Combining this with the potential
estimate yields
\begin{equation}
\norm{\ind{\norm{\bm{k}_{\bm{p}_a}}>\lambda}A\ket{\psi_g(s)}}
\le C\Xi\bigl(\lambda^{-2}+K^{-1/2}+\lambda^{-1/2}\bigr)
\le C\Xi\lambda^{-1/2}.
\end{equation}
Adding the unweighted a priori tail gives the $1+A$ estimate;
dividing by $\lambda^2/(2m_a)$ gives the plain-tail estimate,
as in the preceding proof.
\end{proofx}

The lattice estimates underlying these two lemmas initially use
thresholds above fixed multiples of $2\pi/L$. We have written the
range as $\lambda\ge1$: any intervening bounded interval is
covered by the uniform graph and norm bounds, with an enlarged
constant depending on the fixed cell. The displayed mass factors
retain the analytic form of Eqs.~\eqref{eq:dyn-t}
and~\eqref{eq:dyn-wtail}; the formal uniform bounds may use
coarser powers of $m_{\max}$, as in Appendix~\ref{app:dynamics}.
We do not identify the smaller displayed prefactors with separately
checked sharp Lean constants.

\subsection{Sharpened secular rates by symbol capping}
\label{sec:stopping-cap}

The time-dependent, or secular, error comes from integrating a
source pairing along the evolution. A lower cap makes the backward
state less costly in graph norm, but discards a larger part of the
kinetic observable. We will balance these two effects. The momentum tail
bounds just proved allow both costs to be estimated at the chosen
cap scale, independently of the basis cutoff.

For the retained-space error $\ket{e_P(t)}=P\ket{\psi(t)}-\ket{\psi_N(t)}$,
we will pair its Duhamel source against
$\ket{\varphi_v(s)}:=e^{iH_N(t-s)}\mu_\kappa\ket{v}$, where
$\ket{v}$ is either $P\ket{\psi(t)}$ or $\ket{\psi_N(t)}$.
Thus the backward evolution starts from the capped endpoint
$\mu_\kappa\ket{v}$ and enters the matrix element
$\langle e_P(t)|\mu_\kappa|v\rangle$.
We now quantify how capping reduces the regularity needed for the
backward evolution in the observable pairing. A bound on
$\norm{\mu_\kappa\ket{v}}$ controls the size of that pairing,
whereas a bound on $\Theta(\mu_\kappa\ket{v})$ controls the
high-momentum tail of the backward-evolved state.

\begin{lemma}[Capped smearing]
\label{lem:capsmear}
For $1\le\kappa\le K$, let $\mu_\kappa$ be the Fourier multiplier
with symbol
\begin{equation}
\mu_\kappa(\bm{k}_p):=
\frac{\min\{\norm{\bm{k}_p}^2,\kappa^2\}}{2m_p}.
\end{equation}
Fix a constant $c>0$, independent of $K$ and $\kappa$, and suppose
$\Theta(\ket{v})\le c\Xi$ and
\begin{equation}
\norm{\ind{\norm{\bm{k}_p}>\lambda}(1+A)\ket{v}}
\le c\Xi\lambda^{-1/2},\qquad 1\le\lambda\le\kappa.
\end{equation}
Then
\begin{equation}
\Theta(\mu_\kappa\ket{v})\le C\Xi\kappa^{3/2},
\qquad \norm{\mu_\kappa\ket{v}}\le C\Xi.
\label{eq:capsmear}
\end{equation}
Here $C$ may also depend on $c$.
\end{lemma}

\begin{proofx}
Since $0\le\mu_\kappa\le T_p\le\Ttot$ and these operators
are diagonal in the same basis, the second estimate follows from
$\norm{\mu_\kappa\ket{v}}\le\Theta(\ket{v})$.
To obtain the first estimate, we use the tail bound to improve on
the direct operator-norm estimate, which would cost $\kappa^2$.

We split momentum space into $\norm{\bm{k}_p}\le1$, dyadic
shells between $1$ and $\kappa$, and $\norm{\bm{k}_p}>\kappa$.
On the first region the symbol is at most $(2m_p)^{-1}$, so its
contribution to the squared graph norm is at most $C\Xi^2$.
For a shell
$r<\norm{\bm{k}_p}\le\min\{2r,\kappa\}$ with
$r=1,2,4,\ldots$ and $r<\kappa$, let $S_r$ denote its Fourier
projection. On this shell the symbol is at most $C r^2$, and
the tail bound at the lower endpoint gives
\begin{equation}
\norm{(1+A)\mu_\kappa S_r\ket{v}}
\le C r^2\norm{\ind{\norm{\bm{k}_p}>r}(1+A)\ket{v}}
\le C\Xi r^{3/2}.
\end{equation}
The shells are disjoint and all operators in this expression are
Fourier multipliers. Their squared norm contributions therefore
add, giving at most
$C\Xi^2\sum_{r<\kappa}r^3\le C\Xi^2\kappa^3$.
Truncating the last shell at $\kappa$ covers arbitrary cap scales,
without requiring $\kappa$ to be a power of two.

On the remaining region the capped symbol is constant, so the
single admissible tail estimate at $\lambda=\kappa$ gives
\begin{equation}
\norm{(1+A)\mu_\kappa\ind{\norm{\bm{k}_p}>\kappa}\ket{v}}^2
\le C\kappa^4
\norm{\ind{\norm{\bm{k}_p}>\kappa}(1+A)\ket{v}}^2
\le C\Xi^2\kappa^3.
\end{equation}
Adding the three disjoint contributions and taking a square root
proves the first estimate. In particular, we never use a tail
estimate at a threshold above the cap.
\end{proofx}

The lemma applies to the exact, projected exact, Galerkin, and grid
states by Proposition~\ref{prop:dyn-boot} and
Lemmas~\ref{lem:disc-tails} and~\ref{lem:grid-tails}. Fourier
projection can only decrease their weighted tails. Thus the
capped endpoint has graph norm $C\Xi\kappa^{3/2}$, in place of
the $C\Xi K^{3/2}$ needed for an uncapped register endpoint.

\begin{theorem}[Capped kinetic observable: sublinear secular rates]
\label{thm:mult-cap}
Under the hypotheses of Theorem~\ref{thm:mult-gal}, for every
$t\ge0$ and $K\ge K_0$,
\begin{align}
&\left|\langle\psi(t)|T_p|\psi(t)\rangle
-\langle\psi_N(t)|T_p|\psi_N(t)\rangle\right|\\
&\qquad\le C\Xi^2
 \left[K^{-3}+\min\{tK^{-3/2},\ t^{2/3}K^{-2}\}\right]
=\mathcal O\!\left(N^{-1}
 +\min\{tN^{-1/2},\ t^{2/3}N^{-2/3}\}\right).
\label{eq:mult-cap}
\end{align}
For the grid evolution with any kernel in $\mathcal K_K$,
\begin{align}
&\left|\langle\psi(t)|T_p|\psi(t)\rangle
-\langle\psi_g(t)|T_p|\psi_g(t)\rangle\right|\\
&\qquad\le C\Xi^2
 \left[K^{-3}+\min\{tK^{-1},\ t^{2/3}K^{-5/3}\}\right].
\label{eq:mult-cap-grid}
\end{align}
These bounds use the standing $D(\Hnbo^2)$ assumption on the initial
state without additional regularity. The linear branches recover
Theorem~\ref{thm:mult-gal} and Proposition~\ref{prop:mult-grid}.
\end{theorem}

\begin{proofx}
We first derive a bound for a fixed cap and then choose its scale.
Take $\kappa_0\le\kappa\le K/2$, where $\kappa_0\ge1$ is a
fixed lower threshold large enough to absorb the fixed mass and
lattice constants below. We increase $K_0$ so this interval is
nonempty. The identity
$T_p=\mu_\kappa+(T_p-\mu_\kappa)$ separates the observable
comparison into a capped comparison and the error of making that
replacement at its two endpoints.

\emph{Cost of replacing the observable.}
The multiplier difference is non-negative and satisfies
\begin{equation}
0\le T_p-\mu_\kappa
\le T_p\ind{\norm{\bm{k}_p}>\kappa}.
\end{equation}
For any of the exact or discrete states $\ket{u}$ appearing
in the theorem, the weighted and plain tails therefore give
\begin{align}
0\le\langle u|T_p-\mu_\kappa|u\rangle
&\le C\norm{\ind{\norm{\bm{k}_p}>\kappa}(1+A)\ket{u}}
       \norm{\ind{\norm{\bm{k}_p}>\kappa}\ket{u}}\\
&\le C\Xi^2\kappa^{-1/2}\kappa^{-5/2}
=C\Xi^2\kappa^{-3}.
\end{align}
Applying this to both states in the comparison costs
at most $C\Xi^2\kappa^{-3}$. This error is uniform in time.
The separate projection error for the capped observable is still
at most $C\Xi^2K^{-3}$, because $0\le\mu_\kappa\le T_p$
and Lemma~\ref{lem:mult-basic} applies.

\emph{Galerkin source pairing with the capped observable.}
For either endpoint
$\ket{v}\in\{P\ket{\psi(t)},\ket{\psi_N(t)}\}$, let
$\ket{\varphi_v(s)}=e^{iH_N(t-s)}\mu_\kappa\ket{v}$.
Lemma~\ref{lem:capsmear} gives
\begin{equation}
\Theta(\mu_\kappa\ket{v})\le C\Xi\kappa^{3/2},
\qquad\norm{\mu_\kappa\ket{v}}\le C\Xi.
\end{equation}
To propagate the first bound, we again compare the kinetic and
Hamiltonian graph norms on the retained space. One form of the
needed inverse comparison is
\begin{align}
\Theta(\ket{u})
&\le C\left(\norm{\Hnbo\ket{u}}
                     +m_{\max}\norm{\ket{u}}\right)\\
&\le C\left(\norm{H_N\ket{u}}
       +\norm{(1-P)V_{\mathrm{tot}}\ket{u}}
       +m_{\max}\norm{\ket{u}}\right)\\
&\le C\left(\norm{H_N\ket{u}}
                     +m_{\max}\norm{\ket{u}}\right)
       +\tfrac12\Theta(\ket{u}),
\qquad \ket{u}\in\operatorname{ran}P.
\label{eq:graphN}
\end{align}
In the second line we used $(1-P)\Ttot\ket{u}=0$; in the third
we used the potential-tail bound~\eqref{eq:lam-gen} and chose
$K_0$ large enough to absorb its factor $K^{-1/2}$. The forward
graph bound controls $H_N$ on the capped endpoint. Unitarity and
Hamiltonian-norm conservation then preserve these two input norms,
and the inverse comparison gives
\begin{equation}
\Theta(\ket{\varphi_v(s)})
\le C\Xi(m_{\max}+\kappa^{3/2})
\le C\Xi\kappa^{3/2}.
\end{equation}
The last inequality is ensured by the fixed choice of $\kappa_0$.

We now exploit the location of the omitted exact state, rather
than bounding the entire Duhamel source in norm. Since
$\ket{\delta(s)}\in\operatorname{ran}(1-P)$ and
$\ket{\varphi_v(s)}\in\operatorname{ran}P$, self-adjointness
of the potential gives
\begin{align}
\left|\langle\varphi_v(s)|PV_{\mathrm{tot}}|\delta(s)\rangle\right|
&=\left|\langle\delta(s)|(1-P)V_{\mathrm{tot}}|\varphi_v(s)\rangle\right|\\
&\le\norm{\ket{\delta(s)}}
       \norm{(1-P)V_{\mathrm{tot}}\ket{\varphi_v(s)}}\\
&\le C\Xi K^{-5/2}\cdot
       C\Xi\kappa^{3/2}K^{-1/2}
=C\Xi^2\kappa^{3/2}K^{-3}.
\end{align}
The two factors come from Eq.~\eqref{eq:dyn-t} and the
potential-tail estimate, respectively. Integrating from $0$ to
$t$ bounds each retained-space pairing. Together with replacement
and projection, we obtain the fixed-cap estimate
\begin{equation}
\left|\langle\psi(t)|T_p|\psi(t)\rangle
-\langle\psi_N(t)|T_p|\psi_N(t)\rangle\right|
\le C\Xi^2\left[K^{-3}+\kappa^{-3}
                         +t\kappa^{3/2}K^{-3}\right].
\label{eq:F-fixed-cap-gal}
\end{equation}
This displays the tradeoff: raising the cap reduces the replacement
error and increases the graph-norm cost of the accumulated source.

\emph{Grid source pairing with the capped observable.}
We use the same decomposition, now with
$\ket{v}\in\{\Pp\ket{\psi(t)},\ket{\psi_g(t)}\}$ and
$\ket{\varphi_v(s)}=e^{i\Hgrid(t-s)}\mu_\kappa\ket{v}$.
The grid graph comparison propagates
$\Theta(\ket{\varphi_v(s)})\le C\Xi\kappa^{3/2}$.
Its a priori tail at any fixed fraction of $K$ is therefore
\begin{equation}
t_r(cK;\ket{\varphi_v(s)})
\le C\Xi\kappa^{3/2}K^{-2}.
\end{equation}
The Galerkin-type source pairs by support just as above and costs
$C\Xi^2\kappa^{3/2}K^{-3}$. In the dropped-transfer form,
Eq.~\eqref{eq:Dform}, the two terms are now bounded by
\begin{align}
&CK^{-1/2}\left[
(C\Xi\kappa^{3/2})(C\Xi K^{-5/2})
 +(C\Xi)(C\Xi\kappa^{3/2}K^{-2})\right]\\
&\hspace{2cm}\le C\Xi^2\kappa^{3/2}K^{-5/2}.
\end{align}
The weak-coefficient wrap obeys the same estimate. Its
strong-coefficient part is smaller:
\begin{equation}
CK(C\Xi K^{-5/2})(C\Xi\kappa^{3/2}K^{-2})
\le C\Xi^2\kappa^{3/2}K^{-7/2}.
\end{equation}
Thus the grid analogue of the fixed-cap bound is
\begin{equation}
\left|\langle\psi(t)|T_p|\psi(t)\rangle
-\langle\psi_g(t)|T_p|\psi_g(t)\rangle\right|
\le C\Xi^2\left[K^{-3}+\kappa^{-3}
                         +t\kappa^{3/2}K^{-5/2}\right].
\label{eq:F-fixed-cap-grid}
\end{equation}
Only first-order graph bounds for the capped backward flow are
used here. The second-order grid estimates of
Sec.~\ref{sec:grid-upgrade} provide further information, but we
do not optimise that additional input in this theorem.

\emph{Choosing the cap and covering all times.}
For $t>0$, balancing the replacement term with the accumulated
source in the two fixed-cap bounds gives
\begin{equation}
\kappa_*=
\begin{cases}
(K^3/t)^{2/9},&\text{Galerkin},\\
(K^{5/2}/t)^{2/9},&\text{grid}.
\end{cases}
\end{equation}
Whenever $\kappa_0\le\kappa_*\le K/2$, choosing
$\kappa=\kappa_*$ yields, respectively,
$t^{2/3}K^{-2}$ and $t^{2/3}K^{-5/3}$ for both balanced terms.
The remaining times require the endpoint estimates, since a cap
outside the admitted interval cannot be used.

If $\kappa_*>K/2$, then $t$ is at most a fixed multiple of
$K^{-3/2}$ in the Galerkin case or $K^{-2}$ in the grid case.
The corresponding direct bound already implies the desired
sublinear bound, because in these ranges
\begin{equation}
tK^{-3/2}\le C t^{2/3}K^{-2},\qquad
\text{or, respectively,}\qquad
tK^{-1}\le C t^{2/3}K^{-5/3}.
\end{equation}
Choosing the endpoint cap $K/2$ also recovers the direct powers,
with its replacement cost absorbed into $CK^{-3}$.
At $t=0$ we simply use the projection estimate.

If $\kappa_*<\kappa_0$, then $t$ is at least a fixed positive
multiple of $K^3$ for Galerkin evolution or $K^{5/2}$ for grid
evolution. All states have norm at most one, and
Lemma~\ref{lem:mult-basic} bounds their kinetic images uniformly.
Cauchy--Schwarz therefore gives a uniform energy-error bound
$C\Xi^2$. In this regime $t^{2/3}K^{-2}$, respectively
$t^{2/3}K^{-5/3}$, is bounded below by a positive fixed constant,
so the same sublinear bound follows. No cap below $\kappa_0$ is
needed.

We have proved each sublinear bound for all $t\ge0$. Taking its
minimum with the independently valid direct bound, and enlarging
the common constant, proves Eqs.~\eqref{eq:mult-cap}
and~\eqref{eq:mult-cap-grid}.
\end{proofx}

\begin{corollary}[Stopping power: window law]
\label{cor:stopping2}
In the setting of Corollary~\ref{cor:stopping}, for every $T>0$,
\begin{equation}
|\Delta\widehat S|\le\frac{C\Xi^2}{v_p}
\begin{cases}
\min\{K^{-3/2},\ T^{-1/3}K^{-2}\}+2K^{-3}/T,
 &\text{Galerkin},\\
\min\{K^{-1},\ T^{-1/3}K^{-5/3}\}+2K^{-3}/T,
 &\text{grid}.
\end{cases}
\label{eq:stopping2}
\end{equation}
\end{corollary}

\begin{proofx}
As before, the secant error is at most the sum of the two kinetic
endpoint errors divided by $v_pT$. At time zero only the projection
term remains; at time $T$, apply Theorem~\ref{thm:mult-cap}.
Dividing the capped term $T^{2/3}$ by the window length produces
$T^{-1/3}$, giving Eq.~\eqref{eq:stopping2}.
\end{proofx}

At fixed cell size, $N=\Theta(K^3)$, so the Galerkin result in
particular gives
\begin{equation}
|\Delta\widehat S|
=\mathcal O\bigl(T^{-1/3}N^{-2/3}+T^{-1}N^{-1}\bigr)
\qquad(T>0),
\end{equation}
with the fixed physical parameters and regularity budget understood.
For $T\gtrsim K^{-3/2}$ the endpoint projection contribution is
absorbed into the first term, leaving
$\mathcal O(T^{-1/3}N^{-2/3})$. This is
$\mathcal O(N^{-5/6})$ at $T\sim K^{3/2}$ and
$\mathcal O(N^{-1})$ for $T\gtrsim K^3$.
The grid bound similarly guarantees $\mathcal O(N^{-1})$ for
$T\gtrsim K^4$. These are sufficient window lengths from the
stated estimates, not necessary conditions. All statements compare
the discrete and exact secants on the same window; none requires
an additional assumption about a long-time limit of the stopping
power.

The capped estimates complement the logarithmic theorem because
they bound the accumulated error by a sublinear power of time,
even though their cutoff powers are weaker. Their mechanism is to
bound the omitted high-momentum contribution and balance it against
the graph norm of the capped backward state. Theorem~\ref{thm:kinetic-log}
instead separates low and high Hamiltonian energies to compare the
original kinetic observable directly. Both arguments retain the
same measured observable; capping changes only the analysis.

\section{Projection flux and resonant-transfer estimates}
\label{app:flux}

We seek a bound on the change in kinetic-energy production caused by
removing momenta outside the Galerkin basis. This is one part of the
kinetic-energy error: here we compare the exact state with its projection
at each time, whereas Appendix~\ref{app:source} also compares the projected
state with the state evolved by the finite Hamiltonian. Our task is to
control the time integral of this projection contribution. A pointwise
bound alone loses a power of the cutoff; to recover it, we use cancellation
between transfers with different kinetic energies.

We first establish the pointwise estimate and show how a single
integration by parts exploits that cancellation. We then count transfers
whose kinetic mismatch is small. Finally, we combine these counts with
three successive commutator identities in a finite auxiliary Fourier box.
The resulting bound is
$C_\varepsilon\mathcal D^2(1+t)K^{-3+\varepsilon}$ for every
$\varepsilon>0$ and $t\ge0$, under the fixed-cell and cutoff conditions
of Theorem~\ref{thm:flux3}. The auxiliary box is used only in the proof;
it does not enlarge the simulated basis.

\subsection{The projection flux and the first cancellation}

\paragraph{Momentum and constant conventions.}
We distinguish the shape of the retained basis from the norm used to
measure a momentum. In Appendices~\ref{app:flux} and~\ref{app:source}, an
unadorned vector norm is Euclidean. The Galerkin projector $P$ retains
configurations for which every particle momentum lies in the cube
$\norm{\bm{k}_a}_\infty\le K$, and $R:=1-P$ retains the discarded configurations.
To identify transfers that start well inside the retained region, we use
the auxiliary bulk projector
\begin{equation}
 Q=\prod_a\ind{\norm{\bm{k}_a}_2\le K/4}.
 \label{eq:review-bulk-convention}
\end{equation}
Thus $Q$ is a product of Euclidean balls, with $Q\le P$. This distinction
matters in the gap estimate: leaving a retained cube forces the Euclidean
norm of that register to exceed $K$, so Lemma~\ref{lem:gap} applies to
every bulk-to-exterior crossing. A Euclidean high-momentum projector such as $\ind{>\lambda}$, when
written without a register index, retains configurations in which at
least one register has momentum norm greater than $\lambda$. On $\operatorname{ran}P$ Euclidean tails vanish at
$\lambda\ge\sqrt3K$, whereas cube tails vanish at $\lambda\ge K$.
Put $g=2\pi/L$. The reciprocal lattice is $g\mathbb Z^3$: transfer variables in
kinetic mismatch formulas are physical momenta, and constants in lattice
counts may depend on $L$. We put $m_{\max}=\max_a m_a$ and
$\mu_{ab}^{-1}=m_a^{-1}+m_b^{-1}$. A displayed reduced mass refers to the
interacting pair; sums over pairs are absorbed into a polynomial in the
fixed charges, masses and $\eta_{\rm tot}$. Constants $C,C_\varepsilon$ are
independent of $K,t$ and of the initial state apart from its regularity
budget $\Xi$. Dependence on the positive exponent is allowed in
$C_\varepsilon$. A fixed lower cutoff $K_0$ may depend on the other fixed
data. We retain the homogeneous graph norm of the initial state
$\mathcal D=\|\ket{\psi_0}\|+\|H\ket{\psi_0}\|+\|H^2\ket{\psi_0}\|$;
$\Xi=C_0m_{\max}\mathcal D$ is the regularity budget in
Eq.~\eqref{eq:Xi}. We use the same fixed-system constant $C_0$ in both
appendices, increasing it once if needed for the estimates of the evolved
state below.
In particular, scaling the state scales both budgets by the same amount;
we never impose a positive lower bound on $\Xi$ for unnormalised states.
Let $\ket{\vec{\bm{k}}}$ denote the orthonormal many-particle plane-wave
basis, with any spin labels included implicitly. We write the momentum
amplitudes of a state as
\begin{equation}
 u_{\vec{\bm{k}}}:=\langle\vec{\bm{k}}|u\rangle,\qquad
 \ket{u}=\sum_{\vec{\bm{k}}}u_{\vec{\bm{k}}}\ket{\vec{\bm{k}}}.
\end{equation}
The amplitude of $B\ket{u}$ is $\langle\vec{\bm{k}}|B|u\rangle$.
We write $\mathcal E:=1+A$, where $A$ is the momentum-space
multiplier representing the total kinetic energy, so
$T\ket{\vec{\bm{k}}}=A(\vec{\bm{k}})\ket{\vec{\bm{k}}}$. In this representation,
\begin{equation}
 \Theta(\ket{v}):=\|\mathcal E\ket{v}\|
 =\left(\sum_{\vec{\bm{k}}}[1+A(\vec{\bm{k}})]^2|v_{\vec{\bm{k}}}|^2\right)^{1/2}.
\end{equation}
In finite Fourier sums below, $\vec{\bm{p}}$ may also label a momentum
configuration, with $u_{\vec{\bm{p}}}:=\langle \vec{\bm{p}}|u\rangle$. We write $|u|$ for
the nonnegative amplitude profile $(|u_{\vec{\bm{p}}}|)_{\vec{\bm{p}}}$, with the same
convention for other states. Likewise, $|Bu|$ denotes the profile
$(|\langle \vec{\bm{p}}|B|u\rangle|)_{\vec{\bm{p}}}$.
All estimates below are at fixed cell size.

\label{sec:stopping-flux}

We begin with the rate of change of the measured particle's kinetic
energy $T_p$. For a Coulomb potential, differentiating the potential
formally would introduce a singular force. We can avoid that additional
domain issue by using the following sesquilinear form, which is
conjugate-linear in its first argument and linear in its second. For
$\ket{u},\ket{v}\in D(T)$, set
\begin{equation}
 \mathfrak G(\ket{u},\ket{v})
 :=i\big(\langle V_{\mathrm{tot}}u\,|\,T_pv\rangle-\langle T_pu\,|\,V_{\mathrm{tot}}v\rangle\big),
 \qquad G=i[V_{\mathrm{tot}},T_p]\quad\text{in this form sense}.
 \label{eq:flux-id}
\end{equation}
The form is Hermitian, meaning
$\mathfrak G(\ket{v},\ket{u})=\overline{\mathfrak G(\ket{u},\ket{v})}$. For the exact time-dependent state
$\ket{\psi(s)}=e^{-iHs}\ket{\psi_0}$ with $\ket{\psi_0}\in D(H^2)$, and the finite
Galerkin state $\ket{\psi_N}$, the Heisenberg identity gives
\begin{equation}
 \frac{d}{dt}\langle \psi|T_p|\psi\rangle=\mathfrak G(\ket{\psi},\ket{\psi}),
 \qquad
 \frac{d}{dt}\langle \psi_N|T_p|\psi_N\rangle
       =\mathfrak G(\ket{\psi_N},\ket{\psi_N}).
\end{equation}
In this appendix $\langle u|G|v\rangle$ denotes $\mathfrak G(\ket{u},\ket{v})$.
Likewise $\langle u|i[V_{\mathrm{tot}},Y]|v\rangle$ denotes
$i(\langle V_{\mathrm{tot}}u\,|\,Yv\rangle-\langle Yu\,|\,V_{\mathrm{tot}}v\rangle)$ whenever the indicated
factors exist; no claim that $Y$ preserves $D(T)$ is implicit.
Here a bra such as $\langle V_{\mathrm{tot}}u|$ is the adjoint of
$V_{\mathrm{tot}}\ket{u}$. Keeping the operated states on their
respective sides of the inner product avoids asserting that a
product of the unbounded operators is defined.
The part of this rate that is removed by projecting the exact state
is $\mathfrak G(\ket{\psi},\ket{\psi})-\mathfrak G(P\ket{\psi},P\ket{\psi})$. We therefore define
its time integral by
\begin{equation}
 \mathcal F_K(t)=\int_0^t
 \big[\mathfrak G(\ket{\psi(s)},\ket{\psi(s)})-
           \mathfrak G(P\ket{\psi(s)},P\ket{\psi(s)})\big]ds.
\end{equation}
Notice that $P\ket{\psi(s)}$ is obtained by projecting the exact state at time
$s$; it is generally different from $\ket{\psi_N(s)}$. The quantity
$\mathcal F_K(t)$ therefore controls the projection contribution only.
The remaining comparison between these two retained states is the
source problem treated in Appendix~\ref{app:source}.

\begin{lemma}[Hardy bound for the Coulomb commutator form]
\label{lem:hardy}
For the periodic Coulomb Hamiltonian, including its clamped-nucleus
terms, and $\ket{u},\ket{v}\in D(A)$,
\begin{equation}
 |\mathfrak G(\ket{u},\ket{v})|\le C_G\Big[
   \|(1+A)^{1/2}\ket{u}\|\,\|(1+A)\ket{v}\|
  +\|(1+A)\ket{u}\|\,\|(1+A)^{1/2}\ket{v}\|\Big].
 \label{eq:hardy-a}
\end{equation}
Here $C_G$ depends only on the cell, $\eta_{\rm tot}$, the charges and
the masses, not on the initial-state budget $\mathcal D$.
\end{lemma}
\begin{proofx}
Use the definition \eqref{eq:flux-id} directly. The periodic Coulomb
Hardy estimate, summed over the quantum-pair and clamped-nucleus
interactions, gives $\|V_{\mathrm{tot}}\ket{z}\|\le C_G\|(1+A)^{1/2}\ket{z}\|$.
Also $0\le T_p\le A$ implies $\|T_p\ket{z}\|\le\|(1+A)\ket{z}\|$.
Cauchy--Schwarz in each of the two terms of
$i(\langle V_{\mathrm{tot}}u\,|\,T_pv\rangle-\langle T_pu\,|\,V_{\mathrm{tot}}v\rangle)$ proves
\eqref{eq:hardy-a}. Thus no formula involving separate gradients of
the singular potential is needed.
\end{proofx}

To use the Hardy bound, we separate the retained component
$\ket{a}=P\ket{\psi(s)}$ from the discarded component $\ket{\delta}=R\ket{\psi(s)}$. Hermiticity
gives the exact decomposition
\begin{equation}
 \mathfrak G(\ket{\psi},\ket{\psi})-\mathfrak G(\ket{a},\ket{a})
 =\mathfrak G(\ket{\delta},\ket{\delta})
    +2\operatorname{Re}\mathfrak G(\ket{\delta},\ket{a}).
\end{equation}
The tail--tail term is small because both arguments have large momentum.
The more difficult term couples the discarded component to the part of
$\ket{a}$ well inside the basis. The next theorem separates these two scales.

\begin{theorem}[Projection flux bounds]
\label{thm:flux}
Let $\ket{\psi_0}\in D(H^2)$, with homogeneous graph norm $\mathcal D$, and
put $\ket{a}=P\ket{\psi(s)}$, $\ket{\delta}=R\ket{\psi(s)}$. At every fixed positive fraction
of the cutoff the restricted-state budgets give
\begin{equation}
 |\mathfrak G(\ket{\delta},\ket{\delta})|+
 2|\mathfrak G(\ket{\delta},\ind{>K/2}\ket{a})|\le C\Xi^2K^{-2}.
 \label{eq:flux-nc}
\end{equation}
For the full projection flux,
\begin{equation}
 |\mathfrak G(\ket{\psi},\ket{\psi})-\mathfrak G(\ket{a},\ket{a})|
 \le C\Xi^2K^{-1}.
 \label{eq:flux-bd}
\end{equation}
The constants are uniform in $s$, above a fixed lattice-scale cutoff.
\end{theorem}
\begin{proofx}
The momentum-tail estimates for the evolved state give
$\|\ket{\delta}\|\le C\Xi K^{-5/2}$ and
$\|\mathcal E\ket{\delta}\|\le C\Xi K^{-1/2}$. Interpolating between them,
\begin{equation}
 \|\mathcal E^{1/2}\ket{\delta}\|^2
 \le\|\ket{\delta}\|\,\|\mathcal E\ket{\delta}\|
 \le C\Xi^2K^{-3}.
\end{equation}
The same three estimates hold for the part of $\ket{a}$ restricted beyond any
fixed positive fraction of $K$. Thus each product in the Hardy form
bound has size $\Xi^2K^{-3/2}K^{-1/2}=\Xi^2K^{-2}$, which proves
\eqref{eq:flux-nc}. For the remaining cross term, the two expressions
in the definition of $\mathfrak G(\ket{\delta},\ket{a})$ are bounded by
\begin{equation}
 \|V_{\mathrm{tot}}\ket{\delta}\|\|T_p\ket{a}\|\le C\Xi^2K^{-3/2},\qquad
 \|T_p\ket{\delta}\|\|RV_{\mathrm{tot}}\ket{a}\|\le C\Xi^2K^{-1}.
\end{equation}
In the second expression we inserted $R$ because $T_p\ket{\delta}$ is
supported outside the basis. The potential-tail estimate applied to $\ket{a}$
gives $\|RV_{\mathrm{tot}}\ket{a}\|\le C\Xi K^{-1/2}$, since
$\Theta(\ket{a})\le C\Xi$. Multiplication by the
$K^{-1/2}$ bound on $\|T_p\ket{\delta}\|$ produces the weaker $K^{-1}$
term. Adding the quadratic tail contribution proves \eqref{eq:flux-bd}.
This calculation identifies exactly where pointwise estimation loses a
power of $K$.
\end{proofx}

Integrating \eqref{eq:flux-bd} directly would give a bound proportional
to $tK^{-1}$. That estimate takes absolute values before accounting for
the oscillation generated by a change in kinetic energy. We next show
that a transfer starting in $Q$ and ending outside $P$ has a uniformly
large kinetic mismatch. We will divide by that mismatch to write the
corresponding flux as a time derivative plus a smaller remainder.

\begin{lemma}[Uniform off-resonance of bulk-fed transfers]
\label{lem:gap}
Let a transfer $\bm{\nu}$ act on the pair $(p, b)$ by
$\bm{k}_p \mapsto \bm{k}_p + \bm{\nu}$, $\bm{k}_b \mapsto \bm{k}_b - \bm{\nu}$, with kinetic
mismatch
$\omega_{\bm{\nu}}(\vec{\bm{k}}) = \frac{ 2 \bm{k}_p\!\cdot\!\bm{\nu} + \norm{\bm{\nu}}^2 }{2
m_p} + \frac{ \norm{\bm{\nu}}^2 - 2 \bm{k}_b\!\cdot\!\bm{\nu} }{2 m_b}$. If
$\norm{\bm{k}_p}, \norm{\bm{k}_b} \le K/4$ and the image configuration
leaves the zone ($\norm{\bm{k}_p + \bm{\nu}} > K$ or
$\norm{\bm{k}_b - \bm{\nu}} > K$), then
$\norm{\bm{\nu}} > 3K/4$ and
\begin{equation}
\omega_{\bm{\nu}}(\vec{\bm{k}}) \;\ge\; \frac{3 K^2}{32\, \mu_{pb}}
\qquad \text{and} \qquad
\omega_{\bm{\nu}}(\vec{\bm{k}}) \;\ge\; \frac{\norm{\bm{\nu}}^2}{20\, \mu_{pb}},
\qquad
\mu_{pb}^{-1} := m_p^{-1} + m_b^{-1},
\label{eq:gap}
\end{equation}
and the constant $3/32$ is sharp over configurations and mass ratios.
The two quadratic recoil terms add, so a bulk-fed transfer cannot
cancel its kinetic mismatch by moving the two particles in opposite
directions. Its mismatch is at least of order $K^2/m_{\max}$,
regardless of how far beyond the cutoff it lands. Consequently, a
near-resonant crossing cannot have a bulk endpoint. Such crossings
belong to the part where both configurations have a register tail,
which is controlled by the $K^{-2}$ estimate in
Theorem~\ref{thm:flux}.
\end{lemma}

\begin{proofx}
We keep the exit condition throughout the minimisation, because it
correlates the input momentum with the transfer direction. The substitution
$(\bm{k}_p, m_p) \leftrightarrow (-\bm{k}_b, m_b)$ leaves
$\omega_{\bm{\nu}}$ invariant and exchanges the two exit conditions, so
assume $\norm{\bm{k}_p + \bm{\nu}} > K$. Write $n := \norm{\bm{\nu}}$ and
$a := (\bm{\nu}/\|\bm{\nu}\|)\cdot \bm{k}_p$; since $(\bm{\nu}/\|\bm{\nu}\|)\cdot \bm{k}_b\le K/4$,
\begin{equation}
\omega_{\bm{\nu}} \;\ge\; \frac{n^2}{2 \mu_{pb}}
+ \frac{n\, a}{m_p} - \frac{n K}{4 m_b}.
\end{equation}
The naive worst case $a = -K/4$ is not attainable near the
threshold: the exit condition
$\norm{\bm{k}_p}^2 + 2 n a + n^2 > K^2$ with $\norm{\bm{k}_p} \le K/4$
forces $n>3K/4$ and gives the lower bound
$a\ge a_0(n):=\max\{-K/4,\;
(15K^2/16-n^2)/(2n)\}$, the two branches meeting at
$n = 5K/4$. We first consider $3K/4<n\le5K/4$, where the exit condition gives the
active branch of $a_0(n)$. Substituting that branch and using
$\mu_{pb}^{-1}-m_p^{-1}=m_b^{-1}$, we obtain
\begin{equation}
 \frac{n^2}{2\mu_{pb}}+\frac{na_0(n)}{m_p}-\frac{nK}{4m_b}
 =\frac{n(n-K/2)}{2m_b}+\frac{15K^2}{32m_p}.
\end{equation}
The first term increases with $n$ in this range. Taking its value at the
lower endpoint therefore gives
\begin{equation}
 \omega_{\bm{\nu}}\ge\frac{3K^2}{32m_b}+\frac{15K^2}{32m_p}
 \ge\frac{3K^2}{32\mu_{pb}}.
\end{equation}
This proves the cutoff-scale bound in the first range. To obtain the
transfer-scale bound as well, we use $n^2\le25K^2/16$, which gives
$\omega_{\bm{\nu}}\ge3n^2/(50\mu_{pb})\ge n^2/(20\mu_{pb})$.

We next consider $n\ge5K/4$. Here the constraint $a\ge-K/4$ is the
active branch, and substitution gives
\begin{equation}
 \omega_{\bm{\nu}}\ge\frac{n(2n-K)}{4\mu_{pb}}
 \ge\frac{15K^2}{32\mu_{pb}},\qquad
 \omega_{\bm{\nu}}\ge\frac{3n^2}{10\mu_{pb}}.
\end{equation}
Both are stronger than the corresponding claims in \eqref{eq:gap}.
Together, the two ranges prove the required lower bounds.

It remains to explain why the coefficient $3/32$ cannot be increased
uniformly over configurations and mass ratios. In the first range,
consider the boundary configuration
\begin{equation}
 \bm{k}_p=a_0(n)\bm{\nu}/n+(K^2/16-a_0(n)^2)^{1/2}\bm{e}_\perp,
 \qquad \bm{k}_b=(K/4)\bm{\nu}/n,
\end{equation}
where $\bm{e}_\perp$ is a unit vector perpendicular to $\bm{\nu}$. It has
$\|\bm{k}_p\|=K/4$ and $\|\bm{k}_p+\bm{\nu}\|=K$. For fixed $n>3K/4$ sufficiently
close to $3K/4$, increase the longitudinal component of $\bm{k}_p$ slightly
and decrease its perpendicular component to keep its norm $K/4$.
This makes the exit strict, and the mismatch approaches the boundary
value as the perturbation tends to zero. That value satisfies
\begin{equation}
 \frac{\mu_{pb}\omega_{\bm{\nu}}}{K^2}
 \longrightarrow\frac{3m_p+15m_b}{32(m_p+m_b)}
 \quad\text{as }n\downarrow3K/4.
\end{equation}
Taking $m_b/m_p\to0$ then gives $3/32$. These continuous
configurations also establish the uniform lattice sharpness: first
choose strict interior margins for the input bounds and the strict
exit, then approximate the scaled momenta on the fixed reciprocal
lattice as $K/g\to\infty$. The rounding errors divided by $K$ tend
to zero, so the same limiting ratio is attained. The mirrored
configuration covers the opposite mass ordering.

For comparison, when the masses are equal the constrained infimum is
$9K^2/(32\mu_{pb})$. The difference explains why the exit condition
must be retained when claiming a sharp bound uniform in the mass
ratio: the unconstrained estimate
$\omega\ge\|\bm{\nu}\|^2/(2\mu)-K\|\bm{\nu}\|/(4\mu)$ does not use that
condition and is especially loose for unequal masses.
\end{proofx}

We now turn the gap into an identity for the time integral. In a momentum
basis the commutator with $T$ multiplies the $(\vec{\bm{q}},\vec{\bm{p}})$ matrix element of an
operator by $T(\vec{\bm{q}})-T(\vec{\bm{p}})$. We can therefore recover the bulk-crossing force
by choosing an auxiliary Hermitian operator $Y$ whose matrix elements
are divided by this energy difference. Applying the Heisenberg identity
to $Y$ replaces the original flux by endpoint expectations and a new
commutator with the potential. This exact integration by parts is the
elimination step used below; it is not an approximation to the dynamics.
Here $\vec{\bm{p}},\vec{\bm{q}}$ label complete momentum configurations when used as matrix
indices, whereas the $p$ in $T_p$ continues to label the measured particle.

\begin{proposition}[Elimination of the bulk-fed force]
\label{prop:elim}
Let $G_b=RGQ+QGR$ in the form sense, with
$\omega(\vec{\bm{q}},\vec{\bm{p}}):=T(\vec{\bm{q}})-T(\vec{\bm{p}})$. Let $Y$ have the bulk-crossing matrix
element $G_b(\vec{\bm{q}},\vec{\bm{p}})/(i\omega(\vec{\bm{q}},\vec{\bm{p}}))$ and its Hermitian reverse,
and let its other blocks vanish. Then $i[T,Y]=G_b$ as forms on $D(T)$ and
\begin{equation}
 \int_0^t\mathfrak G_b(\ket{\psi(s)},\ket{\psi(s)})\,ds
 =\big[\langle \psi(s)|Y|\psi(s)\rangle\big]_0^t
   -\int_0^t\mathfrak V_Y(\ket{\psi(s)},\ket{\psi(s)})\,ds,
 \label{eq:elim}
\end{equation}
where $\mathfrak V_Y(\ket{u},\ket{v})=i(\langle V_{\mathrm{tot}}u\,|\,Yv\rangle-
\langle Yu\,|\,V_{\mathrm{tot}}v\rangle)$. In particular
\begin{equation}
 \|YQ\ket{v}\|\le CK^{-1/2}\Theta(\ket{v}),\qquad
 |\langle \psi|Y|\psi\rangle|\le C\Xi^2K^{-3}.
 \label{eq:elim-bdry}
\end{equation}
A coarse bound on the remainder is
\begin{equation}
 |\mathfrak V_Y(\ket{\psi},\ket{\psi})|\le C\Xi^2K^{-1}.
 \label{eq:elim-rem}
\end{equation}
\end{proposition}
\begin{proofx}
On a bulk crossing $\|\bm{\nu}\|\ge3K/4$ and
$\omega\ge\|\bm{\nu}\|^2/(20\mu)$ by Lemma~\ref{lem:gap}. The force
matrix element contains the kinetic change of the measured particle,
which is at most $C\|\bm{\nu}\|^2$ for a bulk input. Dividing by the mismatch
therefore leaves a kernel dominated by $C|\hat w(\bm{\nu})|$. To obtain \eqref{eq:elim-bdry}, one must use this
majorant \emph{on the bulk-to-exterior block}. Fix the conserved total momentum of the interacting pair and all
spectator momenta and spin labels. Only the three-dimensional relative
momentum then varies; we call this set of configurations a channel
fibre. Cauchy--Schwarz in the input with weight $(1+A)^2$ gives
\begin{equation}
 \sum_{\vec{\bm{q}}\in R}\left|\sum_{\vec{\bm{p}}\in Q}Y(\vec{\bm{q}},\vec{\bm{p}})v_{\vec{\bm{p}}}\right|^2
 \le \Theta(\ket{v})^2
       \sum_{\vec{\bm{p}}\in Q}(1+A(\vec{\bm{p}}))^{-2}
       \sum_{\vec{\bm{q}}\in R}|Y(\vec{\bm{q}},\vec{\bm{p}})|^2
 \le CK^{-1}\Theta(\ket{v})^2.
\end{equation}
The input weight is summable on the three-dimensional fibre and the
squared transfer tail is $\sum_{\|\bm{\nu}\|\ge3K/4}\|\bm{\nu}\|^{-4}=O(K^{-1})$.
After summing the squared bounds over spectator and spin variables,
we obtain the first estimate in \eqref{eq:elim-bdry}. Since $Y$ has only
$RQ$ and $QR$ blocks, its expectation is
$2\operatorname{Re}\langle R\psi\,|\,YQ\psi\rangle$. Consequently,
\begin{equation}
 |\langle \psi|Y|\psi\rangle|
 \le2\|R\ket{\psi}\|\,\|YQ\ket{\psi}\|
 \le C\Xi^2K^{-5/2}K^{-1/2}
 =C\Xi^2K^{-3}.
\end{equation}
The endpoint gain thus comes from pairing a small discarded-state tail
with the bulk-to-exterior action of $Y$. We do not require an
operator-norm bound on $Y$ acting on arbitrary high-frequency input.

The same argument includes a clamped-nucleus interaction. Only the
measured register moves in such a channel; the pair-gap calculation
reduces to its one-register version by setting $m_b^{-1}=0$. Its
kinetic numerator equals its mismatch, so division leaves the nuclear
Coulomb coefficient itself, up to sign. A translated nucleus contributes
a phase of modulus one, which does not change the bound. We sum over
the three momentum coordinates of that register, fixing all spectators,
and obtain the same weighted estimate and endpoint bound.

We next justify the commutator identity without imposing a domain
condition on $Y\ket{\psi}$. For each fixed $K$, the space selected by $Q$ is
finite dimensional and $\Theta(Q\ket{v})\le C_K\|\ket{v}\|$. The bound for $YQ$
therefore makes $Y=RYQ+QYR$ a bounded Hermitian operator, although this
argument alone gives no useful cutoff-uniform operator norm. On finite
Fourier sums its matrix elements give
\begin{equation}
 i\big(\langle Tu\,|\,Yv\rangle-\langle Yu\,|\,Tv\rangle\big)
 =\mathfrak G_b(\ket{u},\ket{v}).
\end{equation}
Both sides are continuous in the $D(T)$ graph norm, by boundedness of
$Y$ and the Hardy form estimate. Since finite Fourier sums are dense
in that graph norm, the identity holds for $\ket{u},\ket{v}\in D(T)$.
The exact state is strongly differentiable in time in $L^2$, so boundedness
of $Y$ permits direct differentiation of its expectation:
\begin{equation}
 \frac{d}{ds}\langle \psi|Y|\psi\rangle
 =i\big(\langle H\psi\,|\,Y\psi\rangle-
               \langle Y\psi\,|\,H\psi\rangle\big)
 =\mathfrak G_b(\ket{\psi},\ket{\psi})+\mathfrak V_Y(\ket{\psi},\ket{\psi}).
\end{equation}
Integration gives \eqref{eq:elim}. At no point do we require
$Y\ket{\psi}\in D(T)$.

Finally, we bound the potential remainder. Because $Y$ has only the
$RQ$ and $QR$ blocks,
\begin{equation}
 |\mathfrak V_Y(\ket{\psi},\ket{\psi})|
 \le2\|RV_{\mathrm{tot}}\ket{\psi}\|\,\|YQ\ket{\psi}\|
   +2\|R\ket{\psi}\|\,\|YQV_{\mathrm{tot}}\ket{\psi}\|.
\end{equation}
The first product is at most
$C(\Xi K^{-1/2})(\Xi K^{-1/2})=C\Xi^2K^{-1}$.
For the second, the kinetic weight on $Q$ is bounded by $CK^2$ for
cutoffs above the fixed lower threshold. Hardy therefore gives
\begin{equation}
 \Theta(QV_{\mathrm{tot}}\ket{\psi})
 \le CK^2\|V_{\mathrm{tot}}\ket{\psi}\|\le C\Xi K^2,
 \qquad
 \|YQV_{\mathrm{tot}}\ket{\psi}\|\le C\Xi K^{3/2}.
\end{equation}
Pairing this with $\|R\ket{\psi}\|\le C\Xi K^{-5/2}$ gives the same
$C\Xi^2K^{-1}$ bound. This proves \eqref{eq:elim-rem}. We have
obtained a $K^{-3}$ endpoint, but this coarse remainder would still
accumulate at rate $K^{-1}$. The next argument retains the phase
cancellation within that remainder to improve the rate.
\end{proofx}

\subsection{Cancellation in the first potential remainder}

The coarse estimate above treats the two potential factors separately.
To improve it, we keep the two successive transfers together and use
the fact that only the imaginary part of their expectation contributes
to a commutator. We need two ingredients: a way to estimate a pair
kernel from its transfer coefficients, and the cancellation obtained
by reversing a bulk-to-exterior-to-bulk path. We establish them in
that order, then apply them to $\mathfrak V_Y$.

The first ingredient concerns a three-dimensional momentum sum within
one interacting pair. Fixing the pair total momentum and all spectator
registers leaves just one free momentum variable. Weighted
Cauchy--Schwarz in that variable lets us control a form even when its
transfer coefficients do not decay. This is the content of the next
lemma; it does not require summability over the full many-particle
momentum lattice.

\begin{lemma}[Pair-correlation trace bounds]
\label{lem:paird}
Let $V$ be a pair operator on registers $(\alpha, \beta)$,
\begin{equation}
V \;=\; \sum_{\bm{\tau} \ne 0} c(\bm{\tau})\,
e^{-i \bm{\tau} \cdot (\bm{x}_\alpha - \bm{x}_\beta)}\,\sigma_{\bm{\tau}}(\vec{\bm{k}}),
\end{equation}
with real transfer coefficients $c$ and measurable symbols
$\norm{\sigma_{\bm{\tau}}}_\infty \le 1$. We evaluate the symbol at the
input momentum configuration. With the Fourier convention
\eqref{eq:fourier-sign}, the summand acts as
\begin{equation}
 \ket{\ldots,\bm{k}_\alpha,\ldots,\bm{k}_\beta,\ldots}
 \longmapsto
 c(\bm{\tau})\sigma_{\bm{\tau}}(\vec{\bm{k}})
 \ket{\ldots,\bm{k}_\alpha+\bm{\tau},\ldots,\bm{k}_\beta-\bm{\tau},\ldots}.
\end{equation}
The negative exponential and its position to the left of the symbol
in the operator formula implement exactly this input-to-output
displacement. The following analytic bounds are stated on finite Fourier
blocks with the channel kernel and support displayed here; the
formal flux argument uses the corresponding restricted finite-block
forms, not a universal class of operators. Then for every
$\varepsilon \in (0, \tfrac14]$ and all $\ket{u},\ket{v}$ in these blocks:
\begin{align}
| c(\bm{\tau}) | \le h \ \ \forall \bm{\tau}
\quad &\Longrightarrow \quad
| \langle u|V|v\rangle | \;\le\; C_\varepsilon\, h\,
\norm{(1+A)^{3/4+\varepsilon}\ket{u}}\,
\norm{(1+A)^{3/4+\varepsilon}\ket{v}},
\label{eq:paird-flat}
\\
| c(\bm{\tau}) | \le \frac{h \norm{\bm{\tau}}}{K} \ \ \forall \bm{\tau}
\quad &\Longrightarrow \quad
| \langle u|V|v\rangle | \;\le\; \frac{C_\varepsilon\, h}{K}
\Big[ \norm{(1+A)^{5/4+\varepsilon}\ket{u}}\,
\norm{(1+A)^{3/4+\varepsilon}\ket{v}}
+ (\ket{u} \leftrightarrow \ket{v}) \Big].
\label{eq:paird-grad}
\end{align}
At fixed cell size, the constants depend on $\varepsilon$, the masses
and the reciprocal-lattice spacing, but not on the finite block.
When summing over interaction pairs, we absorb their number and charge
factors into the fixed-system prefactor.
\end{lemma}
The symbol $\sigma_{\bm{\tau}}$ is a momentum-dependent multiplier
accompanying the transfer $\bm{\tau}$: it may vary with the configuration,
whereas $c(\bm{\tau})$ depends only on the transfer.

\begin{proofx}
We first remove the bounded symbols by taking absolute values. Writing
$\rho(\bm{\tau})$ for the sum of $|u_{\vec{\bm{k}}'}||v_{\vec{\bm{k}}}|$ over
configuration pairs related by the transfer $\bm{\tau}$, we have
\begin{equation}
 |\langle u|V|v\rangle|\le\sum_{\bm{\tau}} |c(\bm{\tau})|\rho(\bm{\tau}).
\end{equation}
For each fixed pair total momentum and set of spectators, we may use
the momentum $\bm{q}$ of register $\alpha$ as the free coordinate. The
output coordinate is $\bm{q}'=\bm{q}+\bm{\tau}$; thus summing over $(\bm{q},\bm{\tau})$ is
equivalent to summing independently over $(\bm{q},\bm{q}')$. Consequently,
\begin{equation}
 \sum_{\bm{\tau}\ne0}\rho(\bm{\tau})
 \le\sum_{\text{fibres}}
       \left(\sum_{\bm{q}'}|u_{\bm{q}'}|\right)
       \left(\sum_{\bm{q}}|v_{\bm{q}}|\right).
\end{equation}
The inequality allows the additional nonnegative diagonal terms
$\bm{q}'=\bm{q}$, corresponding to zero transfer. Here $u_{\bm{q}'}$ and $v_{\bm{q}}$
denote the amplitudes of the corresponding
full configurations, with the fixed coordinates labelling the fibre
suppressed. For either factor, weighted Cauchy--Schwarz gives
\begin{equation}
 \sum_{\bm{q}}|v_{\bm{q}}|
 \le\left(\sum_{\bm{q}\in g\Z^3}(1+\|\bm{q}\|)^{-3-4\varepsilon}\right)^{1/2}
    \left(\sum_{\bm{q}}(1+\|\bm{q}\|)^{3+4\varepsilon}|v_{\bm{q}}|^2\right)^{1/2}.
\end{equation}
The first sum converges because the fibre is three dimensional and
$\varepsilon>0$. The square root of the weight in the second sum is
bounded by $C_m(1+A)^{3/4+\varepsilon}$, since the total kinetic
energy controls the momentum of each register. Applying the same
estimate to $\ket{u}$ and then Cauchy--Schwarz over the fibre labels proves
\eqref{eq:paird-flat}.

For \eqref{eq:paird-grad}, the coefficient includes one additional
factor $\|\bm{\tau}\|/K$. We use
$\|\bm{\tau}\|\le\|\bm{q}'\|+\|\bm{q}\|$ before separating the two sums.
The first term places one momentum factor on $\ket{u}$, and the second
places it on $\ket{v}$. One momentum factor is bounded by $C_m(1+A)^{1/2}$,
so it raises the corresponding exponent from $3/4+\varepsilon$ to
$5/4+\varepsilon$. This gives the two terms displayed in
\eqref{eq:paird-grad}.
\end{proofx}

We now apply this trace estimate to a pair of successive Coulomb
transfers. To see the relevant structure, write the bulk-to-exterior
part of the auxiliary operator as
$Y_1:=\sum_jR\widetilde Y^{(j)}Q$, so that $Y=Y_1+Y_1^\dagger$.
Here $j$ ranges over all interaction channels moving the measured
register, including quantum pairs and clamped-nucleus terms, and
$\widetilde Y^{(j)}$ is the channel's transfer sum with its kinetic
numerator divided by its mismatch.
The products $V_{\mathrm{tot}}Y_1$ and $Y_1V_{\mathrm{tot}}$ therefore
describe two successive transfers. Their individual kernels have a
bound independent of the net transfer; their difference gains an
additional small factor when both outer configurations are in $Q$.
The symbol $Y_1$ in this preliminary argument denotes only this
bulk-to-exterior block; it will be reassigned explicitly in the
finite-box argument below.

\begin{lemma}[Flat convolution and the Hermitian bulk pairing]
\label{lem:flat}
With the zero Coulomb mode omitted,
\begin{equation}
 \sup_{\bm{\tau}}\sum_{\substack{\|\bm{\nu}\|\ge3K/4\\\bm{\nu}\ne\bm{\tau}}}
       |\hat w(\bm{\nu})\hat w(\bm{\nu}-\bm{\tau})|\le CK^{-1}.
 \label{eq:flatbd}
\end{equation}
For a two-hop form that starts in $Q$, passes through $R$, and returns
to $Q$, the two hops use the same channel.
If $c_1(\vec{\bm{q}},\vec{\bm{p}})$ is the kernel of $QV_{\mathrm{tot}}RYQ$, its anti-Hermitian part obeys
\begin{equation}
 |c_1(\vec{\bm{q}},\vec{\bm{p}})-\overline{c_1(\vec{\bm{p}},\vec{\bm{q}})}|
       \le CK^{-2}\|\vec{\bm{k}}_{\vec{\bm{q}}}-\vec{\bm{k}}_{\vec{\bm{p}}}\|.
 \label{eq:asymbd}
\end{equation}
These statements require Hermitian channel coefficients, not real
matrix elements. In particular they apply to translated clamped-nucleus
channels. A one-body channel has $\Delta T_p/\omega=1$ when it moves
$p$, so its symbol difference vanishes.
\end{lemma}
\begin{proofx}
We first estimate the absolute convolution in \eqref{eq:flatbd}.
If $\|\bm{\tau}\|\le3K/8$ and $\|\bm{\nu}\|\ge3K/4$, then both
$\|\bm{\nu}\|$ and $\|\bm{\nu}-\bm{\tau}\|$ are comparable. The product of
Coulomb coefficients is therefore bounded by $C\|\bm{\nu}\|^{-4}$;
summing this three-dimensional tail gives $CK^{-1}$.
Otherwise split into $\|\bm{\nu}-\bm{\tau}\|\le\|\bm{\tau}\|/2$, the analogous
ball about zero, and their complement. The first two give
$C\|\bm{\tau}\|^{-2}\sum_{0<\|\bm{r}\|\le C\|\bm{\tau}\|}\|\bm{r}\|^{-2}
\le C/\|\bm{\tau}\|$; dyadic annuli bound the complement by the same
quantity. All constants include the fixed reciprocal-lattice spacing.

The stronger bound concerns the difference between the two orientations
of a closed excursion. A transfer from $Q$ to $R$ has size at least
$3K/4$. Every register it moves then lies outside its bulk ball, even
if only one of them lies outside the retained cube. Returning to $Q$
in one further hop must undo each of those displacements, so the
second hop uses the same channel. For a one-register interaction we
group all clamped centres acting on that register into one convolution
channel. Reversing the two outer configurations leaves the intermediate
set unchanged.
Put $s(\vec{\bm{r}},\vec{\bm{p}})=\Delta T_p(\vec{\bm{r}},\vec{\bm{p}})/\omega(\vec{\bm{r}},\vec{\bm{p}})$; the kernel of $Y$ is
$-w(\vec{\bm{r}},\vec{\bm{p}})s(\vec{\bm{r}},\vec{\bm{p}})$. Consequently
\begin{equation}
 c_1(\vec{\bm{q}},\vec{\bm{p}})-\overline{c_1(\vec{\bm{p}},\vec{\bm{q}})}
 =-\sum_{\vec{\bm{r}}\in R} w(\vec{\bm{q}},\vec{\bm{r}})w(\vec{\bm{r}},\vec{\bm{p}})\,[s(\vec{\bm{r}},\vec{\bm{p}})-s(\vec{\bm{r}},\vec{\bm{q}})].
\end{equation}
The minus sign comes from $G(\vec{\bm{q}},\vec{\bm{p}})=-iw(\vec{\bm{q}},\vec{\bm{p}})\Delta T_p(\vec{\bm{q}},\vec{\bm{p}})$.
We must estimate this difference before taking absolute values. Put
$N_{\vec{\bm{p}}}:=\Delta T_p(\vec{\bm{r}},\vec{\bm{p}})$ and $D_{\vec{\bm{p}}}:=\omega(\vec{\bm{r}},\vec{\bm{p}})$, and use the
corresponding notation at $\vec{\bm{q}}$. The elementary identity
\begin{equation}
 \frac{N_{\vec{\bm{p}}}}{D_{\vec{\bm{p}}}}-\frac{N_{\vec{\bm{q}}}}{D_{\vec{\bm{q}}}}
 =\frac{N_{\vec{\bm{p}}}-N_{\vec{\bm{q}}}}{D_{\vec{\bm{p}}}}
   +\frac{N_{\vec{\bm{q}}}(D_{\vec{\bm{q}}}-D_{\vec{\bm{p}}})}{D_{\vec{\bm{p}}}D_{\vec{\bm{q}}}}
\end{equation}
shows where the gain arises. Both outer configurations are in $Q$,
so the differences of their kinetic energies are bounded by a fixed
multiple of $K\|\vec{\bm{k}}_{\vec{\bm{q}}}-\vec{\bm{k}}_{\vec{\bm{p}}}\|$. The two denominators have the bulk gap,
and the kinetic numerator is at most of the order of the squared
transfer. These facts give
$|s(\vec{\bm{r}},\vec{\bm{p}})-s(\vec{\bm{r}},\vec{\bm{q}})|\le C\|\vec{\bm{k}}_{\vec{\bm{q}}}-\vec{\bm{k}}_{\vec{\bm{p}}}\|/K$.
Multiplying this gain by the $CK^{-1}$ convolution bound proves
\eqref{eq:asymbd}. Hermitian reversal conjugates the coefficients,
so the argument remains valid for the complex phases contributed by
translated clamped nuclei. In a one-register channel the ratio $s$
is identically one, making the symbol difference zero.
\end{proofx}

We can now sharpen the potential remainder. Its bulk--bulk contribution
is an imaginary part, so the anti-Hermitian kernel difference in
\eqref{eq:asymbd}, rather than the absolute convolution alone, is the
quantity to estimate. Contributions with an outer state outside $Q$
already have a momentum-tail gain.

\begin{proposition}[Phase-sensitive first remainder]
\label{prop:phase}
For every $\varepsilon>0$, throughout the exact evolution in $D(H^2)$,
\begin{equation}
 |\mathfrak V_Y(\ket{\psi},\ket{\psi})|\le
 C_\varepsilon\Xi^2K^{-2+\varepsilon}.
 \label{eq:phase}
\end{equation}
\end{proposition}
\begin{proofx}
Only the imaginary part of $z=\langle V_{\mathrm{tot}}\psi\,|\,Y\psi\rangle$ enters,
since $\mathfrak V_Y(\ket{\psi},\ket{\psi})=-2\operatorname{Im}z$. With
$Y=RYQ+QYR$, split exactly
\begin{equation}
 z=\langle V_{\mathrm{tot}}Q\psi\,|\,YQ\psi\rangle
  +\langle V_{\mathrm{tot}}(1-Q)\psi\,|\,YQ\psi\rangle
  +\langle YQV_{\mathrm{tot}}\psi\,|\,R\psi\rangle.
\end{equation}
We estimate the second and third terms first, leaving the bulk--bulk
cancellation for the first. Hardy on the restricted state $(1-Q)\ket{\psi}$
gives $\|V_{\mathrm{tot}}(1-Q)\ket{\psi}\|\le C\Xi K^{-3/2}$.
Pairing this with $\|YQ\ket{\psi}\|\le C\Xi K^{-1/2}$ bounds the
second term by $C\Xi^2K^{-2}$.
For the third term, we refine the Gram estimate by splitting the
input into dyadic momentum shells. Here the Gram estimate refers to
the Cauchy--Schwarz bound in Proposition~\ref{prop:elim}, which uses
a sum of squared kernel entries on each channel fibre.
Let $S_\lambda$ retain
configurations with
$\lambda<\max_a\|\bm{k}_a\|\le2\lambda$, together with a separate
lowest shell. On a fixed channel fibre, $QS_\lambda$ contains at
most $C_L\lambda^3$ input points. For each such point the squared
bulk-to-exterior kernel has sum at most $CK^{-1}$. Bounding the
operator norm by the square root of the sum of squared matrix
entries therefore gives, for each channel $j$,
\begin{equation}
 \|Y_jQS_\lambda\|\le CK^{-1/2}\lambda^{3/2}.
\end{equation}
The bound is uniform in the fixed fibre coordinates, so it also
holds on their orthogonal sum. The full potential-tail estimate gives
$\|S_\lambda V_{\mathrm{tot}}\ket{\psi}\|\le C\Xi\lambda^{-1/2}$.
Their product is $C\Xi K^{-1/2}\lambda$. Only shells with
$\lambda\le CK$ meet $Q$, and the sum of their dyadic radii is
$O(K)$. Including the lowest shell, we conclude that
\begin{equation}
 \|YQV_{\mathrm{tot}}\ket{\psi}\|
 \le C\Xi K^{-1/2}\left(1+\sum_{\lambda\le CK}^{\rm dyadic}\lambda\right)
 \le C\Xi K^{1/2}.
\end{equation}
Because the potential-tail bound is for $V_{\mathrm{tot}}\ket{\psi}$,
this argument already includes channels different from the one in
$Y_j$. It uses the shell support before taking a norm, whereas the
coarse graph bound from the preceding proof loses that information.
Pairing the result with the state tail gives
$(C\Xi K^{1/2})(C\Xi K^{-5/2})=C\Xi^2K^{-2}$.

We return to the first term. For a quadratic form, twice its imaginary
part is determined by the kernel minus its Hermitian reverse. We
therefore use \eqref{eq:asymbd} before applying the linear-transfer
trace bound \eqref{eq:paird-grad}. On each finite $Q$ fibre the linear-transfer
trace estimate follows from weighted Cauchy--Schwarz; one endpoint
has weight $(1+A)^{3/4+\varepsilon'}$ and the other
$(1+A)^{5/4+\varepsilon'}$. The $3/4+\varepsilon'$ moment is bounded uniformly throughout the exact
evolution when $\varepsilon'$ is sufficiently small. The
$5/4+\varepsilon'$ moment of $Q\ket{\psi}$ is at most
$C\Xi K^{2\varepsilon'}$, up to the logarithm at the critical
$5/4$ moment. To see the origin of that logarithm, the
$\mathcal E$-weighted tail decays as $\lambda^{-1/2}$, while the
extra factor $\mathcal E^{1/4}$ grows as $\lambda^{1/2}$ on a
momentum shell. The factors cancel, leaving a constant squared
contribution per dyadic shell up to $K$. We absorb the resulting
logarithm into a further arbitrarily small positive power of $K$. Choose $\varepsilon'>0$
sufficiently small relative to $\varepsilon$. The result is
\eqref{eq:phase}. This proof uses the whole complement $1-Q$,
including high momentum on spectator registers.
\end{proofx}

\begin{theorem}[Integrated projection flux after one elimination]
\label{thm:flux2}
For $\ket{\psi_0}\in D(H^2)$, every $\varepsilon>0$, all $t\ge0$ and
sufficiently large fixed-cell cutoffs,
\begin{equation}
 |\mathcal F_K(t)|\le C_\varepsilon\Xi^2
             \big(K^{-3}+tK^{-2+\varepsilon}\big).
 \label{eq:flux2}
\end{equation}
\end{theorem}
\begin{proofx}
Split $P\ket{\psi}=Q\ket{\psi}+(P-Q)\ket{\psi}$. The bulk-to-exterior term is
\eqref{eq:elim}; its endpoint is $O(\Xi^2K^{-3})$ and its
remainder is bounded by Proposition~\ref{prop:phase}.
All other terms have both arguments restricted by a register tail
at $K/4$, and Lemma~\ref{lem:hardy} gives $C\Xi^2K^{-2}$.
The restriction is the whole complement $1-Q$: a spectator register
can supply the large momentum even when the measured particle does
not. Integrating the $K^{-2+\varepsilon}$ remainder over $[0,t]$
and adding the two $K^{-3}$ endpoint terms proves
\eqref{eq:flux2}. This completes the first cancellation argument;
the refined proof below will improve the accumulated contribution
from $K^{-2+\varepsilon}$ to $K^{-3+\varepsilon}$.
\end{proofx}

\subsection{Lattice counts for resonant transfers}

The next elimination steps cannot divide by a kinetic mismatch that
is close to zero. We instead separate those transfers and estimate
their total contribution by counting how many can occur. For a fixed
input configuration, completing the square turns the small-mismatch
condition into a thin spherical layer in transfer space. We first
count its lattice points, then explain why the same count controls
both sides of the corresponding matrix form.

Throughout these counting lemmas, transfers are physical momenta on
$g\Z^3$, with $g=2\pi/L$. Constants may therefore depend on the
fixed cell size. The notation $N(\vec{\bm{k}};\rho,\omega_0)$ and local counts
abbreviated to $N$ denote lattice multiplicities, not the one-particle
basis size.

\begin{lemma}[Two-sided slab count]
\label{lem:slab}
Fix an interaction pair $(p, b)$ and write
\begin{equation}
\bm{v} \;:=\; \frac{\bm{k}_p}{m_p} - \frac{\bm{k}_b}{m_b},
\qquad
\omega_{\bm{\nu}}(\vec{\bm{k}})
\;=\; \frac{\norm{\bm{\nu}}^2}{2 \mu_{pb}} + \bm{\nu} \cdot \bm{v}
\;=\; \frac{1}{2 \mu_{pb}}
\Big( \norm{ \bm{\nu} + \mu_{pb} \bm{v} }^2 - \mu_{pb}^2 \norm{\bm{v}}^2 \Big),
\label{eq:vrel}
\end{equation}
so that the layer $\{ |\omega_{\bm{\nu}}(\vec{\bm{k}})| \le \omega_0 \}$ is, for
fixed $\vec{\bm{k}}$, a spherical annulus in $\bm{\nu}$ centred at
$-\mu_{pb} \bm{v}$ about the elastic sphere through the origin. Then:

(i) \emph{(Count.)} There is a constant $C$ depending on $L$ such that
for every configuration $\vec{\bm{k}}$, every $\rho \ge 1$, and every
$\omega_0 > 0$,
\begin{equation}
N(\vec{\bm{k}}; \rho, \omega_0) \;:=\;
\#\big\{ \bm{\nu} \in g\Z^3 :\;
\rho \le \norm{\bm{\nu}} \le 2\rho,\;
|\omega_{\bm{\nu}}(\vec{\bm{k}})| \le \omega_0 \big\}
\;\le\; C\, \rho^2
\Big( 1 + \frac{\mu_{pb}\, \omega_0}{\rho} \Big).
\label{eq:slabcount}
\end{equation}

(ii) \emph{(Two-sided form bound.)} On the finite Fourier block
used for the resonant-channel construction, let $V^{(0)}$ be a
transfer operator on the pair with coefficients supported on the
shell $\rho \le \norm{\bm{\nu}} \le 2\rho$ and restricted to the
layer,
\begin{equation}
V^{(0)} \;=\; \sum_{\rho \le \norm{\bm{\nu}} \le 2\rho}
c(\bm{\nu})\,e^{-i\bm{\nu}\cdot(\bm{x}_p-\bm{x}_b)}\,
\sigma_{\bm{\nu}}(\vec{\bm{k}})\,
\ind{ |\omega_{\bm{\nu}}(\vec{\bm{k}})| \le \omega_0 },
\qquad |c(\bm{\nu})| \le c_0, \quad
\norm{\sigma_{\bm{\nu}}}_\infty \le 1 .
\end{equation}
As in Lemma~\ref{lem:paird}, both the symbol and the mismatch
are evaluated at the input configuration, before the displacement
$(\bm{k}_p,\bm{k}_b)\mapsto(\bm{k}_p+\bm{\nu},\bm{k}_b-\bm{\nu})$.
Then $V^{(0)}$ is bounded, with
\begin{equation}
\norm{ V^{(0)} }
\;\le\; c_0\, \sup_{\vec{\bm{k}}}\, N(\vec{\bm{k}}; \rho, \omega_0)
\;\le\; C\, c_0\, \rho^2
\Big( 1 + \frac{\mu_{pb}\, \omega_0}{\rho} \Big),
\label{eq:twoside}
\end{equation}
and in the form $|\langle u|V^{(0)}|v\rangle|$ the supremum may be
restricted to the momentum supports of $\ket{u}$ and $\ket{v}$. One-body
terms are the case $m_b \to \infty$: \eqref{eq:vrel},
\eqref{eq:slabcount}, and \eqref{eq:twoside} hold with
$\mu_{pb} \to m_p$ and $\bm{v} \to \bm{k}_p / m_p$.
\end{lemma}

\begin{proofx}
\emph{(i).} We separate a wide mismatch window, where no counting gain
is needed, from a thin window, where the spherical-layer geometry is
useful. Write $\mu:=\mu_{pb}$. If
$\omega_0\ge\rho^2/(64\mu)$, the full transfer shell contains at
most $C_L\rho^3$ points, and
\begin{equation}
 N\le C_L\rho^3
 \le64C_L\rho^2\frac{\mu\omega_0}{\rho}.
\end{equation}
This is already bounded by the right-hand side of
\eqref{eq:slabcount}.

We may therefore assume $\omega_0<\rho^2/(64\mu)$. The claim is
immediate if the layer contains no lattice points. Otherwise its
nonemptiness gives a useful lower bound on the relative speed.
Indeed, if $\|\bm{v}\|\le\rho/(4\mu)$, then every transfer in the shell
would satisfy
\begin{equation}
 \omega_{\bm{\nu}}\ge\frac{\|\bm{\nu}\|^2}{2\mu}-\|\bm{\nu}\|\|\bm{v}\|
 \ge\frac{\|\bm{\nu}\|^2}{4\mu}
 \ge\frac{\rho^2}{4\mu}>\omega_0,
\end{equation}
which contradicts membership in the layer. Thus the radius
$c:=\mu\|\bm{v}\|$ of the elastic sphere satisfies $c>\rho/4$.

By \eqref{eq:vrel}, the mismatch window is the annulus with radii
$r_-\le\|\bm{\nu}+\mu \bm{v}\|\le r_+$, where
$r_\pm^2=c^2\pm2\mu\omega_0$. Our thin-window assumption gives
$2\mu\omega_0<\rho^2/32<c^2/2$, so both radii are positive and
$r_-\ge c/\sqrt2$. Rationalising their difference now yields
\begin{equation}
 r_+-r_-=\frac{4\mu\omega_0}{r_++r_-}
 \le\frac{2\sqrt2\,\mu\omega_0}{c}
 \le\frac{8\sqrt2\,\mu\omega_0}{\rho}.
\end{equation}
This is the required thickness scale. The remaining task is to show
that the part of this annulus inside the transfer shell has area of
order $\rho^2$, even when its centre lies far away.

To pass from thickness to a lattice count, place a disjoint cube of
side $g$ at each lattice point. The cubes belonging to the layer
inside the transfer shell lie in
\begin{equation}
 \{\|\bm{\nu}\|\le\rho'\}\cap
 \{\max(0,r_- -\sqrt3g)\le\|\bm{\nu}+\mu \bm{v}\|\le r_++\sqrt3g\},
 \qquad \rho':=2\rho+\sqrt3g.
\end{equation}
Their total volume is $Ng^3$. Since $\rho\ge1$ and $g$ is fixed,
we have $\rho'\le C_L\rho$; it remains to bound the volume of
this enlarged intersection. Use spherical coordinates about $-\mu \bm{v}$. The intersection of a
radius-$r$ sphere with the ball of radius $\rho'$ is a spherical
cap, possibly the entire sphere. Its solid angle satisfies
\begin{equation}
 \sigma(r)\le\min\left\{4\pi,\frac{\pi\rho'^2}{rc}\right\},
 \qquad \sigma(r)=0\ \text{if }\|r-c\|>\rho'.
\end{equation}
Indeed the cap formula before clipping at $0$ and $4\pi$ is
$\pi[\rho'^2-(r-c)^2]/(rc)$. The enlarged radial interval has
length at most $2h$, where
\begin{equation}
 h:=\tfrac12(r_+-r_-)+\sqrt3g
 \le C_L(1+\mu\omega_0/\rho).
\end{equation}
We use two cases to bound its cross-sectional area uniformly in the
possibly large displacement $c$ of the sphere's centre. If $c \ge \rho'$, then
$r \le c + \rho' \le 2c$ on the support, the integrand
$r^2 \sigma(r) \le \pi \rho'^2\, r / c \le 2 \pi \rho'^2$, and
the volume is at most $4 \pi \rho'^2 h \le C \rho^2 h$. If instead $\rho/4<c<\rho'$, then
$r_+\le c\sqrt{3/2}\le C_L\rho$. Bounding the solid angle by
$4\pi$ gives volume at most
$4\pi(r_++\sqrt3g)^2\,2h\le C_L\rho^2h$.
Both cases give the same bound. Dividing the volume by $g^3$ and
absorbing this fixed factor into $C_L$ proves
$N\le C_L\rho^2(1+\mu\omega_0/\rho)$.

\emph{(ii).} We now use the geometric count to bound the operator.
Write $\vec{\bm{k}}\oplus\bm{\nu}$ for the configuration obtained by adding
$\bm{\nu}$ to register $p$, subtracting it from register $b$, and leaving
all spectators unchanged. Bounding the symbols by one and applying
Cauchy--Schwarz to the sum over configurations and transfers gives
\begin{equation}
| \langle u|V^{(0)}|v\rangle |
\;\le\; c_0
\Big( \sum_{\bm{\nu}} \sum_{\vec{\bm{k}}}
\ind{|\omega_{\bm{\nu}}(\vec{\bm{k}})| \le \omega_0}\,
| u_{\vec{\bm{k}}\oplus\bm{\nu}} |^2 \Big)^{1/2}
\Big( \sum_{\bm{\nu}} \sum_{\vec{\bm{k}}}
\ind{|\omega_{\bm{\nu}}(\vec{\bm{k}})| \le \omega_0}\,
| v_{\vec{\bm{k}}} |^2 \Big)^{1/2} .
\end{equation}
The second bracket is
$\sum_{\vec{\bm{k}}} N(\vec{\bm{k}})\, |v_{\vec{\bm{k}}}|^2$. In the first, substitute
$\vec{\bm{k}}' = \vec{\bm{k}} \oplus \bm{\nu}$ and use
$\omega_{-\bm{\nu}}(\vec{\bm{k}}') = - \omega_{\bm{\nu}}(\vec{\bm{k}})$ together with the
symmetry of the shell under $\bm{\nu}\to-\bm{\nu}$: after extending the
finite-block vectors by zero, it is at most
$\sum_{\vec{\bm{k}}'} N(\vec{\bm{k}}')\, |u_{\vec{\bm{k}}'}|^2$. If $N_u$ and $N_v$ are the largest counts on the respective Fourier
supports, this gives
$|\langle u|V^{(0)}|v\rangle|\le c_0\sqrt{N_uN_v}\|\ket{u}\|\|\ket{v}\|$.
Thus the maximum count on the union of those supports suffices;
taking the unrestricted maximum proves \eqref{eq:twoside}. Had the restriction
constrained only the input --- a one-sided layer --- the first
bracket would return the full shell cardinality
$\asymp \rho^3$ and the bound would degrade by the square root of
the layer's relative measure; the reversal symmetry is what
delivers the full measure.
\end{proofx}

Later we will fix one transfer in a composite and count the allowed
values of another. The remaining mismatch then lies in a shifted
energy window, and the allowed transfers lie in a ball that need not
be centred at zero. The next lemma supplies this more local version
of the same geometric estimate.

\begin{lemma}[Windowed, ball-restricted slab count]
\label{lem:slab2}
Fix a pair, a configuration $\vec{\bm{k}}$, and a window
$W=[\omega_-,\omega_+]$, with $\omega_-\le\omega_+$ and width
$|W|=\omega_+-\omega_-$. Write $x_+:=\max\{x,0\}$.
With $R_\pm := \mu_{pb}^2 \norm{\bm{v}}^2 + 2 \mu_{pb}\, \omega_\pm$
and $t_W := \sqrt{ (R_+)_+ } - \sqrt{ (R_-)_+ }$, for every
$\bm{\tau} \in \R^3$ and every $d \ge 1$,
\begin{equation}
\# \big\{ \bm{\nu} \in g\Z^3 :\; \norm{\bm{\nu} - \bm{\tau}} \le d, \;
\omega_{\bm{\nu}}(\vec{\bm{k}}) \in W \big\}
\;\le\; C\, (1 + d)^2\, ( 1 + t_W ),
\qquad
t_W \le \min\!\Big( \sqrt{ 2 \mu_{pb} |W| },\;
\frac{ 2 \mu_{pb} |W| }{ \sqrt{ (R_+)_+ } } \Big),
\label{eq:ballcount}
\end{equation}
where $C$ depends only on the fixed lattice spacing. In the
thickness estimate, the quotient is interpreted as $+\infty$ when
$R_+\le0$; then $t_W=0$. The count is empty if $R_+<0$, and at
$R_+=0$ the possible set has radius zero.
This is the geometric estimate underlying Lemma~\ref{lem:slab}(i):
for its centred window, the additional transfer-shell restriction and
nonemptiness argument give the sharper thickness bound
$O(\mu_{pb}\omega_0/\rho)$. Here we permit an arbitrary window and
an arbitrary ball centre.
\end{lemma}

\begin{proofx}
By \eqref{eq:vrel} the window set is the annulus
$\sqrt{(R_-)_+} \le \norm{ \bm{\nu} - \bm{c} } \le \sqrt{(R_+)_+}$ about
$\bm{c} := -\mu_{pb} \bm{v}$ (empty if $R_+ < 0$), of radial thickness $t_W$. If $R_+\le0$, the preceding empty-set or
single-point observation proves the claim, so assume $R_+>0$.
We cover the annulus by at most $t_W+2$ unit shells
$\{ r \le \norm{\bm{\nu} - \bm{c}} < r + 1 \}$. Each unit shell meets $B(\bm{\tau},d)$ in at most $C_L(1+d)^2$ lattice
points. To see why this count grows as area rather than volume, we
count along coordinate columns: for a lattice
point $\bm{\nu}$ of the shell, the three squared distances to the
coordinate axes through $\bm{c}$ sum to $2 \norm{\bm{\nu} - \bm{c}}^2$, so one
axis has $s^2 \le \tfrac{2}{3} (r+1)^2$; along that axis the
column through $\bm{\nu}$ meets the unit shell in at most two
intervals of length at most
$( 2r + 1 ) \big/ \big( \sqrt{(r+1)^2 - s^2} +
\sqrt{ (r^2 - s^2)_+ } \big) \le (2r+1)\sqrt{3}/(r+1) \le
2\sqrt 3$, hence at most $C_L$ points each, because successive lattice points
on a column are spaced by $g$. For each of the three axes, the
columns meeting $B(\bm{\tau},d)$ project into a disc of radius at most
$d+1$, giving at most $C_L(1+d)^2$ columns. Shells with $r\le8$
contain at most $C_L$ points and can be included in the same bound.
Multiplying by the number of radial shells proves the count.

It remains to bound the thickness by the energy-window width. If $R_- \ge 0$ then
$t_W = 2\mu|W| / ( \sqrt{R_+} + \sqrt{R_-} ) \le
2\mu|W|/\sqrt{R_+}$, and otherwise
$t_W = \sqrt{R_+} \le \sqrt{2\mu|W|} \le 2\mu|W|/\sqrt{R_+}$
since $R_+ \le R_+ - R_- = 2\mu|W|$; the bound
$t_W \le \sqrt{2\mu|W|}$ follows in the first case from
$R_+ \ge 2\mu|W|$.
\end{proofx}

Counting limits how many transfers have a small mismatch. We also need
to know where their endpoint states can be supported. A transfer that
changes relative kinetic energy very little has nearly equal input
and output relative-momentum norms. If the transfer itself is large,
the triangle inequality then forces both norms to be large. The next
lemma quantifies this observation; we refer to it as pinning the
endpoints away from zero momentum.

\begin{lemma}[Resonant pinning]
\label{lem:pin}
Let $\bm{\xi}=\mu_{pb}(\bm{k}_p/m_p-\bm{k}_b/m_b)$ be the relative momentum of the
pair and let $\|\bm{\tau}\|\in[\varrho,2\varrho)$. Near resonance means
$|\omega_{\bm{\tau}}|<\varrho^2/(8\mu_{pb})$, equivalently
$|\, \norm{\bm{\xi}+\bm{\tau}}^2 - \norm{\bm{\xi}}^2 \,| < \varrho^2/4$. Then
\begin{equation}
\min\big( \norm{\bm{\xi}},\; \norm{\bm{\xi} + \bm{\tau}} \big) \;\ge\;
\tfrac{3}{8}\, \varrho ,
\label{eq:pin}
\end{equation}
and the constant is sharp. Consequently, in any two-hop
composite whose two kinetic mismatches are near-resonant at
their transfer scales, both outer configurations are pinned:
each outer relative momentum exceeds $3/8$ of the transfer scale
of the hop it terminates, hence so does the larger register of
the corresponding pair.
\end{lemma}

\begin{proofx}
Dividing the window by $\norm{\bm{\xi}+\bm{\tau}} + \norm{\bm{\xi}} \ge
\norm{\bm{\tau}} \ge \varrho$ gives
$|\, \norm{\bm{\xi}+\bm{\tau}} - \norm{\bm{\xi}} \,| < \varrho/4$, and the two norms
sum to at least $\varrho$; the smaller is therefore at least
$(\varrho - \varrho/4)/2$. The anti-parallel configuration
$\bm{\xi} = -\tfrac38 \bm{\tau}$, $\norm{\bm{\tau}} = \varrho$, lies on the window's edge. Replacing $3/8$ by $3/8+\delta$ with
$0<\delta<1/8$ gives strict near resonance and approaches the same
constant as $\delta\downarrow0$. As in the preceding gap argument,
large-cutoff lattice approximations preserve strict inequalities, so
the constant is sharp uniformly on the lattice as well.
For a two-hop composite, the outer input is an endpoint of the first
hop and the outer output is an endpoint of the second, so the same
lower bound applies separately at these two outer endpoints.
Finally,
\begin{equation}
 \bm{\xi}=\frac{m_b}{m_p+m_b}\bm{k}_p-\frac{m_p}{m_p+m_b}\bm{k}_b,
 \qquad
 \|\bm{\xi}\|\le\max\{\|\bm{k}_p\|,\|\bm{k}_b\|\}.
\end{equation}
Thus at least one register at each endpoint has the required large
momentum. The statement does not require both individual registers
to be large.
\end{proofx}

\subsection{A finite-box proof of the energy-rate projection flux}
\label{sec:flux-finite-box}

We now complete the improvement from the first-elimination estimate to
an integrated flux of order $(1+t)K^{-3}$, up to logarithms. The
remaining difficulty is the potential commutator created by an
elimination: although dividing by a kinetic mismatch improves the
endpoint, the new commutator contains further transfers that need not
have a large mismatch. We will first separate a part of the force that
can be removed exactly, then use three successive auxiliary Hermitian operators to remove the
remaining components, starting with the centre-of-mass contribution.
At each stage we use
the kinetic commutator of such an operator to reproduce the particular
flux component being removed. This is the meaning of a dressing in the
argument below. Its expectation gives an endpoint term, and its
potential commutator gives the next remainder.

We carry out these operations in a finite auxiliary Fourier box. This
lets us use exact matrix identities throughout, including the final
reflection of closed excursions. That reflection is also what allows
us to include clamped nuclei with complex Fourier coefficients: the
cancellation uses Hermitian reversal, not reality of the coefficients.
The auxiliary box is a proof device; the retained Galerkin basis remains
$P$.

For this subsection write $g=2\pi/L$, let $n$ be the integer cube
radius, and put $K=gn$; the one-particle basis size is $(2n+1)^3$.
The side length used in Appendix~\ref{app:grid} is therefore $2n+1$
in this local convention. The integer $n$ denotes the cube radius
throughout this subsection, including Theorem~\ref{thm:flux3}.
All momentum indices and transfer vectors below
are dimensionless lattice vectors. A register with index $\bm{p}_a$ has
physical momentum $g\bm{p}_a$, so physical kinetic energies contain $g^2$.
In particular, the bulk projector already denoted by $Q$ now has the
integer-coordinate description $\prod_a\ind{\|\bm{p}_a\|\le n/4}$.
We retain the total kinetic weight $\mathcal E=1+A$ and the homogeneous
initial-state graph norm $\mathcal D$; the fixed factor relating $\Xi$ to
$\mathcal D$ is absorbed into constants in this subsection.

Let $B_M$ be the product cube of radius $M=n^6$, with $n\ge64$,
and compress all operators to $B_M$. Write
$\ket{u(s)}=B_M\ket{\psi(s)}$, $T_M=B_MT B_M$, $V_M=B_MV_{\mathrm{tot}}B_M$ and
$P_M=P$, $R_M=B_M-P$. We suppress the subscript $M$ inside matrix
identities. It is important that $\ket{u}$ is the projection of the exact
state, rather than a state generated solely by the compressed
Hamiltonian. Its equation therefore contains a source from the exterior
of the auxiliary box:
\begin{equation}
 i\ket{\dot u}=(T_M+V_M)\ket{u}+\ket{f_M},
 \qquad \ket{f_M}=B_MV_{\mathrm{tot}}(1-B_M)\ket{\psi}.
 \label{eq:flux-box-evolution}
\end{equation}
For every self-adjoint matrix $J$ on this box define
\begin{equation}
 e_J(s)=i\big(\langle f_M|J|u\rangle-\langle Ju\,|\,f_M\rangle\big).
\end{equation}
Differentiating $\langle u|J|u\rangle$ and substituting
\eqref{eq:flux-box-evolution} gives
\begin{equation}
 \frac{d}{ds}\langle u|J|u\rangle
 =\langle u|i[T_M,J]|u\rangle
  +\langle u|i[V_M,J]|u\rangle+e_J(s).
\end{equation}
Thus, whenever $X=i[T_M,J]$,
\begin{equation}
 \int_0^t\langle u|X|u\rangle ds
 =\big[\langle u|J|u\rangle\big]_0^t
   -\int_0^t\langle u|i[V_M,J]|u\rangle ds-\int_0^te_J(s)ds.
 \label{eq:flux-box-elimination}
\end{equation}
This is the identity we will apply repeatedly. It fixes the minus signs
in front of both remainders, and it shows exactly what must be bounded
after we construct $J$. All these are finite-matrix equalities; we make
no additional continuum domain assertion for the singular force or for
an auxiliary operator acting on it.

\subsubsection{An exact force split and a separated small-momentum region}

We begin by distinguishing the two parts of the measured kinetic change
in a single interaction. A channel is either an unordered quantum pair,
or the one-register convolution obtained by combining all clamped-centre
interactions on that register. We use this grouping for fixed centres
because a path can return through a different centre on the same
register while still belonging to the same convolution channel. Let
$V_j$ denote the corresponding contribution to $V_{\mathrm{tot}}$.
The Fourier coefficients can be complex, but obey Hermitian reversal
and are dominated by a positive Coulomb kernel. As throughout the
appendix, the zero Coulomb transfer is omitted.

For a channel moving the measured register $p$, define
$\lambda_j=\mu_j/m_p$ for a quantum pair and $\lambda_j=1$ for a
one-body channel; set $\lambda_j=0$ on other channels. On a pair hop
$\bm{\theta}$, conservation of pair total momentum gives
\begin{equation}
 \Delta T_p=\lambda_j\omega_j+c_j,\qquad
 c_j=\pm\frac{g^2}{m_a+m_b}\,\bm{\theta}\cdot\bm{\Pi}_{ab},
 \qquad \bm{\Pi}_{ab}=\bm{p}_a+\bm{p}_b.
 \label{eq:com}
\end{equation}
Here $\omega_j$ is the change in total kinetic energy along that hop,
and $\bm{\Pi}_{ab}$ is the conserved integer pair momentum. The sign records
which member is measured. This pair formula applies when the measured
register belongs to the pair; for channels not moving that register,
we set $c_j=0$ as well as $\lambda_j=0$. The term $\lambda_j\omega_j$ is proportional
to the total kinetic mismatch; $c_j$ is the remaining centre-of-mass
contribution. One-body channels also have $c_j=0$. If they move the measured
register, its kinetic change is the entire mismatch; otherwise
$\Delta T_p=0$ and $\lambda_j=0$.

We collect the mismatch-proportional part in the weighted potential
$V_\lambda:=\sum_j\lambda_jV_j$, with compression
$V_{\lambda,M}:=B_MV_\lambda B_M$. The subscript denotes this weighted
sum, not a Fourier transform. Since the $(\vec{\bm{q}},\vec{\bm{p}})$ entry of a commutator
with $T_M$ contains the factor $T(\vec{\bm{q}})-T(\vec{\bm{p}})$, equation~\eqref{eq:com}
gives the exact force decomposition
\begin{equation}
 G_M=-i[T_M,V_{\lambda,M}]+G_M^{\rm cm},\qquad
 (G_M^{\rm cm})_{\vec{\bm{q}}\vec{\bm{p}}}=-i\sum_j c_j(\vec{\bm{q}},\vec{\bm{p}})(V_j)_{\vec{\bm{q}}\vec{\bm{p}}}.
 \label{eq:force-cm-exact}
\end{equation}
As earlier, $\vec{\bm{p}},\vec{\bm{q}}$ as matrix indices label complete momentum
configurations, whereas the $p$ in $T_p$ labels the measured register.
This is an identity, not an asymptotic expansion. It is useful because
the first term already has the kinetic-commutator form required by
\eqref{eq:flux-box-elimination}, without division by a possibly small
mismatch.

Set $V_{\lambda,c}:=V_{\lambda,M}-PV_{\lambda,M}P$ and
$X^{\rm cm}:=G_M^{\rm cm}-PG_M^{\rm cm}P$. The flat, or
mismatch-proportional, part of the projection flux is then
$-i[T_M,V_{\lambda,c}]$, so its auxiliary operator is
$-V_{\lambda,c}$. To identify the potential remainder of this first
exact elimination, we must distinguish full-lattice multiplication
from compressed multiplication. The multipliers $V_{\mathrm{tot}}$
and $V_\lambda$ commute on the full lattice, but their compressions
need not commute. We retain their boundary form as
$e_{\rm bd}(\ket{v}):=\langle v|i[V_M,V_{\lambda,M}]|v\rangle$.
Writing $\ket{a}=P\ket{u}$ and $\ket{d}=R_M\ket{u}$ and expanding the two commutator products
gives
\begin{align}
 \langle u|i[V_M,V_{\lambda,c}]|u\rangle
 &=e_{\rm bd}(\ket{u})-e_{\rm bd}(\ket{a})+n^{\rm fl}(\ket{u}),\\
 n^{\rm fl}(\ket{u})
 &=2\operatorname{Im}\langle V_Md\,|\,PV_{\lambda,M}a\rangle
   -2\operatorname{Im}\langle R_MV_Ma\,|\,R_MV_{\lambda,M}a\rangle.
 \label{eq:flat-flux-identity}
\end{align}
The first term in $n^{\rm fl}$ contains the discarded state component
$\ket{d}$. The second describes a retained state leaving the basis and
returning. Its $Q\times Q$ block is zero: a two-hop
$Q\to R_M\to Q$ excursion must undo every register moved out of the
bulk and therefore uses one channel. Both orders then have the same
coefficient $\lambda_j$. Equivalently, the contribution for that
channel is a real multiple of an inner product of a vector with itself,
so its imaginary part vanishes. This argument requires the Hermitian
channel structure and the grouping of fixed centres, but not real
matrix elements.

For the final elimination we will need a subset of $Q$ whose total
kinetic energy is separated from that of every configuration outside
$Q$. We therefore introduce the small-momentum region
\begin{equation}
 C=\prod_a\ind{\|\bm{p}_a\|\le c_*n},\qquad
 c_*=\frac1{32\sqrt{(\eta_{\rm tot}+1)m_{\max}}},\qquad
 F=B_M-Q,
 \label{eq:flux-cold-region}
\end{equation}
and assume $gc_*n\ge1$. We refer to $C$ as the cold region: every
register in it has momentum at most the specified small fraction of
the cutoff. These projectors satisfy $C\subset Q\subset P\subset B_M$.
The region $Q-C$ lies between the cold region and the boundary of the
bulk, while $F$ contains every configuration in the auxiliary box
outside $Q$, including those that remain in $P$. In particular, $F$
is larger than the discarded region $R_M$.

If $\vec{\bm{f}}\in F$, at least one register has physical momentum greater than
$K/4$, so $T(\vec{\bm{f}})\ge K^2/(32m_{\max})$. If $\vec{\bm{b}}\in C$, the inherited
atomic-unit hypothesis $m_a\ge1$ gives
$T(\vec{\bm{b}})\le\eta_{\rm tot}c_*^2K^2/2$. Our choice of $c_*$ makes this
upper bound at most one quarter of the lower bound for $T(\vec{\bm{f}})$.
Consequently,
\begin{equation}
 T(\vec{\bm{f}})-T(\vec{\bm{b}})\ge\frac{3K^2}{128m_{\max}},\qquad
 T(\vec{\bm{f}})-T(\vec{\bm{b}})\ge\frac34T(\vec{\bm{f}}).
 \label{eq:flux-far-cold-gap}
\end{equation}
We will use the first inequality for a uniform inverse-gap bound and
the second to retain decay when the far endpoint has much larger
energy. The mass-dependent radius is essential to this comparison:
a universal cold radius would not supply these inequalities for
arbitrary fixed mass ratios.

\subsubsection{Removing the centre-of-mass flux and its bulk remainder}

We next treat $X^{\rm cm}$, the part of the projection flux left by
\eqref{eq:force-cm-exact}. Its numerator contains the pair total momentum,
so we keep that conserved quantity in the estimates. In integer units
write $\omega_j=g^2F_j(\bm{\theta})/(2\mu_j)$; for a pair,
$F_j(\bm{\theta})=\|\bm{\theta}\|^2+2\bm{\theta}\cdot\bm{\xi}_j$, where
$\bm{\xi}_j=\mu_j(\bm{p}_a/m_a-\bm{p}_b/m_b)$ is its dimensionless relative
momentum, expressed in the integer register coordinates. We use the transfer-linear resonance window
\begin{equation}
 |F_j(\bm{\theta})|\le2\|\bm{\theta}\|,
 \quad\text{equivalently}\quad
 |\omega_j|\le g^2\|\bm{\theta}\|/\mu_j.
 \label{eq:resset}
\end{equation}
The factor $g^2$ is needed because the transfer is an integer index and
the mismatch is a physical energy. Inside this window we estimate the
flux directly by counting allowed transfers. Outside it we divide by
the nonzero mismatch and use \eqref{eq:flux-box-elimination}.

For the remainder of this subsection, $Y$ and $Y_1$ denote the following
finite-box centre-of-mass operators; they are new local objects, not the
bulk-force operators denoted by these letters in the preliminary
one-elimination argument. Let $Y$ be the Hermitian matrix obtained by
summing the kernels $-c_jw_j/\omega_j$ on the gapped set and setting
them to zero on the resonant set, where $w_j(\vec{\bm{q}},\vec{\bm{p}})=(V_j)_{\vec{\bm{q}}\vec{\bm{p}}}$.
Set $Y_1:=Y-PYP$. For each retained matrix element, multiplication by
$i\omega_j$ recovers $-ic_jw_j$, and hence
$i[T_M,Y_1]=X^{\rm cm,gap}$. Hermitian reversal is consistent with this
definition because reversal changes the signs of both $c_j$ and
$\omega_j$. The complementary component is denoted by
$X^{\rm cm,res}$.

The counting estimates enter at this point in two different ways.
Applying the window counts of Lemmas~\ref{lem:slab}
and~\ref{lem:slab2} on dyadic transfer and mismatch shells gives row
bounds of the form $C(1+\|\bm{\Pi}_j\|)\log^2(2+M)$ for the divided,
gapped kernel of $Y$. To see the source of the logarithms, take
$\|\bm{\theta}\|\asymp\rho$ and $|F_j(\bm{\theta})|\asymp h$. On the
gapped set $h\gtrsim\rho$. The absolute divided kernel is bounded
by $C\|\bm{\Pi}_j\|/(\rho h)$, and the number of entries on this shell is
at most $C\rho^2(1+h/\rho)$. Their product is
\begin{equation}
 C\|\bm{\Pi}_j\|\big(\rho/h+1\big)\le C\|\bm{\Pi}_j\|.
\end{equation}
There are only logarithmically many transfer and mismatch shells in
the finite box, giving the stated double logarithm. This row estimate
uses the inverse mismatch in $Y$; it is not a row bound for the bare
centre-of-mass force. Reversal supplies the corresponding column
bounds, so the count can be used in a two-sided weighted Schur
estimate. Here a weighted Schur estimate bounds a matrix form from
its absolute row and column sums after assigning momentum weights
to the two endpoints. For a bulk-to-exterior block we instead have
the stronger structure of the Coulomb Gram estimate from
Proposition~\ref{prop:elim}: one sums the squared Coulomb kernel
only over transfers that cross out of the bulk. The following lemma
records the resulting bounds for the resonant term and the two
endpoints obtained so far.

\begin{lemma}[Resonant centre-of-mass flux and first endpoints]
\label{cor:reslayer}
Uniformly in $s$, for the projected exact state,
\begin{equation}
 |\langle u|X^{\rm cm,res}|u\rangle|\le C\mathcal D^2K^{-3},
 \qquad
 |\langle u|V_{\lambda,c}|u\rangle|+
 |\langle u|Y_1|u\rangle|\le C\mathcal D^2K^{-3}.
 \label{eq:reslayer}
\end{equation}
\end{lemma}
\begin{proofx}
We first bound the flat endpoint by separating the $RR$, $RQ$ and
$R(P-Q)$ blocks, together with their Hermitian reverses. On the $RR$
block, Hardy acts on the discarded component $\ket{d}$; its plain and
half-kinetic tails give a product of order
$\mathcal D^2K^{-5/2}K^{-3/2}$. On $RQ$, the discarded-state tail is
paired with the restricted Gram injection, giving
$\mathcal D^2K^{-5/2}K^{-1/2}=\mathcal D^2K^{-3}$. Finally, the
intermediate region $P-Q$ already has a register tail, so Hardy gives
\begin{equation}
 \|\ket{d}\|\,\|V_\lambda(P-Q)\ket{u}\|
 =O(\mathcal D^2K^{-4}).
\end{equation}
These are the three block estimates required for $V_{\lambda,c}$.
The weighted potential has only fixed channel coefficients, so the
same Hardy and Coulomb majorants apply to it.

For clarity, we distinguish the row estimate for the undivided resonant
force from the row estimate for the divided, gapped kernel $Y$.
Write
\begin{equation}
 S_j(\vec{\bm{p}})=\sum_{a\text{ moved by }j}\|\bm{p}_a\|^2,
 \qquad X(\vec{\bm{p}})=\sum_a\|\bm{p}_a\|^2.
\end{equation}
For a pair channel, $\|\bm{\Pi}_j(\vec{\bm{p}})\|\le\sqrt{2S_j(\vec{\bm{p}})}$; the
centre-of-mass numerator vanishes for a one-body channel.  The
resonant force kernel $G_j^{\rm res}$ satisfies
\begin{equation}
 |G_j^{\rm res}(\vec{\bm{q}},\vec{\bm{p}})|
 \le C_j\frac{\|\bm{\Pi}_j(\vec{\bm{p}})\|}{\|\bm{\theta}\|}
       \mathbf1_{\{|F_j(\bm{\theta})|\le2\|\bm{\theta}\|\}}.
\end{equation}
Here and in the following channel sums the transfer is nonzero and
$\vec{\bm{q}}$ is determined by $\vec{\bm{p}}$, $j$ and $\bm{\theta}$.  On a dyadic annulus
$\rho\le\|\bm{\theta}\|<2\rho$, the window count gives at most
$C\rho^2$ resonant transfers, so its contribution to
$\sum\|\bm{\theta}\|^{-1}$ is at most $C\rho$.
If $\|\bm{\theta}\|>8$, then
$|F_j(\bm{\theta})|\le2\|\bm{\theta}\|<\|\bm{\theta}\|^2/4$, and pinning gives
\begin{equation}
 \frac9{64}\|\bm{\theta}\|^2\le S_j(\vec{\bm{p}}).
\end{equation}
Consequently only scales $\rho\le C\sqrt{S_j(\vec{\bm{p}})}$ occur beyond
the finitely many scales $\|\bm{\theta}\|\le8$.  Summing the geometric
series, rather than bounding each shell by the largest shell, gives
\begin{equation}
 \sum_{\bm{\theta}\ \text{resonant}}\frac1{\|\bm{\theta}\|}
 \le C\bigl(1+\sqrt{S_j(\vec{\bm{p}})}\bigr).
\end{equation}
The actual row and column bounds are therefore
\begin{equation}
 \sum_{\vec{\bm{q}}}|G_j^{\rm res}(\vec{\bm{q}},\vec{\bm{p}})|\le C_j(1+X(\vec{\bm{p}})),
 \qquad
 \sum_{\vec{\bm{p}}}|G_j^{\rm res}(\vec{\bm{q}},\vec{\bm{p}})|\le C_j(1+X(\vec{\bm{q}})).
 \label{eq:flux-res-explicit-rows}
\end{equation}
The second inequality follows by Hermitian reversal.  Applying
Cauchy--Schwarz to the nonnegative kernel $|G_j^{\rm res}(\vec{\bm{q}},\vec{\bm{p}})|$
and summing the finite number of channels gives
\begin{equation}
 |\langle v|G^{\rm res}|w\rangle|
 \le C\left(\sum_{\vec{\bm{q}}}(1+X(\vec{\bm{q}}))|v_{\vec{\bm{q}}}|^2\right)^{1/2}
       \left(\sum_{\vec{\bm{p}}}(1+X(\vec{\bm{p}}))|w_{\vec{\bm{p}}}|^2\right)^{1/2}
 \le C\|\mathcal E^{1/2}\ket{v}\|\,
       \|\mathcal E^{1/2}\ket{w}\|.
 \label{eq:flux-res-explicit-form}
\end{equation}
The last comparison uses the fixed cell and masses; it includes all
spectator kinetic energies.

We also check the endpoint restriction needed to apply this form.
Suppose a resonant hop goes from $\vec{\bm{p}}\in P$ to $\vec{\bm{q}}\notin P$.
If $\|\bm{\theta}\|\le n/8$, the register which exits the cube had an
input component of magnitude greater than $7n/8$.
If $\|\bm{\theta}\|>n/8$, then $n\ge64$ implies $\|\bm{\theta}\|>8$;
the preceding quadratic pinning bound, in its relative-coordinate
form, forces an input register to have magnitude greater than
$3\|\bm{\theta}\|/8>3n/64$.  The output has a register larger than $n$
because it lies outside $P$.  Reversal supplies the other
orientation, while both endpoints of the $RR$ block already lie
outside $P$.  Thus every block in the resonant flux has a register
tail at both endpoints, with physical threshold $3K/64$.
For any such restriction $S_\ell$, the momentum tails of the evolved state give
\begin{equation}
 \|\mathcal E^{1/2}S_\ell \ket{u}\|^2
 \le\|S_\ell \ket{u}\|\,\|\mathcal E S_\ell \ket{u}\|
 \le C\mathcal D^2\ell^{-3}.
\end{equation}
Using this at both ends of \eqref{eq:flux-res-explicit-form} proves
the $C\mathcal D^2K^{-3}$ resonant-flux bound, with no logarithm.

For $Y$, the divided-kernel layer calculation instead gives
\begin{equation}
 \sum_{\vec{\bm{q}}}|Y_j(\vec{\bm{q}},\vec{\bm{p}})|\le r_j(\vec{\bm{p}}),\qquad
 \sum_{\vec{\bm{p}}}|Y_j(\vec{\bm{q}},\vec{\bm{p}})|\le r_j(\vec{\bm{q}}),\qquad
 r_j(\vec{\bm{p}})=C_j\Lambda_M\|\bm{\Pi}_j(\vec{\bm{p}})\|,
 \quad \Lambda_M\le C\log^2(2+M).
 \label{eq:flux-Y-explicit-rows}
\end{equation}
The corresponding weighted form estimate is
\begin{equation}
 |\langle v|Y_j|w\rangle|
 \le\left(\sum_{\vec{\bm{p}}} r_j(\vec{\bm{p}})|v_{\vec{\bm{p}}}|^2\right)^{1/2}
       \left(\sum_{\vec{\bm{p}}} r_j(\vec{\bm{p}})|w_{\vec{\bm{p}}}|^2\right)^{1/2}.
\end{equation}
Since $r_j\le C\Lambda_M\mathcal E^{1/2}$, a restricted
endpoint satisfies the stronger bound
\begin{equation}
 \sum_{\vec{\bm{p}}} r_j(\vec{\bm{p}})|\langle \vec{\bm{p}}|S_\ell|u\rangle|^2
 \le C\Lambda_M\|S_\ell \ket{u}\|\,
                    \|\mathcal E^{1/2}S_\ell \ket{u}\|
 \le C\Lambda_M\mathcal D^2\ell^{-4}.
 \label{eq:flux-Y-tail-weight}
\end{equation}
To use it, expand $Y_1=Y-PYP$ into $RQ$, $R(P-Q)$ and $RR$
blocks and their Hermitian reverses.  On $RQ$ the bulk gap bounds
$|c_j/\omega_j|$ by a fixed constant.  Thus the kernel is bounded
by a Coulomb majorant, and the restricted Gram injection gives
\begin{equation}
 \|RYQ\ket{u}\|\le C\mathcal D K^{-1/2},\qquad
 |\langle Ru\,|\,YQu\rangle|
 \le C\mathcal D^2K^{-5/2}K^{-1/2}.
\end{equation}
Every endpoint in the remaining blocks has a register tail at least
$K/4$.  Applying \eqref{eq:flux-Y-tail-weight} to both endpoints
therefore gives
\begin{equation}
 |\langle u|Y_1|u\rangle|
 \le C\mathcal D^2\bigl(K^{-3}+\Lambda_M K^{-4}\bigr)
 \le C\mathcal D^2K^{-3}.
\end{equation}
For the last inequality we use $M=n^6$, $K=gn$, and the boundedness
of $\log^2(2+n)/n$ for $n\ge64$.  The gapped row itself has a
double logarithm; the extra inverse power in its two-tail blocks
absorbs it in the endpoint estimate.  None of these bounds requires
a pointwise upper bound on spectator kinetic energy.
\end{proofx}

The first centre-of-mass elimination leaves a potential commutator.
We now combine it with the flat remainder from
\eqref{eq:flat-flux-identity}, because their joint bulk block exposes a
useful difference of symbols. Define the self-adjoint matrix $N_2$
by its quadratic form
\begin{equation}
 \langle u|N_2|u\rangle=n^{\rm fl}(\ket{u})
                 -\langle u|i[V_M,Y_1]|u\rangle.
\end{equation}
Its $Q\times Q$ block is $-Z_a$, where
$Z_a:=Qi(V_MR_MY-YR_MV_M)Q$. The flat contribution has vanished on
this block, and the remaining two orders compare the same
bulk-to-exterior excursion. Put $\sigma=c_j/\omega_j$, so that
$Y_{\vec{\bm{r}}\vec{\bm{p}}}=-w_j(\vec{\bm{r}},\vec{\bm{p}})\sigma(\vec{\bm{r}},\vec{\bm{p}})$. On a same-channel bulk fibre,
\begin{equation}
 (Z_a)_{\vec{\bm{q}}\vec{\bm{p}}}=-i\sum_{\vec{\bm{r}}\in R_M}w_j(\vec{\bm{q}},\vec{\bm{r}})w_j(\vec{\bm{r}},\vec{\bm{p}})
                    [\sigma(\vec{\bm{r}},\vec{\bm{p}})-\sigma(\vec{\bm{r}},\vec{\bm{q}})].
 \label{eq:zker}
\end{equation}
We have used Hermitian reversal before taking absolute values. This
order matters: estimating the two products separately would lose the
difference in square brackets, which measures the change of the
divided symbol between the two nearby bulk endpoints.

Let $\bm{\tau}$ denote their outer displacement in the relative channel
coordinate, and let $\varpi:=\omega(\vec{\bm{q}},\vec{\bm{p}})$ denote their outer kinetic
mismatch. Both inner excursions obey the bulk gap. To see how their
quotient difference is estimated, write
\begin{equation}
 \sigma(\vec{\bm{r}},\vec{\bm{p}})-\sigma(\vec{\bm{r}},\vec{\bm{q}})
 =\frac{c_j(\vec{\bm{r}},\vec{\bm{p}})-c_j(\vec{\bm{r}},\vec{\bm{q}})}{\omega(\vec{\bm{r}},\vec{\bm{p}})}
  +c_j(\vec{\bm{r}},\vec{\bm{q}})\frac{\omega(\vec{\bm{r}},\vec{\bm{q}})-\omega(\vec{\bm{r}},\vec{\bm{p}})}
                  {\omega(\vec{\bm{r}},\vec{\bm{p}})\omega(\vec{\bm{r}},\vec{\bm{q}})}.
\end{equation}
The first numerator changes with the outer displacement; the difference
in the second numerator is the outer kinetic mismatch, up to sign.
Using the gap in both denominators and then the two-Coulomb convolution
tail gives
\begin{equation}
 |(Z_a)_{\vec{\bm{q}}\vec{\bm{p}}}|\le C K^{-3}\|\bm{\Pi}_j\|
                  \big(\|\bm{\tau}\|+|\varpi|/K\big).
\end{equation}
The factor $K^{-3}$ is now attached to an outer-transfer weight and an
outer-mismatch weight. We must either absorb those weights in the
state or divide once more by the outer mismatch.

For this purpose, split $Z_a=Z^{\le}+Z^>$ at the \emph{kinetic}
floor $|F_j(\bm{\tau})|=\|\bm{\tau}\|^2/4$. We assign the equality and the
diagonal to the low part. This quadratic threshold differs from the
transfer-linear threshold in \eqref{eq:resset}: here it makes the
low-mismatch endpoints comparable to the outer transfer, as in
Lemma~\ref{lem:pin}. More precisely, its proof with
$\varrho=\|\bm{\tau}\|$ and non-strict inequalities gives
$\min\{\|\bm{\xi}\|,\|\bm{\xi}+\bm{\tau}\|\}\ge3\|\bm{\tau}\|/8$ on this closed
low set. Thus the low part can be bounded using momentum weights at
both endpoints. On the high part define
$(Y_2)_{\vec{\bm{q}}\vec{\bm{p}}}:=(Z^>)_{\vec{\bm{q}}\vec{\bm{p}}}/(i\omega(\vec{\bm{q}},\vec{\bm{p}}))$. Then
$i[T_M,Y_2]=Z^>$ and
\begin{equation}
 |(Y_2)_{\vec{\bm{q}}\vec{\bm{p}}}|\le CK^{-3}\|\bm{\Pi}_j\|/\|\bm{\tau}\|.
 \label{eq:flux-Y2-kernel}
\end{equation}
Indeed the high-mismatch condition bounds $|\omega(\vec{\bm{q}},\vec{\bm{p}})|$ below by
a fixed multiple of $g^2\|\bm{\tau}\|^2/\mu_j$. Dividing the linear term
$\|\bm{\tau}\|$ gives $C/\|\bm{\tau}\|$, while dividing the $|\varpi|/K$ term
gives $C/K$. Both endpoints lie in $Q$, so $\|\bm{\tau}\|\le C_LK$ and
the latter contribution is also bounded by $C/\|\bm{\tau}\|$.
The bound \eqref{eq:flux-Y2-kernel} is used only for $\bm{\tau}\ne0$.
Every zero mismatch lies in the low part, so no division by zero
occurs.

\begin{lemma}[Channel line estimates]
\label{lem:flux-line-estimates}
Let $x_{\vec{\bm{p}}}=|u_{\vec{\bm{p}}}|$ and, for a channel $j$, let
$L_j(x)^2=\sum_{\mathcal L}(\sum_{\vec{\bm{p}}\in\mathcal L}x_{\vec{\bm{p}}})^2$, where
$\mathcal L$ fixes the channel invariants and spectator registers.
Thus a line here is a three-dimensional channel fibre: its variable
is the pair relative momentum, or the moving register momentum in a
one-body channel. The following estimates hold uniformly on the finite
box:
\begin{align}
 L_j(x)^2&\le C_s\|\mathcal E^s\ket{u}\|^2 &&(s>3/4),\\
 \|\mathcal E^{-1/4}V_M\mathcal E^{-1/4}\|&\le C,
 &\|\mathcal E^{5/4}Q\ket{u}\|^2&\le C\mathcal D^2\log(2+n),
 \label{eq:flux-line-estimates}\\
 |\langle u|Z^{\le}|u\rangle|+
 |\langle u|Y_2|u\rangle|+
 |\langle u|i[V_M,Y_2]|u\rangle|
 &\le C\mathcal D^2K^{-3}\log^{q_2}(2+n).
 \label{eq:flux-second-estimates}
\end{align}
Here and below $q_2$ denotes a fixed finite logarithmic power.
The estimates also hold for spin-valued states after summing the
squared component bounds.
\end{lemma}
\begin{proofx}
We first explain the three weighted estimates that supply the endpoint
control. On a channel fibre with relative coordinate $\bm{\xi}$,
Cauchy--Schwarz gives
\begin{equation}
 \left(\sum_{\vec{\bm{p}}\in\mathcal L}x_{\vec{\bm{p}}}\right)^2
 \le\left(\sum_{\bm{\xi}}(1+\|\bm{\xi}\|^2)^{-2s}\right)
     \sum_{\vec{\bm{p}}\in\mathcal L}(1+\|\bm{\xi}\|^2)^{2s}|u_{\vec{\bm{p}}}|^2.
\end{equation}
The first sum is finite precisely in the stated range $s>3/4$ in
three dimensions. The second is bounded by the corresponding total
kinetic weight, with constants depending on the fixed masses and
cell. Summing over the fibre labels proves the first inequality.
We have taken an $\ell^1$ sum only within a three-dimensional fibre;
the distinct fibres are combined by squared norms.

The weighted Coulomb estimate
$\|\mathcal E^{-1/4}V_M\mathcal E^{-1/4}\|\le C$ follows from Hardy
and the square-root operator inequality, or from the weighted positive
Coulomb kernel. Taking the positive square root preserves the ordering
of positive operators, so the squared Hardy bound
$V_M^2\le C\mathcal E$ gives $|V_M|\le C\mathcal E^{1/2}$.
The weighted estimate distributes the potential's momentum
weight between the two sides of a form. The critical moment on $Q\ket{u}$
comes from the graph tail. On a dyadic momentum shell of scale
$\lambda$, the squared $\mathcal E$-weighted tail is
$O(\mathcal D^2\lambda^{-1})$. Replacing $\mathcal E$ by
$\mathcal E^{5/4}$ multiplies that squared bound by
$O(\lambda)$, leaving $O(\mathcal D^2)$ per shell. There are
$O(\log(2+n))$ shells below the bulk cutoff, which gives the last
estimate in \eqref{eq:flux-line-estimates}. This is the borderline
moment in the proof: it is bounded with a logarithm, not uniformly
without one.

We first give the pinning-to-weight calculation for $Z^{\le}$.
On its closed low-mismatch set,
$|\varpi|\le C\|\bm{\tau}\|^2$ and $\|\bm{\tau}\|\le C_LK$, so the
previous kernel estimate reduces to
\begin{equation}
 |Z^{\le}_j(\vec{\bm{q}},\vec{\bm{p}})|\le CK^{-3}\|\bm{\Pi}_j(\vec{\bm{p}})\|\,\|\bm{\tau}\|.
\end{equation}
The diagonal is zero, as is apparent from the difference of symbols
in \eqref{eq:zker}.  For a nonzero entry, pinning at both ends gives
\begin{equation}
 \|\bm{\tau}\|\le\frac83\sqrt{S_j(\vec{\bm{p}})},\qquad
 \|\bm{\tau}\|\le\frac83\sqrt{S_j(\vec{\bm{q}})}.
\end{equation}
Use the right-hand side of the kernel estimate only on this pinned
support as a symmetric nonnegative majorant $B_j(\vec{\bm{q}},\vec{\bm{p}})$.  It is
symmetric because $\bm{\Pi}_j$ is conserved and $\|\bm{\tau}\|$ is unchanged
by reversal.  With $R_{\vec{\bm{p}}}=(8/3)\sqrt{S_j(\vec{\bm{p}})}$, injectivity of the
channel transfer and the integer-box count give
\begin{align}
 \sum_{\vec{\bm{q}}} B_j(\vec{\bm{q}},\vec{\bm{p}})
 &\le CK^{-3}\|\bm{\Pi}_j(\vec{\bm{p}})\|
             \sum_{\|\bm{\tau}\|\le R_{\vec{\bm{p}}}}\|\bm{\tau}\|\\
 &\le CK^{-3}\|\bm{\Pi}_j(\vec{\bm{p}})\|R_{\vec{\bm{p}}}(2R_{\vec{\bm{p}}}+1)^3
 \le CK^{-3}(1+S_j(\vec{\bm{p}}))^{5/2}
 \le CK^{-3}\mathcal E(\vec{\bm{p}})^{5/2}.
\end{align}
There is also a factor $\mathbf1_Q(\vec{\bm{p}})$, since both endpoints are
in $Q$.  The elementary inequality $2|u_{\vec{\bm{q}}}||u_{\vec{\bm{p}}}|\le|u_{\vec{\bm{q}}}|^2+|u_{\vec{\bm{p}}}|^2$
then proves
\begin{equation}
 |\langle u|Z^{\le}|u\rangle|
 \le CK^{-3}\sum_{\vec{\bm{p}}\in Q}\mathcal E(\vec{\bm{p}})^{5/2}|u_{\vec{\bm{p}}}|^2
 \le C\mathcal D^2K^{-3}\log(2+n).
 \label{eq:flux-low-explicit-form}
\end{equation}
This is precisely where the critical moment enters the low part.

For the $Y_2$ endpoint a simpler fibre trace suffices.  On each
channel fibre choose one moved register momentum $\bm{\xi}_j$ as its
free coordinate and put $\mathfrak w_j(\vec{\bm{p}})=(1+\|\bm{\xi}_j(\vec{\bm{p}})\|^2)^{3/2}$.
There are no zero divided transfers, and a nonzero integer transfer
has $\|\bm{\tau}\|\ge1$.  Consequently
\begin{align}
 \sum_{\substack{\vec{\bm{q}},\vec{\bm{p}}\in Q\text{ on }j\\\vec{\bm{q}}\ne \vec{\bm{p}}}}
 \frac{\|\bm{\Pi}_j(\vec{\bm{p}})\|}{\|\bm{\tau}\|}|u_{\vec{\bm{q}}}||u_{\vec{\bm{p}}}|
 &\le\sum_{\mathcal L}\|\bm{\Pi}_j(\mathcal L)\|
                      \left(\sum_{\vec{\bm{p}}\in Q\cap\mathcal L}|u_{\vec{\bm{p}}}|\right)^2\\
 &\le C\log(2+n)\sum_{\vec{\bm{p}}\in Q}\mathfrak w_j(\vec{\bm{p}})\|\bm{\Pi}_j(\vec{\bm{p}})\||u_{\vec{\bm{p}}}|^2\\
 &\le C\log(2+n)\|\mathcal E Q\ket{u}\|^2.
\end{align}
In the second line, Cauchy--Schwarz uses
$\sum_{\vec{\bm{p}}\in Q\cap\mathcal L}\mathfrak w_j(\vec{\bm{p}})^{-1}\le C\log(2+n)$;
the last line uses $\mathfrak w_j(\vec{\bm{p}})\|\bm{\Pi}_j(\vec{\bm{p}})\|\le C\mathcal E(\vec{\bm{p}})^2$.
Multiplying by $CK^{-3}$ from \eqref{eq:flux-Y2-kernel} proves
\begin{equation}
 |\langle u|Y_2|u\rangle|
 \le C\mathcal D^2K^{-3}\log(2+n).
 \label{eq:flux-Y2-explicit-form}
\end{equation}
Thus the inverse transfer can be bounded by one for this endpoint.
It must be kept in the commutator calculation, to which we now turn.

Let $v_j(\vec{\bm{a}},\vec{\bm{c}})\le C_j\|\bm{\sigma}\|^{-2}$ be the nonnegative Coulomb
majorant for a hop in channel $j$, and let
\begin{equation}
 b_j(\vec{\bm{c}},\vec{\bm{b}})=CK^{-3}\frac{\|\bm{\Pi}_j(\vec{\bm{b}})\|}{\|\bm{\tau}\|}
\end{equation}
be the $Y_2$ majorant on a nonzero $j$-hop with $\vec{\bm{c}},\vec{\bm{b}}\in Q$,
zero elsewhere. We bound the absolute sum
\begin{equation}
 \sum_{\vec{\bm{a}},\vec{\bm{c}},\vec{\bm{b}}}|u_{\vec{\bm{a}}}|v_{j'}(\vec{\bm{a}},\vec{\bm{c}})b_j(\vec{\bm{c}},\vec{\bm{b}})|u_{\vec{\bm{b}}}|.
\end{equation}
If $j'=j$, conservation of $\bm{\Pi}_j$ and
\begin{equation}
 \frac1{\|\bm{\sigma}\|^2\|\bm{\tau}\|}
 \le\frac23\|\bm{\sigma}\|^{-3}+\frac13\|\bm{\tau}\|^{-3}
\end{equation}
show that summing over the intermediate state $\vec{\bm{c}}$ costs at most
$C\log(2+M)$.  Both transfer maps are injective along this fibre;
all transfers have size at most $CM$, and their zero terms are
absent.  A second use of the preceding fibre trace, now over the
auxiliary box, costs at most $C\log(2+M)$.  Hence
\begin{equation}
 \sum_{\vec{\bm{a}},\vec{\bm{c}},\vec{\bm{b}}}|u_{\vec{\bm{a}}}|v_j(\vec{\bm{a}},\vec{\bm{c}})b_j(\vec{\bm{c}},\vec{\bm{b}})|u_{\vec{\bm{b}}}|
 \le CK^{-3}\log^2(2+M)\|\mathcal E \ket{u}\|^2.
 \label{eq:flux-Y2-same-explicit}
\end{equation}

For different channels, we specify the coordinates before taking any
sums.  The $Y_2$ channel $j$ is a quantum pair; a one-body
centre-of-mass dressing is zero.  Choose a register of $j$ which
$j'$ does not move and denote its momentum by $\bm{\xi}$.  If $j'$
moves a register outside $j$, let $\bm{\zeta}$ be that register's
momentum.  This includes two pairs sharing a register, disjoint
pairs, and a one-body channel outside the pair.  In the remaining
case $j'$ moves just one register of $j$: take $\bm{\xi}$ to be the
other register and $\bm{\zeta}=\bm{\Pi}_j$, the pair total momentum.
In every case, with signs $\epsilon_1,\epsilon_2\in\{1,-1\}$,
\begin{equation}
 \bm{\xi}(\vec{\bm{c}})=\bm{\xi}(\vec{\bm{b}})+\epsilon_1\bm{\tau},\quad \bm{\zeta}(\vec{\bm{c}})=\bm{\zeta}(\vec{\bm{b}}),
 \qquad
 \bm{\xi}(\vec{\bm{a}})=\bm{\xi}(\vec{\bm{c}}),\quad \bm{\zeta}(\vec{\bm{a}})=\bm{\zeta}(\vec{\bm{c}})+\epsilon_2\bm{\sigma}.
\end{equation}
For example, for $j=(a_0,b_0)$ and $j'=(a_0,c_0)$, the choices
$\bm{\xi}=\bm{p}_{b_0}$ and $\bm{\zeta}=\bm{p}_{c_0}$ have exactly these properties.
Thus, when one endpoint of either hop is fixed, its changing
coordinate and its transfer each give injective parametrisations of
the intermediate configurations.  Identical channels have already
been treated in \eqref{eq:flux-Y2-same-explicit}; there is no
remaining degenerate two-coordinate case.

Put
\begin{equation}
 X(\vec{\bm{p}})=1+\|\bm{\xi}(\vec{\bm{p}})\|^2,\qquad
 Z(\vec{\bm{p}})=(1+\|\bm{\zeta}(\vec{\bm{p}})\|^2)^{1/2},\qquad
 h(\vec{\bm{p}})=\frac1{X(\vec{\bm{p}})Z(\vec{\bm{p}})},\qquad L_M=\log(2+M).
\end{equation}
Remove the conserved numerator from $b_j$ and call the remaining
kernel $b'_j=CK^{-3}\|\bm{\tau}\|^{-1}$ on its original support.  The
four one-hop sums we need are
\begin{align}
 \sum_{\vec{\bm{b}}}\frac{b'_j(\vec{\bm{c}},\vec{\bm{b}})}{X(\vec{\bm{b}})}&\le CK^{-3}L_M,&
 \sum_{\vec{\bm{c}}}\frac{b'_j(\vec{\bm{c}},\vec{\bm{b}})}{X(\vec{\bm{c}})}&\le CK^{-3}L_M,\\
 \sum_{\vec{\bm{c}}}\frac{v_{j'}(\vec{\bm{a}},\vec{\bm{c}})}{Z(\vec{\bm{c}})}&\le CL_M,&
 \sum_{\vec{\bm{a}}}\frac{v_{j'}(\vec{\bm{a}},\vec{\bm{c}})}{Z(\vec{\bm{a}})}&\le CL_M.
 \label{eq:flux-cross-one-hop}
\end{align}
Here is a direct verification of their weights.  Weighted
arithmetic--geometric mean gives
\begin{align}
 \frac1{\|\bm{\tau}\|(1+\|\bm{\xi}\|^2)}
 &\le\frac13\|\bm{\tau}\|^{-3}
       +\frac23(1+\|\bm{\xi}\|^2)^{-3/2},\\
 \frac1{\|\bm{\sigma}\|^2(1+\|\bm{\zeta}\|^2)^{1/2}}
 &\le\frac23\|\bm{\sigma}\|^{-3}
       +\frac13(1+\|\bm{\zeta}\|^2)^{-3/2}.
\end{align}
Each transfer or coordinate in these sums ranges injectively over
an integer box of radius $CM$.  The two inverse-cubic sums are
$O(L_M)$: a shell of radius $2^r$ has $O(2^{3r})$ points and each
summand is $O(2^{-3r})$.  This proves all four inequalities,
including their reversed orientations.

Define the composite nonnegative kernel
$H(\vec{\bm{a}},\vec{\bm{b}})=\sum_{\vec{\bm{c}}} v_{j'}(\vec{\bm{a}},\vec{\bm{c}})b'_j(\vec{\bm{c}},\vec{\bm{b}})$.  Since $Z(\vec{\bm{c}})=Z(\vec{\bm{b}})$ on
the $j$ hop, the first and third bounds in
\eqref{eq:flux-cross-one-hop} give
\begin{equation}
 \sum_{\vec{\bm{b}}} H(\vec{\bm{a}},\vec{\bm{b}})h(\vec{\bm{b}})
 =\sum_{\vec{\bm{c}}}\frac{v_{j'}(\vec{\bm{a}},\vec{\bm{c}})}{Z(\vec{\bm{c}})}
              \sum_{\vec{\bm{b}}}\frac{b'_j(\vec{\bm{c}},\vec{\bm{b}})}{X(\vec{\bm{b}})}
 \le CK^{-3}L_M^2.
\end{equation}
Similarly $X(\vec{\bm{a}})=X(\vec{\bm{c}})$ on the $j'$ hop, so the other two bounds
give $\sum_{\vec{\bm{a}}} H(\vec{\bm{a}},\vec{\bm{b}})h(\vec{\bm{a}})\le CK^{-3}L_M^2$.
Cauchy--Schwarz applied to the kernel $H(\vec{\bm{a}},\vec{\bm{b}})h(\vec{\bm{a}})h(\vec{\bm{b}})$ now yields
\begin{align}
 &\sum_{\vec{\bm{a}},\vec{\bm{c}},\vec{\bm{b}}}|u_{\vec{\bm{a}}}|v_{j'}(\vec{\bm{a}},\vec{\bm{c}})b_j(\vec{\bm{c}},\vec{\bm{b}})|u_{\vec{\bm{b}}}|\\
 &\quad\le CK^{-3}L_M^2
 \left(\sum_{\vec{\bm{a}}} X(\vec{\bm{a}})Z(\vec{\bm{a}})|u_{\vec{\bm{a}}}|^2\right)^{1/2}
 \left(\sum_{\vec{\bm{b}}\in Q}\|\bm{\Pi}_j(\vec{\bm{b}})\|^2X(\vec{\bm{b}})Z(\vec{\bm{b}})|u_{\vec{\bm{b}}}|^2\right)^{1/2}
 \\
 &\quad\le CK^{-3}L_M^2
       \|\mathcal E^{3/4}\ket{u}\|\,
       \|\mathcal E^{5/4}Q\ket{u}\|.
 \label{eq:flux-Y2-cross-explicit}
\end{align}
Indeed $\bm{\xi}$ and $\bm{\zeta}$ are each sums of at most two register
momenta, so $XZ\le C\mathcal E^{3/2}$ and
$\|\bm{\Pi}_j\|^2XZ\le C\mathcal E^{5/2}$.
The first norm is at most $C\mathcal D$, and the critical-moment
bound gives $C\mathcal D\sqrt{\log(2+n)}$ for the second.
Finally, self-adjointness gives
$|\langle u|i[V_M,Y_2]|u\rangle|\le2|\langle V_Mu\,|\,Y_2u\rangle|$.
Sum the finitely many channel contributions in
\eqref{eq:flux-Y2-same-explicit} and
\eqref{eq:flux-Y2-cross-explicit}, and use $M=n^6$.
Together with \eqref{eq:flux-low-explicit-form} and
\eqref{eq:flux-Y2-explicit-form}, this proves
\eqref{eq:flux-second-estimates} with a fixed finite logarithmic
power.  The estimates use three-dimensional channel sums and retain
the full kinetic weights, including spectators.  For spin-valued
states the same positive-kernel arguments apply to the norm of the
spin vector at each momentum, so no spin-space dimension is lost.
\end{proofx}

\subsubsection{The last far-to-cold block and closed excursions}

We have now dealt with the entire $Q\times Q$ block of $N_2$.
To identify what still needs elimination, put
$X:=FN_2C$, $N^{FC}:=X+X^*$, and
$N^{\rm prod}:=N_2+Z_a-N^{FC}$. The superscript $FC$ records the
block from the cold region to $F$, together with its Hermitian reverse.
The matrix $N^{\rm prod}$ has both endpoints outside $C$:
adding $Z_a$ removes its $Q\times Q$ block, and subtracting
$N^{FC}$ removes the remaining $F\times C$ and $C\times F$ blocks.
Thus the exact remainder decomposition is
\begin{equation}
 N_2=N^{\rm prod}-Z^{\le}-Z^>+N^{FC}.
\end{equation}
We give the restricted products explicitly, because support outside
$C$ alone would not bound an arbitrary two-potential matrix.
Temporarily write
\begin{equation}
 \ket{\alpha}=(Q-C)\ket{u},\qquad \ket{\beta}=(P-Q)\ket{u},\qquad \ket{d}=R\ket{u},
 \qquad V=V_M,\quad V_\lambda=V_{\lambda,M}.
\end{equation}
Expanding the flat remainder and $-i[V,Y_1]$, and removing their
$Q\times Q$ and far-to-cold blocks, gives
\begin{align}
 \langle u|N^{\rm prod}|u\rangle={}&
 2\operatorname{Im}\langle Vd\,|\,PV_\lambda(\alpha+\beta)\rangle
 -2\operatorname{Im}\langle RV\alpha\,|\,RV_\lambda\beta\rangle\\
 &-2\operatorname{Im}\langle RV\beta\,|\,RV_\lambda\alpha\rangle
 -2\operatorname{Im}\langle RV\beta\,|\,RV_\lambda\beta\rangle\\
 &+2\operatorname{Im}\langle V(\alpha+\beta+d)\,|\,Yd\rangle\\
 &+2\operatorname{Im}\langle RV(\beta+d)\,|\,RY\alpha\rangle
 +2\operatorname{Im}\langle RV(\alpha+\beta+d)\,|\,RY\beta\rangle.
\end{align}
All vectors to which $V$, $V_\lambda$ or $Y$ is applied are
restrictions of $\ket{u}$ outside $C$; sums such as $\ket{\alpha}+\ket{\beta}+\ket{d}$
are still single restrictions, because their supports are disjoint.
For any such restriction $S$, put $\ell=c_*K\ge1$.  Hardy and
the interpolated momentum-tail estimate for the evolved state give
\begin{equation}
 \|VS\ket{u}\|+\|V_\lambda S\ket{u}\|
 \le C\|\mathcal E^{1/2}S\ket{u}\|
 \le C\mathcal D\ell^{-3/2}.
\end{equation}
To estimate $YS\ket{u}$, use the conservation of $r_j$ on each channel
fibre in \eqref{eq:flux-Y-explicit-rows}.  Row Cauchy--Schwarz
followed by the column bound gives
\begin{equation}
 \|Y_j\ket{v}\|^2
 \le\sum_{\vec{\bm{p}}} r_j(\vec{\bm{p}})^2|v_{\vec{\bm{p}}}|^2
 \le C\Lambda_M^2\|\mathcal E^{1/2}\ket{v}\|^2.
\end{equation}
For the first inequality, the intermediate coefficient is
$\sum_{\vec{\bm{q}}} r_j(\vec{\bm{q}})|Y_j(\vec{\bm{q}},\vec{\bm{p}})|$; conservation permits replacing
$r_j(\vec{\bm{q}})$ by $r_j(\vec{\bm{p}})$, after which the column sum is at most
$r_j(\vec{\bm{p}})^2$.  Summing the finite number of channels therefore gives
\begin{equation}
 \|YS\ket{u}\|\le C\Lambda_M\mathcal D\ell^{-3/2}.
\end{equation}
The four flat products above are bounded by
$8(C\mathcal D\ell^{-3/2})^2$, and the three centre-of-mass
products by
\begin{equation}
 6(C\mathcal D\ell^{-3/2})(C\Lambda_M\mathcal D\ell^{-3/2}).
\end{equation}
Projections on the outputs only decrease these norms. Hence
\begin{equation}
 |\langle u|N^{\rm prod}|u\rangle|
 \le C\mathcal D^2(1+\Lambda_M)\ell^{-3}
 \le C\mathcal D^2K^{-3}\log^2(2+n),
\end{equation}
which has the asserted form with a fixed logarithmic power.

The remaining term $N^{FC}$ has one cold endpoint, for which such
a tail estimate is unavailable. In exchange, every one of its matrix
elements has the gap \eqref{eq:flux-far-cold-gap}.

We therefore define the Hermitian matrix $Y_3$ by
$Y_{3,\vec{\bm{f}}\vec{\bm{b}}}:=N^{FC}_{\vec{\bm{f}}\vec{\bm{b}}}/(i\omega(\vec{\bm{f}},\vec{\bm{b}}))$ on $F\times C$ and by
Hermitian reversal on $C\times F$, with the other blocks zero.
Then $i[T_M,Y_3]=N^{FC}$. The kinetic mismatches $\omega(\vec{\bm{f}},\vec{\bm{b}})$ are
strictly positive in the displayed orientation, so this third
elimination creates no new resonant remainder.

Before estimating it, we make explicit which two-hop terms this last
operator contains. With $F':=P-Q$, direct expansion of $FN_2C$ gives
\begin{align}
 X&=-iBC,\\
 B={}&RV_MPV_{\lambda,M}-F'V_MRV_{\lambda,M}
       +F'V_{\lambda,M}RV_M\\
    &+FV_MRY-RYV_M-F'YRV_M.
 \label{eq:flux-FC-words}
\end{align}
All $R$ in this formula denote $R_M$. Reading each product from right
to left displays the intermediate support of the two hops starting in
$C$. Expanding each $V_{\mathrm{tot}}$ or $Y$ by channel gives a
finite list of such ordered products, with both orientations retained.
For a same-channel flat contribution, the two $F'$ products cancel,
leaving $\lambda_jRV_jPV_jC$. This remaining product is the one whose
closed return will require reflection. The exact formula ensures that
the estimates cover all channels and all support restrictions, rather
than only a representative pair interaction.

\begin{lemma}[Third dressing and reflected closed excursions]
\label{prop:z3}
For the projected exact states $\ket{u}=B_M\ket{\psi(s)}$, with the uniform
plain, half-kinetic and graph-tail estimates established above,
\begin{equation}
 |\langle u|Y_3|u\rangle|+
 |\langle u|i[V_M,Y_3]|u\rangle|
 \le C\mathcal D^2K^{-3}\log^{q_3}(2+n).
 \label{eq:flux-third-estimates}
\end{equation}
The constants include every channel product in
\eqref{eq:flux-FC-words}.
No reality assumption on the channel coefficients is imposed.
\end{lemma}
\begin{proofx}
Write $\widetilde Y:=FY_3C$, so
$Y_3=\widetilde Y+\widetilde Y^*$. We first establish the two product
bounds needed outside the $C\times C$ block. We then keep the extra
structure needed when both outer endpoints are cold. The distinction
is important: an estimate sufficient for $\|\widetilde Y\ket{u}\|$ does
not by itself give the additional power required for a closed
excursion.

\paragraph{Kernel sums for the far-to-cold products.}
For $\vec{\bm{f}}\in F$, put $w_F(\vec{\bm{f}}):=1/T(\vec{\bm{f}})$, and set $w_F=0$ on $Q$.
This symbol is an inverse-energy weight, unrelated to a Coulomb
coefficient. The far-to-cold gap gives
$[T(\vec{\bm{f}})-T(\vec{\bm{b}})]^{-1}\le(4/3)w_F(\vec{\bm{f}})$ for $\vec{\bm{b}}\in C$. Consequently, if
$\kappa(\vec{\bm{f}},\vec{\bm{b}})$ is any of the two-hop kernels in
\eqref{eq:flux-FC-words}, its contribution to the squared norm of
the divided operator is at most
\begin{equation}
 \frac{16}{9}\sum_{\vec{\bm{f}}} w_F(\vec{\bm{f}})^2
       \left(\sum_{\vec{\bm{b}}\in C}|\kappa(\vec{\bm{f}},\vec{\bm{b}})|\,|v_{\vec{\bm{b}}}|\right)^2.
\end{equation}
The finite sum of channel products costs only a fixed factor by
Cauchy--Schwarz. We will bound the displayed expression for each
product, without discarding the energy dependence of $w_F$.

We use positive kernels to make these estimates independent of the
phases of the Fourier coefficients. Write $w_j(\vec{\bm{q}},\vec{\bm{p}})=(V_j)_{\vec{\bm{q}}\vec{\bm{p}}}$ and
choose fixed constants $h_j$ such that
\begin{equation}
 |w_j(\vec{\bm{q}},\vec{\bm{p}})|\le W_j(\vec{\bm{q}},\vec{\bm{p}}):=
 \begin{cases}
 h_j/\|\bm{\theta}\|^2,&\vec{\bm{q}}=\vec{\bm{p}}\oplus_j\bm{\theta},\quad\bm{\theta}\ne0,\\
 0,&\text{otherwise}.
 \end{cases}
\end{equation}
Here $\vec{\bm{p}}\oplus_j\bm{\theta}$ is the configuration obtained from $\vec{\bm{p}}$ by
a hop $\bm{\theta}$ of channel $j$; equivalently, $\vec{\bm{q}}\sim_j \vec{\bm{p}}$ denotes
the corresponding channel-fibre relation.
Let $y_j$ be the channel kernel of the first centre-of-mass operator
$Y$. Thus $y_j=-c_jw_j/\omega_j$ on its gapped set and zero
elsewhere. Lowercase $y_j$ in this proof always denotes a channel of
$Y$, not the later auxiliary operator $Y_2$.
For a channel $j$, let $\vec{\bm{\ell}}_j(\vec{\bm{p}})$ collect its conserved pair
momentum and spectator registers; in a one-body channel it collects
just the spectator registers. A channel coordinate $\bm{\zeta}$ is a
register momentum, or the sum of at most two register momenta, that
changes by $\pm\bm{\theta}$ along a $j$ hop. On a fixed $j$ fibre it
therefore parametrises the three-dimensional lattice injectively.
We write
\begin{equation}
 S_jx(\vec{\bm{\ell}}):=\sum_{\vec{\bm{p}}:\vec{\bm{\ell}}_j(\vec{\bm{p}})=\vec{\bm{\ell}}}x_{\vec{\bm{p}}},
 \qquad L_j(x)^2=\sum_{\vec{\bm{\ell}}} S_jx(\vec{\bm{\ell}})^2.
\end{equation}
This is the channel line sum already used in
Lemma~\ref{lem:flux-line-estimates}; we introduce its unsquared
version only to display the kernels below.

The elementary radial sums that produce the cutoff powers are
\begin{equation}
 \sum_{\bm{\zeta}\in\mathbb Z^3}
       \frac1{\max(n^2,\|\bm{\zeta}\|^2)^2}\le\frac Cn,
 \qquad
 \sum_{\bm{\zeta}\in\mathbb Z^3}
       \frac1{\max(n^2,\|\bm{\zeta}\|^2)^4}\le\frac C{n^5}.
\end{equation}
For the first sum, the ball $\|\bm{\zeta}\|\le n$ has $O(n^3)$ points,
each contributing $n^{-4}$; a dyadic shell of radius $\rho\ge n$
contributes $O(\rho^{-1})$. Summing the shells gives $O(n^{-1})$.
For the second sum the corresponding contributions are $O(n^{-5})$
and $O(\rho^{-5})$. A finite channel fibre is a subset of this
lattice, so the same bounds hold uniformly in its fixed labels.

Split a first Coulomb hop out of $C$ into $W_j^L$ and $W_j^S$ at
$\|\bm{\theta}\|=c_*n$, retaining the cold-input restriction in both
kernels. If the hop is long and $\bm{\zeta}$ is a coordinate of $j$,
then $\|\bm{\zeta}(\vec{\bm{b}})\|\le2c_*n$ and
$\|\bm{\zeta}(\vec{\bm{d}})\|\le\|\bm{\zeta}(\vec{\bm{b}})\|+\|\bm{\theta}\|\le3\|\bm{\theta}\|$. Hence
\begin{equation}
 W_j^L(\vec{\bm{d}},\vec{\bm{b}})\le
 \frac{C}{\max(n^2,\|\bm{\zeta}(\vec{\bm{d}})\|^2)}
       \ind{\vec{\bm{\ell}}_j(\vec{\bm{d}})=\vec{\bm{\ell}}_j(\vec{\bm{b}})},
 \qquad
 \sum_{\vec{\bm{d}}}\left(\sum_{\vec{\bm{b}}}W_j^L(\vec{\bm{d}},\vec{\bm{b}})x_{\vec{\bm{b}}}\right)^2
       \le\frac Cn L_j(x)^2.
\end{equation}
The second inequality follows by replacing the inner sum with the
first majorant times $S_jx(\vec{\bm{\ell}}_j(\vec{\bm{d}}))$ and then using the first
radial sum on each fibre. A first $Y$ hop from $C$ to $R_M$ has
the more precise majorant
\begin{equation}
 |y_j(\vec{\bm{d}},\vec{\bm{b}})|\ind{\vec{\bm{d}}\in R_M,\vec{\bm{b}}\in C}
 \le\frac{C\|\bm{\Pi}_j(\vec{\bm{b}})\|}{n\max(n^2,\|\bm{\zeta}(\vec{\bm{d}})\|^2)}
       \ind{\vec{\bm{\ell}}_j(\vec{\bm{d}})=\vec{\bm{\ell}}_j(\vec{\bm{b}})}.
\end{equation}
Indeed the cold-to-exterior kinetic gap bounds
$|y_j|$ by $C\|\bm{\Pi}_j(\vec{\bm{b}})\|/(n\|\bm{\theta}\|^2)$, and the same long-hop
comparison applies. Since $\|\bm{\Pi}_j(\vec{\bm{b}})\|\le2c_*n$ on $C$, this also
has the preceding $Cn^{-1}L_j(x)^2$ squared-sum bound. Later we
will retain $\|\bm{\Pi}_j(\vec{\bm{b}})\|/n$ instead of bounding it by a constant.

The far energy controls coordinates as well as the cutoff. For any
channel coordinate $\bm{\zeta}$, and for a pair total momentum $\bm{\Pi}_j$,
\begin{equation}
 \begin{split}
 w_F(\vec{\bm{f}})&\le\frac{C}{\max(n^2,\|\bm{\zeta}(\vec{\bm{f}})\|^2)},\\
 w_F(\vec{\bm{f}})^2&\le\frac{C}{n^2(1+\|\bm{\zeta}(\vec{\bm{f}})\|^2)},\qquad
 w_F(\vec{\bm{f}})^2\le\frac{C}{n^2(n^2+\|\bm{\Pi}_j(\vec{\bm{f}})\|^2)}.
 \end{split}
\end{equation}
These inequalities follow by combining $T(\vec{\bm{f}})\ge c n^2$ on $F$
with the lower bounds on total kinetic energy furnished by the
specified coordinate or pair momentum. Their constants include the
fixed masses and $g$. We now show how they control the possible
second hops.

Suppose first that the first hop has the long-hop squared-sum bound,
and write its nonnegative output as
$v_{\vec{\bm{d}}}:=\sum_{\vec{\bm{b}}}K_1(\vec{\bm{d}},\vec{\bm{b}})x_{\vec{\bm{b}}}$. If the second hop is Coulomb, dual Hardy
on its three-dimensional fibre gives
\begin{equation}
 \begin{split}
 \sum_{\vec{\bm{f}}} w_F(\vec{\bm{f}})^2\left(\sum_{\vec{\bm{d}}}W_{j_2}(\vec{\bm{f}},\vec{\bm{d}})v_{\vec{\bm{d}}}\right)^2
 &\le\frac C{n^2}\sum_{\vec{\bm{f}}}
   \frac{\left(\sum_{\vec{\bm{d}}\sim_{j_2}\vec{\bm{f}}}
                      v_{\vec{\bm{d}}}/\|\bm{\theta}_2\|^2\right)^2}
        {1+\|\bm{\zeta}_2(\vec{\bm{f}})\|^2}\\
 &\le\frac C{n^2}\sum_{\vec{\bm{d}}}v_{\vec{\bm{d}}}^2
 \le\frac C{n^3}L_{j_1}(x)^2.
 \end{split}
\end{equation}
Here the sum is on the same second-channel fibre, and all
displayed singular denominators exclude the zero transfer. Dual Hardy is the adjoint form of the channel
Coulomb inequality: it places the inverse coordinate weight on the
output instead of a coordinate weight on the input.

If the second hop is $y_{j_2}$, we use its conserved pair momentum.
The divided-kernel row and column estimates give, for a nonnegative
sequence $v$,
\begin{equation}
 \sum_{\vec{\bm{f}}}\left(\sum_{\vec{\bm{d}}}|y_{j_2}(\vec{\bm{f}},\vec{\bm{d}})|v_{\vec{\bm{d}}}\right)^2
 \le C\log^q(2+M)\sum_{\vec{\bm{d}}}\|\bm{\Pi}_{j_2}(\vec{\bm{d}})\|^2v_{\vec{\bm{d}}}^2.
\end{equation}
Since $\bm{\Pi}_{j_2}$ is constant on this second-channel fibre, we
can apply this estimate to
$v_{\vec{\bm{d}}}/(n^2+\|\bm{\Pi}_{j_2}(\vec{\bm{d}})\|^2)^{1/2}$ and obtain
\begin{equation}
 \begin{split}
 \sum_{\vec{\bm{f}}} w_F(\vec{\bm{f}})^2\left(\sum_{\vec{\bm{d}}}|y_{j_2}(\vec{\bm{f}},\vec{\bm{d}})|v_{\vec{\bm{d}}}\right)^2
 &\le\frac{C\log^q(2+M)}{n^2}
     \sum_{\vec{\bm{d}}}\frac{\|\bm{\Pi}_{j_2}(\vec{\bm{d}})\|^2}
                   {n^2+\|\bm{\Pi}_{j_2}(\vec{\bm{d}})\|^2}v_{\vec{\bm{d}}}^2\\
 &\le\frac{C\log^q(2+M)}{n^3}L_{j_1}(x)^2.
 \end{split}
\end{equation}
Thus both types of second hop give the same $n^{-3}$ squared-norm
rate after a long first hop.

It remains to check a short first Coulomb hop. Its output
$v^S_{\vec{\bm{d}}}:=\sum_{\vec{\bm{b}}}W_{j_1}^S(\vec{\bm{d}},\vec{\bm{b}})x_{\vec{\bm{b}}}$ lies in the small ball in which
every register has norm at most $2c_*n$. A hop from this ball to
$F$ has $\|\bm{\theta}_2\|\ge n/8$. The same coordinate comparison as
above, together with the far energy weight, therefore gives
\begin{equation}
 w_F(\vec{\bm{f}})W_{j_2}(\vec{\bm{f}},\vec{\bm{d}})\le
       \frac{C}{\max(n^2,\|\bm{\zeta}_2(\vec{\bm{f}})\|^2)^2}.
\end{equation}
When the second hop is $y_{j_2}$, the relevant product has output
in $R_M$. The kinetic energy of $\vec{\bm{d}}$ in the small ball is then at
most half that of $\vec{\bm{f}}$, and the same bound holds for
$w_F(\vec{\bm{f}})|y_{j_2}(\vec{\bm{f}},\vec{\bm{d}})|$. Using the second radial sum, either case
gives
\begin{equation}
 \sum_{\vec{\bm{f}}}\left(\sum_{\vec{\bm{d}}} w_F(\vec{\bm{f}})K_2(\vec{\bm{f}},\vec{\bm{d}})v^S_{\vec{\bm{d}}}\right)^2
 \le\frac C{n^5}L_{j_2}(v^S)^2.
\end{equation}
On the same channel, the short-hop column sum is
$\sum_{\vec{\bm{d}}}W_{j_1}^S(\vec{\bm{d}},\vec{\bm{b}})\le Cn$, because
$\sum_{0<\|\bm{\theta}\|<c_*n}\|\bm{\theta}\|^{-2}\le Cn$. Hence
$L_{j_1}(v^S)^2\le Cn^2L_{j_1}(x)^2$.
For distinct channels choose a coordinate $\bm{\xi}_1$ of $j_1$ that
is fixed along $j_2$. Hardy in that transverse coordinate gives
\begin{equation}
 L_{j_2}(v^S)^2
 \le C L_{j_2}\big((1+\|\bm{\xi}_1\|^2)^{1/2}x\big)^2
 \le Cn^2L_{j_2}(x)^2,
\end{equation}
where the last inequality uses cold support. To see the transverse
step concretely, fix all remaining invariants and use coordinates
$\bm{r},\bm{s}$ moved independently by the two channels. Summing over the
$j_2$ line commutes with the $j_1$ convolution, so the relevant
sum becomes
$\sum_{\bm{r}'}\|\bm{r}-\bm{r}'\|^{-2}\sum_{\bm{s}}x(\bm{r}',\bm{s})$. Squaring, summing in $\bm{r}$,
and applying the three-dimensional Coulomb Hardy inequality gives
$C\sum_{\bm{r}'}(1+\|\bm{r}'\|^2)(\sum_{\bm{s}}x(\bm{r}',\bm{s}))^2$, exactly the first
inequality above. For pairs with a shared register, the two
non-shared register momenta provide these coordinates; for a
one-body channel sharing a register with a pair, use the other
register of the pair and the pair total momentum. For disjoint
channels, use a moved register from each. Thus short first hops also give $n^{-5}n^2=n^{-3}$.

These three calculations cover all terms of
\eqref{eq:flux-FC-words}. The flat products have a Coulomb
second hop; the product $FV_MRYC$ has a first $Y$ hop from $C$
to $R_M$; and the products with $Y$ on the left have a second
$Y$ hop. Whenever their middle projector is $R_M$, a short first
hop is absent. We have therefore proved the useful bound, valid
for an arbitrary finite-box input $\ket{v}$,
\begin{equation}
 \|\widetilde Y\ket{v}\|^2
 \le\frac{C\log^q(2+M)}{n^3}
       \sum_j L_j\big(C|v|\big)^2.
 \label{eq:flux-third-cold-input}
\end{equation}
This is a bound on the action from cold inputs with specified line
sums, not a uniform norm estimate for $Y_3$ on arbitrary states.

To apply it to $\ket{v}=\ket{u}$ and $\ket{v}=V_M\ket{u}$, we record the endpoint weights
explicitly. The channel trace at the borderline exponent gives
\begin{equation}
 L_j(\phi|u|)^2
 \le C\log(2+M)\sum_{\vec{\bm{p}}}\mathcal E(\vec{\bm{p}})^{3/2}\phi(\vec{\bm{p}})^2|u_{\vec{\bm{p}}}|^2.
\end{equation}
Every extra weight needed here satisfies
$\phi^2\le C\mathcal E$. Thus the last sum is at most
$C\|\mathcal E^{5/4}\ket{u}\|^2\le
 C\mathcal D^2\log(2+M)$. The latter follows by the same dyadic
graph-tail calculation as the critical moment on $Q\ket{u}$, now summed
to the auxiliary cutoff $M$. We use this full-box moment only with
logarithmic dependence on $M$.

For the potential input, first take a channel $j_3$ of $V_M$.
On its own line, the sum over an output in $C$ obeys
\begin{equation}
 \sup_{\vec{\bm{a}}}\sum_{\vec{\bm{b}}\in C}|w_{j_3}(\vec{\bm{b}},\vec{\bm{a}})|\le C(c_*n+1),
 \qquad
 L_{j_3}\big(C|V_{j_3}u|\big)^2
 \le Cn^2L_{j_3}(|u|)^2.
\end{equation}
The first estimate is the inverse-square sum on a translated ball
of radius $c_*n$; a ball near the singularity has sum $O(n)$,
and a ball far from it has no larger sum. For $j\ne j_3$, the
transverse Hardy calculation instead gives
\begin{equation}
 L_j\big(C|V_{j_3}u|\big)^2
 \le C L_j\big((1+\|\bm{\xi}_{j_3}\|^2)^{1/2}|u|\big)^2,
\end{equation}
where $\bm{\xi}_{j_3}$ is a coordinate of $j_3$ fixed on $j$ lines.
The preceding critical-moment estimate bounds this weighted line
sum by $C\mathcal D^2\log^q(2+M)$. Summing the finitely many
channels therefore yields
\begin{equation}
 \sum_jL_j\big(C|V_Mu|\big)^2
 \le Cn^2\mathcal D^2\log^q(2+M).
\end{equation}
Substituting these two inputs into the cold-input bound and taking
square roots gives, since $K=gn$ and $M=n^6$,
\begin{equation}
 \|\widetilde Y\ket{u}\|\le C\mathcal D K^{-3/2}\log^q(2+n),\qquad
 \|\widetilde YV_M\ket{u}\|\le C\mathcal D K^{-1/2}\log^q(2+n).
 \label{eq:flux-third-products}
\end{equation}
All estimates used positive channel majorants, so the same bounds
hold with those majorants in place of the original kernels.

The endpoint now follows by pairing with the far state tail:
\begin{equation}
 |\langle u|Y_3|u\rangle|
 \le2\|F\ket{u}\|\,\|\widetilde Y\ket{u}\|,
 \qquad \|F\ket{u}\|=O(\mathcal D K^{-5/2}).
\end{equation}
This gives $CK^{-4}$ times the squared graph norm and logarithms, which
suffices for the stated $K^{-3}$ endpoint. For the non-cold part of
the commutator we have
\begin{equation}
 2\|V_M(1-C)\ket{u}\|\|\widetilde Y\ket{u}\|
       +2\|\widetilde YV_M\ket{u}\|\|F\ket{u}\|
 \le C\mathcal D^2K^{-3}\log^q(2+n).
\end{equation}
Indeed Hardy and the half-kinetic tail at $c_*K$ give
$\|V_M(1-C)\ket{u}\|=O(\mathcal D K^{-3/2})$. The two products have
powers $K^{-3/2}K^{-3/2}$ and $K^{-1/2}K^{-5/2}$, respectively.
We are using finite kernel inequalities throughout; no iteration of
a singular continuum multiplier or extra domain assertion is needed.

\paragraph{The two squared sums in a closed excursion.}
It remains to estimate $\langle V_MCu\,|\,\widetilde YCu\rangle$.
Put $x_{\vec{\bm{b}}}:=\ind{\vec{\bm{b}}\in C}|u_{\vec{\bm{b}}}|$. For a closing channel $j_3$ and a
two-hop product with absolute output $H(\vec{\bm{f}})$, define
\begin{equation}
 G_3(\vec{\bm{f}}):=\ind{\vec{\bm{f}}\in F}\sum_{\vec{\bm{c}}}W_{j_3}(\vec{\bm{f}},\vec{\bm{c}})x_{\vec{\bm{c}}},
 \qquad
 \Phi_H:=\sum_{\vec{\bm{f}}} w_F(\vec{\bm{f}})^2H(\vec{\bm{f}})^2.
\end{equation}
Apart from the fixed factor $4/3$ from the inverse gap, the closed
form is bounded by $\sum_{\vec{\bm{f}}}G_3(\vec{\bm{f}})w_F(\vec{\bm{f}})H(\vec{\bm{f}})$. Cauchy--Schwarz
therefore uses the two \emph{specific} squared sums
\begin{equation}
 \left(\sum_{\vec{\bm{f}}}G_3(\vec{\bm{f}})w_F(\vec{\bm{f}})H(\vec{\bm{f}})\right)^2
 \le\left(\sum_{\vec{\bm{f}}}G_3(\vec{\bm{f}})^2\right)\Phi_H.
\end{equation}
Every closing hop from $C$ to $F$ is long. The first radial estimate
on the closing-channel fibre gives
\begin{equation}
 \sum_{\vec{\bm{f}}}G_3(\vec{\bm{f}})^2\le\frac CnL_{j_3}(x)^2.
\end{equation}
This is the $n^{-1}$ factor. We next establish $n^{-5}$ for the
word-side squared sum, retaining endpoint weights instead of
replacing them by their cutoff maxima.

For two distinct flat channels, split the first hop as before. On
its long part,
$\sum_{\vec{\bm{b}}}W_{j_1}^L(\vec{\bm{d}},\vec{\bm{b}})x_{\vec{\bm{b}}}\le Cn^{-2}S_{j_1}x(\vec{\bm{\ell}}_{j_1}(\vec{\bm{d}}))$.
The second channel has a coordinate $\bm{\xi}_2$ fixed on $j_1$ lines.
The transverse Hardy calculation now gives
\begin{equation}
 \begin{split}
 \sum_{\vec{\bm{f}}} w_F(\vec{\bm{f}})^2
   \left(\sum_{\vec{\bm{d}}}W_{j_2}(\vec{\bm{f}},\vec{\bm{d}})\sum_{\vec{\bm{b}}}W_{j_1}^L(\vec{\bm{d}},\vec{\bm{b}})x_{\vec{\bm{b}}}\right)^2
 &\le C\left[\sup_{\vec{\bm{\ell}}}\sum_{\vec{\bm{f}}:\vec{\bm{\ell}}_{j_1}(\vec{\bm{f}})=\vec{\bm{\ell}}}
                    \big(n^{-2}w_F(\vec{\bm{f}})\big)^2\right]
           L_{j_1}\big((1+\|\bm{\xi}_2\|^2)^{1/2}x\big)^2\\
 &\le\frac C{n^5}
           L_{j_1}\big((1+\|\bm{\xi}_2\|^2)^{1/2}x\big)^2.
 \end{split}
\end{equation}
In the last step, the square of $n^{-2}$ contributes $n^{-4}$,
and the fibre sum of $w_F^2$ contributes $n^{-1}$. More explicitly,
in the transverse coordinates used above the factor
$S_{j_1}x$ depends only on the coordinate moved by $j_2$; summing
$(n^{-2}w_F)^2$ in the other coordinate first leaves precisely the
three-dimensional Hardy form. This order of summation is the extra
gain over the earlier $n^{-3}$ product bound.

For a short first hop, the previously proved second-hop estimate
gives
\begin{equation}
 \Phi_H\le\frac C{n^5}L_{j_2}(W_{j_1}^Sx)^2
 \le\frac C{n^5}
       L_{j_2}\big((1+\|\bm{\xi}_1\|^2)^{1/2}x\big)^2.
\end{equation}
We now keep this endpoint weight; replacing it by $Cn$ would lose
two powers in the squared bound. The flat product whose middle
projector is $R_M$ has only the long first-hop case. These two
calculations cover the distinct-channel flat terms.

If the first hop is $Y$ from $C$ to $R_M$, its sharper majorant
retains $\|\bm{\Pi}_{j_1}(\vec{\bm{b}})\|/n$. Applying the long-first-hop Coulomb
calculation to this weighted input gives
\begin{equation}
 \Phi_H\le\frac C{n^2}\frac Cn\frac1{n^2}
                   L_{j_1}(\|\bm{\Pi}_{j_1}\|x)^2
       \le\frac C{n^5}L_{j_1}(\|\bm{\Pi}_{j_1}\|x)^2.
\end{equation}
The three factors are the far-weighted second-hop Hardy bound,
the long first-hop fibre sum, and the square of the additional
$1/n$ in the first $Y$ hop.

For a short first Coulomb hop followed by $y_{j_2}$, we similarly
retain the pair momentum instead of bounding it immediately. On
the relevant output $\vec{\bm{f}}\in R_M$, the short-ball gap and
$\|\bm{\theta}_2\|\ge n/8$ give
\begin{equation}
 w_F(\vec{\bm{f}})|y_{j_2}(\vec{\bm{f}},\vec{\bm{d}})|
 \le\frac{C}{\max(n^2,\|\bm{\zeta}_2(\vec{\bm{f}})\|^2)^2}
                   \frac{\|\bm{\Pi}_{j_2}(\vec{\bm{d}})\|}{n},
 \qquad
 \Phi_H\le\frac C{n^7}
                  L_{j_2}(\|\bm{\Pi}_{j_2}\|v^S)^2.
\end{equation}
When the two channels coincide, pair momentum is conserved on the
short first hop, so
$L_{j_2}(\|\bm{\Pi}_{j_2}\|v^S)^2\le
 Cn^2L_{j_2}(\|\bm{\Pi}_{j_2}\|x)^2$.
When they differ, $v^S$ is in the small ball and
$\|\bm{\Pi}_{j_2}(\vec{\bm{d}})\|\le4c_*n$, after which transverse Hardy gives
\begin{equation}
 L_{j_2}(\|\bm{\Pi}_{j_2}\|v^S)^2
 \le Cn^2L_{j_2}\big((1+\|\bm{\xi}_1\|^2)^{1/2}x\big)^2.
\end{equation}
Both cases recover the required $n^{-5}$ bound with an explicit
endpoint weight.

Finally, suppose the first Coulomb hop is long and the pair
momentum of the second $Y$ channel is conserved by that first
channel. The conserved-weight Schur estimate used above also gives
\begin{equation}
 \Phi_H\le\frac{C\log^q(2+M)}{n^4}
       \sum_{\vec{\bm{d}}}\|\bm{\Pi}_{j_2}(\vec{\bm{d}})\|^2
                \left(\sum_{\vec{\bm{b}}}W_{j_1}^L(\vec{\bm{d}},\vec{\bm{b}})x_{\vec{\bm{b}}}\right)^2
 \le\frac{C\log^q(2+M)}{n^5}
                L_{j_1}(\|\bm{\Pi}_{j_2}\|x)^2.
\end{equation}
Here we used $n^2+\|\bm{\Pi}_{j_2}\|^2\ge n^2$ in that estimate,
then moved the conserved momentum to the cold input before using
the long-hop fibre sum.

All endpoint weights in these formulas have square bounded by
$C\mathcal E$. The critical-moment line estimate therefore bounds
each weighted line sum by $C\mathcal D^2\log^q(2+M)$.
Consequently the named factors satisfy
\begin{equation}
 \sum_{\vec{\bm{f}}}G_3(\vec{\bm{f}})^2\le C\mathcal D^2n^{-1}\log^q(2+M),
 \qquad
 \Phi_H\le C\mathcal D^2n^{-5}\log^q(2+M).
\end{equation}
Their square-root product is
$C\mathcal D^2n^{-3}\log^q(2+M)$, as required. The remaining
case is a second $Y$ channel whose pair momentum changes along
the first channel.

\paragraph{Shared centre-of-mass channels.}
Write the second channel as the pair $j_2=(a,t)$, with $a$ moved
by $j_1$ and $t$ not moved by $j_1$. The following algebra replaces
the changing pair-momentum weight by a weight on the fixed register
$t$. Let $\bm{\xi}=\mu_{at}(\bm{d}_a/m_a-\bm{d}_t/m_t)$ and
$\bm{\theta}=\bm{\theta}_2$. The pair mismatch satisfies
\begin{equation}
 \bm{\theta}\cdot\bm{\xi}=\frac{\mu_{at}}{g^2}\omega_{j_2}
                              -\frac{\|\bm{\theta}\|^2}{2}.
\end{equation}
Also the centre-of-mass numerator, up to the sign of the measured
member, can be written using either register as
\begin{equation}
 \frac{g^2}{m_a+m_t}\bm{\theta}\cdot(\bm{d}_a+\bm{d}_t)
 =\frac{g^2}{m_a}\bm{\theta}\cdot(\bm{d}_a-\bm{\xi})
 =\frac{g^2}{m_t}\bm{\theta}\cdot(\bm{d}_t+\bm{\xi}).
\end{equation}
Use the last expression, the transfer-linear floor
$|\omega_{j_2}|\ge g^2\|\bm{\theta}\|/\mu_{at}$, and
$\mu_{at}\le m_t$. The ratio of numerator to mismatch is at most
$1+3\|\bm{d}_t\|+\|\bm{\theta}\|/2$. Multiplying by the Coulomb coefficient gives
the explicit decomposition
\begin{equation}
 |y_{j_2}(\vec{\bm{f}},\vec{\bm{d}})|
 \le h_{j_2}\frac{1+3\|\bm{d}_t\|}{\|\bm{\theta}_2\|^2}
                         +\frac{h_{j_2}}{2\|\bm{\theta}_2\|}.
\end{equation}
This implies the earlier form
$C(1+\|\bm{d}_t\|)/\|\bm{\theta}_2\|^2+C/(1+\|\bm{\theta}_2\|)$ because nonzero
integer transfers have norm at least one. The first numerator is
constant along the first-channel lines, since $\bm{d}_t=\bm{b}_t$ there.

We need two finite convolution inequalities for the two terms of
this decomposition. For a coordinate $\bm{\zeta}$ of $j_2$, dual Hardy
gives
\begin{equation}
 \sum_{\vec{\bm{f}}}\frac{\left(\sum_{\vec{\bm{d}}\sim_{j_2}\vec{\bm{f}}}
                     z_{\vec{\bm{d}}}/\|\bm{\theta}_2\|^2\right)^2}
                   {1+\|\bm{\zeta}(\vec{\bm{f}})\|^2}\le C\sum_{\vec{\bm{d}}}z_{\vec{\bm{d}}}^2.
\end{equation}
The inverse-first-power kernel satisfies the companion bound
\begin{equation}
 \sum_{\vec{\bm{f}}}\frac{\left(\sum_{\vec{\bm{d}}\sim_{j_2}\vec{\bm{f}}}
                     z_{\vec{\bm{d}}}/\|\bm{\theta}_2\|\right)^2}
                   {1+\|\bm{\zeta}(\vec{\bm{f}})\|^2}
 \le C\sum_{\vec{\bm{d}}}(1+\|\bm{\zeta}(\vec{\bm{d}})\|^2)z_{\vec{\bm{d}}}^2.
\end{equation}
To justify the latter, the positive lattice kernels obey
\begin{equation}
 \frac1{\|\bm{r}-\bm{s}\|}\le C\sum_{\bm{z}\ne \bm{r},\bm{s}}
                \frac1{\|\bm{r}-\bm{z}\|^2\|\bm{z}-\bm{s}\|^2},\qquad \bm{r}\ne \bm{s},
\end{equation}
with the intermediate sum restricted to a box of radius
$O(1+\|\bm{r}\|+\|\bm{s}\|)$. A box about the midpoint with $O(\|\bm{r}-\bm{s}\|^3)$
points, each at distance $O(\|\bm{r}-\bm{s}\|)$ from both endpoints, supplies
the lower bound of order $\|\bm{r}-\bm{s}\|^{-1}$; the finitely many shortest
separations are covered by increasing the constant. Apply dual
Hardy to the outer Coulomb convolution and Hardy to the inner
one. This proves the displayed finite-sum inequality. It is the
precise meaning of applying Hardy twice here, and does not assert
that a singular continuum potential preserves a Sobolev domain.

Suppose first that there is a coordinate $\bm{\zeta}$ of $j_2$ fixed
by both $j_1$ and the closing channel $j_3$. Set
$v_{\vec{\bm{d}}}:=\sum_{\vec{\bm{b}}}W_{j_1}^L(\vec{\bm{d}},\vec{\bm{b}})x_{\vec{\bm{b}}}$ and
$\Omega(\vec{\bm{f}}):=1+\|\bm{\zeta}(\vec{\bm{f}})\|^2$. Since $w_F^2\le Cn^{-4}$,
the two convolution inequalities give
\begin{equation}
 \begin{split}
 \sum_{\vec{\bm{f}}}\frac{w_F(\vec{\bm{f}})^2}{\Omega(\vec{\bm{f}})}
           \left(\sum_{\vec{\bm{d}}}|y_{j_2}(\vec{\bm{f}},\vec{\bm{d}})|v_{\vec{\bm{d}}}\right)^2
 &\le\frac C{n^4}\left[
       \sum_{\vec{\bm{d}}}(1+\|\bm{d}_t\|^2)v_{\vec{\bm{d}}}^2+\sum_{\vec{\bm{d}}}\Omega(\vec{\bm{d}})v_{\vec{\bm{d}}}^2\right]\\
 &\le\frac C{n^5}\left[
       L_{j_1}\big((1+\|\bm{p}_t\|^2)^{1/2}x\big)^2
       +L_{j_1}(\Omega^{1/2}x)^2\right].
 \end{split}
\end{equation}
The second step moves both weights through the first hop, using
their invariance on $j_1$ lines, and then applies the long-hop
$n^{-1}$ fibre bound. On the closing side, invariance on $j_3$
lines gives
\begin{equation}
 \sum_{\vec{\bm{f}}}\Omega(\vec{\bm{f}})G_3(\vec{\bm{f}})^2
 \le\frac CnL_{j_3}(\Omega^{1/2}x)^2.
\end{equation}
Weighted Cauchy--Schwarz pairs these two sums, with factors
$\Omega^{1/2}$ and $\Omega^{-1/2}$. They again have the powers
$n^{-1}$ and $n^{-5}$, and their endpoint weights are precisely
those controlled by the critical moment.

The common-coordinate condition covers all but the following
geometries. If $j_3$ does not move $t$, take $\bm{\zeta}=\bm{p}_t$.
If $j_3=(t,u)$ and $u$ is not moved by $j_1$, take
$\bm{\zeta}=\bm{p}_t+\bm{p}_u$. If $j_1=(a,s)$ and $j_3$ is the one-body
channel on $t$, take $\bm{\zeta}=\bm{p}_a+\bm{p}_s$. The remaining triples
are $j_3=j_2$, the two one-body outer channels $j_1=(a)$ and
$j_3=(t)$, and $j_1=(a,s)$ with $j_3=(t,s)$.
In these cases the closing channel moves $t$, and we use an
absolute kernel bound on their two-coordinate fibres.

We can also see directly why a sufficiently short second hop
cannot close through a channel moving $t$. A hop from a cold
$\vec{\bm{c}}$ to $\vec{\bm{f}}\in F$ has
$\|\bm{\theta}_3\|>n/4-c_*n$. Because $\bm{d}_t=\bm{b}_t$ and both $\vec{\bm{b}},\vec{\bm{c}}$ are
cold,
\begin{equation}
 \|\bm{\theta}_2\|=\|\bm{f}_t-\bm{d}_t\|
 \ge\|\bm{f}_t-\bm{c}_t\|-\|\bm{b}_t-\bm{c}_t\|
 \ge\|\bm{\theta}_3\|-2c_*n
 >n/4-3c_*n\ge5n/32>n/8.
\end{equation}
Thus $\|\bm{\theta}_2\|<n/8$ rules out such a closing channel. The
common-coordinate argument above handles the cases where this
exclusion is useful without introducing a new mismatch division.

For one of the remaining triples, write its positive closed kernel
as
\begin{equation}
 \mathcal K_{321}(\vec{\bm{c}},\vec{\bm{b}}):=
 \sum_{\vec{\bm{f}}\in F}W_{j_3}(\vec{\bm{f}},\vec{\bm{c}})w_F(\vec{\bm{f}})
                    \sum_{\vec{\bm{d}}}|y_{j_2}(\vec{\bm{f}},\vec{\bm{d}})|W_{j_1}^L(\vec{\bm{d}},\vec{\bm{b}}),
 \qquad \vec{\bm{b}},\vec{\bm{c}}\in C.
\end{equation}
For fixed $\vec{\bm{f}},\vec{\bm{b}}$ there is at most one admissible intermediate $\vec{\bm{d}}$.
Indeed $\bm{d}_t=\bm{b}_t$, while the $j_2$ hop changes register $t$ by
$\pm\bm{\theta}_2$; hence $\bm{f}_t-\bm{b}_t$ determines $\bm{\theta}_2$, which
then determines $\vec{\bm{d}}$. The sum over $\vec{\bm{d}}$ therefore has no hidden
multiplicity. The closing hop satisfies $\|\bm{\theta}_3\|\ge n/8$,
and the preceding triangle inequality gives
$\|\bm{\theta}_2\|\ge\|\bm{\theta}_3\|/2$, using $c_*\le1/32$.
The symbol decomposition, $\bm{d}_t=\bm{b}_t$ and cold support imply
\begin{equation}
 \sum_{\vec{\bm{d}}}|y_{j_2}(\vec{\bm{f}},\vec{\bm{d}})|W_{j_1}^L(\vec{\bm{d}},\vec{\bm{b}})
 \le\frac C{n^2}
       \left(\frac{1+n}{\|\bm{\theta}_3\|^2}
                         +\frac1{\|\bm{\theta}_3\|}\right).
\end{equation}
Here the first $n^{-2}$ is the long first-hop Coulomb bound.
Multiplying by $w_F(\vec{\bm{f}})\le Cn^{-2}$ and
$W_{j_3}(\vec{\bm{f}},\vec{\bm{c}})\le C\|\bm{\theta}_3\|^{-2}$, then summing in the
closing-channel transfer, gives
\begin{equation}
 \begin{split}
 \mathcal K_{321}(\vec{\bm{c}},\vec{\bm{b}})
 &\le\frac C{n^4}\left[
      (1+n)\sum_{\|\bm{\theta}_3\|\ge n/8}\|\bm{\theta}_3\|^{-4}
       +\sum_{n/8\le\|\bm{\theta}_3\|\le CM}\|\bm{\theta}_3\|^{-3}\right]\\
 &\le\frac C{n^4}\log(2+M).
 \end{split}
\end{equation}
The inverse-fourth tail contributes $O(n^{-1})$, cancelling the
factor $1+n$ up to a constant; the inverse-cubic sum contributes
one logarithm. This is the required $n^{-4}$ closed-kernel bound.

To turn it into a form bound, fix the invariants common to the
three channels. In a triple involving one-body channels, these
are the untouched registers. For three pair channels on
$a,t,s$, also fix the total momentum of those three registers.
In either case the remaining configuration is determined by
$(\bm{p}_a,\bm{p}_t)$ on its fibre $\mathcal F$. Both coordinates of a
cold endpoint have norm at most $c_*n$. Weighted Cauchy--Schwarz
therefore gives
\begin{equation}
 \begin{split}
 \left(\sum_{\vec{\bm{p}}\in\mathcal F\cap C}x_{\vec{\bm{p}}}\right)^2
 &\le\left[\sum_{\|\bm{r}\|\le c_*n}(1+\|\bm{r}\|^2)^{-3/2}\right]
       \left[\sum_{\|\bm{s}\|\le c_*n}(1+\|\bm{s}\|^2)^{-1}\right]\\
 &\hspace{1cm}\times
       \sum_{\vec{\bm{p}}\in\mathcal F\cap C}
           (1+\|\bm{p}_a\|^2)^{3/2}(1+\|\bm{p}_t\|^2)x_{\vec{\bm{p}}}^2\\
 &\le Cn\log(2+n)
       \sum_{\vec{\bm{p}}\in\mathcal F\cap C}
           (1+\|\bm{p}_a\|^2)^{3/2}(1+\|\bm{p}_t\|^2)x_{\vec{\bm{p}}}^2.
 \end{split}
\end{equation}
The first reciprocal-weight sum is logarithmic and the second is
$O(n)$ in three dimensions. The product weight has degree five
and is bounded by $C\mathcal E^{5/2}$. Summing the squared
bounds over fibres and using the critical moment proves a closed
form bound of order
$n^{-4}\cdot n\,\mathcal D^2\log^q(2+M)
=\mathcal D^2n^{-3}\log^q(2+M)$.
We have now covered the distinct flat channels, first-$Y$ products,
short second-$Y$ products, and both geometries of long shared
centre-of-mass products. The same-channel flat contribution remains;
for that term we must use its two orientations before taking absolute
values, as follows.

The final class is the same-channel flat product
$\lambda_jRV_jPV_jC$. A direct absolute estimate for its closed
kernel gives only $n^{-2}$, so it does not suffice. We now write the
two orientations explicitly, retaining their common Coulomb product
until after subtraction. Fix cold configurations $\vec{\bm{b}},\vec{\bm{c}}$ on one
$j$-channel fibre. Let $\tau_j\bm{\theta}$ denote the configuration
transfer that adds $\bm{\theta}$ to the moving register of a one-body
channel, or adds $\bm{\theta}$ and $-\bm{\theta}$ to the two registers of a
pair channel. Thus $\vec{\bm{c}}=\vec{\bm{b}}+\tau_j\bm{\theta}_{\vec{\bm{c}}}$ for a unique
$\bm{\theta}_{\vec{\bm{c}}}\in\mathbb Z^3$. Let $\gamma_j$ be the full channel convolution
coefficient, so that whenever the two endpoints are in the auxiliary
box,
\begin{equation}
 \gamma_j(\bm{\theta})=(V_j)_{\vec{\bm{b}}+\tau_j\bm{\theta},\vec{\bm{b}}},\qquad
 s_j(\vec{\bm{k}})=\sum_{a\text{ moved by }j}\|\bm{k}_a\|.
\end{equation}
Convolution makes $\gamma_j$ independent of $\vec{\bm{b}}$, and Hermiticity gives
$\gamma_j(-\bm{\theta})=\overline{\gamma_j(\bm{\theta})}$. As before,
$|\gamma_j(\bm{\theta})|\le C_j\|\bm{\theta}\|^{-2}$ for nonzero transfer and
$\gamma_j(0)=0$. The weight $s_j$ records all moving-register momenta,
including any fixed pair centre-of-mass momentum; it does not replace
those momenta by the single free fibre coordinate.

Let $\widetilde Y_j$ be the contribution of this flat product to
$\widetilde Y$. The signs in \eqref{eq:flux-FC-words} and division
by $i\omega(\vec{\bm{f}},\vec{\bm{b}})$ give
\begin{equation}
 (\widetilde Y_j)_{\vec{\bm{f}}\vec{\bm{b}}}
 =-\lambda_j\ind{\vec{\bm{f}}\in R}\ind{\vec{\bm{b}}\in C}
     \sum_{\vec{\bm{d}}\in P}\frac{(V_j)_{\vec{\bm{f}}\vec{\bm{d}}}(V_j)_{\vec{\bm{d}}\vec{\bm{b}}}}{T(\vec{\bm{f}})-T(\vec{\bm{b}})}.
\end{equation}
Here $\vec{\bm{f}},\vec{\bm{d}}$ are in the auxiliary box; $\vec{\bm{f}}\in R$ already implies
$\vec{\bm{f}}\in F$. Consequently the oriented closed kernel, for $\vec{\bm{b}},\vec{\bm{c}}\in C$,
is
\begin{equation}
 \mathcal A(\vec{\bm{c}},\vec{\bm{b}})
 :=\sum_{\vec{\bm{f}}\in B_M}\overline{(V_j)_{\vec{\bm{f}}\vec{\bm{c}}}}(\widetilde Y_j)_{\vec{\bm{f}}\vec{\bm{b}}}
 =-\lambda_j\sum_{\substack{\vec{\bm{f}}\in B_M\setminus P\\\vec{\bm{d}}\in P}}
 \frac{(V_j)_{\vec{\bm{c}}\vec{\bm{f}}}(V_j)_{\vec{\bm{f}}\vec{\bm{d}}}(V_j)_{\vec{\bm{d}}\vec{\bm{b}}}}{T(\vec{\bm{f}})-T(\vec{\bm{b}})}.
 \label{eq:flux-reflection-kernel}
\end{equation}
It is zero when $\vec{\bm{b}},\vec{\bm{c}}$ are not on the same channel fibre. The support
in this formula is the support of the actual matrix product, including
the auxiliary-box restriction.

Put $\vec{\bm{f}}=\vec{\bm{b}}+\tau_j\bm{\theta}$ and $\vec{\bm{d}}=\vec{\bm{b}}+\tau_j\bm{\phi}$. Let $p_s(\vec{\bm{k}})$
be the indicator of the product cube of radius $s$, so $p_n=P$ and
$p_M=B_M$, and define
\begin{align}
 \chi_{\vec{\bm{b}}}(\bm{\theta},\bm{\phi})
 &=p_M(\vec{\bm{b}}+\tau_j\bm{\theta})
   [1-p_n(\vec{\bm{b}}+\tau_j\bm{\theta})]p_n(\vec{\bm{b}}+\tau_j\bm{\phi}),\\
 \chi_{\vec{\bm{c}}}^-(\bm{\theta},\bm{\phi})
 &=p_M(\vec{\bm{c}}-\tau_j\bm{\theta})
   [1-p_n(\vec{\bm{c}}-\tau_j\bm{\theta})]p_n(\vec{\bm{c}}-\tau_j\bm{\phi}),\\
 H(\bm{\theta},\bm{\phi})
 &=\gamma_j(\bm{\theta}_{\vec{\bm{c}}}-\bm{\theta})\gamma_j(\bm{\theta}-\bm{\phi})\gamma_j(\bm{\phi}),\\
 e_{\vec{\bm{k}}}(\bm{\theta})&=T(\vec{\bm{k}}+\tau_j\bm{\theta})-T(\vec{\bm{k}})
 =\frac{g^2}{2\mu_j}
       \bigl(\|\bm{\theta}\|^2+2\bm{\theta}\cdot\bm{\xi}_j(\vec{\bm{k}})\bigr).
\end{align}
For a one-body channel take $\mu_j=m_a$ and $\bm{\xi}_j(\vec{\bm{k}})=\bm{k}_a$;
for a pair, $\mu_j$ and $\bm{\xi}_j$ are the reduced mass and relative
momentum already used above. All transfer sums in the following
identities can be taken over the common finite cube
$\{-2M,\ldots,2M\}^3$ in each variable. A quotient with a vanishing
support indicator is defined to be zero, so it is evaluated only
where its energy denominator is strictly positive.

Reflect the intermediate configurations by
\begin{equation}
 (\vec{\bm{d}},\vec{\bm{f}})\longmapsto(\vec{\bm{b}}+\vec{\bm{c}}-\vec{\bm{d}},\vec{\bm{b}}+\vec{\bm{c}}-\vec{\bm{f}}).
 \label{eq:flux-reflection}
\end{equation}
The reflected path starts at $\vec{\bm{c}}$, with intermediate configurations
$\vec{\bm{c}}-\tau_j\bm{\phi}$ and $\vec{\bm{c}}-\tau_j\bm{\theta}$, and ends at $\vec{\bm{b}}$.
Its three-factor product after Hermitian conjugation is
\begin{equation}
 \overline{\gamma_j(\bm{\theta}-\bm{\theta}_{\vec{\bm{c}}})
              \gamma_j(\bm{\phi}-\bm{\theta})\gamma_j(-\bm{\phi})}
 =\gamma_j(\bm{\theta}_{\vec{\bm{c}}}-\bm{\theta})\gamma_j(\bm{\theta}-\bm{\phi})\gamma_j(\bm{\phi})
 =H(\bm{\theta},\bm{\phi}).
\end{equation}
Thus the two exact finite sums are
\begin{equation}
 \begin{split}
 \mathcal A(\vec{\bm{c}},\vec{\bm{b}})&=-\lambda_j\sum_{\bm{\theta},\bm{\phi}}
       \frac{\chi_{\vec{\bm{b}}}(\bm{\theta},\bm{\phi})H(\bm{\theta},\bm{\phi})}{e_{\vec{\bm{b}}}(\bm{\theta})},\\
 \overline{\mathcal A(\vec{\bm{b}},\vec{\bm{c}})}&=-\lambda_j\sum_{\bm{\theta},\bm{\phi}}
       \frac{\chi_{\vec{\bm{c}}}^-(\bm{\theta},\bm{\phi})H(\bm{\theta},\bm{\phi})}{e_{\vec{\bm{c}}}(-\bm{\theta})}.
 \end{split}
 \label{eq:flux-reflection-pair}
\end{equation}
Each oriented kernel has one inverse kinetic gap. Both denominators
appear together only when these two inverses are subtracted below.
In particular, reflection has not been assumed to preserve either
support indicator. The equality of the products uses conjugation,
not reality of $\gamma_j$. It therefore retains all phases from translated
clamped nuclei, including the cross terms between centres grouped
in a one-register channel.

We first estimate the part where both indicators are one. Cold
support and either far endpoint give $\|\bm{\theta}\|\ge n/2$, while
$\|\bm{\theta}_{\vec{\bm{c}}}\|\le2c_*n$ gives
$\|\bm{\theta}_{\vec{\bm{c}}}-\bm{\theta}\|\ge\|\bm{\theta}\|/2$. Moreover,
$\sum_i|\bm{\xi}_j(\vec{\bm{k}})_i|\le3s_j(\vec{\bm{k}})$ and
$s_j(\vec{\bm{k}})\le2c_*n$ for $\vec{\bm{k}}=\vec{\bm{b}},\vec{\bm{c}}$. Hence
\begin{equation}
 2|\bm{\theta}\cdot\bm{\xi}_j(\vec{\bm{k}})|
 \le12c_*n\|\bm{\theta}\|
 \le24c_*\|\bm{\theta}\|^2\le\tfrac34\|\bm{\theta}\|^2.
\end{equation}
Using $1/2\le\mu_j\le m_{\max}$, both denominators obey
\begin{equation}
 e_{\vec{\bm{b}}}(\bm{\theta}),\ e_{\vec{\bm{c}}}(-\bm{\theta})
 \ge\frac{g^2}{8m_{\max}}\|\bm{\theta}\|^2.
\end{equation}
Their quadratic terms cancel in the difference:
\begin{equation}
 e_{\vec{\bm{c}}}(-\bm{\theta})-e_{\vec{\bm{b}}}(\bm{\theta})
 =-\frac{g^2}{\mu_j}\,
       \bm{\theta}\cdot\bigl(\bm{\xi}_j(\vec{\bm{b}})+\bm{\xi}_j(\vec{\bm{c}})\bigr).
\end{equation}
It follows that
\begin{equation}
 \left|\frac1{e_{\vec{\bm{b}}}(\bm{\theta})}-\frac1{e_{\vec{\bm{c}}}(-\bm{\theta})}\right|
 =\frac{|e_{\vec{\bm{c}}}(-\bm{\theta})-e_{\vec{\bm{b}}}(\bm{\theta})|}
        {e_{\vec{\bm{b}}}(\bm{\theta})e_{\vec{\bm{c}}}(-\bm{\theta})}
 \le\frac{384m_{\max}^2}{g^2}
       \frac{s_j(\vec{\bm{b}})+s_j(\vec{\bm{c}})}{\|\bm{\theta}\|^3}.
 \label{eq:flux-reflection-denominators}
\end{equation}
This is the extra inverse power of the long transfer. The common
product also satisfies
\begin{equation}
 |H(\bm{\theta},\bm{\phi})|
 \le\frac{C}{\|\bm{\theta}\|^2\|\bm{\theta}-\bm{\phi}\|^2\|\bm{\phi}\|^2}.
\end{equation}
As in the earlier Coulomb convolution estimate, zero transfers are
omitted and
\begin{equation}
 \sum_{\bm{\phi}}\frac1{\|\bm{\theta}-\bm{\phi}\|^2\|\bm{\phi}\|^2}
 \le\frac C{\|\bm{\theta}\|}.
\end{equation}
The common-support contribution to the difference in
\eqref{eq:flux-reflection-pair} is therefore at most
\begin{equation}
 C\bigl(s_j(\vec{\bm{b}})+s_j(\vec{\bm{c}})\bigr)
 \sum_{\|\bm{\theta}\|\ge n/2}\|\bm{\theta}\|^{-6}
 \le Cn^{-3}\bigl(s_j(\vec{\bm{b}})+s_j(\vec{\bm{c}})\bigr).
\end{equation}
Without subtracting the denominators, the same calculation contains
$\|\bm{\theta}\|^{-5}$ in the last sum and gives only $n^{-2}$.

We next estimate the paths belonging to just one of the two supports.
Define the cube-disagreement set
\begin{equation}
 D_s=\{\bm{t}:\ p_s(\vec{\bm{b}}+\tau_j\bm{t})\ne p_s(\vec{\bm{c}}-\tau_j\bm{t})\},
 \qquad s\in\{n,M\}.
\end{equation}
If $\chi_{\vec{\bm{b}}}\ne\chi_{\vec{\bm{c}}}^-$, then $\bm{\theta}\in D_M$, or
$\bm{\theta}\in D_n$, or $\bm{\phi}\in D_n$. To locate these sets, let
$\epsilon_a\in\{1,-1\}$ be the sign with which the channel moves
register $a$. A disagreement at register $a$ and component $i$
requires
\begin{equation}
 s-|b_{a,i}+c_{a,i}|
 <|b_{a,i}+\epsilon_a t_i|
 \le s+|b_{a,i}+c_{a,i}|.
\end{equation}
Indeed the two tested coordinates are $x=b_{a,i}+\epsilon_at_i$
and $b_{a,i}+c_{a,i}-x$; their absolute values differ by at most
$|b_{a,i}+c_{a,i}|$. Spectator coordinates agree and cannot cause
the disagreement. Each displayed condition confines $t_i$ to two
integer intervals with at most $4|b_{a,i}+c_{a,i}|+2$ points.
There are at most six register-component choices, and their total
number of allowed coordinate values is at most
\begin{equation}
 C\bigl(1+s_j(\vec{\bm{b}})+s_j(\vec{\bm{c}})\bigr).
\end{equation}
Also $|b_{a,i}|+|b_{a,i}+c_{a,i}|\le3c_*n\le s/2$.
The same layer condition consequently forces $\|\bm{t}\|\ge s/2$.

For completeness, the lattice estimates used on these layers are
\begin{equation}
 \sum_{\substack{\|\bm{t}\|\ge s/2\\t_i\in I}}\|\bm{t}\|^{-4}
 \le C|I|s^{-2},\qquad
 \sum_{\substack{\|\bm{t}\|\ge s/2\\t_i\in I}}\|\bm{t}\|^{-5}
 \le C|I|s^{-3}.
 \label{eq:flux-reflection-slabs}
\end{equation}
Here $I$ is any finite set of integer coordinate values. In an
annulus of radius $r\ge s/2$, each fixed-coordinate slice contains
at most $Cr^2$ lattice points. Summing $C|I|r^{2-4}$ over the
dyadic radii $r=2^\ell s/2$ proves the first bound. Multiplying
each summand by $\|\bm{t}\|^{-1}\le2/s$ proves the second. This gives
a bound uniform in the outer box radius.

On a path belonging to just one support, its active denominator
still has the quadratic lower bound above. Thus its absolute
summand is bounded by
\begin{equation}
 \frac C{\|\bm{\theta}\|^4\|\bm{\theta}-\bm{\phi}\|^2\|\bm{\phi}\|^2},
 \qquad \|\bm{\theta}\|\ge n/2.
\end{equation}
For $\bm{\theta}\in D_s$, first sum in $\bm{\phi}$ using the two-Coulomb
convolution. The remaining $\|\bm{\theta}\|^{-5}$ layer sum is bounded
by \eqref{eq:flux-reflection-slabs}, giving
$Cs^{-3}(1+s_j(\vec{\bm{b}})+s_j(\vec{\bm{c}}))$ for each $s=n,M$. Since $M\ge n$,
both are bounded by the required $n^{-3}$ expression.

For $\bm{\phi}\in D_n$ we instead sum in $\bm{\theta}$ first. The needed
inverse-fourth convolution tail is
\begin{equation}
 \sum_{\|\bm{\theta}\|\ge R}
 \frac1{\|\bm{\theta}\|^4\|\bm{\theta}-\bm{\phi}\|^2}
 \le\frac C{R\|\bm{\phi}\|^2},\qquad \|\bm{\phi}\|\ge R\ge1.
\end{equation}
To verify it, split at $\|\bm{\theta}-\bm{\phi}\|=\|\bm{\phi}\|/2$. In the near
part $\|\bm{\theta}\|\ge\|\bm{\phi}\|/2$, so the bound
$\sum_{0<\|\bm{z}\|\le r}\|\bm{z}\|^{-2}\le Cr$ gives $C\|\bm{\phi}\|^{-3}$.
In the other part $\|\bm{\theta}-\bm{\phi}\|^{-2}\le4\|\bm{\phi}\|^{-2}$, and
the radial tail $\sum_{\|\bm{\theta}\|\ge R}\|\bm{\theta}\|^{-4}\le C/R$
gives $C/(R\|\bm{\phi}\|^2)$. The first result is bounded by the second
because $\|\bm{\phi}\|\ge R$. Both radial estimates follow by counting
$O(r^3)$ points in each dyadic annulus.
Taking $R=n/2$ and using the remaining factor $\|\bm{\phi}\|^{-2}$,
then the first slab bound, yields
\begin{equation}
 \sum_{\bm{\phi}\in D_n}\frac1{\|\bm{\phi}\|^2}
    \sum_{\|\bm{\theta}\|\ge n/2}
       \frac C{\|\bm{\theta}\|^4\|\bm{\theta}-\bm{\phi}\|^2}
 \le\frac Cn\sum_{\bm{\phi}\in D_n}\|\bm{\phi}\|^{-4}
 \le Cn^{-3}\bigl(1+s_j(\vec{\bm{b}})+s_j(\vec{\bm{c}})\bigr).
\end{equation}
This accounts for every support disagreement, including the
auxiliary-box boundary. Combining it with the common-support bound
proves the log-free kernel estimate
\begin{equation}
 |\mathcal A(\vec{\bm{c}},\vec{\bm{b}})-\overline{\mathcal A(\vec{\bm{b}},\vec{\bm{c}})}|
 \le Cn^{-3}\bigl(1+s_j(\vec{\bm{b}})+s_j(\vec{\bm{c}})\bigr).
 \label{eq:flux-reflection-defect}
\end{equation}
In particular it gives the
$Cn^{-3}(1+\|\vec{\bm{b}}\|+\|\vec{\bm{c}}\|)\log^q(2+n)$ bound, with $\|\vec{\bm{b}}\|,\|\vec{\bm{c}}\|$
understood as endpoint momentum sizes on the moved registers.

Finally set $\ket{v}=C\ket{u}$ and $x_{\vec{\bm{p}}}=|v_{\vec{\bm{p}}}|$. Only the anti-Hermitian part
of $\mathcal A$ contributes to the imaginary part of its form:
\begin{equation}
 |\operatorname{Im}\langle V_jv\,|\,\widetilde Y_jv\rangle|
 \le\frac12\sum_{\vec{\bm{c}}\sim_j \vec{\bm{b}}}x_{\vec{\bm{c}}}x_{\vec{\bm{b}}}
       |\mathcal A(\vec{\bm{c}},\vec{\bm{b}})-\overline{\mathcal A(\vec{\bm{b}},\vec{\bm{c}})}|.
\end{equation}
On each channel fibre let $S=\sum x_{\vec{\bm{p}}}$ and
$W=\sum s_j(\vec{\bm{p}})x_{\vec{\bm{p}}}$. Then
$S^2+2SW\le2S^2+W^2$, and therefore
\begin{equation}
 \sum_{\vec{\bm{c}}\sim_j \vec{\bm{b}}}x_{\vec{\bm{c}}}x_{\vec{\bm{b}}}[1+s_j(\vec{\bm{b}})+s_j(\vec{\bm{c}})]
 \le2L_j(x)^2+L_j(s_jx)^2.
\end{equation}
The weighted line square is also explicit. Choose one moving
register whose momentum $\bm{\zeta}$ parametrises the fibre. Weighted
Cauchy--Schwarz, $s_j(\vec{\bm{p}})^2\le C\mathcal E(\vec{\bm{p}})$, and
$\sum_{\|\bm{\zeta}\|\le\sqrt3 M}(1+\|\bm{\zeta}\|^2)^{-3/2}
\le C\log(2+M)$ give
\begin{equation}
 L_j(s_jx)^2
 \le C\log(2+M)\sum_{\vec{\bm{p}}}
    (1+\|\bm{\zeta}(\vec{\bm{p}})\|^2)^{3/2}s_j(\vec{\bm{p}})^2|v_{\vec{\bm{p}}}|^2
 \le C\log(2+M)\|\mathcal E^{5/4}C\ket{u}\|^2.
\end{equation}
Since $C\subset Q$ and $M=n^6$, \eqref{eq:flux-line-estimates}
bounds this by $C\mathcal D^2\log^2(2+n)$ and also controls
$L_j(x)^2$. The resulting closed-form bound is
$C\mathcal D^2n^{-3}\log^q(2+n)$, equivalently the required
$K^{-3}$ bound at fixed $g$. A different closing channel cannot
meet this same-channel word: a configuration joined by different
channels to two cold endpoints has every register bounded by
$3c_*n<n/4$, whereas the word's far endpoint lies in $R$.
Thus the channelwise bounds cover its full closed return.
Combining the three classes with the non-cold contributions proves
\eqref{eq:flux-third-estimates}. The leading same-channel
contribution has been cancelled by the paired reflection; we have
not declared a complex kernel to be real symmetric or its
expectation to vanish.

\end{proofx}

\subsubsection{Boundary terms and the integrated flux bound}

We have bounded the endpoints of all four auxiliary operators and
the remainders that will remain under the time integral. We still
need to control the errors introduced by the auxiliary box, since
its projected exact state satisfies the forced equation
\eqref{eq:flux-box-evolution}. These errors are estimated at each
time, uniformly in that time; integrating them incurs a factor $t$.

We separate three errors, since they use different parts of the
estimates for the evolved state. First put $\ket{d_M}:=(1-B_M)\ket{\psi}$, so that
$\ket{\psi}=\ket{u}+\ket{d_M}$ and $P\ket{\psi}=P\ket{u}$. The force form of two vectors
supported in the box agrees exactly with the compressed matrix
form. Therefore the instantaneous flux replacement error is
\begin{equation}
 \mathfrak G(\ket{\psi},\ket{\psi})-\mathfrak G(\ket{u},\ket{u})
 =\mathfrak G(\ket{u},\ket{d_M})+\mathfrak G(\ket{d_M},\ket{\psi}).
\end{equation}
Each of the two forms is a sum of potential--kinetic pairings. The
graph and half-kinetic tails give
\begin{equation}
 \|T_p\ket{d_M}\|\le C\mathcal D(gM)^{-1/2},\qquad
 \|V_{\mathrm{tot}}\ket{d_M}\|
 \le C\|\mathcal E^{1/2}\ket{d_M}\|
 \le C\mathcal D(gM)^{-3/2}.
\end{equation}
The corresponding norms of $\ket{u}$ and $\ket{\psi}$ without tail projections are bounded
by $C\mathcal D$. Consequently the replacement error is at most
$C\mathcal D^2(gM)^{-1/2}$. In particular, no operator norm of
the compressed force, which would grow with the box, is used here.

Second, compression prevents us from simply commuting two
multiplication operators. Reflecting the intermediate index by
$\vec{\bm{r}}\mapsto \vec{\bm{p}}+\vec{\bm{q}}-\vec{\bm{r}}$ cancels the matched terms and leaves the boundary
majorant
\begin{equation}
 |e_{\rm bd}(\ket{v})|\le
 C\sum_{\vec{\bm{y}}\in B_{3M}\setminus B_M}
      \left(\sum_{\vec{\bm{q}}\in B_M}|v_{\vec{\bm{q}}}|\,\mathcal M(\vec{\bm{q}}-\vec{\bm{y}})\right)^2,
\end{equation}
where $\mathcal M$ is the sum of the positive Coulomb channel
majorants. We need this estimate for $\ket{v}=\ket{u}$ and $\ket{v}=P\ket{u}$.
Split each inner sum into inputs in $B_{M/2}$ and its complement.
A hop from $B_{M/2}$ to the exterior of $B_M$ has transfer
$\|\bm{\theta}\|>M/2$. On a channel fibre, the squared Coulomb kernel
therefore has the two bounds
\begin{equation}
 \sum_{\bm{\theta}\ne0}\|\bm{\theta}\|^{-4}\le C,\qquad
 \sum_{\|\bm{\theta}\|>M/2}\|\bm{\theta}\|^{-4}\le CM^{-1}.
\end{equation}
Young's convolution inequality from $\ell^1*\ell^2$ to $\ell^2$,
followed by weighted Cauchy--Schwarz on the input fibre, gives
\begin{equation}
 \|k*a\|_{\ell^2}^2
 \le\|k\|_{\ell^2}^2
    \left(\sum_{\bm{\xi}}(1+\|\bm{\xi}\|^2)^{-2}\right)
    \sum_{\bm{\xi}}(1+\|\bm{\xi}\|^2)^2|a_{\bm{\xi}}|^2.
\end{equation}
Apply the tail kernel bound to the inner inputs and the full kernel
bound to the outer inputs, then sum the fibre labels and channels.
Since an input outside $B_{M/2}$ has some register momentum larger
than $gM/2$, this gives the explicit estimate
\begin{equation}
 |e_{\rm bd}(\ket{v})|\le C\left[
 M^{-1}\|\mathcal E \ket{v}\|^2+
 \sum_a\|\mathcal E\ind{\{\|\bm{k}_a\|>gM/2\}}\ket{v}\|^2\right]
 \le C\mathcal D^2\big(M^{-1}+(gM)^{-1}\big).
\end{equation}
The sum is over all quantum registers, including spectators. The
last step uses the squared graph-tail bound, and is uniform for
both projected states $\ket{v}$ under consideration. It is stronger
than the $O(\mathcal D^2(gM)^{-1/2})$ error needed in the ledger.

Finally consider the four forcing forms. The same half-kinetic
tail used above bounds their common source by
\begin{equation}
 \|\ket{f_M}\|\le\|V_{\mathrm{tot}}\ket{d_M}\|
 \le C\mathcal D(gM)^{-3/2},\qquad
 |e_J|\le2\|\ket{f_M}\|\,\|J\ket{u}\|.
\end{equation}
We only need crude bounds on $J\ket{u}$ for this purpose. Keeping the
auxiliary radius $M$ visible, the four bounds are
\begin{align}
 \|V_{\lambda,c}\ket{u}\|&\le C\mathcal D,&
 \|Y_1\ket{u}\|&\le C\mathcal D\log^2(2+M),\\
 \|Y_2\ket{u}\|&\le C\mathcal D n,&
 \|Y_3\ket{u}\|&\le C\mathcal D n^{1/2}\log^q(2+M).
\end{align}
Here the first follows from Hardy applied to $\ket{u}$ and $P\ket{u}$. For the
second, the gapped row bound proved above has the conserved channel
factor $\|\bm{\Pi}_j\|$. Weighted Schur therefore bounds $\|Y\ket{u}\|$ by
$C\log^2(2+M)\|\mathcal E^{1/2}\ket{u}\|$; the subtraction $PYP$
has the same bound.

For the third, the support of $Y_2$ lies in $Q\times Q$. Each
channel row has at most $C(n+1)^3$ entries, $\|\bm{\Pi}_j\|\le Cn$,
and every nonzero integer transfer satisfies $\|\bm{\tau}\|\ge1$.
The kernel bound \eqref{eq:flux-Y2-kernel} thus gives the crude
row and column bound $Cn^{-3}n(n+1)^3\le Cn$, as claimed.
For the fourth, use the cold-input squared-norm estimate
\eqref{eq:flux-third-cold-input} for $\widetilde Y$. The
critical channel-line Cauchy--Schwarz bound on an arbitrary cold
vector $\ket{z}$ is
\begin{equation}
 \sum_j L_j(C|z|)^2
 \le C\log(2+M)
       \sum_j\sum_{\vec{\bm{p}}\in C}(1+\|\bm{p}_{\mathrm{lead}(j)}\|^2)^{3/2}|z_{\vec{\bm{p}}}|^2
 \le Cn^4\log(2+M)\|\ket{z}\|^2.
\end{equation}
$\mathrm{lead}(j)$ denotes the moving register chosen as the
fibre coordinate for channel $j$. We used the loose bound
$(1+\|\bm{p}_{\mathrm{lead}(j)}\|^2)^{3/2}\le Cn^4$ on $C$.
Combining this with the $n^{-3}$ cold-input estimate gives
$\|\widetilde Y\|\le Cn^{1/2}\log^q(2+M)$. Its adjoint has
the same norm, proving the stated bound for $Y_3$.

The four forcing errors are consequently bounded, respectively,
by $C\mathcal D^2(gM)^{-3/2}$ times
\begin{equation}
 1,\qquad \log^2(2+M),\qquad n,\qquad
 n^{1/2}\log^q(2+M).
\end{equation}
With $M=n^6$, these have powers $n^{-9}$, $n^{-9}$, $n^{-8}$
and $n^{-17/2}$, up to fixed-system constants and logarithms.
Each is smaller than the required $K^{-3}\log^q(2+n)$.
This calculation makes explicit that no positive power of the
auxiliary radius is hidden in the forcing estimates.

This choice of $M$ matches the box error to the target power:
$(gM)^{-1/2}=g^{-1/2}n^{-3}=O_L(K^{-3})$, while $M^{-1}=n^{-6}$
is smaller. Moreover, logarithms of $M=n^6$ are bounded by fixed
multiples of logarithms of $n$. Thus the auxiliary box depends only
on the basis cutoff, not on the evolution time. The time dependence
enters only when the uniformly bounded error forms are integrated.

We can now assemble the exact identities. Apply
\eqref{eq:flux-box-elimination} first with $J=-V_{\lambda,c}$,
then with $J=Y_1$, $J=-Y_2$ and $J=Y_3$. The minus sign for $Y_2$
comes from the term $-Z^>$ in the decomposition of $N_2$.
The result is the following identity, with the explicitly bounded
box corrections collected in $\mathcal R_M(t)$:
\begin{align}
 \mathcal F_K(t)
 ={}&\big[-\langle u|V_{\lambda,c}|u\rangle+
          \langle u|Y_1|u\rangle-\langle u|Y_2|u\rangle+
          \langle u|Y_3|u\rangle\big]_0^t\\
 &+\int_0^t\left[\begin{aligned}
       &\langle u|X^{\rm cm,res}|u\rangle+
        \langle u|N^{\rm prod}|u\rangle-\langle u|Z^{\le}|u\rangle\\
       &+\langle u|i[V_M,Y_2]|u\rangle-
        \langle u|i[V_M,Y_3]|u\rangle
       \end{aligned}\right]ds+\mathcal R_M(t),
 \label{eq:ledger}
\end{align}
where all expectations are evaluated in $\ket{u}=\ket{u(s)}$ at the corresponding time.
The boundary corrections are the flux replacement error,
$e_{\rm bd}(\ket{u})-e_{\rm bd}(P\ket{u})$, and the four forcing terms with
signs $+,-,+,-$, respectively. Their integrated magnitude is bounded
by $C\mathcal D^2tK^{-3}\log^q(2+n)$ under the preceding choice
of $M$.

Each term in \eqref{eq:ledger} now has an assigned estimate.
Lemma~\ref{cor:reslayer} controls the flat and first endpoints and
the resonant centre-of-mass flux. Lemma~\ref{lem:flux-line-estimates}
controls the second endpoint, its low-mismatch remainder and its
potential commutator. The restricted-endpoint estimate controls
$N^{\rm prod}$, and Lemma~\ref{prop:z3} controls the third endpoint
and commutator. The endpoint bounds contribute at the two times
$0,t$, while the remaining uniform form bounds contribute a factor
$t$. Increasing the fixed logarithmic power to cover all terms gives
\begin{equation}
 |\mathcal F_K(t)|\le
 C\mathcal D^2(1+t)K^{-3}\log^q(2+n).
 \label{eq:ledger2}
\end{equation}
The endpoint of $Y_2$ carries a logarithm; it cannot be replaced by
an unqualified $CK^{-3}$ endpoint in this argument.

\begin{corollary}[Completed flux ledger]
\label{cor:flux3m}
For every $\varepsilon>0$,
\begin{equation}
 |\mathcal F_K(t)|\le C_\varepsilon\mathcal D^2
                   (1+t)K^{-3+\varepsilon}.
 \label{eq:flux3m}
\end{equation}
\end{corollary}
\begin{proofx}
At fixed $L$, the relation $K=gn$ gives
$\log^q(2+n)\le C_{\varepsilon,L}K^\varepsilon$ for every
$\varepsilon>0$. Substitution in \eqref{eq:ledger2} absorbs the
finite logarithmic loss into an arbitrarily small cutoff power and
proves the claim.
\end{proofx}

We finally restate the bound for the original force form and exact
evolution. This also makes explicit the dependence on the initial state and the
cutoff conditions used in choosing the cold region and auxiliary box.

\begin{theorem}[Integrated projection flux at the energy rate]
\label{thm:flux3}
For the periodic Coulomb model, including clamped nuclei, let
$\ket{\psi_0}\in D(H^2)$ and
$\mathcal D=\|\ket{\psi_0}\|+\|H\ket{\psi_0}\|+\|H^2\ket{\psi_0}\|$.
For every $\varepsilon>0$ there is a constant $C_\varepsilon$ depending
on the fixed system and $L$, but not on $t$, the state or the cutoff,
such that for every cube radius $n\ge64$ with $K=gn$ and
$gc_*n\ge1$, and every $t\ge0$,
\begin{equation}
 \left|\int_0^t[\mathfrak G(\ket{\psi(s)},\ket{\psi(s)})-
                  \mathfrak G(P\ket{\psi(s)},P\ket{\psi(s)})]ds\right|
 \le C_\varepsilon\mathcal D^2
          \big(K^{-3+\varepsilon}+tK^{-3+\varepsilon}\big).
 \label{eq:flux3}
\end{equation}
No normalisation or exchange-sector assumption on the initial state
is required for this estimate.
\end{theorem}
\begin{proofx}
The integral on the left is $\mathcal F_K(t)$, so
\eqref{eq:flux3m} gives the required estimate. To verify the homogeneous
dependence on the initial state explicitly, take a nonzero initial state and scale it by
$1/\mathcal D$. The scaled state has unit graph norm, and the
force form and every term of \eqref{eq:ledger} are quadratic in the
state. Scaling back therefore multiplies the bound by
$\mathcal D^2$. A zero initial state remains zero. No normalisation
or symmetry-sector condition enters this argument.
\end{proofx}

\section{Logarithmic kinetic-energy bounds and auxiliary source estimates}
\label{app:source}

We prove that the error in a particle's kinetic-energy expectation is
$O(K^{-3}[1+t+\log(2+K)])$ for both plane-wave Galerkin evolution and
the truncated-kernel grid. The main difficulty is that kinetic energy
is an unbounded observable: a small state-norm error alone need not
give a comparably small error in its expectation. We overcome this by
comparing the evolved states at low Hamiltonian energies and estimating
separately how their higher-energy components enter the kinetic
pairing. At fixed time and cell size, the resulting bound is
$O(\log N/N)$.

We use two different kinds of projection. The Fourier projections
define the finite discretisations, whereas the Hamiltonian-energy
projections separate parts of the proof. The latter do not change
the simulated Hamiltonians or require filtering the prepared initial
state. We first prove the grid result, where the higher-energy
pairing needs particular care, and then obtain the simpler Galerkin
comparison. The stopping-power bounds follow from the kinetic
theorem. The direct source, thin-shell and stability estimates at the
end of this appendix are independent results, rather than additional
hypotheses of the logarithmic bound.

Fix the periodic system of Eq.~\eqref{eq:overview-hamiltonian}, with
$\eta_{\rm tot}$ quantum particles of positive masses and any clamped
nuclei. Write $H=T+V_{\mathrm{tot}}$ and
$T=\sum_{r=1}^{\eta_{\rm tot}}T_r$. All constants in this appendix may
depend on this fixed system and cell. The Galerkin projector $P$
restricts each register to $\|\bm{k}\|_\infty\le K$, and $H_N=PHP$ on
$\operatorname{ran}P$. For the grid we use the padded parametrisation
$n'=2n-1$, $K=\pi(n-1)/L$, and $P'$ restricts each register to
$\|\bm{k}\|_\infty\le2K$. Its Hamiltonian is $\Hgrid$, with any kernel in
Eq.~\eqref{eq:kernelclass}: the Fourier coefficients agree with the
Coulomb kernel below transfer momentum $K$, are dominated in magnitude
by its coefficients, and vanish outside the transfer cube of radius
$2K$, with the stated zero-mode and reality conventions. The letter
$A$ retains its meaning as the Fourier multiplier representing the
total kinetic operator $T$. We denote positively shifted Hamiltonians
by $\mathsf H$, $\mathsf H_N$ and $\mathsf H_g$, so that their energy
projections remain distinct from momentum restrictions.

\begin{theorem}[Logarithmic kinetic-energy error for both discretisations]
\label{thm:kinetic-log}
There are constants $C,K_0$, depending only on the fixed periodic system
and cell, such that every admissible $K\ge K_0$, every
$\ket{\psi_0}\in D(H^2)$, every measured register $p$, and every $t\ge0$
satisfy the following bound. Here $\ket{\psi_0}\in D(H^2)$ means that
$\ket{\psi_0}\in D(H)$ and $H\ket{\psi_0}\in D(H)$, so both $H\ket{\psi_0}$ and
$H^2\ket{\psi_0}$ have finite norm:
\begin{equation}
 \left|\langle \psi(t)|T_p|\psi(t)\rangle
       -\langle \psi_{\rm d}(t)|T_p|\psi_{\rm d}(t)\rangle\right|
 \le C\mathcal D^2K^{-3}[1+t+\log(2+K)],
 \qquad
 \mathcal D=\norm{\ket{\psi_0}}+\norm{H\ket{\psi_0}}+\norm{H^2\ket{\psi_0}},
 \label{eq:kinetic-log}
\end{equation}
where $\ket{\psi(t)}=e^{-iHt}\ket{\psi_0}$ and either
\begin{equation}
 \ket{\psi_{\rm d}(t)}=\ket{\psi_N(t)}=e^{-iH_Nt}P\ket{\psi_0}
 \quad\hbox{or}\quad
 \ket{\psi_{\rm d}(t)}=\ket{\psi_g(t)}=e^{-i\Hgrid t}P'\ket{\psi_0}.
\end{equation}
The grid constants are uniform over the full kernel
class~\eqref{eq:kernelclass}. The projected initial states are not
renormalised. No normalisation or physical spectral-gap assumption is required.
The estimate holds on the full particle space and remains valid in a
fixed particle-exchange symmetry sector, such as the electronic
antisymmetric subspace. For Galerkin evolution we also
have the stronger retained-state comparison
\begin{equation}
 \norm{P\ket{\psi(t)}-\ket{\psi_N(t)}}
 \le C\mathcal D K^{-3}[1+t+\log(2+K)].
 \label{eq:kinetic-retained-galerkin}
\end{equation}
At fixed cell size, $N=\Theta(K^3)$ for either discretisation, so
Eq.~\eqref{eq:kinetic-log} is
$C\mathcal D^2[1+t+\log(2+N)]/N$ after changing the constant.
\end{theorem}

The retained-state estimate compares the finite-basis evolved state with
$P\ket{\psi(t)}$. The full continuum state also has a discarded Fourier tail,
whose norm is bounded here by $C\mathcal D K^{-5/2}$; we do not claim
Eq.~\eqref{eq:kinetic-retained-galerkin} for that full state difference.
For kinetic energy the discarded contribution is smaller because it
pairs two tails and has no cross term with the retained modes.

The proof compares the two Hamiltonians through their consistency
residual: the difference between applying the finite Hamiltonian
after Fourier projection and projecting after applying the continuum
Hamiltonian. We obtain the useful $K^{-3}$ bound by testing this
residual between low-energy states, each weighted by the inverse of
its shifted Hamiltonian. Those inverse weights keep their kinetic
norms bounded; the low-energy restriction then improves their
momentum tails to $O(K^{-5/2})$. Pairing a state tail with a potential
tail of size $O(K^{-1/2})$ gives $K^{-3}$.

We convert this residual estimate into a dynamical comparison by
dividing Hamiltonian energies into bands whose endpoints grow by
factors of two. The sum over separated bands produces the logarithm.
For the grid, the remaining state difference at high Hamiltonian
energies is too large to finish with its norm alone. We therefore
use the kinetic pairing itself: spectral orthogonality supplies an
inverse energy cutoff, while the momentum tails control an additional
kinetic weight on the vectors being paired. This gives the same
$K^{-3}$ rate without another logarithm.

The proof can be reduced to masses at least one without restricting the
physical system. Set
$\lambda=1+\sum_r m_r^{-1}$, replace each moving mass by
$m'_r=\lambda m_r\ge1$, and divide every particle charge,
including the charges of clamped nuclei, by $\sqrt\lambda$. The resulting operators satisfy
$H=\lambda H'$ and $T_p=\lambda T'_p$, so the original evolution at
time $t$ is the rescaled evolution at time $\lambda t$. The graph-norm bound
for the initial state obeys $\mathcal D'\le\mathcal D$. Thus a kinetic-error constant
$C'$ for the rescaled system gives a constant $\lambda^2C'$ for the
original one; the retained-state constant becomes $\lambda C'$.
All these factors depend only on the fixed system. We may therefore
establish the estimates with masses at least one and restore arbitrary
positive masses at the end.

\subsection{Uniform estimates and the full kernel class}
\label{sec:kinetic-inputs}

We first establish the bounds that will be available for both the
continuum and grid states throughout evolution. We need to control their kinetic
norms, the part of those norms carried by large momenta, and the
residual relating their equations of motion. Throughout, the
constants must be independent of the cutoff and of the particular
kernel in the class~\eqref{eq:kernelclass}.

Let $\ket{\vec{\bm{k}}}$ denote the orthonormal many-particle plane-wave
basis, with any spin labels included implicitly. We write the momentum
amplitudes of a state as
\begin{equation}
 u_{\vec{\bm{k}}}:=\langle\vec{\bm{k}}|u\rangle,\qquad
 \ket{u}=\sum_{\vec{\bm{k}}}u_{\vec{\bm{k}}}\ket{\vec{\bm{k}}}.
\end{equation}
The amplitude of $B\ket{u}$ is $\langle\vec{\bm{k}}|B|u\rangle$.

Put $\E:=1+T$, so that $\|\E \ket{u}\|$ controls both the norm of $\ket{u}$ and
the norm of its kinetic image $T\ket{u}$. Let $\bm{p}_r\in\mathbb Z^3$ label the Fourier mode of
register $r$, with momentum $\bm{k}_{\bm{p}_r}=2\pi\bm{p}_r/L$, and define
\begin{equation}
 Q_{r,\lambda}=\mathbf 1_{\|\bm{k}_{\bm{p}_r}\|_\infty>\lambda},\qquad
 Q_\lambda=1-\prod_{r=1}^{\eta_{\rm tot}}(1-Q_{r,\lambda}).
 \label{eq:kinetic-momentum-projectors}
\end{equation}
Thus $Q_\lambda$ selects configurations in which at least one particle
momentum exceeds $\lambda$ in the maximum norm. Since
$\|\bm{k}\|_\infty\le\|\bm{k}\|_2$, each cube tail is contained in the corresponding
Euclidean tail. The Euclidean tail bounds of
Appendix~\ref{app:dynamics} therefore also bound these projectors,
with unchanged constants. These projectors commute with $T$ and
$\E$, so the same inclusion applies to the kinetically weighted tails.

The Coulomb relative bounds and the grid graph equivalence
(Lemma~\ref{lem:grid-KR}, especially \eqref{eq:grid-graph-equivalence}) allow a
common $K$-independent shift $c$ for which
\begin{equation}
 \mathsf H=H+c\ge1,\quad \mathsf H_g=\Hgrid+c\ge1,\quad
 \norm{\mathsf H \ket{u}}\asymp\norm{\E \ket{u}},\quad
 \norm{\mathsf H_g\ket{v}}\asymp\norm{\E \ket{v}}.
 \label{eq:kinetic-graph}
\end{equation}
Here $\ket{u}\in D(H)=D(T)$ and $\ket{v}\in\ran P'$.
These graph-norm equivalences let us transfer bounds between the
Hamiltonian, whose evolution conserves its own powers, and the
kinetic operator, whose Fourier weights directly control momentum
tails. We increase $c$ if necessary and keep it fixed throughout.
In particular, $\norm{\mathsf H^2\ket{\psi_0}}\le C\mathcal D$. On the
finite register, every particle momentum component is at most $2K$,
so
\begin{equation}
 \norm{\E|_{\ran P'}}\le CK^2.
 \label{eq:kinetic-grid-bandwidth}
\end{equation}
This operator-norm bound grows with the cutoff. We will use momentum
tails to obtain a better bound when an additional kinetic weight acts
on the evolved states. No commutation between $\mathsf H_g$ and
$\E$ is asserted or needed.

Write $V_g:=\Hgrid-T|_{\ran P'}$ for the grid potential. We will use
the following bounds on the part of a potential image lying above
momentum $\lambda$:
\begin{equation}
 \norm{Q_{r,\lambda}V_{\mathrm{tot}}\ket{u}}\le C\lambda^{-1/2}\norm{\E \ket{u}},\qquad
 \norm{Q_{r,\lambda}V_g\ket{v}}\le C\lambda^{-1/2}\norm{\E \ket{v}}
 \quad(\lambda\ge1).
 \label{eq:kinetic-generic-tails}
\end{equation}
The first is Eq.~\eqref{eq:lam-gen}. For the second,
\eqref{eq:grid-tail} and \eqref{eq:grid-potential-tail} give the grid
version, and their proof applies at every threshold. To see explicitly
why the kernel choice and smaller thresholds cause no difficulty,
let $w$ denote the periodic Coulomb kernel and $g$ a kernel in the
allowed class. Taking absolute values in their Fourier convolution
sums, the inequality $|\widehat g|\le\widehat w$ bounds the $g$
convolution by the $w$ convolution acting on the componentwise
absolute values $(|v_{\vec{\bm{k}}}|)_{\vec{\bm{k}}}$. The replacement
$v_{\vec{\bm{k}}}\mapsto|v_{\vec{\bm{k}}}|$ preserves the weighted norm
$\|\E\ket{v}\|$ because $\E$ is diagonal in the momentum basis. We can therefore apply the Coulomb tail estimate with
the same norm on its right-hand side. The grid additionally
folds exterior Fourier coefficients into the register. Each target has
a bounded number of preimages, and a centred representative modulo
$n'$ has no larger absolute momentum component than its preimage.
Consequently, a folded output above any threshold $\lambda$ can only
come from an unfolded coefficient above that same threshold. The
bounded-folding estimates \eqref{eq:alias-mult} and
\eqref{eq:alias-restricted} therefore preserve~\eqref{eq:kinetic-generic-tails},
including at fixed fractions of $K$.

The same domination and folding argument gives the uniform relative
bound $\norm{V_g\ket{v}}\le\epsilon\norm{T\ket{v}}+C_\epsilon\norm{\ket{v}}$ and the
uniform lower bound used in~\eqref{eq:kinetic-graph}. For the dropped-transfer
part, the elementary facts
\begin{equation}
 \widehat w-\widehat g=0\quad(0<\|\bm{k}_{\bm{\nu}}\|\le K),\qquad
 |\widehat w-\widehat g|\le2\widehat w
 \label{eq:kinetic-dropped-domination}
\end{equation}
are sufficient. Thus the square root of the sum of squared dropped coefficients is
$O(K^{-1/2})$. A matrix element of this operator connects an input
momentum configuration to an output configuration. Because its
transfer has magnitude greater than $K$, at least one of those two
configurations has a moved register above $\theta K$, where we fix
$\theta=1/4$. The wrap terms arise when a transfer leaves the grid's
Fourier cube and is brought back by modular reduction. They also
have a large-momentum input or output; in the part where the Coulomb
coefficient is not small, both are large in a folding register.

These support properties let us use a momentum tail on at least one
side of each matrix element. The form bounds
\eqref{eq:Dform} and \eqref{eq:Rform} then give, uniformly over the
kernel class,
\begin{align}
 |\langle v|D_{\mathrm{drop}}|u\rangle|+|\langle v|R_{\mathrm{wrap}}|u\rangle|
 &\le CK^{-1/2}\sum_r
 \left[\norm{\E \ket{v}}\norm{Q_{r,\theta K}\ket{u}}
       +\norm{\E \ket{u}}\norm{Q_{r,\theta K}\ket{v}}\right]\\
 &\quad+CK\sum_r\norm{Q_{r,\theta K}\ket{u}}
                     \norm{Q_{r,\theta K}\ket{v}},
 \qquad \ket{u},\ket{v}\in\ran P'.
 \label{eq:kinetic-forms}
\end{align}
Here $D_{\mathrm{drop}}$ and $R_{\mathrm{wrap}}$ are the operators denoted
$D$ and $R$ in Eq.~\eqref{eq:grid-structure}: the omitted-transfer
Galerkin operator and the wrap operator, respectively. The latter
includes all modular-total-momentum terms. We use these subscripts to
distinguish the wrap operator from the exterior Fourier projector
$R=1-P$ used in Appendix~\ref{app:flux} and in the auxiliary estimates below. To see the two contributions in Eq.~\eqref{eq:kinetic-forms}, first
consider the matrix elements whose squared row and column sums are
$O(K^{-1})$. Lemma~\ref{lem:paircorr} applies weighted
Cauchy--Schwarz on each interaction slice: fixing the spectator
momenta and, for a pair interaction, the pair's total momentum leaves
one three-dimensional momentum variable to sum. The kinetic weight
on one vector makes that sum finite. Splitting the bilinear sum into
terms with a large-momentum input and terms with a large-momentum
output therefore gives
$CK^{-1/2}(\|\E \ket{v}\|\|Q_{r,\theta K}\ket{u}\|
+\|\E \ket{u}\|\|Q_{r,\theta K}\ket{v}\|)$.

In the remaining wrap terms, both the input and output lie in a
large-momentum region of a folding register. Their absolute row and
column sums are $O(K)$. Schur's test bounds the matrix norm by the
geometric mean of these two maximal sums, hence by $CK$. Both
vectors can now be restricted to their momentum tails, giving the
last term in Eq.~\eqref{eq:kinetic-forms}. For cyclic pair slices,
the fixed total momentum is understood modulo the register size;
the same count and bounded folding multiplicity apply. Coefficient
domination changes neither support argument. This explains why the
estimate holds for the full kernel class without positivity or
radiality of $\widehat g$.

The momentum tails of the exact evolved state follow from Proposition~\ref{prop:dyn-boot},
specifically
\eqref{eq:dyn-wtail} and \eqref{eq:dyn-t}. For
$\ket{\phi}\in D(H^2)$, let $\mathcal D_\phi=\norm{\ket{\phi}}+\norm{H\ket{\phi}}
+\norm{H^2\ket{\phi}}$. Uniformly in $t$ and $\lambda\ge1$,
\begin{equation}
 \begin{aligned}
 \norm{\E e^{-iHt}\ket{\phi}}&\le C\mathcal D_\phi,\\
 \norm{Q_{r,\lambda}\E e^{-iHt}\ket{\phi}}
       &\le C\mathcal D_\phi\lambda^{-1/2},&
 \norm{Q_{r,\lambda}e^{-iHt}\ket{\phi}}
       &\le C\mathcal D_\phi\lambda^{-5/2}.
 \end{aligned}
 \label{eq:kinetic-exact-flow-tails}
\end{equation}
Sums over the fixed number of registers give the same estimates with
$Q_\lambda$ in place of $Q_{r,\lambda}$.

\begin{lemma}[Initial-state consistency and momentum tails during grid evolution]
\label{lem:kinetic-strong}
Define the consistency residual, as a map from $D(H)$ to $\ran P'$, by
\begin{equation}
 \mathcal C_g:=\Hgrid P'-P'H
 =-D_{\mathrm{drop}}P'+R_{\mathrm{wrap}}P'-P'V_{\mathrm{tot}}(1-P').
 \label{eq:kinetic-consistency}
\end{equation}
For every $\ket{\phi}\in D(H^2)$,
\begin{equation}
 \norm{\mathcal C_g\ket{\phi}}\le C\mathcal D_\phi K^{-1/2}.
 \label{eq:kinetic-data-consistency}
\end{equation}
The evolved grid state $\ket{v(t)}=e^{-i\Hgrid t}P'\ket{\phi}$ satisfies
\begin{equation}
 \norm{\E \ket{v(t)}}\le C\mathcal D_\phi,\qquad
 \norm{Q_{r,\lambda}\E \ket{v(t)}}
      \le C\mathcal D_\phi\lambda^{-1/2}
      \quad(1\le\lambda\le K).
 \label{eq:kinetic-grid-flow-tails}
\end{equation}
\end{lemma}

\begin{proofx}
We first bound the residual by quantities that the continuum
momentum-tail estimates control. Its displayed decomposition follows
from the grid structure formula~\eqref{eq:grid-structure}. To estimate
its norm, test it against a unit vector $\ket{v}\in\ran P'$. The finite
bandwidth gives $\norm{\E \ket{v}}\le CK^2$, while
$\norm{Q_{r,\theta K}\ket{v}}\le1$. Therefore
Eq.~\eqref{eq:kinetic-forms}, with $P'\ket{u}$ in place of its second
argument, gives
\begin{equation}
 |\langle v|(-D_{\mathrm{drop}}+R_{\mathrm{wrap}})P'|u\rangle|
 \le C K^{-1/2}\norm{\E \ket{u}}
       +C(K^{3/2}+K)\norm{Q_{\theta K}\ket{u}}.
\end{equation}
The sum over the fixed number of registers has been absorbed into
$C$. For the exterior term, self-adjointness of the potential and
Eq.~\eqref{eq:kinetic-generic-tails} yield
\begin{equation}
 |\langle v|P'V_{\mathrm{tot}}(1-P')|u\rangle|
 =|\langle (1-P')V_{\mathrm{tot}}v\,|\,(1-P')u\rangle|
 \le CK^{-1/2}\norm{\E \ket{v}}\norm{(1-P')\ket{u}}.
\end{equation}
Here $\langle Bu|$ denotes the adjoint of $B\ket{u}$; such a pairing
requires only that its two displayed vectors exist.
The exterior of $P'$ is contained in the momentum tail
$Q_{\theta K}$, and $\norm{\E \ket{v}}\le CK^2$. Taking the supremum
over unit $\ket{v}$, and using $K\ge1$, consequently gives the useful
residual estimate
\begin{equation}
 \norm{\mathcal C_g \ket{u}}
 \le C\bigl(K^{-1/2}\norm{\E \ket{u}}
             +K^{3/2}\norm{Q_{\theta K}\ket{u}}\bigr),\qquad \ket{u}\in D(H).
 \label{eq:kinetic-residual-tail}
\end{equation}
For $\ket{u}=\ket{\phi}\in D(H^2)$, the exact-state estimates
\eqref{eq:kinetic-exact-flow-tails} at time zero bound the two norms
by $C\mathcal D_\phi$ and $C\mathcal D_\phi K^{-5/2}$.
This proves Eq.~\eqref{eq:kinetic-data-consistency} uniformly over
the stated kernel class. We will also apply
Eq.~\eqref{eq:kinetic-residual-tail} to inverse-weighted low-energy
states after establishing their momentum tails.

We next propagate the weighted tails on the grid. Graph equivalence
and conservation of $\norm{\mathsf H_g\ket{v(t)}}$ first give
$\norm{\E \ket{v(t)}}\le C\mathcal D_\phi$. Split the Hamiltonian image as
\begin{equation}
 \Hgrid \ket{v(t)}=\ket{\chi(t)}+\ket{\rho(t)},\qquad
 \ket{\chi(t)}:=e^{-i\Hgrid t}P'H\ket{\phi},\qquad
 \ket{\rho(t)}:=e^{-i\Hgrid t}\mathcal C_g\ket{\phi}.
\end{equation}
We control these two terms differently. For $\ket{\chi}$, the assumption
$H\ket{\phi}\in D(H)$, Fourier contraction and the forward grid graph
bound give
\begin{equation}
 \|\mathsf H_gP'H\ket{\phi}\|
 \le C\|\E P'H\ket{\phi}\|
 \le C\|\E H\ket{\phi}\|
 \le C\mathcal D_\phi.
\end{equation}
This Hamiltonian norm is conserved by the grid flow, so graph
equivalence yields $\|\E\ket{\chi(t)}\|\le C\mathcal D_\phi$. For
$\ket{\rho}$, we need only its ordinary norm: unitarity and
Eq.~\eqref{eq:kinetic-data-consistency} give
$\|\ket{\rho(t)}\|\le C\mathcal D_\phi K^{-1/2}$. Separating these terms
avoids asking for a uniform second grid-Hamiltonian norm of the
projected initial state. Restricting
$T\ket{v}=\Hgrid \ket{v}-V_g\ket{v}$ to a momentum tail now gives
\begin{equation}
 \norm{Q_{r,\lambda}\E \ket{v(t)}}
 \le C\mathcal D_\phi\bigl[\lambda^{-2}+K^{-1/2}
                                  +\lambda^{-1/2}\bigr]
 \le C\mathcal D_\phi\lambda^{-1/2}
 \quad(1\le\lambda\le K).
\end{equation}
The $\lambda^{-2}$ term bounds the tails of $\ket{v}$ and $\ket{\chi}$ from
their first graph norms; the $\lambda^{-1/2}$ term bounds the
potential image using Eq.~\eqref{eq:kinetic-generic-tails}. We used
$K^{-1/2}\le\lambda^{-1/2}$ only in the final step. This proves the
bounds during grid evolution without requiring a cutoff-independent
second grid-Hamiltonian norm for $P'\ket{\phi}$. The remaining finite
shell above $K$ can be bounded by its containment in the tail at $K$.
\end{proofx}

\subsection{The low-energy comparison}
\label{sec:kinetic-low}

The preceding residual bound is $O(K^{-1/2})$ in state norm. To
obtain the kinetic theorem, we now exploit the fact that the residual
will be tested between states with controlled Hamiltonian energies.
We apply one inverse shifted Hamiltonian on each side, which bounds
the two kinetic norms by graph equivalence. Restricting those states
to sufficiently low Hamiltonian energies also suppresses their
large-momentum components. We can then use
Eq.~\eqref{eq:kinetic-forms} to obtain a $K^{-3}$ pairing, even though
the residual itself need not be that small in norm.

The relevant momentum threshold is $\theta K$. A vector supported
above that threshold has kinetic energy at least a fixed multiple
of $K^2$. We therefore choose the Hamiltonian-energy cutoffs below
that scale, with a sufficiently small fixed $a>0$, and set
\begin{equation}
 \Lambda=aK^2,\qquad S=\mathbf 1_{\mathsf H\le\Lambda},\qquad
 S_g=\mathbf 1_{\mathsf H_g\le2\Lambda}.
 \label{eq:kinetic-spectral-filters}
\end{equation}
We take $K_0$ large enough that $\Lambda\ge2$. The projectors $S$
and $S_g$ select eigenvectors of the shifted continuum and grid
Hamiltonians with energies at most $\Lambda$ and $2\Lambda$,
respectively. The larger grid cutoff leaves a separation of at least
$\Lambda$ between the continuum energies retained by $S$ and the
grid energies omitted by $S_g$. We will use this chosen separation
to bound the omitted grid component; it is not a gap assumption on
either physical Hamiltonian.

\begin{lemma}[Momentum tails and consistency below the spectral cutoffs]
\label{lem:kinetic-low}
With the fixed choice of $a$ sufficiently small, for every register $r$,
\begin{equation}
 \norm{Q_{r,\theta K}\mathsf H^{-1}S}
 +\norm{Q_{r,\theta K}\mathsf H_g^{-1}S_g}\le CK^{-5/2},
 \label{eq:kinetic-spectral-momentum-tail}
\end{equation}
and
\begin{equation}
 \norm{S_g\mathsf H_g^{-1}\mathcal C_g \mathsf H^{-1}S}\le CK^{-3}.
 \label{eq:kinetic-low-consistency}
\end{equation}
The residual with only its input restricted satisfies
\begin{equation}
 \norm{\mathcal C_g\mathsf H^{-1}S}\le CK^{-1/2}.
 \label{eq:kinetic-strong-consistency}
\end{equation}
\end{lemma}

\begin{proofx}
We first show that a low-Hamiltonian-energy state, after one inverse
Hamiltonian weight, has little support above momentum $\theta K$.
Let $Q=Q_{r,\theta K}$ and $B=\mathsf H^{-1}S$. On $\ran Q$, the
kinetic energy is at least $bK^2$, with a fixed $b>0$ determined by
$\theta$ and the masses. On the other hand, $B$ only has input
energies at most $\Lambda$. Writing $T=H-V_{\mathrm{tot}}$ relates
these two facts; since $Q$ commutes with $T$, we have
\begin{align}
 bK^2\norm{QB}
 &\le\norm{QTB}
 \le\norm{QHB}+\norm{QV_{\mathrm{tot}}B}\\
 &\le(\Lambda+c)\norm{QB}+CK^{-1/2}.
 \label{eq:kinetic-absorption}
\end{align}
For the Hamiltonian term, $HB=BHS$ gives $QHB=QBHS$ and hence
$\|QHB\|\le(\Lambda+c)\|QB\|$. We have used commutation of the
spectral functions $B,H,S$, without commuting the momentum projector
$Q$ through $H$. For the potential term, graph equivalence gives
$\|\E B\|\le C$, so Eq.~\eqref{eq:kinetic-generic-tails} supplies
$CK^{-1/2}$. The same argument with $B=\mathsf H_g^{-1}S_g$ gives
$2\Lambda+c$ in place of $\Lambda+c$.

We now choose $a$ so that $2a<b/2$, then increase $K_0$ so that
$c\le bK^2/4$. In both cases, moving the term proportional to
$\|QB\|$ to the left leaves a positive multiple of $K^2\|QB\|$.
Dividing the remaining $CK^{-1/2}$ by that $K^2$ proves
Eq.~\eqref{eq:kinetic-spectral-momentum-tail}. We choose the constants
using the smallest $b$ over the fixed set of registers, so the result
holds for all of them. A tail at the larger threshold $2K$ is contained
in the tail just bounded.

We next use the improved momentum tails on both sides of the
consistency residual. For arbitrary vectors $\ket{f}$ and $\ket{h}$ in the
continuum and grid spaces, respectively, take
$\ket{u}=\mathsf H^{-1}S\ket{f}$ and $\ket{v}=\mathsf H_g^{-1}S_g\ket{h}$. Their graph norms are bounded by
$C\norm{\ket{f}}$ and $C\norm{\ket{h}}$, and their high-momentum norms are bounded
by~\eqref{eq:kinetic-spectral-momentum-tail}. Since $P'$ preserves these
Fourier-weighted bounds, the dropped and wrap terms in~\eqref{eq:kinetic-forms}
are at most
\begin{equation}
 C\bigl(K^{-1/2}K^{-5/2}+K K^{-5}\bigr)
       \norm{\ket{f}}\norm{\ket{h}}\le CK^{-3}\norm{\ket{f}}\norm{\ket{h}}.
\end{equation}
For the exterior part of the residual, self-adjointness of $V_{\mathrm{tot}}$ gives
\begin{align}
 |\langle v|P'V_{\mathrm{tot}}(1-P')|u\rangle|
 &=|\langle (1-P')V_{\mathrm{tot}}v\,|\,(1-P')u\rangle|\\
 &\le CK^{-1/2}\norm{\E \ket{v}}\,
                 CK^{-5/2}\norm{\ket{f}}
 \le CK^{-3}\norm{\ket{f}}\norm{\ket{h}}.
\end{align}
We used~\eqref{eq:kinetic-generic-tails} and the union over the fixed number
of registers. Taking the supremum over $\ket{f},\ket{h}$ proves
Eq.~\eqref{eq:kinetic-low-consistency}.

Finally, take $\ket{u}=\mathsf H^{-1}S\ket{f}$ in
Eq.~\eqref{eq:kinetic-residual-tail}. Graph equivalence bounds
$\norm{\E \ket{u}}$ by $C\norm{\ket{f}}$, and the momentum-tail estimate just
proved bounds $\norm{Q_{\theta K}\ket{u}}$ by
$CK^{-5/2}\norm{\ket{f}}$. Consequently
\begin{equation}
 \norm{\mathcal C_g\mathsf H^{-1}S\ket{f}}
 \le C(K^{-1/2}+K^{-1})\norm{\ket{f}}
 \le CK^{-1/2}\norm{\ket{f}},
\end{equation}
which establishes Eq.~\eqref{eq:kinetic-strong-consistency}.
Both consistency estimates follow from the same low-energy
momentum tails. The two-sided, inverse-weighted estimate controls
the difference of the low-energy evolutions. The strong estimate,
with only its input restricted, will instead control transfer from
the low-energy continuum space into grid energies above $2\Lambda$.
\end{proofx}

We have obtained a $K^{-3}$ estimate with one inverse Hamiltonian
on each side of the residual. The initial state instead provides two
Hamiltonian powers on the input side through
$\|\mathsf H^2\ket{\psi_0}\|\le C\mathcal D$. To relate these two forms,
we divide the input and output energies into dyadic bands, intervals
of the form $[2^j,2^{j+1})$. Within neighbouring bands, the ratio of
the two energies stays bounded. Between separated bands, integration
of their relative oscillation supplies an inverse energy difference.
The next lemma makes this conversion precise and shows where the
logarithm enters.

\begin{lemma}[Duhamel comparison in dyadic energy bands]
\label{lem:kinetic-dyadic}
Let $B_1,B_2$ be positive self-adjoint operators on possibly different
finite-dimensional Hilbert spaces, with spectra in $[1,M]$, $M\ge2$.
Let $F$ map the second
space to the first, and suppose $\|B_1^{-1}FB_2^{-1}\|\le\delta$.
Then, for every $t\ge0$,
\begin{equation}
 \left\|\int_0^t e^{-iB_1(t-s)}F e^{-iB_2s}B_2^{-2}\,ds\right\|
 \le C\delta[t+\log(2+M)].
 \label{eq:kinetic-dyadic}
\end{equation}
\end{lemma}
\begin{proofx}
We separate input and output energies according to their size. If they
are comparable, the inverse weights already match and direct time
integration suffices. If they differ by more than a fixed factor, the
relative oscillation supplies an inverse energy separation. We will
sum the resulting estimates without using a physical spectral gap.

Let $E_j$ and $F_k$ be the spectral projectors of $B_1$ and $B_2$ onto
$[2^j,2^{j+1})$, respectively, with $0\le j,k\le J$ and
$J=O(\log(2+M))$. Empty bands cause no difficulty. Write
$B=B_1^{-1}FB_2^{-1}$. The operator $F$ is the residual being propagated; $F_k$ is a
spectral projector on its input space. The contribution from input
band $k$ to output band $j$ is
$L_{jk}:=E_jFB_2^{-2}F_k$. Substituting $F=B_1BB_2$ gives
\begin{equation}
 \|L_{jk}\|=\|B_1E_jBF_kB_2^{-1}\|
       \le 2^{j-k+1}\delta.
\end{equation}
For neighbouring bands, $|j-k|\le1$, the ratio is at most four, so
time integration costs at most $4t\delta$ per block. Crucially, for
each fixed offset $j-k$, different blocks have mutually orthogonal
input spaces and mutually orthogonal output spaces. Their sum is a
direct sum of operators, whose norm is the largest block norm. The
three neighbouring offsets therefore contribute at most $12t\delta$,
without multiplying the time-dependent term by the number of bands.

For $j\ge k+2$, the output energies exceed the input energies by at
least $d_{jk}=2^{j-1}$. On these two spectral subspaces define
\begin{equation}
 X_{jk}=\int_0^\infty e^{-sB_1}L_{jk}e^{sB_2}\,ds.
\end{equation}
The integrand has norm at most $e^{-d_{jk}s}\|L_{jk}\|$, so the
integral converges in operator norm and
$\|X_{jk}\|\le\|L_{jk}\|/d_{jk}$. Differentiating the integrand and
integrating its endpoint values gives the Sylvester identity
$B_1X_{jk}-X_{jk}B_2=L_{jk}$. This identity is useful because
\begin{equation}
 \frac{d}{ds}\big(e^{-iB_1(t-s)}X_{jk}e^{-iB_2s}\big)
 =i e^{-iB_1(t-s)}L_{jk}e^{-iB_2s}.
\end{equation}
The Duhamel block is therefore a difference of two unitary translates
of $X_{jk}$, up to a scalar phase. Its norm is at most
$2\|X_{jk}\|\le C\delta2^{-k}$. Summing over the separated bands gives
\begin{equation}
 \sum_{k=0}^J\sum_{j=k+2}^J C\delta2^{-k}
 \le C\delta(J+1).
\end{equation}
The factor $J+1$ is the source of the logarithm. When $j\le k-2$,
the input energies exceed the output energies by at least $2^{k-1}$.
In this case we define
\begin{equation}
 X_{jk}=-\int_0^\infty e^{sB_1}L_{jk}e^{-sB_2}\,ds.
\end{equation}
The negative sign gives the same identity
$B_1X_{jk}-X_{jk}B_2=L_{jk}$. The integrand decays at least as
$e^{-2^{k-1}s}$, so the endpoint argument bounds the Duhamel block by
$C\delta2^{j-2k}$. These blocks have a bounded total contribution,
since
\begin{equation}
 \sum_{k=2}^J\sum_{j=0}^{k-2}2^{j-2k}
 \le \sum_{k=2}^J2^{-k-1}\le1.
\end{equation}
Adding the neighbouring and separated contributions proves
Eq.~\eqref{eq:kinetic-dyadic}. Spectral multiplicities and spacings
within a band never enter the estimate.
\end{proofx}

We apply this lemma to the low-energy continuum and grid spaces in
Eq.~\eqref{eq:kinetic-spectral-filters}. Both are finite-dimensional:
the grid register is finite, and the periodic Coulomb Hamiltonian has
compact resolvent, so only finitely many eigenvalues, counted with
multiplicity, lie below $\Lambda$. Thus we can choose orthonormal
eigenbases for the two restricted Hamiltonians and apply the lemma
without any bound depending on the dimensions of these spaces.

Define the full register vectors
\begin{equation}
 \ket{u(t)}:=e^{-iHt}S\ket{\psi_0},\qquad
 \ket{x(t)}:=P'\ket{u(t)},\qquad
 \ket{y(t)}:=e^{-i\Hgrid t}P'S\ket{\psi_0}.
 \label{eq:kinetic-low-trajectories}
\end{equation}
Only the continuum initial state has been spectrally filtered; $\ket{x}$ and $\ket{y}$
have not been restricted by $S_g$. Their initial states agree.
The definition of $\mathcal C_g$ gives
\begin{equation}
 i\frac{d}{dt}(S_g\ket{x})=\Hgrid S_g\ket{x}-S_g\mathcal C_g\ket{u},
 \qquad
 i\frac{d}{dt}(S_g\ket{y})=\Hgrid S_g\ket{y}.
\end{equation}
Here $S_g$ commutes with the grid Hamiltonian. The Duhamel identity
for their low-energy difference is therefore
\begin{equation}
 S_g(\ket{x(t)}-\ket{y(t)})
 =i\int_0^t e^{-i\Hgrid(t-s)}S_g\mathcal C_g e^{-iHs}S\ket{\psi_0}\,ds.
\end{equation}
The common shift $c$ changes only a scalar phase in this integral.
In Lemma~\ref{lem:kinetic-dyadic}, take $B_1$ to be $\mathsf H_g$ on
$\operatorname{ran}S_g$, $B_2$ to be $\mathsf H$ on
$\operatorname{ran}S$, and $F=S_g\mathcal C_gS$. Their spectra lie in
$[1,2\Lambda]$, so $M=2\Lambda$, and
Eq.~\eqref{eq:kinetic-low-consistency} supplies $\delta=CK^{-3}$.
The initial state is $S\ket{\psi_0}=B_2^{-2}(S\mathsf H^2\ket{\psi_0})$, with
$\|S\mathsf H^2\ket{\psi_0}\|\le C\mathcal D$. Matching these two
Hamiltonian powers to the two inverse weights in the lemma
gives
\begin{equation}
 \norm{S_g(\ket{x(t)}-\ket{y(t)})}
 \le C\mathcal D K^{-3}\bigl[t+\log(2+K)\bigr].
 \label{eq:kinetic-low-flow-error}
\end{equation}
Both full register vectors have first graph norm $O(\mathcal D)$:
$P'$ does not increase the $\E$-norm, since it commutes with $\E$
and is an orthogonal projection, and the grid flow conserves the
$\mathsf H_g$-norm, which is equivalent to the $\E$-norm. In
particular, $\norm{T_p\ket{x(t)}}+\norm{T_p\ket{y(t)}}\le C\mathcal D$.
The identity
\begin{equation}
 \langle z|T_p|z\rangle-\langle w|T_p|w\rangle
   =\langle z-w\,|\,T_pz\rangle+\langle T_pw\,|\,z-w\rangle
 \label{eq:kinetic-pairing}
\end{equation}
explains why these norms matter. To estimate the full kinetic
expectation difference, we must pair both the low- and high-energy
parts of $\ket{x}-\ket{y}$ with $T_p\ket{x}$ and $T_p\ket{y}$. Equation~\eqref{eq:kinetic-low-flow-error}
controls the low-energy part; we now obtain the complementary bound.

\subsection{The high-energy contribution to the grid kinetic error}
\label{sec:kinetic-high}

We have controlled $S_g(\ket{x}-\ket{y})$, but the kinetic pairing in
Eq.~\eqref{eq:kinetic-pairing} also contains $(1-S_g)(\ket{x}-\ket{y})$. Our next
two estimates address this remaining contribution. First we show
that a continuum state below energy $\Lambda$ has only a small
component above grid energy $2\Lambda$ after Fourier projection.
We then show that $T_p\ket{x}$ and $T_p\ket{y}$, the vectors paired with the
state difference, also have controlled components at those
higher grid energies. We keep the full vectors $\ket{x}$ and $\ket{y}$ in both
kinetic expectations throughout; replacing them by $S_g\ket{x}$ and $S_g\ket{y}$
would introduce cross terms because $T_p$ need not commute with $S_g$.

\begin{lemma}[Separated spectral transfer]
\label{lem:kinetic-leakage}
The map $X:=(1-S_g)P'\mathsf H^{-1}S$ takes a continuum input below
energy $\Lambda$, applies the inverse Hamiltonian weight and Fourier
projection, and retains only the grid energies above $2\Lambda$.
As an operator from the continuum space to the grid register, it obeys
\begin{equation}
 \|X\|\le CK^{-5/2}.
 \label{eq:kinetic-transfer}
\end{equation}
Consequently,
\begin{equation}
 \|(1-S_g)(\ket{x(t)}-\ket{y(t)})\|\le C\mathcal D K^{-5/2}.
 \label{eq:kinetic-leakage-state}
\end{equation}
\end{lemma}
\begin{proofx}
The two energy ranges in $X$ are separated by at least $\Lambda$.
We use this separation to turn the strong residual bound into a
smaller bound on $X$ itself. Substituting
$\mathcal C_g=\Hgrid P'-P'H$ and commuting each spectral projector
with its own Hamiltonian gives
\begin{equation}
 \mathsf H_gX-X\mathsf H=F,\qquad
 F:=(1-S_g)\mathcal C_g\mathsf H^{-1}S.
\end{equation}
Equation~\eqref{eq:kinetic-strong-consistency} gives
$\|F\|\le CK^{-1/2}$. Since the output and input energies are
separated by at least $\Lambda$, we can recover $X$ from this
identity by
\begin{equation}
 X=\int_0^\infty e^{-s\mathsf H_g}F e^{s\mathsf H}S\,ds,
 \qquad \|X\|\le\frac{\|F\|}{\Lambda}\le CK^{-5/2}.
 \label{eq:kinetic-transfer-integral}
\end{equation}
Indeed, the integrand is bounded by $e^{-\Lambda s}\|F\|$.
Differentiating $e^{-s\mathsf H_g}Xe^{s\mathsf H}S$ and integrating
from zero to infinity verifies the identity, because its norm tends
to zero at the upper limit. This separation is introduced by our
chosen spectral cutoffs and does not assume a gap in the physical
spectrum.

The high-energy part of $\ket{x(t)}$ is $X\mathsf H \ket{u(t)}$, whose norm is
at most $C\mathcal D K^{-5/2}$. The high-energy part of $\ket{y(t)}$ is
$e^{-i\Hgrid t}X\mathsf H S\ket{\psi_0}$, because $S_g$ commutes
with the grid evolution. Its norm has the same bound. The triangle
inequality gives Eq.~\eqref{eq:kinetic-leakage-state}. We need only
these state norms; the next lemma supplies the additional kinetic
control on the vectors paired with their difference.
\end{proofx}

Equation~\eqref{eq:kinetic-leakage-state} gives only a $K^{-5/2}$
state-norm bound. To recover $K^{-3}$ in the kinetic expectation,
we also need decay on the other side of the high-energy pairing.
We obtain it by bounding $\mathsf H_gT_p\ket{w}$ for $\ket{w}=\ket{x},\ket{y}$.
The filtered initial state $S\ket{\psi_0}$ still has
$\|S\ket{\psi_0}\|+\|HS\ket{\psi_0}\|+\|H^2S\ket{\psi_0}\|\le\mathcal D$,
since $S$ commutes with $H$. The momentum-tail estimates therefore
apply to both full register vectors:
\begin{equation}
 \|Q_\lambda\E \ket{w(t)}\|\le C\mathcal D\lambda^{-1/2},
 \qquad 1\le\lambda\le K,\qquad \ket{w}\in\{\ket{x},\ket{y}\}.
\end{equation}
For $\ket{x}$, Fourier projection can only decrease the weighted norm;
for $\ket{y}$, we use the tail estimate for the grid evolution. The next lemma sums
these tails over momentum shells, adding one more kinetic weight
while keeping track of its dependence on $K$.

\begin{lemma}[Higher kinetic moment on the finite register]
\label{lem:kinetic-high}
For $\ket{w(t)}=\ket{x(t)}$ or $\ket{w(t)}=\ket{y(t)}$ and every measured particle $p$,
\begin{equation}
 \|\mathsf H_gT_p\ket{w(t)}\|\le C\mathcal D K^{3/2}.
 \label{eq:kinetic-high-moment}
\end{equation}
The constants are uniform in time and over the allowed grid kernels.
\end{lemma}
\begin{proofx}
We first bound $\|\E T_p\ket{w}\|$ using the momentum tails. Although
$\E T_p$ applies two kinetic weights, the vector lies in a finite
register, so the higher moment is finite. We must determine how it
grows with $K$, rather than assume an ordinary fourth derivative for
the continuum state.

Decompose the register into disjoint momentum shells according to
$\max_r\|\bm{k}_{\bm{p}_r}\|_\infty$. On a shell with radius between
$\lambda/2$ and $\lambda$, the Fourier multipliers satisfy
$T_p\le T\le C\lambda^2$. If $F_\lambda$ is its projector and
$2\le\lambda\le K$, then commutation of these multipliers and
Eqs.~\eqref{eq:kinetic-exact-flow-tails} and~\eqref{eq:kinetic-grid-flow-tails} give
\begin{equation}
 \|\E T_pF_\lambda \ket{w}\|^2
 \le C\lambda^4\|\E F_\lambda \ket{w}\|^2
 \le C\mathcal D^2\lambda^3.
 \label{eq:kinetic-one-shell}
\end{equation}
The low shell, where all momentum components have magnitude at most
one, costs at most $C\mathcal D^2$. Choose disjoint dyadic shells
up to the largest dyadic radius $\lambda_*\le K$, so that
$K/2<\lambda_*\le K$. The remaining top shell lies above
$\lambda_*$ and below the register cutoff $2K$. There
$T_p\le CK^2$, and its weighted norm is controlled by the tail at
$\lambda_*$; its squared contribution is therefore at most
$CK^4\mathcal D^2\lambda_*^{-1}\le C\mathcal D^2K^3$.
No tail estimate at a threshold larger than $K$ is needed.

The shell components are orthogonal after applying $\E T_p$, since
this operator is diagonal in the same Fourier basis. Thus we add
their squared norms. The dyadic sum is geometric:
\begin{equation}
 \sum_{2^j\le K}(2^j)^3\le CK^3,
 \qquad
 \|\E T_p\ket{w}\|^2\le C\mathcal D^2K^3.
\end{equation}
Taking a square root and using graph equivalence on the grid vector
$T_p\ket{w}$ proves Eq.~\eqref{eq:kinetic-high-moment}. This argument
bounds the action of the individual $T_p$ and does not assume that
an individual particle's kinetic energy is conserved.
\end{proofx}

We now combine the two estimates in the kinetic pairing. The
high-energy state difference is orthogonal to all grid states
below energy $2\Lambda$. Thus only the high-energy part of the test
vector can contribute: for every register vector $\ket{z}$,
\begin{equation}
 \big|\langle (1-S_g)(x-y)\,|\,z\rangle\big|
 =\big|\langle (1-S_g)(x-y)\,|\,(1-S_g)z\rangle\big|.
\end{equation}
On the range of $1-S_g$, the eigenvalues of $\mathsf H_g$ exceed
$2\Lambda$. Expanding $\ket{z}$ in the grid Hamiltonian's orthonormal
eigenbasis gives
\begin{equation}
 \|(1-S_g)\ket{z}\|^2
 \le (2\Lambda)^{-2}\|\mathsf H_g(1-S_g)\ket{z}\|^2
 \le (2\Lambda)^{-2}\|\mathsf H_g\ket{z}\|^2.
\end{equation}
Taking a square root supplies the inverse energy cutoff that we
need on this side of the pairing.
Combining this fact with the low- and high-energy difference bounds
above yields
\begin{equation}
 \begin{aligned}
 |\langle x(t)-y(t)\,|\,z\rangle|
 &\le C\mathcal D K^{-3}[t+\log(2+K)]\|\ket{z}\|\\
 &\quad+C\mathcal D K^{-5/2}(2\Lambda)^{-1}
                      \|\mathsf H_g\ket{z}\|.
 \end{aligned}
 \label{eq:kinetic-grid-pairing}
\end{equation}
We apply this inequality with $\ket{z}=T_p\ket{x}$ and $\ket{z}=T_p\ket{y}$. The first
kinetic graph bounds give $\|T_p\ket{x}\|+\|T_p\ket{y}\|\le C\mathcal D$,
and Lemma~\ref{lem:kinetic-high} bounds the two higher moments by
$C\mathcal D K^{3/2}$. Since $\Lambda=aK^2$, the high-energy term
has the required size
\begin{equation}
 (\mathcal D K^{-5/2})K^{-2}(\mathcal D K^{3/2})
 =\mathcal D^2K^{-3}.
\end{equation}
Finally, self-adjointness of $T_p$ gives
\begin{equation}
 \langle x|T_p|x\rangle-\langle y|T_p|y\rangle
 =\langle x-y\,|\,T_px\rangle+\langle T_py\,|\,x-y\rangle.
\end{equation}
The two applications of Eq.~\eqref{eq:kinetic-grid-pairing} therefore
prove
\begin{equation}
 \left|\langle x(t)|T_p|x(t)\rangle
             -\langle y(t)|T_p|y(t)\rangle\right|
 \le C\mathcal D^2K^{-3}[1+t+\log(2+K)].
 \label{eq:kinetic-high-kinetic-error}
\end{equation}
Only the low-energy dynamical comparison contributes the logarithm.
The high-energy contribution uses the full grid states and
spectral orthogonality; it does not require $S_g$ to commute with
$T_p$ or to preserve their momentum tails. To finish the theorem,
we will restore the continuum initial state $\ket{\psi_0}$ and its omitted Fourier
modes, whose errors are estimated separately.

\subsection{Restoring the original initial state and Fourier modes}

We have compared the two full register expectations starting from
the filtered continuum initial state $S\ket{\psi_0}$. Two differences remain
between that comparison and the theorem: the original initial state
is $\ket{\psi_0}$, and the continuum expectation includes Fourier modes
outside $P'$. We estimate these contributions separately, using
Hamiltonian regularity for the first and momentum tails for the second.

\begin{proofx}[Grid case of Theorem~\ref{thm:kinetic-log}]
We first remove the initial spectral filter used in the proof. Since
$\mathsf H>\Lambda$ on $\ran(1-S)$,
\begin{equation}
 \norm{(1-S)\ket{\psi_0}}\le\Lambda^{-2}\norm{\mathsf H^2\ket{\psi_0}}
                         \le C\mathcal D K^{-4}.
 \label{eq:kinetic-initial-high}
\end{equation}
The exact evolutions from $\ket{\psi_0}$ and $S\ket{\psi_0}$ both have first graph
norm $O(\mathcal D)$. The same is true for the grid evolutions from $P'\ket{\psi_0}$
and $P'S\ket{\psi_0}$, by~\eqref{eq:kinetic-graph} and conservation of the grid graph
norm. Unitarity bounds the difference within each pair of states
by Eq.~\eqref{eq:kinetic-initial-high}. Applying the pairing
identity~\eqref{eq:kinetic-pairing} therefore gives, for example,
\begin{equation}
 \big|\langle \psi|T_p|\psi\rangle-\langle u|T_p|u\rangle\big|
 \le\|\ket{\psi}-\ket{u}\|\bigl(\|T_p\ket{\psi}\|+\|T_p\ket{u}\|\bigr)
 \le C\mathcal D^2K^{-4}.
\end{equation}
The same argument applies to the two grid states. Hence
removing the initial filter costs at most $C\mathcal D^2K^{-4}$
in each expectation, uniformly in time.

For the exact filtered state $\ket{u(t)}=e^{-iHt}S\ket{\psi_0}$, the
ordinary momentum projection contributes
\begin{align}
 0\le\langle u|T_p|u\rangle-\langle P'u\,|\,T_pP'u\rangle
 &=\langle (1-P')u\,|\,T_p(1-P')u\rangle\\
 &\le\norm{(1-P')\ket{u}}\norm{\E(1-P')\ket{u}}
 \le C\mathcal D^2K^{-3}.
 \label{eq:kinetic-projection-error}
\end{align}
The cross terms vanish because $P'$ commutes with $T_p$, and the last
line uses the two tails in~\eqref{eq:kinetic-exact-flow-tails} at cutoff
$2K$. This is the projection contribution isolated in
Eq.~\eqref{eq:overview-kinetic-projection}.

The two full register vectors in Eq.~\eqref{eq:kinetic-low-trajectories}
are exactly $P'\ket{u(t)}$ and the grid state evolved from $P'S\ket{\psi_0}$.
Equation~\eqref{eq:kinetic-high-kinetic-error} compares their kinetic
expectations directly. Adding the initial-filter costs and the
Fourier-projection contribution gives
\begin{equation}
 \begin{aligned}
 &\left|\langle \psi(t)|T_p|\psi(t)\rangle-
              \langle \psi_g(t)|T_p|\psi_g(t)\rangle\right|\\
 &\qquad\le C\mathcal D^2
       \left\{K^{-4}+K^{-3}[1+t+\log(2+K)]\right\}.
 \end{aligned}
\end{equation}
For $K\ge K_0\ge1$, the $K^{-4}$ term is absorbed into the constant
multiplying $K^{-3}$. This proves the grid case of
Eq.~\eqref{eq:kinetic-log}.
\end{proofx}

\subsection{The simpler Galerkin comparison}
\label{sec:kinetic-galerkin}

For Galerkin evolution we can use the same energy-band comparison,
but the residual is smaller: it contains only the coupling to
omitted Fourier modes, with no dropped-transfer or wrap terms.
The resulting high-energy components are already $O(K^{-3})$ in
norm. Their kinetic contribution can therefore be bounded using
the first kinetic norms, without the extra moment needed on the
grid. We give the comparison explicitly because it also proves
the retained-state estimate~\eqref{eq:kinetic-retained-galerkin}.

Let $\mathsf H_N=H_N+c\ge1$, increasing the common fixed shift if
needed. The Coulomb relative bound and Fourier restriction give
\begin{equation}
 \norm{\mathsf H_N\ket{v}}\asymp\norm{\E \ket{v}},\qquad
 \norm{\mathsf H_NP\ket{u}}\le C\norm{\E \ket{u}},
 \quad \ket{v}\in\operatorname{ran}P,\quad \ket{u}\in D(H).
 \label{eq:kinetic-galerkin-graph}
\end{equation}
Consequently the exact, projected and Galerkin states have first
graph norms bounded by $C\mathcal D$, uniformly in time. The residual
is
\begin{equation}
 \mathcal C_N=H_NP-PH=-PV_{\mathrm{tot}}(1-P),
 \qquad
 \norm{\mathcal C_N\mathsf H^{-1}}\le CK^{-1}.
 \label{eq:kinetic-galerkin-strong}
\end{equation}
Indeed, Hardy's inequality gives
$\|V_{\mathrm{tot}}\ket{v}\|\le C\|\E^{1/2}\ket{v}\|$, and on the Fourier
complement $\E\ge bK^2$. Therefore
\begin{equation}
 \|\mathcal C_N\mathsf H^{-1}\|
 \le C\|\E^{1/2}(1-P)\mathsf H^{-1}\|
 \le CK^{-1}\|\E\mathsf H^{-1}\|\le CK^{-1}.
\end{equation}
This gains a half power of $K$ over the strong grid residual in
Eq.~\eqref{eq:kinetic-strong-consistency}.

Choose $\Lambda=aK^2$, $S=\ind{\mathsf H\le\Lambda}$ and
$S_N=\ind{\mathsf H_N\le2\Lambda}$, with a sufficiently small fixed
$a$ and $\Lambda\ge2$. The absorption argument in
Eq.~\eqref{eq:kinetic-absorption}, now on the Fourier complement of
$P$, gives
\begin{equation}
 \norm{(1-P)\mathsf H^{-1}S}\le CK^{-5/2},\qquad
 \norm{S_N\mathsf H_N^{-1}\mathcal C_N\mathsf H^{-1}S}
       \le CK^{-3}.
 \label{eq:kinetic-galerkin-low}
\end{equation}
For clarity, the first inequality follows by setting
$B=\mathsf H^{-1}S$ and using
\begin{equation}
 bK^2\|(1-P)B\|
 \le\|(1-P)TB\|
 \le(\Lambda+c)\|(1-P)B\|+CK^{-1/2}.
\end{equation}
The potential term uses the generic tail in
Eq.~\eqref{eq:kinetic-generic-tails}; the Hamiltonian term uses
$HB=BHS$, without commuting $P$ and $H$. Choosing $a$ small and then
$K$ large absorbs the first term on the right. To prove the second
inequality in Eq.~\eqref{eq:kinetic-galerkin-low}, test it on
$\ket{u}=\mathsf H^{-1}S\ket{f}$ and $\ket{v}=\mathsf H_N^{-1}S_N\ket{h}$. Self-adjointness
of $V_{\mathrm{tot}}$ gives
\begin{equation}
 |\langle v|\mathcal C_N|u\rangle|
 =|\langle (1-P)V_{\mathrm{tot}}v\,|\,(1-P)u\rangle|
 \le CK^{-1/2}\|\E \ket{v}\|\,CK^{-5/2}\|\ket{f}\|
 \le CK^{-3}\|\ket{h}\|\|\ket{f}\|.
\end{equation}
Unlike the grid proof, this step needs no improved momentum tail for
the finite test vector $\ket{v}$.

Put
\begin{equation}
 \ket{u(t)}=e^{-iHt}S\ket{\psi_0},\qquad
 \ket{x_N(t)}=S_NP\ket{u(t)},\qquad
 \ket{y_N(t)}=e^{-iH_Nt}S_NPS\ket{\psi_0}.
\end{equation}
Here $\ket{x_N}$ and $\ket{y_N}$ already include the Galerkin energy projector
$S_N$, unlike the full register vectors $\ket{x}$ and $\ket{y}$ used in the grid
proof. Their initial states agree, and $S_N$ commutes with $H_N$.
The exact residual identity gives
\begin{equation}
 \ket{x_N(t)}-\ket{y_N(t)}
 =i\int_0^t e^{-iH_N(t-s)}S_N\mathcal C_Ne^{-iHs}S\ket{\psi_0}\,ds.
\end{equation}
Apply Lemma~\ref{lem:kinetic-dyadic} with $B_1=\mathsf H_N$ on
$\operatorname{ran}S_N$, $B_2=\mathsf H$ on
$\operatorname{ran}S$, $F=S_N\mathcal C_NS$, $M=2\Lambda$ and input
$S\mathsf H^2\ket{\psi_0}$. The common shift changes only a scalar phase.
Equation~\eqref{eq:kinetic-galerkin-low} then yields
\begin{equation}
 \norm{\ket{x_N(t)}-\ket{y_N(t)}}
 \le C\mathcal D K^{-3}[t+\log(2+K)].
 \label{eq:kinetic-galerkin-low-flow}
\end{equation}

To restore the full Galerkin register vectors, we next bound their
components above energy $2\Lambda$. As in the grid comparison, the
relevant transfer map and its residual are
\begin{equation}
 X_N=(1-S_N)PS\mathsf H^{-1}S,\qquad
 F_N=(1-S_N)\mathcal C_N\mathsf H^{-1}S.
\end{equation}
They satisfy $\mathsf H_NX_N-X_N\mathsf H=F_N$. The output energies
are above $2\Lambda$ and the input energies are at most $\Lambda$, so
\begin{equation}
 X_N=\int_0^\infty e^{-s\mathsf H_N}F_Ne^{s\mathsf H}S\,ds,
 \qquad
 \norm{X_N}\le\Lambda^{-1}\norm{F_N}\le CK^{-3}.
 \label{eq:kinetic-galerkin-transfer}
\end{equation}
The integral converges in norm across this chosen spectral separation,
and its norm is bounded by $\|F_N\|/\Lambda$. In particular,
$(1-S_N)P\ket{u(t)}=X_N\mathsf H \ket{u(t)}$ has norm at most
$C\mathcal D K^{-3}$. The high component of
$e^{-iH_Nt}PS\ket{\psi_0}$ is the unitary evolution of its initial high
component and has the same bound. Adding these two components to
Eq.~\eqref{eq:kinetic-galerkin-low-flow} proves
\begin{equation}
 \norm{P\ket{u(t)}-e^{-iH_Nt}PS\ket{\psi_0}}
 \le C\mathcal D K^{-3}[1+t+\log(2+K)].
 \label{eq:kinetic-galerkin-filtered}
\end{equation}
Thus the high components already have the required norm rate.
Pairing this state difference with the first kinetic graph norms is
sufficient; the higher-moment estimate used for the grid is unnecessary.

\begin{proofx}[Galerkin case of Theorem~\ref{thm:kinetic-log}]
The initial filter error satisfies
$\|(1-S)\ket{\psi_0}\|\le C\mathcal D K^{-4}$ by
Eq.~\eqref{eq:kinetic-initial-high}. Unitarity and contraction of $P$
therefore remove the filter from Eq.~\eqref{eq:kinetic-galerkin-filtered}
at a cost of at most $2C\mathcal D K^{-4}$. This proves
Eq.~\eqref{eq:kinetic-retained-galerkin} for the original projected
initial state.

For the observable, $P$ commutes with $T_p$, so
\begin{equation}
 \langle \psi|T_p|\psi\rangle-\langle P\psi\,|\,T_pP\psi\rangle
 =\langle (1-P)\psi\,|\,T_p(1-P)\psi\rangle
 \le\|(1-P)\ket{\psi}\|\,\|\E(1-P)\ket{\psi}\|
 \le C\mathcal D^2K^{-3}.
\end{equation}
The last step uses the momentum tails of the exact evolved state at $K$. Meanwhile,
Eq.~\eqref{eq:kinetic-pairing}, the first graph bounds in
Eq.~\eqref{eq:kinetic-galerkin-graph}, and
Eq.~\eqref{eq:kinetic-retained-galerkin} bound the difference between
the expectations in $P\ket{\psi(t)}$ and $\ket{\psi_N(t)}$ by
$C\mathcal D^2K^{-3}[1+t+\log(2+K)]$. Adding the projection error
completes the Galerkin case. Finally, $N=\Theta(K^3)$ at fixed cell
size gives the stated form in terms of $N$ for both methods.
\end{proofx}

The logarithm in the final estimate comes from summing separated
Hamiltonian-energy bands in Lemma~\ref{lem:kinetic-dyadic}. The
high-grid-energy contribution is $O(\mathcal D^2K^{-3})$ without a
logarithmic factor. All graph-norm
bounds used in the proof follow from the continuum initial state in $D(H^2)$;
we never assume a cutoff-independent bound on
$\|\Hgrid^2P'\ket{\psi_0}\|$ or $\|H_N^2P\ket{\psi_0}\|$.

\subsection{Stopping power and the earlier power-law endpoints}
\label{sec:kinetic-stopping}

For stopping power we compare the kinetic-energy change across a
finite observation window, divided by the distance $v_p\tau$
travelled at the prescribed speed $v_p$. This is the signed secant
used in Corollary~\ref{cor:stopping}. Its discretisation error is
controlled by the kinetic errors at the two endpoints, so the
kinetic theorem immediately gives a window estimate. Any difference
between the exact finite-window quantity and an instantaneous or
asymptotic stopping law remains a separate bias.

\begin{corollary}[Logarithmic stopping-power error]
\label{cor:stopping-log}
In the setting of Corollary~\ref{cor:stopping}, use either the Galerkin
or truncated-kernel grid evolution from the projected initial state.
For every window $[0,\tau]$ with $\tau>0$ and every sufficiently large
admissible $K$,
\begin{equation}
 |\Delta\widehat S_\tau|
 \le\frac{C\mathcal D^2}{v_p}K^{-3}
       \left[1+\frac{1+\log(2+K)}{\tau}\right].
 \label{eq:stopping-log}
\end{equation}
The grid bound is uniform over the kernel
class~\eqref{eq:kernelclass}. Here $v_p>0$ is the projectile speed used
to convert elapsed time to travelled distance.
\end{corollary}
\begin{proofx}
Write
$E(t)=\langle \psi(t)|T_p|\psi(t)\rangle$ and
$E_{\rm d}(t)=\langle \psi_{\rm d}(t)|T_p|\psi_{\rm d}(t)\rangle$.
The exact and discretised window functionals are
\begin{equation}
 \widehat S_\tau=\frac{E(\tau)-E(0)}{v_p\tau},\qquad
 \widehat S_{{\rm d},\tau}
       =\frac{E_{\rm d}(\tau)-E_{\rm d}(0)}{v_p\tau}.
\end{equation}
Taking their difference bounds its magnitude by the sum of the two
endpoint kinetic errors divided by $v_p\tau$. Applying
Theorem~\ref{thm:kinetic-log} at $t=\tau$ and $t=0$ gives
\begin{equation}
 |\Delta\widehat S_\tau|
 \le\frac{C\mathcal D^2}{v_p}K^{-3}
       \left[1+\frac{2(1+\log(2+K))}{\tau}\right].
\end{equation}
Absorbing the factor two into $C$ proves
Eq.~\eqref{eq:stopping-log}. The sharper projection-only estimate at
time zero is available from Eq.~\eqref{eq:kinetic-projection-error},
but is not needed for this rate. Both evolutions start at the
common initial time, so the stated functional uses the window
$[0,\tau]$.
\end{proofx}

The following two statements preserve the earlier Galerkin endpoints
and their original time ranges. They now follow from the logarithmic
bounds by absorbing the logarithm into an arbitrarily small power.
The same bounds in fact hold for every $t\ge0$ and every window
$\tau>0$, respectively, by Theorem~\ref{thm:kinetic-log} and
Corollary~\ref{cor:stopping-log}: we use
$1+t+\log(2+K)\le(1+t)[1+\log(2+K)]$ and absorb the remaining
logarithm into $K^\varepsilon$.

\begin{theorem}[Kinetic-energy error at the energy rate]
\label{thm:src3}
Fix the periodic Coulomb system, including optional clamped nuclei,
a measured register $p$ and $\varepsilon>0$. There exist $C_\varepsilon$
and a cutoff threshold $K_0$, independent of $t$ and the initial state,
such that for every sufficiently large admissible $K\ge K_0$,
every $\ket{\psi_0}\in D(H^2)$ and every $t\ge1$,
\begin{equation}
 \left|\langle \psi(t)|T_p|\psi(t)\rangle
       -\langle \psi_N(t)|T_p|\psi_N(t)\rangle\right|
 \le C_\varepsilon\mathcal D^2(1+t)K^{-3+\varepsilon},
 \quad \mathcal D=\|\ket{\psi_0}\|+\|H\ket{\psi_0}\|+\|H^2\ket{\psi_0}\|.
 \label{eq:src3}
\end{equation}
The constants and $K_0$ may depend on the fixed system, cell and
$\varepsilon$, but not on $K,t$ or $\ket{\psi_0}$. No normalisation
or exchange-sector restriction is required; such restrictions may
be imposed without changing the estimate.
\end{theorem}
\begin{proofx}
For fixed $\varepsilon>0$,
$1+\log(2+K)\le C_\varepsilon K^\varepsilon$. The Galerkin case of
Eq.~\eqref{eq:kinetic-log} therefore implies Eq.~\eqref{eq:src3}.
\end{proofx}

\begin{corollary}[Stopping power at the energy rate]
\label{cor:stopping3}
In the setting of Corollary~\ref{cor:stopping}, for the Galerkin
evolution and a window $[0,\tau]$ with $\tau\ge1$,
\begin{equation}
 |\Delta\widehat S|
 \le\frac{C_\varepsilon\Xi^2}{v_p}
       \left[K^{-3+\varepsilon}
                      +\frac{K^{-3+\varepsilon}}{\tau}\right],
 \label{eq:stopping3}
\end{equation}
where $\Xi$ is the homogeneous regularity budget of
Proposition~\ref{prop:dyn-reg}. At fixed cell size this gives
$\mathcal O(N^{-1+\varepsilon})$ discretisation error in the window
functional, without choosing an optimal window length.
\end{corollary}
\begin{proofx}
Use Eq.~\eqref{eq:stopping-log}, the fixed-system relation
$\Xi=C_0m_{\max}\mathcal D$, and
$1+\log(2+K)\le C_\varepsilon K^\varepsilon$.
\end{proofx}

\subsection{Direct kinetic source bounds}
\label{sec:source}

We now give a second way to compare the exact and Galerkin kinetic
expectations. We express their difference directly in terms of the
potential acting on the omitted Fourier component of the exact state.
This gives a useful short-time estimate proportional to $tK^{-2}$;
the capped estimates retain their sublinear long-time growth. The
source estimate, the shell estimates that follow, and their stability
statement are auxiliary results: none is a hypothesis of the spectral
proof of the logarithmic kinetic-energy rate above. In particular,
the direct source argument concerns the Galerkin Hamiltonian $H_N$;
it does not add an assumption to the truncated-kernel grid result.

\paragraph{Regularity used by the auxiliary estimates.}
We retain the momentum and constant conventions of
Appendix~\ref{app:flux}. Unless a stronger threshold is stated,
arguments involving a nonempty dyadic lattice shell use
$K\ge\max\{1,4\pi/L\}$ and
$\lambda\ge\max\{1,8\pi/L\}$. We absorb the finitely many smaller
shells into a constant depending on the fixed cell. Algebraic
estimates stated for $K\ge1$ do not require a shell to be nonempty.
The multiplier $A(\vec{\bm{k}})$ is the total kinetic energy of the
momentum configuration $\vec{\bm{k}}$, so
$T\ket{\vec{\bm{k}}}=A(\vec{\bm{k}})\ket{\vec{\bm{k}}}$. Thus the operator
$\mathcal E=1+T$ used earlier gives the weighted norm
\begin{equation}
 \Theta(\ket{u})
 =\left(\sum_{\vec{\bm{k}}}[1+A(\vec{\bm{k}})]^2|u_{\vec{\bm{k}}}|^2\right)^{1/2}
 =\|\mathcal E\ket{u}\|.
\end{equation}
When working in the momentum basis below, we also write
$\mathcal E=1+A$. This is the same norm in the two representations,
not an additional regularity condition. We identify Fourier
projectors with their conjugates on the state space.

The direct source argument needs control of all momenta beyond a
threshold, rather than only of the total kinetic norm. For
$q\in\{2,\infty\}$, let
\begin{equation}
 Q^{(q)}_{a,\lambda}:=\ind{\|\bm{k}_a\|_q>\lambda}
\end{equation}
be the projector onto that whole tail in register $a$. We call a
state $\ket{u}$ \emph{good with budget $\Xi$} when
\begin{equation}
 \|\ket{u}\|+\Theta(\ket{u})\le C\Xi,\qquad
 \|Q^{(2)}_{a,\lambda}\ket{u}\|\le C\Xi\lambda^{-5/2},\qquad
 \|\mathcal E Q^{(2)}_{a,\lambda}\ket{u}\|\le C\Xi\lambda^{-1/2}
 \label{eq:good-state}
\end{equation}
for every register $a$ and every $\lambda\ge1$. A time-dependent state is good
when the same constants work at every time under consideration. The
Euclidean bounds also imply the corresponding cube-tail bounds:
$\|\bm{k}_a\|_\infty>\lambda$ implies $\|\bm{k}_a\|_2>\lambda$, and all these
projectors commute with $\mathcal E$.

Later we also need to distinguish a thin layer from the whole tail.
We use
\begin{equation}
 S^{(q)}_{a,\lambda,\rho}
 :=\{\lambda\le\|\bm{k}_a\|_q<\lambda+\rho\}
\end{equation}
for a shell of radius $\lambda$ and thickness $\rho$. We call a
state \emph{band-good} when, in addition to
Eq.~\eqref{eq:good-state}, it satisfies the two bounds in
Eq.~\eqref{eq:band} for both shell geometries, every register,
$\lambda\ge\lambda_{\rm band}$ and $1\le\rho\le\lambda$.
Those bounds resolve the dependence on the thickness $\rho$; a
first graph norm alone does not supply them. A cube-supported state
has no modes beyond Euclidean radius $\sqrt3K$ or cube radius $K$
in any register, so an empty shell contributes zero. A shell whose
lower radius equals the support boundary can still contain boundary
modes, because its lower endpoint is included.

Fourier projectors are contractions for the state norm and every
$\mathcal E$-weighted norm used here. Proposition~\ref{prop:dyn-boot}
and Lemma~\ref{lem:disc-tails} therefore give
Eq.~\eqref{eq:good-state} for the exact, projected and Galerkin
states. We will use only these whole-tail estimates in the direct
source proof. The budget $\Xi$ is the homogeneous budget of
Proposition~\ref{prop:dyn-reg}; multiplying it by a fixed system
constant places no new condition on the initial state.

\paragraph{Moving the observable into the source pairing.}
The retained error $\ket{e_P}=P\ket{\psi}-\ket{\psi_N}$ is driven by the omitted exact
component $(1-P)\ket{\psi}$. To estimate a kinetic expectation, we could
try to control a kinetic norm of $\ket{e_P}$ itself. Instead, we use its
Duhamel formula and move the finite-dimensional propagator to the
other argument of the inner product. This makes the test vector a
backward evolution of $T_p\ket{v}$, where $\ket{v}$ is one of two terminal states.
We first establish the exact identity, including both terminal
states, and then bound the energy norm of this test vector.

\begin{proposition}[Exact decomposition through the backward
observable flow]
\label{prop:src-decomp}
Let $\ket{a}:=P\ket{\psi}$, $\ket{\delta}:=(1-P)\ket{\psi}$, $\ket{e_P}:=\ket{a}-\ket{\psi_N}$, and
$V^{RP}:=(1-P)V_{\mathrm{tot}}P$. Then
\begin{equation}
\langle \psi(t)|T_p|\psi(t)\rangle-\langle \psi_N(t)|T_p|\psi_N(t)\rangle
=\langle \delta(t)|T_p|\delta(t)\rangle
+\langle e_P(t)|T_p|a(t)\rangle
+\overline{\langle e_P(t)|T_p|\psi_N(t)\rangle} ,
\label{eq:src-decomp}
\end{equation}
and, for either $\ket{v}=\ket{a(t)}$ or $\ket{v}=\ket{\psi_N(t)}$,
\begin{equation}
 \langle e_P(t)|T_p|v\rangle
 =+i\int_0^t\langle \psi(s)|V^{RP}|\phi_v(s)\rangle\,ds,
 \qquad
 \ket{\phi_v(s)}:=e^{iH_N(t-s)}T_p\ket{v}\in\operatorname{ran}P.
 \label{eq:src-pair}
\end{equation}
\end{proposition}

\begin{proofx}
We first separate the omitted component from the retained one.
Since $T_p$ and $P$ are diagonal in momentum, $[T_p,P]=0$ and the
cross terms between $\ket{a}$ and $\ket{\delta}$ vanish. Hence
\begin{align}
 \langle \psi|T_p|\psi\rangle
 &=\langle a|T_p|a\rangle+\langle \delta|T_p|\delta\rangle,\\
 \langle a|T_p|a\rangle-\langle \psi_N|T_p|\psi_N\rangle
 &=\langle e_P|T_p|a\rangle+
       \overline{\langle e_P|T_p|\psi_N\rangle}.
\end{align}
The second line follows by substituting $\ket{e_P}=\ket{a}-\ket{\psi_N}$ and using
self-adjointness of $T_p$. This proves the decomposition without
introducing a quadratic term in $\ket{e_P}$.

For each of its two remaining pairings, Eq.~\eqref{eq:duhamel} gives
\begin{equation}
 \ket{e_P(t)}=-i\int_0^t e^{-iH_N(t-s)}PV_{\mathrm{tot}}\ket{\delta(s)}\,ds.
\end{equation}
Our inner product is antilinear in its first argument. The scalar
$-i$ therefore becomes $+i$, and moving the propagator to the second
argument gives
\begin{equation}
 \langle e_P(t)|T_p|v\rangle
 =+i\int_0^t
   \langle PV_{\mathrm{tot}}\delta(s)\,|\,e^{iH_N(t-s)}T_pv\rangle\,ds.
\end{equation}
Finally, $\ket{\phi_v(s)}\in\operatorname{ran}P$ and
$\ket{\delta(s)}=(1-P)\ket{\psi(s)}$, so self-adjointness of the potential yields
\begin{equation}
 \langle PV_{\mathrm{tot}}\delta\,|\,\phi_v\rangle
 =\langle \delta|V_{\mathrm{tot}}|\phi_v\rangle
 =\langle \psi|(1-P)V_{\mathrm{tot}}P|\phi_v\rangle.
\end{equation}
This proves Eq.~\eqref{eq:src-pair} with the stated sign.
\end{proofx}

The vector $\ket{\phi_v}$ in this identity is a backward \emph{observable}
flow: we apply $T_p$ to the terminal state before evolving it. It is
different from a backward state such as
$e^{iH_N(t-s)}P\ket{\psi(t)}$. Applying $T_p$ increases the weight of high
momenta, so goodness of $\ket{v}$ does not assert goodness of $\ket{\phi_v}$.
The next lemma supplies exactly the weaker, cutoff-dependent norms
that we need for this observable flow.

\begin{lemma}[Regularity of the backward observable flow]
\label{lem:phi}
Let $\ket{v}\in\operatorname{ran}P$ be good with budget $B$, and let
$H_{\rm ret}=H_N+D_K$, where $D_K$ is self-adjoint and uniformly
bounded. Take $O=T_p$, or more generally $O=T_p\Pi$ with $\Pi$ a
Fourier projector, and put $\ket{\phi(\tau)}=e^{iH_{\rm ret}\tau}O\ket{v}$.
The additional restriction $\Pi$ is allowed because it can only
decrease the weighted norms of the initial observable vector $O\ket{v}$.
Uniformly for all real $\tau$,
\begin{equation}
 \|\ket{\phi}\|\le CB,\qquad
 \|\mathcal E^{1/2}\ket{\phi}\|\le CBK^{1/2},\qquad
 \|\mathcal E\ket{\phi}\|\le CBK^{3/2}.
 \label{eq:phi-norms}
\end{equation}
For either choice of $O$, interpolation gives
\begin{equation}
 \|\mathcal E^\alpha\ket{\phi}\|
 \le
 \begin{cases}
 C_\alpha BK^\alpha,&0\le\alpha\le1/2,\\
 C_\alpha BK^{2\alpha-1/2},&1/2\le\alpha\le1.
 \end{cases}
\end{equation}
In particular, for $0\le\alpha\le1/2$ one may use the conservative
bound $C_\alpha BK^\alpha\log^{q_\alpha}(2+K)$ with a fixed
nonnegative logarithmic power, although the displayed bound is
log-free. Also,
$\|\mathcal E^{3/4+\varepsilon}\ket{\phi}\|
\le C_\varepsilon BK^{1+2\varepsilon}$ for
$0<\varepsilon\le1/4$. No normalisation or exchange-symmetry
condition is required.
\end{lemma}

\begin{proofx}
We first estimate the terminal vector $O\ket{v}$ and then transport its
norms by the retained Hamiltonian. For the terminal estimate, we
partition momentum space into a bounded low-momentum block and
disjoint dyadic shells of the largest Euclidean register momentum.
Here ``dyadic'' means that the shell radii double successively.
Because $\ket{v}$ is supported in the product of cubes defining $P$, this
largest momentum is at most $\sqrt3K$.

On a shell with largest register momentum comparable to $\lambda$,
we have $\mathcal E\le C(1+\lambda^2)$. The shell is contained in
the union of the register tails at a comparable radius, so
Eq.~\eqref{eq:good-state} gives
$\|\mathcal E\ind{\lambda}\ket{v}\|\le CB\lambda^{-1/2}$; here
$\ind{\lambda}$ denotes just this dyadic shell. Since $O$ is
diagonal and $0\le O\le T_p\le A$, we obtain
\begin{align}
 \|\mathcal E^{1/2}\ind{\lambda}O\ket{v}\|
 &\le C\lambda\|\mathcal E\ind{\lambda}\ket{v}\|
 \le CB\lambda^{1/2},\\
 \|\mathcal E\ind{\lambda}O\ket{v}\|
 &\le C\lambda^2\|\mathcal E\ind{\lambda}\ket{v}\|
 \le CB\lambda^{3/2}.
\end{align}
The disjoint shells are orthogonal, including after multiplication
by $\mathcal E$. We therefore sum squared norms, giving
\begin{equation}
 \|\mathcal E^{1/2}O\ket{v}\|^2
 \le CB^2\left(1+\sum_{\lambda\lesssim K}\lambda\right)
 \le CB^2K,
 \qquad
 \|\mathcal E O\ket{v}\|^2
 \le CB^2\left(1+\sum_{\lambda\lesssim K}\lambda^3\right)
 \le CB^2K^3.
\end{equation}
Both sums run over dyadic radii, so each is bounded by a constant
multiple of its largest term. The low-momentum block contributes
$CB^2$, and $\|O\ket{v}\|\le\|\mathcal E\ket{v}\|\le CB$.

To propagate these estimates, we use a positive shift
$H_{\rm ret}+C_{\rm low}+1$. Here $C_{\rm low}$ is a constant independent
of $K$ chosen so that $H_{\rm ret}\ge-C_{\rm low}$, making the shifted
Hamiltonian at least one. Hardy's inequality and uniform
boundedness of $D_K$ make its first graph norm uniformly equivalent
to the $\mathcal E$ norm. Interpolation gives the corresponding
equivalence for powers between zero and one. Powers of the shifted
Hamiltonian commute with $e^{iH_{\rm ret}\tau}$, and their norms
are conserved. We may thus convert each terminal
$\mathcal E$-weighted norm to the shifted-Hamiltonian norm,
propagate it, and convert back with constants independent of $K$
and $\tau$. This proves Eq.~\eqref{eq:phi-norms}; it does not require
$\mathcal E$ itself to commute with the propagator.

Finally, H\"older's inequality for the spectral measure of
$\mathcal E$ in the vector $\ket{\phi}$, which assigns to each eigenvalue
the squared norm of the corresponding component of $\ket{\phi}$, gives,
for $0\le\alpha\le1/2$,
\begin{equation}
 \|\mathcal E^\alpha\ket{\phi}\|
 \le\|\ket{\phi}\|^{1-2\alpha}
      \|\mathcal E^{1/2}\ket{\phi}\|^{2\alpha}
 \le C_\alpha BK^\alpha.
\end{equation}
Interpolating between the half-power and first-power bounds instead
gives
\begin{equation}
 \|\mathcal E^\alpha\ket{\phi}\|
 \le\|\mathcal E^{1/2}\ket{\phi}\|^{2-2\alpha}
      \|\mathcal E\ket{\phi}\|^{2\alpha-1}
 \le C_\alpha BK^{2\alpha-1/2},
 \qquad 1/2\le\alpha\le1.
\end{equation}
The special cases in the statement follow by substitution.
\end{proofx}

\begin{proposition}[A direct bound on the complete source]
\label{prop:src1}
Let $\ket{v}\in\operatorname{ran}P$ be good with budget $C\Xi$,
$\ket{\delta(s)}=(1-P)\ket{\psi(s)}$ and
$\ket{\phi_v(s)}=e^{iH_N(t-s)}T_p\ket{v}$. For $t\ge0$ and $K\ge1$,
\begin{equation}
 \left|\int_0^t
   \langle PV_{\mathrm{tot}}\delta(s)\,|\,\phi_v(s)\rangle\,ds\right|
 \le C\Xi^2tK^{-2}.
 \label{eq:src1}
\end{equation}
The bound includes the complete potential source; it requires no
split into bulk, resonant and gapped crossings.
\end{proposition}

\begin{proofx}
We place the potential on the observable vector, where Hardy's
inequality bounds it by a half-power energy norm. Self-adjointness
of the Coulomb multiplication form gives
\begin{equation}
 |\langle PV_{\mathrm{tot}}\delta\,|\,\phi_v\rangle|
 =|\langle \delta|V_{\mathrm{tot}}|\phi_v\rangle|
 \le C\|\ket{\delta}\|\|\mathcal E^{1/2}\ket{\phi_v}\|.
\end{equation}
The exact-state tail is $\|\ket{\delta}\|\le C\Xi K^{-5/2}$, by taking
the union of the register tails, while Lemma~\ref{lem:phi} gives
$\|\mathcal E^{1/2}\ket{\phi_v}\|\le C\Xi K^{1/2}$. Consequently
\begin{equation}
 |\langle PV_{\mathrm{tot}}\delta\,|\,\phi_v\rangle|
 \le C\Xi K^{-5/2}\cdot\Xi K^{1/2}
 =C\Xi^2K^{-2}.
\end{equation}
This estimate is uniform in the integration variable $s$, so
integration gives Eq.~\eqref{eq:src1}. Every transfer size remains
in $V_{\mathrm{tot}}$ throughout the argument; there is no upper
cutoff on a crossing hop.
\end{proofx}

\begin{theorem}[Unconditional kinetic-energy error]
\label{thm:src1}
Let $\ket{\psi_0}\in D(H^2)$ and
$\ket{\psi_N(t)}=e^{-iH_Nt}P\ket{\psi_0}$. For every measured register $p$,
all $t\ge0$ and $K\ge1$,
\begin{equation}
 \left|\langle \psi(t)|T_p|\psi(t)\rangle
       -\langle \psi_N(t)|T_p|\psi_N(t)\rangle\right|
 \le C\mathcal D^2\bigl(K^{-3}+tK^{-2}\bigr),
 \label{eq:src1thm}
\end{equation}
where $C$ depends only on the fixed system and cell and
$\mathcal D=\|\ket{\psi_0}\|+\|H\ket{\psi_0}\|+\|H^2\ket{\psi_0}\|$.
\end{theorem}

\begin{proofx}
We apply the three-term identity of
Proposition~\ref{prop:src-decomp}. Its first term is the kinetic
expectation of the omitted Fourier component. Since $T_p$ commutes
with $P$, it is bounded by the product of the plain and weighted
tails:
\begin{equation}
 |\langle \delta|T_p|\delta\rangle|
 \le\|\ket{\delta}\|\|\mathcal E\ket{\delta}\|
 \le C\Xi^2K^{-5/2}K^{-1/2}
 \le C\mathcal D^2K^{-3}.
\end{equation}
For the two retained pairings, the terminal states are exactly
$\ket{v}=P\ket{\psi(t)}$ and $\ket{v}=\ket{\psi_N(t)}$. Both satisfy
Eq.~\eqref{eq:good-state} by the dynamical and Galerkin tail bounds,
so Proposition~\ref{prop:src1} bounds each pairing by
$C\mathcal D^2tK^{-2}$. Adding the three contributions proves the
claim. Every step is homogeneous in $\ket{\psi_0}$, and the uniform
state bounds impose neither a small-time restriction nor an
upper bound on $t$.
\end{proofx}

\subsection{Band-resolved tails}

The direct source bound used all momenta above a threshold together.
We now ask how much of that tail can lie in a shell of thickness
$\rho$. This gives information about individual crossing terms
beyond the total kinetic expectation. We retain both Euclidean and
cube shells: the Euclidean geometry enters the regularity estimates,
whereas the product of cubes defines the simulation projector.

\begin{lemma}[Band-resolved tails]
\label{lem:band}
Assume $\ket{\psi_0}\in D(H^2)$ and use the homogeneous budget $\Xi$ of
Proposition~\ref{prop:dyn-reg}. Let $\ket{u}$ be any of
\begin{equation}
 \ket{\psi(t)},\qquad P\ket{\psi(t)},\qquad\ket{\psi_N(t)},\qquad
 e^{iH_N(t-s)}P\ket{\psi(t)}.
\end{equation}
For $q\in\{2,\infty\}$, set
$S=S^{(q)}_{a,\lambda,\rho}
=\{\lambda\le\|\bm{k}_a\|_q<\lambda+\rho\}$.
Let $C_H$ bound the sum of the particle-pair and clamped-nucleus
Hardy estimates by $C_H\|\mathcal E^{1/2}\ket{u}\|$, and take
\begin{equation}
 \lambda_{\rm band}
 :=\max\{1,2\pi/L,8C_H\sqrt{2m_{\max}}\}.
\end{equation}
For $\lambda\ge\lambda_{\rm band}$ and $1\le\rho\le\lambda$,
\begin{equation}
 \|\ind S\ket{u}\|\le C\Xi\lambda^{-3}\rho^{1/2},
 \qquad
 \|\ind S\mathcal E\ket{u}\|
 \le C\Xi\lambda^{-1}\rho^{1/2}.
 \label{eq:band}
\end{equation}
For the discrete states we need only consider
$\lambda\le\sqrt3K$; all shells beyond the support are empty.
The constants are uniform in time.
The backward Galerkin state in the list is also good in the sense of
Eq.~\eqref{eq:good-state}. For evolution under a perturbed retained Hamiltonian, the
corresponding conclusion requires the additional hypotheses of
Lemma~\ref{lem:good-stab}, rather than only a first graph bound.
\end{lemma}

\begin{proofx}
We first prove the shell estimate for the exact state and explain
how each interaction can feed the selected shell. We then identify
the additional term that must be controlled for the Galerkin
states. The equation
\begin{equation}
 A(\vec{\bm{k}})u_{\vec{\bm{k}}}
 =\langle\vec{\bm{k}}|H|u\rangle
  -\langle\vec{\bm{k}}|V_{\mathrm{tot}}|u\rangle
\end{equation}
relates the weighted state on the shell to the Hamiltonian-applied
state and to the potential. Since $\Theta(H\ket{u})\le C\Xi$ and
$A\ge\lambda^2/(2m_a)$ on $S$, the first contribution is
\begin{equation}
 \|\ind S H\ket{u}\|\le C\Xi\lambda^{-2}.
\end{equation}
Our remaining task is to control the potential on $S$ without
assuming bounds on the spectator momenta.

Consider first a pair $(b,c)$ that does not contain the shell
register $a$. This interaction leaves $\bm{k}_a$ unchanged and therefore
commutes with $\ind S$. Hardy's inequality gives
\begin{equation}
 \|V_{bc}\ind S \ket{u}\|
 \le C_H\|\mathcal E^{1/2}\ind S\ket{u}\|
 \le C_H\sqrt{2m_a}\lambda^{-1}
           \|\mathcal E\ind S\ket{u}\|.
\end{equation}
The last step uses only the lower bound on $A$ supplied by register
$a$; every other register may have arbitrarily large momentum.
After summing all interactions that leave $a$ unchanged, the choice
of $\lambda_{\rm band}$ makes the coefficient of the desired
weighted shell norm small enough to move this contribution to the
left-hand side.

For a pair containing $a$, we split the input at momentum
$\lambda/2$ in the chosen norm. An input with
$\|\bm{k}_a\|_q\le\lambda/2$ can reach $S$ only through a transfer
whose Euclidean size is at least $\lambda/2$. Its Coulomb coefficient
is consequently bounded in absolute value by $C\lambda^{-2}$.
We apply this observation separately on each three-dimensional
pair slice: fix the total momentum of the interacting pair and
all spectator momenta, and let the momentum of one member of the
pair vary. On such a slice, the target shell contains at most
$C_L\lambda^2\rho$ lattice modes for either shell geometry.

To see why the input sum on a slice is controlled, write $\bm{j}$ for its
variable three-dimensional momentum and $\vec{\bm{k}}(\bm{j})$ for the full
configuration on that slice. Put $u_{\bm{j}}:=\langle\vec{\bm{k}}(\bm{j})|u\rangle$
and $A_{\bm{j}}:=A(\vec{\bm{k}}(\bm{j}))$, with the other slice labels fixed.
Weighted Cauchy--Schwarz gives
\begin{equation}
 \sum_{\bm{j}}|u_{\bm{j}}|
 \le
 \left(\sum_{\bm{j}}(1+A_{\bm{j}})^{-2}\right)^{1/2}
 \left(\sum_{\bm{j}}|(1+A_{\bm{j}})u_{\bm{j}}|^2\right)^{1/2}.
\end{equation}
The first factor is uniformly bounded: the kinetic energy includes
a positive multiple of $\|\bm{j}\|^2$, and
$\sum_{\bm{j}\in(2\pi/L)\mathbb Z^3}(1+c\|\bm{j}\|^2)^{-2}<\infty$.
Thus each output coefficient is at most $C\lambda^{-2}$ times
the weighted norm of the input on its slice. Taking the norm over
the $C_L\lambda^2\rho$ possible outputs gives the factor
\begin{equation}
 C\lambda^{-2}(\lambda^2\rho)^{1/2}
 =C\lambda^{-1}\sqrt\rho.
\end{equation}
Finally we sum the squared bounds over the mutually orthogonal
slices. Their squared weighted input norms add to at most
$\Theta(\ket{u})^2$, yielding
$C\Xi\lambda^{-1}\sqrt\rho$ for the low-input part.
This uses a three-dimensional weighted sum on each slice, never
the full many-particle Fourier $\ell^1$ norm.

For the remaining input, Hardy's inequality and the two whole-tail
bounds give
\begin{align}
 \|\mathcal E^{1/2}Q_{a,\lambda/2}^{(q)}\ket{u}\|
 &\le
 \bigl(\|Q_{a,\lambda/2}^{(q)}\ket{u}\|
       \|\mathcal E Q_{a,\lambda/2}^{(q)}\ket{u}\|\bigr)^{1/2}\\
 &\le C\Xi\lambda^{-3/2}.
\end{align}
For cube tails we use their inclusion in Euclidean tails. Any fixed
small thresholds in this use of $\lambda/2$ are absorbed into the
system constant. This controls the high-input part of every pair
that moves $a$.

The clamped-nucleus terms obey the same two alternatives, and it is
useful to spell out their application. A potential acting on a
register other than $a$ commutes with $\ind S$, so its Hardy
constant is included in the contribution absorbed above. For a
potential acting on $a$, we fix all spectator momenta and work on
the three-dimensional slice of $\bm{k}_a$ itself. A low input again needs
a transfer of size at least $\lambda/2$ to reach $S$. Translating
the nucleus multiplies each Coulomb coefficient by a phase of
modulus one, which leaves both the $C\lambda^{-2}$ coefficient
bound and the weighted Cauchy--Schwarz estimate unchanged. The
$C_L\lambda^2\rho$ target count therefore gives the same
$C\Xi\lambda^{-1}\sqrt\rho$ bound. The high-input
part uses the one-register Hardy inequality and the interpolated
tail $C\Xi\lambda^{-3/2}$. The number of nuclei is fixed, so
summing these terms changes only the system constant.

We can now collect the estimates. In passing from $A$ to
$\mathcal E=1+A$, the additional unweighted shell norm is already
bounded by the whole-tail estimate. After absorbing the commuting
potential terms, we obtain
\begin{equation}
 \|\mathcal E\ind S\ket{u}\|
 \le C\Xi
       \bigl(\lambda^{-2}
             +\lambda^{-1}\sqrt\rho+\lambda^{-3/2}\bigr)
 \le C\Xi\lambda^{-1}\sqrt\rho.
\end{equation}
Here $\lambda\ge1$ and $\rho\ge1$ ensure that the middle term
dominates the other displayed powers. Dividing by
$A\ge\lambda^2/(2m_a)$ on $S$ gives the unweighted shell estimate.
Projection by $P$ is a contraction for both norms, so the result for
$P\ket{\psi(t)}$ follows as well.

For the forward Galerkin state, Lemma~\ref{lem:disc-tails}
replaces $H\ket{u}$ by a sum of a graph-bounded part and a part of norm
at most $C\Xi K^{-3/2}$. The potential estimates just proved are
unchanged, because the additional output projection $P$ can only
decrease the norm. On the retained support,
\begin{equation}
 K^{-3/2}\le C\lambda^{-1}\sqrt\rho,
 \qquad 1\le\lambda\le\sqrt3K,\quad\rho\ge1.
\end{equation}
The norm-small part therefore fits within the required weighted
shell bound.

For completeness, the backward evolution in the statement starts
from $P\ket{\psi(t)}$, with no observable $T_p$ applied to that initial vector.
At its terminal time,
\begin{equation}
 H_NP\ket{\psi(t)}
 =PH\ket{\psi(t)}-PV_{\mathrm{tot}}(1-P)\ket{\psi(t)}.
\end{equation}
We propagate these two vectors separately by
$e^{iH_N(t-s)}$. The first remains graph bounded by
Eq.~\eqref{eq:graphN}, because
$\Theta(PH\ket{\psi(t)})\le C\Xi$. The second has norm
$C\Xi K^{-3/2}$ by Hardy's inequality and the exact-state tails,
and unitarity preserves this norm. Together with the first graph
bound for the backward state itself, this decomposition gives its
whole-tail estimates by the same argument as
Lemma~\ref{lem:disc-tails}: apply a register tail projector to
$A\ket{u}=H_N\ket{u}-PV_{\mathrm{tot}}\ket{u}$, use the graph bound on the first
part, the norm bound on the second, and the potential-tail estimate
on the last term. We may then apply the shell argument above.
This proves the claimed backward-state estimates without a uniform
second Galerkin graph norm and without asserting them for the
backward observable flow $\ket{\phi_v}$.
\end{proofx}

\paragraph{Resolving the momentum transfers into a shell.}
We next refine the contribution from high input momenta. The useful
geometric fact is that a transfer shell can meet a thin output shell
in only a small fraction of its modes. Schur's test turns this count
into an operator norm: if the sums of absolute matrix entries in
every row and column are bounded by $M_{\rm row}$ and $M_{\rm col}$, respectively, the norm is at most
$\sqrt{M_{\rm row}M_{\rm col}}$.

\begin{lemma}[Thin-shell Schur bound]
\label{lem:thinschur}
Let $V_{\varrho'}$ be the part of a pair potential with transfers
$\|\bm{\sigma}\|_2\in[\varrho',2\varrho')$, where $\varrho'\ge1$.
Let $S$ be a Euclidean or cube shell in a register moved by the
transfer, of radius $\lambda\ge4\varrho'$ and thickness
$1\le\varrho\le\lambda$. Then
\begin{equation}
 \|\ind S V_{\varrho'}\|
 \le C\sqrt{\varrho'\min(\varrho,\varrho')}.
\end{equation}
\end{lemma}

\begin{proofx}
We apply Schur's test on each pair slice, with total pair momentum
and all spectator momenta fixed. At a fixed output, the row sum is
bounded by the sum over the whole transfer shell:
\begin{equation}
 \sum_{\|\bm{\sigma}\|_2\sim\varrho'}|\hat w(\bm{\sigma})|
 \le C\varrho'^3\varrho'^{-2}=C\varrho'.
\end{equation}
For a fixed input, only transfers whose output lies in $S$
contribute to the column sum. Across a ball of transfer radius
$2\varrho'$, a Euclidean shell at radius $\lambda\ge4\varrho'$
occupies a spherical cap with normal thickness $C\varrho$.
For a cube shell, we use the union of its at most six face slabs.
The available two-dimensional extent is $O(\varrho'^2)$, and the
remaining thickness is bounded by both $C\varrho$ and
$C\varrho'$. Covering this region by lattice cells therefore gives
at most
$C\varrho'^2\min(\varrho,\varrho')$ transfers. Since each
coefficient is at most $C\varrho'^{-2}$, the column sum is at most
$C\min(\varrho,\varrho')$. The geometric mean of the row and
column bounds proves the claim on each slice, and hence on their
orthogonal sum.
\end{proofx}

The same argument covers a clamped-nucleus potential acting on the
shell register. Its slice consists of that register's momentum with
all spectator momenta fixed, instead of relative momentum at fixed
pair total momentum. The row and column counts are unchanged, and
the nuclear translation phase disappears when we take absolute
values. We can therefore use the transfer split below for every
interaction that moves the shell register. Interactions that leave
it unchanged continue to use the commuting Hardy estimate.

For a band-good state $\ket{u}$, we have the full potential-shell bound
\begin{equation}
 \|\ind{S_{a,\lambda,\rho}}V_{\mathrm{tot}}\ket{u}\|
 \le C\Xi\lambda^{-1}\sqrt\rho
\end{equation}
in either shell geometry. Input below $\lambda/2$ in the shell
register is controlled by the slice-wise estimate used in
Lemma~\ref{lem:band}. For an interaction moving $a$, the part with
input beyond $\lambda/2$ in at least one register moved by that
interaction has the stronger bound
$C\Xi\lambda^{-2}\sqrt\rho$. We now verify this
improvement by sorting transfers into dyadic radii $r$. The point of
the split is to use a thin input shell for small transfers and a
whole input tail for larger transfers.

If $r<\lambda/4$, an input reaching the output shell must lie in
a shell of width at most $\rho+4r$, at a radius comparable to
$\lambda$. Combining its band bound with
Lemma~\ref{lem:thinschur} gives
\begin{equation}
 C\Xi\lambda^{-3}
       \sqrt{r\min(\rho,r)}\sqrt{\rho+4r}.
\end{equation}
If the input shell is wider than its lower radius, we partition it
into a bounded number of shells whose widths are no greater than
their lower radii; all these radii remain comparable to $\lambda$.
For $r\ge\rho$, the two square-root factors are bounded by
$C\sqrt{r\rho}\sqrt r=C\sqrt\rho\,r$. For $r<\rho$, they
are bounded by $Cr\sqrt\rho$. Thus both cases give
$C\Xi\lambda^{-3}\sqrt\rho\,r$, and the dyadic sum is
\begin{equation}
 C\Xi\lambda^{-3}\sqrt\rho
       \sum_{r<\lambda/4}r
 \le C\Xi\lambda^{-2}\sqrt\rho.
\end{equation}
When a comparable input radius lies below the fixed threshold
$\lambda_{\rm band}$, there are only fixed-radius contributions;
we absorb them using the first graph bound and a fixed system
constant.

For $\lambda/4\le r\le4\lambda$, the row sum is at most $Cr$.
There are at most $C\lambda^2\rho$ output modes on a slice, so
the column sum is at most
$Cr^{-2}\lambda^2\rho\le C\rho$ in this range. Schur's test
therefore gives $C\sqrt{r\rho}$. The high-input restriction is a
tail in either member of the interacting pair, so the union of
their two whole-tail bounds gives $C\Xi\lambda^{-5/2}$.
For a clamped-nucleus channel there is only the one moved register.
Multiplying by this input bound yields
$C\Xi\lambda^{-2}\sqrt\rho$ for each such transfer shell.
Only a bounded number of dyadic shells lie in this interval.

For $r>4\lambda$, the same counts give a row sum $Cr$ and a
column sum $Cr^{-2}\lambda^2\rho$, hence the Schur norm
$C\lambda\sqrt\rho\,r^{-1/2}$. An output in $S$ has Euclidean
momentum at most $2\sqrt3\lambda$, for either geometry. A transfer
of Euclidean size at least $r$ can therefore reach it only from
an input of Euclidean size at least $c r$, for a fixed $c>0$.
The input norm is consequently at most $C\Xi r^{-5/2}$, and
\begin{equation}
 C\Xi\lambda\sqrt\rho\sum_{r>4\lambda}r^{-3}
 \le C\Xi\lambda^{-2}\sqrt\rho.
\end{equation}
This completes the high-input estimate for interactions moving
$a$. For an interaction that leaves $a$ unchanged, we instead
combine its commuting Hardy bound with the two shell norms:
\begin{equation}
 \|\mathcal E^{1/2}\ind S \ket{u}\|
 \le\bigl(\|\ind S \ket{u}\|\|\mathcal E\ind S \ket{u}\|\bigr)^{1/2}
 \le C\Xi\lambda^{-2}\sqrt\rho.
\end{equation}
Adding these contributions proves the stated full
$C\Xi\lambda^{-1}\sqrt\rho$ bound. All input
estimates were slice-wise weighted estimates or $\ell^2$ tail
estimates; no full many-particle Fourier $\ell^1$ norm was used.

\paragraph{Resonant crossings near the cutoff.}
The preceding shell estimates describe states and potential images.
We now record a separate bound for crossings whose kinetic-energy
change is small. Recall the bulk projector from
Eq.~\eqref{eq:review-bulk-convention},
\begin{equation}
 Q=\prod_a\ind{\|\bm{k}_a\|_2\le K/4},\qquad R:=1-P,
\end{equation}
and define
\begin{equation}
 V_{\mathrm{nc}}
 :=RV_{\mathrm{tot}}(P-Q)+(P-Q)V_{\mathrm{tot}}R.
\end{equation}
Here $Q$ requires every register to lie in the Euclidean bulk;
it is distinct from the one-register tail projectors
$Q^{(q)}_{a,\lambda}$. Since $Q\le P$, the projector $P-Q$
selects retained modes with at least one register beyond $K/4$.
Thus $V_{\mathrm{nc}}$ couples those modes to the discarded
space $\operatorname{ran}R$ in both directions.

We impose the resonance condition separately on each interaction
channel. For a particle pair $(a,b)$, a transfer $\bm{\nu}$ changes the
two momenta by $+\bm{\nu}$ and $-\bm{\nu}$, and its kinetic mismatch is
\begin{equation}
 \omega=\frac{\|\bm{\nu}\|_2^2}{2\mu_{ab}}
       +\bm{\nu}\cdot\left(\frac{\bm{k}_a}{m_a}-\frac{\bm{k}_b}{m_b}\right),
 \qquad
 \mu_{ab}:=\frac{m_am_b}{m_a+m_b}.
\end{equation}
For a clamped-nucleus channel acting on $a$, the corresponding
formula uses $\mu=m_a$ and relative velocity $\bm{k}_a/m_a$.
These are the channel conventions of Lemma~\ref{lem:slab};
the pair indices need not include the measured register $p$.

\begin{proposition}[Resonant near-cutoff coupling]
\label{prop:src-res}
Let $V_{\mathrm{nc}}^{\mathrm{res}}$ retain, channel by channel,
the crossings in $V_{\mathrm{nc}}$ with transfer $\bm{\nu}$ of size
$\varrho$ and kinetic mismatch
$|\omega|\le g\varrho/\mu$, where $g=2\pi/L$ and $\mu$ is the
reduced mass of that channel, or the moving mass for a
clamped-nucleus channel. For the backward observable flow
$\ket{\phi_v}$ of Proposition~\ref{prop:src1}, pointwise in time,
\begin{equation}
 |\langle \psi|V_{\mathrm{nc}}^{\mathrm{res}}|\phi_v\rangle|
 \le C\Xi^2K^{-3}\log(2+K).
\end{equation}
At large fixed-cell cutoffs, this is the same
$K^{-3}\log K$ rate; the form $\log(2+K)$ also retains a positive
bound at the allowed endpoint $K=1$.
\end{proposition}

\begin{proofx}
We first show that only finitely many dyadic transfer scales can
contribute. Put $M:=K/g$, the integer cube radius in lattice units,
and $\bm{\ell}:=\bm{\nu}/g\in\mathbb Z^3$. For a pair channel the
dimensionless relative momentum
$\mu_{ab}(\bm{k}_a/m_a-\bm{k}_b/m_b)/g$ has norm at most $\sqrt3M$
when the input lies in the retained cube: each component is a
weighted difference of two numbers of absolute value at most
$M$, and $\mu_{ab}/m_a+\mu_{ab}/m_b=1$.
The same bound holds for $\bm{k}_a/g$ in a clamped-nucleus channel.
The resonance inequality consequently implies
\begin{equation}
 \|\bm{\ell}\|^2-2\sqrt3M\|\bm{\ell}\|\le2\|\bm{\ell}\|,
 \qquad\text{and hence}\qquad
 \|\bm{\ell}\|\le2\sqrt3M+2
\end{equation}
for nonzero transfers. If $\varrho$ is the lower radius of the
dyadic transfer shell, we use $\varrho\le\|\bm{\nu}\|$ in obtaining
the same inequality. Thus at most $C\log(2+M)$ dyadic transfer
shells occur. At fixed $L$, this is bounded by
$C\log(2+K)$, so we will not sum a constant shell bound over
infinitely many scales.

Fix one transfer shell of radius $\varrho$. At a fixed input,
Lemma~\ref{lem:slab} counts at most $C\varrho^2$ resonant
outputs when $\mu\omega_0=g\varrho$. Its two-sided version
gives the same count at a fixed output. Each Coulomb coefficient
has magnitude at most $c_w\varrho^{-2}$, so both absolute row
and column sums are bounded by a constant. Schur's test therefore
bounds the restricted shell operator by $C$. Translation phases
in clamped-nucleus channels do not affect this estimate.

Because $\ket{\phi_v}\in\operatorname{ran}P$, only the term mapping
$P-Q$ to $R$ contributes to the pairing. Its left argument is
$\ket{\delta}=R\ket{\psi}$, and its right argument is $(P-Q)\ket{\phi_v}$. The
exact-state tail gives
$\|\ket{\delta}\|\le C\Xi K^{-5/2}$. On the support of $P-Q$,
some register has Euclidean momentum greater than $K/4$, so
$\mathcal E^{1/2}\ge cK$. Lemma~\ref{lem:phi} then gives
\begin{equation}
 \|(P-Q)\ket{\phi_v}\|
 \le CK^{-1}\|\mathcal E^{1/2}\ket{\phi_v}\|
 \le C\Xi K^{-1/2}.
\end{equation}
Each transfer shell consequently contributes at most
\begin{equation}
 C\|\ket{\delta}\|\|(P-Q)\ket{\phi_v}\|
 \le C\Xi K^{-5/2}\cdot\Xi K^{-1/2}
 =C\Xi^2K^{-3}.
\end{equation}
Summing the $C\log(2+K)$ possible scales and the fixed set of
interaction channels proves the claim.
\end{proofx}

It is useful to distinguish this whole-tail estimate for $\ket{\phi_v}$
from a thinner-shell estimate that is available only at its
terminal time. If $\ket{v}$ is band-good, then $\ket{\phi_v(t)}=T_p\ket{v}$ obeys
\begin{equation}
 \|P\ind{K-\rho<\|\bm{k}_a\|_\infty\le K}\ket{\phi_v(t)}\|
 \le C\Xi K^{-1}\sqrt\rho,
 \qquad 1\le\rho\le K/2.
 \label{eq:phi-band}
\end{equation}
Indeed, the shell radius is comparable to $K$ and
$T_p\le\mathcal E$, so the weighted shell estimate for $\ket{v}$
gives this bound after covering the retained layer by a bounded
number of shells of radius comparable to $K$ and total thickness
$O_L(\rho)$. This covering includes the upper boundary modes.
We do not assert propagation of Eq.~\eqref{eq:phi-band} from
$s=t$ to earlier times. The resonant-coupling proof instead uses
the norm of the whole high-momentum component $(P-Q)\ket{\phi_v(s)}$,
which Lemma~\ref{lem:phi} controls at every time. This is why that
proof does not require the backward observable flow to be
band-good.

\subsection{Stability of momentum tails under evolution}

We finally identify perturbations of the retained Hamiltonian for
which the bounds on evolved states remain valid. A uniformly
bounded self-adjoint perturbation preserves the first graph norm,
but a first graph bound alone gives less momentum decay than
Eq.~\eqref{eq:good-state}. We therefore keep track of the size of
the perturbation relative to that graph norm. The assumptions for
the whole-tail conclusion and the stronger assumptions for thin
shells are stated separately.

\begin{lemma}[Stability of regularity under evolution]
\label{lem:good-stab}
Let $D_K$ be self-adjoint on $\operatorname{ran}P$ and uniformly
bounded in $K$. We use the following homogeneous estimates, where
each applicable bound holds for all $\ket{u}\in\operatorname{ran}P$:
\begin{equation}
 \|D_K\ket{u}\|\le\alpha_1\Theta(\ket{u}),\qquad
 \|Q_{a,\lambda}^{(q)}D_K\ket{u}\|
       \le\alpha_2\lambda^{-1/2}\Theta(\ket{u}),\qquad
 \|\ind{S^{(q)}_{a,\lambda,\rho}}D_K\ket{u}\|
       \le\alpha_3\lambda^{-1}\sqrt\rho\,\Theta(\ket{u}).
 \label{eq:stab-hyp}
\end{equation}
For the conclusion that states remain good throughout evolution, assume the first bound with
$\alpha_1\le CK^{-1/2}$, and increase $K_0$ so that
$\alpha_1\le1/4$. The second and third bounds are optional
sufficient refinements; they are not required for this conclusion.
Let
\begin{equation}
 \ket{\chi_0}=P\ket{\psi_0}+\ket{r_0},\qquad
 \ket{\psi_0}\in D(H^2),\qquad \ket{r_0}\in\operatorname{ran}P,
 \qquad
 \|\ket{r_0}\|+\Theta(\ket{r_0})\le C\Xi K^{-1/2}.
\end{equation}
Then $\ket{\chi(s)}=e^{-i(H_N+D_K)s}\ket{\chi_0}$ is good with budget
$C\Xi$, uniformly for all real $s$.

For the band-good conclusion, strengthen the initial-state assumptions to
\begin{equation}
 \Theta(\ket{r_0})\le C\Xi K^{-1},\qquad
 \|D_K\ket{\chi_0}\|\le C\Xi K^{-1}.
 \label{eq:stab-band-datum}
\end{equation}
In addition, assume either $\alpha_1\le CK^{-1}$, or the third
bound of Eq.~\eqref{eq:stab-hyp} with $\alpha_3\le C$ for the
\emph{whole} range
$\lambda\ge\lambda_{\rm band}$, $1\le\rho\le\lambda$,
and $q\in\{2,\infty\}$, for every register $a$.
A shell bound only for $\lambda>K/8$ does not suffice for this
second alternative. All constants may depend on the fixed system,
but not on $K$, $s$, or the initial state beyond $\Xi$.
\end{lemma}

\begin{proofx}
Put $H':=H_N+D_K$. Hardy's inequality and the uniform bound on
$D_K$ give uniform equivalence of the first graph norms of $H'$
and $\mathcal E$. Since $\Theta(\ket{\chi_0})\le C\Xi$, conservation
of the shifted $H'$ graph norm yields
\begin{equation}
 \Theta(\ket{\chi(s)})\le C\Xi
\end{equation}
for all real $s$. To obtain the sharper tails, we need more than
this estimate: we separate $H'\ket{\chi(s)}$ into a graph-bounded part
and a small remainder.

At the initial time, expand $H'\ket{\chi_0}$ using
$H_NP\ket{\psi_0}=PH\ket{\psi_0}-PV_{\mathrm{tot}}(1-P)\ket{\psi_0}$.
Commuting $H'$ with its own propagator then gives the exact
decomposition
\begin{equation}
 H'\ket{\chi(s)}=\ket{f(s)}+\ket{e(s)},\qquad
 \ket{f(s)}:=e^{-iH's}PH\ket{\psi_0},\qquad
 \ket{e(s)}:=e^{-iH's}\ket{e(0)},
\end{equation}
where
\begin{equation}
 \ket{e(0)}=-PV_{\mathrm{tot}}(1-P)\ket{\psi_0}
       +H_N\ket{r_0}+D_K\ket{\chi_0}.
\end{equation}
The condition $\ket{\psi_0}\in D(H^2)$ gives
$\Theta(PH\ket{\psi_0})\le C\Xi$, and the same first graph-norm
argument used for $\ket{\chi}$ therefore gives
$\Theta(\ket{f(s)})\le C\Xi$. We estimate the three parts of $\ket{e(0)}$
in norm. Hardy's inequality and the exact-state tails give
$C\Xi K^{-3/2}$ for the omitted-potential term; graph-norm
equivalence gives $C\Theta(\ket{r_0})\le C\Xi K^{-1/2}$ for
$H_N\ket{r_0}$; and the first bound of Eq.~\eqref{eq:stab-hyp} gives
$\|D_K\ket{\chi_0}\|\le C\Xi\alpha_1$. Unitarity hence implies
\begin{equation}
 \|\ket{e(s)}\|
 \le C\Xi\bigl(K^{-3/2}+K^{-1/2}+\alpha_1\bigr)
 \le C\Xi K^{-1/2}.
\end{equation}

We now apply a register tail projector to
\begin{equation}
 A\ket{\chi}=\ket{f}+\ket{e}-PV_{\mathrm{tot}}\ket{\chi}-D_K\ket{\chi},
\end{equation}
where we use the Fourier representation of the vectors. On
$\|\bm{k}_a\|_q>\lambda$, the four terms on the right have bounds
\begin{equation}
 C\Xi\lambda^{-2},\qquad C\Xi K^{-1/2},\qquad
 C\Xi\lambda^{-1/2},\qquad C\Xi\alpha_1,
\end{equation}
respectively. The first comes from $\Theta(\ket{f})\le C\Xi$.
The second uses only the norm of $\ket{e}$. The third is the
potential-tail estimate applied to the graph-bounded state
$\ket{\chi}$, and the fourth uses the assumed size of $D_K$; no
tail-specific bound on $D_K$ is needed here. For
$1\le\lambda\le\sqrt3K$, both $K^{-1/2}$ and $\alpha_1$
are at most $C\lambda^{-1/2}$, so
\begin{equation}
 \|A Q^{(q)}_{a,\lambda}\ket{\chi}\|
 \le C\Xi\lambda^{-1/2}.
\end{equation}
The elementary graph-norm tail of $\ket{\chi}$ adds at most
$C\Xi\lambda^{-2}$ when we replace $A$ by $\mathcal E=1+A$.
We thus obtain the weighted-tail bound in
Eq.~\eqref{eq:good-state}. Since
$A\ge\lambda^2/(2m_a)$ on the tail, division gives
\begin{equation}
 \|Q^{(q)}_{a,\lambda}\ket{\chi}\|
 \le C\lambda^{-2}
       \|A Q^{(q)}_{a,\lambda}\ket{\chi}\|
 \le C\Xi\lambda^{-5/2}.
\end{equation}
Beyond the retained support the tail is zero. Together with the
uniform first graph norm, this proves goodness for all times.

For the stronger shell conclusion, the initial-state assumptions in
Eq.~\eqref{eq:stab-band-datum} improve the same remainder to
$\|\ket{e(s)}\|\le C\Xi K^{-1}$. Throughout the retained support,
\begin{equation}
 K^{-1}\le C\lambda^{-1}\sqrt\rho,
 \qquad \lambda\le\sqrt3K,\quad\rho\ge1,
\end{equation}
so this remainder fits within the weighted shell estimate of
Lemma~\ref{lem:band}. That lemma's graph-bounded contribution
applies to $\ket{f}$, and its potential estimates apply to $\ket{\chi}$
because we have already proved goodness.

The only new shell contribution is $D_K\ket{\chi(s)}$. Under the first
alternative, $\alpha_1\le CK^{-1}$ gives
\begin{equation}
 \|D_K\ket{\chi(s)}\|\le C\Xi K^{-1}
 \le C\Xi\lambda^{-1}\sqrt\rho,
\end{equation}
which controls every shell by its full norm. Under the second
alternative, the third bound in Eq.~\eqref{eq:stab-hyp} supplies
the required estimate directly on each shell. We need that bound
at every relevant radius, since goodness alone does not resolve
thin shells below $K/8$. The absorption and division steps in
Lemma~\ref{lem:band} now give both shell bounds. This establishes
the two alternatives exactly as stated, without using the optional
$\alpha_2$ bound or assuming any stronger control of $D_K$.
\end{proofx}

\end{document}